\documentclass[11pt,a4paper,parskip=half]{scrartcl}

\usepackage[utf8]{inputenc}
\usepackage{LCVision}
\usepackage{threeparttable}
\usepackage{eurosym}
\usepackage{multicol, multirow}
\usepackage{epstopdf}
\usepackage[version=4]{mhchem}

\usepackage{LCVision_definitions}

\usepackage{url,color}
\definecolor{orange}{rgb}{1,0.5,0}
\definecolor{gray}{rgb}{0.5,0.5,0.5}
\definecolor{roetlich}{rgb}{1, .7, .7}
\definecolor{camel}{rgb}{0.76, 0.6, 0.42}
\definecolor{britishracinggreen}{rgb}{0.0, 0.26, 0.15}

\usepackage{xpatch}
\usepackage{numprint}

\makeatletter
\xpatchcmd\@HepConStyle
 {\edef\@upcode{\updefault}}
 {\ifdefined\shapedefault\edef\@upcode{\shapedefault}\else\edef\@upcode{\updefault}\fi}
 {}{}
\makeatother
\usepackage{soul}

\RedeclareSectionCommand[
  beforeskip=1.\baselineskip,
  afterskip=.5\baselineskip]{subsection}
\RedeclareSectionCommand[
  beforeskip=1.\baselineskip,
  afterskip=.5\baselineskip]{subsubsection}

\usepackage{todonotes}
\usepackage{xpatch}
\usepackage{bm}
\makeatletter
\xpretocmd{\todo}{\@bsphack}{}{}
\xapptocmd{\todo}{\@esphack}{}{}
\makeatother

\usepackage{booktabs}
\usepackage{csquotes}

\usepackage{graphicx} 
\usepackage{appendix}

\def\beq{\begin{equation}}
\def\eeq#1{\label{#1}\end{equation}}
\def\eeqn{\end{equation}}

\newenvironment{Eqnarray}%
   {\arraycolsep 0.14em\begin{eqnarray}}{\end{eqnarray}}
\def\beqa{\begin{Eqnarray}}
\def\eeqa#1{\label{#1}\end{Eqnarray}}
\def\eeqan{\end{Eqnarray}}

\def\leqn#1{(\ref{#1})}

\def\lsim{\mathrel{\raise.3ex\hbox{$<$\kern-.75em\lower1ex\hbox{$\sim$}}}}
\def\gsim{\mathrel{\raise.3ex\hbox{$>$\kern-.75em\lower1ex\hbox{$\sim$}}}}

\def\L{{\cal L}}

\def\L{{\cal L}}

\def\del{\partial}
\def\Dslash{\not{\hbox{\kern-4pt $D$}}}
\def\dslash{\not{\hbox{\kern-2pt $\del$}}}

\def\Dlr{\mathrel{\raise1.5ex\hbox{$\leftrightarrow$\kern-1em\lower1.5ex\hbox{$D$}}}}

\def\Pl{{\mbox{\scriptsize Pl}}}

\def\ee{\ensuremath{\Pep\Pem}}

\def\mw{m_W}

\def\alphas{\alpha_s}
\def\msb{{\bar{\scriptsize M \kern -1pt S}}}

\def\drb{{\bar{\scriptsize D \kern -1pt R}}}

\def\eps{\epsilon}

\usepackage[italic]{mathastext}

\DeclareCiteCommand{\citejournal}[\mkbibbrackets]
  {\usebibmacro{prenote}}
  {\usebibmacro{citeindex}%
   \printtext[bibhyperref]{\printfield{journaltitle}}%
   \iffieldundef{volume}
     {}%
     {\setunit{\addspace}%
     \printtext[bibhyperref]{\printfield{volume}}}%
   \setunit{\addspace}%
   \printtext[bibhyperref]{(\printdate)}%
   \iffieldundef{pages}
     {}
     {\setunit{\addspace}%
     \printtext[bibhyperref]{\printfield{pages}}%
     }%
     }
  {\multicitedelim}
  {\usebibmacro{postnote}}
  
\DeclareCiteCommand{\citesubmit}[\mkbibbrackets]
  {\usebibmacro{prenote}}
  {\usebibmacro{citeindex}%
   \printtext[bibhyperref]{\printfield{journaltitle}}%
   \setunit{\addspace}%
   \printtext[bibhyperref]{(\printdate)}}
  {\multicitedelim}
  {\usebibmacro{postnote}}
  
  \DeclareCiteCommand{\citeconf}[\mkbibbrackets]
  {\usebibmacro{prenote}}
  {\usebibmacro{citeindex}%
   \printtext[bibhyperref]{\printfield{howpublished}}%
   \setunit{\addspace}%
   \printtext[bibhyperref]{(\printdate)}}
  {\multicitedelim}
  {\usebibmacro{postnote}}

\title{A Linear Collider Vision for the Future of Particle Physics} 

\nolicence 

\date{\today}

\abstract{\noindent
In this paper we review the physics opportunities at linear $\ee$ colliders with a special focus on high centre-of-mass energies and beam polarisation, take a fresh look at the various accelerator technologies available or under development and, for the first time, discuss how a facility first equipped with a technology that is mature today could be upgraded with technologies of tomorrow to reach much higher energies and/or luminosities. In addition, we discuss detectors, alternative collider modes, as well as opportunities for beyond-collider experiments and R\&D facilities as part of a linear collider facility (LCF). The material of this paper supports all plans for $\boldmath\ee$ linear colliders and the additional opportunities they offer, independently of technology choice or proposed site, as well as R\&D for advanced accelerator technologies.
This joint perspective on the physics goals, early technologies and upgrade strategies has been developed by the LCVision team based on an initial discussion at LCWS2024 in Tokyo and a follow-up at the LCVision Community Event at CERN in January 2025. It heavily builds on decades of achievements of the global linear collider community, in particular in the context of CLIC and ILC.
}

\addauthor{
C.~Balazs\\
\emph{Monash University, Melbourne 3800 Victoria, Australia}
}

\addauthor{
G.~Taylor\\
\emph{University of Melbourne, Victoria 3010, Australia}
}

\addauthor{
W.~Mitaroff\\
\emph{Institute of High Energy Physics, Austrian Academy of Sciences, 1050 Vienna, Austria}
}

\addauthor{
A.H.~Hoang,
S.~Pl\"{a}tzer\\
\emph{University of Vienna, Faculty of Physics, 1090 Vienna, Austria}
}

\addauthor{
G.~Durieux\\
\emph{Centre for Cosmology, Particle Physics and Phenomenology (CP3), Universit\'e catholique de Louvain, 1348 Louvain-la-Neuve, Belgium}
}

\addauthor{
M.~Tytgat\\
\emph{Inter-University Institute for High Energies, Vrije Universiteit Brussel, 1050 Ixelles, Belgium}
}

\addauthor{
C.~Hensel\\
\emph{Centro Brasileiro de Pesquisas F\'isicas, Rio de Janeiro 22290-180, Brazil}
}

\addauthor{
A.B.~Bellerive\\
\emph{Carleton University, Ottawa, Ontario K1S5B6, Canada}
}

\addauthor{
F.~Corriveau\\
\emph{Centre for High-Energy Physics, McGill University, Montreal, Quebec H3A2T8, Canada}
}

\addauthor{
D.~Tuckler\\
\emph{TRIUMF, Wesbrook Mall, Vancouver, Canada}
}

\addauthor{
C.D.~Fu,
Q.~Ouyang,
H.R.~Qi,
M.Q.~Ruan\\
\emph{Institute of High Energy Physics, Chinese Academy of Sciences, Beijing, China}
}

\addauthor{
I.P.~Ivanov\\
\emph{Sun Yat-sen University, Guangzhou, Guangdong, China}
}

\addauthor{
M.~Ouchemhou,
T.~Robens\\
\emph{Rudjer Boskovic Institute, 10000 Zagreb, Croatia}
}

\addauthor{
J.~Cvach,
J.~Kvasni\v{c}ka,
I.~Pol\'{a}k,
J.~Z\'{a}le\v{s}\'{a}k\\
\emph{FZU - Institute of Physics of the Czech Academy of Sciences, 18200 Prague 8, Czechia}
% ...
}

\addauthor{
J.D.~Wells\\
\emph{Physics Dept, Ann Arbor, MI 48109, USA}
}

\addauthor{
H.~Baer\\
\emph{University of Oklahoma, Norman, OK 73019, USA}
}

\addauthor{

J.E.~Brau,
C.T.~Potter\footnote[2]{deceased on July 12, 2025},
J.~Strube\\
\emph{University of Oregon, Eugene, OR 97403, USA}

}

\addauthor{
K.~Sugizaki\\
\emph{University of Pennsylvania, Philadelphia, PA 19104, USA}
}

\addauthor{
A.~Freitas\\
\emph{University of Pittsburgh, Pittsburgh, PA 15260, USA}
}

\addauthor{
A.P.~White\\
\emph{University of Texas at Arlington, Arlington, TX 76019, USA}
}

\addauthor{
Y.~Bai,
S.~Dasu\\
\emph{University of Wisconsin--Madison, Madison, WI 53706, USA}
}

\addauthor{
\begin{flushleft}
{$^{0}$}Also at: King Abdulaziz City for Science and Technology (KACST), Riyadh, Saudi Arabia\\
{$^{1}$}Also at: NSC KIPT National Science Center ``Kharkiv Institute of Physics and Technology'' of The National Academy of Sciences of Ukraine, Kharkiv, Ukraine\\
{$^{2}$}Also at: University of Hamburg\\
{$^{3}$}Also at: Institute for Theoretical Physics I, Heinrich Heine University, D\"{u}sseldorf\\
{$^{4}$}Also at: St. Joseph's University, Bangalore 560027, India\\
{$^{5}$}Also at: Indian Institute of Science Education and Research Kolkata, Mohanpur, West Bengal, India\\
{$^{6}$}Also at: Nambu Yoichiro Institute of Theoretical and Experimental Physics (NITEP)\\
{$^{7}$}Also at: ICEPP\\
{$^{8}$}Also at: Kavli IPMU\\
{$^{9}$}Also at: Ludwig-Maximilians-Universit\"at M\"unchen, 85748 Garching, Germany\\
{$^{10}$}Also at: Deutsches Elektronen-Synchrotron DESY, Germany\\
{$^{11}$}Also at: CERN, Geneva, Switzerland\\
{$^{12}$}Also at: Physics and Astronomy, Stony Brook University, NY 11794, USA\\
{$^{13}$}Also at: Argonne National Laboratory, Lemont, IL 60439, USA
\end{flushleft}
}

\graphicspath{ {./logos/}{./figures/} }

\DeclareTOCStyleEntry[beforeskip=.2cm]{section}{section}

\begin{document}

\titlepage
\pagenumbering{arabic}\setcounter{page}{2}

\tableofcontents
\clearpage

\section{Introduction} 

The 2020 update of the European Strategy for Particle Physics made a firm statement about its priorities in its first line under ,,High-priority future initiatives'':  ,,An electron-positron Higgs factory is the highest-priority next collider.'' The importance of a Higgs factory was re-emphasized by the Snowmass Study and the P5 report in the US. There are good reasons for this statement.   

The key insight that we have gained from the Large Hadron Collider (LHC) is that the Standard Model of particle physics gives an excellent description of the interactions of elementary particles in all of the processes that it probes.  This is true throughout studies of the strong, weak, and electromagnetic interactions. The Standard Model (SM) requires the Higgs boson, and, indeed, the LHC experiments were able to discover the Higgs boson and to show that it has properties very close to those that the SM predicts. 

On the other hand, the SM is manifestly incomplete.  It does not include the cosmic dark matter or dark energy, and it does not address the issues of formulating a quantum theory of gravity.  But it is also incomplete in its description of particle physics.  For the interactions mediated by gauge bosons, it does describe a wealth of elementary processes by a minimal set of gauge theory equations of motion with only three coupling constants that must be specified.  For interactions mediated by the Higgs boson, the reverse is true.  Every outcome of the theory --- the spectrum of quark and lepton masses, their mixing under the weak interactions,  the presence of CP violation, and the masses of neutrinos --- comes from a specific number put into the Higgs couplings with no underlying explanation. Even the most basic element of the Higgs field's behaviour, the fact that it has spontaneous symmetry breaking, is put into the SM by hand.  

Dynamical explanations of these features must come from new 
couplings of the Higgs boson.  Over the past decades, we have searched for new physics beyond the SM in every possible low-energy setting, seeking high-precision measurements and extremely rare processes. At the energy frontier, the LHC experiments have carried out extensive searches and have made great progress in measuring the properties of the Higgs boson.

Many important  opportunities remain for studies at higher precision with the potential to reveal the influence of new physics on the Higgs boson.  We must continue this search, since this is the one place where the imprint of these new interactions must be found.

An $\ee$ collider is the ideal machine for precision measurements.  Because electron and positrons are elementary particles, their reactions are simple and display the structure of the underlying interactions directly.  Backgrounds are dominated by electroweak processes, and these are also simple and --- more importantly --- precisely understood at the part-permil level.  The low event rates relative to proton collisions allow the construction of low-material-budget, high-precision detectors and of trigger-less data taking. All of these features minimize systematic uncertainties.  This makes it possible to measure small deviations from the SM with high confidence and credibility.

The choice of a linear $\ee$ collider redoubles these advantages.  Electroweak physics is intrinsically chiral, with, for example, the left- and right-handed electron giving different information.  Linear colliders will have high beam polarisation to take advantage of the new observables that this produces.  Higgs bosons are produced in a number of different reactions that complement one another, as is well appreciated at the LHC.  Because linear collider designs have luminosities that increase with the centre-of-mass energy, they can study directly not only the reaction $\ee\to \PZ\PH$ that is most important at low energies but also $\PW\PW$ fusion, reactions involving the top quark Yukawa coupling, multiple reactions with Higgs boson pair production, and at the highest energies even constrain triple-production of Higgs bosons.  They can also carry out a programme of precision measurements on the top quark,  the quark for which consequences of non-Standard Higgs physics would be most visible. Many of these measurements see a tremendous gain in sensitivity at higher centre-of-mass energies, significantly above the $\PQt\PAQt$ production threshold. This applies likewise to processes with multiple electroweak gauge bosons. Linear colliders thus provide a large number of distinct observables covering the full range of interactions of the Higgs boson and its closest relatives in the SM. These observables will be crucial to discover deviations from the SM through precision measurement, and also to re-discover it in a variety of processes.   This programme will allow us to confirm the presence of new physics and to fully understand the pattern of the effects it has on SM particles.

The purpose of this paper is to review our knowledge of the physics opportunities at linear $\ee$ colliders with a special focus on high centre-of-mass energies and beam polarisation, to take a fresh look at the various accelerator technologies available or under development and, for the first time, discuss how a facility first equipped with a technology mature today could be upgraded with technologies of tomorrow to reach much higher energies and/or luminosities. In addition, we will discuss detectors and alternative collider modes, as well as opportunities for beyond-collider experiments and R\&D facilities as part of a generic linear collider facility (LCF).

Substantial documentation exists already on the more mature linear collider proposals with their detectors and physics potential, including the Technical Design Report of the International Linear Collider (ILC)~\cite{Behnke:2013xla, ILC:2013jhg, Adolphsen:2013jya, Adolphsen:2013kya, Behnke:2013lya} and its staging report~\cite{Evans:2017rvt}, the Conceptual Design Report of the Compact Linear Collider (CLIC)~\cite{Aicheler:2012bya, Lebrun:2012hj, Linssen:2012hp}, its updated baseline~\cite{CLIC:2016zwp} and its Project Implementation Plan~\cite{Aicheler:2018arh}, the ILC and CLIC reports for the previous update of the European Strategy for Particle Physics (ESPP)~\cite{Bambade:2019fyw, CLICdp:2018cto, CLIC:2018fvx}, the ILC and CLIC reports to the Snowmass 2021 study in the US~\cite{ILCInternationalDevelopmentTeam:2022izu,Brunner:2022usy} and the current update of the ESPP~\cite{ILC-EPPSU:2025, CLIC-EPPSU:2025, Adli:ESU25RDR}. Here, we will build on those papers, provide references to the previous literature and update the conclusions from new studies wherever possible. Based on the detailed designs for ILC and CLIC, the Linear Collider Facility at CERN~\cite{LCF:EPPSU} is now being proposed as a future flagship project for CERN.

In recent years, even more proposals or concepts for linear colliders based on technologies under development today have been proposed, including for instance --- but not being restricted to --- C$^3$~\cite{vernieri2023cool,CCC:ESPPU}, HELEN~\cite{Belomestnykh:2023uon, Belomestnykh:2023naf}, HALHF~\cite{HALHF:EPPSU, Foster:HALHF2025}, ALEGRO~\cite{ALEGRO:EPPSU, Cros:2019tns, Proceedings:2024ncv}, the  Design Initiative for a \SI{10}{TeV} pCM Wakefield Collider~\cite{10TeV_AAC}, the ERLC~\cite{Telnov_2021} or ReLiC~\cite{Litvinenko:2022qbd}. We will discuss how these technologies, or a photon collider like for instance XCC~\cite{Barklow_2023} could boost the performance of the linear collider facility in later upgrades. From the beginning, a linear collider could also host a large variety of beyond collider experiments and R\&D facilities.

{\bfseries The material of this paper will support all plans for $\boldmath\ee$ linear colliders and additional opportunities they offer, independently of technology choice or proposed site, as well as R\&D for advanced accelerator technologies.} 

An outline of this paper is as follows: in Section~\ref{sec:RunScenarios} we will outline potential running scenarios and the experimental environment as the foundation for the following sections. Section~\ref{sec:physics} reviews the science opportunities offered by a linear collider facility, Section~\ref{sec:acc} discusses the available baseline designs and future upgrade strategies for a linear collider facility.   
Section~\ref{sec:beyond} is dedicated to the facilities that a linear collider facility would make available for beam dump and fixed target experiments, while Section~\ref{sec:det} will discuss linear collider detectors with their current technologies and future developments.  Section~\ref{sec:conclusions} will present our conclusions.

\section{Operating scenarios, experimental environment and algorithmic developments} 
\label{sec:RunScenarios}

In this section we summarise the operating scenarios discussed for linear colliders as well as the experimental environment at these machines. For a comprehensive introduction to the energy and polarisation dependence of various physics processes in $\ee$ collisions, as well as the physics impact of beam polarisation and the measurement strategies for luminosity and polarisation we refer to Chapter~5 of~\cite{ILCInternationalDevelopmentTeam:2022izu} and to~\cite{Moortgat-Pick:2005jsx}.

Quantitative studies of the physics potential of future colliders are necessarily based on some reference assumptions on the centre-of-mass energy, integrated luminosity as well as the beam polarisation (absolute values and luminosity sharing between the different sign configurations). In electron-positron collisions, different physics observables often have their own preferred centre-of-mass energy range. Therefore the overall physics performance of any $\ee$ collider depends on the choice of a full operating or running scenario, combining a set of energies with the respective luminosity and polarisation information. Different choices made in the running scenarios often reflect different choices on which part of the physics programme to emphasize. We will introduce some of current standard running scenarios in Sec.~\ref{sec:operating}, while stressing that one of the key assets of a Linear Collider remains the flexibility to adjust the operation mode as scientific priorities demand.

We also note that in case of detailed Geant4-based detector simulations, i.e.\ the de-facto standard for ILC and CLIC projections, analyses in some cases have been performed on MC data sets corresponding to much lower integrated luminosities and then extrapolated to the full running scenario due to resource limitations.

\subsection{Beam polarisation}
\label{sec:polarisation}
Beam polarisation is an important aspect of linear collider physics. Hence, the beam polarisations to be used must be specified as part of the operating plan. For the convenience of the reader, we give a brief introduction here, referring to the literature, e.g.~\cite{ILCInternationalDevelopmentTeam:2022izu, Moortgat-Pick:2005jsx} for a more in-depth discussion. Longitudinal beam polarisation quantifies the number asymmetry between the $N_L$ left- and the $N_R$ right-handed particles in an ensemble of particles, like a bunch of a collider beam\footnote{Here, we assume a highly relativistic beam, for which the helicity is for all practical purposes equivalent to the chirality.}: 
\begin{equation}
                     P =    \frac{N_R - N_L}{N_L + N_R}
\end{equation}
Vice-versa, a beam of polarisation $P$ contains the fractions of particles of each helicity
\begin{equation}
                 f_L = \frac{1 - P}{2}           \qquad     f_R = \frac{1 +  P}{2} \ .
\end{equation}
A collider with both beams polarised will have four different possible longitudinal polarisation sign configurations, plus the options to take parts of the data with zero polarisation or transverse polarisation~\cite{Moortgat-Pick:2005jsx}. Since  $\PemR$ and $\PemL$ have different $SU(2)\times U(1)$ quantum numbers, each of the four polarisation settings is effectively a different scattering experiment.  
The ILC as well as the LCF at \SI{250}{GeV} assume absolute values of $80\%$ electron polarisation and $30\%$ positron polarisation, with $60\%$ positron polarisation as upgrade. Taking the case of P($\Pem,\Pep$)=$(-80\%/+30\%)$ as example, the content in terms of the electron and positron chiral states is
\begin{equation}
         f_L(\Pem) = 90\%  \quad   f_R(\Pem) = 10\%  \quad  ; \quad
         f_R(\Pep) =  65\%  \quad     f_L(\Pep) = 35\%  \   ,
\end{equation}
while $\Pep\Pem$ collisions with non-polarised beams lead to equally populated helicity combinations LR, LL, RR, and RL, i.e.\ all $f$'s equal to 50\%. 

The cross-section for any given polarisation choice is the sum of the chiral, i.e.\ fully polarised cross-sections with the product of the respective $f$'s as prefactors:
\begin{eqnarray}
        \upsigma_{P_{\Pem},\,P_{\Pep}} &=& f_L(\Pem) f_R(\Pep) \upsigma_{LR} \,+\, f_R(\Pem) f_L(\Pep) \upsigma_{RL} \,+\, f_L(\Pem) f_L(\Pep) \upsigma_{LL} \,+\, f_R(\Pem) f_R(\Pep) \upsigma_{RR} \\
        &=& \frac{(1 - P_{\Pem})(1 + P_{\Pep})}{4} \upsigma_{LR}
        + \frac{(1 + P_{\Pem})(1 - P_{\Pep})}{4} \upsigma_{RL} \\
       & &  + \frac{(1 - P_{\Pem})(1 - P_{\Pep})}{4} \upsigma_{LL}
        + \frac{(1 + P_{\Pem})(1 + P_{\Pep})}{4} \upsigma_{RR} \nonumber
\end{eqnarray}

For a purely vector-boson mediated $s$-channel process, $\upsigma_{LL}$ and $\upsigma_{RR}$ are zero due to angular momentum conservation, and the polarised cross-section simplifies to
\begin{equation}
    \upsigma_{P_{\Pem},\,P_{\Pep}} = \frac{1}{4} \left[  
    \left( 1-P_{\Pem} P_{\Pep} \right) 
    \left( \upsigma_{LR} + \upsigma_{RL}  \right) +
    \left( P_{\Pem} - P_{\Pep} \right)
    \left( \upsigma_{RL} - \upsigma_{LR}  \right)
    \right],
\end{equation}
This means that in case of an unpolarised collider, half of the collisions do not contribute to these processes, thus polarised beams enhance the effective luminosity of an $\Pep\Pem$ collider by a factor
\begin{equation}
     {\cal L}_{\mathrm{eff}}/{\cal L}   =  (1 - P_{\Pem}P_{\Pep}) .
\end{equation}  
For $t$-channel processes involving $\PW$ bosons, the enhancement is even more dramatic, as only left-handed fermions and right-handed anti-fermions take part in the charged weak interaction.
Besides the enhancement of the effective luminosity, different beam polarisation settings can be used to selectively enhance and suppress individual processes w.r.t.\ others, thereby increasing the signal-to-background ratio in one data-set, while creating otherwise identical control samples with suppressed signal in the data-set with the opposite polarisation signs. This is an extremely effective way to minimise the impact of experimental systematic uncertainties on physics observables, in particular when the beam polarisation signs can be flipped at high frequency, i.e.\ faster than the typical timescales of drifting calibrations etc~\cite{Beyer:2022xyz, Habermehl:2020njb, Karl:2019hes}.

The most powerful benefit of beam polarisation, however, is the provision of additional observables in form of polarised cross-sections and left-right asymmetries. While they can already be accessed with only one beam polarised, the presence of polarisation also for the second beam does not only add redundancy and control of systematics, but also enhances the analysing power, quantified in analogy to the effective luminosity in terms of the effective polarisation:
\begin{equation}
     {P}_{\mathrm{eff}}   =   \frac{P_{\Pem} - P_{\Pep}}{1 - P_{\Pep}P_{\Pem}}.
\end{equation}  

The ratio of the polarised cross-section to the unpolarised one, refered to as enhancement factor in Sec.~\ref{sec:singleHiggs:pol}, can be written in terms of the effective luminosity, effective polarisation and the left-right asymmetry $\asym{LR} =  (\upsigma_{RL} - \upsigma_{LR}) /  (\upsigma_{RL} + \upsigma_{LR})$ as 
\begin{equation}
\frac{\upsigma_{P_{\Pem},\,P_{\Pep}}}{\upsigma_{0}}
       =   \frac{{\cal L}_{\mathrm{eff}}}{\cal L} \left( 1 + {P}_{\mathrm{eff}} \asym{LR} \right).
\label{eqn:pol_enhance}
\end{equation}  

For a more information on the role of beam polarisation we refer to~\cite{ILCInternationalDevelopmentTeam:2022izu, Fujii:2018mli, Moortgat-Pick:2005jsx}.

In the scenarios below,  polarisations of 80\% for electrons and 30\% for positrons are assumed in most cases. A row of each table shows the assumed splitting among the 
$-+$, $-\,-$, $++$, and $+-$ operating modes. 
The luminosity sharing between the beam polarisation sign configurations can be adjusted depending on the physics needs. This is especially true for the ILC and LCF, where the configurations can be adjusted arbitrarily at run time.

\subsection{Operating scenarios}
\label{sec:operating}

For the ILC, a canonical straw-man running scenario has been worked out by the LCC Joint Working Group on Beam Parameters~\cite{Barklow:2015tja}, based on a comprehensive physics-driven optimisation, which aimed to choose the optimal energy for each of the targeted measurements, leading to several desired operation energies below \SI{1}{TeV}. Capitalising on the fact that a linear accelerator based on super-conducting radio frequency cavities (SCRF) can easily be run at lower gradients, operation at centre-of-mass energies below the maximal installed energy is almost always possible. While an extended running period at \SI{250}{GeV} is optimal for measuring the total Higgs\-strahlung cross-section, giving access to the Higgs boson's total width and its coupling to the \PZ boson, most measurements of top-quark properties benefit from a centre-of-mass energy significantly above the $\PQt\PAQt$ production threshold. 
A theoretically proper determination of the top quark mass requires a threshold scan at a centre-of-mass energy around twice the top quark mass. 
The experimental uncertainties become negligible compared to the theory uncertainties already with about \SI{100}{\fbinv}. Thus, the ILC run plan foresees only a brief run at the $\PQt\PAQt$ production threshold. 

The time ordering of the centre-of-mass energies was later adjusted to the staged machine starting at a centre-of-mass energy of \SI{250}{GeV}~\cite{Fujii:2017vwa}.
At the same time, the importance of the left-right asymmetry $A_{LR}$ in the Higgs-strahlung process for disentangling certain SMEFT operators was recognized, leading to an adjustment of the luminosity sharing between the polarisation sign configurations at \SI{250}{GeV}. At the $\PQt\PAQt$ threshold, however, the majority of the luminosity is foreseen to be taken in the $(-,+)$ configuration. 
We'd like to stress, however, that the ILC design foresees fast helicity reversal for both beams, which allows the luminosity sharing between the polarisation sign configurations to be adjusted arbitrarily at run time, depending on physics needs.

Table~\ref{tab:ILC-runplan} summarises the integrated luminosities envisioned at each energy stage, along with the beam polarisations. The Cool Copper Collider~\cite{vernieri2023cool} envisions a similar scenario, adopting though \SI{550}{GeV} instead of \SI{500}{GeV} due to the significantly higher $\PQt\PAQt\PH$ cross-section. The actual time needed for any of the stages will depend on the assumed instantaneous luminosity as well as on the assumptions on operating efficiency and learning curves when starting to operate the collider in a new mode.
 
\begin{table}[htb]
\begin{center}
\begin{tabular}{lccccc}
           &  \SI{91}{\GeV}    &    \SI{250}{\GeV}      &   \SI{350}{\GeV}   &   \SI{500}{\GeV}  &   \SI{1000}{\GeV} \\ \hline
$\int  \L$ (ab$^{-1}$) &     0.1    &       2          &   0.2       &     4      &      8    \\     
beam polarisation ($\Pem/\Pep$; \%)   &   80/30  &  80/30    &  80/30  &   80/30  &  80/20  \\
($-\,-$, $-+$, $+-$, $++$) (\%)      &  (10,40,40,10) & (5,45,45,5) &  (5,68,22,5) &  (10,40,40,10)   & (10,40,40,10) \\
\end{tabular}
\end{center}
\caption{Centre-of-mass energy, integrated luminosity and beam polarisation of the various stages of the straw-man ILC operating scenario~\cite{Barklow:2015tja}. Running at the $\PW\PW$ threshold is technically possible, however currently not prioritized, c.f.~Sec.~\ref{sec:Wprec} for a discussion of the role of the threshold running for the $\PW$ mass determination.}
\label{tab:ILC-runplan}
\end{table}

The main objective for CLIC, on the other hand, is to reach the TeV-regime as quickly as possible. Thus, the optimisation of the running scenario was conducted in a very different way by considering which single centre-of-mass energy offers the best compromise covering the majority of the physics programme below \SI{1}{TeV}, as detailed in~\cite{CLIC:2016zwp}. This concluded that a centre-of-mass energy of \SI{380}{GeV} is the optimal choice before proceeding to the TeV-regime, since it allows to perform both top-quark and Higgs boson measurements. In particular the total Higgs\-strahlung cross-section measurement via the hadronic recoil benefits at this energy from a much cleaner kinematic separation from important background processes compared to \SI{250}{GeV}, minimizing model-dependencies due to decay-mode-dependent selection efficiencies~\cite{Thomson:2015jda}.

Building on the CLIC report to Snowmass~\cite{Brunner:2022usy}, the CLIC luminosity has been updated in December 2024~\cite{Robson:2025clic} for all three stages: \SI{380}{GeV}, \SI{1.5}{TeV} and \SI{3}{TeV}. For the lowest energy point, two options
are available, differing only in the repetition rate of \SI{50}{Hz} or \SI{100}{Hz}, i.e., doubling the luminosity. The envisaged integrated luminosities and beam polarisations are summarised in Table~\ref{tab:CLIC-runplan}.

\begin{table}[htb]
\centering{
\begin{tabular}{ lccc } 
   &  \SI{380}{\GeV}  &  \SI{1500}{\GeV}    & \SI{3000}{\GeV} \\  \hline
  $\int  \L$ (ab$^{-1}$) & 2.2 (4.3) & 4     & 5 \\
beam polarisation (\Pem,\Pep; \%)   &   80/0  &  80/0   &  80/0   \\
($-0$, $+0$) (\%)     &  (50,50) & (80,20) &  (80,20)  \\
\end{tabular}
}
\caption{The updated baseline CLIC operation model (December 2024). Two options for \SI{380}{GeV} running are given, with \SI{50}{Hz} and \SI{100}{Hz} repetition rates, respectively.  
Running at $\roots=$\SI{91}{GeV} is an option.
\label{tab:CLIC-runplan}}
\end{table}

Many of the physics projections in the following section will be based on these ``classic'' operating scenarios for ILC and CLIC.

In the context of the LCVision discussions, however, the physics-driven optimisation for a generic Linear Collider has been revisited, considering also that up to now no direct or indirect signals of new particles in the TeV regime have been found --- giving more emphasis to precision measurements of the Higgs boson, the top quark, and the electroweak gauge bosons, while keeping the flexibility to target energies of up to \SI{3}{TeV} in the future. 
The design of the ILC as a linear collider based on superconducting technologies was frozen more than ten years ago. Meanwhile there is major progress in SCRF (R\&D as well as operational experience with XFELs), ultra-low emittance storage rings and other individual components (e.g.\ klystron efficiency). Implications will be discussed in Sec.~\ref{sec:acc}, where we lay out suggestions for an updated SCRF-based linear collider and discuss the option to upgrade an initial facility constructed with today's technology with higher-gradient developments in the future. Based on a significantly higher instantaneous luminosity compared to ILC as considered in Japan, the LCVision team proposes the run plan summarised in Table~\ref{tab:LCF-runplan}. Note for completeness that up to 550\,GeV this plan is identical to the plan presented in the LCF proposal for CERN~\cite{LCF:EPPSU}. Thus we will refer to this operating scenario as the LCF scenario. It differs from the ILC scenario (Table~\ref{tab:ILC-runplan}) by planning for 3 instead of 2\,\abinv at \SI{250}{GeV}, and by doubling the integrated luminosities of the top threshold scan and the \SI{550}{GeV} run. It also assumes that after the  operation of the lower energy stages, the helical undulator of the positron source can be prolonged and equipped with a tighter collimation system for the undulator photons to increase the absolute value of the positron polarisation to \SI{60}{\%}~\cite{Alharbi:2021ctf,Riemann:2014hia,Clarke:2008zz}.
Example upgrade scenarios and timelines for collecting these data sets will be discussed in Sec.~\ref{sec:acc:scenarios}.

\begin{table}[htb]
\begin{center}
\begin{tabular}{lccccc}
           &  \SI{91}{\GeV}    &     \SI{250}{\GeV}      &   \SI{350}{\GeV}   &   \SI{550}{\GeV}  &   1-3 TeV \\ \hline
$\int  \L$ (ab$^{-1}$) &     0.1    &       3          &   0.4       &     8      &      8    \\     
beam polarisation (\Pem,\Pep; \%)   &   80/30  &  80/30    &  80/60  &   80/60  &  80/20  \\
($-\,-$, $-+$, $+-$, $++$) (\%)     &  (10,40,40,10) & (5,45,45,5) &  (5,68,22,5) &  (10,40,40,10)   & (10,40,40,10) \\
\end{tabular}
\end{center}
\caption{Centre-of-mass energy, integrated luminosity and beam polarisation of the various stages of the strawman LCVision -- or LCF -- scenario. Running at the $\PW\PW$ threshold is technically possible, however currently not prioritized, c.f.~Sec.~\ref{sec:Wprec} for a discussion of the role of the threshold running for the $\PW$ mass determination.}
\label{tab:LCF-runplan}
\end{table}


\subsection{Expected number of events in the LCF operating scenario \label{sec:nevts}}
For illustration purposes, we present here the number of events produced from selected processes during the LCF operating scenario as defined in Table~\ref{tab:LCF-runplan}, based on the comprehensive MC event generation campaigns done in the past for the ILC, in particular by the LCC Generator Group~\cite{Berggren:2021sju}. In these campaigns, the full set of Standard Model processes is covered systematically by  grouping tree-level processes according to the sum of fermions and anti-fermions in the final-state of the hard interaction. Thereby, the intermediate particles are not restricted, thus including the full interference for fermion-level final-states reachable via different intermediate particles (vector bosons, top quark, etc). These MC productions were performed for 100\% polarised beams, in all four helicity configurations, allowing easy re-weighting to any desired polarisation scenario as described in Sec.~\ref{sec:polarisation}. 

On this basis, Table~\ref{tab:nevts} gives expected numbers of events from some selected SM processes at the 250, 350 and 550\,GeV runs of the LCF operating scenario.  In total, the LCF program is expected to produce about 2.3 million Higgs bosons, 11 million top- or anti-top quarks, and 300 million $\PW$ bosons.

When comparing event numbers, it is important to realise that in many cases, different production modes for the same particle carry qualitatively different information. For example, the ``total number of Higgs bosons produced'' can have a very different physics impact depending on how it is distributed on individual processes like $\Pep\Pem\to \PZ\PH$, $\Pep\Pem\to \PGne\PAGne\PH$, $\Pep\Pem\to \PQt\PAQt\PH$ or $\Pep\Pem\to \PZ\PH\PH$.
At a polarised collider, this goes one step further: Due to the chiral nature of weak interactions, even events from initial-states of different chirality carry qualitatively different information. Thus e.g.\ a certain number of Higgs bosons from the unpolarised $\Pep\Pem\to \PZ\PH$ process carry less information then the same number of Higgs bosons separated\footnote{Of course these cannot be separated at the level of individual events, but data-taking with all four different polarisation sign combinations allows a statistical separation and measurements of the chiral cross-sections.} into $\PepR\PemL\to \PZ\PH$ and $\PepL\PemR\to \PZ\PH$. Therefore, the processes and chiral initial-states are listed separately in Tab~\ref{tab:nevts}, illustrating also how some processes hugely profit from higher centre-of-mass energies.

\begin{table}[htb]
\begin{center}
\begin{tabular}{l|ccc|c}
           &  \SI{250}{\GeV}      &   \SI{350}{\GeV}   &   \SI{550}{\GeV}  &   total \\ \hline
$\int  \L$ (ab$^{-1}$) &      3       &   0.4       &     8      &         \\     
beam polarisation (\Pem,\Pep; \%)   &  80/30    &  80/60  &   80/60  &    \\
($-\,-$, $-+$, $+-$, $++$) (\%)     &  (5,45,45,5) &  (5,68,22,5) &  (10,40,40,10)    \\
\hline
$\PepR\PemL \to \PZ\PH$              & 4.6E+05  &  6.7E+04 & 3.5E+05  &  8.7E+05 \\
$\PepL\PemR \to \PZ\PH$              & 2.9E+05  &  1.6E+04 & 2.2E+05  &  5.3E+05 \\
$\PepR\PemL \to \PGne\PAGne \PH$     & 2.3E+04  &  2.1E+04 & 8.2E+05  &  8.6E+05 \\ 
$\PepR\PemL \to \PZ\PH\PH$           & --  &  -- & 9.6E+02  &  9.6E+02 \\
$\PepL\PemR \to \PZ\PH\PH$           & --  &  -- & 6.2E+02  &  6.2E+02 \\
$\PepR\PemL \to \PGne\PAGne \PH \PH$ & --  &  -- & 3.9E+01  &  3.9E+01 \\ 
$\PepR\PemL \to \PQt\PAQt\PH  $      & --  &  -- & 1.1E+04  &  1.1E+04 \\
$\PepL\PemR \to \PQt\PAQt\PH $       & --  &  -- & 4.4E+03  &  4.4E+03 \\\hline
\bfseries{N (Higgs bosons)}                   &  \bfseries{7.7E+05} & \bfseries{1.0E+05}  & \bfseries{1.4E+06}  &  \bfseries{2.3E+06} \\\hline
$\PepR\PemL \to \PQt\PAQt$   & --  & 1.1E+05  & 3.6E+06  &  3.7E+06 \\
$\PepL\PemR \to \PQt\PAQt$   & --  & 1.6E+04  & 1.5E+06  &  1.5E+06 \\ \hline
\bfseries{N (top (anti-)quarks)}   & \bfseries{--}  & \bfseries{2.5E+05}  & \bfseries{1.0E+07}  & \bfseries{1.1E+07}  \\ \hline
$\PepR\PemL \to \PW\PW$        & 5.8E+07  & 9.5E+06  & 7.2E+07  &  1.4E+08 \\
$\PepL\PemR \to \PW\PW$        & 5.2E+05  & 2.3E+04  & 2.8E+05  &  8.3E+05 \\
$\PepL\PemL \to \Pe\PGne\PW$   & 1.1E+05  & 2.2E+04  & 1.5E+06  &  1.6E+06 \\
$\PepR\PemL \to \Pe\PGne\PW$   & 5.4E+06  & 1.0E+06  & 1.1E+07  &  1.7E+07 \\
$\PepR\PemR \to \Pe\PGne\PW$   & 9.9E+04  & 4.5E+03  & 1.2E+06  &  1.3E+06 \\
$\PepL\PemR \to \Pe\PGne\PW$   & 4.9E+05  & 2.2E+04  & 3.1E+05  &  8.2E+05 \\\hline
\bfseries{N (W bosons)}   & \bfseries{1.2E+08}  & \bfseries{2.0E+07}  & \bfseries{1.6E+08} &  \bfseries{3.0E+08} \\\hline
\end{tabular}
\end{center}
\caption{Number of events expected for selected processes in the LCF operating scenario. Fermion combinations in the final state refer explicitly to the non-resonant contributions. Numbers derived from the LCC Generator Group MC production campaigns.}
\label{tab:nevts}
\end{table}

\subsection{Instantaneous luminosity and power consumption of linear colliders \label{sec:lumipower}}

\begin{figure}[htbp]
    \centering
    \begin{subfigure}{.5\textwidth}
    \centering
        \includegraphics[width=0.95\textwidth]{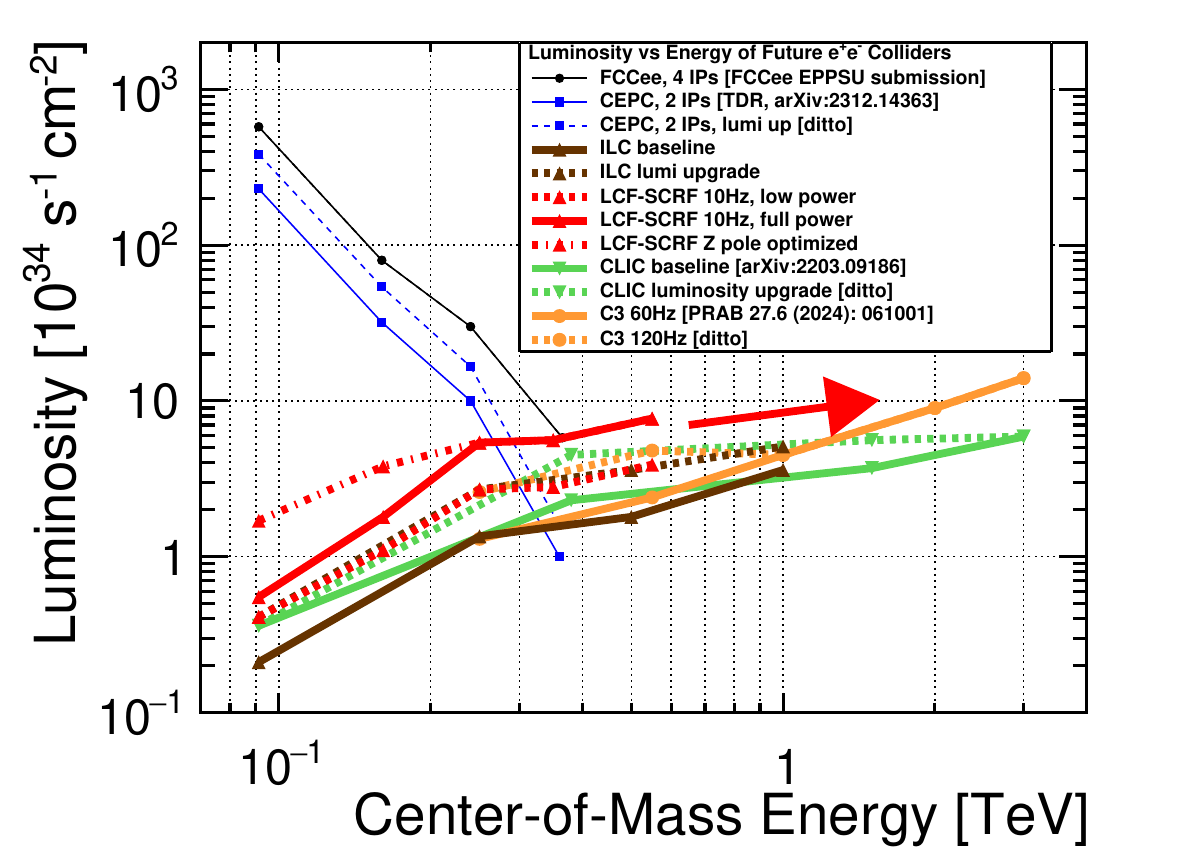}
        \caption{}
        \label{fig:lep:lumi}    
    \end{subfigure}\hfill%
    \begin{subfigure}{.5\textwidth}
        \centering
        \includegraphics[width=0.95\textwidth]{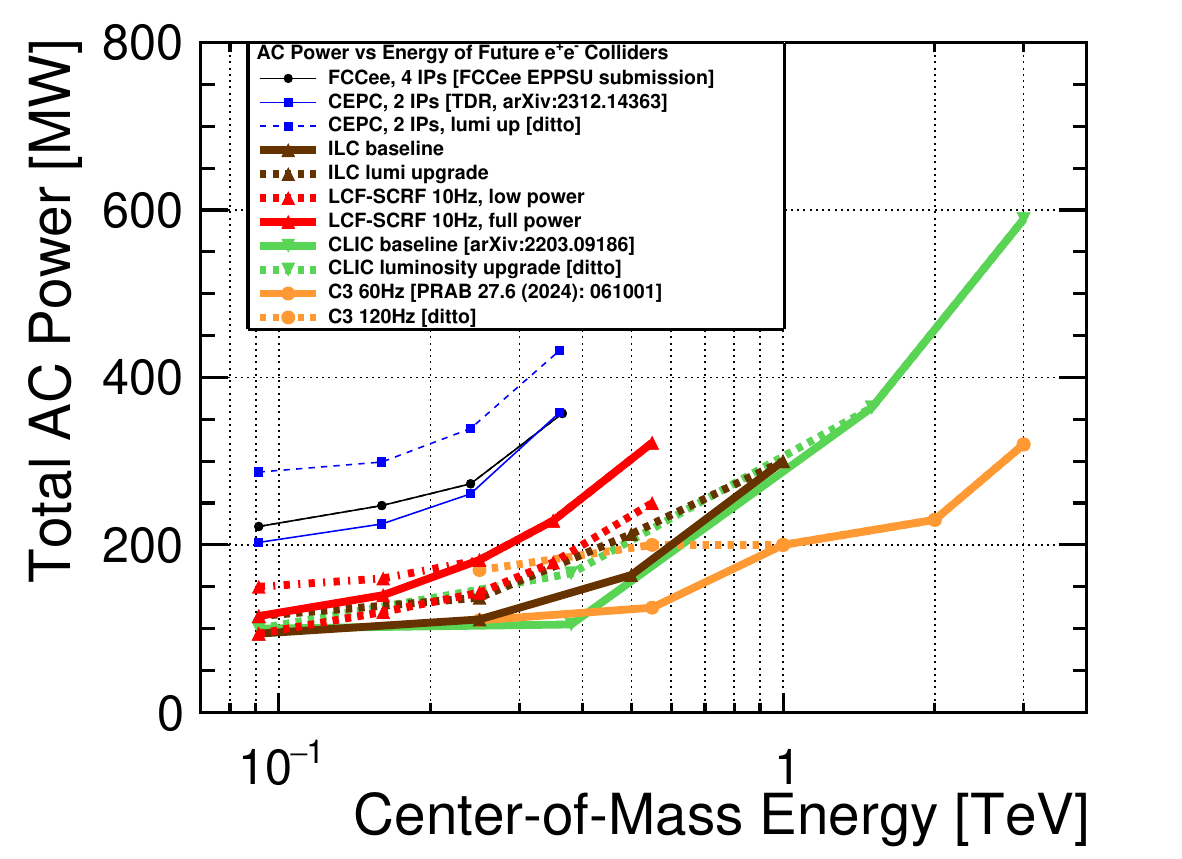}
        \caption{}
        \label{fig:lep:power}    
    \end{subfigure}%
    \vspace{0.1cm}
    \begin{subfigure}{.5\textwidth}
    \centering
        \includegraphics[width=0.95\textwidth]{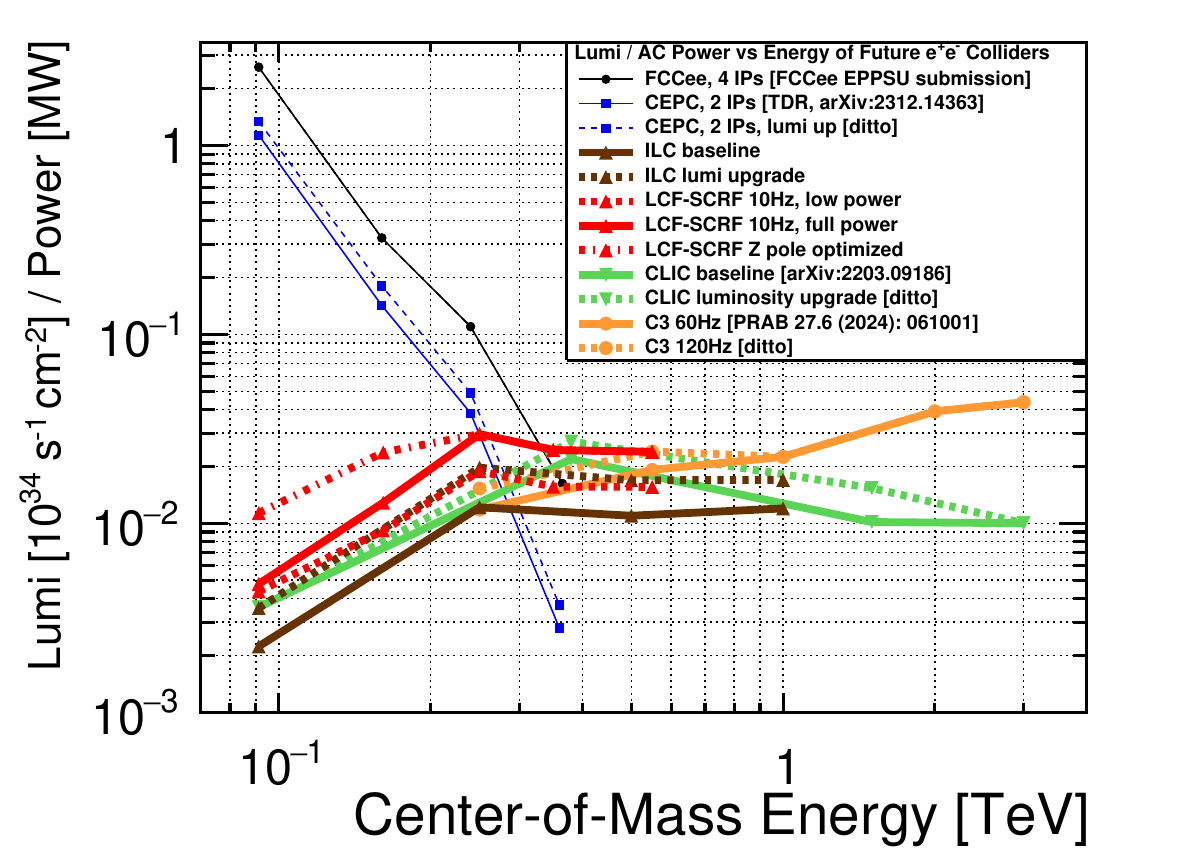}
        \caption{}
        \label{fig:lep:lumipower}    
    \end{subfigure}\hfill%
    \begin{subfigure}{.5\textwidth}
        \centering
        \includegraphics[width=0.95\textwidth]{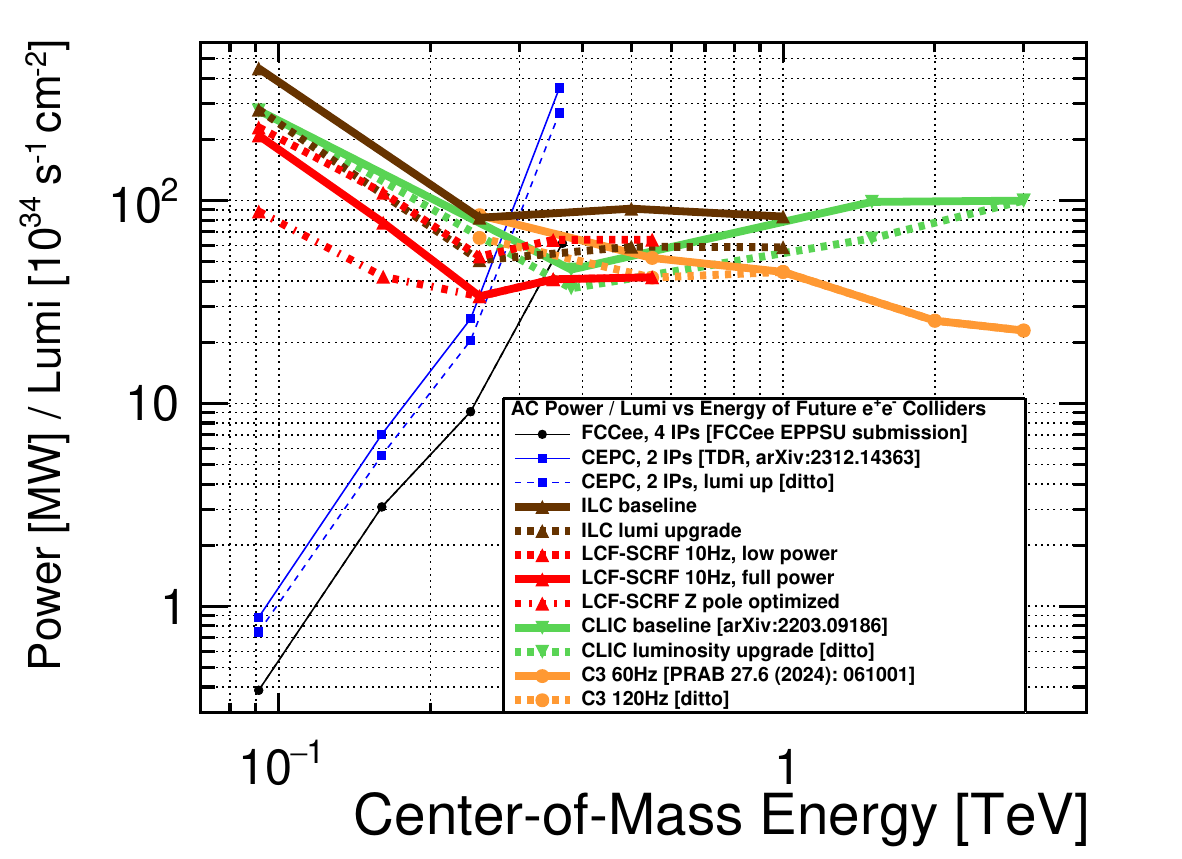}
        \caption{}
        \label{fig:lep:powerlumi}    
    \end{subfigure}%
    \caption{Instantaneous luminosity (a), total site power (b), and their ratios (c), (d) as a function of the centre-of-mass energy for various $\ee$ colliders. The LCF is drawn only up to \SI{550}{GeV}, since the luminosity and power consumption of higher energies will depend on the yet-to-be-chosen technology for these energies. However, they can be expected to be similar to the high-energy stages of CLIC and \CCC, as indicated by the red arrow.}
    \label{fig:lep}
\end{figure}

The instantaneous luminosity is intimately connected to the electrical power budget of a collider. Linear collider proposals have traditionally been designed against self-imposed limits on the power consumption. For instance ILC aimed to stay around \SI{100}{MW} for operation at \SI{250}{GeV}, and not to exceed \SI{300}{MW} for operation at \SI{1}{TeV}. This has led to some misconceptions about the technical possibilities of linear colliders. In Sec.~\ref{sec:acc}, we will come back to this theme and discuss in more detail the possibilities to increase the luminosity w.r.t.\ the classic linear collider proposals and their respective implications on the power consumption. Here, we give already a summary in Fig.~\ref{fig:lep}, where LCF (red lines) refers to an ILC-like collider with some of the improvements discussed in Sec.~\ref{sec:acc} applied, as proposed in~\cite{LCF:EPPSU}.

Figure~\ref{fig:lep:lumi} gives the instantaneous luminosity of various proposed circular and linear colliders as a function of the centre-of-mass energy, clearly showing the complementary performance of circular and linear colliders. The corresponding total AC power consumption is presented in Fig.~\ref{fig:lep:power}, indicating that the higher luminosity of LCF w.r.t.\ ILC comes at the price of a higher power consumption -- although still significantly below that of circular colliders. Nevertheless, the luminosity-to-power ratio (c.f.\ Fig.~\ref{fig:lep:lumipower}) for LCF is more favourable than for ILC, thus the moderate increase in power consumption enables a more efficient operation of the machine.

At centre-of-mass energies above 1\ or \SI{1.5}{TeV}, both SCRF and drive-beam based linear colliders become very power hungry (c.f.\ Fig.~\ref{fig:lep:power}) -- in case of CLIC at \SI{3}{TeV} in particular due to the required second drive-beam complex. The cool copper technology promises a more benign growth of the AC power with centre-of-mass energy. Therefore, the LCF proposal chose to not fix the technology for centre-mass-energies above \SI{550}{GeV} at this point in time, and instead proposes that the initial facility should be designed maximising the compatibility with various different upgrade options which are expected to become production ready in the next decade.

\subsection{Experimental environment and detector requirements\label{sec:expenv}}

A common feature of most linear colliders is the pulsed operation, i.e.\ the grouping of bunches into trains or pulses, with a time structure determined by the accelerator technology. In case of the ILC, the beam structure features \SI{0.73}{ms} long bunch trains being delivered to the interaction point 
with a rate of \SI{5}{Hz}/\SI{10}{Hz}. The bunches are spaced equally every \SI{554}{ns} or \SI{336}{ns}, c.f.\ Table~\ref{tab:ILCbaseline}. At CLIC, the bunch separation is reduced to \SI{0.5}{ns}, with a train repetition rate of \num{50} or \SI{100}{Hz}, c.f.\ Table~\ref{tab:CLIC380}, and at C$^3$ the bunch separation ranges between \SI{1.65}{ns} and \SI{5.26}{ns}, with a train repetition rate of \num{60} or \SI{120}{Hz}, c.f.\ Table~\ref{tab:C3baseline}. 

The detectors at a future linear collider will be optimized for precision measurements of the Higgs boson, electroweak bosons, the top quark, and other particles, with requirements significantly exceeding those of hadron colliders due to the cleaner \ee\ collision environment and lower radiation levels. The linear collider detectors must achieve very precise and highly efficient tracking, exceptional jet energy resolution, and trigger-less operation, leveraging the lower complexity and rates of \ee\ collisions. The considerations in the previous paragraphs lead to following set of requirements:

\begin{description}
    \item[Impact parameter resolution:]  An impact parameter resolution of $ 5\,\mu \mathrm{m} \oplus 10\,\mu \mathrm{m} / [ p~[{\mathrm{GeV}/c}]\sin^{3/2}\theta$] has been defined as a goal in order to enable efficient identification of charm-quark decays. 
    \item[Momentum resolution:] An inverse momentum resolution of $\Delta (1 / p) = 2 \times 10^{-5}\,\mathrm{[(GeV/c)}^{-1}]$ asymptotically at high momenta should be reached in order not to limit the precision of the Higgs recoil measurement in the $\PZ\to\PGm\PGm$ channel. Maintaining excellent tracking efficiency and very good momentum resolution at lower momenta will be achieved by an aggressive design to minimise the detector's material budget.
    \item[Jet energy resolution:] A jet energy resolution $\Delta E/ E =$\SI{3}{-}\SI{4}{\%} is required for the statistical separation of hadronically decaying \PW, \PZ, and Higgs bosons. Particle flow-based detectors and corresponding reconstruction algorithms have been shown to meet this requirement. 
    \item[Readout:] The detector readout will not use a hardware trigger, ensuring full efficiency for all possible event topologies.
    \item[Powering:] The power of major systems will be cycled between bunch trains. This will reduce the power consumption of the detectors and minimise the amount of material in the detector. The corresponding reduced service requirements will also be beneficial for the acceptance and hermeticity of the detectors.
\end{description}

The detectors must also handle radiation exposure, with most regions experiencing modest doses (about $10^{11}\,\Pn/\mathrm{cm}^2/\mathrm{year}$ NIEL), except for the forward calorimeters near the beamline. 

Common to all linear machines is the need for strong beam focusing resulting in beam sizes in the nanometer range at the interaction point. This causes a challenge -- but also brings a benefit:

\begin{itemize}

\item The strong electromagnetic fields generated by the tiny beams lead to beamstrahlung and the creation of \ee\ pair background, posing a potential challenge to the detectors. However all linear collider detector concepts demonstrated that the expected impact is manageable, with detector occupancies kept under control, see e.g.~\cite{Behnke:2013lya} for the ILC detector concepts. 
A recent example for C$^3$ is presented in~\cite{Ntounis:2024gjw}. CLIC at \SI{3}{TeV} constitutes the most challenging environment for the detectors in view of the high beam-induced background levels, requiring a slightly larger radius for the innermost layer of the vertex detector as well as timing information to separate signals from different bunch crossings.

\item For the identification of $\PQb$- and $\PQc$-jets, the tiny beam spot is a clear advantage, as the transverse size of the luminous region is negligible and does not contribute visibly to the impact parameter and vertex significance. This enables a full exploitation of the vertex detector resolution in high-level reconstruction.

\end{itemize}

More details on the linear collider detector concepts, the R\&D supporting these designs and new ideas will be discussed in Sec.~\ref{sec:det}. Comprehensive descriptions of the detector concepts have been given e.g.\ in~\cite{ILCInternationalDevelopmentTeam:2022izu, Breidenbach:2021sdo, ILDConceptGroup:2020sfq,  Dannheim:2019rcr, CLICdp:2018vnx, Behnke:2013lya}

\subsection{Event reconstruction --- a brief overview and outlook}
\label{sec:hlreco}

The physics performance of modern highly-granular particle-flow detectors decisively depends on both excellent hardware, discussed in more detail in Sec.~\ref{sec:det}, and sophisticated reconstruction algorithms. This concerns track reconstruction and particle flow as well as particle identification, jet flavour tagging and many other aspects. Linear collider physics studies have a long-standing tradition of using detailed and realistic detector simulations, based upon performance of prototypes observed in test beam campaigns. Reconstruction and analysis tools leveraging machine-learning and other advanced techniques are however only now being developed, and the vast majority of physics projections use more traditional algorithms. A deeper discussion of the state of the art in this respect can be found in e.g.\ the ILD Interim Design Report~\cite{ILDConceptGroup:2020sfq}, the ILC Report to Snowmass~\cite{ILCInternationalDevelopmentTeam:2022izu} and, even more recently, in the report of the ECFA Study on Higgs / Top / Electroweak Factories~\cite{ECFA-HF-Report}.

The recent and very dynamically ongoing developments in jet flavour tagging deserve special highlighting:
Since jets (especially with heavy bosons) are contained in most of final states including Higgs and gauge bosons, the identification of the jet flavour is an essential tool to distinguish such final states to derive various coupling constants. LCFIPlus~\cite{Suehara:2015ura} is a successful software tool (released in 2013) which consists of vertex finding, jet clustering and BDT-based jet flavour tagging. Nearly all of the currently available physics projections are based on LCFIPlus.
More recently, jet flavour tagging based on deep-learning technologies have been rapidly advanced and significant improvement has been observed for hadron collider experiments. Among many algorithms, ParticleNet~\cite{FlavourTagger_Mareike} and ParticleTransformer (ParT)~\cite{Tagami:2024gtc} have been applied to linear collider studies and in particular ParT shows 5-10 times better rejection ratio for both $\PQb$-tag and $\PQc$-tag with (ILD) full simulation. 
Among the first analyses leveraging these tools as well as the improved kinematic reconstruction discussed below is the recent update of the $\PZ\PH\PH$ analysis by ILD~\cite{Berggren:2025fpw}, showing a significant impact on the expected  precision for the Higgs self-coupling measurement.

These techniques also enable the tagging of strange and gluon jets with appropriate particle identification (PID) features implemented in detectors such as dE/dx in gaseous trackers, time-of-flight in outer detectors (calorimeters or outer trackers) or dedicated Cherenkov detectors. 
The PID information is provided by the new tool CPID~\cite{Einhaus:2023amy} and both this and the inference part of the flavour tagger have been recently implemented in the ILD reconstruction chain. 

Beyond the tagging of jet flavour, another important development is a decay-by-decay correction for the missing four-momentum of neutrinos from semi-leptonic $\PQb$-decays, along with a bottom-up estimate of energy and angular uncertainties for each individual jet based on the measurement uncertainties of its constituent particle flow objects~\cite{Radkhorrami:2024bbf}. 

The impact of these and many other ongoing improvements on physics analyses is expected to be broad: Similar developments are ongoing in the context of FCC-ee and CEPC, and the shared software framework key4hep~\cite{Carceller:2025ydg, Key4hep:2021rub} allows to leverage the obvious synergies between circular and linear collider studies.  

\subsection{Measurement of luminosity, beam energy and polarisation 
\label{sec:LEPmeas}}
The overall strategy for precise control of the top-level beam parameters always combines fast, time-resolved measurements with dedicated instrumentation up- and downstream of the \ee\ interaction point with an absolute scale calibration from collision data themselves. For instance in the case of the beam energy, the up- and downstream spectrometers have been design to deliver precisions of \num{1e-4} (\SI{100}{ppm})~\cite{Boogert:2009ir}, while $\ee\to\PGmp\PGmm(\PGg)$ events will allow for an absolute scale calibration to a few ppm at the $\PZ$ pole and to \SI{10}{ppm} at higher energies~\cite{Wilson:2023mll,Madison:2022spc,Wilson:2021mG}.

A recent summary of luminosity, beam energy and beam polarisation measurements and their expected precision at the ILC can be found in Sec.~5.4 of~\cite{ILCInternationalDevelopmentTeam:2022izu}. In the following we highlight recent progress in particular w.r.t.\ the control of the integrated luminosity and beam polarisation.

\subsubsection{Integrated luminosity}
There are numerous previous studies on the design and analysis of measuring integrated luminosity at future \ee\ linear colliders ~\cite{Wilson:2024wck,Madison:2025myj,Sadeh:2010ey,Bozovic-Jelisavcic:2010rbh}, and a recent discussion of luminosity measurements at future \ee\ colliders in general, including a discussion of the involved theoretical challenges, can be found in~\cite{Altmann:2025feg}. 

Small angle Bhabha scattering (SABS) has been proposed as the primary integrated luminosity method for similar reasons it was used at previous \ee\ colliders. SABS is simple from a theoretical perspective, as it is dominantly t-channel and QED at small angles. SABS has a huge cross-section compared to other processes, often hundreds to thousands of times larger, and therefore can offer both small statistical uncertainty and comparatively small backgrounds. More recent studies have indicated that SABS may not be as simple due to final state beam electromagnetic deflection effects being more severe, which can affect integrated luminosity precision at the percent level~\cite{Madison:2025myj,Rimbault:2007ork,Rimbault:2007zz}. There are investigations into ways to correct this by reconstructing the collision position along the beam axis, typically referred to as $z_0$. This correction favours more granular forward detectors that can extend to smaller values of the inner acceptance, such as \SI{20}{mrad} or smaller~\cite{Wilson:2025LumiDefl}. However, the optimisation of the position resolution  in the luminosity calorimeters, as well as more detailed studies of the effectiveness and feasibility of this correction, are on-going~\cite{Madison:2024jak}.

Recent work proposes an alternative integrated luminosity process by using di-photons ($\ee\to\PGg\PGg$), which avoids beam electromagnetic deflection due to the neutral final state~\cite{Madison:2025myj,Wilson:2024wck}. Estimations for integrated luminosity uncertainty sources for SABS and di-photons from the $\PZ$ pole to \SI{1}{TeV} can be found in a recent study~\cite{Madison:2025myj}. These estimates found that both SABS and di-photons have paths to precision of \num{1e-4} on integrated luminosity if a careful choice of forward detectors and methodology is made. In particular, the uncertainties for both Bhabhas and di-photons improve if one deploys more granular forward detectors that extend to smaller values of inner acceptance~\cite{Madison:2025myj}. Beyond detector choice and analysis methodology, using di-photons for integrated luminosity at a polarised \ee\ collider relies significantly on the precision of the beam polarisation measurement. The precision in the di-photon cross-section is, at leading-order, linearly dependent on the precision of beam polarisation~\cite{Madison:2025myj}. A recent study has found that, once NLO QED+Weak helicity corrections are included, this dependence becomes less severe than linear~\cite{Madison:2025jsn}. The choice of taking a fraction of data at same-sign beam polarisations, such as the four polarisation combinations of (\SI{10}{\%}, \SI{40}{\%}, \SI{40}{\%}, \SI{10}{\%}) discussed in Sec.~\ref{sec:singleHiggs}, also significantly lowers the strictness of the requirements on beam polarisation precision for measuring the di-photon cross-section~\cite{Madison:2025jsn}. 

\subsubsection{Beam polarisation}
A recent study found that the precision on beam polarisation needed for using di-photons for integrated luminosity at centre-of-mass energies from the $\PZ$ pole to \SI{3}{TeV}, or for the $\PZ$ pole $\asym{\Pe}$ measurement, was both similar and near $1\times10^{-3}$~\cite{Madison:2025jsn}. Collider designs with polarised positron beams were found to have less strict requirements on beam polarisation for measuring $\asym{\Pe}$ and stricter requirements for measuring the di-photon cross-section. A precision of \num{2.5e-3} for beam polarisation has been shown to be plausible for polarimeters~\cite{Boogert:2009ir}. This can be improved to \num{1e-3}  when combining with \ee\ collision data, where rates or angles of well-known polarisation dependent processes can reconstruct the beam polarisation~\cite{Beyer:2022xyz,Karl:2019hes,Wilson:2012SingleBosonPolarimetry}. 

At the level of $10^{-3}$ to $10^{-4}$ in the precision of beam polarisation, the prediction of the luminosity-weighted average polarisation at the \ee\ interaction point from the polarimeter measurements become sensitive to beam transport and beam-beam effects~\cite{Beckmann:2014mka}, while the extraction from collision data becomes sensitive to higher order theory modelling and statistical uncertainties~\cite{Beyer:2022xyz,Karl:2019hes,Marchesini:2011aka,Moortgat-Pick:2008zjf}. 
The combination of these methods with their fully complimentary systematic uncertainties, as well as the redundancy offered by both beams being polarised, will allow to over-constrain the system and to detect and resolve any bias possibly present in any of the single methods individually.

\section{The portal to new physics}

\label{sec:physics}

As we have discussed already in the introduction, precision measurements of the Higgs boson offer unique opportunities to discover deviations from the SM that point to solutions to the many questions the SM cannot answer. In this section, we will review the many signals of new physics that could appear in the precision study of the Higgs boson and other heavy particles of the SM, as well as in direct searches for new particles.

The programme requires several stages, at different $\ee$ centre-of-mass energies that are listed again here for convenience and match the stages sketched in Table~\ref{tab:LCF-runplan}. 

\begin{description}
  \item[An initial stage at $\mathbf{\SI{250}{\GeV}}$: ]
  The cross section of the Higgs\-strahlung process $\ee \to \PH\PZ$ is largest close to the $\PH\PZ$ production threshold. Therefore, this energy allows for recording a large dataset for determining the Higgs branching fractions as well as the absolute normalisation of Higgs couplings and the Higgs boson mass. The clean environment of \ee\ collisions is also an ideal situation in which to search for non-standard Higgs boson decays, including invisible ones. 

    This data set will also give most precise measurements of the $\PW$ boson non-linear interactions and of electroweak fermion pair production, providing further opportunities for discovery. The $\PW$ boson's mass can be measured 
    using the \SI{250}{GeV} data or in an optional one-year threshold scan. 
 Also, at this stage, one can collect a few $10^9$ events on the $\PZ$ pole with polarised beams, to test the SM electroweak sector with high precision. In this regard, the use of beam polarisation  compensates almost three orders of magnitude in integrated luminosity with respect to unpolarised colliders.
 
  \item[A second stage at $\mathbf{\SI{550}{\GeV}}$:] At this energy the $\PW\PW$-fusion reaction $\ee\to \PGn \PAGn \PH$ has become the dominant production mechanism for Higgs bosons. 
  This will provide an independent setting for the measurement of Higgs branching fractions and searches for exotic decays.  
  It also allows the measurement of the top-quark Yukawa coupling and the trilinear Higgs self-coupling through the processes $\ee\to \PQt\PAQt\PH$ and $\ee\to  \PZ\PH\PH$, respectively. 

    The top quark is the SM particle most strongly
    coupled to the Higgs boson and thus most likely  to be affected by
    new Higgs interactions.  This stage, well above the top-quark
    threshold, will also provide precision measurements of the top-quark electroweak couplings 
    and include a short run around the top pair production 
    threshold for a precise mass determination. 

    \item[A third stage of at least $\mathbf{\SI{1}{\TeV}}$:] Here the reaction of $\PW\PW$
      fusion to $\PH\PH$ is the dominant Higgs pair production
      mechanism.   This energy stage thus provides a second method to
      determine the trilinear Higgs self-coupling, and can put constraints on the quartic Higgs self-coupling~\cite{Stylianou:2023xit}. 
      The two Higgs pair production processes, $\PZ\PH\PH$ and $\PW\PW$ fusion, are strikingly complementary; as the self-coupling is increased from its SM value, the $\PZ\PH\PH$ cross section increases while the $\PW\PW$-fusion cross section decreases.  Observing these deviations in the same experiment would give the most persuasive evidence for a non-standard value of the Higgs self-coupling.
      
      Anomalies in top-quark or multi-gauge-boson couplings suggested in earlier stages will be
      dramatically larger, proportionally to $E_{CM}^2$ in typical cases.
      The ample production of (single) Higgs bosons in $\PW\PW$ fusion and $\PQt\PAQt\PH$ will allow us to further scrutinise the couplings of the Higgs boson to SM particles including the top quark and the $\PW$ boson.
    
    \end{description}

At any of these stages, but increasingly so at the higher energies, there is additional discovery potential for new particles. We also stress that in case a discovery -- at the HL-LHC, at any other current experiment, or at the first stage of a linear collider itself -- points to a specific energy scale, the run plan and upgrade schedule of a linear collider should be -- and can be -- adjusted.


\subsection{Scrutinizing the Higgs boson via single-Higgs production}
\label{sec:singleHiggs}

The precise measurement of Higgs boson properties is critical for probing effects of new physics in the electroweak sector of the SM. At a parametris operating at a centre-of-mass energy of \SI{250}{GeV}, precision studies of the Higgs boson will focus on the dominant Higgs\-strahlung process, $\ee \to \PZ\PH$. At this energy there is also some sensitivity to the  $\PW\PW$-fusion process, which however increases significantly at higher energy stages. The Higgs\-strahlung production mechanism allows for model-independent measurements of Higgs boson properties by reconstructing the Higgs mass from the recoil mass spectrum of the $\PZ$ boson, independent of the Higgs decay mode. The $\PW\PW$-fusion process contributes to the determination of the Higgs boson width.

For CLIC the first stage has a nominal centre-of-mass energy of \SI{380}{GeV}, aiming for a balanced sensitivity to both Higgs production modes. The physics projections, however, have been performed with Monte-Carlo samples generated at a centre-of-mass energy of \SI{350}{GeV}, which is thus used as reference energy in this section.

Above \SI{500}{GeV}, also di-Higgs production, both in double Higgs\-strahlung and $\PW\PW$ fusion, as well as $\ttbar\PH$ production enter the scene. They will be discussed in Secs.~\ref{sec:phys:highEHiggs} and~\ref{sec:Higgstop}, respectively.

\subsubsection{Impact of beam polarisation on Higgs production}
\label{sec:singleHiggs:pol}

As discussed in Sec.~\ref{sec:polarisation}, polarised beams are an important characteristic of a linear collider. Specifically for Higgs production, both opposite-sign helicity configurations contribute to Higgs\-strahlung, while for $\PW\PW$ fusion, due to the pure left-handed coupling of the \PW\ bosons, the ($-\,+$) configuration is by far the most relevant one.

In the following some polarisation enhancement factors (as defined in Eq.~\ref{eqn:pol_enhance} in Sec.~\ref{sec:polarisation}) for these processes are given at leading order. Radiative corrections will lead to changes of these enhancement factors at the level of about \SI{5}{\%} for processes mediated through the \PZ\ boson, mainly via changes to the value of $\sin^2\theta_W$, while other radiative corrections are not polarisation dependent. Thus the enhancement factors for the $\PW\PW$-fusion process are not affected. 

For $P(\Pem,\Pep)=(-1,1)$ the Higgs\-strahlung cross section would be enhanced by a factor \num{2.43} and for $P(\Pem,\Pep)=(1,-1)$ by a factor \num{1.57} compared to the unpolarised cross section. This difference in the enhancement factor is due to the different couplings for left- an right-handed fermions to the $\PZ$ boson, corresponding to the left-right asymmetry in Eq.~\ref{eqn:pol_enhance}. 

For $P(\Pem,\Pep)=(-0.8,0.3)$, the enhancement factor is still \num{1.48}, whereas $P(\Pem,\Pep)=(0.8,-0.3)$ leads to a cross section roughly equal to the unpolarised cross section.
For the like-sign polarisation combinations, the factors are \num{0.87} and \num{0.65}, thus suppressing the Higgs signal, thus serving readily as signal-depleted control samples with otherwise identical conditions.

If the positron beam is not polarised, the enhancement for $P(\Pem,\Pep)=(-0.8,0)$ is only \num{1.17} and for $P(\Pem,\Pep)=(0.8,0)$ the cross section is even reduced by a factor of \num{0.83} compared to the unpolarised cross section \cite{Moortgat-Pick:2005jsx}.

For $\PW\PW$ fusion only the LR initial state leads to a signal. Running with $P(\Pem,\Pep)=(-0.8,0.3)$, the effective fraction
of LR states increases from \SI{25}{\%} to \SI{58}{\%}, resulting in an effective enhancement 
of the cross section by a factor \num{2.34} compared to the unpolarised cross section. Without positron polarisation, this factor is only \num{1.8}.

In the ILC and LCF running scenarios (c.f.\, Tables~\ref{tab:ILC-runplan} and~\ref{tab:LCF-runplan}, the integrated luminosity at \SI{250}{GeV} is divided among the four polarisation combinations as (\SI{5}{\%}, \SI{45}{\%}, \SI{45}{\%}, \SI{5}{\%}). Running of the collider with the two polarisation combinations that enhance the cross section corresponds 
to \SI{90}{\%} of the total expected integrated luminosity. This leads to a sample with about \SI{20}{\%} more Higgs bosons from both Higgs\-strahlung and $\PW\PW$ fusion
than for a machine without beam polarisation delivering the same integrated luminosity. Similarly, for \SI{550}{GeV} and \SI{1}{TeV} running, the division of polarisation components is (\SI{10}{\%}, \SI{40}{\%}, \SI{40}{\%}, \SI{10}{\%}). If the goal were to  
maximize the number of Higgs bosons, running purely in $P(\Pem,\Pep)=(-0.8,0.3)$ is clearly favoured. However, the absolute number of Higgs bosons is not the only consideration.
The SM backgrounds to the Higgs measurements result from processes mediated through photon, \PZ boson and \PW boson exchanges.
The polarisation will therefore affect the background composition, since the photon exchange is agnostic to the helicities, the \PZ couplings are modified as in  single \PZ production, and the \PW boson pairs only contribute in the pure $\Pem\Pep$ LR initial
state, resulting in a significant reduction of this background
for the $+-$ polarisation choice. The different background compositions of the four data sets with different beam helicity combinations also provide an additional handle to control systematic uncertainties.
Another important consideration is the additional sensitivity to the EFT operators via the left-right asymmetries of the Higgs\-strahlung cross section,
which is discussed in Sec.~\ref{sec:glob}. 

The disfavoured helicity combinations provide 
further information for the determination of the background. These combinations provide also a significant 
sample of Higgs bosons, between \SI{65}{\%} and \SI{87}{\%} of the unpolarised cross section for the same integrated luminosity. 
Furthermore, the two \SI{5}{\%}-components of the integrated luminosity 
also provide data for a closure test, testing the compatibility of the measured polarisation and integrated
luminosity under the assumption of the SM couplings for the Higgs boson.

With the equal sharing of the integrated luminosity at \SI{380}{GeV}, CLIC will de facto produce the same number of Higgs bosons as an unpolarised machine for the same total integrated luminosity. The shift from Higgs\-strahlung to $\PW\PW$ fusion as the predominant production mode at the higher energy stages motivates the (\SI{80}{\%}, \SI{20}{\%}) sharing between the $P(\Pem,\Pep)=(-0.8,0)$ and $P(\Pem,\Pep)=(0.8,0)$ configurations. With only two polarisation configurations available, the redundancy for closure tests and control of systematic uncertainties is strongly reduced. Therefore we stress that also for CLIC, positron polarisation is a technically possible option~\cite{Zang:2010kya}.

\subsubsection{Higgs mass and total cross section}
\label{sec:Higgs_mass_xs}

The most precise and least model-dependent Higgs boson mass measurement can be performed with the recoil mass technique at energies not too high above the $\PZ\PH$ production threshold, exploiting the high precision measurements of the leptonic decays of the \PZ boson. Full detector simulation studies of the recoil method have been performed for ILC~\cite{Yan:2016xyx} and CLIC~\cite{Thomson:2015jda}. 

\begin{figure}
    \begin{subfigure}{0.6\textwidth}
        \includegraphics[width=\textwidth]{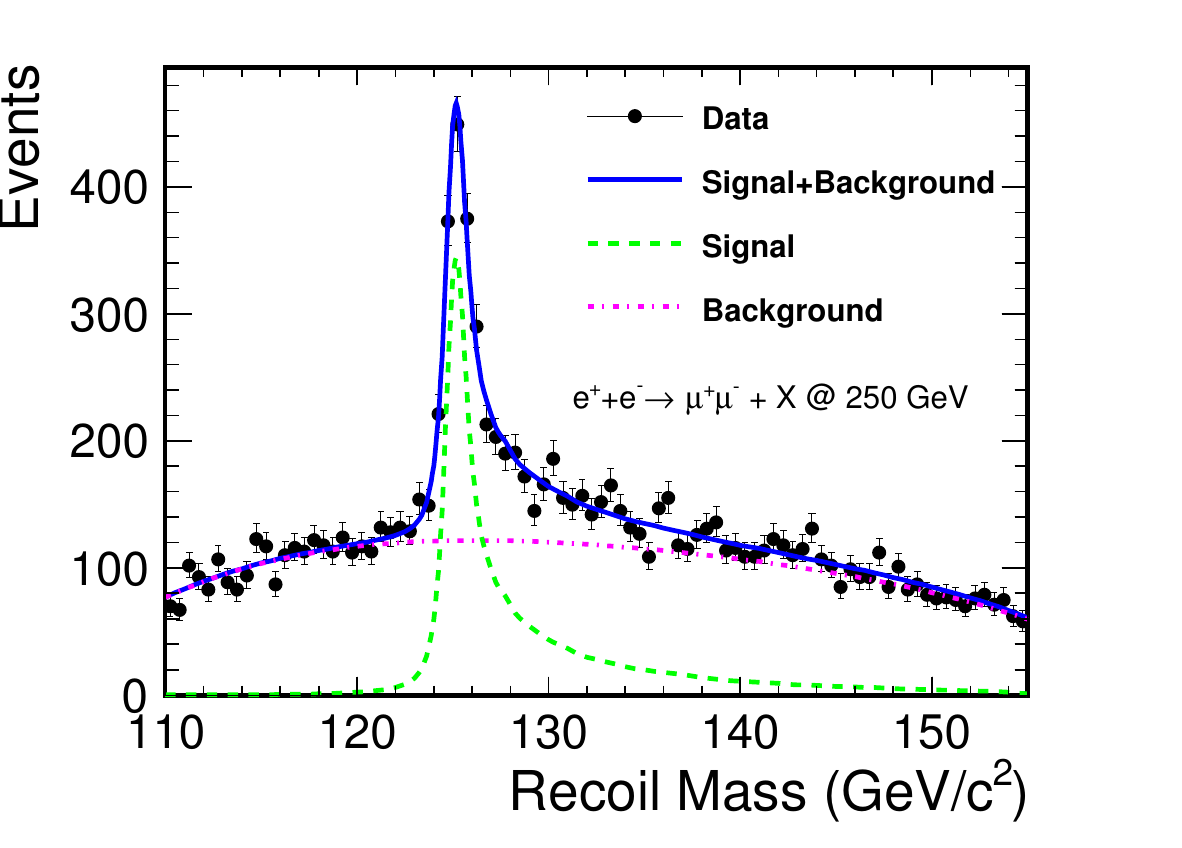} 
        \caption{}
        \label{fig:HiggsMassILC}
    \end{subfigure}
    \quad
    \begin{subfigure}{0.4\textwidth}
        \includegraphics[width=\textwidth]{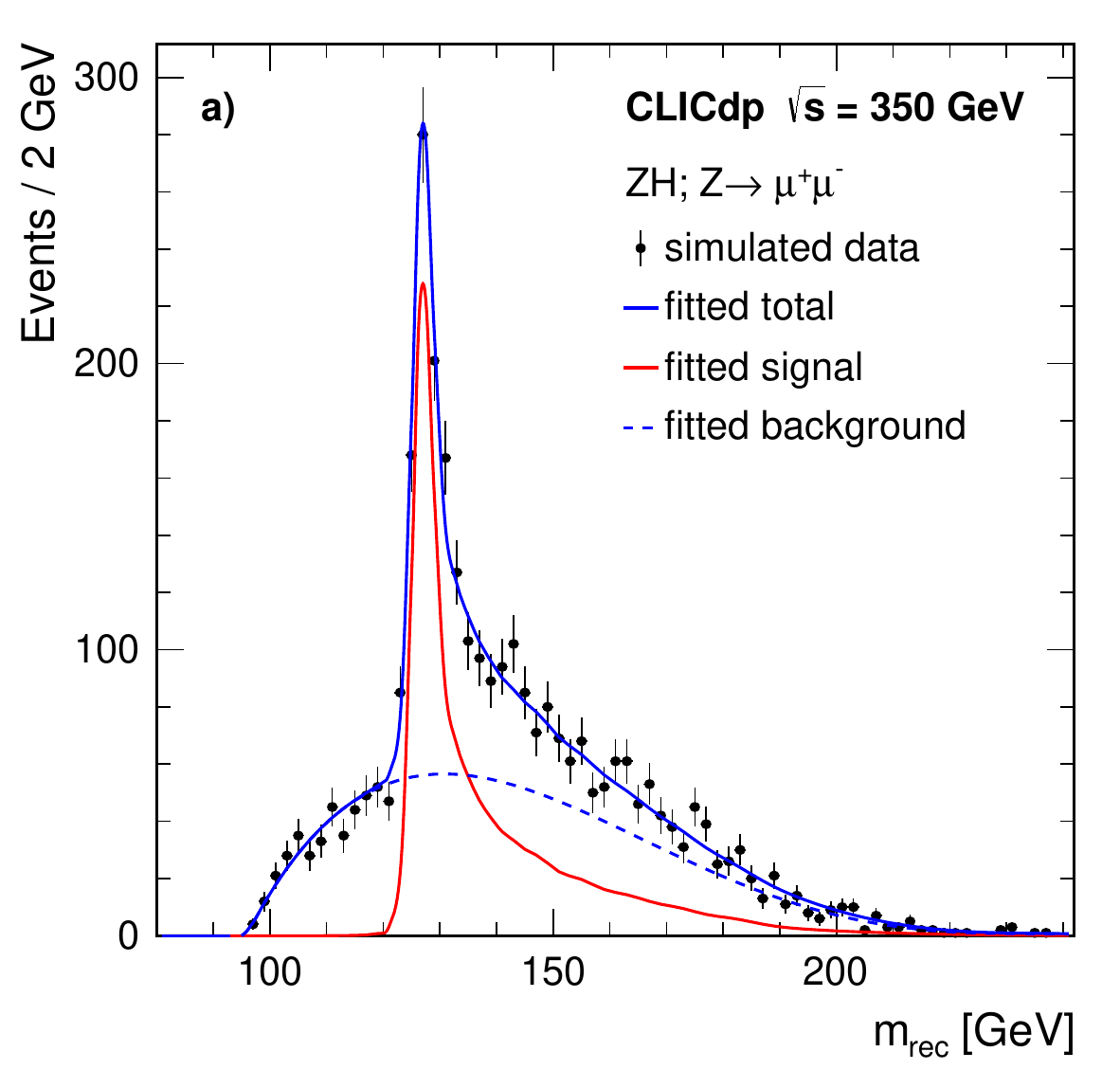}
        \caption{}
    \label{fig:HiggsMassCLIC}
    \end{subfigure}
    \caption{(a) Determination of the Higgs mass at the ILC at a centre-of-mass energy of \SI{250}{GeV}~\cite{Yan:2016xyx}. The statistics of the data points correspond to an integrated luminosity of \SI{250}{\fbinv} with a polarisation of (-0.8,+0.3). (b) Determination of the Higgs mass at CLIC at a centre-of-mass energy of \SI{350}{GeV}~\cite{Abramowicz:2016zbo}. The statistics of the data points correspond to an integrated luminosity of \SI{500}{\fbinv} with unpolarised beams.}
\end{figure}

\begin{table}[htb]
    \centering
    \begin{tabular}{c|ccccc}
    Accelerator & ILC/LCF    & CLIC     & CLIC     & CLIC    & CLIC \\
    \roots      & 250 GeV & 350 GeV  & 350 GeV  & 1.4 TeV & 3 TeV \\
    \lumi       & 2.7\abinv & 2.2\abinv & 4.3\abinv & 4\abinv & 5\abinv \\
    $\Delta m_{\PH}$ & [MeV] & [MeV]  & [MeV] & [MeV] & [MeV] \\
    \hline
    $\PH\PZ$ Recoil                     &  12 & 52 & 38 &    &   \\
    $\PGne\PAGne\PH\to\PGne\PAGne\bb$   &     &    &    & 29 & 28 \\ 
    \end{tabular}
    \caption{\label{tab:mHPrec} Summary of the expected precision of the Higgs mass measurements at various centre-of-mass energies from~\cite{Robson:2025clic} and scaled statistically from~\cite{Yan:2016xyx, deBlas:2022ofj}. If operation at \SI{250}{GeV} is available (like for ILC or LCF), the recoil measurement is more precise than any direct reconstruction at higher energies. }
\end{table}

The determination of the Higgs boson mass is shown in Fig.~\ref{fig:HiggsMassILC} for the ILC 
for the dominant helicity combination $P(\Pem,\Pep)=(-0.8,0.3)$ and in Fig.~\ref{fig:HiggsMassCLIC} 
for CLIC as discussed in~\cite{Yan:2016xyx,Abramowicz:2016zbo,Robson:2025clic}. Away from threshold, the sensitivity of the recoil measurement to the mass of the Higgs boson is significantly reduced by initial-state radiation and the worse precision of the muon momentum measurement, clearly seen in the more pronounced tail to higher recoil masses in Fig.~\ref{fig:HiggsMassCLIC} compared to Fig.~\ref{fig:HiggsMassILC}. Thus if no operation at \SI{250}{GeV} is considered, the direct measurement of the reconstructed invariant $\PQb\PAQb$ invariant mass becomes competitive, in particular when leveraging the $\PW\PW$-fusion 
process~\cite{Abramowicz:2016zbo} at higher energies. With CLIC running at \SI{350}{GeV}, the expected
precision is \SI{52}{MeV} with \SI{2.2}{\abinv} and even \SI{38}{MeV} with \SI{4.3}{\abinv}. This measurement can be further improved 
at \SI{1.4}{TeV} and \SI{3}{TeV} to about \SI{28}{MeV}~\cite{Robson:2025clic}.

Table~\ref{tab:mHPrec} compares the Higgs mass precisions expected from the major polarisation combinations of various ILC, LCF and CLIC runs based on the above-mentioned full detector simulation studies and~\cite{deBlas:2022ofj}.

Precise knowledge of the Higgs mass is particularly important for accurate predictions of branching fractions to $\PW\PW$ and $\PZ\PZ$, which depend strongly on the Higgs mass. The achievable precision at the ILC or the LCF is sufficient to reduce the parametric error on the
branching fractions to the permil level.

The inclusive cross section $\sigma_{\zhsm}$ is determined in a joint fit with the Higgs boson mass. Its value in the two major polarisation combinations has been determined in full
simulation with \SI{250}{\fbinv}. Translating this estimation into a precision on the measurement with
the full expected statistics, the expected error for one of the 
major polarisation combinations is \SI{0.87}{\%} as shown in Table~\ref{tab:HiggsBRPrec}, giving an error of \SI{0.6}{\%} for the combination
of the two major polarisation combinations. This means that an error on 
the absolute coupling measurement $g_{\PH\PZ\PZ}$ of \SI{0.3}{\%} is reachable, also see Sec.~\ref{sec:glob}.

For CLIC, operating at an energy of \SI{350}{GeV}, the expected precision on the cross section measurement has been discussed in~\cite{Abramowicz:2016zbo}. Translating the numbers to the new CLIC 
running scenarios with increased integrated luminosity, a precision of \SI{0.79}{\%} or even \SI{0.56}{\%} can be
reached as listed in Table~\ref{tab:HiggsBRPrec}, thus reaching the several permil level precision on the measurement of the absolute coupling value $g_{\PH\PZ\PZ}$.

\subsubsection{Branching fraction measurements}

\begin{table}[tb]
    \centering
    \begin{tabular}{l|cccccccc}
    Collider                                  & LCF                & LCF      & CLIC & CLIC  & LCF & LCF & CLIC & CLIC \\
    \roots                                    & 250 GeV            & 350 GeV  & 350 GeV & 350 GeV & 500 & 1 TeV & 1.4 TeV & 3 TeV \\
    \lumi                                     & 2.7\abinv          & 0.135\abinv & 2.2\abinv & 4.3\abinv & 6.4\abinv & 6.4\abinv & 4\abinv & 5\abinv \\
                                               & [\%]               & [\%]    & [\%] & [\%]  & [\%] & [\%] & [\%] \\
    \hline
    $\sigma_{\zhsm}$                          &    0.62    & 2.5 & 0.79 & 0.56 & 0.8 &  &2.0 & 4.3\\
    \hline
      $\sigma_{\zhsm}\cdot\br{\bb}$           &    0.41   & 2.1 & 0.41 & 0.29 & 0.5 & &&\\
      $\sigma_{\zhsm}\cdot\br{\cc}$           &    2.5    & 15  & 7    &  5   & 3.6  &&&\\
      $\sigma_{\zhsm}\cdot\br{\Pg\Pg}$        &    2.1    & 11.4 & 2.9  &  2.1 & 3.0 &&&\\
      \hline
      $\sigma_{\zhsm}\cdot\br{\PGt\PGt}$      &    0.98   & 5.5 & 3.0 &  2.1 & 1.2 &&&\\
      \hline
      $\sigma_{\zhsm}\cdot\br{\zz}$           &    5.5    & 34   &     &      & 6.9  &&&\\
      $\sigma_{\zhsm}\cdot\br{\PW\PW}$        &    1.4    & 7.6  & 2.4 &  1.7 & 1.6 &&&\\
      $\sigma_{\zhsm}\cdot\br{\PGg\PGg}$      &    10     &      &     &      & 10  &&&\\
      \hline
      $\sigma_{\PGne\PAGne\PH}\cdot\br{\bb}$          &  2.5   & 2.5 & 0.9 & 0.6 &  0.30& 0.30 & 0.2 & 0.2\\
      $\sigma_{\PGne\PAGne\PH}\cdot\br{\cc}$          &        & 26  & 12  &  9  &  2.5 & 1.6 & 3.7 & 4.4\\
      $\sigma_{\PGne\PAGne\PH}\cdot\br{\Pg\Pg}$       &        & 11  & 4.8 & 3.4 &  1.6 & 1.3 & 3.6 & 2.7 \\
      \hline
      $\sigma_{\PGne\PAGne\PH}\cdot\br{\PGt\PGt}$     &        & 22  &  & &  2.8 & 1.5 & 2.6 & 2.8 \\
      $\sigma_{\PGne\PAGne\PH}\cdot\br{\mu\mu}$       &        &  &  & &  28 & 16 & 23  & 16 \\
      \hline
      $\sigma_{\PGne\PAGne\PH}\cdot\br{\zz}$          &        & 27  &   & &  3.4 & 2.2 & 3.4  & 2.5 \\
      $\sigma_{\PGne\PAGne\PH}\cdot\br{\PW\PW}$       &        & 7.8 &   & &  0.96 & 0.88 & 0.6  & 0.4 \\
      $\sigma_{\PGne\PAGne\PH}\cdot\br{\PGg\PGg}$ &        &   &   & &  7.6  & 4.6 &  9   & 6\\
      \hline
      $\sigma_{\ee\PH}\cdot\br{\bb}$ &&&& & & &1.1 & 1.5\\
      \hline
      \hline
     $\Gamma(\PH \to$ inv), \SI{95}{\%} CL &   0.21    & 1.6 & 0.6 &  0.4  & 0.55 & &&  \\      
      \end{tabular}
    \caption{Expected measurement precision of $\sigma\cdot\text{BR}$ at different centre-of-mass energies, extrapolated from~\cite{deBlas:2022ofj} for some of the running scenarios discussed in Sec.~\ref{sec:RunScenarios}. For the LCF the precision is based on detailed ILC studies extrapolated to the LCF running scenario, not accounting for the \SI{10}{\%} increase in centre-of-mass energy from \num{500} to \SI{550}{GeV}.
     The precision for the two CLIC running scenario is based on~\cite{Abramowicz:2016zbo} extrapolated for the increased luminosity in~\cite{Robson:2025clic}. The precision on the total cross section is given for the subsample of events with hadronic \PZ boson decay. For CLIC not all of the exclusive final states have been studied. In case of LCF, only the data-sets  with opposite helicities are included. For simplicity, in this table, the two results from each accelerator configuration are combined to a precision on the unpolarised cross section. In practice, the two results are used separately, to provide the maximum information to the global fits described in Sec.~\ref{sec:glob}.}
    \label{tab:HiggsBRPrec}
\end{table}

The measurement of Higgs couplings at a linear collider is projected to reach a precision at the permil level, making it highly sensitive to deviations from the SM caused by new physics at the TeV scale. The total cross sections as well as cross section times branching fractions will be measured in all beam polarisation combinations, though current studies often only evaluate the two major, opposite-sign data-sets. 
The asymmetry between these measurements in different polarisations offers an important additional input to the global understanding of Higgs couplings, which we be discussed in Sec.~\ref{sec:glob}.

The expected precisions for the inclusive cross section measurements
as well as the exclusive final states are listed in Table~\ref{tab:HiggsBRPrec}, including only measurements with an expected relative precision better
than \SI{25}{\%} are listed. The numbers are based on original work and extrapolated to the running scenarios~\cite{deBlas:2022ofj,Abramowicz:2016zbo,Robson:2025clic}.
For the LCF running scenarios, the measurement precisions listed for ILC in~\cite{deBlas:2022ofj} have
been scaled to the LCF running scenario presented in Sec.~\ref{sec:RunScenarios}.

The ILC results are based on full simulation studies for the dominant polarisation mode $P(\Pem,\Pep)=(-0.8,0.3)$. Several final states and their backgrounds 
have also been studied for $P(\Pem,\Pep)=(0.8,-0.3)$ achieving comparable precision within \SI{20}{\%}. The lower cross sections for $+-$ beams are approximately compensated by lower levels of background processes.  Therefore similar precision is expected for the
Higgs\-strahlung process for the major polarisation combinations, independent of the final state. 
Dominated by the statistical error, the combined precision was obtained as scaling with
the statistics increase from~\cite{deBlas:2022ofj} and combining the two major polarisation
combinations, de facto a three-fold increase in statistics for \SI{250}{GeV}. According to the running scenarios in Sec.~\ref{sec:RunScenarios}, at the data  at \SI{350}{GeV}  will be taken mainly in $P(\Pem,\Pep)=(-0.8,0.3)$ mode, therefore the expected
precision is given for this polarisation combination as in~\cite{deBlas:2022ofj}. For Higgs\-strahlung, the combined 
error for the two major polarisation combinations at this energy is improved by about \SI{14}{\%}. It is basically unchanged
for $\PW\PW$-fusion processes. For  \SI{550}{GeV}, the analyses 
performed at \SI{500}{GeV} are listed in the table. For this energy  and for \SI{1}{TeV} running, the Higgs\-strahlung processes are enhanced in  \SI{80}{\%} of the total integrated luminosity, and so this is the luminosity quoted in the table. For the $\PW\PW$-fusion processes, the dominant $-+$ mode comprises \SI{40}{\%} of the total integrated luminosity. The results for CLIC are quoted for unpolarised cross sections for all energies.  The individual contributions of the two polarisation running modes can be calculated using the cross section scaling factors and the fraction of the integrated luminosity. The optimized running scenario for CLIC in Sec.~\ref{sec:RunScenarios} differs in energy by about \SI{10}{\%} with respect to the energy at which the analyses were performed, therefore the analysis energy is quoted in Sec.~\ref{tab:HiggsBRPrec}.
 
The cross section times branching fraction for key decay modes such as
the gauge bosons $\PH\to\Pg\Pg$ and  $\PH\to\PW\PW$ will be measured with high precision. 
Due to the smallness of its branching fraction, the mode $\PH\to\PZ\PZ$ is less precisely measured.

Decays to fermions, $\PH\to\PGtm\PGtp$, $\PH\to\bb$, $\PH\to\cc$ will be measured at the
(sub)percent level. The study of hadronic decays, particularly $\PH\to\ssbar$, presents a significant experimental challenge due to small branching fractions and large multi-jet backgrounds. Nonetheless, novel experimental techniques, including  $K/\pi$ particle identification and $dE/dx$ measurements, are expected to enable the observation of this process and corresponding analyses are underway.

Rare decays of the Higgs boson, such as $\PH\to\PGmm\PGmp$ and loop induced processes such as $\PH\to\PGg\PGg$ and $\PH\to\PZ\PGg$ , will be explored 
as well, though less precisely than at the HL-LHC as their small branching fractions limit the achievable precision.  
The $\PH\PGg\PZ$ coupling can also be probed via the $\epem\to\PGg\PH$ process, whose cross section is also maximal around \SI{250}{GeV}. Upper limits at \SI{95}{\%} CL on the production cross sections in the different polarisation scenarios can be set at \SI{1.8}{fb} for the same beam polarisation~\cite{Aoki:2021khh}.  The polarisation-dependence of this limit and the implications for limits on SMEFT parameters are discussed in~\cite{Aoki:2022dxg}.

On the partial width of the Higgs decaying invisibly,  a \SI{95}{\%} CL limit of  several permil is expected. The achievable sensitivity is dominated by the hadronic decay mode of the \PZ boson, with a performance intimately depending on the quality of the jet reconstruction. Thus, full detector simulation and reconstruction was necessary to make these projections reliable.

To extract the total width of the Higgs boson ($\Gamma_{\PH}$), the inclusive cross section measurement, electroweak decays to $\PW\PW$ and $\PZ\PZ$ and the decay to \bb will play a pivotal role in constraining the total Higgs decay width. 
Using the inclusive cross section (at tree-level proportional to $g_{\PH\PZ\PZ}^2$) and the exclusive Higgs\-strahlung final state $\sigma_{\PH\PZ}\cdot\br{\zz}$ (proportional to $g_{\PH\PZ\PZ}^4/\Gamma_{\PH}$) allows to extract the total width with a precision dominated by the exclusive decay channel which suffers from a low branching fraction. At \SI{250}{GeV} this will result in a precision of \SI{5}{-}\SI{6}{\%}. 

A more precise estimate can be achieved by putting together four measurements: the Higgs\-strahlung exclusive final states $\PW\PW$ (proportional to $g_{\PH\PZ\PZ}^2g_{\PH\PW\PW}^2/\Gamma_{\PH}$), \bb (proportional to $g_{\PH\PZ\PZ}^2g_{\PH\bb}^2/\Gamma_{\PH}$), the inclusive cross section and the $\PW\PW$-fusion
channel to \bb. The fusion channel divided by the product of the ratios of Higgs\-strahlung
channel to inclusive cross section is the total width. With the precisions given in Table~\ref{tab:HiggsBRPrec} the measurement precision is dominated by the $\PW$-fusion channel, improving
the precision significantly.

The results summarized in Table~\ref{tab:HiggsBRPrec} are based on the standard flavour tagging algorithms. 
The performance of the LCFIPlus package has been improved significantly using modern machine learning techniques as described in Sec.~\ref{sec:hlreco}.
A task for the future is to evaluate the impact of new working points which promise improved rejection for the same efficiency
or improved tagging efficiency for the same background rejection. Additionally the introduction of particle
identification may open the measurement of the $\PH\to\ssbar$ final state, albeit with large uncertainties.

\subsubsection{Interpretation}

Once deviations of the Higgs boson couplings from the SM predictions are discovered in future experiments, we can perform ``fingerprinting'' of the Higgs sector and can extract the mass scale of new particles such as a second Higgs boson mass. Such analyses can generally be done by using the SMEFT approach because non-zero deviations come from the coefficients of higher-order operators, and the new mass scale typically corresponds to the cut-off parameter $\Lambda$ which appears in the denominator of such an operator. We will employ this approach in Sec.~\ref{sec:glob}. 

Although the SMEFT approach captures the general patterns of the deviation, we here discuss some concrete examples of renormalizable models giving rise to the coupling deviations, especially focusing on the models with extended Higgs sectors.  
\begin{itemize}
\item In models with iso-singlet scalars, the coupling deviation only comes from the Higgs boson mixing $\alpha$. In this case, all the Higgs boson couplings are universally suppressed by the factor of $\cos\alpha$, i.e., $\kappa_V^{} = \kappa_f^{} = \cos\alpha$.  
\item In the 2 Higgs doublet models (2HDMs), the Yukawa couplings ($\PH\mathrm{f}\bar{\mathrm{f}}$) and the gauge couplings $\PH\mathrm{VV}$ ($\mathrm{V}=\PWpm,\PZ$) can differently be modified from the SM predictions. It has been shown that independent of the Higgs decays, the NLO EW corrections enable us to distinguish 2HDMs from the SM by just looking at the Higgs production, inclusively and differentially~\cite{Bredt:2025zxc}.
The pattern of the deviations of $\PH\mathrm{f}\bar{\mathrm{f}}$ couplings depends on the scenarios of the 2HDM. In the most general case, i.e., without imposing any additional symmetries, Yukawa couplings for $\PH$ depend on an arbitrary $3\times 3$ matrix in flavour space. Therefore, any values of $\PH\mathrm{f}\bar{\mathrm{f}}'$ couplings (even for flavour-violating cases $\mathrm{f}\neq \mathrm{f}'$) can be taken in principle. 
This general scenario, however, introduces dangerous flavour-changing neutral currents (FCNCs) at tree level, so that some assumptions to avoid such FCNCs are often imposed on the 2HDM. 
For instance, the Yukawa-alignment scenario has been proposed in~\cite{Pich:2009sp}, in which two Yukawa matrices are assumed to be proportional with each other, and then the flavour-violating Yukawa couplings for neutral Higgs bosons are forbidden. In this case, $\kappa_\mathrm{f}
$ take flavour universal values, and are expressed as $\kappa_\mathrm{f}^{}= \kappa_V^{} + \zeta_\mathrm{f}\sqrt{1 - \kappa_V^2}$ with $\zeta_\mathrm{f}$ being  constants (generally complex and $\zeta_u\neq \zeta_d \neq \zeta_e$).  
The 2HDMs with a softly-broken $Z_2$ symmetry correspond to the special case of the Yukawa-alignment scenario, where three $\zeta_\mathrm{f}$ parameters are determined either $\cot\beta$ or $-\tan\beta$ depending on the four types of the Yukawa interactions (Type-I, Type-II, Type-X and Type-Y)~\cite{Barger:1989fj,Grossman:1994jb,Aoki:2009ha}.  

\item In models with higher isospin representation fields such as $SU(2)_L$ triplet Higgs field, we expect the situation that cannot be realized in the previous two examples. Namely, the $\PH\PW\PW$ and $\PH\PZ\PZ$ couplings can differently be modified from the SM predictions, and also $\kappa_{\PW}$ and $\kappa_{\PZ}$ can exceed unity. 
The size of deviations in $\PH\PV\PV$ couplings are often constrained by the electroweak $\rho$ parameter because the deviation depends on vacuum expectation values of such exotic Higgs fields. However, one can consider models with $\rho = 1$ at tree level by imposing the custodial symmetry in the Higgs sector. The simplest example has been known as the Georgi-Machacek model~\cite{Georgi:1985nv,Chanowitz:1985ug}, where a real and a complex triplet Higgs fields are added to the SM.  

\end{itemize}


\subsubsection{CP properties}
The search for CP violation is an important research direction of future experiments in particle physics, as CP violation is required for baryogenesis and cannot be sufficiently explained with present knowledge. 
CP violation can be tested in interactions of the Higgs boson with either fermions or bosons, see e.g.~\cite{Gritsan:2022php} for a recent overview.
At \ee\ linear colliders, Higgs CP properties can be tested in $\PH \to \PGt\PGt$ decays~\cite{Jeans:2018anq}, in $\PZ\PZ$-fusion production~\cite{Vukasinovic:2023ewd} as well as in $\PQt\PAQt\PH$ associated production~\cite{Godbole:2011hw}. 

The amount of CP violation is characterized by the quantity
\begin{equation}
f_{\mathrm{CP}}^{\PH \PX}\equiv
\frac{\Gamma^{\mathrm{CP\,odd}}_{\PH\to \PX}
}{\Gamma^{\mathrm{CP\,odd}}_{\PH\to \PX}
+\Gamma^{\mathrm{CP\,even}}_{\PH\to \PX}}
\,.
\label{eq:fCP} 
\end{equation}

Table~\ref{tab:cpscenarios} lists estimates of the sensitivity on $f_{\mathrm{CP}}^{\PH \PX}$ in unpolarised \ee\ collisions at various centre-of-mass energies and for various processes. It should be noted that for the case of $\PH\PQt\PAQt$, a significant improvement is expected from the increase of the centre-of-mass energy from \num{500} to \SI{550}{GeV}, as well as from the higher luminosity at this energy in the LCF running scenario.

\begin{table}[ht]
\renewcommand{\arraystretch}{1.5}
\begin{center}
\begin{tabular}{|l|cccc|c|}
Collider   &  \epem    &   \epem    &  \epem    &  \epem    & target \\
ECM [GeV]   &  250          &  350  & 500    &   1000   &   (theory) \\
\lumi [\abinv] & 2.5          &  3.5                  & 5              &    10     &     \\ 
\hline
$\PH\zz/\PW\PW$ &  $3.9\!\cdot\!10^{-5}$ & $2.9\! \cdot\!10^{-5}$ & $1.3\!\cdot \!10^{-5}$ & $3.0\!\cdot\!10^{-6}$   & $<10^{-5}$  \\
 $\PH\PQt\PAQt$ & -- & -- & 0.29 & 0.08  & $<10^{-2}$  \\
 $\PH\PGt\PGt$ &   $0.01$ & $0.01$  &  ${0.02}$  & ${0.06}$  &  $<10^{-2}$  \\

\end{tabular}
\end{center}
\caption{
List of expected precision (at \SI{68}{\%} CL) of CP-sensitive measurements of the parameters $f_{\mathrm{CP}}^{\PH \PX}$ defined in Eq.~(\ref{eq:fCP}).
The $\epem\to\PZ\PH$ projections are performed with $\PZ\to\ell\ell$  
but scaled to a ten times larger luminosity to account for $\PZ\to\PQq\PAQq$. Table adapted from~\cite{gritsan2022snowmasswhitepaperprospects,Li:2025ouv}. 
\label{tab:cpscenarios}
}
\end{table}

Polarisation of the colliding beams may give additional sensitivity in these
measurements. Using the major polarisation mode $P(\Pem,\Pep)=(0.8,-0.3)$ the 
background cross section is decreased while the signal cross section is increased.
However negative interference compensates partially this improvement.
In total up to \SI{10}{\%} improvement in the expected constraints on the $f^{\PH\PZ\PZ}_{\mathrm{CP}}$ parameter \cite{gritsan2022snowmasswhitepaperprospects,Li:2025ouv} are expected from polarisation.


\subsubsection{The Higgs boson and the running of quark masses}

The Higgs and $\PZ$-pole data collected at an \ee\ Higgs/top/EW factory offer an experimental window on the scale evolution of quark masses. Just like the strong coupling constant $\alpha_{\mathrm{s}}$, quark masses are renormalized, scheme-dependent quantities. In the $\overline{MS}$ scheme the value of the quark mass depends on the renormalization scale $\mu$, that is identified with the energy scale of the process at hand. Even if quark mass values themselves are not predicted by the SM, the Renormalization Group Equations (RGE) give a precise prediction for the scale evolution, or ``running'', of quark masses. Calculations for the the running quark masses are available at five loops~\cite{Vermaseren:1997fq,Chetyrkin:1997dh,Baikov:2014qja}. 

A large number of measurements over a broad range of energies characterize the evolution of the strong coupling $\alphas(\mu)$~\cite{ParticleDataGroup:2022pth}. Experimental tests of the running of quark masses have been performed for the charm~\cite{Gizhko:2017fiu} and top~\cite{Defranchis:2022nqb} quarks. In this Section, the focus is on the potential of the linear collider facility at CERN for the study of the scale evolution of the bottom quark mass.

\begin{figure}[htbp!]
\begin{center}
\includegraphics[width=0.6\columnwidth]{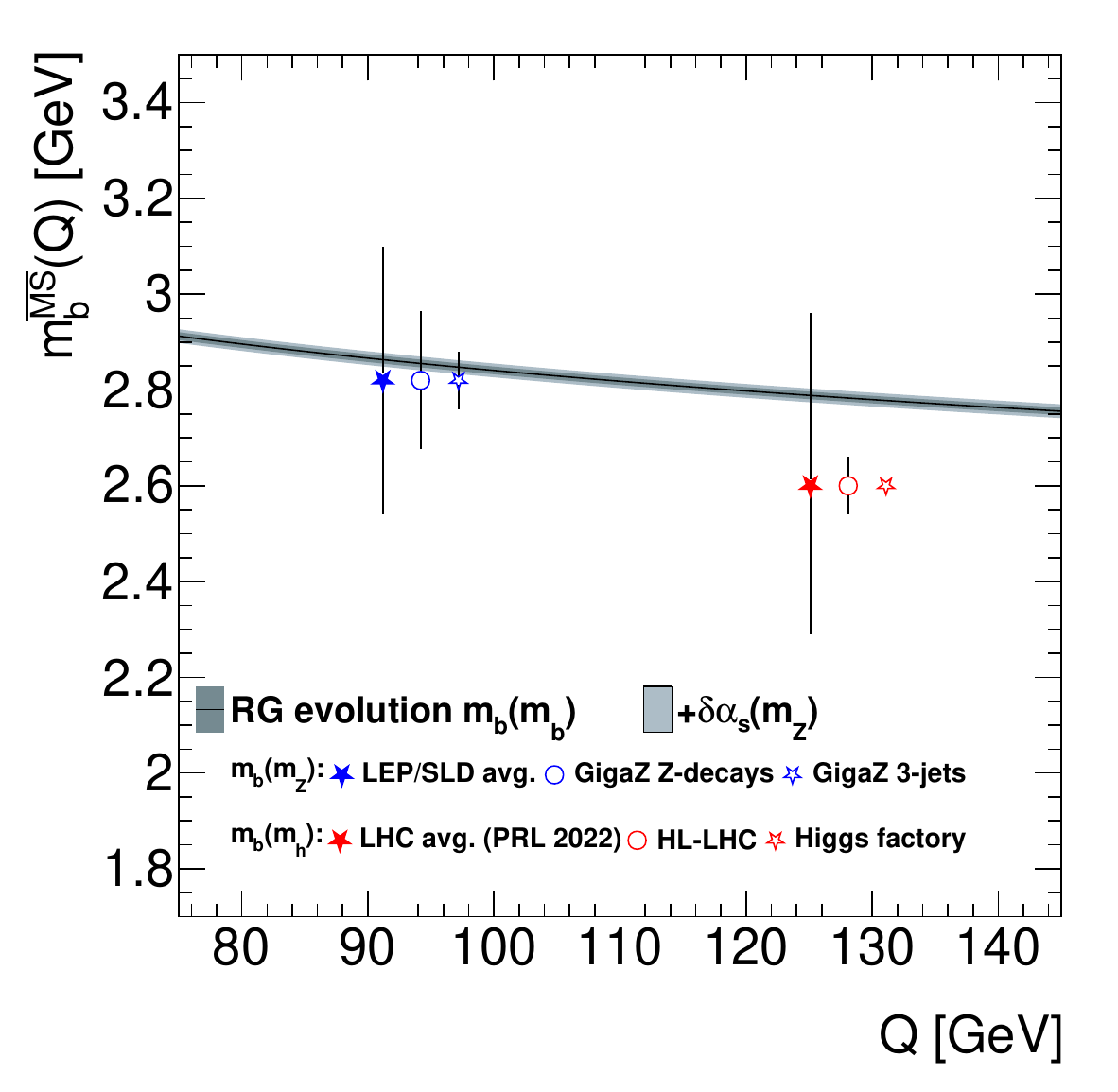}
\end{center}
\caption{\label{fig:running_mass_projection} The scale evolution of the bottom quark $\overline{MS}$ mass. The markers are projections for $m_{\PQb}(m_{\PZ})$ from three-jet rates at the \PZ pole and for $m_{\PQb}(m_{\PH})$ from Higgs boson branching fractions. The prediction of the evolution of the mass is calculated at five-loop precision with REvolver~\cite{Hoang:2021fhn}. The grey error band includes the effect of missing higher orders and the projected parametric uncertainties from $m_{\PQb}(m_{\PQb})$ and $\alpha_{\mathrm{s}}$. 
Figure from~\cite{Snowmass:bprospects}.
}
\end{figure}

The most promising avenue to measuring the bottom quark mass at high scale is based on Higgs decay rate measurements~\cite{Aparisi:2021tym}. The Higgs decay width to bottom quarks has a very strong dependence on the bottom quark mass and can be measured very precisely. Importantly, the strong coupling plays a minor role in this electroweak process, and the scale of the process is clearly identified with the Higgs boson mass. These properties make Higgs decays the ideal laboratory to determine $m_{\PQb}(m_{\PH})$, the bottom quark mass at the scale of the Higgs boson mass.    Figure~\ref{fig:running_mass_projection} compares the precision that can be achieved on $m_{\PQb}(m_{\PH})$: from \SI{14}{\%} using LHC Run-2 measurements to \SI{2}{\%} after the complete high-luminosity program, to sub-percent precision using the Higgs factory stage of the linear collider facility.

Figure~\ref{fig:running_mass_projection} also shows projections for $m_{\PQb}(m_{\PZ})$ from Z-boson decay rates to bottom quarks. Also these can be improved well beyond the precision of the current LEP and SLD results~\cite{Kluth:2022ucw}.
A measurement of three-jet rates in a high-luminosity ``GigaZ/TeraZ'' $\PZ$-pole run can improve the measurement of $m_{\PQb}(m_{\PZ})$, reaching a factor of two~\cite{ILDnote2020} with respect to today's precision. The determination of $m_{\PQb}(m_\mathrm{Z})$ from the partial width of the $\PZ\,\to \PQb\PAQb$ decay, as proposed in~\cite{Kluth:2022ucw}, is currently not competitive, but could reach an interesting precision (\SI{5}{\%}) with a new $\PZ$-pole run. 

Combined with the high-precision bottom quark mass measurements at low scale (that dominate the world average~\cite{ParticleDataGroup:2022pth} for $m_{\PQb}(m_{\PQb})$) the high-scale measurements constitute a stringent test of the scale evolution of quark masses. The scale evolution is potentially affected by unknown massive states that carry colour charge~\cite{Llorente:2018wup, Jezabek:1992sq}. 
A joint analysis of the scale evolution of the strong coupling and the quark masses provides a powerful and model-independent handle on new coloured states in the mass range between $m_{\PQb}$ and $m_{\PH}$.

\subsection{Two and four fermion production -- Z pole and above}
\label{sec:2f4f}

The programme described here serves several purposes. Di-fermion production at the $\PZ$ pole allows for the measurement of electroweak precision observables to at least an order of magnitude better than this was possible at LEP/SLC (and than this is possible at the LHC). These updated electroweak precision observables are the backbone of the measurements at higher energies. Di-fermion production might also see the onset of new physics and the interpretation of the data must not be compromised by an insufficient knowledge of electroweak precision observables. $\PW$\,pair production leads to four-fermion final states. These are thus at the heart of the determination of the $\PW$\,mass and of flavour physics by the determination of CKM matrix elements. Multi-boson production in general, including triple and quartic gauge couplings, will be discussed in Sec.~\ref{sec:ewhigh}.

\subsubsection{Z pole}
\label{sec:Zpole}

A linear $\ee$ collider designed for Higgs boson production can be run at the $\PZ$ resonance. The programme for this mode of operation, and the expected precision measurements of SM parameters, have been described in some detail in Sec.~9.1 of the ILC Snowmass report~\cite{ILCInternationalDevelopmentTeam:2022izu}. That discussion envisioned a programme of about 2 years to collect 
an event sample of $5\times 10^9$ $\PZ$-boson decays. In the LCF scenario, this would be a 1-year program, assuming that in such a short run only about \SI{50}{\%} of the peak luminosity possible in a $\PZ$ pole optimized machine configuration (c.f.\ Fig.~\ref{fig:lep:lumi} and  Sec.~\ref{sec:acc:lumiup:zpole}) would be reached. For comparison, the number of $\PZ$ bosons assumed in the FCC-ee programme correspond to 3 years of operation at full peak luminosity~\cite{deBlas:2025gyz}. 

However, even the short $\PZ$-pole run currently foreseen in the LCF programme will bring typical precisions on SM observables to the $10^{-4}$ level.  Also, since the most important electroweak observables at the $\PZ$ are chiral and a linear collider will provide highly polarised beams, these observables provide highly effective tests of the SM electroweak theory.  Precision theory for the $\PZ$-pole observables is presented in the review~\cite{Freitas:2019bre} and 
in \cite{Blumlein:2020jrf, Ablinger:2020qvo, Blumlein:2021jdl}.

\begin{table}[t]
\begin{center}
  \begin{tabular}{cc|c|cc|cc}
    \hline
    Quantity & Value &  current &  \multicolumn{2}{c}{Z pole}     & 
  \multicolumn{2}{c}{LCF250}    \\ 
 &   &    $\delta[10^{-4}]$&
                                                 $\delta_{stat}[10^{-4}]$&
               $\delta_{sys}[10^{-4}]$ & $\delta_{stat}[10^{-4}]$ & $\delta_{sys}[10^{-4}]$ \\ 
   \hline
boson properties \\ \hline
   $m_{\PW}${[}GeV{]}           & 80.379  & 1.5  &  -                       &  -    & 0.08 &  0.2  \\
   $m_{\PW}${[}GeV{]}$^\dagger$ &         &      &  -                       &  -    &      &  0.09 \\
   $m_{\PZ}${[}GeV{]}           & 91.1876 & 0.23 & 0.006 (0.002)$^\ddagger$ & 0.022 & 0.08 &  -    \\
  $\Upgamma_{\PZ}${[}GeV{]}     &  2.4952 & 9.4  & 0.5 (0.07)$^\ddagger$    &  -    & 6    &  -    \\
\hline
$\PZ$-$\Pe$ couplings \\  \hline
    $1/\brar{\Pe}$    &   0.0482   &  24    & 2                    &  5   &  5.5  & 10 \\
    $\asym{\Pe}$      &   0.1513   & 139    & 1.5 (0.8)$^\ddagger$ &  1.2 & 12    &  9 \\ \hline    
$\PZ$-$\ell$ couplings \\  \hline 
   $1/\brar{\PGm}$  & 0.0482 & 16    & 0.9 &  0.9               & 5.5 & 10 \\
   $1/\brar{\PGt}$  & 0.0482 & 22    & 0.9 &  0.9               & 5.7 & 10 \\
    $\asym{\PGm}$   & 0.1515 & 991   & 7   &  2.7 (1.3)$^\ast$  & 54  &  3 \\
   $\asym{\PGt}$    & 0.1515 & 271   & 7   &  2.7 (1.3)$^\ast$  & 57  &  3 \\  \hline
$\PZ$-$\PQb$ couplings  \\ \hline
   $\brar{\PQb}$  & 0.2163 &  31  &  0.7 & 0.6              & 3.5  & 10 \\ 
   $\asym{\PQb}$  & 0.935  & 214  &  0.9 & 2.7 (1.3)$^\ast$ & 5.7  &  3 \\ \hline
$\PZ$-$\PQc$ couplings \\ \hline  
  $\brar{\PQc}$ & 0.1721 & 174  & 1.4 &  2.5               &   5.8 & 50  \\
  $\asym{\PQc}$ & 0.668  & 404  & 4   &  2.7 (1.3)$^\ast$  &  21   &  3  \\ \hline
$\PZ$-$\PQs$ couplings \\ \hline  
   $\brar{\PQs}$ & $\approx R_b$ & N.A.  & 1.6  &  0.3              &   -  & -   \\
   $\asym{\PQs}$ & 0.895         & 1011  & 6    &  2.7 (1.3)$^\ast$ &  - &  -  \\ \hline
  \end{tabular}
\end{center}
\caption{Projected precision of electroweak quantities
  expected from the LCF.  Precisions are given as {\em relative}
  errors ($\delta A = \Delta A/A$) in units of $10^{-4}$. The column labelled ``$\PZ$ pole'' refers to 
  the dedicated $\PZ$-pole run; the column labelled ``LCF250'' refers to values that 
  can be obtained from radiative return events at \SI{250}{GeV}. The $\PW$ mass is also determined in runs at 250\,GeV. The second line for the $\PW$ mass, marked by the $^\dagger$ symbol, excludes hadronisation and radiative corrections from the systematic error calculation.
  By default \num{5e9} $\PZ$ bosons collected in \SI{100}{\fbinv} are assumed. Numbers in parentheses are optimistic estimations supposing either \SI{800}{\fbinv} in case of $m_{\PZ}$, $\Upgamma_{\PZ}$ and $\asym{\Pe}$ (all marked by $^\ddagger$) or 60\% positron polarisation (marked by $^\ast$), respectively.
}

   \label{tab:PEWresults}
 \end{table}

The precisions expected from this programme in electroweak observables at the $\PZ$ are given in Table~\ref{tab:PEWresults}. For some observables, the impact of a larger data-set or higher positron polarisation is indicated in addition.
The table also gives the update of precision electroweak measurements that would be expected from the analysis of radiative return events at \SI{250}{GeV}.  This already offers a sizeable improvement from the current situation.  According to Table~12.3 of~\cite{ILCInternationalDevelopmentTeam:2022izu}, this is already quite sufficient input to the analysis of Higgs boson couplings to reach the precisions on Higgs properties discussed in Section~\ref{sec:singleHiggs}.  This section will deal with the improvement of precision electroweak measurements for its own sake.  In the following, we will quote all uncertainties in terms of relative errors $\delta A = \Delta A/A$, as is done in Table~\ref{tab:PEWresults}.

The strongest test of the SM available from precision electroweak interactions is the prediction of $\sin^2\theta_w$ from the SM input parameters $\alpha(m_{\PZ}^2)$, $m_{\PZ}$, and $G_F$.   The dependence on these SM inputs is
\beq
\delta \sin^2\theta_w =   1.4 \,\delta\alpha^{-1} \oplus 1.4\, \delta G_F  \oplus 2.9\, \delta m_{\PZ} \ . 
\eeq{RHSuncertainty}
Using the uncertainty in $m_{\PZ}$ from the table, the last two terms are negligible.   We
expect that $\alpha^{-1}$ will be known sufficiently well from lattice QCD that the uncertainty in the direct measurement of $\sin^2\theta_w$ will be the dominant source of uncertainty.    From the uncertainty  in $A_e$ listed in the table, we find
\beq
\delta \sin^2\theta_w = 1.6 \times 10^{-5},     
\eeqn
which is 
comparable with the strongest precision electroweak tests available 
in a TeraZ programme with unpolarised beams, for example, the error is identical to that obtained from the $A_e$ determination using $\tau$ polarisation~\cite{ESPPU-ID217}.  

Polarisation asymmetries are an excellent way to measure precision electroweak quantities, for two reasons.  First, the asymmetries are typically larger, allowing measurements of the same absolute precision to reach lower relative uncertainties.  For example, the asymmetry $A_\ell$ is about \SI{15}{\%}, while at unpolarised colliders $A_\ell$ is extracted from the $A_{FB,b}$ and $A_{FB,\mu}$, with values of \SI{11}{\%} and \SI{2}{\%}, respectively.  Second, polarisation asymmetries are measured through very robust observables. For example, $A_e$ is measured as the change in the total hadronic and leptonic (excluding Bhabha) cross sections as the beam polarisation is flipped.  In contrast, $A_{FB,b}$ has substantial uncertainties from the modelling of the hadronic final state.  Some of the issues are presented in the discussion of $R_b$ below. $A_{FB,\mu}$ has substantial uncertainties due to non-resonant contributions to the $\ee\to \mu^+\mu^-$ cross section with their own forward-backward asymmetries.  This is sometimes described as the uncertainty of converting measured asymmetries into ``ideal pseudo-observables'' whose asymmetries are calculated by theorists.  In both cases, these uncertainties are often labelled as ``theory uncertainties''.  Given the achievements of precision QCD calculation for LHC, we anticipate that, in era of Higgs factories, theory predictions from electroweak diagrams will be computed at the N$^3$LO level.  But there will still remain uncertainties will arise from nonperturbative effects in hadronization and event modelling, whose understanding has improved very little in the LHC era.  For polarisation asymmetries, the dominant uncertainty comes from the measurement of the beam polarisation.  This has recently advanced very much with new strategies and the use of both electron and positron polarisation; see also Sec.~\ref{sec:LEPmeas} and the review in Section 5.4 of~\cite{ILCInternationalDevelopmentTeam:2022izu}.

Beyond $A_e$, Table~\ref{tab:PEWresults} lists many more precision electroweak tests. In the following we summarise the reasoning behind the numbers given in the table. 

\paragraph[Z mass and width]{Z mass and width}
The main issue for the measurement of $m_{\PZ}$ is the centre-of-mass energy scale, which can be controlled statistically to much better than the statistical 
precision on the observable itself using a momentum-based 
estimator in di-lepton events at the $\PZ$ pole. The primary systematic uncertainty is consequently driven by the uncertainty on the tracker momentum-scale; this should be controlled to better 
than \SI{10}{ppm} given the low material budget of the envisaged trackers and excellent momentum resolution~\cite{Wilson:2021}, with an ultimate target of \SI{2}{ppm} that matches the uncertainties from knowledge of the J/$\psi$ meson mass. This limit could potentially be pushed further using techniques relying on the much better knowledge (sub-ppm) of the pion 
and proton masses as discussed in ~\cite{Rodriguez:2020qhf, Wilson:2021}. 
For $\Gamma_{\PZ}$, the main systematic uncertainty is 
from point-to-point variation in the centre-of-mass energy scale which should be under good control.

\paragraph[$1/R_{\mathrm{e}},\,1/R_{\mathrm{\mu}},\,1/R_{\mathrm{\tau}}$]{$\mathbf{ 1/R_{\Pe},\,1/R_{\PGm},\,1/R_{\PGt}}$}

The calculation of the statistical uncertainty assumes an angular acceptance of $|\cos{\theta}|<$ \num{0.95} and a selection efficiency of \SI{100}{\%} for $\PGm$ and $\Pe$ pairs (for the sake of simplicity).
The selection efficiency is assumed to be \SI{90}{\%} in case of $\PGt$ pairs. This acceptance is larger than the one obtained at LEP (c.f.\ Table~2.1 of~\cite{ALEPH:2005ab}), but should be well achievable with modern detectors.
The selection efficiencies are compatible with the ones measured in~\cite{ALEPH:2005ab}\footnote{... and for the projections it is irrelevant whether it is going to be \SI{95}{\%}, \SI{97}{\%} or \SI{98}{\%} in case of $\PGm$ and $\Pe$ pairs} and found also in~\cite{Kawada:2020lxr} for muon pairs. 
It is further assumed that the systematic uncertainties can be brought down to the level of the statistical ones.

\paragraph[$R_{ \mathrm{b}}\,(R_{\mathrm{c}})$]{$\mathbf{R_{\PQb}\,(R_{\PQc})}$}
The contamination of $\PQb$ by $\PQc$ quarks and vice versa is one of the leading systematic uncertainties for the determination of $R_{\PQb}$ and $R_{\PQc}$.
This observation has been for example confirmed in the detailed simulation study presented in~\cite{Irles:2023ojs}.
The DELPHI study~\cite{DELPHI:1998cnd} was at the origin of the estimation of the systematic uncertainty for the TESLA physics report~\cite{Hawkings:1999ac, ECFADESYLCPhysicsWorkingGroup:2001igx}. At the time an improvement of the systematic uncertainty of $\brar{\PQb}$ by a factor of five w.r.t.\ the results obtained by DELPHI seemed reasonable. This estimation was taken into account in the assessments for the ESPPU 2020 and Snowmass 2021 studies~\cite{LCCPhysicsWorkingGroup:2019fvj, ILCInternationalDevelopmentTeam:2022izu}.  
DELPHI had chosen a working point with a $\PQb$ tagging efficiency of \SI{30}{\%} and a $\PQc$ mistagging efficiency of around 0.004. With modern flavour tagging algorithms, the mistagging rate  can be reduced to effectively zero. The study in~\cite{FlavourTagger_Mareike} shows that this can already be achieved at a $\PQb$ tagging efficiency of \SI{60}{\%}.  
The same study suggests that in case of $\PQc$ quarks the contamination by $\PQb$ quarks is effectively zero for a $\PQc$ tagging efficiency of \SI{15}{\%}. It is further assumed that also the mistag of lighter quarks can be reduced to zero at the given working points. 

The efficiencies at the chosen working points together with a general pre-selection efficiency of \SI{70}{\%} and an acceptance cut $\mathrm{| \cos \theta |}$ < 0.95 following the detailed simulation study presented in~\cite{Irles:2023ojs} determine the calculation of the statistical uncertainties.

 The authors of~\cite{Hawkings:1999ac} have identified two further sources of systematic uncertainty.
\begin{itemize}
     \item  Hemisphere correlations: Hemisphere correlations due to particle assignment to jets, gluon radiation and/or a non-homogeneous detector were a source of systematic error at DELPHI (and to a much lesser extent at SLC due to the much smaller beam spot). The contribution to the hemisphere correlations at DELPHI was $\uprho\approx0.03$. In~\cite{Irles:2023ojs} it was shown that hemisphere correlations were compatible with zero 
     for ILC detectors and ILC beam spot and that at \SI{250}{GeV} they can be neglected w.r.t.\ to other uncertainty sources. Reference~\cite{Rohrig:2025bea} 
     finds a residual hemisphere correlation of $\uprho=0.001\pm0.003$ for FCCee detectors, still compatible with zero. Even $\uprho=0.001$ would be a factor 30 below the value found for DELPHI. Correspondingly, the influence of the hemisphere correlations on the systematic uncertainty could be expected to shrink by a factor of 30 without improving on the sources of its own systematic uncertainty. In addition, the authors of~\cite{Hawkings:1999ac} suggested an improvement by a factor of five due to a better understanding of the hemisphere correlations. The projected reduction of the hemisphere correlations by a factor 30 alone, i.e.\ without the factor 5 from~\cite{Hawkings:1999ac}, would allow for reducing the systematic uncertainty from hemisphere correlations by at least a factor 6 compared with Snowmass 2021 and ESPPU 2020. It can be assumed that either the residual hemisphere correlations itself or the sources of its systematic uncertainty will be further improved. An improvement of either of the two by a factor of two is a safe and conservative estimate and is included in the results in Table~\ref{tab:PEWresults}.    
     \item Gluon splitting \Pg $\to$ \bb, \cc: As in previous works~\cite{LCCPhysicsWorkingGroup:2019fvj, ILCInternationalDevelopmentTeam:2022izu} is assumed that gluon splitting will be controlled at least to the level of the statistical uncertainty. The uncertainty from the modelling of gluon splitting  mainly introduces an uncertainty on the  mistagging of light jets~\cite{SLD:2005zyw}. With the recent and further future improvements in flavour tagging, also the uncertainty due to gluon splitting is expected to be drastically reduced.
\end{itemize}

The previous reasoning together with the anticipated reduction of the contamination by quarks of other flavours to effectively zero allows supposing that the systematic uncertainty estimated in~\cite{Hawkings:1999ac} and adopted for ESPPU2020 and Snowmass 2021 can be reduced by at least a factor of 12. 

All estimations for future detectors are based on simulation studies assuming a symmetric detector. A fictive asymmetry of \SI{1}{\%} in the tagging efficiency of $\PQb$ quarks would yield a contribution to $\uprho$ of $\approx2.5\cdot 10^{-5}$, which would be below the systematic uncertainty estimated above. 

\paragraph[$R_{\mathrm{s}}$]{$\mathbf{\brar{\PQs}}$\,} The estimation of the branching fraction into $\PQs \bar{\PQs}$ pairs (and the asymmetry $\asym{\PQs}$, see below) is new w.r.t.\ to earlier assessments. It is based on the analysis carried out in~\cite{Okugawa:2024dks} that achieved a selection efficiency of around \SI{5}{\%} for $\PQs \bar{\PQs}$ pairs. This efficiency has been used for the estimations of $\brar{\PQs}$ in Table~\ref{tab:PEWresults}. 
The cut-based analysis targets the selection of ``leading'' charged kaons, resulting in a signal-to-background ratio of about one. The main background arises from light quarks ($\PQu$ and $\PQd$) and the size of this background was considered as the main source of systematic uncertainty. 
Modern flavour tagging algorithms promise to reduce the background rate by at least a factor of 100 at a similar signal efficiency. As sketched above, the analysis in~\cite{Okugawa:2024dks} followed an exclusive ansatz. Therefore, typical uncertainties due to jet formation are minimised and the estimation of the systematic uncertainty is based on the reduction of the background. 
The exclusive ansatz may also lead in the future to a relatively low efficiency. However, due to the fact that also the light quarks may be well isolated by a given flavour tagging algorithm, the background will be measured as well. Once the background is controlled the working point could be chosen such that an adequate efficiency is ensured.

\paragraph[Asymmetries other than $A_{\mathrm{e}}$]{Asymmetries other than $\mathbf{\asym{\Pe}}$} The statistical uncertainties for $\asym{\PGm}$ and $\asym{\PGt}$ were obtained for the pair selection efficiencies and the acceptances introduced above. 
While it can be safely assumed that most of the systematic uncertainties cited by e.g.\ SLD (c.f.\ Table 3.2 in~\cite{ALEPH:2005ab}) can be controlled to a level below the uncertainty on the polarisation (see below), we comment here on the dominant systematic uncertainty on $\asym{\PGt}$ of $18\cdot10^{-4}$ found by SLD. This uncertainty is generated by a bias in the reconstructed event mass in case of $\PGt \to \PGp \PGn$ decays due to the V-A nature of the $\PGt$ decay. Modern $\PGt$ reconstruction algorithms like the one presented in~\cite{Jeans:2015vaa} will allow for a complete event reconstruction, therefore alleviating any bias of the event mass due to the missing neutrino momenta. It is therefore assumed that this uncertainty source can be controlled at the LCF at least to the level of the statistical uncertainty.

In the case of asymmetries of $\PQb$ and $\PQc$ quarks, today's analyses~\cite{Irles:2023ojs} apply double charge tagging, suppressing therefore systematic uncertainties due to hadronisation and decay of the heavy quarks. The successful suppression of mutual contamination discussed before leaves the uncertainty on the beam polarisation as the main remaining source of systematic uncertainties.\footnote{Uncertainties calculated for the ESPPU 2020 and Snowmass 2021 were calculated supposing only a \SI{80}{\%} polarised electron beam and disregarding the polarisation of the positron beam.}

In the case of $\asym{\PQs}$, the selection efficiency found in~\cite{Okugawa:2024dks} is \SI{0.5}{\%} at a purity of \SI{80}{\%} against charge flip for the proper measurement of the differential cross section. This selection efficiency has been used for the calculation of the statistical uncertainty.
The uncertainty due to the background will fall below the systematic uncertainty due to beam polarisation that is discussed next.

The systematic uncertainties on all the asymmetries have been calculated under the default assumption of \SI{80}{\%} electron and \SI{30}{\%} positron polarisation but assuming $\updelta \mathrm{P}/ \mathrm{P}$=\SI{0.05}{\%} for both beams (as before). For the optimistic estimate in Table~\ref{tab:PEWresults}, \SI{60}{\%} positron polarisation has been assumed but no improvement on the uncertainty of the beam polarisations. The systematic uncertainty of the asymmetries can be further reduced by taking both polarisation and the asymmetries as parameters in a fit to the polar angle spectrum, as demonstrated for the case of \SI{250}{GeV}~\cite{Beyer:2022xyz}. 

It is important to note that the linear collider measurements discussed here 
do not require an 
exceptionally high-rate environment, which would compromise 
the detector capabilities both at the $\PZ$ pole and for 
the precision Higgs boson programme at higher energy.

Finally, we comment on the SM gauge couplings $\alpha_s$ and $\alpha$.  It is already true that the most precise determinations of $\alpha_s$ do not come from collider measurements but rather from lattice QCD. The current best determination of the MSbar  $\alpha_s(m_{\PZ}^2)$, with $\delta\alpha_s = 5.7 \times 10^{-3}$, comes from 10 lattice QCD groups using 5 different methods, all in agreement within statistics.  In all cases, these uncertainties are improvable with greater computer power. 
In \cite{Lepage:2014fla}, it is explained in Table 1 how the error can be improved to $0.8\times 10^{-3}$ with a factor 100 in computer power. The improvement over the next 20 years is expected to be more than 1,000; from 1997 to 2017, the speed of the
world's fastest computers increased by $10^6$.  
We expect a similar improvement in the lattice determination of $\alpha(m_{\PZ}^2)$.  We note the recent improvement by a factor 10 in the lattice determination of the hadronic vacuum polarisation contribution to the muon g-2~\cite{Borsanyi:2020mff}, now confirmed by 4 other lattice groups~\cite{Bazavov:2024eou}.

The uncertainty of $0.8\times 10^{-3}$  in $\alpha_s$ corresponds to an error in the theoretical prediction for $\Upgamma_{\PZ}$ of $\delta\Upgamma_{\PZ} = 2\times 10^{-5}$.  This is comparable to the statistical error in $\Upgamma_{\PZ}$ expected from Table~\ref{tab:PEWresults}.  For more precise measurements of $\Gamma_{\PZ}$ and $\upsigma^0_{\mathrm{had}}$, 
in their use for tests of the SM,  the uncertainty in $\alpha_s$ will be the dominant source of systematic uncertainty.

In fact, we expect to do even better in the precision electroweak closure test than the estimate given above,  since in the 
intervening time there have been significant updates to the 
linear collider $\PZ$-pole program.  These are:
\begin{enumerate}
   \item Much higher luminosities at the $\PZ$ pole than originally envisaged 
for ILC~\cite{Yokoya:2019rhx} can be realized by accommodating higher beamstrahlung 
with an ILC-like accelerator. Studies reported in Sec.~\ref{sec:acc:lumiup:zpole}
indicate that an eight-fold increase in 
luminosity to~$1.7\times 10^{34}$~cm$^{-2}$s$^{-1}$ is 
potentially within reach, allowing polarised data-sets of \SI{0.8}{\abinv} to 
be collected in a reasonable time frame. 

  \item Results from polarised $\PZ$ scans that include the statistical effects of the luminosity spectrum indicate statistical precisions with 
  such an \SI{800}{\fbinv} data-set with (\SI{80}{\%}/\SI{30}{\%}) beam polarisations 
  of \SI{14}{keV}, \SI{18}{keV}, and \num{1.0e-6} on $m_{\mathrm{Z}}$, 
  $\Gamma_{\mathrm{Z}}$ and $\sin^{2}\theta_{\mathrm{eff}}$, respectively, using simply the hadronic and leptonic polarisation-averaged lineshapes and left-right asymmetries. Higher luminosities still can be envisaged particularly 
  if the accelerator can provide even higher bunch charges. 
  \item 
  The eventual precision on observables such as $A_{\mathrm{b}}$, $R_{\ell}$, and $\sigma^{0}_{\mathrm{had}}$ depends partly on how well one can measure  hadronic events, and do b-tagging at very forward angles. Current estimates for the $\PZ$ pole are based on full simulation studies at \SI{250}{GeV} as presented in Sec.~\ref{sec:qprod}.
A key part will be being able to instrument the tracking to low angles where the linear collider design and the associated backgrounds may permit tracking 
instrumentation down to the \SI{40}{mrad} regime.
  \item Measurements rely on precise knowledge of the initial-state conditions. One recent development is a new approach to precision luminosity where the process $\ee \to \PGg\PGg$ is also considered~\cite{Wilson:2024wck}. The measurement techniques for 
  precision centre-of-mass energy measurement using leptonic final states discussed 
  in~\cite{Madison:2022spc} have been further refined including a method for measuring statistically the energy of each colliding beam~\cite{Wilson:2023mll}.
\end{enumerate}

\begin{figure}
   \begin{center}
        \includegraphics[width=0.7\textwidth]{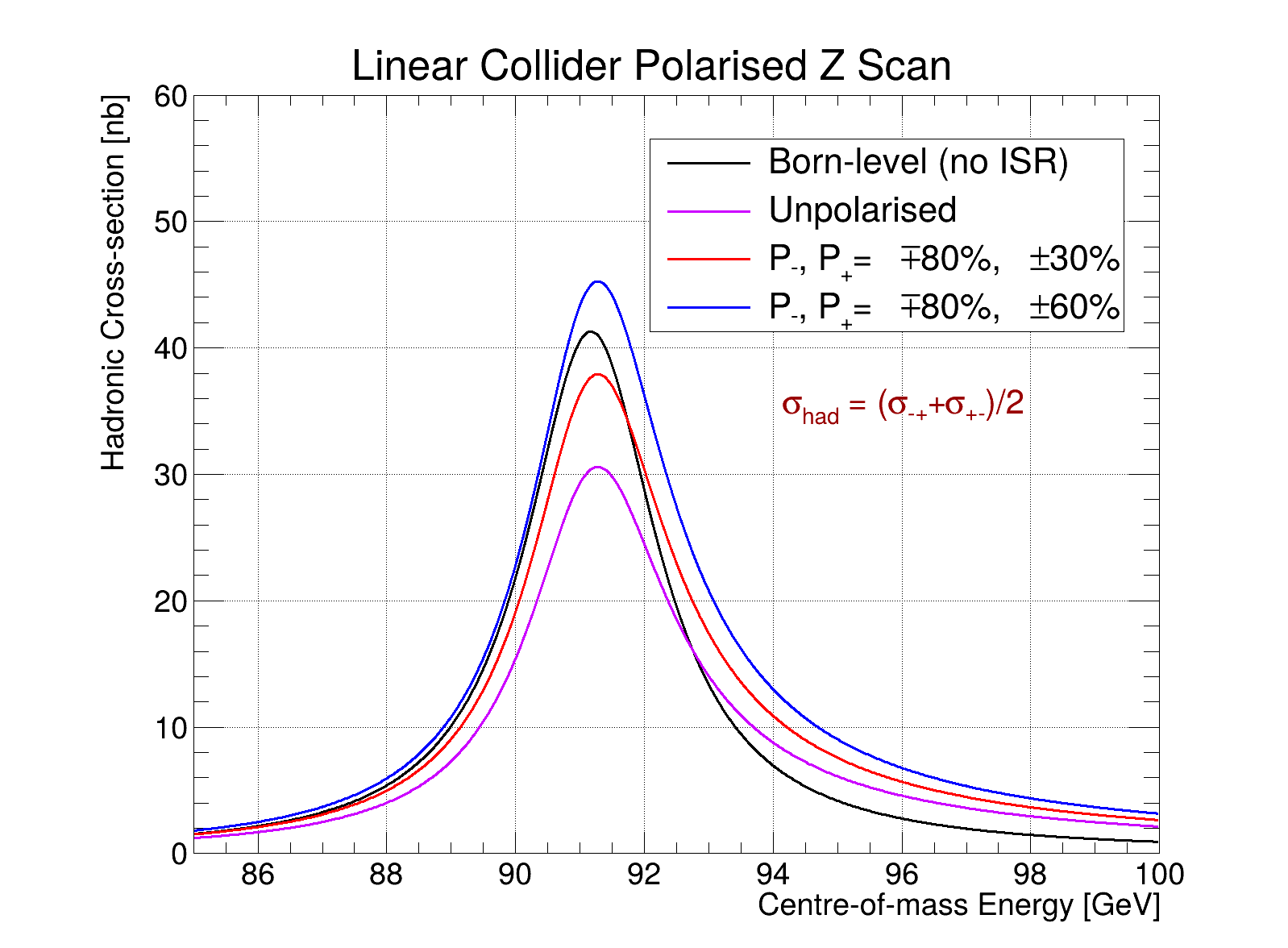} 
   \end{center}
        \caption{$\PZ$ lineshape with polarised beams. The blue, red, and violet curves are the measurable averaged hadronic cross section after including ISR for various polarisation values of the electron and positron beams. The black curve is with unpolarised beams and no QED radiative corrections.}
        \label{fig:Zscan}
\end{figure}

The potential for highly longitudinally polarised electron 
and positron beams, and the ability to instrument well the detector to 
very low angles are significant advantages for the linear collider 
approach. Polarisation adds unique observables, improves statistical precision, and allows for control of backgrounds. Forward acceptance and the freedom to 
design appropriate detectors may be determinative in setting the scale 
for realistic systematic uncertainties both absolute and relative 
that play a key role in measuring the $\PZ$ lineshape. 
Figure~\ref{fig:Zscan} illustrates that polarised beams in the $-+$ 
and $+-$ configurations result in a significant increase in the 
averaged cross section over a facility with no polarisation, while 
providing a simple way of 
simultaneously measuring $A_{\mathrm{LR}}$ which is 
proportional to $(\sigma_{-+} - \sigma_{+-})/(\sigma_{-+} + \sigma_{+-}) $. 
By taking some data with all four polarisation sign combinations 
and so measuring also $\sigma_{-\,-}$ and $\sigma_{++}$, the polarisation magnitudes can be reliably reconstructed~\cite{Moortgat-Pick:2005jsx}.

For ultimate luminosity at the $\PZ$ pole and $\PW\PW$ threshold, while retaining  
the polarisation advantages inherent to a LC, luminosities 
in the $10^{36}$~cm$^{-2}$s$^{-1}$ regime may be realizable in the future 
using energy and particle recovery technologies such as those discussed in Sec.~\ref{sec:acc:lumiup:erl}.


\subsubsection{W mass measurements}
\label{sec:Wprec}

Any $\ee$ collider designed to explore the Higgs boson with high precision will collect hundreds 
of millions of $\PW$ bosons and has the opportunity to make significant improvements 
to the measurement of the $\PW$ mass. Prospects for $\mw$ measurement with ILC are documented in some 
detail in~\cite{ILCInternationalDevelopmentTeam:2022izu} and references therein, based partly on earlier studies 
documented in~\cite{Baak:2013fwa}. The measurements are likely to be primarily 
systematics limited.
As was done at LEP, there are four well established techniques\footnote{It is expected that kinematically-constrained reconstruction of fully-hadronic $\PW\PW$ events 
will not be useful for a precision $\mw$ measurement given 
current and projected uncertainties from colour 
reconnection and Bose-Einstein effects. Future $\ee$ 
colliders can help to model better these phenomena~\cite{Christiansen:2015yca}.} for 
measuring the $\PW$ mass that can be utilized at a future linear collider operating at centre-of-mass 
energies above the $\PZ\PH$ threshold. This would be synergistic with the envisaged 
overall physics programme. The techniques include: 
\begin{enumerate} 
\item kinematically-constrained reconstruction of semi-leptonic $\PW\PW$ events (denoted K) 
\item direct measurement of the hadronic mass in single-$\PW$ and semi-leptonic events (H)
\item measurement of the lepton energy spectra in particular the kinematic endpoints in both fully-leptonic (FL) and semi-leptonic $\PW\PW$ events
\item measurement of the pseudo-masses in fully-leptonic $\PW\PW$ events with electrons and muons (FL)
\end{enumerate}
The estimated uncertainty on $\mw$ from these techniques is best for data taking 
at $\sqrt{s}=$\SI{250}{GeV} given the higher cross section and in general somewhat smaller systematic 
uncertainties. Quantitatively we estimate total uncertainties of \SI{2.2}{MeV} (K), \SI{2.6}{MeV} (H) and 
\SI{5.0}{MeV} (FL) for the subsets of these measurements that are most easily separable, 
where statistical uncertainties are \SI{0.6}{MeV}, \SI{0.6}{MeV} and \SI{4.8}{MeV} respectively for 
\SI{3.0}{\abinv} at \SI{250}{GeV}. Combined, the overall uncertainty on $\mw$ is estimated to be \SI{1.8}{MeV} taking into 
account (positive) correlations from effects such as beam energy, luminosity spectrum, and hadronisation modelling. As 
the dominant measurements are foreseen to be systematics limited, the additional data taken at higher centre-of-mass energies will give further cross-checks with two detectors, but is not expected to add 
significantly in overall precision.

\begin{figure}
   \centering
    \begin{subfigure}{0.49\textwidth}
        \includegraphics[width=1\textwidth]{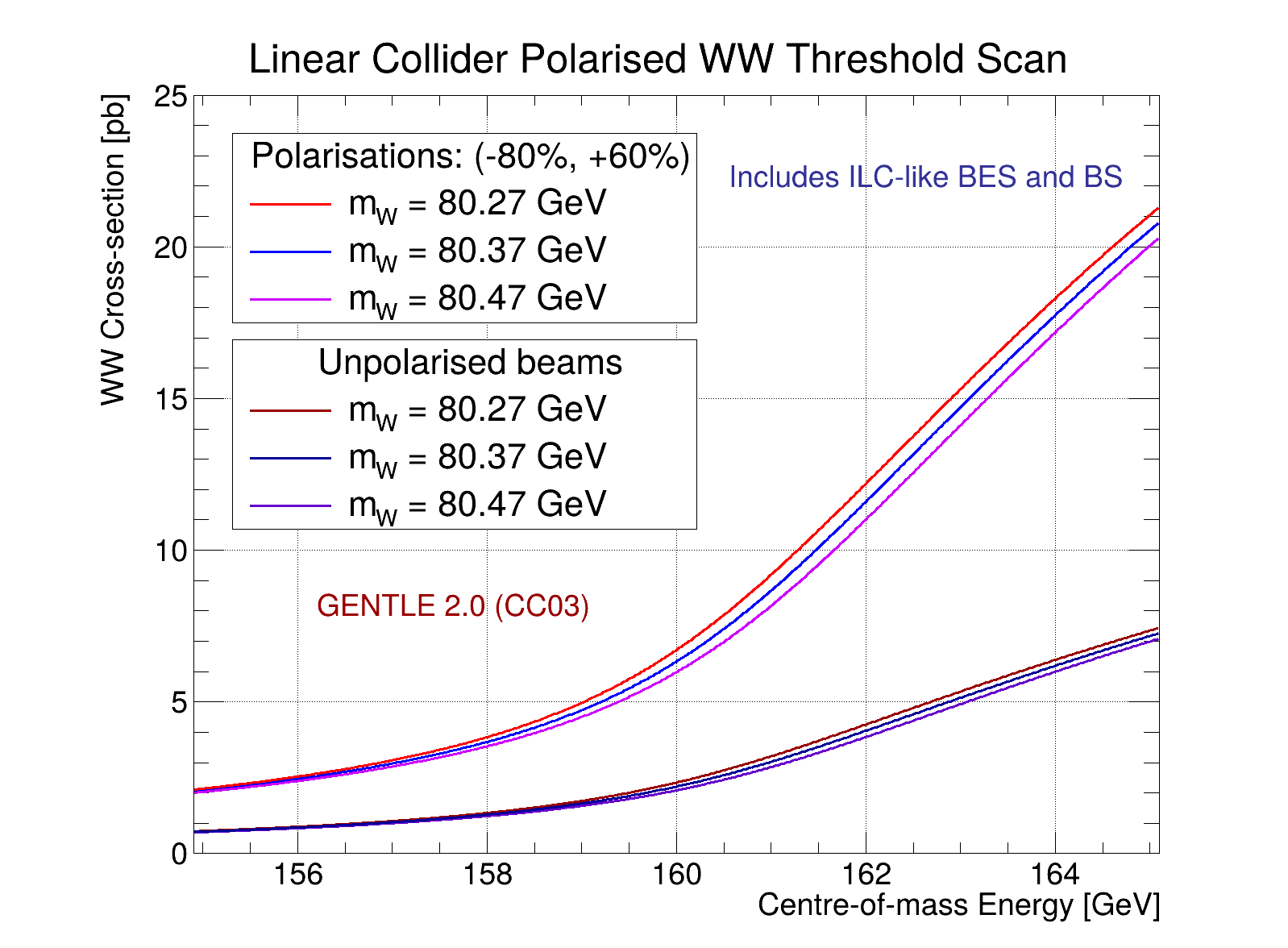} 
        \caption{\label{fig:WWScan:xsec}}
    \end{subfigure}
    \begin{subfigure}{0.49\textwidth}
        \includegraphics[width=1\textwidth]{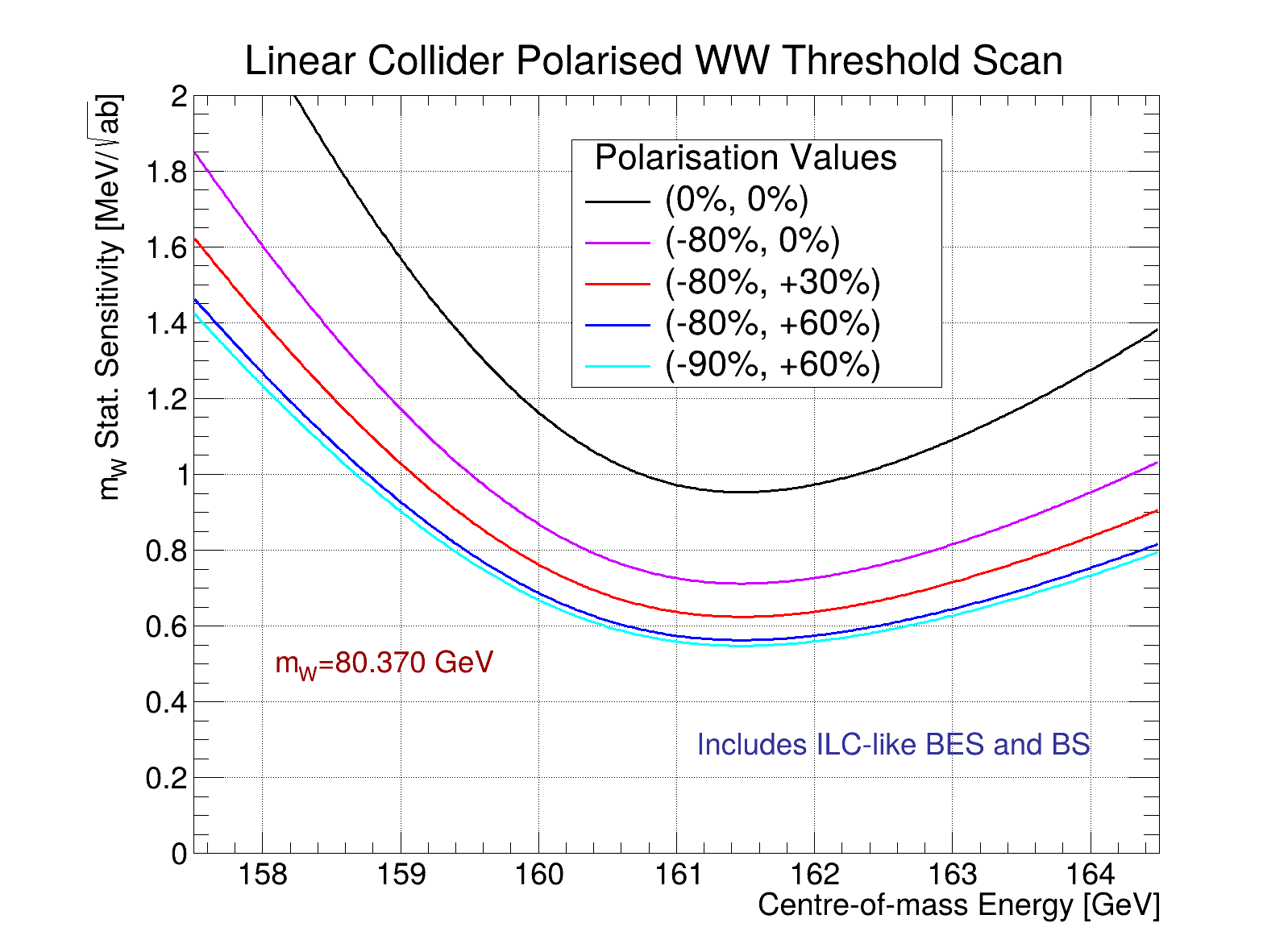}
        \caption{\label{fig:WWScan:Kplot}}
    \end{subfigure}
        \caption{(a) Illustration of the centre-of-mass energy dependence of the $\PW\PW$ cross section on $\mw$ and the level of polarisation. (b) The intrinsic sensitivity, $K$, 
        of polarised cross section measurements to $\mw$ near threshold assuming an efficiency-purity product of unity.}
    \label{fig:WWScan}
\end{figure}

A similar overall precision of \SI{2.0}{MeV} could also be reached 
with a dedicated \SI{0.5}{\abinv} polarised scan\footnote{For illustration these numbers 
use (\SI{90}{\%}, \SI{60}{\%}) polarisation and a 6-point scan with all four helicity combinations.} near the $\PW\PW$ threshold where all $\PW\PW$ final states contribute to the cross section measurement. Together with the 
baseline programme discussed above the estimated combined uncertainty on $\mw$ would be \SI{1.3}{MeV}. 
The availability of longitudinal polarisation significantly increases the cross section 
near threshold when using left-handed electrons and right-handed positrons given that 
the $t$-channel $\PGne$ exchange dominates as illustrated 
in Fig.~\ref{fig:WWScan:xsec}, and it is this data-set that dominates the sensitivity. 
This is a significant advantage over circular colliders. It is not just an increase in 
signal rate and consequent $\mw$ sensitivity, but a capability of switching helicities 
and turning off $\PW\PW$ production to assist in measuring directly 
the background at the same centre-of-mass energy.
Figure~\ref{fig:WWScan:Kplot} shows the centre-of-mass energy dependence of the basic sensitivity factor of the polarised cross section to $\mw$, defined as $K$ below, including the effects of beam energy spread (BES) and beamstrahlung (BS)  
with beam parameters consistent with the ILC250 baseline. 
The relationship~\cite{Stirling:1995xp} is 
\[  \Delta m_{\mathrm{stat}} = \left| \frac{\mathrm{d}\sigma}{\mathrm{d}m} \right| ^{-1} \Delta \sigma  
= 
   \left| \frac{\mathrm{d}\sigma}{\mathrm{d}m} \right| ^{-1} \sqrt{\frac{\sigma}{\varepsilon p \cal{L}}
}  
   = \frac{K}{\sqrt{Q \cal{L}}} \; \mathrm{with} \;  Q \equiv \varepsilon p \: \: .
   \]
Clearly high polarisations can improve the sensitivity dramatically 
over the unpolarised case.
Values for the quality factor, $Q$, vary by $\PW\PW$ final state and polarisation assumption. 
With the current scan assumptions the systematic uncertainty of \SI{1.7}{MeV} is dominated 
by the statistical uncertainty on the background measurement while the overall statistical uncertainty of the polarised data-set of \SI{0.5}{\abinv} is \SI{1.1}{MeV} equivalent to an effective $K/\sqrt{Q}=\SI{0.77}{MeV/\sqrt{\mathrm{ab}}}$. 
Further improvements in the background uncertainty 
can be envisaged by reducing hadronic event backgrounds for example 
by rejecting/de-weighting  events with one 
or more $\PQb$-tagged jets or those with hadronic \PW candidates consisting of two $\PQc$-tagged jets. There is 
also potential for improving the statistical uncertainty 
by devoting more time to the most favourable helicity combination, namely the one illustrated in Fig.~\ref{fig:WWScan:xsec}.

In summary, there are excellent prospects for a major advance in the $\mw$ precision 
by an order of magnitude from today's PDG uncertainty of \SI{13.3}{MeV} with both the 
data collected above the production threshold for Higgs bosons and from  
a dedicated run of at least \SI{0.5}{\abinv} with a polarised $\PW\PW$ threshold scan. 
The projected precisions in the two regimes are similar but 
running at $\PW\PW$ threshold precludes many other measurements of prime interest. 
The current accelerator designs retain accelerator compatibility with running 
at $\PW\PW$ threshold with polarised beams, recognizing the complementary nature of 
such a measurement, but do not include such a dedicated run in the running scenarios.
$\PW\PW$ threshold running is retained as an option. 
Further improvements in the instantaneous luminosity near the $\PW\PW$ threshold as discussed in Sec.~\ref{sec:acc:lumiup:zpole} and potentially different future 
physics perspectives can lead to such running 
becoming part of the actual run plan whenever it becomes the most interesting thing to do.

\subsubsection{Quark pair production}
\label{sec:qprod}

Precise measurements of the $\PZ$ boson's couplings to quarks may expose deviations from Standard Model predictions. Jets originating from \PQb and \PQc-quarks can be identified via secondary and tertiary vertices, while jets from \PQs-quarks often  feature leading kaons. For  constraining the $\PZ$ boson's couplings to first-generation quarks, radiative events can be used: Since the probability of photon emission is proportional to electric charge, up-type quarks produced in the $\PZ$-boson decays are significantly more likely to radiate photons than down-type quarks. On the other hand, the non-radiative decay sample is dominated by down-type quarks, reflecting their stronger electroweak coupling in the Standard Model. These facts can be used to constrain light-quark electroweak couplings from the hadronic decays of the $\PZ$ boson by simultaneously examining both radiative and non-radiative signatures. With these techniques, the couplings of the $\PZ$ boson to each quark flavour can be measured both  at the $\PZ$ pole and at higher energies, providing complementary SM tests and BSM sensitivities.

\paragraph[At the Z pole]{At the Z pole} A run at the $\PZ$ pole provides an ideal environment for such studies by comparing radiative and non-radiative $\PZ$ boson decays~\cite{Mekala:2024xal}. Similar measurements at LEP~\cite{DELPHI:1991utk, L3:1992kcg, L3:1992ukp, DELPHI:1995okj, OPAL:2003gdu} were statistically limited and yielded precision of up to 3\% for $\PQd$ and 6\% for $\PQu$ quarks. 

The concept of the measurement is based on the observation that the hadronic cross section at the $\PZ$ pole, $\epem \to \PZ \to \PQq\PAQq$, $\PQq = \PQu, \PQd, \PQs, \PQc, \PQb$, can be expressed as
\begin{equation}
    \sigma_{had} = \sum_{q} \sigma_q  = \mathcal{C}_1 \cdot \sum_{q} c_q, 
\end{equation}
where $c_q \equiv c_V^2 + c_A^2$ is the $\PZ$ boson's coupling to the quark $\PQq$ and $\mathcal{C}_1$ is a constant. The radiative hadronic cross section with exactly one photon identified in the final state (the one-photon inclusive cross section) can be parametrised as
\begin{equation}
       \sigma_{had+\gamma} = \mathcal{C}_2 \cdot \sum_{q} c_q Q_q^2,
\end{equation}
where $\mathcal{C}_2$ is another constant (for a given set of cuts and isolation criteria for photons). The values of $\mathcal{C}_1$ and $\mathcal{C}_2$ can be estimated from theoretical calculations and simulations.

\begin{figure}[hbt]
    \centering
    \includegraphics[width=0.52\textwidth]{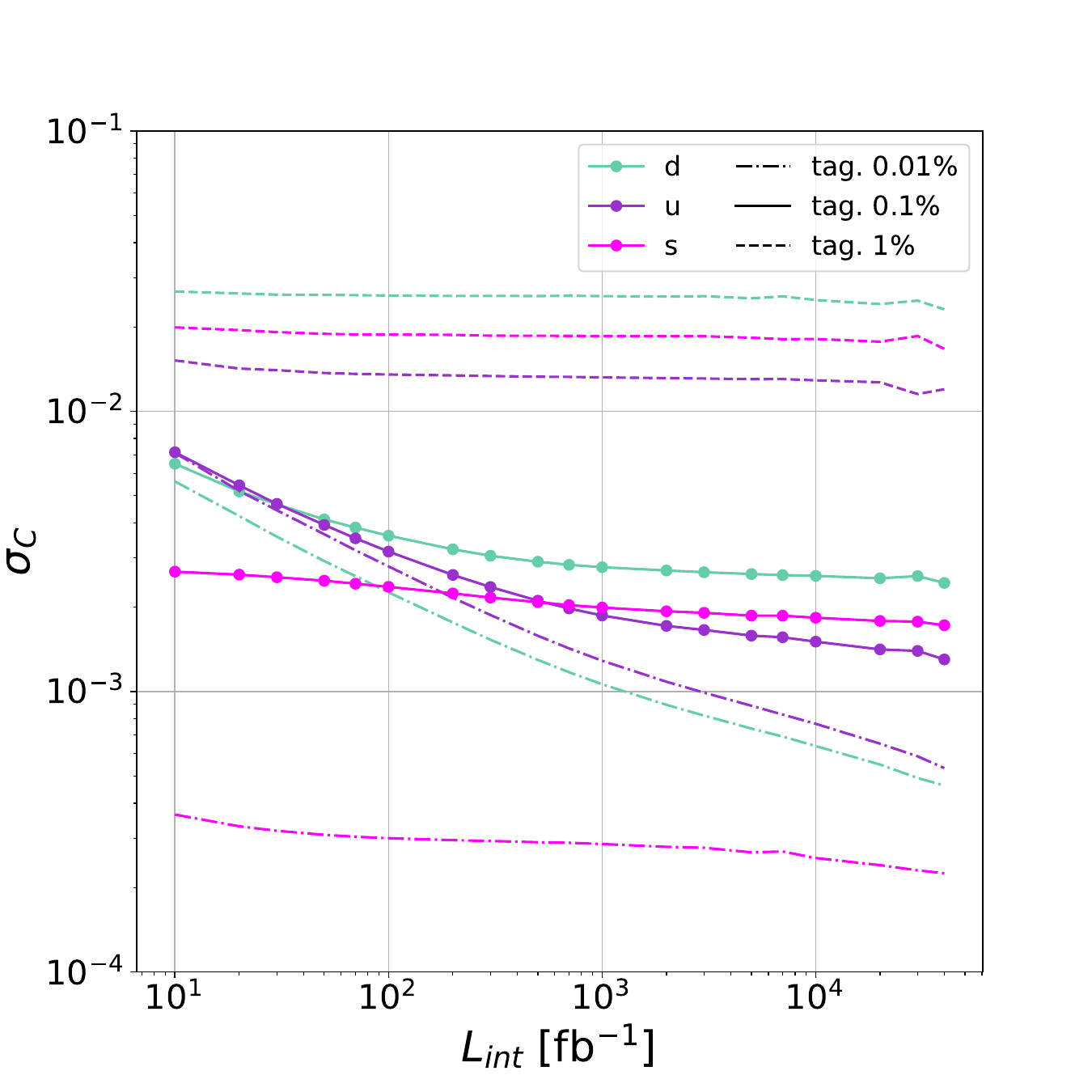}
    
    \caption{The uncertainty of the $d$ (aquamarine), $u$ (purple) and $s$ (pink) coupling measurement as a function of the integrated luminosity collected at the $\PZ$ pole for unpolarised beams. Three assumptions on the jet-flavour tagging uncertainty are shown: \SI{0.1}{\%} (solid lines), \SI{1}{\%} (dashed lines) and \SI{0.01}{\%}. Other systematic uncertainties are fixed at \SI{0.1}{\%}, except for the luminosity which is assumed to be known to 10$^{-4}$.
    Taken from \cite{Mekala:2025rlk}. }
    \label{fig:5flav_unc_lumi}
\end{figure}

One of the advantages of the measurement at the $\PZ$ pole is the suppression of the ISR background contribution due to the small phase space for photon emissions. However, background photons originating from hadronic decays must be considered in the analysis. Photon reconstruction criteria should be optimised at the detector level to enhance the contribution from events with the actual hard emissions from the final state. The systematic uncertainties on the acceptance of radiative events and the mis-acceptance of hadronisation photons into the radiative sample should be included in the measurement. Similarly, the flavour tagging uncertainty has a crucial impact on the results.

Figure~\ref{fig:5flav_unc_lumi} shows the precision of the light-quark coupling measurement as a function of the integrated luminosity for different assumptions on the tagging uncertainty~\cite{Mekala:2025rlk}. For the typical $\PZ$ pole programme of a linear collider, comprising a few hundred $\fbinv$ of data,  sub-percent precision on the light-quark couplings to the $Z$ boson can be achieved, provided that the jet-flavour tagging uncertainty is reduced below \SI{0.3}{\%} and the relative uncertainty on the background contribution to the radiative sample from hadronisation photons remains at or below \SI{1}{\%}. The effect of the uncertainty on the radiative event selection efficiency is found to be marginal. The figure also demonstrates that in order to benefit from much higher data statistics, sub-permil uncertainty on the jet-flavour tagging would need to be achieved.

\begin{figure}
    \begin{subfigure}{0.44\textwidth}
        \includegraphics[width=\textwidth]{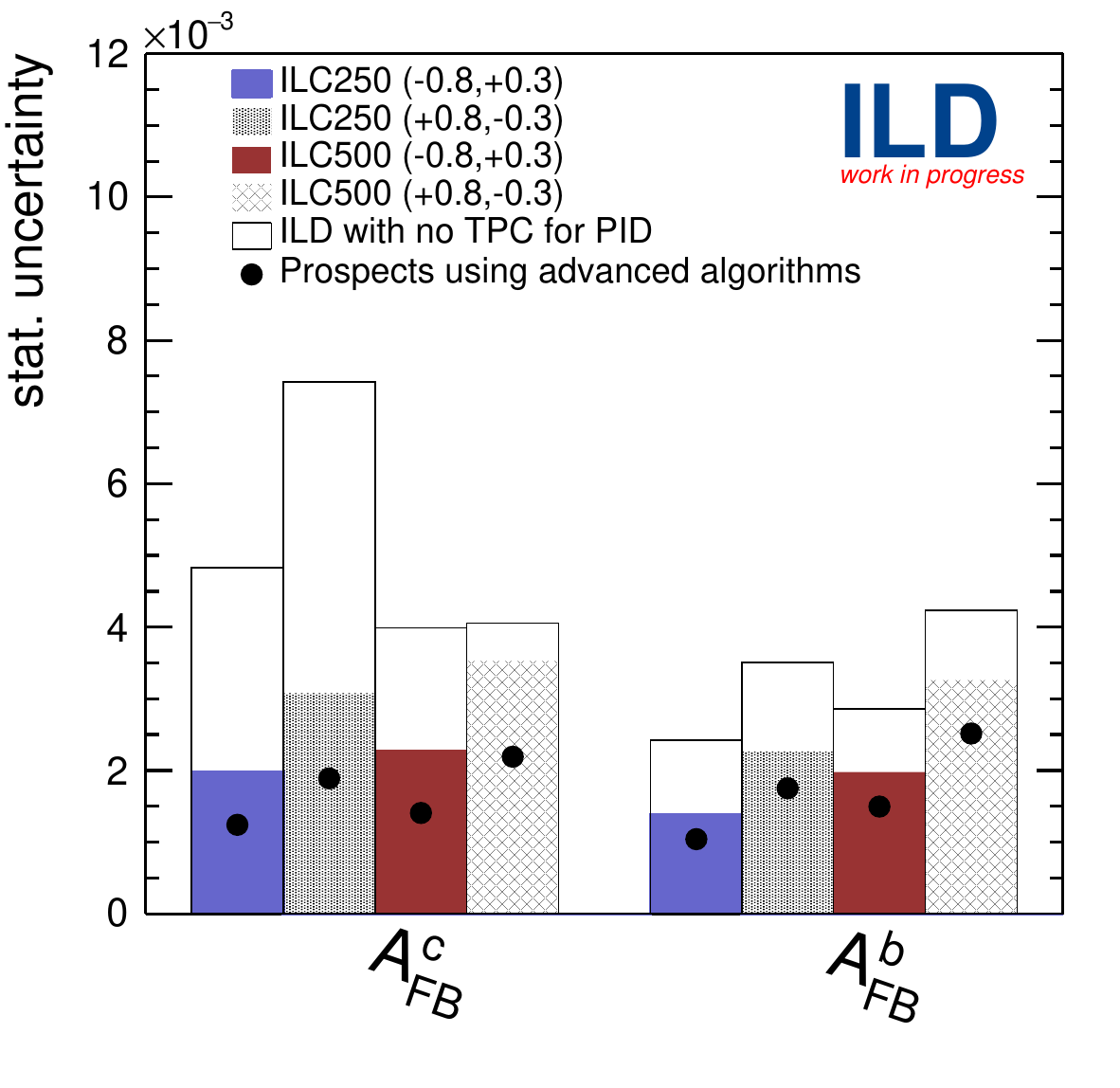} 
        \caption{}
        \label{fig:afbbafbc}
    \end{subfigure}
    \quad
    \begin{subfigure}{0.52\textwidth}
        \includegraphics[width=\textwidth]{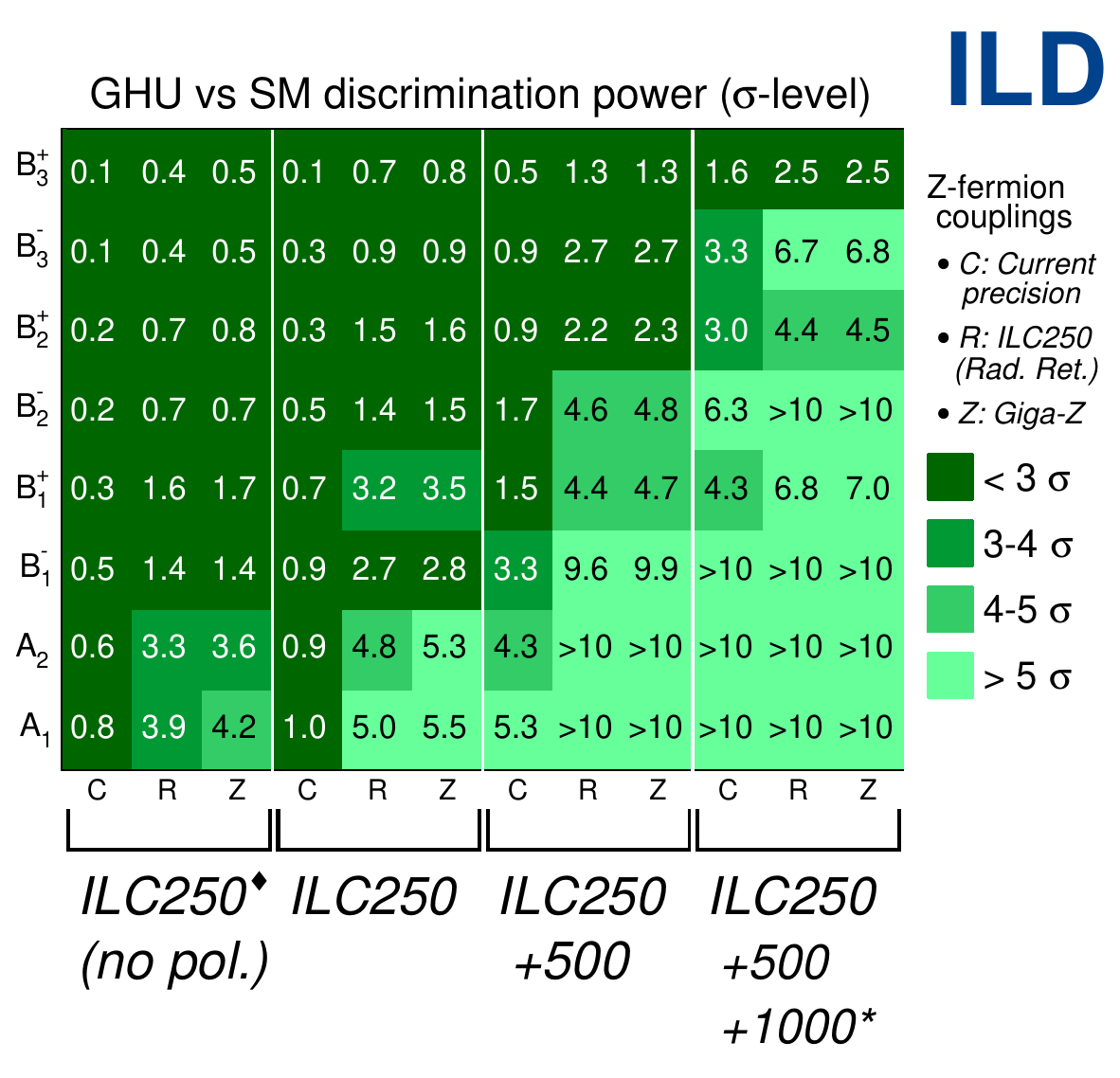}
        \caption{}
    \label{fig:ghuvssm}
    \end{subfigure}
\caption{Results on heavy quark production at $\roots=$\SI{250}{GeV}: (a) Statistical precisions on \afbf{b} and \afbf{c} for integrated luminosities of \SI{2}{\abinv} at \SI{250}{GeV} and \SI{4}{\abinv} at \SI{500}{GeV}~\cite{ECFA-HF-Report}. The figure also shows the impact of particle identification and updated flavour tag. (b) Statistical discrimination power between the GHU models~\cite{Funatsu:2017nfm,Funatsu:2020haj,Funatsu:2023jng} and the SM. Figure updates that in~\cite{Irles:2024ipg}. }
\end{figure}

\paragraph[Above the Z pole]{Above the Z pole} The physics potential of two fermion processes above the $\PZ$ pole is illustrated, for example, by recent studies of quark pair production.
The publications~\cite{Irles:2023ojs, Irles:2024ipg} describe comprehensive experimental studies on viability and prospects for the measurement of electroweak observables in $\epem \to \bb$ and $\epem \to \cc$ processes at the ILC operating at $\roots= $\SI{250}{GeV} and \SI{500}{GeV} of centre-of-mass energy. The studies are based on a detailed simulation of the ILD concept, leveraging its particle flow approach, based on excellent vertexing and tracking capabilities, as well as its charged hadron identification thanks to the \dedx measurements in the TPC. Polarised beams allow for inspecting in detail the four independent chirality combinations of the electroweak couplings to electrons and other fermions with an almost background-free analysis. 
A classical key observable is the forward-backward asymmetry $\afb$. Figure~\ref{fig:afbbafbc} shows the expected statistical precision at \SI{250}{GeV} and \SI{500}{GeV} for integrated luminosities of \SI{2}{\abinv} and \SI{4}{\abinv}, respectively. 
It illustrates the successive improvements obtained by particle identification and an updated flavour tagging that was introduced in Sec.~\ref{sec:hlreco}.
The results can be interpreted in the frame of so-called Grand Higgs Unification Models. In these models that include {\em compactified} warped extra dimensions the Higgs appears as a fluctuation mode of the Aharonov-Bohm phase $\theta_H$ and can be associated as the fifth component of the known SM vector potentials. For an introduction see~\cite{10.1093/ptep/ptv153} and references therein. The model predicts massive Kaluza-Klein excitation of neutral vector bosons that influence observables like \afb. A statistical analysis carried out in~\cite{Irles:2024ipg} demonstrates the discovery potential of a linear collider. The result is shown in Fig.~\ref{fig:ghuvssm}. 
Different running scenarios of ILC are compared: ILC (no pol.) refers to a hypothetical case with no beam polarisation and \SI{2}{\abinv} of integrated luminosity, ILC250 (\SI{2}{\abinv}), ILC500 (\SI{4}{\abinv}), and ILC1000* (\SI{8}{\abinv}), the last one not using full simulation studies but extrapolations of uncertainties from ILC500. Three different assumptions for the $\PZ$-fermion couplings uncertainties are considered~\cite{ILCInternationalDevelopmentTeam:2022izu}: \textit{C} for current knowledge; \textit{R} for expected knowledge after the full ILC250 programme and the study of $\PZ$-fermion couplings from radiative return events, and \PZ for expected knowledge after a full ILC \PZ-pole program.~

Neutral boson masses \mass{Z^{1}} of up to \SI{19.6}{GeV} have been tested. The results show that already at \SI{250}{GeV} different GHU predictions could be distinguished from the SM prediction. The separation power is amplified at higher energies as expected. Beam polarisation increases the separation power, as illustrated for the \SI{250}{GeV} case.

\subsubsection{Tau-lepton pair production}

$\PGt$ lepton decays provide sensitivity to the $\PGt$ polarisation through the distribution of decay products. The momenta of the $\PGt$'s daughters can be used to define an optimal ``polarimeter'' whose orientation is correlated with the parent $\PGt$ lepton's polarisation. The $\PGt$-pair final state of the process
$\epem \to \PGtp\PGtm$ 
allows unique tests of the chiral nature of electroweak interactions, complemented at higher centre-of-mass energies by similar analyses of the top-quark pair final state.
The combination of a linear collider's polarised initial state and high statistics distributed over a wide energy range promises a suite of intriguing measurements to test the validity of the SM in a distinctive way. 
Such measurements can be made both at the nominal centre-of-mass energy of the collider, and at lower effective energies by identifying events with significant initial state radiation, in particular the frequent return to the $\PZ$ pole.
The current status of ongoing studies is discussed in more detail Sec.~5.3.5 of \cite{ECFA-HF-Report}. 

The influence of relevant SMEFT coefficients on different components of the $\PGt$ lepton's polarimeters, and also on the correlations between polarimeters within tau pairs is being studied. The effects of new Lagrangian structures are often rather subtle, and tend to become more significant with increasing centre-of-mass energy.

On the experimental side, the focus is on leveraging the expected performance of detectors to extract maximal information about event kinematics, recovering as much as possible the lost neutrino momenta by the use of constraints on the $\PGt$ mass, four-momentum conservation, charged $\PGt$ daughter trajectories, and the small, well-defined interaction region. 
The key performance metrics relevant in such an approach are the vertex detector's impact parameter resolution and the ability to identify $\PGt$ decay modes, principally using the calorimeter to identify and measure neutral pions. 

For the case of the ILC, the $\epem\to\PGtp\PGtm$ process at \SI{500}{GeV} has been studied in full simulation of the ILD detector concept~\cite{Jeans:2019brt}, achieving a precision of \SI{0.5}{\%} for the $P_{e^-e^+}=(-0.8,+0.3)$ sub-set of the total \SI{4}{\abinv} foreseen at the ILC. Further improvements to the reconstruction of the $\PGt$ polarisation have been investigated since~\cite{Yumino:2022vqt}, but not yet applied in the full simulation analysis. With the analysis improvements and the larger data-set envisioned for LCF, permil-level precision can be expected.

More complete studies of spin correlations in $\PGt$ pair systems are also underway, with the aim to reconstruct the full spin density matrix, or``quantum tomography'', potentially allowing tests of quantum entanglement between the $\PGt$~\cite{Barr:2024djo}. This topic will be discussed in some more detail in Sec.~\ref{sec:top:quantum} in the context of top-quark pair production.

\subsubsection{Lepton flavour violation at high energies}
\label{sec:LFV} 

The observation of lepton flavour violation (LFV) would constitute an unambiguous sign of physics beyond the SM, and it is therefore highly motivated to pursue dedicated searches for LFV processes.
Traditionally, such searches have focused on rare decays of muons and tau leptons. 
Complementary to these low-energy probes, high-energy colliders provide a powerful alternative avenue for exploring LFV interactions. 
At high energies, one can search for lepton flavour-violating decays of heavy resonances such as the $\PZ$, Higgs, or top quark, or alternatively, one can directly search for the non-resonant production of lepton-antilepton pairs with different flavours in high-energy collisions, for example $\epem \to \PGt\PGm$. 
The latter method takes advantage of the possible energy growth of the new physics signal cross section compared to SM backgrounds that typically fall with the centre-of-mass energy. 
Linear $\epem$ colliders offer high centre-of-mass energies, large integrated luminosities, and the ability to polarise the electron and positron beams, giving them unique capabilities to test new physics in $\epem \to \PGt\PGm$. 

The main source of background for such searches arises from $\epem \to \PGtp \PGtm$ where one tau decays leptonically to a muon and neutrinos, producing a continuous spectrum of muon momenta with a kinematic endpoint at the beam momentum $p_\text{beam}$. 
In contrast, the $\epem \to \PGt\PGm$ signal is characterized by a sharp peak at this endpoint, with the width of the peak determined by the beam energy spread, the momentum resolution of the detector, and effects of initial state radiation. 
Sensitivity estimates are obtained by applying a cut $p_{\PGm} /p_\text{beam} \gtrsim 1$ which removes almost the entire background while retaining an $\mathcal O(1)$ fraction of the signal~\cite{Dam:2018rfz, Altmannshofer:2023tsa, Altmannshofer:2025nbp}. 

\begin{figure}[tb]
\centering
\includegraphics[width = \linewidth]{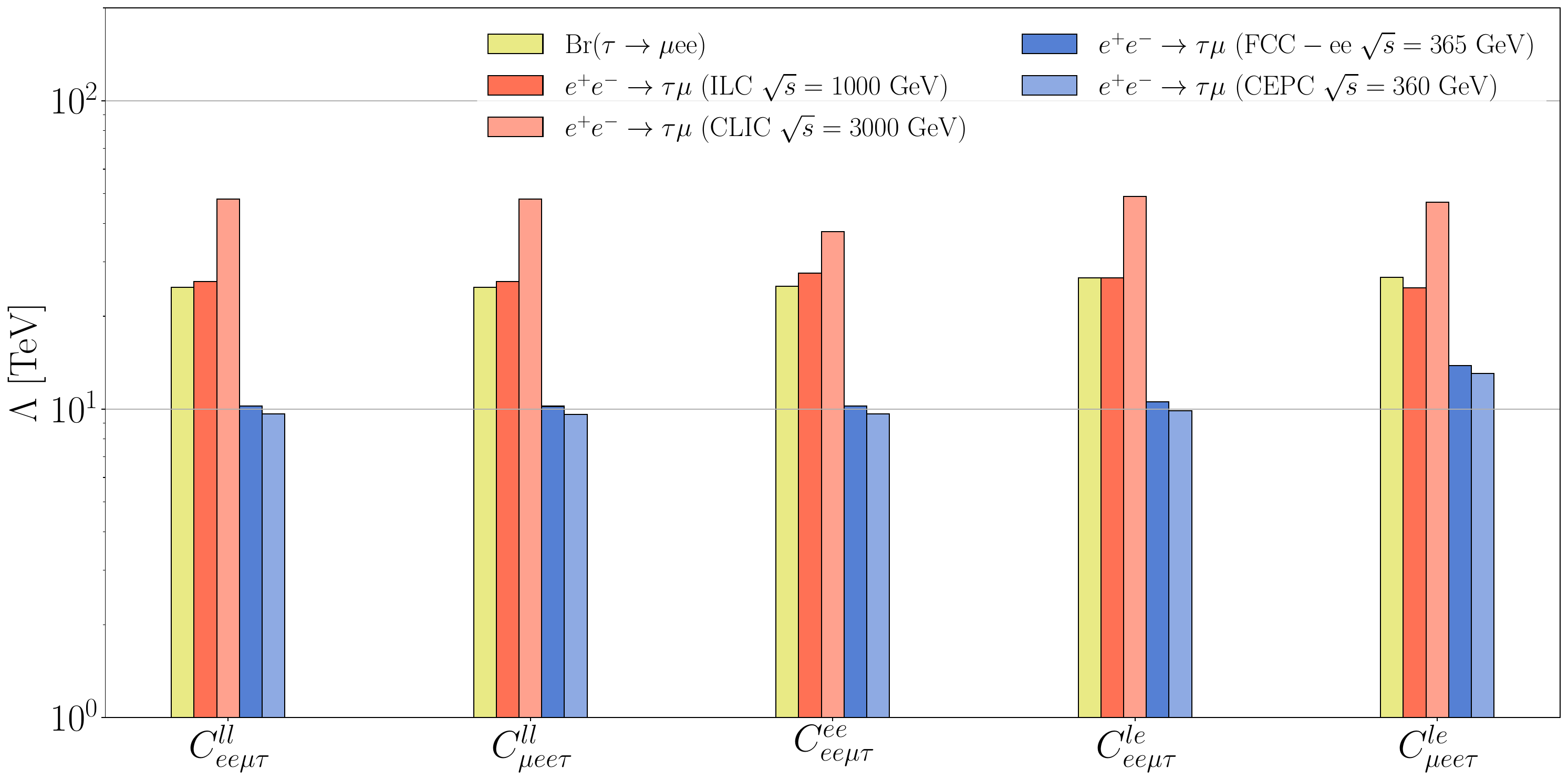}
\caption{Expected sensitivity to the new physics scale of lepton flavour-violating 4-fermion interactions in SMEFT. The Belle~II sensitivity of tau decays, $\PGt\to\PGm \epem$, is shown in yellow~\cite{Belle-II:2018jsg}. The sensitivity of searches for $\PGt\PGm$ production in $\epem$ collisions is shown in red for ILC and CLIC and in blue for FCC-ee and CEPC~\cite{Altmannshofer:2023tsa}. Figure from~\cite{Altmannshofer:2025nbp}.} 
\label{fig:bar_chart}
\end{figure}

The effects of heavy new physics from a scale $\Lambda \gtrsim \sqrt{s}$ can be systematically parameterised in the SMEFT framework. 
Figure~\ref{fig:bar_chart} shows the expected \SI{95}{\%} CL sensitivities to the new physics scale of lepton flavour-violating 4-fermion interactions from $\epem \to \PGt\PGm$~\cite{Altmannshofer:2025nbp}. The sensitivities achievable at a \SI{1}{TeV} run of ILC with an integrated luminosity of \SI{8}{\abinv} and a \SI{3}{TeV} run of CLIC with \SI{5}{\abinv} are shown in red and compared to the projections for FCC-ee and CEPC (blue)~\cite{Altmannshofer:2023tsa} and the expected reach of the rare tau decay $\PGt\to \PGm \epem$ at Belle~II (yellow)~\cite{Belle-II:2018jsg}.
Since the 4-fermion operator contributions to the $\epem \to \PGt\PGm$ cross section grow with the squared centre-of-mass energy $s$, one finds exceptional sensitivity at linear colliders, exceeding both Belle~II and future circular $\epem$ machines. ILC could probe new physics scales up to $\Lambda \simeq $\SI{25}{TeV} while CLIC could reach $\Lambda \simeq $\SI{50}{TeV}.

If a $\epem \to \PGt\PGm$ signal were to be observed, the availability of various electron and positron beam polarisation settings would offer valuable diagnostics about the underlying new physics.
Different polarisation configurations provide complementary sensitivity to operators involving left-handed or right-handed leptons, enabling one to disentangle SMEFT operators with distinct chirality structures~\cite{Altmannshofer:2025nbp}.

\subsubsection[Precision QCD at $\mathrm{e}^+\mathrm{e}^-$ Higgs factories]{Precision QCD at $\ee$ Higgs factories}
\label{sec:QCD}

Although $\ee$ colliders are thought of primarily as machines to study the electroweak interactions, they have had a profound influence on our understanding of Quantum Chromodynamics (QCD). Electron-positron annihilation to hadronic final states is the simplest QCD process, resulting typically in two well-defined hadronic jets.  This simplicity allowed the discovery of jets at a time when hadronic experiments struggled to find direct evidence of quark interactions~\cite{Hanson:1975fe} and, later, the discovery of the gluon through the observation of 3-jet events~\cite{Barber:1979yr,PLUTO:1979dxn,JADE:1979rke,TASSO:1980lqw}.

The understanding of QCD has advanced dramatically since those energy measurements.  In the LHC era, the study of jets has become a central tool in all aspects of the experimental program.  New theoretical  methods such as anti-$k_T$ clustering~\cite{Cacciari:2008gp}, 
 N-jettiness~\cite{Stewart:2010tn} and jet grooming methods such as soft-drop~\cite{Larkoski:2014wba}  have offered refined jet observables that have enabled higher precision QCD tests and also the use of jet substructure for important physics 
discoveries~\cite{ATLAS:2018kot,CMS:2018nsn}. These theoretical development up to 2020 are reviewed in ~\cite{Larkoski:2017jix}.

Still, many issues remain open the study of QCD jets. The final stage of quark and gluon hadronisation is poorly understood.  In simulations, it is still represented by ad-hoc models.    Non-perturbative effects in jet physics fall of slowly with momentum transfer, as $Q^{-1}$. These facts limit the precision in the modelling of hadronic final states, to the extent that this will be a leading source of systematic error for the precision measurements at the HL-LHC. To overcome these problems, we need to go back to basics and make high-precision measurements of final states in $\ee$ annihilation, in a setting unaffected by an underlying event or pile-up.  To better understand the boundary between perturbative and non-perturbative physics, these measurements should be carried out over as large a range of centre-of-mass energies as possible, always with 
up-to-date precision detectors.  A Higgs factory programme with 21st-century trackers and calorimeters beginning at the $\PZ$ pole and at \SI{250}{GeV} and extending eventually up to \SI{1}{TeV} in the centre of mass will provide the setting for the measurements that are needed.

New theoretical tools are now available for this study.  The first of these is the energy-energy correlation function~\cite{Basham:1978bw}, or, in more modern language, the energy flow operators~\cite{Hofman:2008ar}
\beq
     {\cal E}(\vec n) = \lim_{r\to \infty} \int dt r^2 n^i T_{0i} (t, r\vec n) \ .
\eeqn
where $\vec n$ is a light-like unit vector emerging from the particle collision.  Recently, these objects have been intensively studied in precision perturbative QCD~\cite{Dixon:2018qgp,Chen:2020vvp}, leading to a highly sophisticated theoretical framework for discussing jets structure and substructure.   In particular, measurement of higher-point correlation functions of these operators gives detailed information on the energy-flow pattern inside a jet.
The second is the use of ``track functions'', energy correlation functions constructed entirely from charged hadron track information~\cite{Li:2021zcf}. Using the superb trackers that Higgs factories need, for example, for the precision measurement of the Higgs boson mass, experiments at $\ee$ Higgs factories will be able to map the internal structure of QCD jets in unprecedented detail.  Track functions require their own precision formalism for theoretical analysis, but this formalism  is already being built in preparation for this new level of experimental data~\cite{Jaarsma:2023ell,Jaarsma:2024ngl}. 

Hadronic Higgs boson decays that do not involve heavy quarks are dominated by events with $\PH\to \PGg\PGg$, giving an almost pure sample of gluon jets.  Analysis of this sample will greatly improve our understanding of gluon jets and will contribute to strategies for quark-gluon discrimination in final states at hadron colliders~\cite{Gao:2019mlt,Luo:2019nig}.   Additional unique aspects of QCD at $\ee$ Higgs factories are discussed in Section 8.4 of~\cite{ILCInternationalDevelopmentTeam:2022izu}.

\subsubsection{Flavour physics with W bosons: CKM matrix elements}
\label{sec:phys:WWCKM}

A linear collider already at its energy stages below \SI{1}{TeV} typically produces a few $10^8$ $\PW$ bosons in a clean \ee\ environment. The hadronic decays of these real \PW\ bosons offer an opportunity to measure CKM matrix elements in a complementary approach to \PB\ factories.
In particular, there is at the moment an intriguing discrepancy at the $3\sigma$ level in the value of $|V_{\PQc\PQb}|$ from \PB\ decays, being at $(42.19 \pm 0.78) \times 10^{-3}$ from inclusive $\PB$ decays and at $(39.10 \pm 0.50) \times 10^{-3}$ from exclusive \PB, $\PBs$ and $\PGLb$ decays.
While this discrepancy is difficult to resolve in $\PB$ decays due to inherent hadronic uncertainties, these are absent in decays of real \PW bosons.
With the large QCD background, the LHC prospects for this measurement are only at the level of \SI{10}{\%}. 

Moreover, $|V_{\PQc\PQb}|$ is not the only point of interest, rather all six CKM matrix elements that do not involve the top quark can be determined via hadronic $\PW$ decays in  a manner complementary to $\PB$ and $\PD$ meson decays at corresponding factories.
Table~\ref{tab:CKMprecision} summarises the branching fractions, expected number of $\PW$ decays for $10^8$ $\PW$s and the following ideal determination uncertainty for these six elements.

\begin{table}[h!]
\begin{center}
\begin{tabular}{c|c c c c c c }
 \PWm & $\PAQu\PQd$ & $\PAQu\PQs$ & $\PAQu\PQb$ & $\PAQc\PQd$ & $\PAQc\PQs$ & $\PAQc\PQb$\\\hline
 BR & 31.8$\%$ & 1.7$\%$ & $4.5 \times 10^{-6}$ & 1.7$\%$ & 31.7$\%$ & $5.9 \times 10^{-4}$ \\
 $N_{ev}$ & $32 \times 10^6$ & $1.7 \times 10^6$ & 450 & $1.7 \times 10^6$ & $32 \times 10^6$ & $59 \times 10^3$ \\
 $\delta^{stat}_{V_{ij}}$ & 0.018$\%$ & 0.077$\%$ & 4.7$\%$ & 0.077$\%$ & 0.018$\%$ & 0.41$\%$
\end{tabular}
\caption{$\PW$ decay branching fractions in the SM, corresponding occurrence for $10^8$ $\PWm$ bosons and CKM matrix element determination precision assuming an efficiency-purity product of unity; adapted from~\cite{deBlas:2024bmz}.}
 \label{tab:CKMprecision}
\end{center}
\end{table}

A first study~\cite{Marzocca:2024mkc} considering 2-fermion processes as background, a parametrised flavour tagging inspired by the IDEA detector concept for FCC-ee and potentially limiting systematic effects showed that for $|V_{\PQc\PQb}|$ a control of the flavour tagging efficiency at the level of about \SI{0.1}{\%} would be sufficient, while in case of $|V_{\PQc\PQs}|$ the efficiency would need to be controlled to better than \SI{0.01}{\%} in order not to limit the achievable precision. Based on this study, the report of the ECFA Higgs/Electroweak/Top Factory study~\cite{ECFA-HF-Report} concluded that if the flavour tagging efficiency could be controlled to a precision of \SI{0.1}{\%},  $|V_{\PQc\PQb}|$ could reach a precision of \SI{0.15}{\%}, and $|V_{\PQc\PQs}|$ of \SI{0.05}{\%}, to be compared to the current precisions of \SI{3.4}{\%} and \SI{0.6}{\%}, respectively.

A more comprehensive set of background processes and systematic uncertainties has been considered in a full-simulation study based on the ILD version for CEPC~\cite{Liang:2024hox}. This study showed that it is not sufficient to consider only 2-fermion backgrounds, since also 4-fermion processes contribute significantly to the final selection. It also showed that systematic uncertainties should improve by about a factor 4 over LEP systematics in order not to limit the final result, and that in terms of statistical precision, \SI{5}{\abinv} of unpolarised data give about the same result as \SI{2}{\abinv} of polarised data. 

ILD and CLD are working towards a full analysis of all $\PW\PW$ and single-$\PW$ channels. Detailed detector simulation studies are of particular importance given the significant impact of the flavour tagging, which will be discussed in the following section. As the CEPC study showed, the statistical precision at a linear collider is expected to be comparable to those at circular collider due to the strong enhancement of $\PW$ production processes from beam polarisation. The much smaller beam spot at a linear collider gives an additional advantage in flavour tagging. $\PZ$-pole operation will be an important ingredient to minimising the systematic uncertainties related to flavour tagging. The final results of the ongoing studies are needed for a reliable quantitative assessment of the potential of CKM matrix element determinations from hadronic $\PW$ decays at future $\ee$ colliders.


\subsection{Higgs at high(est) energies}
\label{sec:phys:highEHiggs}

The higher energy stages are of crucial importance to a comprehensive Higgs program: they add many more produced Higgs bosons and thereby typically improve precisions on the Higgs couplings to gauge bosons and fermions by a factor two or more, but -- even more importantly -- they do so by opening different production modes of the Higgs boson, allowing a complementary scrutiny of any observed deviation. The impact on the Higgs couplings to fermion and gauge bosons is discussed in Secs.~\ref{sec:singleHiggs} and~\ref{sec:glob}. 

Here, we will focus on illuminating the Higgs potential, the origin of electroweak symmetry breaking (EWSB) and whether it provides the strong first-order phase transition required to explain the observed baryon-antibaryon asymmetry in the universe via electroweak baryogenesis. We will also comment on the interplay with gravitational wave experiments and primordial black hole searches.

\subsubsection{The Higgs self-coupling}
\label{sec:phys:selfcoup}
Many of the open questions of particle physics are related to the Higgs sector and in particular to the Higgs potential, which for this reason is often referred to as the ``holy grail'' of particle physics. 
While the Higgs discovery in 2012 has confirmed the existence of the Higgs potential, its actual form realised in nature as well as its physical origin remain until now largely unknown. Since the Higgs potential triggers EWSB, obtaining information about its shape is crucial to understand how the electroweak phase transition (EWPT) took place in the early universe. 

The SM postulates a minimal form of the Higgs potential, with a single Higgs boson that is an elementary particle. While the location of the electroweak minimum and curvature of the Higgs potential around this minimum are known, the trilinear and quartic Higgs self-couplings $\trlh$ and $\qtlh$ --- i.e.\ the coefficients of the $\PH^3$ and $\PH^4$ terms in the Higgs potential, respectively --- are only loosely constrained so far.

\begin{figure}[ht]
\centering
  \includegraphics[width=0.6\textwidth]{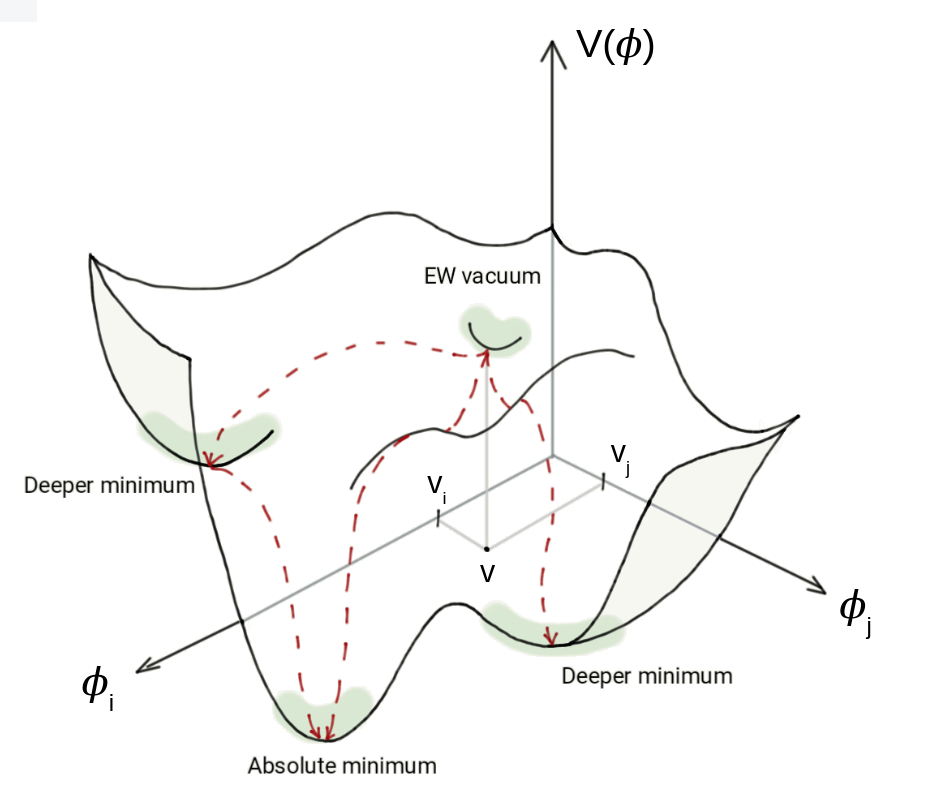}
\caption{Simplified illustration of the form of the Higgs potential $V$ in a 
two-dimensional field space (fields $\phi_i$ and $\phi_j$) for a theory with an extended Higgs sector~\cite{Radchenko:2025}. The electroweak vacuum and 
possible tunnelling directions into deeper minima are indicated.}
\label{fig:higgspot}
\end{figure}

Beyond the Standard Model (BSM) theories often feature extended scalar sectors, resulting in a complicated form of the Higgs potential in the multi-dimensional field space of all contributing scalar fields. 
In particular, the Higgs potential receives contributions not only from the detected Higgs boson\footnote{We continue to denote the 125-GeV Higgs boson here with \PH, independently of its mass ordering with other neutral, CP-even scalars in extended Higgs sectors.} $\PH$, but also from all additional BSM scalars that may be present but have escaped detection so far. 
As a consequence, the Higgs potential can have a rich structure of minima, as illustrated conceptually in Fig.~\ref{fig:higgspot}. Understanding the Higgs potential and the mechanism of the EWPT is especially important when investigating the scenario of electroweak baryogenesis (EWBG)~\cite{Kuzmin:1985mm,Cohen:1993nk} as a possible explanation for the observed baryon-antibaryon asymmetry in the universe (BAU). 

In the present section we focus on the self-couplings of the detected Higgs boson, while Higgs self-couplings involving additional Higgs fields are discussed in Sec.~\ref{sec:phys:bsm:BSMHiggs}. 

\paragraph[Di-Higgs measurements at LHC and HL-LHC]{Di-Higgs measurements at LHC and HL-LHC} Existing constraints on $\trlh$ from the LHC have mainly been obtained from the searches for the Higgs pair production process. Here gluon fusion is the dominant channel, and the dependence on $\trlh$ appears at leading order. A large destructive interference between the SM-type diagrams with and without $\trlh$ implies that the total cross section for di-Higgs production changes very substantially -- by up to two orders of magnitude -- if $\trlh$ is varied around the SM value. The current upper bound on the di-Higgs production cross section from ATLAS and CMS translates into an upper limit on $\trlh$ that is about 7 times larger than the SM value (while a lower limit can also be set around $-1$), assuming SM-like values for all other Higgs couplings besides $\trlh$~\cite{ATLAS:2024ish,CMS:2022dwd} 
(this assumption can be relaxed by incorporating data from single Higgs production~\cite{ATLAS:2022jtk}). The current projections for the HL-LHC -- using the combined integrated luminosity that is expected to be collected by both ATLAS and CMS -- yield a \SI{68}{\%} CL of $[0.74, 1.29]$ for $\trlh$ if the SM value is realised in nature ($\kappa_\lambda = 1$)~\cite{ATL-PHYS-PUB-2025-018}. The di-Higgs cross-section measurement also dominates the precision on $\trlh$ in the global interpretation in NLO SMEFT, which projects a precision on $\trlh$ very similar to the single-parameter extraction as discussed in Sec.~\ref{sec:glob:NLO}.

\begin{figure}[htb]
\begin{center}
  \begin{subfigure}{.5\textwidth}
    \centering
    \includegraphics[width=0.96\hsize]{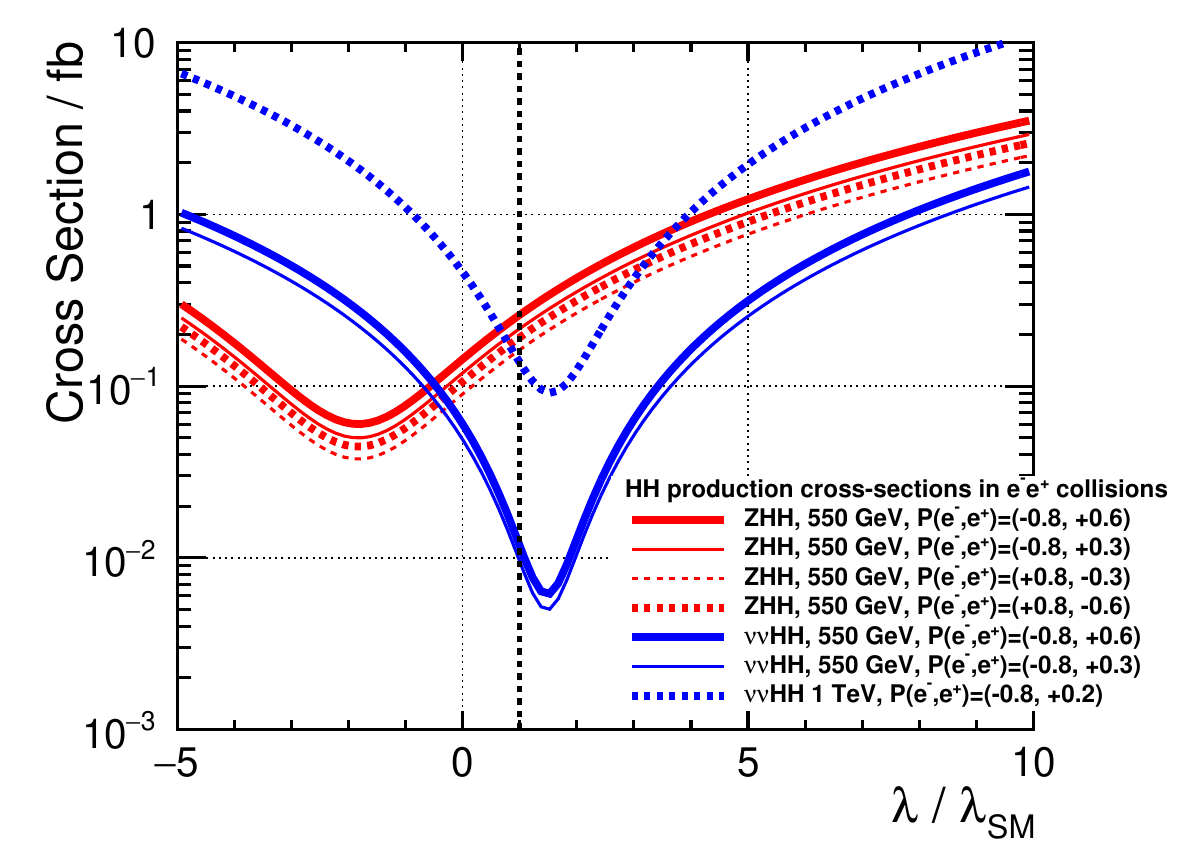}
    \caption{}
    \label{fig:HHxsec:xsec_vs_kala}    
  \end{subfigure}\hfill%
  \begin{subfigure}{.5\textwidth}
    \centering
    \includegraphics[width=0.96\hsize]{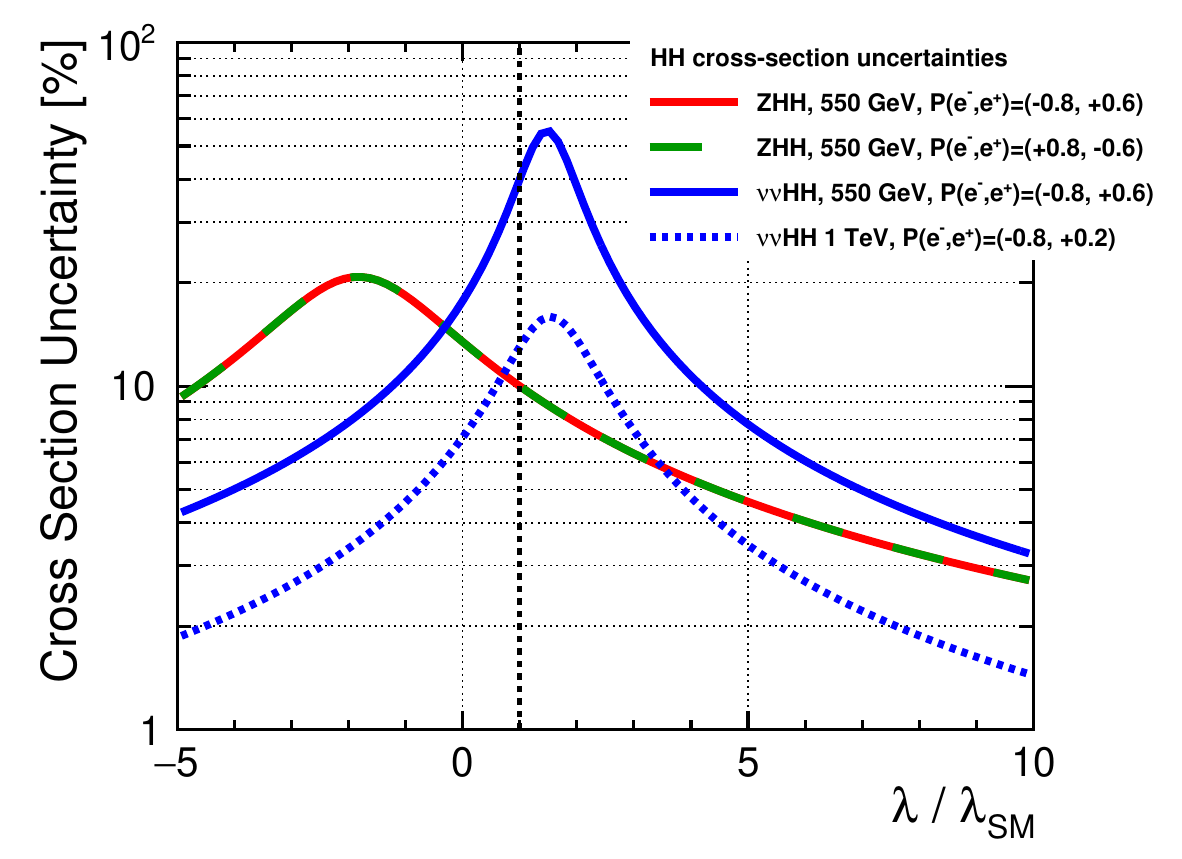}
    \caption{}
    \label{fig:HHxsec:dxsec_vs_kala}    
    \end{subfigure}\hfill
\end{center}
\caption{(a) Cross sections of di-Higgs production in $\ee$ collisions as function of $\kappa_\lambda = \lambda / \lambda_{\mathrm{SM}}$:  $\ee \to \PZ\PH\PH$ at \SI{550}{GeV} for different beam polarisation configurations (red lines) and di-Higgs production from $\PW\PW$ fusion at \SI{550}{GeV} (blue, solid lines) and at \SI{1}{TeV} (blue, dashed line). (b) Projected cross-section-measurement precisions at LCF for $\PZ\PH\PH$ at \SI{550}{GeV} (red / green) and $\PW\PW$ fusion (blue) at \SI{550}{GeV} (solid) and \SI{1}{TeV} (dashed). In (a) and (b), the vertical black dashed lines indicate $\kappa_\lambda = 1$, i.e.\ the SM value, both from~\cite{Berggren:2025fpw}.\label{fig:HHxsec}}
\end{figure}

\paragraph[Di-Higgs measurements at linear colliders]{Di-Higgs measurements at linear colliders} At an $\ee$ linear collider with a centre-of-mass energy of \SI{500}{GeV} or higher, di-Higgs production can be directly measured in two processes, $\ee\to \PZ\PH\PH$ and the $\PW\PW$-fusion process $\ee\to \PGn\PAGn \PH\PH$, in a largely model-independent way. This is a qualitatively new feature of high-energy running of a linear collider, distinguishing its physics capabilities from those of $\ee$ colliders at lower energies. The direct measurement of the two di-Higgs production processes allows the determination of the self-coupling $\trlh$ from its lowest-order contribution to these processes  
in a way that is independent of other possible BSM effects on Higgs couplings~\cite{Djouadi:1999gv,Barklow:2017awn} that affect the indirect determination of $\trlh$ from its loop contributions to other observables at lower energies.  
The cross sections of both di-Higgs production processes are shown in Fig.~\ref{fig:HHxsec:xsec_vs_kala} as a function of $\kappa_\lambda = 1$ for several choices of beam polarisations and centre-of-mass energy.
Remarkably, for values of $\kappa_\lambda$ around unity, the two reactions $\ee\to \PZ\PH\PH$ and $\ee\to \nu\bar \nu \PH\PH$ have opposite dependence on $\lambda_{\PH\PH\PH}$, so that an enhancement in $\PZ\PH\PH$ can be checked by observation of a deficit in the $\PW\PW$-fusion reaction.

With recent improvements in the full-simulation studies of $\PH\PH$ pair production, the projected accuracy provided by a \SI{550}{GeV} linear \ee\ collider is now \SI{15}{\%} for $\kappa_\lambda = 1$, based on an integrated luminosity of \SI{4}{\abinv} assuming a positron polarisation of \SI{60}{\%} -- or \SI{4.4}{\abinv} if only \SI{30}{\%} would be available~\cite{Berggren:2025fpw}.  This value is an update (superseding the values of \SI{20}{\%} from~\cite{List:2024ukv} and \SI{27}{\%} from~\cite{Fujii:2015jha}), reflecting analysis improvements including the combination of results from the $\PZ\PH\PH$ and $\nu\bar\nu \PH\PH$ production channels as well as significant progress made on flavour tagging and jet reconstruction~\cite{Berggren:2025fpw}; see also Sec.~\ref{sec:hlreco}. 
The LCF, thanks to operating at \SI{10}{Hz}, aims to collect \SI{8}{\abinv} with \SI{60}{\%} positron polarisation, and would thus reach even a precision of \SI{11}{\%} for $\kappa_\lambda = 1$.
It should be noted that all other projections, i.e.\ for \SI{1}{TeV}, \SI{1.5}{TeV} or \SI{3}{TeV} have not yet been updated with modern high-level reconstruction and analysis algorithms. Therefore, the projected accuracy for $\kappa_\lambda = 1$ for e.g.\ CLIC operating at \SI{3}{TeV} remains for now at the previously obtained $[\SI{-8}{\%},\SI{+11}{\%}]$~\cite{Roloff:2019crr}, while it is expected that all the analyses will see similar improvements as already demonstrated for \SI{550}{GeV}.

\begin{figure}[htb]
\begin{center}
  \begin{subfigure}{.5\textwidth}
    \centering
    \includegraphics[width=0.96\hsize]{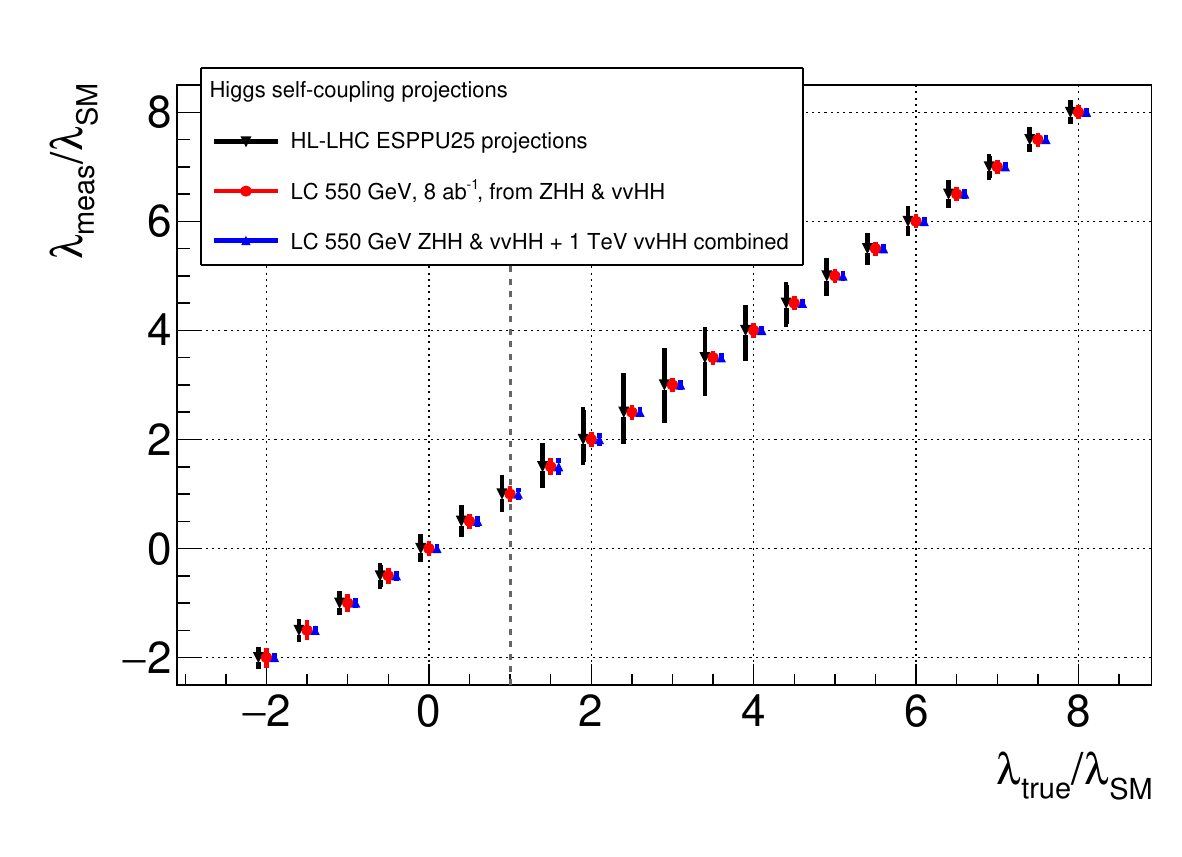}
    \caption{}
    \label{fig:kappalambda550all}    
  \end{subfigure}\hfill%
  \begin{subfigure}{.5\textwidth}
    \centering
    \includegraphics[width=0.96\hsize]{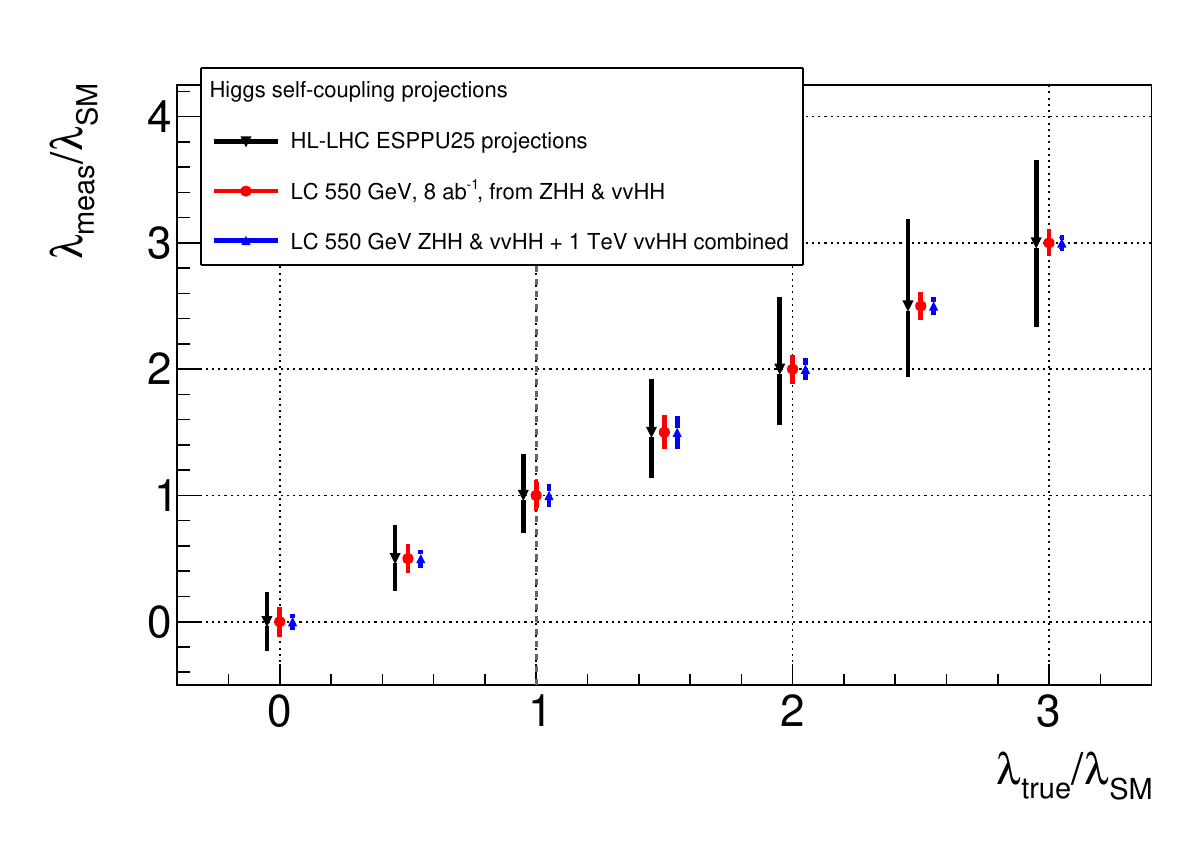}
    \caption{}
    \label{fig:kappalambda550zoom}    
    \end{subfigure}\hfill
\end{center}
\caption{(a) Projected accuracies on $\trlh$ at the HL-LHC (black)~\cite{ATL-PHYS-PUB-2025-018} and an $\ee$ collider at a centre-of-mass energy of \SI{550}{GeV} alone (red) and combined with \SI{1}{TeV} (blue) in dependence of the actual value of $\trlh$ that is realised in nature. (b) zoomed version. \label{fig:kappalambda550}}
\end{figure}

\begin{figure}[htb]
\begin{center}
  \begin{subfigure}{.5\textwidth}
    \centering
    \includegraphics[width=0.96\hsize]{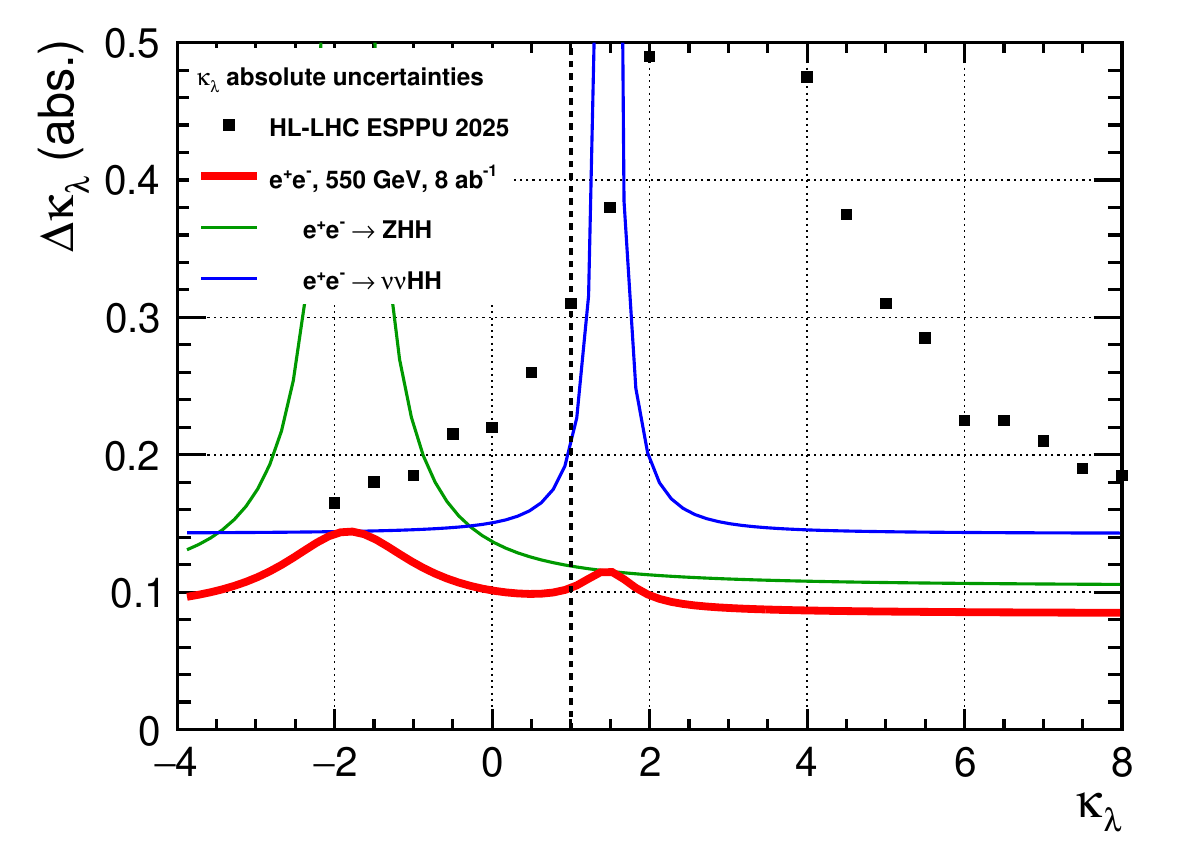}
    \caption{}
    \label{fig:dkala_vs_kala_LCF550_only}    
  \end{subfigure}\hfill%
  \begin{subfigure}{.5\textwidth}
    \centering
    \includegraphics[width=0.96\hsize]{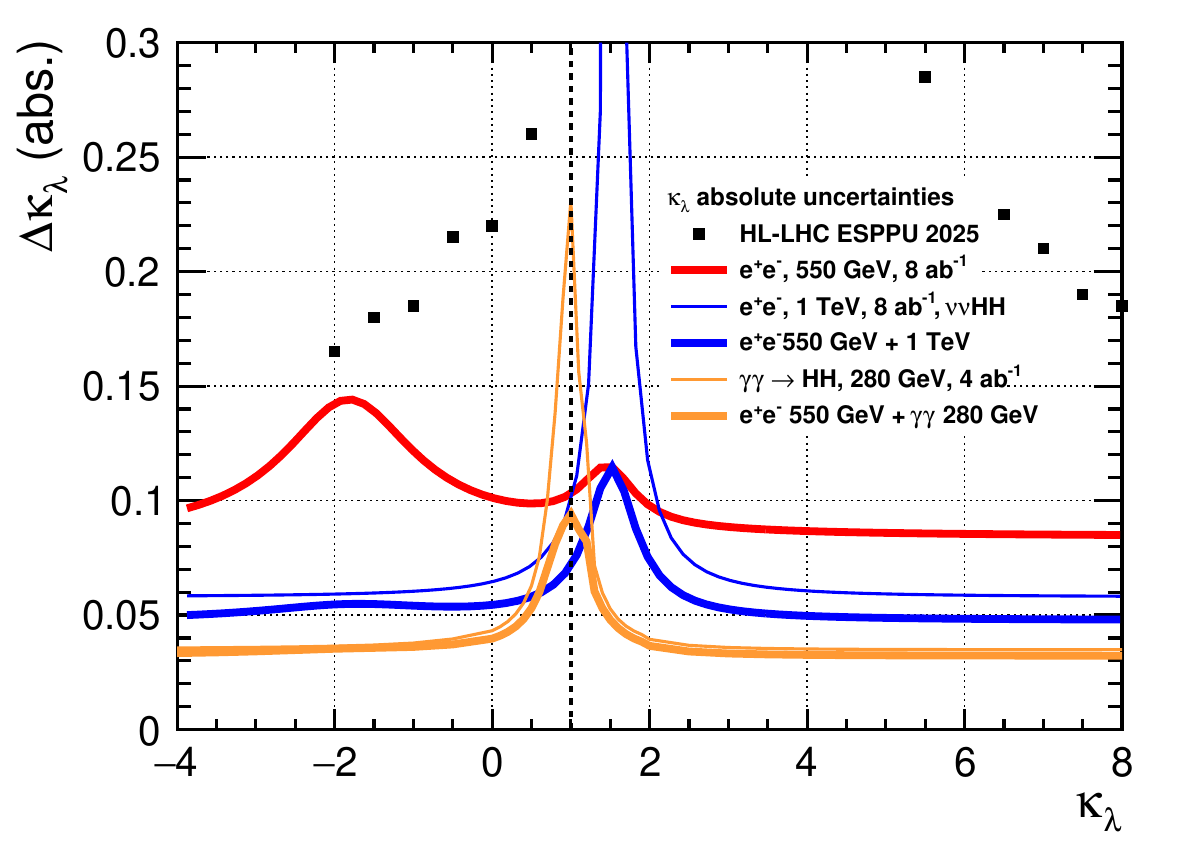}
    \caption{}
    \label{fig:dkala_vs_kala_LCF550_upgrades}    
    \end{subfigure}\hfill
\end{center}
\caption{(a)  Absolute precision on ]$\kappa_{\lambda}$ as a function of $\kappa_{\lambda}$ from the $\PZ\PH\PH$ and $\PW\PW$-fusion processes at the LCF at \SI{550}{GeV}, showing the complementary information from the two production modes. (b) The LCF550 precision from (a) and the potential of further  upgrades to \SI{1}{TeV} or a $\gamma\gamma$-collider at a centre-of-mass energy of \SI{280}{GeV} as discussed in Secs.~\ref{sec:phys:altmodes:gg_HH} and~\ref{sec:acc:altmodes}. Modified from~\cite{Berggren:2025fpw}.    \label{fig:dkala_vs_kala_LCF550}    
}
\end{figure}

\paragraph[BSM projections and comparison with HL-LHC]{BSM projections and comparison with HL-LHC} However, the expected precisions for both, the cross-section measurements and the resulting precision on $\kappa_\lambda$, depend strongly on the value of $\trlh$ realised in nature. 
Figure~\ref{fig:HHxsec:dxsec_vs_kala} shows this dependency for the projections of the cross-section precision achievable at LCF, while Figs.~\ref{fig:kappalambda550} and~\ref{fig:dkala_vs_kala_LCF550} illustrate the corresponding effect on the projected precision on $\kappa_\lambda$.
For instance, if the value of $\trlh$ realised in nature corresponds to $\kappa_\lambda =2$ (which would be favoured in scenarios giving rise to a strong first-order electroweak phase transition, as discussed below), the projected uncertainty on $\trlh$ at the HL-LHC grows to about $\pm 0.6$ (i.e.\ \SI{30}{\%} relative). 
Owing to the different interference pattern in the $\PZ\PH\PH$ channel, the accuracy at the LCF at \SI{550}{GeV} improves to around $\kappa_\lambda =2\pm 0.1$  (i.e.\ \SI{5}{\%} relative).

Figure~\ref{fig:kappalambda550} compares the current projections for the HL-LHC~\cite{ATL-PHYS-PUB-2025-018} and for an LCF, as a function of $\kappa_\lambda$. 
It is expected that future projections for the HL-LHC will be improved over the present case, but this is also true for the case of the \ee\ projections, which by far are not yet fully exploiting the capabilities of the proposed detectors and modern data analysis techniques. 

The interplay of the two di-Higgs production modes accessible at \SI{550}{GeV} and their complementary sensitivity to BSM value of $\trlh$ is illustrated in Fig.~\ref{fig:dkala_vs_kala_LCF550_only}, which shows the absolute uncertainty on $\kappa_\lambda$ as function of the value of $\kappa_lambda$, again in comparison to the HL-LHC projections. Figure~\ref{fig:dkala_vs_kala_LCF550_upgrades} illustrates the further improvements in accuracy expected from additional $\ee$ data at \SI{1}{TeV}\footnote{Analysis not yet updated in view of new high-level reconstruction tools.} or a $\PGg\PGg$ collider at an $\PGg\PGg$ centre-of-mass energy of \SI{280}{GeV} (c.f.\ Secs.~\ref{sec:phys:altmodes:gg_HH} and~\ref{sec:acc:altmodes}.).

\begin{figure}[htb]
    \centering
    \begin{subfigure}{0.52\textwidth}
    \includegraphics[width=\textwidth]{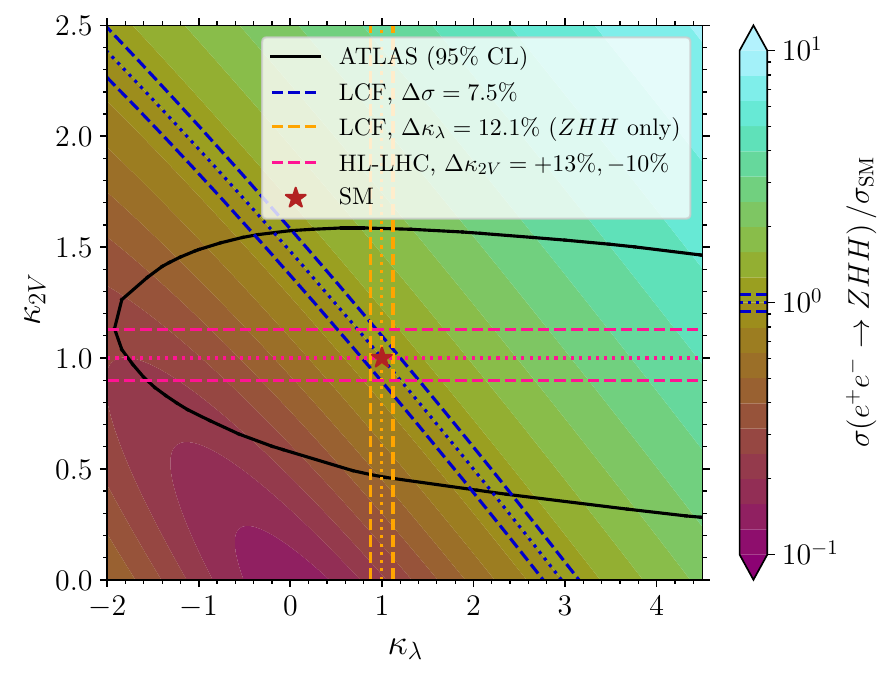}
    \caption{}
    \label{fig:kala_vs_k2v_ee_LCF}
    \end{subfigure}
    \begin{subfigure}{0.47\textwidth}
    \includegraphics[width=\textwidth]{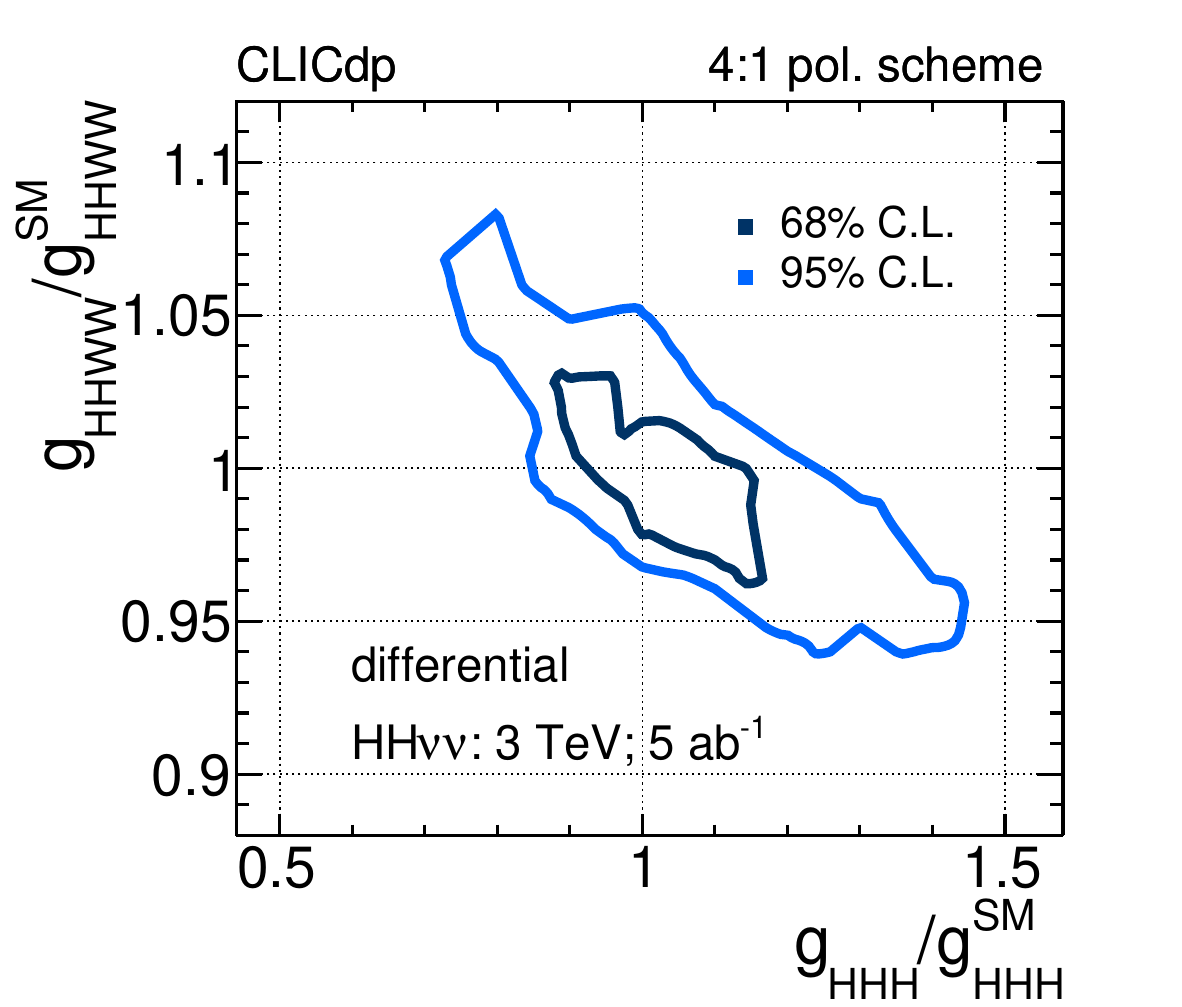}
    \caption{}
    \label{fig:kala_vs_k2v_ee_CLIC}
    \end{subfigure}
    \caption{(a) $\ee\to \PZ\PH\PH$ cross section, normalised to its SM value, shown as contours in the plane of $\kappa_\lambda$ and $\kappa_{2V}$~\cite{FranArcoPriv}. Also shown are current limits from ATLAS at \SI{95}{\%} CL~\cite{ATLAS:2024ish} (black) and the HL-LHC projection for $\kappa_{2V}$ (pink), as well as the cross-section (blue) and $\kappa_\lambda$ accuracy (orange, note that the \SI{12.1}{\%} here correspond to only \SI{30}{\%} positron polarisation, single-parameter extraction) from LCF550. (b) Joint fit of $\kappa_{2W}$ and $\kappa_\lambda$ to differential di-Higgs cross-section measurements at CLIC at \SI{3}{TeV}~\cite{Roloff:2019crr}. \label{fig:kala_vs_k2v_ee}} 
\end{figure}

\paragraph[Interplay with $\kappa_{2V}$]{Interplay with $\kappa_{2V}$} In the SMEFT interpretations discussed in Sec.~\ref{sec:glob}, $\kappa_{2V}$ closely related to $\kappa_{W}$ and $\kappa_{Z}$, which will be constrained to the permil-level from single-Higgs production. However this close relation is a model choice, and in principle $\kappa_{2V}$ can vary independently of the $\kappa_{V}$'s, like it is the case in Higgs Effective Field Theory (HEFT, c.f.~\ref{sec:glob}), or any other model considering a non-linear realisation of EWSB. 
The effects on the $\ee\to ZHH$ cross section of simultaneously varying $\kappa_\lambda$ and $\kappa_{2V}$, where the latter is the coupling modifier of the $HHVV$ interactions ($V=W^\pm,\,Z$), are displayed in Fig.~\ref{fig:kala_vs_k2v_ee_LCF} for the LCF with $\sqrt{s}=550\text{ GeV}$~\cite{FranArcoPriv}. 
This shows that with the anticipated precision on $\kappa_{2V}$ from HL-LHC, the impact on the accuracy of $\kappa_\lambda$ is non-negligible. However, this can be reduced substantially when including differential information. This has been pioneered for CLIC, as shown in Fig.~\ref{fig:kala_vs_k2v_ee_CLIC}. A combination of HL-LHC and all di-Higgs measurements at a linear collider at several centre-of-mass energies, including differential ones, is expected to improve the  constraints on $\kappa_{2V}$ significantly.


\paragraph[Quartic coupling]{Quartic coupling}
Finally, a qualitatively new feature at centre-of-mass energies of about \SI{1}{TeV} and above is the sensitivity to the triple Higgs-boson production processes 
$\ee \to \PZ\PH\PH\PH$ and $\ee \to \PGn\PAGn \PH\PH\PH$, which provide 
experimental access to the quartic Higgs-boson self-coupling $\qtlh$. Because of their dependence also on the trilinear Higgs-boson self-coupling $\trlh$ (some contributions to the triple Higgs-boson production processes even involve its square), the triple Higgs-boson production processes also provide complementary information on $\trlh$ that can be combined with the results that are obtained from the di-Higgs production processes.

\begin{figure}[ht]
\centering
  \includegraphics[width=0.6\textwidth]{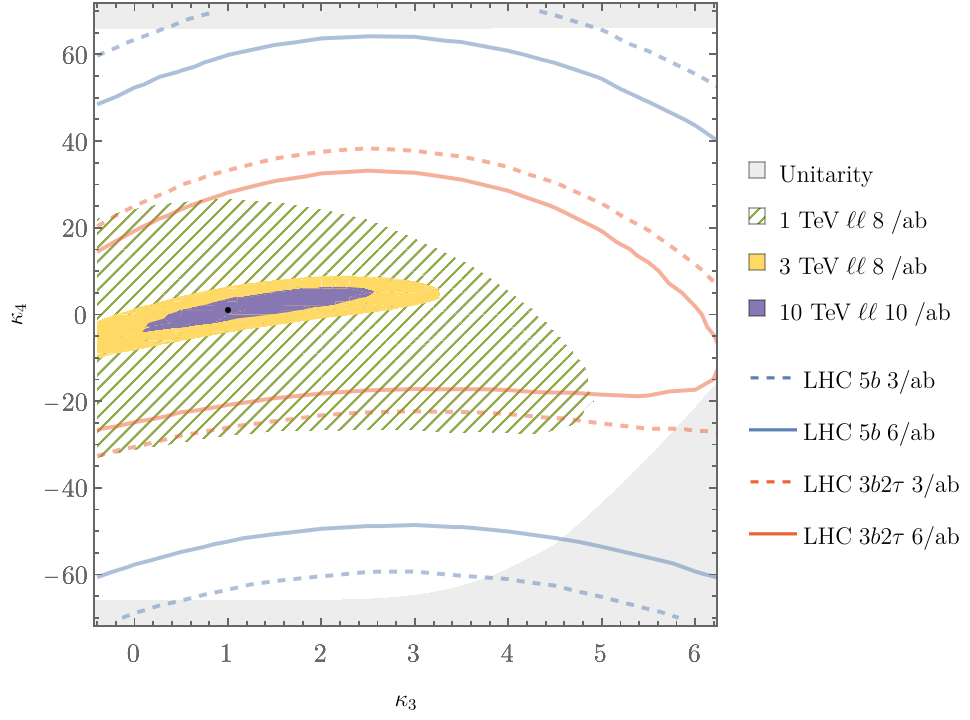}
\caption{Prospects of future lepton colliders with a centre-of-mass energy of \SI{1}{TeV}, \SI{3}{TeV} and \SI{10}{TeV} for constraining the trilinear (horizontal axis) and quartic (vertical axis) Higgs self-couplings, both normalised to their tree-level values in the SM, at the \SI{95}{\%} CL using exclusively searches for triple-Higgs production~\cite{Stylianou:2023xit}. These prospects are presented in comparison to the projected \SI{95}{\%} CL\ contours for the $5\PQb$ and $3\PQb 2\PGt$ analyses at the HL-LHC. The shaded gray area indicates the region that is excluded by the bound from (tree-level) perturbative unitarity. }
\label{fig:kappa3kappa4}
\end{figure}

Figure~\ref{fig:kappa3kappa4} illustrates the sensitivity of future lepton colliders with centre-of-mass energies of \SI{1}{TeV}, \SI{3}{TeV} and \SI{10}{TeV} to set constraints on the trilinear (horizontal axis) and quartic (vertical axis) Higgs-boson self-couplings, both normalised to the tree-level SM values~\cite{Stylianou:2023xit}. The prospective sensitivities shown here are obtained only from the triple Higgs-boson production processes, without additional information from single- and di-Higgs production. The prospects for the lepton colliders are compared to the ones for the HL-LHC obtained in a recent exploratory study for the $5\PQb$ channel and the $3\PQb2\PGt$ channel~\cite{Stylianou:2023xit}. These prospective bounds go significantly beyond the current theoretical constraints from tree-level perturbative unitarity, shown in gray in Fig.~\ref{fig:kappa3kappa4}. 
The displayed results indicate that for a SM-like value of $\trlh$, a \SI{1}{TeV} lepton collider gives similar constraints on $\lambda_{\PH\PH\PH\PH}$ as HL-LHC, while for a larger value of $\trlh$, it would improve significantly over HL-LHC bounds. Higher-energetic lepton colliders (see also~\cite{Maltoni:2018ttu,Chiesa:2020awd,Gonzalez-Lopez:2020lpd,Papaefstathiou:2023uum}) can drastically improve on the HL-LHC capabilities for any value of $\trlh$.

\subsubsection{Expected deviations in and constraints on extended Higgs sectors} 
\label{sec:phys:highEHiggs:BSM}

While the existing bounds on $\trlh$ are rather weak, they nevertheless already probe so far untested parameter regions of physics beyond the SM~\cite{Abouabid:2021yvw,Bahl:2022jnx}, because $\trlh$ can receive large mixing effects and/or very large loop corrections from the additional scalars (which we will generically denote $\Phi$ in the following) of models with extended Higgs sectors. These large radiative contributions to $\trlh$ --- often referred to as mass-splitting or non-decoupling effects --- were first pointed out in~\cite{Kanemura:2002vm,Kanemura:2004mg}, but have now been shown to occur in various BSM models with extra scalars (see~\cite{Aoki:2012jj,Kanemura:2015fra,Kanemura:2015mxa,Arhrib:2015hoa,Kanemura:2016sos,Kanemura:2016lkz,He:2016sqr,Krause:2016xku,Kanemura:2017wtm,Kanemura:2017wtm,Kanemura:2017gbi,Chiang:2018xpl,Basler:2018cwe,Senaha:2018xek,Braathen:2019pxr,Braathen:2019zoh,Kanemura:2019slf,Basler:2019iuu,Basler:2020nrq,Braathen:2020vwo,Bahl:2022jnx,Bahl:2022gqg,Bahl:2023eau,Aiko:2023xui,Aiko:2023nqj,Basler:2024aaf, Bahl:2025wzj,Braathen:2025qxf,Braathen:2025svl} for studies in models with additional singlets, doublets, triplets, etc.\ or combinations thereof). Moreover, the physical nature of these effects was confirmed by explicit two-loop calculations~\cite{Braathen:2019pxr,Braathen:2019zoh}. These corrections are driven by couplings of the generic form $g_{\PH\PH\Phi\Phi}\propto(m_\Phi^2-\mathcal{M}^2)/v^2$, where $m_\Phi$ denotes the physical mass of a BSM scalar $\Phi$, $\mathcal{M}$ is the mass scale that controls the decoupling of the BSM scalar(s), and $v$ is the EW vacuum expectation value. In scenarios (away from the decoupling limit) where a splitting occurs between $\mathcal{M}$ and $m_\Phi$, the $g_{\PH\PH\Phi\Phi}$ couplings can be substantial, resulting in large BSM contributions to $\kappa_\lambda$. It should be emphasised that these loop effects involving BSM scalars and $g_{\PH\PH\Phi\Phi}$ couplings are not a perturbation of the tree-level expression of the trilinear Higgs coupling, but rather a new class of contributions only entering at the loop level. They can therefore become larger than the tree-level contribution, without being associated with a violation of perturbative unitarity --- this situation is analogous to that of loop-induced decays of the Higgs boson like e.g.\ $\PH\to\PGg\PGg$ or $\PH\to \PZ\PGg$. A very significant upward shift in $\trlh$ is also motivated in many scenarios giving rise to a strong first-order EWPT which is required for electroweak baryogenesis, see~\cite{Grojean:2004xa,Kanemura:2004ch} or for more recent works e.g.~\cite{Basler:2017uxn,Basler:2019iuu,Biekotter:2021ysx,Biekotter:2022kgf} for the cases of the (C)2HDM and N2HDM. In these latter cases, the parameter region featuring a strong first-order EWPT and a potentially detectable gravitational wave signal at the future space-based observatory LISA is correlated with an enhancement of $\lambda_{\PH\PH\PH}$ compared to the SM value by about a factor of 2 (see also Fig.~\ref{fig:FOEWPT:2HDM} below).

The existence of scenarios where $\trlh$ receives large radiative corrections naturally raises the question of how sizeable higher-order corrections to other Higgs properties (e.g.\ single-Higgs couplings) involving $g_{\PH\PH\Phi\Phi}$ couplings can become. In turn, one may wonder in which observable, or which Higgs coupling extracted from the measurement(s), one would first observe a deviation from the SM prediction in these types of scenarios. It should be noted that deviations in $\trlh$ may also be small or of the same order as those in single-Higgs couplings: this is for instance the case in decoupling scenarios of supersymmetric models \cite{Hollik:2001px,Dobado:2002jz,Brucherseifer:2013qva,Nhung:2013lpa,Muhlleitner:2015dua,Borschensky:2022pfc} or in composite Higgs models \cite{Grober:2010yv,Gillioz:2012se,Grober:2013fpx,Grober:2016wmf,DeCurtis:2023pus}.

On the other hand, various studies have found that there exist scenarios of new physics in which the size of the BSM deviations in $\trlh$ can be significantly larger, by more than two orders of magnitude, than those in the couplings of $\PH$ to gauge bosons and fermions, as will be discussed in more detail below --- see for instance~\cite{Gupta:2012mi,Gupta:2013zza,DiLuzio:2017tfn,Abouabid:2021yvw,Aoki:2012jj,Kanemura:2015fra,Kanemura:2015mxa,Arhrib:2015hoa,Kanemura:2016sos,Kanemura:2016lkz,He:2016sqr,Krause:2016xku,Kanemura:2017wtm,Kanemura:2017wtm,Kanemura:2017gbi,Chiang:2018xpl,Basler:2018cwe,Senaha:2018xek,Braathen:2019pxr,Braathen:2019zoh,Kanemura:2019slf,Basler:2019iuu,Basler:2020nrq,Braathen:2020vwo,Bahl:2022jnx,Bahl:2022gqg,Bahl:2023eau,Aiko:2023xui,Aiko:2023nqj,Basler:2024aaf, Bahl:2025wzj} for investigations with explicit calculations in extended Higgs sectors and e.g.~\cite{Durieux:2022hbu} for analyses based on EFT arguments. To be more specific, two main types of situations can be distinguished in which corrections to $\trlh$ can be significantly larger than those in other (single) Higgs couplings: \textit{(i)} models with some particular structure of the BSM sector, and \textit{(ii)} scenarios, away from the decoupling limit, with relatively light new physics with significant couplings to the detected Higgs boson (as described above in terms of generic $g_{\PH\PH\Phi\Phi}$ couplings).

BSM models of the first case, which were systematically investigated in~\cite{Durieux:2022hbu}, include for instance the extension of the SM with a custodial EW quadruplet --- a single field extension of the SM, where the enhancement of the effects in $\trlh$ occurs already at the tree level --- or the Gegenbauer's Twin models~\cite{Durieux:2022sgm} where the Higgs boson is a pseudo Nambu-Goldstone boson. Turning next to the second case, power counting arguments~\cite{Bahl:2022jnx,Bahl:2025xyz} in the limit of large $g_{\PH\PH\Phi\Phi}$ couplings show that the leading one-loop BSM contributions to the trilinear Higgs coupling are of $\mathcal{O}(g_{\PH\PH\Phi\Phi}^2)$, while those in Higgs couplings to fermions or gauge bosons grow at most linearly with $g_{\PH\PH\Phi\Phi}$.

\begin{figure}[tb]
\centering
  \begin{subfigure}{0.48\textwidth}
        \includegraphics[width=\textwidth]{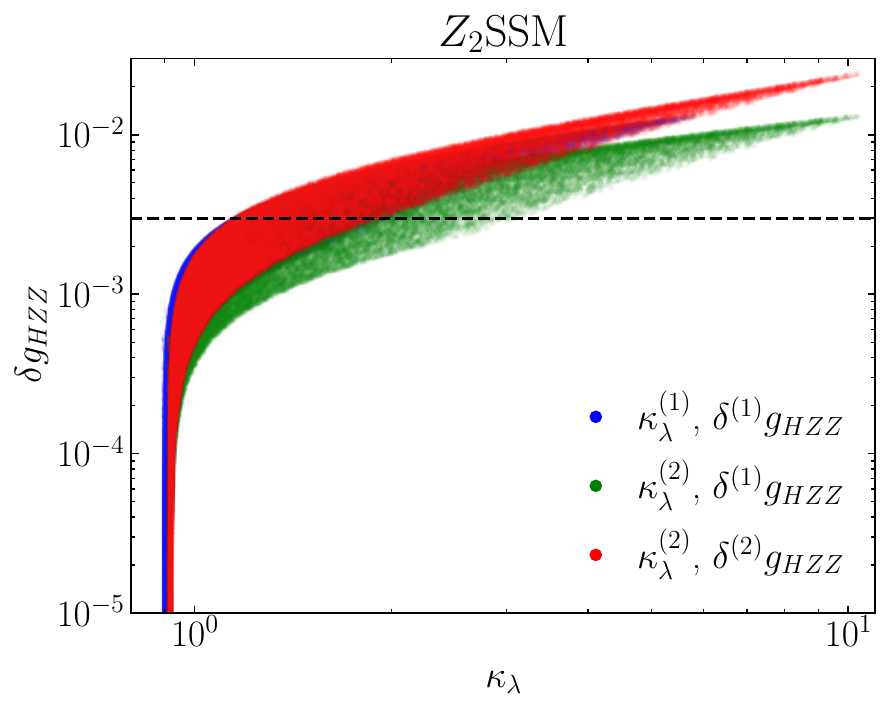}
        \caption{}
        \label{fig:kala_vs_hZZ:Z2SSM}
  \end{subfigure}
  \begin{subfigure}{0.50\textwidth}
        \includegraphics[width=\textwidth]{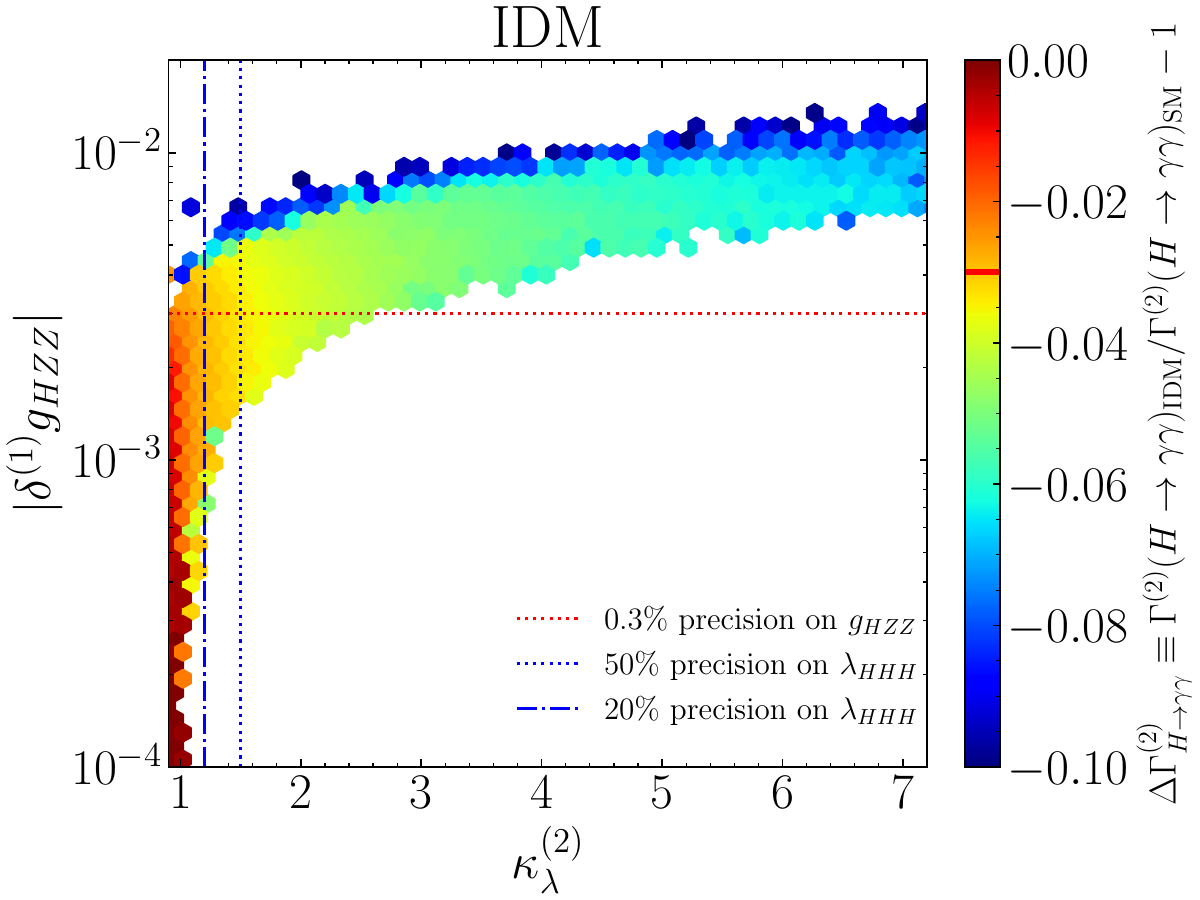}
        \caption{}
        \label{fig:kala_vs_hZZ:IDM}
  \end{subfigure}
\caption{(a) Parameter scan in the $Z_2$-SSM, shown in the plane of $\kappa_\lambda$ and $\delta \ghzz$. The colour of the points indicates the order (one or two loops) at which these two quantities are computed, as explained in the legend of the plot. The black dashed line corresponds a precision of \SI{0.3}{\%} on the determination of the $\ghzz$ coupling (derived to the expected $1\sigma$ accuracy of FCC-ee on the $\ee \to \PZ \PH$ cross section at 240 GeV~\cite{FCCICHEP:2024}). (b) Parameter scan in the IDM, in the plane of $\kappa_\lambda^{(2)}$ (at two loops) and $\delta^{(1)}\ghzz$ (at one loop). The points are grouped in hexagonal bins, and the colour of each of these bins represents the minimal BSM deviation among the points of the bin. The red solid line in the colour bar corresponds to the expected $1\sigma$ accuracy of FCC-ee on $\PH\to\upgamma\upgamma$~\cite{FCCICHEP:2024}, while the horizontal and vertical lines are described in the legend and the text.}
\label{fig:kala_vs_hZZ}
\end{figure}

\paragraph[Singlet extension]{Singlet extension} A first simple but illustrative example is provided by the $Z_2$-SSM\footnote{We refer the reader to, e.g.,~\cite{Braathen:2020vwo} for an overview of the notations and conventions employed here (considering the $N=1$ case of the $O(N)$-symmetric SSM discussed in~\cite{Braathen:2020vwo}).}, i.e.\ a real-singlet extension of the SM with an unbroken global $Z_2$ symmetry. Due to this symmetry, the BSM scalar $S$ does not mix with the detected Higgs boson at 125 GeV. Its mass takes the form $m_S^2=\mu_S^2+\lambda_{\PH\Phi}v^2$, where $\mu_S$ is the singlet Lagrangian mass term and $\lambda_{\PH\Phi}$ the portal quartic coupling between the singlet and the (SM-like) doublet. In the $Z_2$-SSM, the coupling $\lambda_{\PH\Phi}$ plays the exact role of the generic $g_{\PH\PH\Phi\Phi}$ coupling in the discussion above. In this model, single-Higgs couplings $g_{\PH XX}$, where $X$ can be a gauge boson or fermion, only receive BSM contributions via external-leg corrections --- there are no mixing effects, and moreover vertex-type corrections do not appear because of the singlet nature of the BSM scalar and of the unbroken $Z_2$ symmetry. This allows obtaining compact expressions for the single-Higgs coupling modifier $c_\text{eff}\equiv g_{\PH XX}/g_{\PH XX}^\text{SM}$ at one and two loops~\cite{Bahl:2025xyz}. 
Figure~\ref{fig:kala_vs_hZZ:Z2SSM} presents parameter scan results for the $Z_2$-SSM, in the plane of $\kappa_\lambda$ and the BSM deviation in the $g_{h\PZ\PZ}$ coupling, denoted $\delta g_{h\PZ\PZ}$, calculated at the one- and two-loop levels. Large shifts in $\kappa_\lambda$ are found to be possible, e.g.\ at two loops up to $\kappa_\lambda^{(2)} \approx 3 (2)$ for the case where one-loop (two-loop) corrections to the $\ghzz$ coupling are taken into account, while the effect in $\ghzz$ remains below the $1\,\sigma$ sensitivity of single-Higgs measurements (derived from~\cite{FCCICHEP:2024}, using as example the quoted precision on the measurement of the $\ee \to \PZ\PH$ cross section at the FCC-ee at \SI{240}{GeV}). 

\paragraph[Inert Doublet Model]{Inert Doublet Model} The Inert Doublet Model (IDM) is a second example of a BSM theory in which large BSM deviations can occur in $\trlh$ while effects in other Higgs couplings would not be detectable, even with precision measurements. The IDM is a variant of the 2HDM in which the second doublet is charged under an unbroken $\mathbb{Z}_2$ symmetry --- overviews of the model can be found e.g.\ in~\cite{Deshpande:1977rw,Barbieri:2006dq,Braathen:2024lyl}. Similarly to the case of the $Z_2$-SSM, the masses of the BSM scalars $\Phi=\PSh,\PSA,\ \PSHpm$ of the IDM take the form $m_\Phi^2=\mu_2^2 +\frac{1}{2}\lambda_\Phi v^2$, with $\lambda_\Phi$ some combination of the Lagrangian quartic couplings and $\mu_2$ the BSM mass scale of the IDM. However, different from the previous model, single Higgs couplings in the IDM receive genuine BSM vertex corrections, which have here been evaluated consistently at the one-loop level (with also leading two-loop corrections from~\cite{Aiko:2023nqj} included for the partial decay width of the Higgs to two photons). Results of parameter scans in the IDM are shown in Fig.~\ref{fig:kala_vs_hZZ:IDM}, with predictions for $\kappa_\lambda$, including complete one-loop corrections --- computed with the public tool \texttt{anyH3}~\cite{Bahl:2023eau} --- together with leading two-loop corrections from~\cite{Braathen:2019pxr,Braathen:2019zoh,Aiko:2023nqj}, on the horizontal axis and the one-loop BSM deviation in the $\ghzz$ coupling on the vertical axis. For the latter, BSM vertex and external-leg corrections are included, with vanishing external momenta in order to obtain a quantity corresponding to $\kappa_{\PZ}$~\cite{LHCHiggsCrossSectionWorkingGroup:2012nn, LHCHiggsCrossSectionWorkingGroup:2013rie}. The colour of the hexagonal bins corresponds to the minimal BSM deviation in the Higgs partial decay width to two photons, evaluated at two loops, among the points within the bin --- this additional observable is relevant in the context of the IDM, due to the presence of a charged BSM Higgs boson in the model. Vertical blue lines correspond to \SI{20}{\%} (dot-dashed) and \SI{50}{\%} (dotted) levels of accuracy on the determination of $\trlh$~(taking $\kappa_\lambda = 1$ as central value), while the red dotted horizontal corresponds to the $1\sigma$ precision expected for $\ghzz$ at FCC-ee. Interestingly, we observe that a significant population of scan points exhibit significant BSM deviations in $\kappa_\lambda$, with values up to $\kappa_\lambda\sim 2.8$, while maintaining $\delta^{(1)}g_{\PH\PZ\PZ}$ below the $1\sigma$ line of FCC-ee. Meanwhile, for the majority of these points a measurement of $\Gamma(\PH\to\PGg\PGg)$ at the FCC-ee would also remain compatible with the SM up to the $1.5\sigma$ level. Finally, the impact of including both momentum dependence, as well as diagrams with an insertion of the modified value of $\kappa_\lambda$ (while these are formally of two-loop order, the question of their size becomes relevant for large $\kappa_\lambda$) in the calculation of the $\ghzz$ coupling has been checked. Importantly, it was found for these IDM scan points that both effects lead overall to a reduction of the BSM deviation in $\ghzz$ --- in other words, the results shown in Fig.~\ref{fig:kala_vs_hZZ:IDM} constitute a conservative representation of the possible deviations in $\kappa_\lambda$ for IDM scenarios with BSM effects in $\ghzz$ below the $1\sigma$ sensitivity of the FCC-ee. 

While the discussion above was only done in two specific models --- the $\mathbb{Z}_2$-SSM and the IDM --- these cases offer a clear illustration of the important complementarity of measurements of the trilinear Higgs self-coupling with those of single-Higgs couplings, in order to probe the parameter space of BSM scenarios as thoroughly and completely as possible.

\subsubsection{Interplay with other fields of science --- gravitational waves and primordial black holes}
\label{sec:gravwaves}
The determination of the trilinear Higgs self-coupling provides insights in the cosmological evolution of the Higgs potential. Its phase history is closely connected to the generation of the baryon-antibaryon asymmetry. In particular, the requirement of a strong first-order EWPT (FOEWPT), i.e.~departure from the thermal equilibrium, is one of the three Sakharov conditions \cite{Sakharov:1967dj} to be fulfilled for successful EWBG. While the SM cannot reproduce quantitatively the EWBG scenario, models with extended Higgs sectors may feature strong first-order EWPTs. A strong FOEWPT typically comes along with a large trilinear Higgs self-coupling \cite{Grojean:2004xa,Kanemura:2004ch,Basler:2017uxn,Basler:2019iuu,Biekotter:2022kgf,Enomoto:2021dkl,Kanemura:2022ozv}. This in turn implies in general a mass gap between the lighter and heavier Higgs bosons of the Higgs spectrum, leading to typical signatures to be probed at present and future colliders \cite{Asakawa:2010xj, Dorsch:2013wja,Dorsch:2014qja, Dorsch:2017nza,Basler:2016obg,Biekotter:2022kgf,Biekotter:2023eil}.  
More generally, new physics effects in the scalar and/or Higgs-Yukawa sectors, as required for a strong first-order phase transition, have impact not only on BSM Higgs signatures \cite{Basler:2019nas,Anisha:2022hgv,Anisha:2023vvu,Biekotter:2023eil} but also on the phenomenology of di- and triple Higgs production \cite{Bahl:2022jnx,Bahl:2023eau,Stylianou:2023xit,Biermann:2024oyy}. There is hence a strong interplay between the size of the Higgs self-interaction, the actual shape of the Higgs potential and EWBG, such that exploiting synergies between particle physics and cosmology will largely advance our knowledge about the underlying physics that is realised in nature.

A fully complementary approach is offered by gravitational wave detection: Strong FOPTs can provoke gravitational waves that can be tested at a future space-based gravitational waves observatory such as LISA, see e.g.~\cite{
Kakizaki:2015wua,Hashino:2016rvx,Hashino:2018wee,Basler:2024aaf,Biekotter:2022kgf,Biekotter:2023eil}. 
Another possibility to observe indications of a FOPT are primordial black holes (PBHs): 
In general, in a FOPT of the vacuum, a spatial and temporal contrast of vacuum energy density occurs. 
If the magnitude of the contrast exceeds a certain critical value, the spatial portion of the phase transition that is delayed can cause gravitational collapse and become a PBH. 
The mass of a PBH derived from the FOEWPT is expected to be about $10^{-5}$ of the solar mass. 
Searching for a PBH of this mass using micro-lensing experiments in the present or near future such as Subaru HSC~\cite{Niikura:2017zjd}, OGLE~\cite{Niikura:2019kqi}, PRIME~\cite{Kondo_2023} and Roman Space Telescope~\cite{Fardeen:2023euf} may provide an early insight into the physics of the FOEWPT,  complementary to future Higgs self-coupling measurements and gravitational wave observations. 

In the following we will give two examples for the interplay of these methods in an two-Higgs-doublet model (2HDM) and an aligned Higgs EFT scenario.

\begin{figure}[thb]
\centering
  \begin{subfigure}{0.48\textwidth}
        \includegraphics[width=\textwidth]{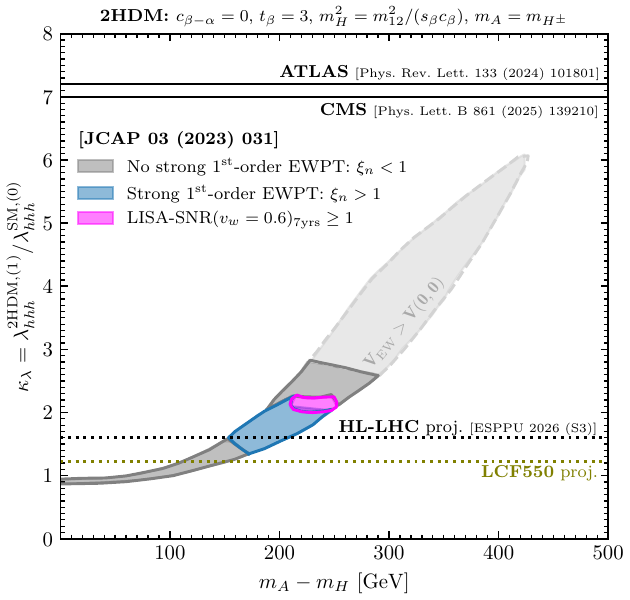}
        \caption{}
        \label{fig:FOEWPT:2HDM}
  \end{subfigure}
  \begin{subfigure}{0.50\textwidth}
        \includegraphics[width=\textwidth]{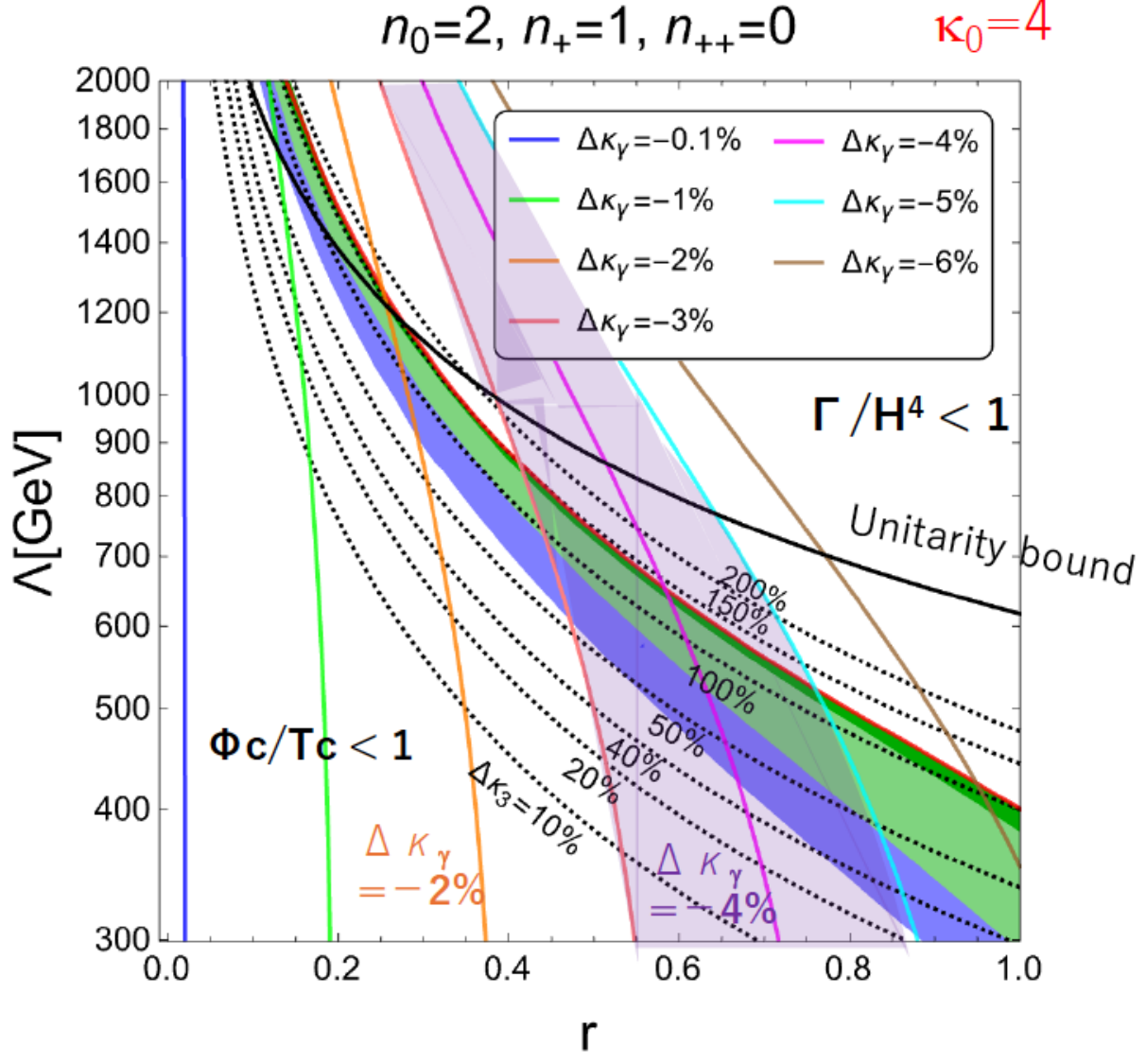}
        \caption{}
        \label{fig:FOEWPT:aHEFT}
  \end{subfigure}
\caption{(a) Scan in the plane of the mass difference $m_A - m_H$ and $\kappa_\lambda$ of the type-II 2HDM ~\cite{Biekotter:2022kgf}. 
The parameter region giving rise to a strong first-order electroweak phase transition is displayed by the blue  and pink regions, where in the latter region a gravitational wave signal is potentially detectable at the space-based gravitational wave observatory LISA.
The projected \SI{95}{\%} CL upper limits on $\kappa_\lambda$ from the HL-LHC and a \SI{550}{GeV} linear collider are indicated by the horizontal black and olive-green dotted lines, respectively. 
(b) First-order phase transition in the $r$-$\Lambda$ plane of the aligned Higgs EFT~\cite{Florentino:2024kkf} ($r$: degree of non-decoupling, $\Lambda$: scale of new physics,  $n^0$, $n^+$, and $n^{++}$: degrees of freedom of the electrically neutral, singly-charged and doubly-charged scalars in the model, respectively): The coloured bands indicate areas consistent with FOEWPT and how they can be probed experimentally (red: PBH searches \& colliders,  dark (light) green: DECIGO (LISA) \& colliders, blue: colliders only). As collider probes, $g_{\PH\PGg\PGg}$ and $\trlh$ measurements are considered, their coverage for various precisions indicated as lines. The purple shaded region corresponds to the case that $\Delta\kappa_\gamma$ is determined to be $\SI{-4}{\%}\pm \SI{1}{\%}$ (using HL-LHC and ILC). 
}
\label{fig:FOEWPT}
\end{figure}
\paragraph[Colliders and gravitational waves scrutinizing the 2HDM]{Colliders and gravitational waves scrutinizing the 2HDM} The 2HDM contains a Higgs spectrum that could trigger a FOEWPT. A case study involving measurements of the addition Higgs bosons will be discussed in Sec.~\ref{sec:phys:bsm:BSMHiggs}, while we focus here on constraints from the trilinear coupling of the 125-GeV Higgs boson and gravitional wave detection.
Figure~\ref{fig:FOEWPT:2HDM} shows a scan in the 2HDM parameter space. 
The ranges of the masses that were scanned are \SI{200}{GeV} $< m_{\PH} < $ \SI{1}{TeV}\footnote{In this section, $\PH$ refers to the additional neutral scalar of the 2HDM, while the 125-GeV Higgs boson is denoted as $\Ph$.} and \SI{600}{GeV} $< m_{\PSA} = m_{\PSHpm} <$ \SI{1.2}{TeV}, with $\tan{\beta} = 3$. 
Exclusion limits from LHC searches were applied, which excluded most of the points with $m_{\PH}$ below the $\PQt\PAQt$ threshold. 
For larger values of $m_{\PH}$, there was no sensitivity from LHC searches. 
Shown in pink are the points in the $\kappa_\lambda$ versus $m_{\PSA}-m_{\PH}$ plane that feature a FOEWPT that leads to a gravitational wave signal which is potentially detectable at LISA \cite{Biekotter:2022kgf}. 
For probing the full parameter region giving rise to a strong FOEWPT (blue and pink areas), however, collider measurements of $\kappa_\lambda$ are indispensable. The precision expected at LCF550 (indicated as \SI{95}{\%} upper limit by the olive-green dotted line in Fig.~\ref{fig:FOEWPT:2HDM}) is sufficient to probe the full area consistent with FOEWPT in this model.

\paragraph[Collider, gravitational wave and primordial black hole signatures in aligned HEFT]{Collider, gravitational wave and primordial black hole signatures in aligned HEFT} The formation of PBHs from a FOEWPT was studied in the context of SMEFT~\cite{Hashino:2021qoq}. Subsequently, an analysis using Higgs EFT, an effective theory that is more compatible with the physics causing strong FOEWPTs, was performed, and a methodology to comprehensively verify the physics of a FOEWPT using Higgs self-coupling measurements, gravitational wave observations, and PBH searches was proposed~\cite{Hashino:2022tcs}. Furthermore, the analysis was extended to include correlations with the di-photon coupling of the Higgs boson~\cite{Florentino:2024kkf}.

In Fig.~\ref{fig:FOEWPT:aHEFT}, the region providing a FOEWPT is indicated in the plane of the scale of new physics $\Lambda$ vs the degree of non-decoupling $r$. 
The coloured lines indicate the expected deviations in $\kappa_\gamma$, while the dotted lines indicate the deviations in $\kappa_\lambda$ (denoted $\Delta\kappa_{3}$ in the figure). The coloured bands show the region in consistent with a FOEWPT, coloured by their accessibility with different experimental probes: the narrow red band can be tested by PBH searches and Higgs properties, and the (dark and light) green bands can be examined by gravitational wave observation at DECIGO and LISA, respectively, as well as Higgs properties -- but the blue band can only be probed by Higgs properties, in particular $\kappa_\gamma$ and  $\kappa_\lambda$. The purple-shaded band corresponds to the case where $\Delta\kappa_{\gamma}$ is determined to be $\SI{-4}{\%} \pm \SI{1}{\%}$ by combining HL-LHC and linear collider data, c.f.\ Sec.~\ref{sec:singleHiggs}. 

PBH production resulting from FOEWPT has also been studied in specific models, see e.g.~\cite{Kanemura:2024pae}. These studies highlight that the parameter region in which the FOEWPT can be verified through PBH searches and gravitational waves alone is limited, and that precise measurements of the Higgs self-coupling and Higgs di-photon coupling by future collider experiments are essential for a broader verification of the FOEWPT.


\subsection{Electroweak physics at highest energies: multi-boson processes}

Compared to QCD, our understanding of electroweak (EW) interactions as a quantum field theory is still relatively poor. The EW scale is three orders of magnitude above the QCD scale ($v=$\SI{246}{GeV} compared to $\Lambda_{\text{QCD}}$), such that non-Abelian collinear phenomena like EW jets, EW showers, and EW (Sudakov) suppression of exclusive final states become visible only in the regime of a TeV and above. Exploring the non-Abelian EW structure is of utmost importance to understand this fundamental force of nature. A first glimpse showed up in $\ee \to \PW\PW / \PZ\PZ$ at LEP.  In the recent years more insights could be gained at the LHC in di-bosons and the first-ever observed EW triple-boson production and vector-boson scattering processes.  In the following, we will first introducing the opportunities from multi-EW-boson production in the TeV regime, before commenting more specifically on constraining triple-gauge boson couplings from $\ee \to \PW\PW$ at linear colliders and prospects w.r.t.\ vector-boson induced processes and quartic-gauge couplings.


\subsubsection{Scrutinizing electroweak interactions at high(est) energies}
\label{sec:ewhigh}

\begin{figure}[htb]
\begin{center}
\begin{subfigure}{0.49\textwidth}
    \includegraphics[width=\textwidth]{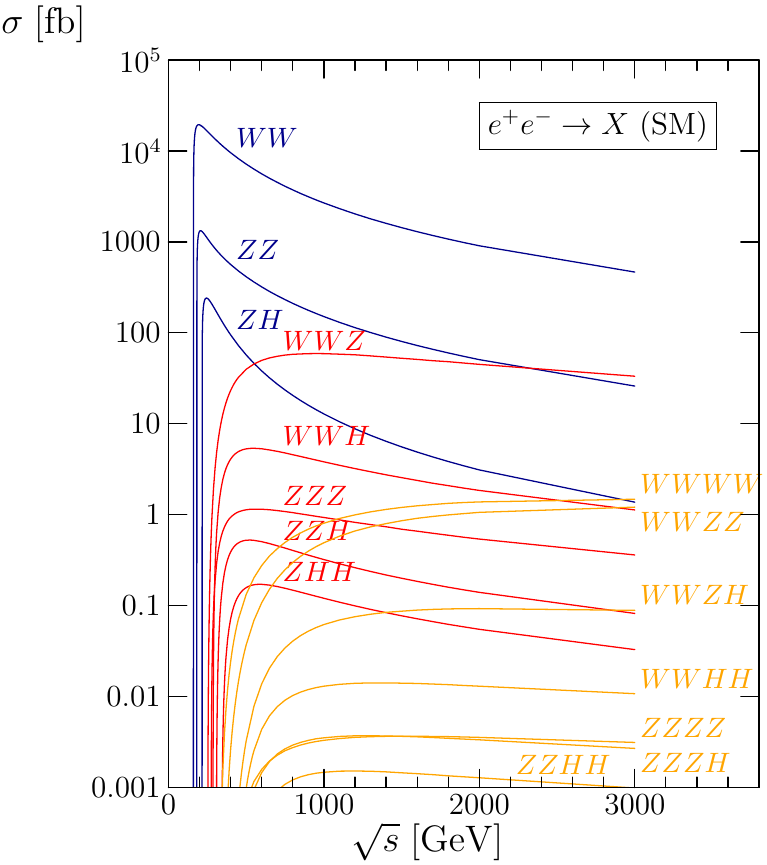}
     \caption{}
     \label{fig:multibosons:wp}
\end{subfigure}
\begin{subfigure}{0.49\textwidth}
\includegraphics[width=\textwidth]{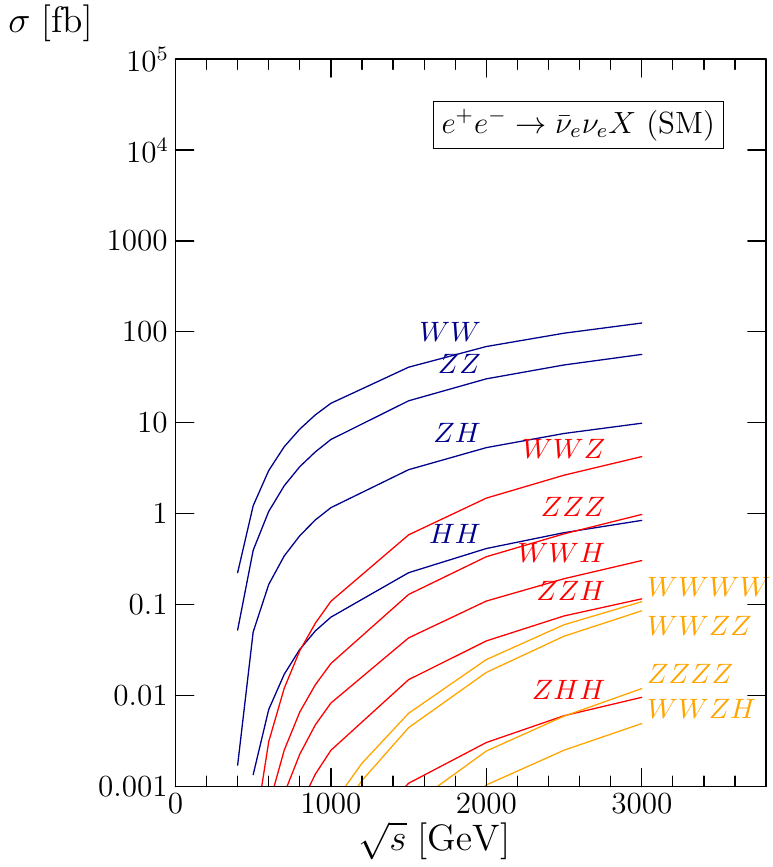}
     \caption{}
     \label{fig:multibosons:vbs}
\end{subfigure}
\end{center}
\caption{Unpolarised born-level cross sections~\cite{CLIC:2018fvx} for (a) multi-boson production (LO) in annihilation and (b) vector-boson-fusion processes.
\label{fig:multibosons}}
\end{figure}

In general, measurements of multi-(EW)-boson processes are needed to access the non-Abelian gauge structure and to ultimately disentangle transversal from longitudinal-scalar degrees of freedom of the EW gauge bosons, which are assigned to distinct sectors of the SM. In the SM, we expect to observe the restoration of electroweak symmetry at high energy such that transversal $\PW$ and $\PZ$ bosons become genuine gauge degrees of freedom, while longitudinal $\PW$ and $\PZ$ bosons combine with the Higgs boson to form a scalar multiplet which complements the fermionic matter of the SM. Multi-particle production and scattering in the multi-TeV regime can then be understood as electroweak-jet dynamics analogous to QCD.  In extensions of the SM, the asymptotic symmetry pattern could be different, which would typically result in a systematic enhancement of specific multi-boson final states as compared to the SM prediction shown in Fig.~\ref{fig:multibosons}. Gauge boson polarisation measurements at the LHC opened a first door to this. However, at a hadron collider EW effects are shadowed because QCD radiation and virtual corrections largely dominate over EW contributions. A detailed understanding of EW interactions in the complex structure of initial- and final-state radiation can be achieved at a TeV-scale lepton collider~\cite{Bredt:2022dmm,Denner:2024yut}.

Electroweak jets, most notably also neutrino jets can be studied in high-energetic lepton collisions~\cite{Ma:2024ayr}. Again, weak production and low-background environments make these precision studies more favourable than at a high-energy hadron collider. Besides electroweak multi-particle production these processes would also comprise electroweak fragmentation.  It is essential that at a high-energy $\ee$ collider the analysis of fully hadronic final states provides a detailed picture of exclusive multi-boson states without significant loss of efficiency.

The observables here are 6-, 8-, 10-fermion final states which are complementary to corresponding processes at hadron colliders, limited mostly by statistics but with very low systematic uncertainties. The thresholds for the simplest multi-boson production processes are listed in Table~\ref{tab:multiboson-thresholds}, and the overall increase in particle multiplicity towards higher energies can be read off from Fig.~\ref{fig:multibosons}.
Some of these processes have been studied in simulations as  discussed in the following subsections, however a comprehensive study of EW processes at high energies remains future work.

\begin{table}
\centering
\begin{tabular}{ |c|c|c| } 
 \hline
 $ee\to$ & \text{threshold [GeV]} & \text{maximum [GeV]} \\\hline
 $\PZ\PH$ & 160.8 & 240 \\ 
 $\PZ\PZ$ & 182.4& 200 \\ 
 $\PW\PW$ & 216.3 & 195 \\ 
 \hline 
 $\PW\PW\PZ$ & 252.0 & 950 \\ 
 $\PZ\PZ\PZ$ & 273.6 & 550 \\ 
 $\PW\PW\PH$ & 285.9 & 550 \\
 $\PZ\PZ\PH$ & 307.5 & 520 \\ 
 $\PZ\PH\PH$ & 341.4 & 590 \\
 \hline
 $\PW\PW\PW\PW$ & 321.5 & 3000 \\
 $\PW\PW\PZ\PZ$ & 343.1 & 4000 \\
 $\PW\PW\PZ\PH$ & 377.0 & 2000 \\
 $\PW\PW\PH\PH$ & 410.9 & 1400 \\
 \hline
\end{tabular}
\caption{Production thresholds and peak cross section positions for 2-, 3- and 4 EW boson processes in $\ee$ collisions. From~\cite{CLIC:2018fvx}.}
\label{tab:multiboson-thresholds}
\end{table}

\subsubsection{Triple and quartic gauge couplings}
\label{sec:Wboson500}

\begin{figure}[htb]
\begin{center}
\includegraphics[width=0.75\hsize]{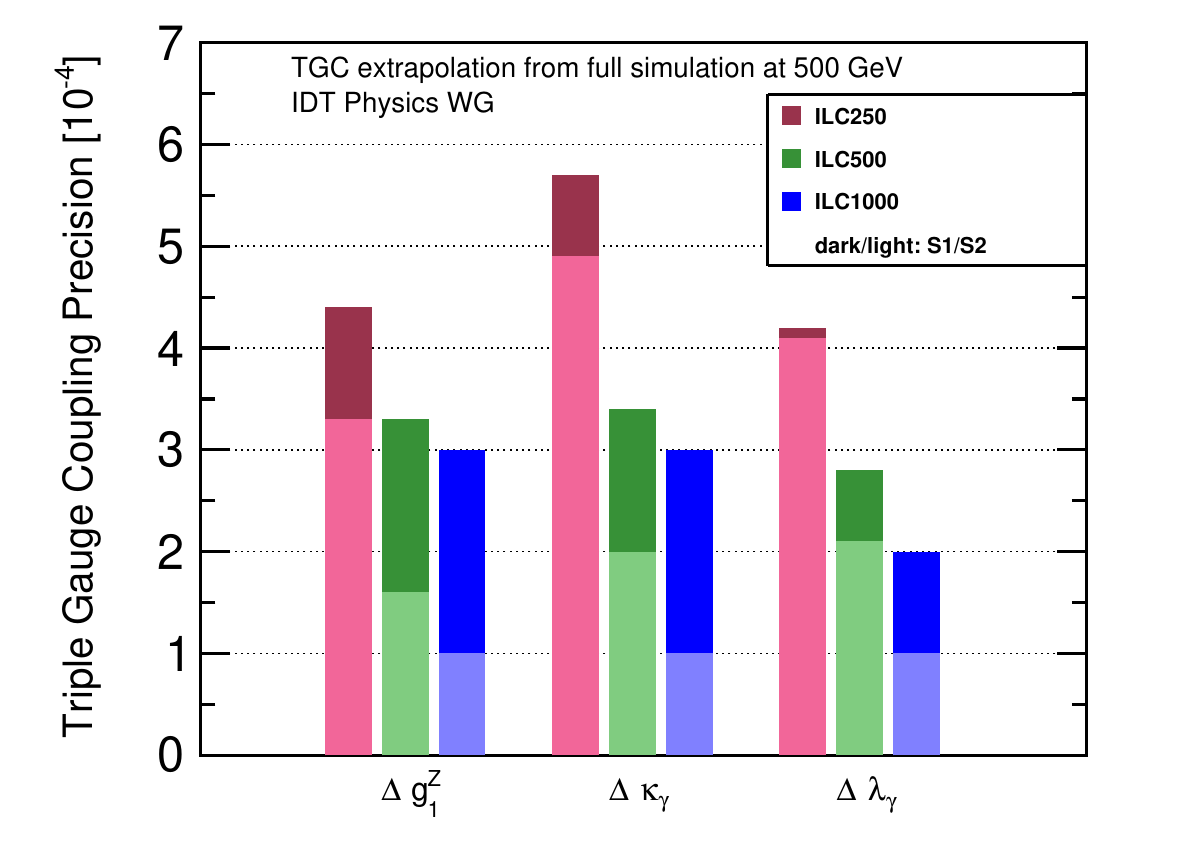}
\end{center}
\caption{Expected precisions on the three triple gauge coupling parameters at the three energy stages of ILC. The results at \SI{500}{GeV} and at \SI{1}{TeV} are based on the ILD full simulation analyses of semi-leptonic $\PW$ pair production, extrapolated to include improvements from the fully hadronic channel and single-$\PW$ production as well as for upgrading from a binned analysis of three angles to an optimal observable technique~\cite{Karl:2019hes}. The S1 scenario assumes the systematic uncertainties from~\cite{Karl:2019hes}, the S2 illustrates the hypothetical reduction by a further factor 2--3 to the level of $1 \times 10^{-4}$.
\label{fig:TGC_ILC_allECM}}
\end{figure}

The ILC prospects for triple gauge coupling measurements 
at $\sqrt{s}=$\SI{500}{GeV} and \SI{1}{TeV} have been studied based on full simulation of the ILD detector concept~\cite{Marchesini:2011aka, Rosca:2016hcq}, and extrapolated to \SI{250}{GeV}~\cite{Karl:2019hes}. With increasing energies, the relative effect on the differential cross section of the three TGC parameters $g_{\PZ}$, $\kappa_\gamma$, and  $\lambda_\gamma$ grows proportional to $s/m_{\PW}^2$. There are  a  number of experimental effects that become more challenging at higher energies --- more forward-boosted event topologies, higher pile-up from beamstrahlung pairs and photoproduction of low-$p_T$ hadrons ---  the fundamental gain in sensitivity with $s$ dominates by far. Figure~\ref{fig:TGC_ILC_allECM} summarizes the current state of the expected precisions, as discussed in more detail in the following paragraphs.

The full simulation study at \SI{500}{GeV}~\cite{Marchesini:2011aka} was limited to a binned analysis of three (out of five) angles in the $\PW\PW \to \PGm\PGn \PQq\PAQq$ and $\PW\PW \to \Pe\PGn \PQq\PAQq$ channels. For an integrated luminosity of \SI{500}{\fbinv}, this study found statistical uncertainties of $(6.1, 6.4, 7.2) \times 10^{-4}$ for $g^Z_1$, $\kappa_{\gamma}$ and $\lambda_{\gamma}$, respectively. 
An unbinned likelihood or optimal observable analysis of all five angles, including also fully hadronic $\PW\PW$ events as well as single-$\PW$ events has been estimated~\cite{Barklow:2018priv} to improve these numbers by a factor of 2.4 for $g^{\PZ}_1$ and by a factor of 1.9 for $\kappa_{\gamma}$ and $\lambda_{\gamma}$. Assuming the full integrated luminosity of ILC500 instead of only \SI{500}{\fbinv} gives another factor of 2 improvement to $(1.3, 1.7, 1.9) \times 10^{-4}$. At this level of precision, systematic uncertainties need to be considered. As shown in~\cite{Beyer:2022xyz}, the effects of a finite knowledge of the luminosity and the beam polarisations are negligible when including them as nuisance parameters in a global fit. The effect of different permil level uncertainties on the selection efficiency and percent-level uncertainties on the residual background has been evaluated in~\cite{Marchesini:2011aka} by propagation through the whole analysis chain, thereby treating them as fully uncorrelated between data sets and observables, obviously a very pessimistic assumption. Based on considerations of correlated uncertainties and nuisance parameters in global fits, more recent studies expect that systematic uncertainties of $(3, 3, 2) \times 10^{-4}$ can be reached~\cite{Karl:2019hes}. In total, the expected precisions on the three couplings thus reach $(3.3, 3.4, 2.8) \times 10^{-4}$ for ILC500. The LCF run plan increases the statistics wrt ILC by \SI{50}{\%} and \SI{100}{\%}, c.f.\ Sec.~\ref{sec:RunScenarios}. Quantifying the corresponding gain in TGC precision requires a fresh and more detailed look at the involved systematic uncertainties.

The full simulation study at \SI{1}{TeV}~\cite{Rosca:2016hcq} found statistical precisions of $(1.9, 1.7, 2.7) \times 10^{-4}$ for a luminosity of \SI{1}{\abinv} with the same analysis technique as at \SI{500}{GeV} (semi-leptonic $\PW$ pairs, binned analysis using three angles). A simple scaling to the full luminosity of \SI{8}{\abinv} renders the statistical uncertainty negligible with respect to the systematic uncertainties as given above. Thus, an adequate estimate of the \SI{1}{TeV} prospects requires a thorough re-analysis of the systematic effects. It has already been shown that any global scaling as well as the variation of a simple angular cut-off can be determined from the data without any loss of precision on the TGCs~\cite{Beyer:2022xyz}.  But a  complete treatment of the remaining backgrounds in a multivariate fit still remains to be done.  In Fig.~\ref{fig:TGC_ILC_allECM} we present the expected precisions on the TGC parameters assuming the currently understood level of systematic uncertainties and the result of possible improvements by a factor 2-3 to the level of \num{1e-4}.

\subsubsection{Vector-boson fusion and scattering}
\label{sec:VBS}

Vector boson scattering (VBS) is one of the most important processes to scrutinize the non-Abelian gauge structure of EW interactions, to connect to 
\begin{figure}
    \centering
    \begin{subfigure}{0.49\textwidth}
       \includegraphics[width=\textwidth]{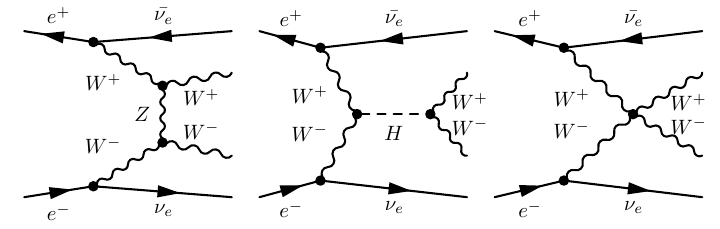}
       \caption{}
       \label{fig:VBS_feyn:sig}
    \end{subfigure}
    \begin{subfigure}{0.49\textwidth}
       \includegraphics[width=\textwidth]{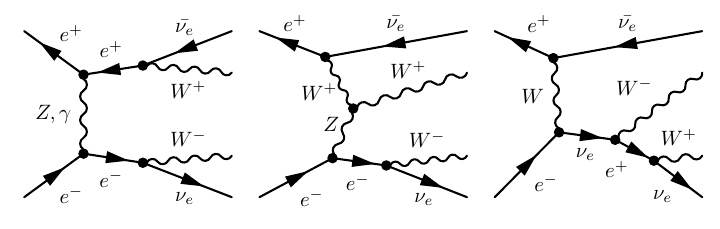}
       \caption{}
       \label{fig:VBS_feyn:bkg}
    \end{subfigure}
    \caption{Feynman diagrams relevant for vector boson scattering (VBS)~\cite{Fleper:2016frz}: (a) VBS signal, (b) ``irreducible'' background processes with identical particle content in the final state.}
    \label{fig:VBS_feyn}
\end{figure}
the Higgs mechanism and EWSB and to compare transverse and longitudinal degrees of freedom. In addition, VBS (the  serves as a vehicle to potentially discover resonances that couple dominantly to either the gauge or Goldstone sector of particle physics. The kinematic phase space of typical VBS Feynman diagrams, shown in Fig.~\ref{fig:VBS_feyn:sig}, opens up at collider energies of about \SI{500}{GeV} and therefore VBS is only accessible at linear $\ee$ colliders, not at circular ones. The advantage compared to VBS at hadron colliders is that fully hadronic final states are experimentally accessible; these allow the full reconstruction of the invariant mass of the di-boson system. Also the reconstruction of polarisations of the intermediate vector bosons from fully hadronic states is possible at a linear collider. VBS can be used to search for deviations from the SM in terms of dim-6 or dim-8 operators in SMEFT or HEFT; indeed, VBS is one of the most sensitive processes to discriminate between the two EFT frameworks. Previously this has been studied as searches for deviations of quartic gauge couplings (QGCs) from the SM (aQGCs)~\cite{Boos:1997gw,Boos:1999kj,Beyer:2006hx,Fleper:2016frz}. To get physically meaningful results that obey the unitarity of the scattering matrix it is important to automatically implement such constraints in the signal modelling, see e.g.~\cite{Alboteanu:2008my,Kilian:2007gr,Kilian:2014zja}. Figure~\ref{fig:VBS_dim8_res:LCreach} illustrates the huge improvement of high centre-of-mass energies  w.r.t.\ constraining dim-8 operators with measurements of VBS processes in \ee\ collisions.

\begin{figure}
    \centering
    \begin{subfigure}{0.45\textwidth}
            \includegraphics[width=\linewidth]{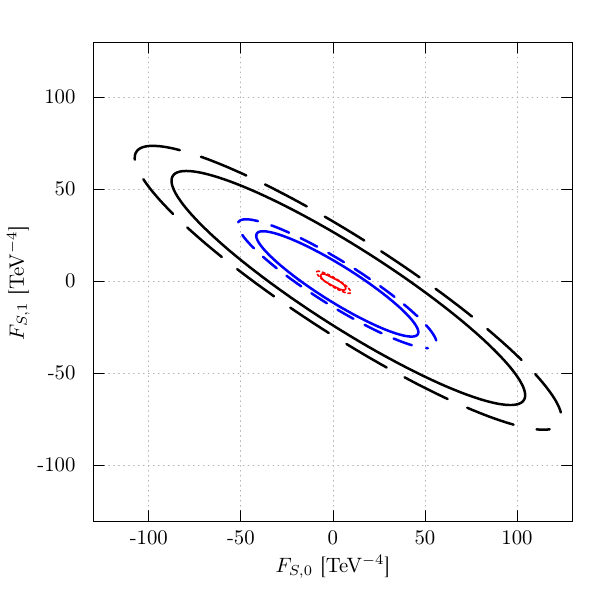}
            \caption{}
            \label{fig:VBS_dim8_res:LCreach}
    \end{subfigure}
    \begin{subfigure}{0.54\textwidth}
            \includegraphics[width=\linewidth]{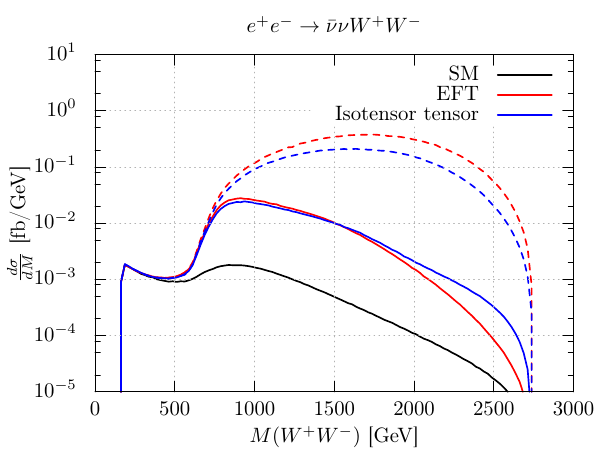}
            \caption{}
            \label{fig:VBS_dim8_res:3TeV}
    \end{subfigure}
    \caption{(a) Reach in dim-8 operator coefficients for  quartic longitudinal vector boson couplings from VBS measurements at ILC1000, CLIC1.5 and \SI{3}{TeV} in black, blue and red, respectively. Dashed curves are unpolarised, full lines with design polarisation fractions.
    (b) Effect of an isotensor tensor resonance in $\PW\PW$ on the $\PW\PW$ invariant mass spectrum at \SI{3}{TeV} CLIC. Full lines are the signal projected to a UV-complete scenario. Both figures from~\cite{Fleper:2016frz}.}
    \label{fig:VBS_dim8_res}
\end{figure}

Vector-boson-scattering processes and multi-boson prompt production processes do, in general, probe the same physics, however, in kinematically vastly different regions and are, hence, complementary to each other~\cite{Beyer:2006hx}. Prompt multi-boson production processes can be used to constrain deviations in triple and more importantly quartic gauge couplings and prove (or disprove) the equivalence of gauge and Goldstone degrees of freedom.

Beyond EFTs, VBS can be used for searches as generic di-boson resonances with spin and isospin quantum numbers of $S,I = 0, 1, 2$ each~\cite{Alboteanu:2008my,Kilian:2015opv,Fleper:2016frz}. Figure~\ref{fig:VBS_dim8_res:3TeV} shows a tensor resonance that could appear in a composite Higgs model. The invariant mass of the di-boson spectrum shows a broad excess that in this special case is difficult to distinguish from a dimension-8 EFT description.


\subsection{Top physics programme -- from threshold to highest energies}
\label{sec:Top_physics}
As the heaviest particle of the SM, the top quark is most intimately linked to the Higgs boson and the mechanism of EWSB. Thus, the study of the top quark is not only important for providing precision inputs to the SM tests, but it is even more important as a probe of the mechanism of mass generation, since any modification of the Higgs sector and any new interactions involved in mass generation will be reflected in the properties of the top quark. 

Apart from the top-quark mass and width, best measured in a scan of the production threshold as will be discussed in Sec.~\ref{sec:Top_mass}, top-quark measurements profit tremendously from the higher centre-of-mass energies only accessible to linear colliders. 
We will illustrate for the top Yukawa coupling (Sec.~\ref{sec:Higgstop}), the top-quark electroweak couplings (Sec.~\ref{sec:top:ewcoup}, complemented by the EFT perspective in Sec.~\ref{sec:top:EFT}), the constraints on 4-top operators (Sec.~\ref{sec:top:4topop}), spin correlations and possible quantum entanglement in the top sector (Sec.~\ref{sec:top:quantum}), FCNC interactions of the top quark (Sec.~\ref{sec:top:fcnc}), as well as for constraining top-axion interactions (Sec.~\ref{sec:top:axions}).

\subsubsection{The top-quark mass and width} \label{sec:Top_mass}

The top-quark mass is one of the fundamental parameters of the SM that must be determined experimentally. The TeVatron experiments left a legacy top-quark mass determination~\cite{CDF:2016vzt}: $m_{\PQt}= 174.34 \pm 0.65$\,GeV, combining CDF and D0 results in several decay channels. Run 1 of the LHC yields a combination of 15 ATLAS and CMS measurements with an experimental precision of two permil~\cite{ATLAS:2024dxp}: $m_{\PQt}= 172.52 \pm 0.33 $\,GeV. These results correspond to {\em direct} measurements, where templates from Monte Carlo generators are compared to the observed distributions of top-quark decay products. The Monte Carlo mass parameter is identified with the top-quark pole mass, within an uncertainty of about 500 MeV~\cite{Hoang:2020iah}. Extractions of the top-quark mass from (differential) cross sections allow for a better control over the mass scheme, and have reached a precision of approximately \SI{1}{GeV}~\cite{ATLAS:2024kxj,CMS:2024irj}. The HL-LHC is expected to improve the experimental precision of direct mass measurements to several \SI{100}{MeV}~\cite{Azzi:2019yne}. 

A threshold scan at an \ee\ collider, scanning the centre-of-mass energy through the top-quark pair production threshold at $\sqrt{s} \sim 2 m_{\PQt}$ and mapping out the sharp rise of the cross section, provides a ``golden'' measurement of the top-quark mass~\cite{Gusken:1985nf, Fadin:1987wz, Fadin:1988fn, Strassler:1990nw, Guth:1991ab, Bach:2017ggt}. The location of the threshold is highly sensitive to the top-quark mass. A comparison of cross-section measurements at several centre-of-mass energies to first-principle predictions gives access to the top-quark mass in a well-controlled mass scheme. Typically threshold masses, such as the ``1S'' or ``PS'' mass are determined and then converted to the $\overline{\text{MS}}$ scheme. Non-relativistic QCD (NR-QCD) calculations have reached N$^3$LO precision~\cite{Hoang:2000yr, Hoang:2013uda, Beneke:2007zg, Beneke:2015kwa, Beneke:2016kkb} and the conversion between mass schemes is known to to four-loop precision~\cite{Marquard:2015qpa, Hoang:2017suc}. 

Beyond the top-quark mass, the shape of the threshold provides a direct handle on the top-quark width. The cross section in the threshold region is moreover sensitive to the exchange of soft gluons and virtual Higgs bosons, leading to a strong dependence on the strong coupling and the top-quark Yukawa coupling. These parameters can be extracted from a precise measurement of the differential top-quark pair production cross section versus centre-of-mass energy in a narrow window around the threshold.

Experimental studies have been performed by a number of groups~\cite{Martinez:2002st, Horiguchi:2013wra, Seidel:2013sqa,CLICdp:2018esa,Li:2022iav,Defranchis:2025auz}. The statistical uncertainty on the top-quark threshold mass can be reduced to below \SI{20}{MeV} with approximately \SI{200}{\fbinv} of data~\cite{Horiguchi:2013wra,CLICdp:2018esa}. The calibration of the beam energy and luminosity have an impact of a few MeV, provided the centre-of-mass energy is calibrated to \SI{5}{MeV} and the luminosity to one permil~\cite{Defranchis:2025auz}. Both of these requirements are expected to be within reach. The luminosity spectrum must be reconstructed precisely, as demonstrated in~\cite{Poss:2013oea}, to benefit fully from the potential of the threshold scan. 

The limiting factor in the extraction of the top-quark mass is expected to be of a theoretical nature. Missing higher orders in today's state-of-the-art prediction of the cross section amount to approximately \SI{35}{MeV}~\cite{Defranchis:2025auz}. The conversion from the threshold mass (1S, MSR or PS mass) to the $\overline{MS}$ scheme~\cite{Marquard:2015qpa, Hoang:2017suc} contributes another \SI{10}{-}\SI{20}{MeV}.\footnote{Subtle issues in the use of the PS scheme are discussed in~\cite{Hoang:2017suc}.} The parametric uncertainties due to the current world average uncertainty in the strong coupling $\alpha_s (m_{\PZ})$ has a non-negligible impact, through the impact on the prediction for the cross section and on the conversion to the $\overline{MS}$ mass. The impact of the uncertainty of the top-quark Yukawa coupling is expected to be below \SI{10}{MeV} assuming the uncertainty of $\delta y_{\PQt}/y_{\PQt}=$\SI{3}{\%} projected by the HL-LHC~\cite{Azzi:2019yne}, and can be further reduced with high-energy $\ee$ data, as discussed in the next section. The top-quark $\overline{MS}$ mass can then be determined to approximately \SI{50}{MeV} with the theory tools currently on the market and under reasonable assumptions for the impact of experimental systematic uncertainties. 

The top-quark width is extracted in the same fit, providing a direct measurement of the top-quark width. A precision of about \SI{50}{MeV} can be achieved, where statistical uncertainties, scale uncertainties in the prediction and the parametric uncertainty from the top-quark Yukawa coupling all have a non-negligible impact.

The Yukawa coupling and $\alpha_s$ can be determined with a small statistical uncertainty from the threshold scan~\cite{Horiguchi:2013wra}, but these measurements cannot compete with the most precise determinations, for instance from lattice QCD calculations (see discussion in Sec.~\ref{sec:Zpole}) once systematic uncertainties are fully accounted for.
A very important reduction of the uncertainties from missing higher orders in the NR-QCD predictions is required to enable precision measurements of these quantities.

While the threshold scan is definitely the {\em golden} top-quark mass determination, high-luminosity runs at centre-of-mass energies above the top-quark pair production threshold yield further opportunities. Very good statistical precision, below \SI{100}{MeV}, can be achieved in a {\em direct} determination of the top-quark mass from fits of MC templates to mass-sensitive observables formed with the top-quark decay products~\cite{CLICdp:2018esa}.
These measurements may be very valuable to connect the results of the threshold scan with the hadron collider measurements. Radiative events, where the top-quark pair is produced in association with an ISR photon, enable a measurement with a precision of approximately \SI{150}{MeV} using \SI{4}{\abinv} of data at \SI{500}{GeV}~\cite{Boronat:2019cgt}.
This analysis maintains full flexibility in the choice of the top-quark mass scheme and provides access to the top-quark mass at several scales, thus probing the scale evolution (``running'') of the top-quark mass. 

\subsubsection{The top-quark Yukawa coupling} 
\label{sec:Higgstop}

The value of the top-quark Yukawa coupling is expected to be close to unity. In the SM and other weakly coupled theories of EWSB, the top-quark Yukawa coupling enters in perturbation theory in the parameter $\alpha_t = y_t^2/4\pi \sim 1/12$. 
In these models, the effects of the top-quark Yukawa coupling are relatively small, in a well-defined perturbation series. 
However, models in which the Higgs boson is composite can produce large deviations from the SM in the Yukawa coupling. Large deviations can also appear in other top-quark interactions, as will be described in the following sections. 
It is important to search for all of these effects with high precision.  
They provide searches for new physics associated with the Higgs boson that are independent of the measurements of Higgs decay described in Sec.~\ref{sec:singleHiggs}.  
In some classes of models, these are the strongest probes of new physics associated with the Higgs field.   
This requires, and gives special importance to, running of a Higgs factory at centre-of-mass energies well above the top-quark threshold.

In this section, we will discuss the direct measurement of the top-quark interaction with the Higgs field through the Yukawa coupling.

The top-quark Yukawa coupling is already constrained by the  study of the Higgs boson at the LHC, in two different ways.
The Higgs-boson discovery channels at the LHC are sensitive to this coupling indirectly, through Higgs production and decay channels such as $\Pg\Pg \rightarrow \PH$ and $\PH\rightarrow \PGg\PGg$; in the SM, these proceed primarily through top-quark loops. 
Under certain assumptions, the Higgs production and decay rates can yield a precise bound on the top-quark Yukawa coupling. 
A more direct, and more robust, measurement is possible in the associated $\Pp\Pp \rightarrow \PQt\PAQt\PH$ production process, observed in 2018~\cite{ATLAS:2018mme,CMS:2018uxb}. 
The projection for the HL-LHC envisages an uncertainty of approximately \SI{3}{\%} on the signal multiplier $\kappa_{\PQt}$ dominated by theory uncertainties~\cite{Cepeda:2019klc}. 
Several groups have studied the interplay between measurements in top-quark and Higgs-boson production processes in the framework of SMEFT fits~\cite{Ellis:2020unq, Ethier:2021bye}.

At an \ee\ collider, indirect probes are also available: the $\PH\rightarrow \PGg\PGg$, $\PH\rightarrow \Pg\Pg$ and $\PH\rightarrow \PZ\PGg$ channels provide sensitivity to the top-quark Yukawa coupling already in \SI{250}{GeV} data~\cite{Durieux:2018ggn,Boselli:2018zxr,Jung:2020uzh}. 
These measurements can determine the top-quark Yukawa coupling with $\sim$\SI{1}{\%} precision, under the assumption that no new particles enter in the loops. 
These measurements may therefore provide an early indication of new physics, but a deviation of the SM cannot be unambiguously pinpointed. 
In more general EFT fits, the constraint on the coefficient $C_{\PQt}\phi$ of the operator that shifts the top-quark Yukawa coupling obtained from these indirect probes is not robust, as its effect is degenerate with poorly bounded degrees of freedom~\cite{Jung:2020uzh}.

The $\PQt\PAQt$ threshold scan offers an indirect determination that is more specific for the top-quark Yukawa coupling. 
The production rate close to threshold is sensitive to Higgs-exchange effects and can yield a competitive precision of \SI{4}{\%} on the top-quark Yukawa coupling~\cite{Yonamine:2011jg}. 
However, this extraction from the $\PQt\PAQt$ threshold scan currently suffers from large uncertainties in state-of-the-art calculation, with no clear perspective to reduce or circumvent this uncertainty to a level matching the experimental uncertainty~\cite{Beneke:2015lwa}.

The direct measurement in $\ee \rightarrow \PQt\PAQt\PH$ production requires a centre-of-mass energy of at least \SI{500}{GeV}.
The cross section rises sharply around that energy; raising the centre-of-mass energy to \SI{550}{GeV} enhances the production rate by a factor of approximately four and the measurement of the $\PQt\PAQt\PH$ coupling by a factor two.
Several groups have performed detailed full-simulation studies at centre-of-mass energies ranging from \SI{500}{GeV} to \SI{1.4}{TeV}~\cite{Abramowicz:2016zbo, Price:2014oca, Yonamine:2011jg}.
With \SI{8}{\abinv} at \SI{550}{GeV}, a precision of \SI{1.9}{\%} is expected on the top-quark Yukawa coupling, which could improve to \SI{1}{\%} with \SI{8}{\abinv} at \SI{1}{TeV}.
Measurements at multiple centre-of-mass energies and with different beam polarisations can further characterize the $\PQt\PAQt\PH$ coupling~\cite{Han:1999xd}. 

Another important target requiring centre-of-mass energies between \SI{600}{GeV} and \SI{1}{TeV} are the CP properties of the $\PQt\PAQt\PH$ coupling. 
Achievable constraints have been studied at the cross-section level~\cite{Godbole:2011hw}, showing a significant improvement due to polarised beams. 
A detailed detector-level study of the relevant observables remains an interesting task for future studies.

\subsubsection{The top-quark electroweak couplings}
\label{sec:top:ewcoup}

In models in which the Higgs boson is composite, the heavy top quark can also participate in forces that create this compositeness or can be affected by them. 
This is called ``partial top-quark  compositeness''~\cite{Kaplan:1991dc}.
Its effects include modification of the top-quark Yukawa coupling, but they go beyond this to processes that do not directly include the Higgs boson. 
The effects of partial top-quark compositeness can arise from three sources that are present in composite Higgs models: 
(1) top-quark mixing with heavy vector-like fermions, (2) new four-fermion interactions between top quarks and electrons mediated by new vector bosons, and (3) new four-fermion interactions among top and bottom quarks due to their intrinsic compositeness. The mechanism (1) appears in Little Higgs Models, e.g.~\cite{Berger:2005ht}; the mechanism (2) appears in models with extra space dimensions, including Randall-Sundrum warped-space models, e.g.~\cite{Djouadi:2006rk}; the mechanism (3) is also present in effective field 
descriptions~\cite{Grojean:2013qca}.

We first discuss the  effects of partial top compositeness on the gauge couplings of the top quark. 
The effects are limited in the top-quark QCD couplings due to constraints from QCD gauge invariance, but they can be manifest in the top-quark couplings to the massive electroweak gauge bosons $\PW$ and $\PZ$. 
The search for modification of the top-quark electroweak couplings thus nicely complements the searches for deviations from the SM in the Higgs couplings to $\PW$ and $\PZ$ and in the top-quark Yukawa coupling.   
It is important to note that since,  at \ee\ colliders, top quark pairs are produced from s-channel $\PGg$ and $\PZ$ exchange, the $\PZ$ coupling to the top quark appears in leading order in the production cross section and the forward-backward asymmetry.  
Both of these quantities can be measured to high precision at \ee\ colliders. Both depend strongly on beam polarisation, adding another pair of precision observables.

Figure~\ref{fig:models-rp} shows the predictions from a large number of models of partial top compositeness for modifications of the left- and right-handed vector couplings of the top quark to the $\PZ$ boson. 
The points shown here correspond to individual parameter sets from the parameter spaces of the various models.  We have rescaled the masses of new bosons to \SI{5}{TeV} and the masses of vector-like fermions to \SI{2}{TeV}, so all of these points lie well beyond the current LHC exclusion limits. 
The expected uncertainties reflect a 2-parameter fit to the two $\PZ$ couplings.   
More general fits to effective field theory are described in the following section.

\begin{figure}[tbp]
  \centering
\setlength{\unitlength}{1.6mm}

\begin{picture}(120,70)
\linethickness{0.3mm}
  \put(9,33){\line(0,1){2}}
  \put(5,34){\line(1,0){8}}  
  \put(15,34){...} 
  \put(21,34){\vector(1,0){60}} 
  \put(82,33){\Large{$\delta g^Z_R / g^Z_R$}}
  \multiput(31,33)(10,0){5}{\line(0,1){2}} 
\put(6.5,31){\footnotesize{-120\%}} 
  \put(29,31){\footnotesize{-10\%}} 
  \put(39,31){\footnotesize{-5\%}}
  \put(60,31){\footnotesize{5\%}}
  \put(70,31){\footnotesize{10\%}}
  \put(51,4){\vector(0,1){60}} 
  \put(45,66){\Large{$\delta g^Z_L / g^Z_L$}}
  \multiput(50,14)(0,10){5}{\line(1,0){2}} 
\put(45,13.5){\footnotesize{-10\%}} 
  \put(45,23.5){\footnotesize{-5\%}} 
  \put(45.5,43.5){\footnotesize{5\%}} 
  \put(45.5,53.5){\footnotesize{10\%}}
  \put(51,34){\color{red}\circle*{2}} 
  \put(53,36){\color{red}SM}
%
\put(51,26.8){\color{britishracinggreen}\circle*{1.5}}
\put(53,25.8){\color{britishracinggreen}Light top partners~\cite{Grojean:2013qca}}

%
\put(43.8,26.8){\color{britishracinggreen}\circle*{1.5}} 
\put(23.8,28.8){\color{britishracinggreen} Light top partners}
\put(23.8,25.4){\color{britishracinggreen} Alternative 1~\cite{bib:panico-priv}}

%
  \put(69,52){\color{britishracinggreen}\circle*{1.5}} 
  \put(52,54.5){\color{britishracinggreen} Light top partners Alternative 2~\cite{bib:panico-priv}}

%
 \put(51,18){\color{cyan}\circle*{1.5}} 
 \put(53,17){\color{cyan}Little Higgs~\cite{Berger:2005ht}}
%
  \put(51,19.6){\color{gray}\circle*{1.5}} 
  \put(53,19.0){\color{gray}RS with Custodial SU(2)~\cite{Carena:2006bn}}
%
 \put(51,16){\color{orange}\circle*{1.5}} 
 \put(53,15){\color{orange}Composite Top~\cite{Pomarol:2008bh}}

%
  \put(36.6,19.6){\color{magenta}\circle*{1.5}} 
  \put(25.6,21.8){\color{magenta}5D Emergent~\cite{Cui:2010ds}}

%
\put(54.6,30.4){\color{camel}\circle*{1.5}} 
\put(56.6,29.0){\color{camel}4D Composite Higgs Models~\cite{Barducci:2015aoa}}

  \put(9,34){\color{blue}\circle*{1.5}} 
  \put(1,36){\color{blue}RS with Z-Z' Mixing~\cite{Djouadi:2006rk}}
\multiput(56,40)(25,0){2}{\line(0,1){8}} 
\multiput(56,40)(0,8){2}{\line(1,0){25}}
\put(60,44){\large{ILC Precision}}
\color{red}
\input{figures/topphysics/ellipse4ab}
\end{picture}

\caption{
Predictions of a large number models of different types that predict deviations of the left- and right-handed couplings of the $\PQt$~quark to the $\PZ$ boson. The ellipse in the frame in the upper right part indicates the precision that can be expected for the ILC running at a centre-of-mass energy of \SI{500}{GeV} after having accumulated ${\mathcal L}=4\,\abinv$ of integrated luminosity shared equally between the beam polarisations  $P(\Pem,\Pep)=(\pm 0.8,\mp 0.3)$. Figure updated from~\cite{Richard:2014upa}, with  predictions rescaled to masses of new bosons or compositeness scales of \SI{5}{TeV} and the masses of vector-like fermions to \SI{2}{TeV}, so that all points are well beyond current LHC limits.
}
\label{fig:models-rp}
\end{figure}

The projected uncertainty contours presented in these figures are supported by full-simulation studies of the selection and reconstruction of $\PQt\PAQt$ final states~\cite{Amjad:2015mma,Bernreuther:2017cyi,CLICdp:2018esa,Okugawa:2019ycm}. 
The conspicuous six-fermion final state is readily isolated from backgrounds and the charged lepton yields a tag to distinguish the top and anti-top-quark candidates up to ambiguities in the event reconstruction. 
The most recent projections are based on the analysis of optimal observables in~\cite{Durieux:2018tev}, that propagates statistical uncertainties, taking into account a conservative energy dependent signal acceptance
from the full-simulation studies. 
For bottom-quark production, the production cross section and forward-backward asymmetry are considered, based on~\cite{Irles:2024ipg}, see also Sec.~\ref{sec:qprod}. 

\begin{figure}\centering
\includegraphics[width=.48\textwidth]{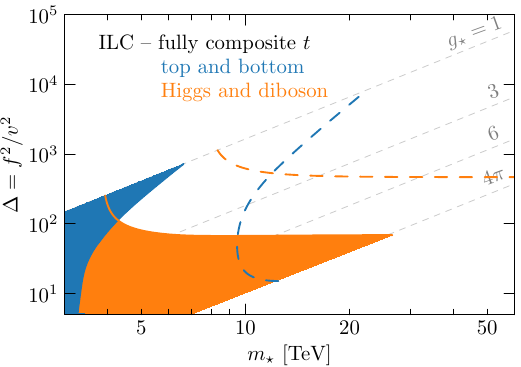}\hfill%
\includegraphics[width=.48\textwidth]{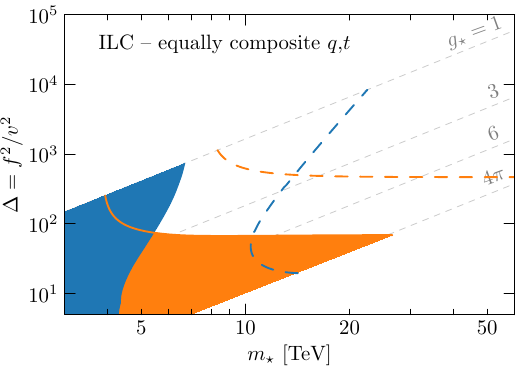}
\caption{\SI{5}{\sigma} reach of top- and bottom-quark measurements in the parameter space of generic composite Higgs models.
The model parameters on the axes are explained in the text.
Runs at \SI{500}{GeV} and \SI{1}{TeV} are included, with \num{0.5} and \SI{1}{\abinv}, respectively, and polarised beams.
Note that this is 1/16 and 1/8, respectively, of the luminosity samples expected for the LCF.
Blue and orange regions represent, respectively, the regions probed by measurements of $\PQt$ and $\PQb$ and regions probed by measurements of Higgs and di-boson processes.
The solid areas represent the potential reach in a pessimistic case with cancellations between the various contributions; the dashes lines are obtained in the optimistic case where no such cancellations occur.
Figure reproduced from~\cite{Durieux:2018ekg}.
}
\label{fig:topcompositness}
\end{figure}

New four-fermion interactions can also effect the structure and centre-of-mass energy dependence of the top-quark pair-production cross section. 
We illustrate this with two analyses from the literature. Top-quark compositeness was studied in an effective field theory description  in~\cite{Durieux:2018ekg}. The authors found the regions of sensitivity in the model space shown in Fig.~\ref{fig:topcompositness}. The horizontal axis is the mass scale of composite resonances $m_*$; note that the sensitivity extends into the 10's of TeV region. The vertical axis shows the hierarchy between the compositeness scale $f$ and the Higgs boson vacuum expectation value. Measurements of top- and bottom-quark interactions probe the blue regions of the parameters space, while Higgs and di-boson measurements probe the orange regions. The analysis nicely illustrates the complementarity  with studies of the Higgs couplings to vector bosons.

\begin{figure}[h!]
    \centering
\includegraphics[width=0.46\linewidth]{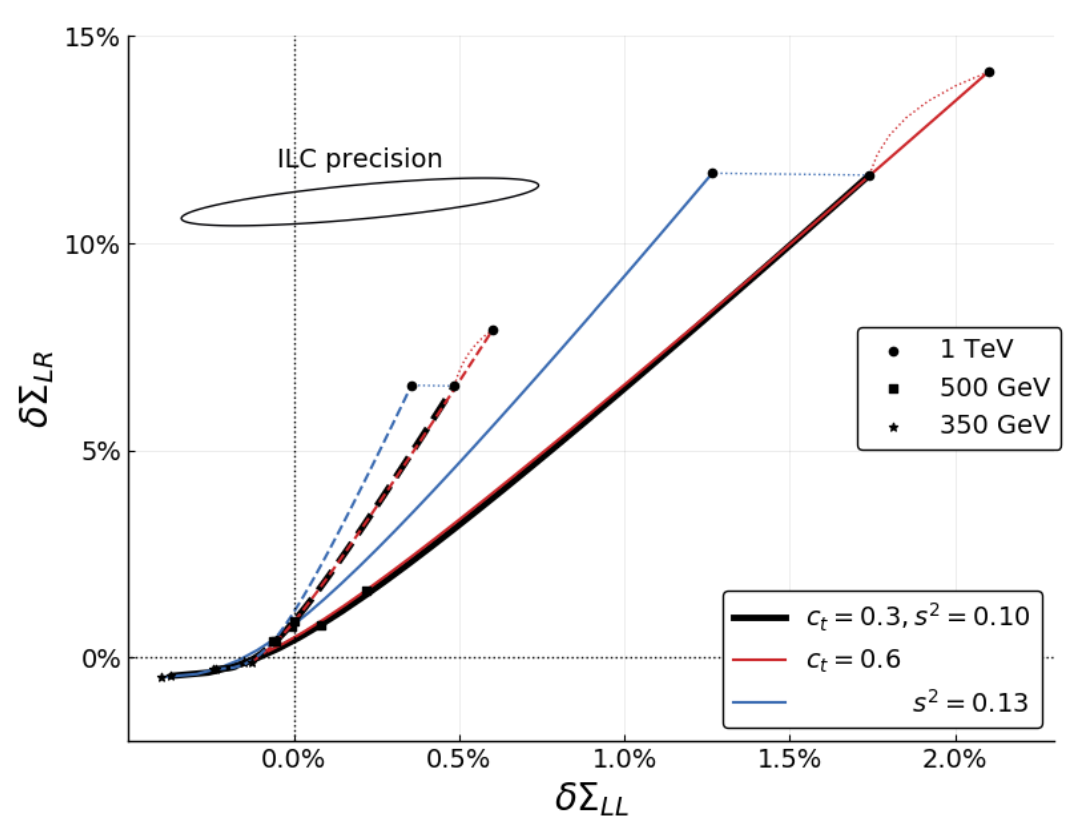} 
    \ \ \includegraphics[width=0.46\linewidth]{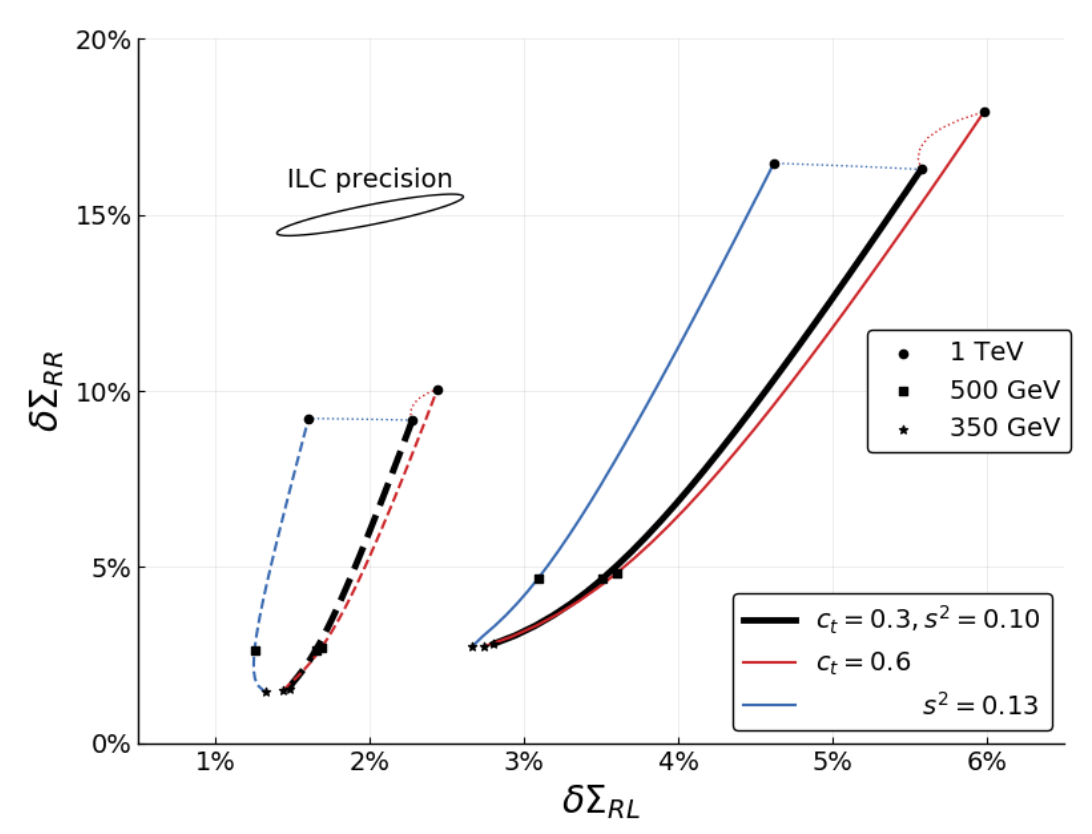} 
\caption{ Deviations from the SM in helicity-dependent cross sections for $\ee\to \PQt\PAQt$ in two examples of Randall-Sundrum theories, for centre-of-mass energies up to \SI{1}{TeV}, from~\cite{Yoon:2018xud}. The ILC contour is the \SI{68}{\%} 2-d uncertainty contour at \SI{550}{GeV}, from~\cite{Amjad:2015mma}.
}
    \label{fig:ttbarexamples}
\end{figure}

The polarisation-dependent cross sections for $\ee\to \PQt\PAQt$ were computed in~\cite{Yoon:2018xud} as a function of centre-of-mass energy in a class of models with a Randall-Sundrum fifth dimension and electroweak gauge bosons in the bulk.  \autoref{fig:ttbarexamples} shows the relative deviations from the SM in the four helicity-dependent cross sections $\Pem(L,R)\Pep(R,L) \to \PQt(L,R)\PAQt(R,L)$ in two of these models.  The figure shows the power using initial- and final-state polarisation in recognizing the BSM effects, and also the 
dramatic  increase in the size of these effects with centre-of-mass energy.  

\subsubsection{Effective field theory analysis of top-quark couplings}
\label{sec:top:EFT}

From the results of the previous section, we see that the search for an observed violation of the SM predictions for the polarised-beam cross sections and forward-backward asymmetries in top quark pair production will be a strong probe of Higgs composite interactions reflected in the top quark properties.
However, since we have discussed two different possible sources for these effects, we will need a more general analysis to understand the origin of any deviations. 
 To gain more information, it is best to analyse the full data set for $\ee\to \PQt\PAQt$ and $\ee\to \PQb\PAQb$ parametrising deviations from the SM  as the coefficients of SMEFT dimension-6 operators.  There are 30 dimension-6 SMEFT operators that couple to the top and bottom quarks, of which 14 are particularly important for $\ee$ reactions.
For orientation, the Wilson coefficient $c_{t\phi}$  modifies the  top-quark Yukawa coupling, the coefficients $c^{(1)}_{\phi Q}, c^{(3)}_{\phi Q},c_{\phi t}$ modify the top-quark vector and axial vector couplings to the \PZ\ boson, and the coefficients $c^{(1)}_{\ell Q}, c^{(3)}_{\ell Q},c_{\ell t},c_{e t}$  give new four-fermion interactions.  

\begin{figure}[h!]
    \centering
    \includegraphics[width=0.9\linewidth]{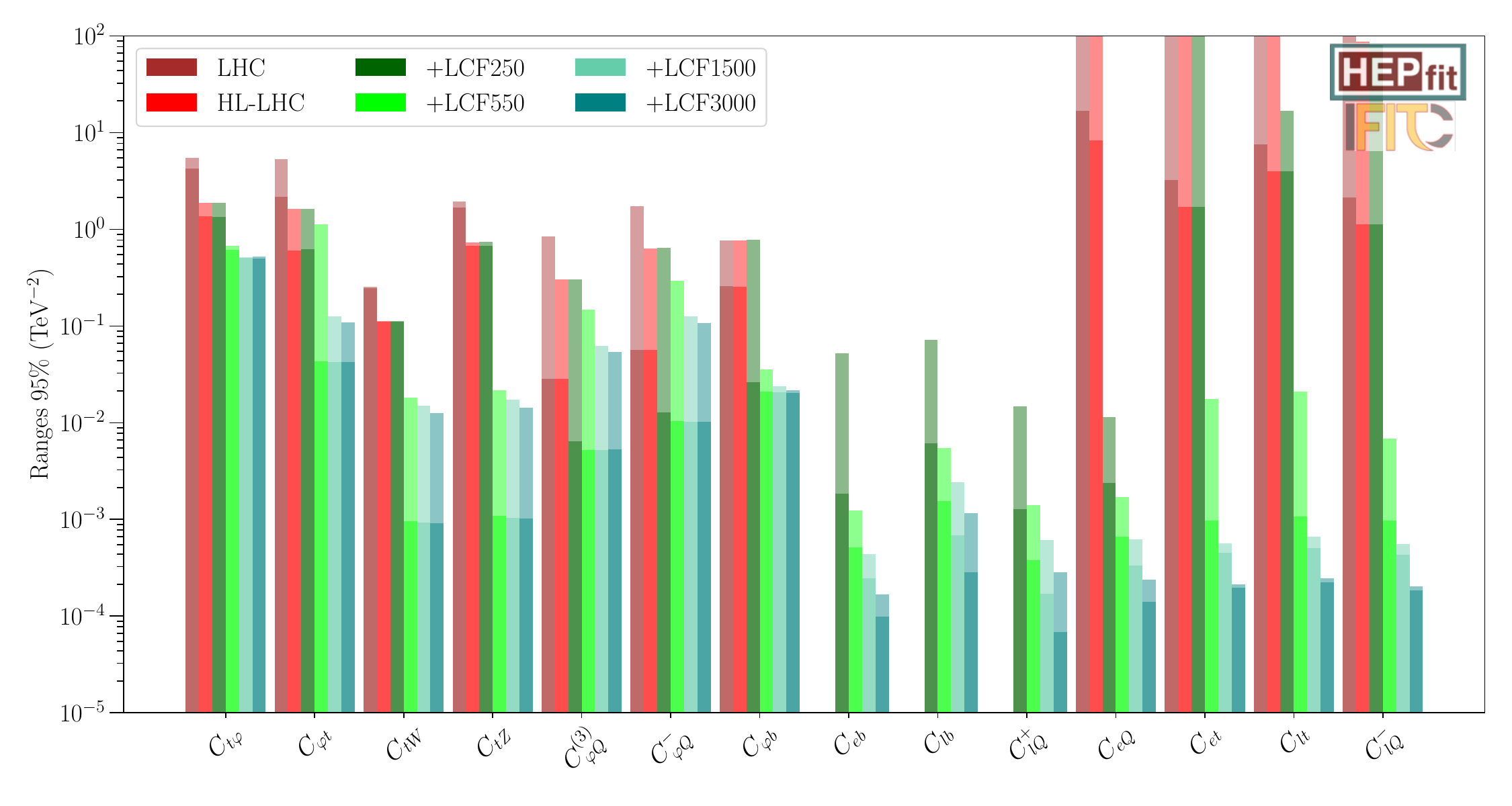}
    \caption{The \SI{95}{\%} CL bounds from a global fit to 30 Wilson coefficients of SMEFT operators with top and bottom quarks.
Coefficients are shown of two-fermion operators and four-fermion operators with charged leptons and heavy quarks.
The fit includes also $C_{tG}$ and fourteen $q\bar{q}\PQt\PAQt$ operators that are not shown.
The four sets of results include the HL-LHC projection based on an extrapolation of current bounds in the S2 scenario~\cite{Azzi:2019yne} (indicated in red).
Subsequent bars add the different stages of a linear collider facility, with $\ee$ data at increasing $\sqrt{t}$.
Here, \SI{3}{\abinv} at \SI{250}{GeV}, \SI{5}{\abinv} at \SI{550}{GeV}, \SI{3}{\abinv} at \SI{1.5}{TeV} and \SI{5}{\abinv} at \SI{3}{TeV} are assumed.
The dependence of the observables on the Wilson coefficients is parametrised at order ${\cal{O}}(\Lambda^{-2})$. Updated from~\cite{Cornet-Gomez:2025jot}. 
}
\label{fig:smeft_top_sector}
\end{figure}

We now report the projected uncertainties from such a fit to the full-simulation data~\cite{deBlas:2022ofj, Cornet-Gomez:2025jot}. 
The \SI{95}{\%} confidence level bounds on the SMEFT operator coefficients are shown in Fig.~\ref{fig:smeft_top_sector}.
The solid-coloured bars show the limits from fits for individual Wilson coefficients; the lightly-shaded bars show the limits from a global fit including all operators.
The HL-LHC projection (indicated in red) is based on an extrapolation of the current bounds of~\cite{Cornet-Gomez:2025jot} to an integrated luminosity of \SI{3}{\abinv}, assuming that statistical and experimental uncertainties scale with the inverse square root of the integrated luminosity, while theory and modelling uncertainties are reduced to half of the current value.
Global bounds are around ${\cal{O}}(0.1-1)$ TeV$^{-2}$ for the two-fermion operator coefficients. 
No meaningful global bounds are found on the two-charged-lepton-two-top-quark operators, even if some LHC analyses of $\PQt\PAQt l^+l^-$ production start to have some sensitivity~\cite{CMS:2021aly}.
The electron-positron data at $\sqrt{s}=$\SI{250}{GeV} sharpen the bounds on the bottom operators, and coefficients that affect both bottom and top-quark production such as $C_{\varphi Q}^{(3)}$, $C_{\varphi Q}^{-}$ and $C_{lQ}^{-}$, but bring no additional sensitivity for pure top-quark operators.

The \SI{550}{GeV} data yield an important improvement of all top operators.
However, the global bounds remain an order of magnitude behind the individual bounds.  The problem here is that there is a degeneracy between the effects of the two-fermion operators that modify the electroweak couplings and the four-fermion operators that generate contact interactions. Little Higgs models
generate only the operators of the first type, while extra-dimensional models primarily generate operators of the second type, so it is important to break this degeneracy.

A second run above the $\PQt\PAQt$ threshold accomplishes this, since then the two-fermion and four-fermion operators can be disentangled using the strong energy dependence of the four-fermion operators.
Operation at even higher energy yields stronger bounds still on the four-fermion operators.
Note that the $1/s$ decrease of the top-quark pair production cross section in the SM is compensated by the energy-growth of the sensitivity and the linear increase in instantaneous luminosity.

The study of top-quark production at $\ee$ colliders can characterize the electroweak interactions of the top quark and a complementary set of four-fermion operators to a precision that cannot be achieved at hadron colliders. High-energy operation and beam polarisation are key to robust global bounds. The stringent bounds on the top sector of the SMEFT correspond to a sensitivity well beyond the collider centre-of-mass energy in BSM scenarios where the top quark and the Higgs boson are (partially) composite.

Several studies in~\cite{Jung:2020uzh, Celada:2024mcf} explore the interplay between the top and Higgs and electroweak sectors of the SMEFT, while~\cite{terHoeve:2025gey} studies the impact of renormalization group evolution of operators. These studies include the $\ee$ projections from~\cite{Durieux:2018tev}. While these effects can be sizeable, and should be part of future global analyses,
the qualitative conclusions of this section remain unaltered.

The discussion of this section includes only possible deviations from the SM seen in top quark observables. 
A truly complete SMEFT analysis would also include the full set of dimension-6 SMEFT operators in a global fit to Higgs factory data.  
We will discuss such a fit in Sec.~\ref{sec:glob:NLO}.  
See especially Fig.~\ref{fig:eft:top-eft} and the accompanying discussion for the great increase in the sensitivity to top quark effective interactions in this more general fit with measurements at centre-of-mass energies of \SI{550}{GeV} and \SI{1}{TeV}. 

\subsubsection{Sensitivity to the four-top operator}
\label{sec:top:4topop}

A partial composite top quark will also exhibit compositeness effects from four-top contact interactions.   These interactions do not appear in Higgs factory processes in effective field analyses carried out the leading order, but they do induce corrections to the $\ee \to \ttbar$ cross section from  NLO SMEFT effects.  In this section, we discuss the sensitivity of Higgs factories operating at high energy to these corrections.

\begin{figure}[!t]
    \centering
    \includegraphics[width=0.55\textwidth]{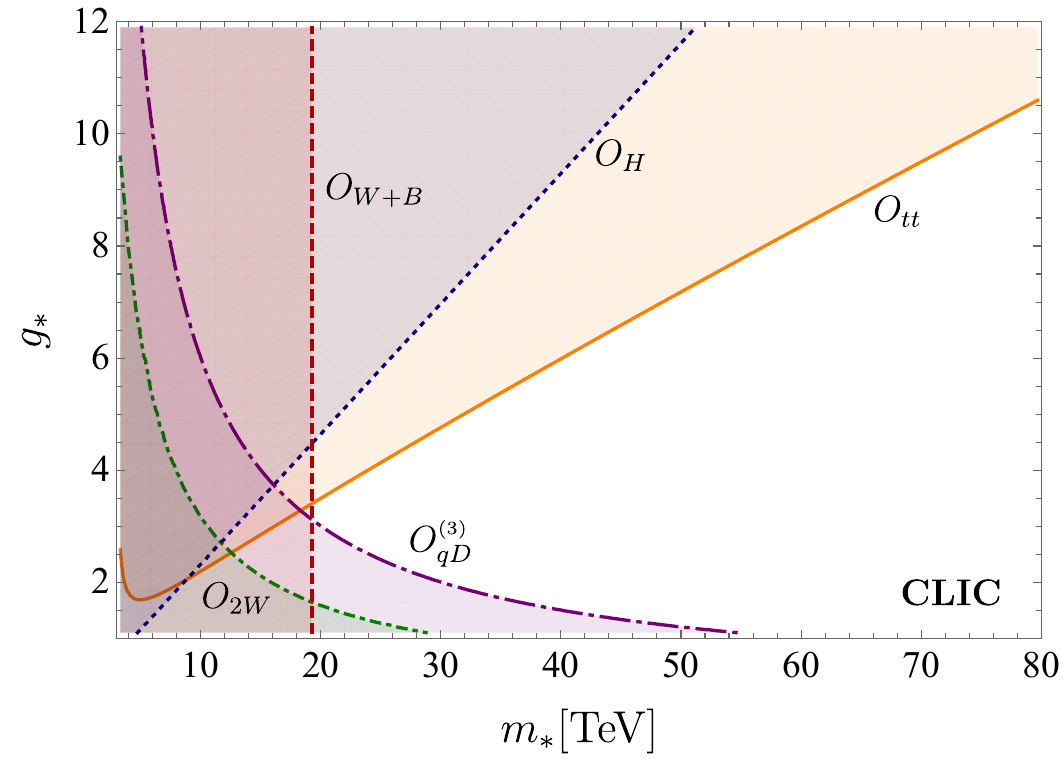}
    \vspace{0.15cm}
    \includegraphics[width=0.55\textwidth]{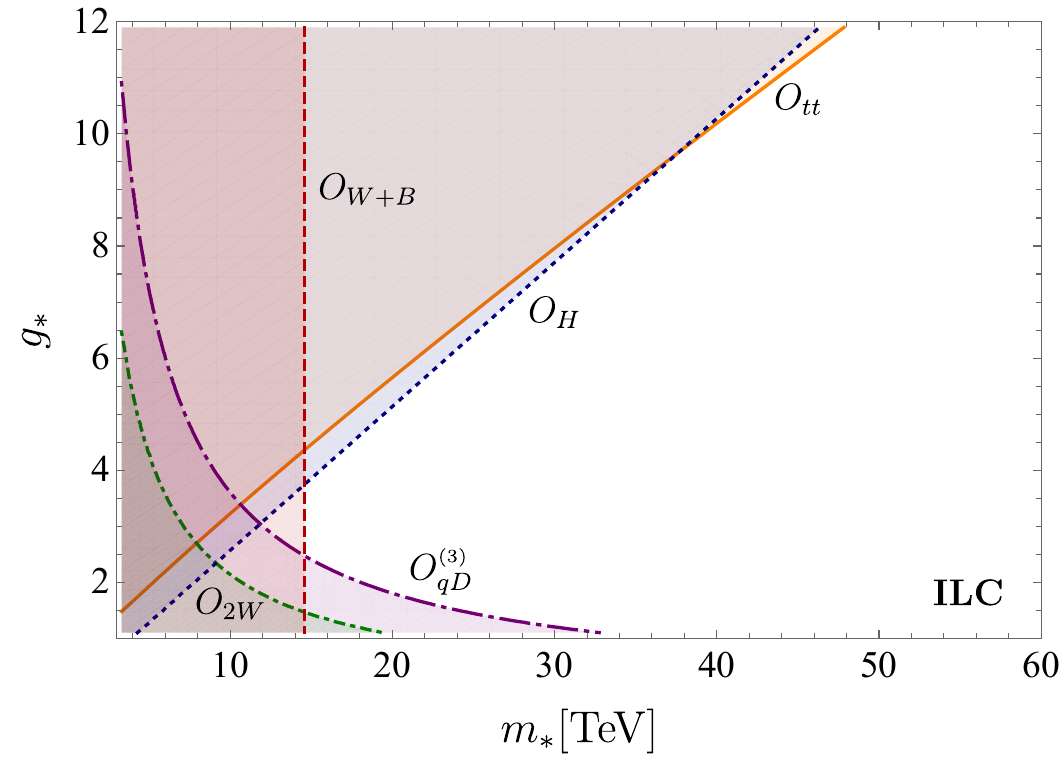}
    \vspace{0.15cm}
    \includegraphics[width=0.55\textwidth]{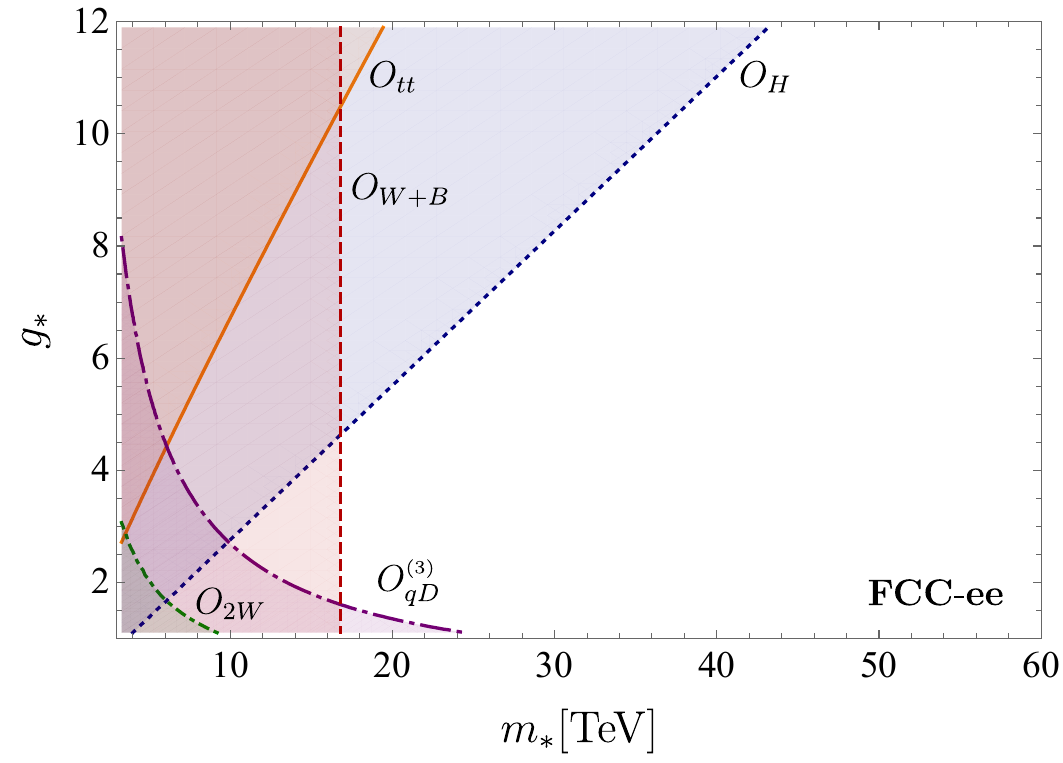}
    \caption{Future sensitivities at lepton colliders: CLIC (top), ILC (middle), and FCC-ee (bottom), at \SI{95}{\%} CL in the $(m_*,g_*)$ plane of scenarios featuring a strongly-interacting Higgs and (right-handed) top quark. The different limits are associated with constraints on individual operators, each dominating the corresponding observables in a certain region of parameter space. Reproduced from~\cite{Banelli:2020iau}.}
    \label{fig:LCbounds}
\end{figure}

The authors of~\cite{Banelli:2020iau} investigate a strongly interacting top-quark scenario\footnote{``Strong" means here that all new couplings are stronger than any of the known Standard Model couplings.}. Corresponding toy models can be characterised by a mass scale $m_*$ and a generic coupling constant $g_*$. The studies are solely 
based on projections for precisions of electroweak top (and bottom) couplings. This direct test is preferable to indirect probing by electroweak precision observables where a comparatively small number of observables has to cover a comparatively large number of EFT operators. The results of the studies are shown in Fig.~\ref{fig:LCbounds}. The following scenarios have been assumed: CLIC \SI{3}{TeV} and \SI{3}{\abinv}, ILC \SI{1}{TeV} and \SI{1}{\abinv}, FCCee \SI{0.365}{TeV} and \SI{1.5}{\abinv}. For linear colliders, their results are essentially an extension of the results shown in Fig.~\ref{fig:topcompositness}.

One sees that quite a large surface of the $(m_*,g_*)$ plane would be covered by the physics (top) programme at a linear collider already for the described scenarios. 
In general, Higgs precision measurements represented by the operator $\mathcal{O}_{H}$ dominate the sensitivity and are superior to four-fermion operators represented by $\mathcal{O}_{2W}$. This is in qualitative agreement with the results in Fig.~\ref{fig:topcompositness}. Finally, the operator $\mathcal{O}_{W+B}$ represents the expected sensitivity to the scale $m_*$ from electroweak precision observables. 

Let us emphasise that the sensitivity to the important four-top operator $\mathcal{O}_{\ttbar}$ increases with centre-of-mass energy. 
The operator $\mathcal{O}_{\ttbar}$ enters the typical four-fermion operators
\begin{equation}
\frac{\mathrm{c_{te}}}{\Lambda^2} (\bar e_R \gamma_\mu e_R) (\bar t_R \gamma^\mu t_R) + \frac{\mathrm{c_{t\ell}}}{\Lambda^2} (\bar \ell_L \gamma_\mu \ell_L) (\bar t_R \gamma^\mu t_R)\,
\label{eett}    
\end{equation}
via a one-loop effect and RGE running (again for details the reader is referred to~\cite{Banelli:2020iau}). A strongly-interacting (right-handed) top quark leads to a new-physics amplitude that scales like
\begin{equation}
\mathcal{M}_{e^+ e^- \to t \bar t} \sim \frac{g'^2}{16\pi^2} \frac{s}{f^2} \log \Big(\frac{m_*^2}{s}\Big) \,.
\label{Aeettt}    
\end{equation}
Here $f$ is a decay constant characterising composite states in that sector. 
Defining $f=m_*/g_*$, the \SI{95}{\%} CL bounds for $\mathrm{f_{\ttbar}}$ are \SI{7.7}{TeV} for CLIC, \SI{4.1}{TeV} for ILC and \SI{1.6}{TeV} for FCC-ee.   

Clearly, the higher the energy the bigger the sensitivity to $\mathcal{O}_{\ttbar}$. 
The growth of the sensitivity to $\mathcal{O}_{\ttbar}$ can be tested at various centre-of-mass energies (e.g.\ already in the step from \SI{550}{GeV} to \SI{1}{TeV}). 
As we will discuss in Sec.~\ref{sec:acc}, LCF is planned to be compatible with upgrades for centre-of-mass energies between \num{1} and \SI{3}{TeV}. The testing of $\mathcal{O}_{\ttbar}$ in the controlled environment of an $\epem$ facility is a unique opportunity, even outperforming an \SI{85}{TeV} hadron collider, as can be seen e.g.
in Fig.~3.8 of~\cite{deBlas:2025gyz}.
As already suggested in Figs.~\ref{fig:ttbarexamples} and~\ref{fig:models-rp}, the (custodial) symmetry of the new strong sector would lead to characteristic patterns in the energy dependence of the electroweak couplings or equivalently of the corresponding EFT operators. 

Note finally that at the LCF the bounds are expected to become much stronger than those in Fig.~\ref{fig:LCbounds} due to two effects: 
\begin{itemize}
    \item The running scenario at the LCF foresees \SI{8}{\abinv} instead of only \SI{1}{\abinv} at \SI{1}{TeV};
    \item All analyses taken into account in e.g.~\cite{Durieux:2018tev} use only the semi-leptonic final state of $\ttbar$ production. It can be expected that with improved reconstruction techniques also the fully hadronic final state will become available in the future. Potentially, this adds another factor two of statistics (however, systematic uncertainties will play an ever increasing role).
\end{itemize}

\subsubsection{Quantum observables in top-quark pair production}
\label{sec:top:quantum}

Following the proposal in~\cite{Afik:2020onf}, ATLAS and CMS have demonstrated that in top-quark pair production at the LHC the top quark and top anti-quark are in a quantum-entangled state in regions of phase space close to the $\PQt\PAQt$ threshold and in the boosted regime~\cite{ATLAS:2023fsd,CMS:2024pts,CMS:2024zkc}. These results establish colliders as laboratories for the study of quantum phenomena at the
highest energies. A steady flow of new ideas has been put forward since (see e.g.~\cite{Barr:2024djo} and~\cite{Afik:2025ejh} for recent reviews). 

Top-quark pairs at \ee\ colliders offer an interesting potential for the study of quantum observables. Unlike the LHC, top-quark pairs are expected to be entangled in the entire phase space.
The controlled initial state and clean final state allow for precise measurements.
The entanglement marker $D_n$, as defined in~\cite{Maltoni:2024csn}, has been evaluated for the SM prediction. The $\epem \to \PQt\PAQt$ process was generated at leading order in MadGraph and the top decays were simulated with MadSpin, and $D_n$ was reconstructed from the charged leptons produced in the $\PQt\rightarrow \PW\PQb \rightarrow \Pl\PGn \PQb$ decay. 

\begin{figure}
    \centering
    \includegraphics[width=0.9\linewidth]{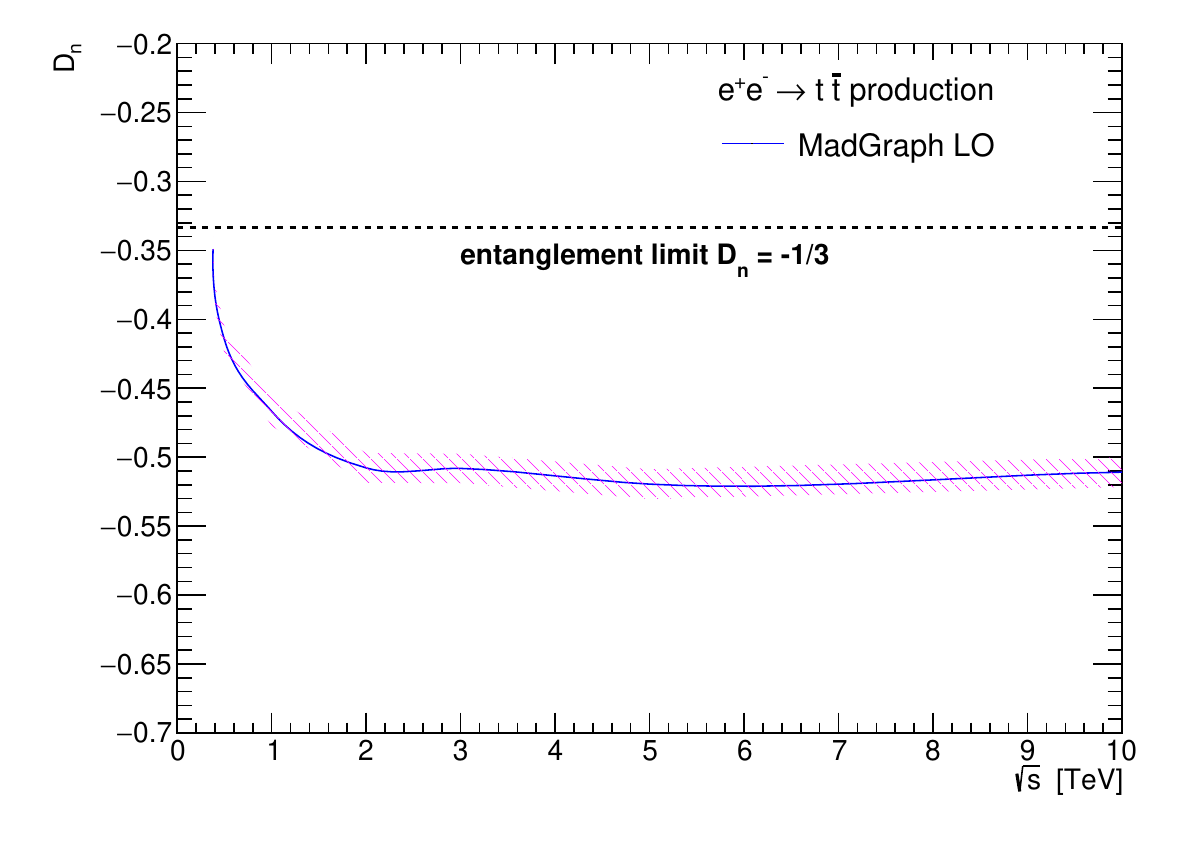}
\caption{The SM prediction for the observable $D_{n}$ in $\epem \to \PQt\PAQt$ as defined in~\cite{Maltoni:2024csn} versus the centre-of-mass energy $\sqrt{s}$, for unpolarised beams~\cite{topentanglement}.
Its value is used as an entanglement marker; a values below -$1/3$ indicate the top-quark pair forms an entangled system.
}
    \label{fig:top_entanglement}
\end{figure}

The result is shown in Fig.~\ref{fig:top_entanglement} as function of the \ee\ centre-of-mass energy for  unpolarised beams.
The entanglement marker $D_n$ exceeds the entanglement limit of $-1/3$ at all energies above the top-quark pair production threshold.
$|D_n|$ initially grows strongly with the centre-of-mass energy and reaches a value of approximately $-0.5$ at energies above \SI{1}{-}\SI{1.5}{TeV}.

A detailed quantum tomography of the $\ee \to \PQt\PAQt$ process is possible at \ee\ colliders with polarised beams~\cite{Altomonte:2024upf}.
The full Choi matrix ---~that determines the input-output transitions of the quantum process~--- can be experimentally reconstructed from measurements of top-quark spin observables with different beam polarisation configurations.
Note that this requires, in addition to the standard data-sets with the different longitudinal beam polarisation configurations also data-sets with transverse beam polarisation. In case of the ILC, spin rotators~\cite{Moffeit:2005pb} have been designed to enable the choice of any polarisation vector orientation at the interaction point.

The study of quantum observables in $\ee$ collisions offers a new approach to probe potential extensions of the SM, but also tests the foundations of quantum mechanics at the highest energies.

\subsubsection{Top-quark FCNC interactions}
\label{sec:top:fcnc}

Single top-quark production in association with an up or charm quark, $\ee \to \PQt\PQu / \PQc$ , can  be searched for already below the $\ttbar$ production threshold.
Any such a quark-flavour-changing neutral current interaction (commonly denoted FCNC) is particularly suppressed in the SM, so that its observation would constitute an unambiguous sign of new physics.

If the underlying dynamics is heavier than the centre-of-mass energy, the SMEFT can be employed to model such FCNCs.
They are first generated by dimension-six operators.
At tree-level, those which contribute to $\ee \to \PQt \PQu / \PQc$ generate $\PQt\PQq\PGg$, $\PQt\PQq\PZ$, and $\epem\PQt\PQq$ interactions.
A new $\PQt\PQq\PSh$ interaction also generates top-quark decays, while a $\PQt\PQq\Pg$ one is efficiently probed in single-top-quark production at hadron colliders.

Predictions for top-quark FCNC processes, accurate to next-to-leading order in QCD, were computed in~\cite{Zhang:2014rja, Degrande:2014tta} and a global analysis of existing constraints ---~including LEP2 ones~--- was performed in~\cite{Durieux:2014xla}.
Based on this, HL-LHC prospects were provided in Sec.~8 of~\cite{Cerri:2018ypt}.
Single top-quark FCNC production at future lepton colliders was notably investigated in~\cite{Aguilar-Saavedra:2001ajk, Khanpour:2014xla, Shi:2019epw}, in Sec.~3.1.2 of~\cite{CLIC:2018fvx} for CLIC and in Sec.~10.1.4 of~\cite{ILCInternationalDevelopmentTeam:2022izu} for the ILC.

\begin{figure}\centering
\includegraphics[width=.75\textwidth]{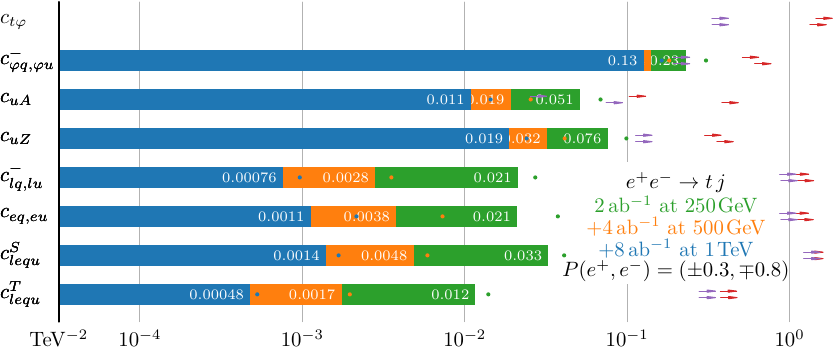}
\caption{Global projected \SI{95}{\%} CL constraints on dimension-six operator coefficients (shorter bars are better) leading to $\ee\to \PQt\PQc / \PQu$ production at increasing centre-of-mass energies assuming the ILC running scenario.
Horizontal bars account for longitudinal beam polarisation, while the coloured dots would be obtained with unpolarised beams.
For comparison, the red and purple arrows show present constraints and HL-LHC prospects respectively, for top-up (upper arrow) and top-charm (lower-arrow) flavour changes. Reproduced from Sec.~10.1.4 of~\cite{ILCInternationalDevelopmentTeam:2022izu}.}
\label{fig:eetj}
\end{figure}

Figure~\ref{fig:eetj}, reproduced from~\cite{ILCInternationalDevelopmentTeam:2022izu}, shows the constraints expected from collisions accumulated at centre-of-mass energies of \num{250}, \num{500}, and \SI{1000}{GeV} (coloured bars).
Compared to HL-LHC prospects (purple arrows), the improvement in sensitivity mostly concerns four-fermion $\epem\PQt\PQq$ interactions (lower four bars).
It reaches almost two orders of magnitude with \SI{2}{\abinv} collected at \SI{250}{GeV}.
Higher-energy runs lead to even tighter constraints.
This family of contact operators indeed generates amplitudes which grow quadratically with the centre-of-mass energy.
Three BSM benchmarks for top-quark FCNC interactions will be provided in Sec.~5.3 of the ECFA focus topic study for Higgs, top-quark, and electroweak factories~\cite{ECFA-HF-Report}.
Their contributions to four-fermion operators are however not estimated.
It would be interesting to understand which new-physics scenarios generate sizeable such operators and therefore benefit the most from lepton collisions.
The complementarity between FCNC and flavour-conserving channels (e.g.\ top-, bottom- and charm-quark pair production) also remains to be examined in specific scenarios where they would be correlated.

\subsubsection{Top quarks and axions}
\label{sec:top:axions}

The subject of axions and axion-like particles (ALPs) has become an important 
focus topic in the high energy physics community. This section is dedicated to ALP searches via their coupling to fermions, like the top quark. We will comment on additional opportunities to search for ALPs at linear colliders, e.g.\ leveraging the beam dumps, in Sec.~\ref{sec:phys:PBC:beamdumps}.

ALPs represent a well-motivated, 
testable scenario of new physics hypothesising a new, relatively light pseudo-scalar 
particle, \PXXA, that is a singlet under the SM gauge group. The lightness of this 
particle with respect to other putative new physics states is explained by the 
fact that the ALP is assumed to be the pseudo-Nambu-Goldstone boson (pNGB) of a 
spontaneously broken global symmetry in the UV. The pNGB nature manifests 
itself in a shift-symmetry of the ALP field, $a\to a+\textit{c}$  where $c$ is a 
constant, under which the ALP interaction Lagrangian should be approximately invariant. 
This means that the ALP couples preferentially via derivative interactions, 
built out of shift-invariants like  $(\partial^\mu a)$. This fact, along with the ALP quantum 
numbers, dictates that its interactions with SM particles are necessarily described by an EFT, 
suppressed by a generic new physics scale, $f_{a}$. The leading, non-redundant 
set of interactions is described by the following dimension-5 effective Lagrangian,
\begin{align}
\label{eq:ALP_lag}
   \mathcal{L}^{D=5}_{\text{ALP}} = & \frac{1}{2}\,(\partial_\mu a ) (\partial^\mu a ) - \frac{m_a^2}{2}\,a^2 + 
   \,c_{GG}\,\frac{a}{f_a}\,\frac{\alpha_s}{4\pi}\,G_{\mu\nu}^A\,\tilde G^{\mu\nu\,A} 
   + c_{\PW\PW}\,\frac{\alpha_L}{4\pi}\,\frac{a}{f_a}\,W_{\mu\nu}^I\,\tilde W^{\mu\nu\,I}        
  \nonumber\\
   &+ c_{\PQb\PQb}\,\frac{\alpha_Y}{4\pi}\,\frac{a}{f_a}\,B_{\mu\nu}\,\tilde B^{\mu\nu} 
    + \frac{\partial^\mu a}{f_a}\,\sum_F \bar F\, \mathbf{c}_F \,\gamma_\mu\, F,
\end{align}
where $\alpha_s$, $\alpha_L$ and $\alpha_Y$ denote the $SU(3)_C$, $SU(2)_L$ and $U(1)_Y$ gauge couplings of the SM,
$\tilde{X}^{\mu\nu}~=~\frac{1}{2}\epsilon_{\mu\nu\rho\sigma}X^{\rho\sigma}$ 
is the dual field strength tensor, and $F$ runs over the five fermionic 
representations of the SM, $\{Q,L,u,d,e\}$. The Wilson coefficients, 
$\mathbf{c}_F$, are $3\times3$ Hermitian matrices in flavour space.

The study of ALPs at colliders is a rich topic. 
ALP couplings to the heavier fermions are well motivated as derivative couplings to fermions can be shown to yield effects proportional to the fermion mass. 
However in particular the ALP coupling to the top quark has received relatively little  attention to date. 
The term in the Lagrangian describing the the ALP-top coupling is
\begin{align}
  \mathcal{L}_{\text{ALP}}^{\text{top}}=c_{\PQt\PQt}\frac{\partial_\mu a}{f_a}\, \bar{t}\gamma^\mu\gamma^5 t,\quad c_{\PQt\PQt}\equiv [c_u]_{33}- [c_Q]_{33}.
\end{align}
Several works have considered the sensitivity of the LHC to this coupling in 
top-rich final states such as in the direct production of an ALP in association 
with a $\PQt\PAQt$ pair, four-top production, as well as virtual ALP effects in $\PQt\PAQt$ 
production~\cite{Esser:2023fdo, Blasi:2023hvb, Phan:2023dqw, Anuar:2024qsz}. The broad conclusion 
is that this coupling is quite challenging to test, and new physics scales 
on the order of a few hundred GeV are still permissible, with the sensitivity
currently driven by precision measurements of $\PQt\PAQt$ production. The precision 
measurement of the spin-density matrix has the potential to improve on this sensitivity~\cite{Hosseini:2024kuh}.

At $\ee$ colliders, there are very few explicit studies of the sensitivity to 
$c_{\PQt\PQt}$ or investigations of ALP signatures in top-quark final states.
However, several ALP production mechanisms have been studied in this context. 
For example, it is known that ALP of mass $m_{a}<m_{\PZ}$ can be efficiently produced 
via decays of the $\PZ$ bosons which leads to bounds from LEP and LHC 
data~\cite{Jaeckel:2012yz,Mimasu:2014nea,Jaeckel:2015jla} and 
promising sensitivity to its gauge boson interactions at a future collider 
$\PZ$-pole run~\cite{Steinberg:2021iay,Polesello:2025gwj}. Several other 
production mechanisms, mostly focused on the gauge bosons couplings, have been 
investigated for future $\ee$ colliders~\cite{Mimasu:2014nea,CLIC:2018fvx,Bauer:2018uxu,Inan:2020aal,Zhang:2021sio,Yue:2022ash,Wang:2022ock,Lu:2022zbe,Cheung:2023nzg}.

In top-quark final states, the direct sensitivity to $c_{\PQt\PQt}$ can primarily 
be probed via direct production in the $\PQt\PAQt a$ process or via indirect effects impacting $\PQt\PAQt$ cross-section measurements.
These would become accessible at energies above the top-pair threshold and have the potential to 
improve on existing bounds, although no explicit studies exist to date. Precise measurements of 
spin correlations and scans across the top-pair threshold would be particularly suited to testing these indirect effects.
Associated production sensitivity would depend on the dominant ALP decay channel, 
and it is crucial to also consider loop-induced decay modes of the ALP~\cite{Bonilla:2021ufe}. 
Indeed, the clean environment of a future $\ee$ collider would facilitate the identification of the ALP decay products and therefore provide access to many of the ALPs couplings. 
For example, in the top-philic case, where $c_{\PQt\PQt}$ is taken to dominate, the ALP preferentially decays at one-loop into a $\PQb\PAQb$ pair when its mass is below the top-pair threshold.

An interesting feature of ALP production is that its derivative couplings lead to 
enhanced off-shell effects, in which non-resonant diagrams with intermediate ALP 
contributions are comparatively important with respect to non derivatively-coupled particles~\cite{Gavela:2019cmq}. These allow the possibility of studying 
$\PQt\PAQt$ production mediated by off-shell ALPs, probing combinations of 
$c_{\PQt\PQt}$ with other ALP couplings that mediate its production at $\ee$ colliders. 
Since derivative couplings to fermions can be shown to yield effects proportional 
to the fermion mass, direct production from $\ee$ annihilation is expected to be suppressed.
However, at higher energy $\ee$ colliders, production via photon or vector-boson fusion is a promising candidate. These channels have mostly been 
used to determine the sensitivity to gauge bosons couplings via gauge boson 
final states~\cite{Yue:2021iiu,RebelloTeles:2023uig}, demonstrating substantial improvements over LHC bounds.
A study of fermionic couplings, apart from the top quark, was also conducted in~\cite{Yue:2023mjm}, considering vector-boson fusion at the \SI{1}{TeV} ILC, 
and assuming ALP decay via a universal coupling to fermions weighted by their mass. 
A promising sensitivity was determined for higher mass ALPs above a few tens of GeV for the $\PQb\PAQb$ decay mode.
In the case of the 
$\PQt\PAQt$ final state, such searches would yield sensitivity to 
combinations of $c_{\PQt\PQt}$ and $c_{\PW\PW}$ or $c_{\PQb\PQb}$. The photon-fusion case was 
investigated in~\cite{Inan:2025bdw}, for $\ee$ centre-of-mass energies of \num{1.5} and \SI{3}{TeV}. 
The study determined the best projected sensitivities of $(c_{\PQb\PQb} c_{\PQt\PQt})^{\frac{1}{2}}<$\SI{2.8}{TeV^{-1}}
and $(\sqrt{c_{\PW\PW} c_{\PQt\PQt}})^{\frac{1}{2}}<$\SI{4.3}{TeV^{-1}} for $m_{a}=$ \SI{10}{GeV}, which
decrease by about an order of magnitude for ultra-heavy ALPs with $m_{a}=$ \SI{10}{TeV}. 
This channel would benefit from further investigation.

Finally, it is worth noting that beyond dimension-5, the ALP EFT should 
strictly be considered alongside other potential dimension-6 deformations of the SM, 
encapsulated by, \emph{e.g.}, the SMEFT. It is known, for instance, that at this order, 
the ALP EFT operators mix with the usual dimension-6 SMEFT operators under 
renormalisation group evolution~\cite{Chala:2020wvs,Galda:2021hbr}. This means that 
ALP couplings can be probed in a global manner using SMEFT interpretations of collider 
data~\cite{Biekotter:2023mpd}, and the precision SMEFT programme enabled by a 
future $\ee$ machine in the top sector (c.f.\ Secs.~\ref{sec:top:ewcoup} and~\ref{sec:glob:NLO}) would likely lead to better sensitivity to some ALP couplings. 
These should be complemented by the full future $\ee$ measurement programme, 
including EW precision tests, Higgs physics, $\PWp\PWm$ production, etc.


\subsection{Direct searches for BSM including SUSY}
\label{sec:directBSM}

The previous sections highlighted the discovery potential of a linear collider facility by precision measurements, indirectly in most cases. 
It is evident that within the kinematic limit a linear collider facility can also directly detect new particles, example of ALP production has been discussed in the previous section. 
The following sections will focus on the search opportunities offered by the energy upgrade of a linear collider facility (see Sec.~\ref{sec:acc:upgrade} for details on the upgrade options), the region between \SI{500}{GeV} and $\sim$\SI{1}{-}\SI{3}{TeV} in particular.
Extensive discussion of the discovery potential at the lower energies can be found e.g.\ in~\cite{ILCInternationalDevelopmentTeam:2022izu,ECFA-HF-Report}.

When discussing discovery prospects at a linear collider, one has to stress that lepton colliders are complementary to the hadron ones.
Probed with the highest sensitivity are electroweak production channels and, thanks to the clean environment, the sensitivity extends down to the smallest masses or mass splittings, covering also the invisible and nearly invisible final states in many cases.
In addition to the direct BSM discovery potential in model parameter range not accessible at LHC, follow-up precision studies are possible in a much wider domain if any new state is observed at HL-LHC or the linear collider itself. In these cases, energy upgrades of a linear collider facility can be prioritized as soon as sufficient funding can be made available.
Presented below are selected studies addressing discovery prospects at different energy stages of a linear collider: searches for light electroweak SUSY  scenarios (Sec.~\ref{sec:phys:bsm:light_susy}), light exotic scalars (Sec.~\ref{sec:phys:bsm:exscalar}), Heavy Neutral Leptons (Sec.~\ref{sec:phys:bsm:hnl}), dark matter particles (Sec.~\ref{sec:phys:bsm:dm}), $\PZ'$ particles (Sec.~\ref{sec:phys:bsm:zprime}, BSM Higgs bosons, (Sec.~\ref{sec:phys:bsm:BSMHiggs}), as well as unexpected signatures (Sec.~\ref{sec:phys:bsm:novelBSM}).

\subsubsection{Linear collider reach for light electroweak SUSY}
\label{sec:phys:bsm:light_susy}

Supersymmetry (SUSY), i.e.\ the symmetry between fermions and bosons, predicts the existence of spin-1/2 partners of the SM gauge bosons as well as for the Higgs bosons (which in the minimal version, the MSSM, correspond to a 2HDM-like Higgs sector), the electroweakinos (EWinos). 
These states play a crucial role in gauge coupling unification, they are intimately tied to an understanding of the electroweak scale through the Higgsino mass parameter $\mu$, and they are of primordial interest in the context of SUSY dark matter (DM). 
Naturally, EWinos are one of main  goals of SUSY searches at the LHC, but limits are still relatively weak and strongly model dependent.
Recently, full scans of the R-parity- and CP-conserving parameter space of the MSSM -- the so-called pMSSM -- have been done both by ATLAS \cite{ATLAS:2024qmx} and by CMS \cite{CMS-PAS-SUS-24-004}.
These show only a quite modest progress in extending the excluded model-space with respect to what has been excluded by the experiments at LEP. 
For further discussion regarding the interpretation of LHC SUSY limits we refer to~\cite{Berggren:2020tle}.

The nature of the EWinos is determined by the hierarchies of the underlying mass parameters.
Within the MSSM, the relevant mass parameter are $M_1$ and $M_2$, the soft SUSY-breaking parameters corresponding to 
the $U(1)$ and $SU(2)$, respectively, as well as the Higgsino mass parameter $\mu$.
The motivation for light EWinos can be two-fold. There are interesting theoretical arguments that favour
$\mu \ll M_1, M_2$, corresponding to light higgsinos. On the other hand, there are some interesting
anomalies in the LHC searches for light EWinos, possibly favouring $M_1, M_2 \ll \mu$, corresponding to
a light wino/bino scenario. Both scenarios have their own implications for the searches at the LC, as will 
be briefly described below, focusing on the MSSM as the benchmark model.

\paragraph[Natural SUSY: linear collider reach for light higgsinos]{Natural SUSY: linear collider reach for light higgsinos}

While supersymmetry offers a 't Hooft technically natural solution to the
Big Hierarchy problem, cumulative results from
LHC Run 2~\cite{Canepa:2019hph,ATLAS:2024lda}
and WIMP direct detection searches have brought out a
Little Hierarchy problem (LHP):
why is there a gap between the weak scale $m_{weak}\sim m_{\PW,\PZ,\PSh}\sim$ \SI{100}{GeV}
and the apparently lowest-possible soft SUSY breaking scale $m_{\mathrm{SUSY}} >$ \SI{1}{-}\SI{2}{TeV}?

The LHP is a problem of practical naturalness~\cite{Baer:2015rja}:
an observable is natural
if all independent contributions to the observable are comparable
to or less than the observable in question.\footnote{this is how naturalness was successfully applied
for instance by Gaillard and Lee in predicting the charm quark mass shortly
before it was discovered~\cite{Gaillard:1974mw}.} In the case of SUSY, the magnitude of the weak scale
is related to the soft SUSY breaking terms and $\mu$ parameter by the
electroweak minimization conditions:
\begin{equation}
  m_{\PZ}^2/2=\frac{m_{H_d}^2+\Sigma_d^d-(m_{H_u}^2+\Sigma_u^u)\tan^2\beta}{\tan^2\beta -1}-\mu^2\simeq -m_{H_u}^2-\Sigma_u^u(\tilde{t}_{1,2})-\mu^2 
\label{eq:mzs}
\end{equation}
where $m_{H_{u,d}}^2$ are soft SUSY breaking Higgs mass terms,
the $\Sigma_{u,d}^{u,d}$ contain a large assortment of SUSY loop
corrections
and $\mu$ is the SUSY conserving $\mu$ parameter which feeds mass to
$\PW,\PZ,\PSh$ and also to the SUSY higgsinos (the superpartners of the Higgs bosons).
Practical naturalness requires all terms on the
right-hand-side of Eq.~\eqref{eq:mzs} to be comparable
(within a factor of a few) to $m_{\PZ}^2/2$.
Note this is the most conservative and non-optional application of naturalness
(plausibility) within the MSSM.
Gluinos, top squarks and other SUSY particles have loop-suppressed
contributions to the weak scale and so can live at the multi-TeV scale with
little cost to naturalness, but higgsinos, with mass $m(higgsino)\sim \mu$,
must lie close to the weak scale, with $\mu \lesssim 200-400$ GeV.
Thus, higgsino pair production is a key target for an $\ee$ collider
operating with $\sqrt{s}>2m(higgsino)$~\cite{Baer:2014yta}.
Indeed, there are at present some
slight $2\sigma$ excesses for both ATLAS~\cite{ATLAS:2019lng}
and CMS~\cite{CMS:2021edw,CMS:2025ttk} Run~2 data
for the soft dilepton+jet+missing energy signature which could arise from
light higgsino pair production at LHC. 

\begin{figure}[!htbp]
\begin{center}
\includegraphics[height=0.4\textheight]{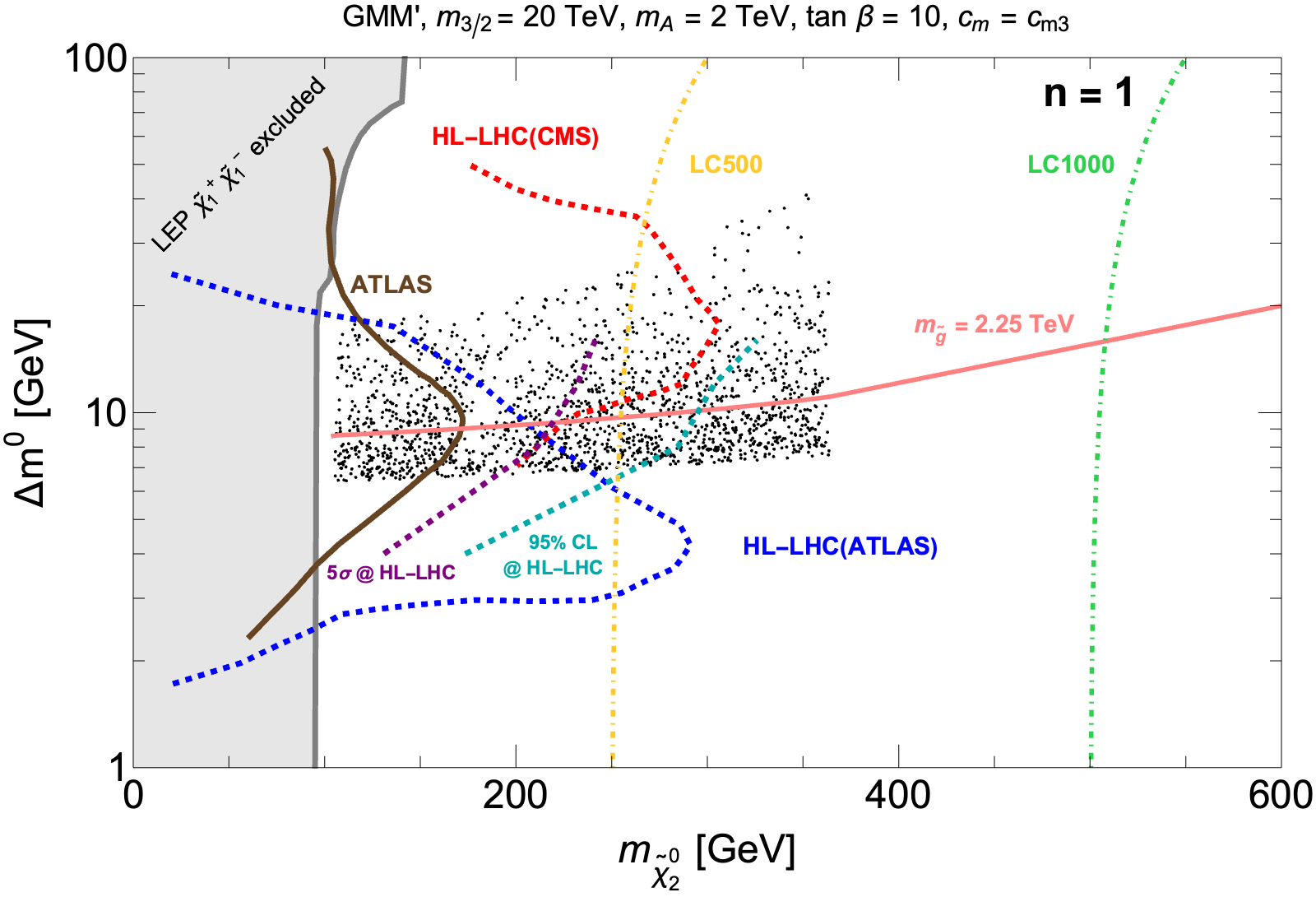}
\caption{The higgsino discovery plane~\cite{Baer:2020sgm} along with predictions from
  a stringy natural landscape SUSY model (black dots:
  the greater the density, the more stringy natural it is)
  along with projected HL-LHC reach and
  the reach of a linear collider at \SI{500}{GeV} and \SI{1}{TeV}.
  The various colliders can probe the left-side of the reach contours.
\label{fig:LCsusy}}
\end{center}
\end{figure}

In Fig.~\ref{fig:LCsusy}, we show the higgsino discovery
plane~\cite{Baer:2020sgm} $m_{\tilde{\chi}_2^0}$ vs.
$\Delta m^0\equiv m_{\tilde{\chi}_2^0}-m_{\tilde{\chi}_1^0}$
along with theoretically preferred dots from (stringy) natural SUSY as
expected from the string landscape: these prefer mass gaps
$\Delta m^0\sim$\SI{5}{-}\SI{10}{GeV}, making discovery at LHC difficult in searching
for dileptons from $\tilde{\chi}_2^0\rightarrow \ell^+\ell^-\tilde{\chi}_1^0$
decay since kinematically $m(\ell^+\ell^-)< m_{\tilde{\chi}_2^0}-m_{\tilde{\chi}_1^0}$.
Several HL-LHC reach contours are shown which probe the
left-side of the considered mass plane \cite{Baer:2021srt}.
Whereas HL-LHC can see much of the theoretically-favoured region,
it cannot see all of it.
We also show the reach contours of a linear collider operating at \SI{500}{GeV} and \SI{1}{TeV}.
In particular, once $\sqrt{s}>m_{\tilde{\chi}_2^0}+m_{\tilde{\chi}_1^0}$ or
$2m_{\tilde{\chi}_1^\pm}$, then a linear collider should have no problem picking up the
soft decay debris from light higgsino pair production, and ultimately
reconstructing the various higgsino masses and underlying SUSY
parameters~\cite{Baer:2019gvu}.

\begin{figure}[!htbp]
\begin{center}
\begin{subfigure}{0.49\textwidth}
    \includegraphics[width=\textwidth]{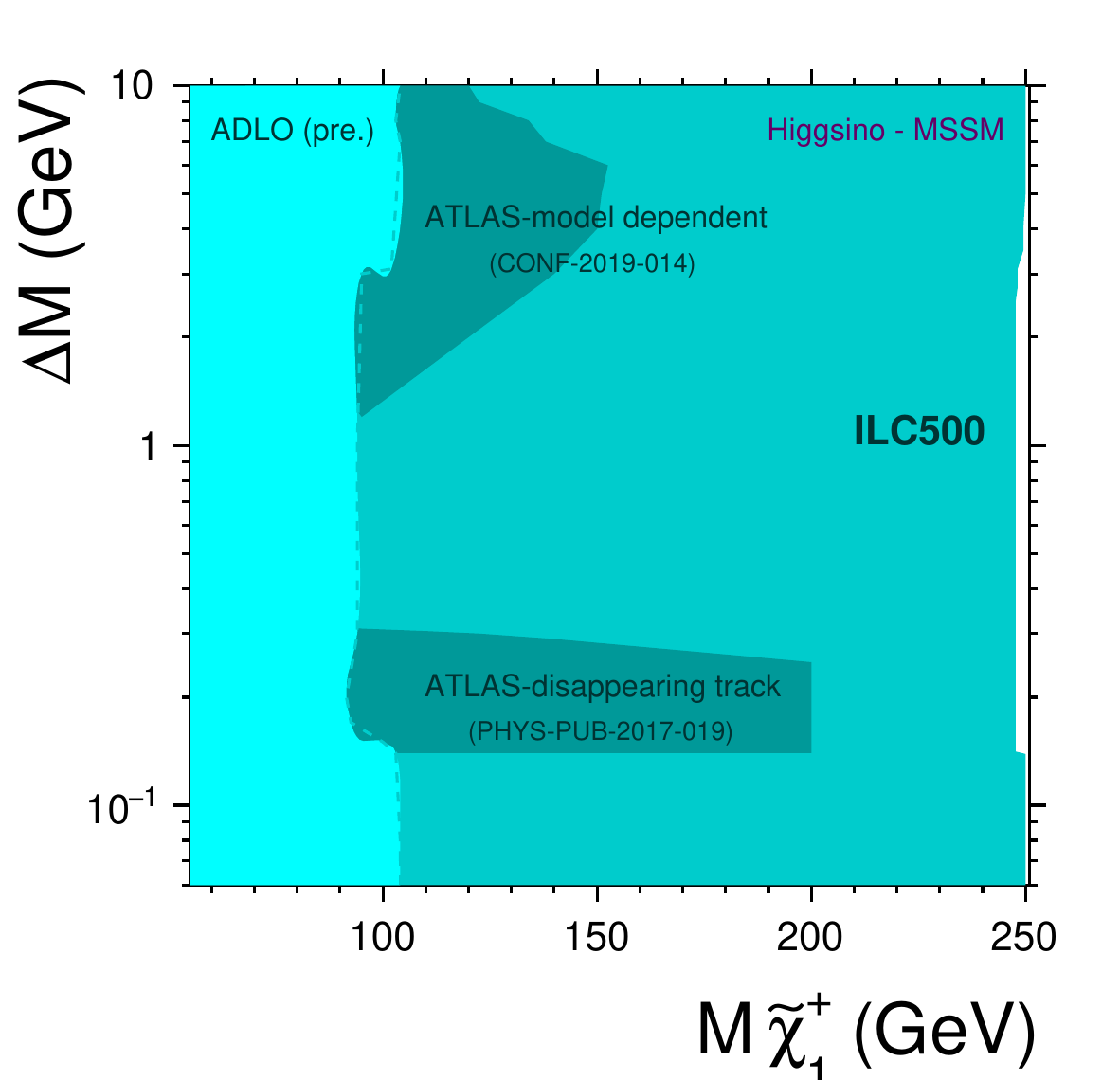}
    \caption{}
    \label{fig:LCsusy2:higgsino}
\end{subfigure}
\begin{subfigure}{0.49\textwidth}
    \includegraphics[width=\textwidth,trim=0 0 1.5cm 1.5cm,clip]{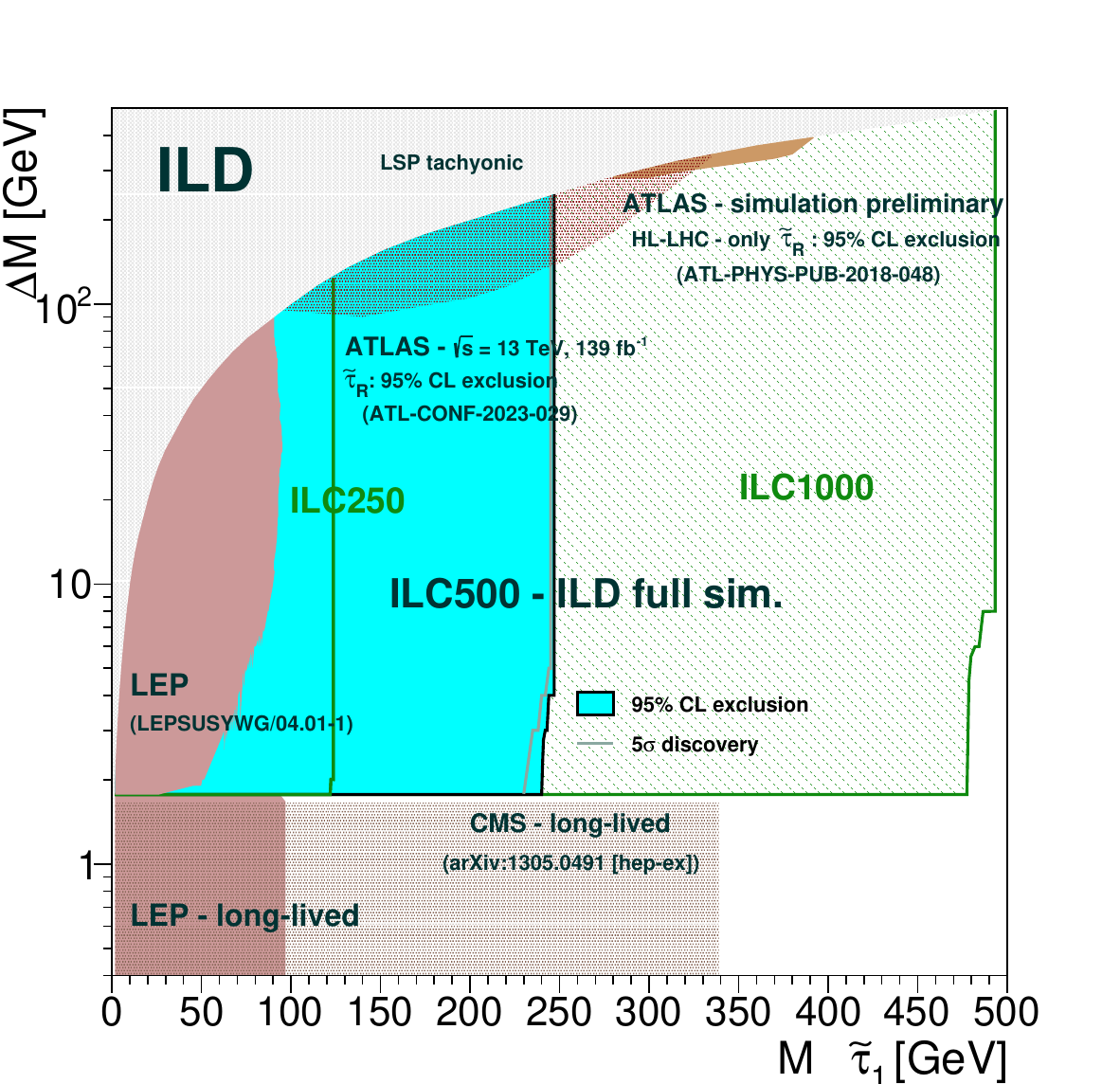}
    \caption{}
    \label{fig:LCsusy2:stau}
\end{subfigure}
\caption{Comparison of the existing limits from LEP and LHC with the expected ILC limits for (a) $\tilde{\chi}^\pm_1$ search in the higgsino-like scenario~\cite{PardodeVera:2020zlr} and (b) $\PSGt$ search in the worst-case scenario, modified from~\cite{Berggren:2024ckz}.
\label{fig:LCsusy2}
}
\end{center}
\end{figure}

SUSY searches at the linear \ee\ colliders profit from the well defined initial state, electron and positron beam polarisations, hermetic detectors, clean running environment and trigger-less operation, which is a huge advantage when looking for unexpected and most likely very soft final states. 
Prospects for SUSY particle discovery, identification and measurements were investigated in great detail, based on the full detector simulation, for selected scenarios~\cite{Berggren:2013vfa,Baer:2016new}. As example, Fig.~\ref{fig:LCsusy2:higgsino} shows the coverage in the $\Delta m$-vs-$m$ plane for higgsinos expected from the ILC~\cite{PardodeVera:2020zlr}, showing that a linear collider facility can explore the full kinematic space of the considered models up to the kinematic limit.

\paragraph[Linear collider reach for light sleptons]{Linear collider reach for  light sleptons}

Another example for the complementarity of \ee\ and hadron colliders regarding SUSY searches are sleptons. Often, in particular when they are the next-to-lightest SUSY particle, sleptons decay into the SM partner lepton and the LSP. In case of the $\PSe$ or the $\PSGm$, the clean environment of a lepton collider allows comfortable discovery up to the kinematic limits~\cite{Berggren:2013vna}, and actually precision measurements of their masses and cross sections~\cite{Berggren:2015qua, Berggren:2020cjs}. 
The experimentally most challenging case is the one of the $\PSGt$, due to the $\PGt$-lepton decaying further with only partial visibility. Therefore this case was studied more intensely in detailed simulations of the ILD concept~\cite{Berggren:2024ckz, NunezPardodeVera:2022izz, NunezPardodeVera:2021cdw}. However Fig.~\ref{fig:LCsusy2:stau} shows that even under the most challenging circumstances w.r.t\ the mixing of the $\PSGt$'s and the nature of the LSP, a linear collider facility can explore the full kinematic space of the considered models up to the kinematic limit~\cite{NunezPardodeVera:2021cdw}.


\paragraph[Linear collider reach for the light wino/bino scenario]{Linear collider reach for the light wino/bino scenario}

Recent searches for the ``golden channel'' for EWinos, $\Pp\Pp \to \PSGczDt \PSGcpmDo \to \PSGczDo Z^{(*)} \, \PSGczDo \PW^{(*)}$
show consistent excesses between ATLAS and CMS in the 2-lepton and the 3-lepton 
searches, assuming $m_{\PSGczDt} \approx m_{\PSGcpmDo} \ge$ \SI{200}{GeV} and 
$\Delta m := m_{\PSGczDt} - m_{\PSGczDo} \approx$ \SI{25}{GeV}~\cite{ATLAS:2019lng,ATLAS:2021moa,CMS:2021edw,CMS:2024gyw,CMS:2025ttk}.
This mass configuration arises naturally in SUSY scenarios with wino/bino DM. 
In these scenarios the lightest supersymmetric particle (LSP),
assumed to be the lightest neutralino, $\PSGczDo$, serves as a DM candidate. 

\begin{figure}[!htbp]

\centering
\begin{subfigure}{0.49\textwidth}
\includegraphics[width=\textwidth]{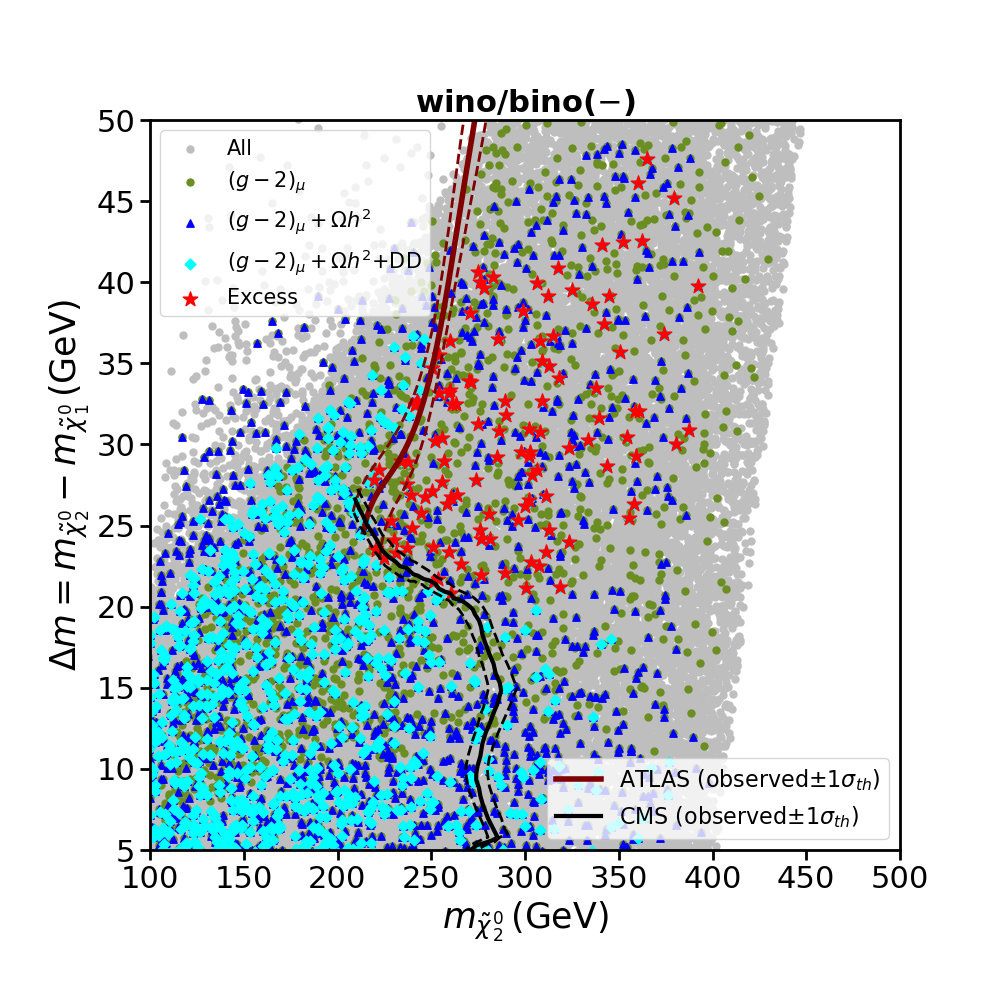}
\caption{}
\label{fig:neu2-cha1:a}
\end{subfigure}
\begin{subfigure}{0.49\textwidth}
\includegraphics[width=\textwidth, trim=0 0 6cm 0,clip]{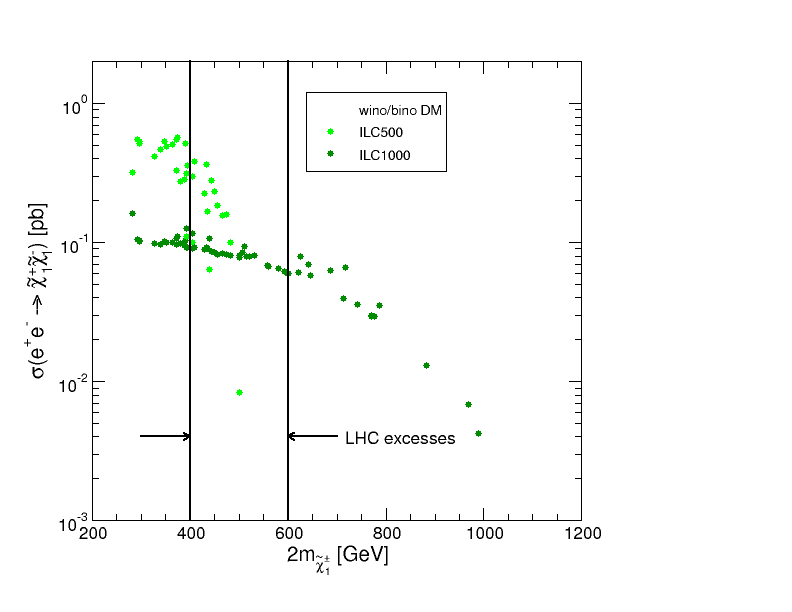}
\caption{}
\label{fig:neu2-cha1:b}
\end{subfigure}
\caption{(a) Results of a parameter scan in the wino/bino DM scenario in the 
$m_{\PSGczDt}$--$\Delta m (:= m_{\PSGczDt} - m_{\PSGczDo})$ plane of the MSSM.
For the colour coding see text. The red stars show the points that are in agreement with all
experimental constraints. Taken from~\cite{Chakraborti:2024pdn}.
(b) Cross section predictions for $\Pep\Pem \to \PSGcpmDo\PSGcmpDo$ in the wino/bino DM scenario
for two centre-of-mass energies, $\sqrt{s} = $\SI{500}{GeV}, \SI{1}{TeV}.
The range that roughly corresponds to the observed LHC excesses is marked by two vertical lines.
Adapted from~\cite{Chakraborti:2020vjp}.
}
\end{figure}

Taking into account the relevant LHC search limits, the DM relic density~\cite{Planck:2018vyg} 
as well as the DM direct detection limits from LZ~\cite{LZ:2022lsv}, it was shown in \cite{Chakraborti:2024pdn} 
that indeed wino/bino DM in the MSSM can fit the observed excesses. To this end, a parameter scan in the wino/bino scenario was performed with $|M_1|$ varied between \SI{100}{GeV} and \SI{400}{GeV}, 
$|M_1| \le M_2 \le 1.1 |M_1| \le \mu$ and $2 \le \tan\beta \le 60$. The slepton masses were assumed to be 
heavier than the LSP. The result of the scan (in the case $M_1 \times \mu < 0$) is shown in Fig.~\ref{fig:neu2-cha1:a} in the $m_{\PSGczDt}$--$\Delta m$ plane of the MSSM. 
The gray points show all scan points, and the various colours indicate the effects of the experimental  constraints taken into account.\footnote{Light slepton masses yield a large contribution for the anomalous magnetic moment of the muon, whereas heavier slepton masses yield a negligible contribution.}
The solid dark red (black) line show the search limits from ATLAS~\cite{ATLAS:2019lng,ATLAS:2021moa} 
(CMS~\cite{CMS:2021edw,CMS:2024gyw}) in the two search channels discussed above, together with the corresponding theory uncertainties indicated by  dashed lines. Shown in only the stronger of the two experimental limits. 
Consequently, the red stars indicate the scan points that fulfill all experimental constraints, including the LHC searches. 
One can observe that the part of the parameter space where ATLAS and CMS observe an excess in both their search channels is well populated. 
In \cite{Chakraborti:2024pdn} it was furthermore shown that for $m_{\PSGczDt} \approx m_{\PSGcpmDo} \le $\SI{250}{GeV} and $\Delta m \approx $\SI{25}{GeV} the LHC production cross section for $pp \to \PSGczDt \PSGcpmDo \to \PSGczDo Z^{(*)} \, \PSGczDo W^{(*)}$ has roughly the size to produce the observed excesses. 
Going a step further, \cite{Agin:2024yfs} discussed the complementarity of the soft lepton excesses with other  excesses in LHC monojet searches~\cite{ATLAS:2021kxv,CMS:2021far}, and showed that non-SUSY interpretations fit these excesses less well.

This opens up very interesting prospects for the production of light SUSY particles at a LC. In a previous
analysis~\cite{Chakraborti:2020vjp} it was shown that the production cross section for 
$\Pep\Pem \to \PSGcpmDo\PSGcmpDo$ 
is in the range of ${\cal O}$(\SI{100}{fb}) in this kind of scenarios. This is demonstrated in Fig.~\ref{fig:neu2-cha1:b} (adapted from~\cite{Chakraborti:2020vjp}), 
where for wino/bino DM the chargino production cross section is shown as a 
function of $2 m_{\PSGcpmDo}$ for two centre-of-mass energies, $\sqrt{s} = $\SI{500}{GeV}, \SI{1}{TeV}.
The range that rougly corresponds to the observed LHC excesses is marked by two vertical lines.
This prominent example demonstrates the possibility that BSM particles discovered by the LHC that are 
within the kinematic reach of a linear collider can be produced abundantly, and thus can also be studied in detail.


\subsubsection{Linear collider reach for light exotic scalars}
\label{sec:phys:bsm:exscalar}

\begin{figure}[htbp]
    \centering
    \includegraphics[width=0.8\linewidth]{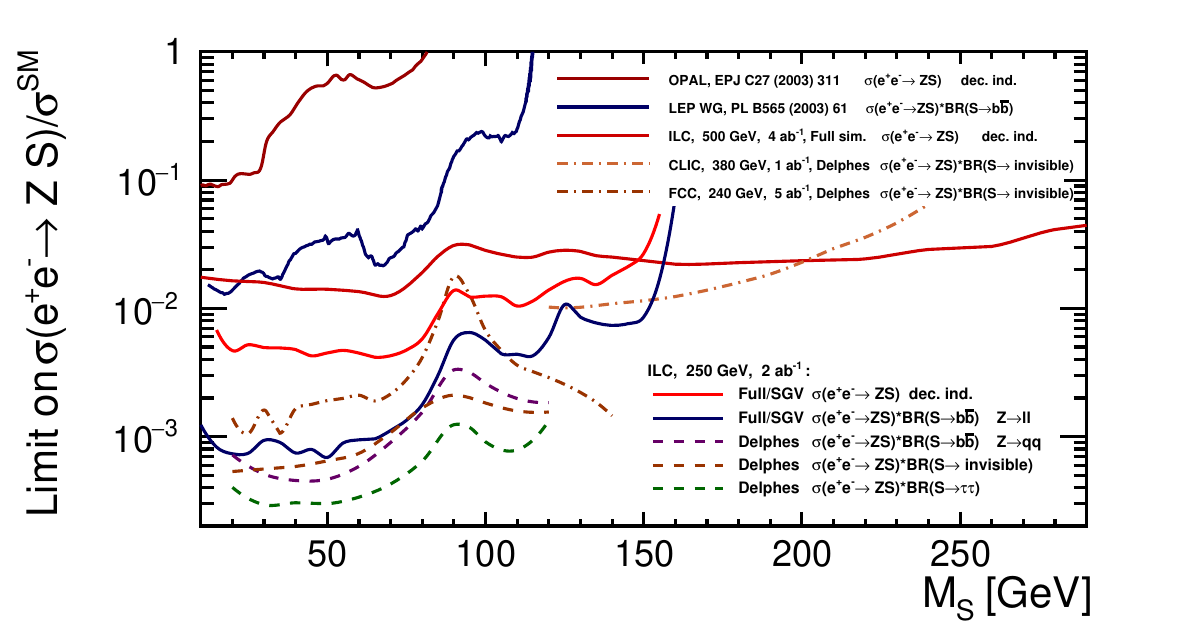}
    \caption{Comparison of expected exclusion limits for exotic light scalar production searches in the scalar-strahlung processes, for different search strategies and scalar decay channels considered. When particular decay channel is indicated, limits are set on the cross section ratio times the branching fraction. Figure updated from~\cite{ECFA-HF-Report}.}
    \label{fig:EXscalar}
\end{figure}

BSM scenarios 
involving light scalars, with masses accessible at future lepton colliders, are not excluded even with the latest experimental data. 
Sizeable production cross sections for new scalars can also coincide with non-standard decay patterns, so a range of decay channels should be considered.
Lepton colliders can be sensitive to exotic scalar production even for very light scalars, thanks to the clean
environment, and precision and hermeticity of the detectors. 
Different production and decay channels, including invisible decays of a new scalar, S, can be considered. The highest sensitivity is expected to come from the scalar-strahlung production process, $\ee \to \PZ$S, where production of the new scalar can be tagged, independent of its decay, based on the recoil mass technique (similar to the approach used for the \SI{125}{GeV} Higgs produced in the Higgs\-strahlung process).
Recent results on light exotic scalar searches in the scalar-strahlung process, submitted as input to the report of the ECFA Higgs/Electroweak/Top Factory Study~\cite{ECFA-HF-Report}, are summarized in Fig.~\ref{fig:EXscalar}.
Strong limits on the light scalar production cross section are already expected from decay-mode-independent searches, based on the recoil mass reconstruction technique.
If a light scalar preferentially decays to tau pairs, an improvement in search sensitivity by more than an order of magnitude is expected, making this channel the most sensitive one in this type of searches (in terms of limits on production cross section times corresponding branching fraction).
Expected exclusion limits in the invisible scalar channel as well as limit estimates for the scalar decay to \bb with hadronic \PZ boson decays, are slightly weaker due to much higher backgrounds, while constraints on scalar decays to \bb with leptonic signature are limited by the leptonic branching fraction of the \PZ boson.  

\begin{figure}[htbp]
    \centering
    \includegraphics[width=0.6\linewidth]{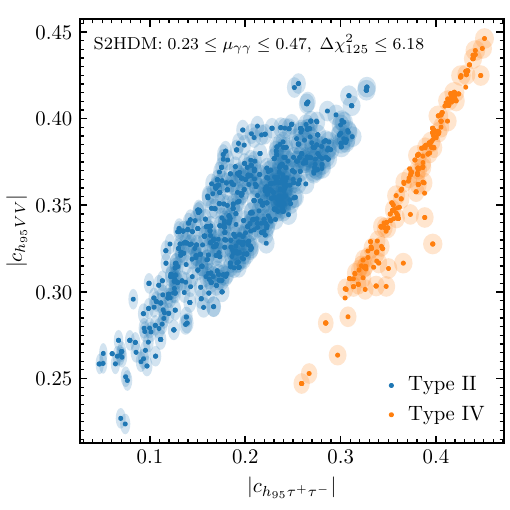}
    \caption{S2HDM parameter points passing the applied constraints that predict a di-photon signal strength in the
preferred range by CMS~\cite{CMS:2024yhz} in the ($|c_{\PSh_{95}\PGt\PGt}|$--$|c_{\PSh_{95}\PV\PV}|$) plane (see text)
The type II and the type IV parameter points are shown in blue and orange, respectively. The shaded ellipses around
the dots indicate the projected experimental precision with which the couplings of $\PSh_{95}$ could be measured at the ILC250
with \SI{2}{\abinv} of integrated luminosity. Taken from \cite{Biekotter:2023jld}.}
    \label{fig:h95-coup}
\end{figure}

As a concrete example we take the hints of a new Higgs boson at $\sim$ \SI{95.4}{GeV} ($\PSh_{95}$) in the 
$\Pg\Pg \to \PSh_{95} \to \PGg\PGg$ channel as reported by CMS~\cite{CMS:2024yhz}
and ATLAS~\cite{ATLAS:2024itc};  see~\cite{Biekotter:2023oen} for a combination leading to an excess of \SI{3.1}{\sigma}. 
In the same mass range LEP reported a $\sim$ \SI{2}{\sigma} excess in the $\ee \to \PZ \to \PZ \PSh_{95} \to \PZ \PQb\PAQb$ 
channel~\cite{LEPWorkingGroupforHiggsbosonsearches:2003ing}. 
These two types of excesses can simultaneously be described in a variety of BSM models. 
It was shown in~\cite{Biekotter:2019kde},
that models extending the SM Higgs sector by one additional doublet and one additional singlet 
are 
an interesting option in this context.
The interpretation of the $\PSh_{95}$ as a new light CP-even Higgs boson,
receiving its couplings to SM particles from a mixing with the SM Higgs doublet, ensures a non-negligible coupling
to $\PZ$~bosons, falling into the regime accessible at future $\ee$ colliders as shown in Fig.~\ref{fig:EXscalar}. 
This results in an abundant production of the $\PSh_{95}$, allowing an analysis of its decay patterns and couplings.
The result of such an analysis is shown in Fig.~\ref{fig:h95-coup}, based on an analysis within the S2HDM (the SM extended
by a second complex doublet and one complex singlet)~\cite{Biekotter:2021ovi, Biekotter:2023jld}. 
Shown are points passing all relevant theoretical
and experimental constraints that predict a di-photon signal strength $\PGm{\PGg\PGg}$ in the
preferred range by CMS~\cite{CMS:2024yhz} in the ($|c_{\PSh_{95}\PGt\PGt}|$--$|c_{\PSh_{95}\PV\PV}|$) plane. Here we denote with 
$c_{\PSh_{95} xx}$ the coupling strength of the $\PSh_{95}$ to the particle $x$ (where $\PV$ denotes $\PWpm$ or $\PZ$)
relative to the coupling strength of the SM Higgs boson.
The Yukawa type~II and the type~IV parameter points are shown in blue and orange, respectively. The shaded ellipses around
the dots indicate the projected experimental precision with which the couplings of $\PSh_{95}$ could be measured at the ILC250
with \SI{2}{\abinv} of integrated luminosity following \cite{Heinemeyer:2021msz} (and references therein). 
One can observe the anticipated 
high precision of the coupling determination of the $\PSh_{95}$ that will allow to discriminate between various Yukawa
types, identify the model realised in nature and pin down its parameters.

\subsubsection{Linear collider reach for heavy neutral leptons}
\label{sec:phys:bsm:hnl}

The observation of neutrino oscillations provides direct evidence that neutrinos have non-zero masses. This indicates the existence of right-handed neutrinos, which are however not a part of the SM. Models with Heavy Neutral Leptons (HNLs) could explain a number of SM ``mysteries'', like the tiny neutrino masses (via see-saw mechanism), the baryon asymmetry of the universe (with additional source of CP violation) or existence of dark matter (with stable sterile neutrinos).

\begin{figure}[!htbp]
\centering
\begin{subfigure}{0.49\textwidth}
\includegraphics[width=\linewidth]{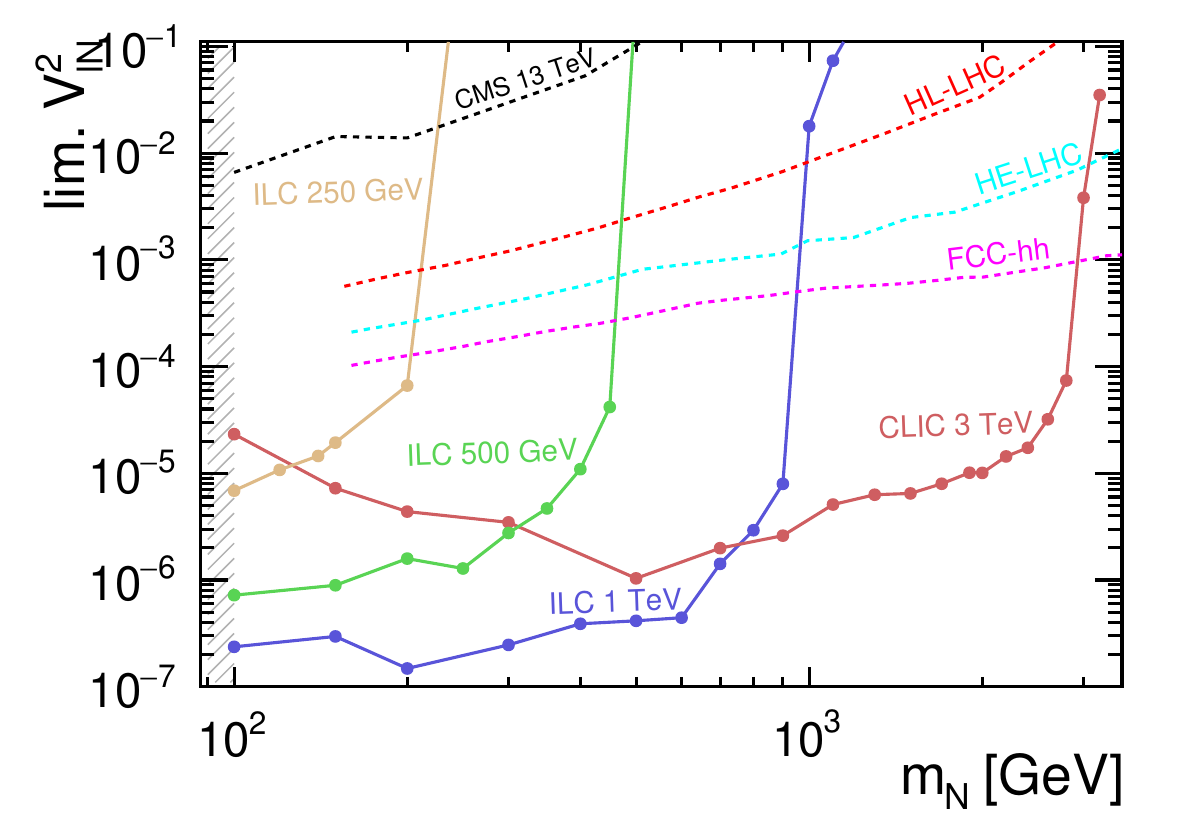}
\caption{}
\label{fig:HNL:a}
\end{subfigure}
\begin{subfigure}{0.49\textwidth}
\includegraphics[width=\linewidth]{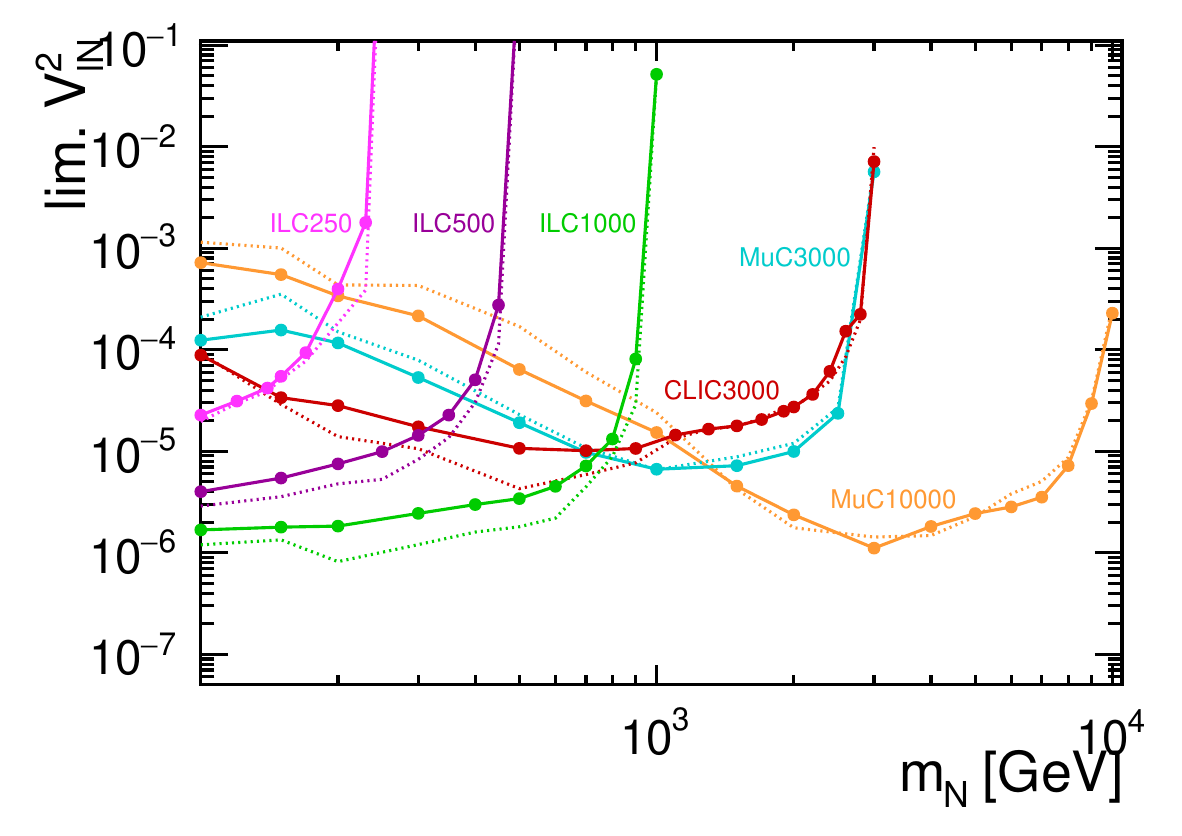}
\caption{}
\label{fig:HNL:b}
\end{subfigure}

\caption{(a) Expected \SI{95}{\%} CL exclusion limits on the coupling $V^2_{\ell N}$ of the Dirac neutrino and (b) \SI{95}{\%} CL discrimination limits between Majorana and Dirac nature, as a function of the heavy neutrino mass, m$_\text{N}$, for different lepton collider setups, as indicated in the plot. Dashed lines in (a) indicate limits from current and future hadron colliders based on~\cite{Sirunyan:2018mtv,Pascoli:2018heg}. Expected \SI{5}{\sigma} discovery sensitivities are indicated by the dotted lines in (b). From~\cite{Mekala:2023kzo}.}
\end{figure}

Lepton colliders are clearly the preferred option for searching for exotic leptons. Depending on the HNL's mass, width and coupling to the SM particles, as well as additional model details (for UV-complete models), many different experimental signatures can be considered.
For light HNL states, up to the masses of the order of m$_{\PW}$, they could show up in rare or invisible decays of the $\PZ$, $\PW$ or Higgs bosons.
With very small SM couplings, light states could also be long-lived, resulting in the striking signature of displaced vertices.
However, for masses above  m$_{\PW}$, only prompt decays of HNLs are expected.
The sensitivity of future lepton colliders to the production of heavy neutrinos was considered in a series of studies~\cite{Mekala:2022cmm,Mekala:2023diu,Mekala:2023kzo}, and their results are summarized in Fig.~\ref{fig:HNL:a}.
The study was based on an effective extension of the SM with three flavours of right-handed neutrinos~\cite{Pascoli:2018heg}, assuming only one of them is kinematically accessible at the studied colliders.
All the heavy-neutrino couplings to the SM leptons were assumed to be the same, $|V_{\Pe N}|^{2} = |V_{\PGm N}|^{2} = |V_{\PGt N}|^{2} \equiv V_{\ell N}^{2}$.
For each of the considered running scenarios, sensitivity range extends almost up to the centre-of-mass energy and the expected limits on the heavy neutrino coupling to the SM leptons, $V^2_{\ell N}$, are much more stringent than any results for high-energy hadron colliders.
This is because, for the considered mass and energy range, the cross section for single HNL production in \epem collisions is dominated by the contribution from t-channel $\PW$-boson exchange and depends only very weakly on the heavy neutrino mass almost up to the kinematic limit. At $\Pp\Pp$ colliders, only s-channel production processes (Drell-Yan) contribute, decreasing fast with the invariant mass of the produced final state.

Heavy Neutral Leptons can be of Dirac nature with only lepton-number conserving (LNC) decays or of Majorana nature, including also lepton-number violating (LNV) production and decay processes. 
If heavy neutrinos are discovered (at \SI{5}{\sigma} level) at a future lepton collider, we should also be able to identify their nature (with \SI{95}{\%} CL), as shown in Fig.~\ref{fig:HNL:b}.

Experiments at the beam dumps of a linear collider facility offer interesting complementary sensitivity to HNLs at lower masses, as discussed in Sec.~\ref{sec:phys:PBC:beamdumps}.


\subsubsection{Linear collider reach for dark matter scenarios}
\label{sec:phys:bsm:dm}

High energy \epem colliders offer a unique possibility for the most general search of dark matter (DM) particles based on the mono-photon signature. The advantage of this approach is that the photon radiation from the initial state is fully described within the SM and depends only indirectly on the DM production mechanism. Prospects for DM searches were
studied for ILC and CLIC, both in the heavy mediator approximation (using the EFT-based approach)~\cite{Blaising:2021vhh,Habermehl:2020njb} as well as for the finite mediator masses, when contributions from resonant production of light mediators have to be taken into account~\cite{Kalinowski:2021tyr}. 

\begin{figure}[!tbp]
\centering
\begin{subfigure}{0.49\textwidth}
\includegraphics[width=\textwidth]{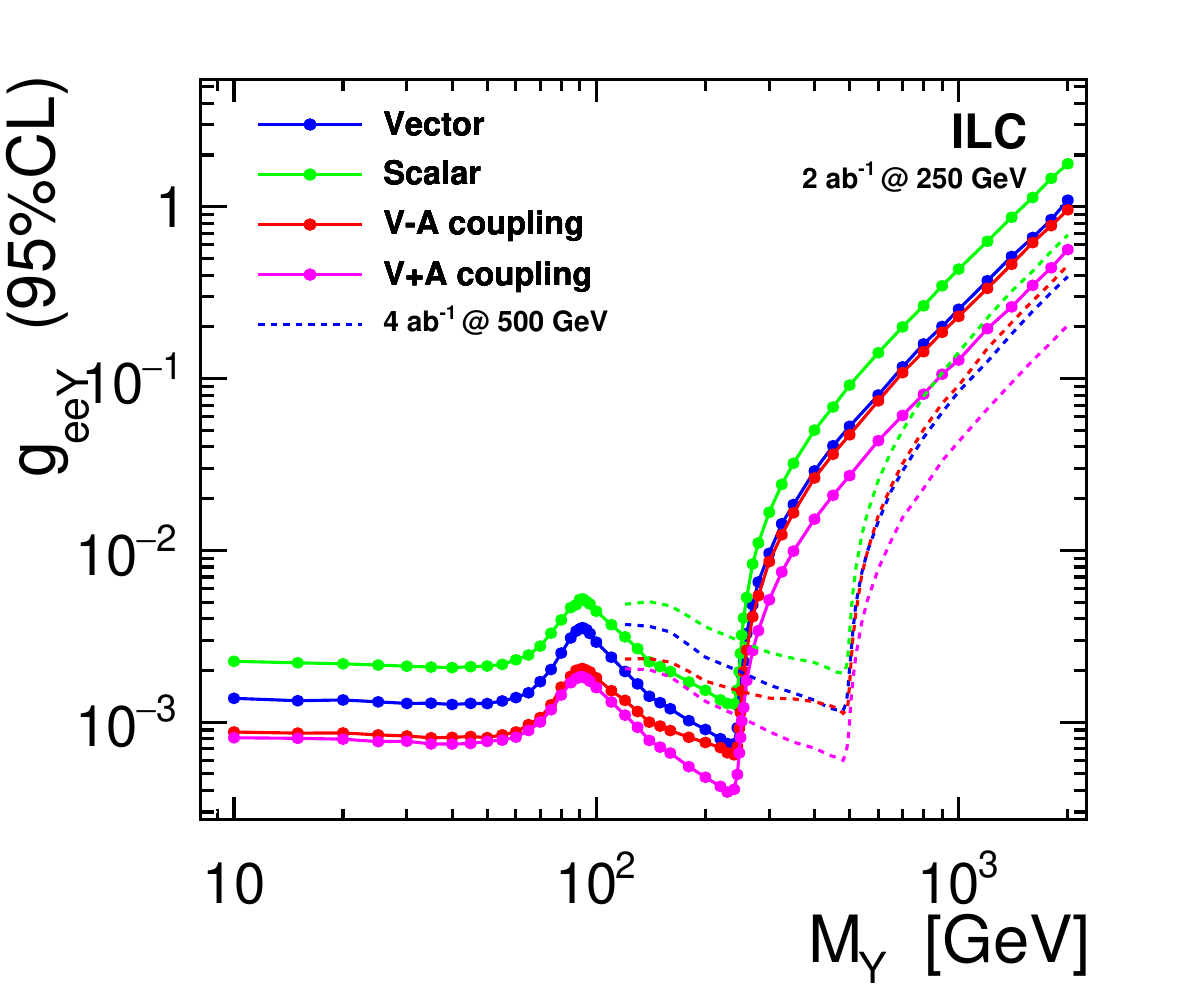}
\caption{}
\label{fig:DM:a}
\end{subfigure}
\hspace{0.001cm}
\begin{subfigure}{0.49\textwidth}
\includegraphics[width=\textwidth]{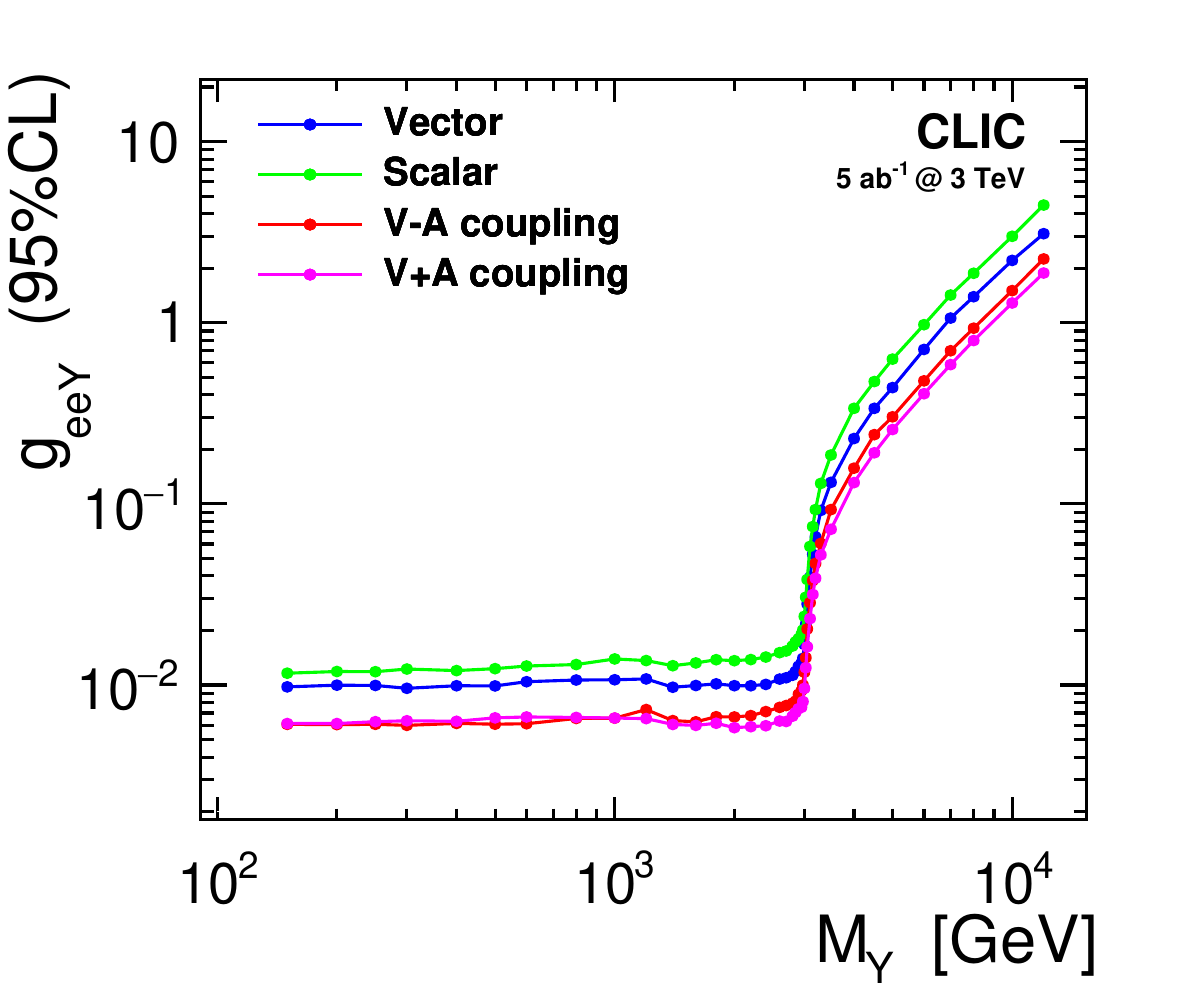}
\caption{}
\label{fig:DM:b}
\end{subfigure}
\caption{(a) Expected limits on mediator coupling to electrons for the ILC running at \num{250} and \SI{500}{GeV} (based on~\cite{Altmann:2025feg}) and (b) CLIC running at \SI{3}{TeV} (taken from~\cite{Kalinowski:2021tyr}), for a relative mediator width $\Gamma$/M = 0.03 and different mediator coupling scenarios, as indicated in the plots. Presented are combined limits corresponding to the baseline running scenarios, with systematic uncertainties taken into account.
\label{fig:DM}
}
\end{figure}

Example limits on the mediator coupling to electrons, $g_{\Pe\Pe Y}$, for the ILC running at \num{250} and \SI{500}{GeV}, and CLIC running at \SI{3}{TeV} are presented in Fig.~\ref{fig:DM}. Considered are scenarios with light Dirac DM pair-production, m$_\chi$=\SI{1}{GeV} for ILC and \SI{50}{GeV} for CLIC, and different structures of mediator coupling to electrons. 
As mediator decays to DM particles are expected to dominate, the mediator coupling to DM is fixed by setting its width to \SI{3}{\%} of its mass.
One has to note that for the coupling range considered, the analysis of mono-photon spectra gives higher sensitivity to processes with light mediator exchange than their direct searches in SM decay channels.

\begin{figure}[!tbp]
\centering

\begin{subfigure}{0.43\textwidth}
\includegraphics[width=\linewidth]{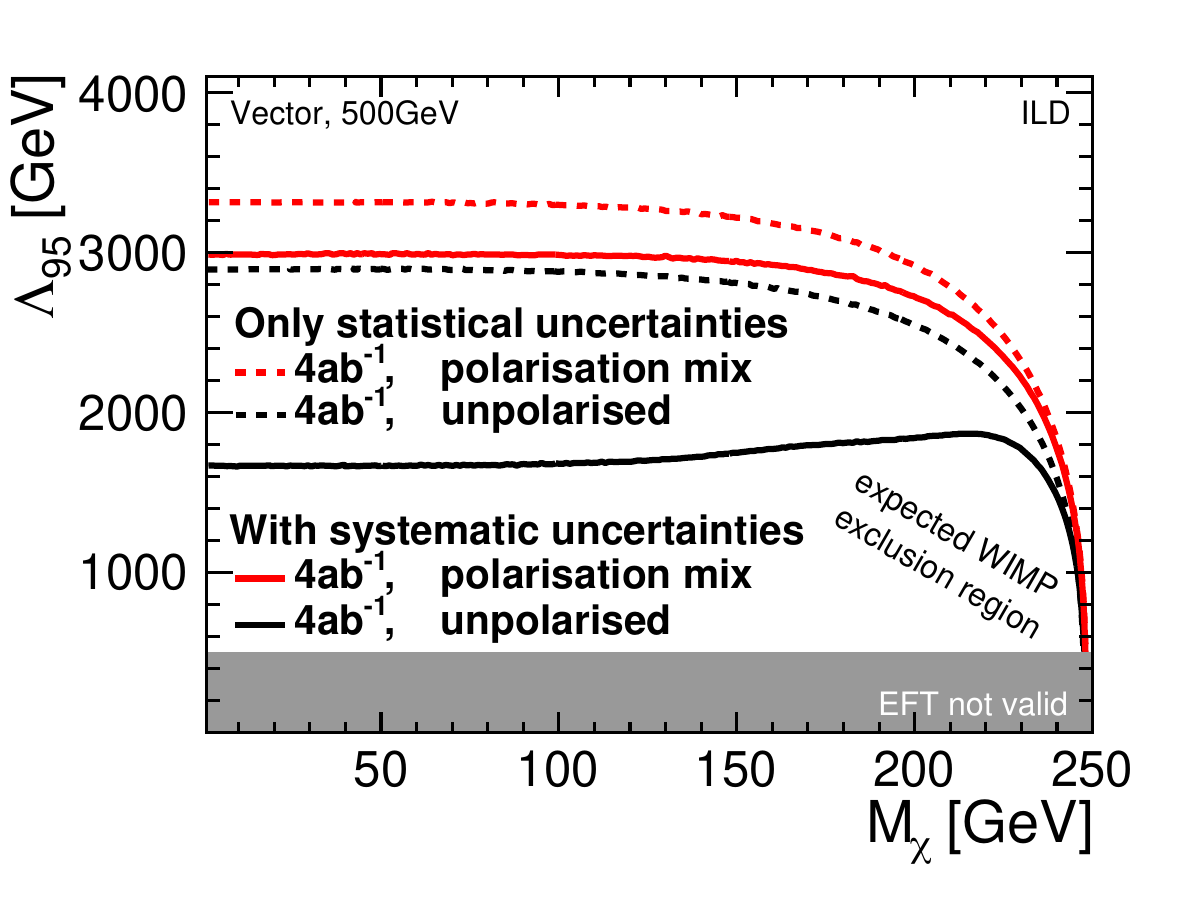}
\caption{}
\label{fig:WIMP:a}
\end{subfigure}
\hspace{0.0001cm}
\begin{subfigure}{0.46\textwidth}
\includegraphics[width=\linewidth, trim=0 0 0 1cm]{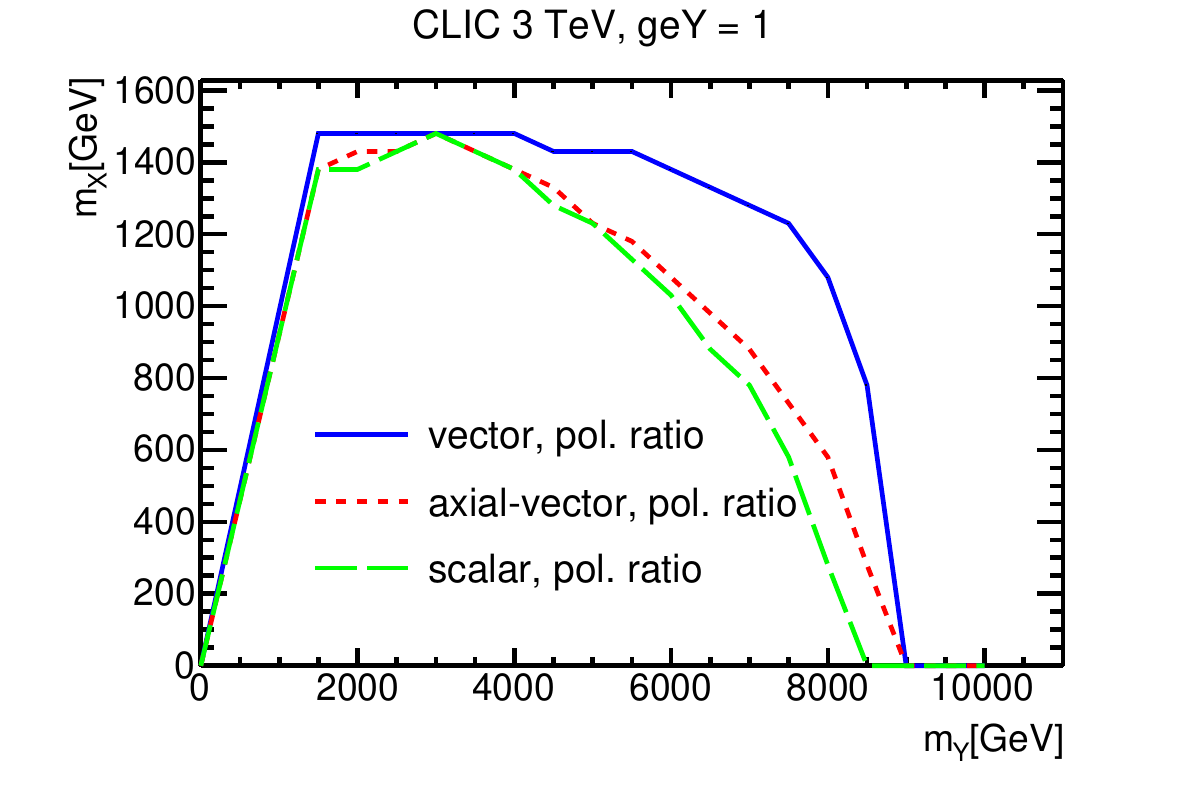}
\caption{}
\label{fig:WIMP:b}
\end{subfigure}

\caption{Expected exclusion limits in the dark matter mass vs mediator mass scale plane, from the search for dark matter pair-production with the mono-photon signature. (a) impact of systematic uncertainties and beam polarisation on the expected limits from the ILC running at \SI{500}{GeV}, assuming vector mediator couplings \cite{Habermehl:2020njb}.
(b) expected sensitivity of CLIC running at \SI{3}{TeV}, based on the measured polarisation ratio, for different mediator coupling scenarios~\cite{Blaising:2021vhh}.
\label{fig:WIMP}}

\end{figure}

The results presented in Fig.~\ref{fig:DM} indicate also that, assuming couplings of the order of one, precision measurements in the mono-photon channel are sensitive to mediator mass scales significantly above the collision energy.
In this domain, the EFT approach is safely applicable, unlike at hadron colliders, which often probe only the mass scales below the collision energy.
In the EFT limit, for high mediator masses above the collision energy, the quantity constrained from the analysis of mono-photon events is the ratio of the mediator mass to the square root of the product of its couplings to electrons $g_{\Pe\Pe Y}$ and DM particles $g_{\chi\chi Y}$: $\Lambda = m_Y/\sqrt{g_{\Pe\Pe Y} g_{\chi\chi Y}}$. Expected exclusion limits in the dark matter mass vs mediator mass plane can be calculated for the selected values of the couplings. Figures~\ref{fig:WIMP:a} and~\ref{fig:WIMP:b} show the expected limit contours resulting from the ILD~\cite{Habermehl:2020njb} and CLIC~\cite{Blaising:2021vhh} studies, respectively, the latter assuming the product of the two couplings to equal unity ($g_{\Pe\Pe Y}\cdot g_{\chi\chi Y} = 1$). These results confirm that the mono-photon signature allow \epem colliders to be sensitive to DM pair-production also for mediator mass scales much higher than the collision energy.

\begin{figure}[htb]
\begin{center}
  \includegraphics[width=0.8\textwidth]{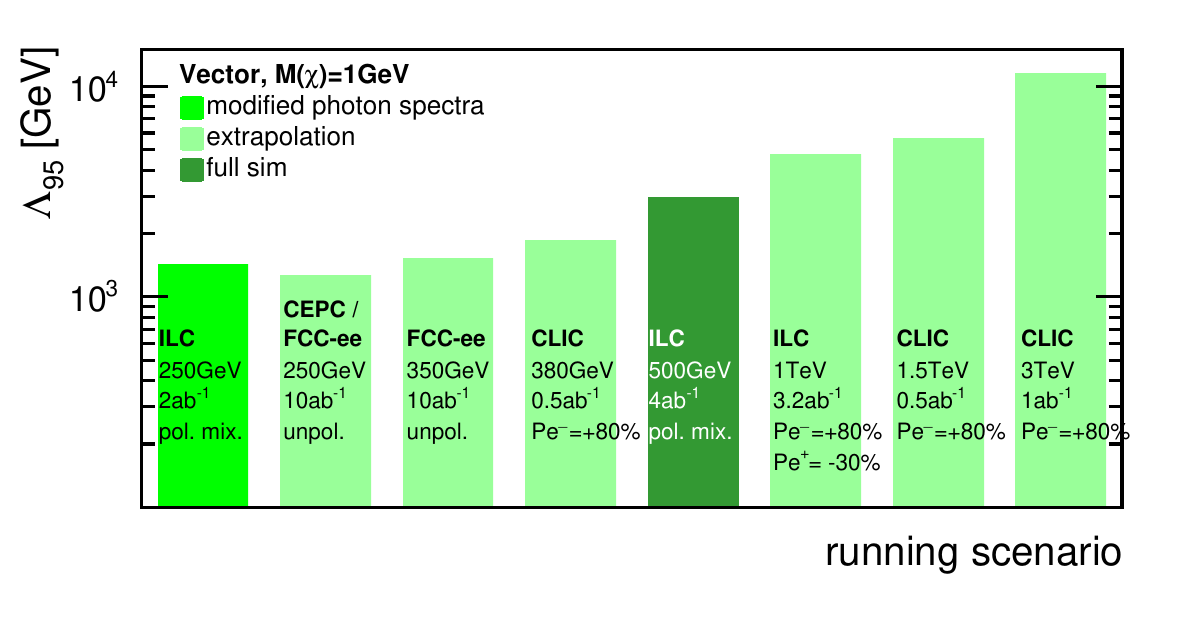} 
\end{center}
 \caption{Results from the search for dark matter pair-production with the mono-photon signature, for vector mediator couplings and dark matter mass of \SI{1}{GeV}.  Compared are expected exclusion limits for different settings of centre-of-mass energy, integrated luminosity and polarisation combination~\cite{Habermehl:2020njb}. }
 \label{fig:manhattan}
\end{figure}

It is important to stress the key role of the beam polarisation in the mono-photon analysis. It is not only essential to reduce SM backgrounds,
but also to control systematic uncertainties, which become important for large data sets expected.
Figure~\ref{fig:manhattan} compares the exclusion limits expected for different running scenarios of future Higgs factories~\cite{Habermehl:2020njb}.
Even with \SI{10}{\abinv}, CEPC and FCC-ee -- circular colliders running without beam polarisation -- cannot compete with \SI{2}{\abinv} of polarised data collected at \SI{250}{GeV} ILC. 

Dark matter charged under electroweak interactions can also be explored at $\epem$ colliders. In this scenario, DM corresponds to the neutral component of an electroweak multiplet. For a recent review on this class of DM candidates, see~\cite{Cirelli:2024ssz}. As it turns out, the thermal relic abundance of electroweak-charged DM agrees with the observed DM density for a mass $\gtrsim \mathcal{O}$(\SI{1}{TeV}); e.g.\ \ee $\simeq$\,\num{1} and \SI{3}{TeV} for an $\mathrm{SU}(2)_L$ doublet and triplet fermion DM, respectively~\cite{Hisano:2006nn, Cirelli:2007xd}. Lighter DM masses can also be viable if supplemented by other DM species (such as axions) or non-thermal production mechanisms~\cite{Baer:2014eja}. At $\epem$ colliders, electroweak-charged DM can be probed for masses up to $\simeq \sqrt{s}/2$.
The search strategy corresponds to that presented in Sec.~\ref{sec:phys:bsm:light_susy} for light SUSY searches -- results presented in Fig.~\ref{fig:LCsusy2} apply for the search of any $\mathrm{SU}(2)_L$ doublet fermion. %
Precision measurements of di-fermion production cross sections could extend the mass reach by $\simeq \mathcal{O}$(\SI{100}{GeV})~\cite{Harigaya:2015yaa, DiLuzio:2018jwd}. High energy $\epem$ colliders thus provide a crucial testing ground for electroweak-charged DM. In particular, CLIC at \SI{3}{TeV} can fully test both the thermal and non-thermal scenarios for the $\mathrm{SU}(2)_L$ doublet case. 

\begin{figure}[bth]
\begin{center}
  \includegraphics[width=0.5\textwidth]{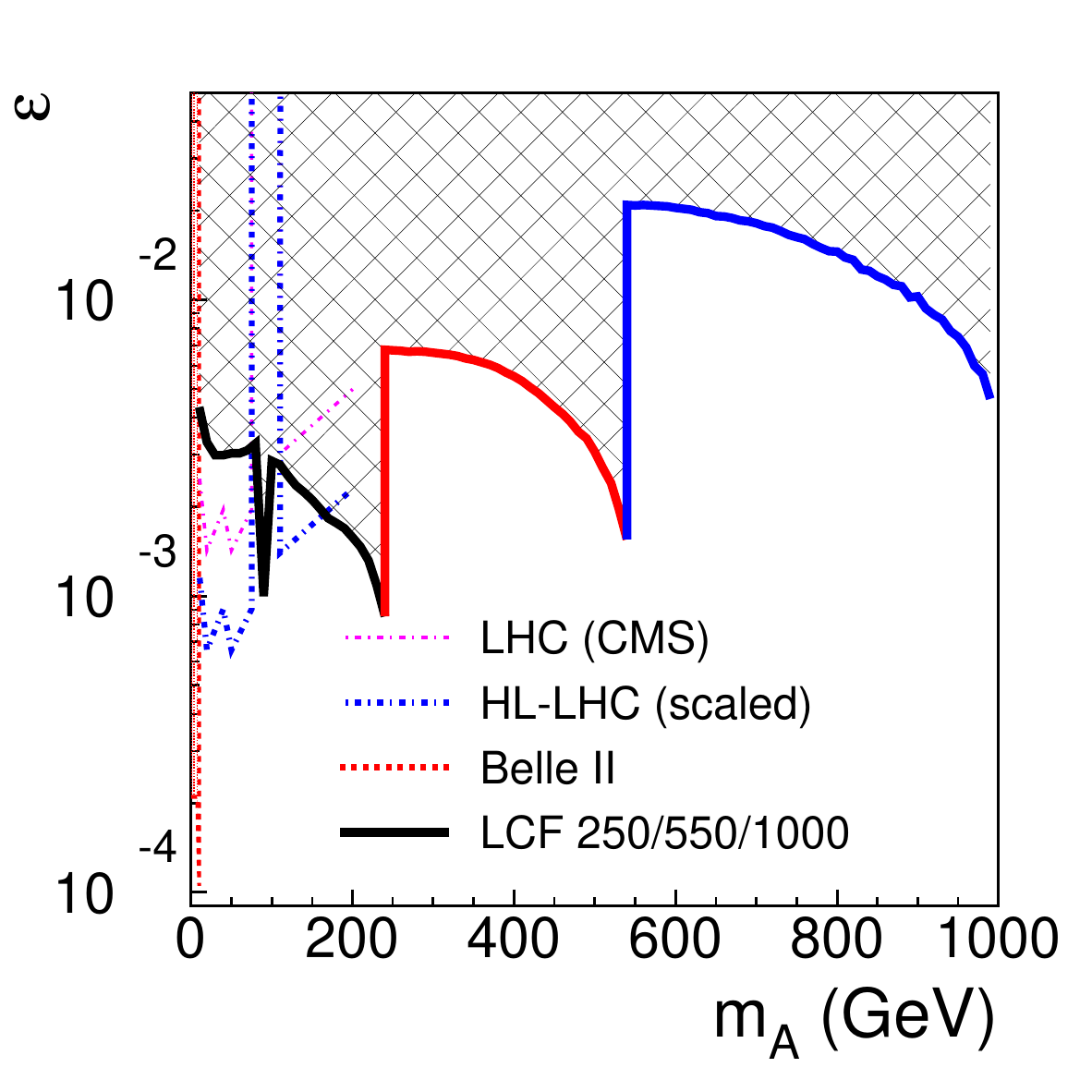} 
\end{center}
\caption{The exclusion reach for the kinetic mixing parameter $\eps$  as a function of the dark photon mass. 
The expected reaches of LCF250, LCF550, and LCF1000 obtained from the full simulation study of ILD~\cite{Hosseini-Senvan:2024,Berggren:2024zed} are shown in the black, red and blue portions of the solid line, respectively, covering together the whole hatched area.
Also shown are the current upper limit from CMS~\cite{CMS:2019buh} as dashed-dotted magenta line, the HL-LHC expectation obtained from re-scaling the CMS result to the expected size of the data-set at HL-LHC (blue dash-dotted), as well as the expectation for Belle II (red dotted, from Fig.~8.16 of~\cite{EuropeanStrategyforParticlePhysicsPreparatoryGroup:2019qin}).}
 \label{fig:darkphoton}
\end{figure}

In a further class of theories, dark matter is explained by postulating the existence of a `dark sector', which interacts gravitationally with ordinary matter. 
If this dark sector contains a U(1) symmetry,
and a corresponding `dark' photon ($A_{D}$), it is natural to expect that this particle kinetically mix with the ordinary photon, and hence become a `portal' through which the dark sector can be studied~\cite{Curtin:2014cca}.
The strength of the mixing is given by a mixing parameter $(\epsilon)$. This same parameter governs both the production and the decay of the $A_{D}$ back to SM particles, and for values of  $\epsilon$ not already excluded, the signal would be a quite small, and quite narrow resonance.
In~\cite{Hosseini-Senvan:2024,Berggren:2024zed} this process was studied using the $A_{D}\rightarrow\PGmp\PGmm$ decay mode in the presence of SM background, using fully simulated signal and background events in the ILD detector~\cite{ILDConceptGroup:2020sfq} at the ILC operating at \SI{250}{GeV}.
It was found that full detector simulation is essential, and that previous theory-based studies~\cite{Karliner:2015tga} are much too optimistic, in that they did not properly take into account the variation of the uncertainty of the determination of the di-muon mass with the momentum and polar angle of the muons.
The analysis at ILC250 was recast to the higher integrated luminosity expected at LCF250, as well as to LCF550 and LCF1000.
The expected exclusion reach in $\epsilon$ is shown in Fig.~\ref{fig:darkphoton}, where also the expected reach of Belle II and HL-LHC are indicated.

\subsubsection{Linear collider sensitivity for the leptophilic Z’ boson}
\label{sec:phys:bsm:zprime}

Many BSM models include new gauge bosons, and the corresponding collider phenomenology strongly depends on their couplings to the SM fields.
One of the interesting scenarios, selected for consideration in the recently published Physics Briefing Book~\cite{deBlas:2025gyz}
is a model with a leptophilic \PZpr boson,
assuming the $L_{\Pe} - L_{\PGm}$ U(1) gauge group and a vector coupling to SM leptons.
While avoiding gauge anomaly and limits from EWPO, this scenario is very challenging for hadron colliders, where only associated production can be considered, and the projected HL-LHC bounds are not expected to be stronger than the existing LEP2 limits, which are also relatively weak.
On the other hand, future lepton colliders offer excellent \PZpr search sensitivity in multiple complementary channels \cite{GonzalezSuarez:2024dsp,Yue:2024kwo}.

As one third of \PZpr decays is expected to be invisible (decays to SM neutrino pairs), coupling limits from the mono-photon search for DM production by exchange of a light mediator~\cite{Kalinowski:2021tyr} shown in Fig.~\ref{fig:DM} can be directly scaled to obtain bounds on the leptophilic \PZpr coupling in the considered scenario.
However, as shown in an FCC-ee study~\cite{GonzalezSuarez:2024dsp}, the best sensitivity to light \PZpr scenarios is expected from the 
direct search in the $\PGmp\PGmm\PGg$ final state.
This is why we selected the di-muon channel for a dedicated LCF sensitivity study.

For the production of light \PZpr bosons with $m_{\PZpr} < \sqrt{s}$, the leading-order signal process considered is $\ee \to \PZpr \, \PGg   \to \PGmp \PGmm \, \PGg $.
For a proper description of the kinematics we include processes with up to three photons on the matrix element (ME) level as well as effects of soft and collinear ISR emissions, with proper ISR-ME matching to avoid double-counting, as described in~\cite{Kalinowski:2020lhp}.
Including these effects is important, as they result in up to \SI{40}{\%} increase in the signal cross section compared to the LO approximation without ISR.
However, higher order effects result also in significant tails in the reconstructed energy distributions of the di-muon system (\PZpr candidate) and of the radiated photons.
We thus decided to look for alternative signal selection criteria:
For radiative \PZpr-return events, the momentum of the radiated photon should be balanced by the momentum of the \PZpr.
The value of the longitudinal momentum of the di-muon system,
$|P^{(\PGm\PGm)}_Z|$ can be thus used as an estimate of the radiated photon energy -- even if photon itself is lost in the beam pipe. 
This allows for efficient selection of radiative di-muon events with a cut on the energy and longitudinal momentum sum, $E^{(\PGm\PGm)} + |P^{(\PGm\PGm)}_Z|$. 
With an additional cut on the total transverse momentum of the event, $P^{(tot)}_T$, suppressing the contribution of processes with neutrinos in the final state, the dominant background comes from the SM production of di-muon pairs, $\ee \to \PGmp \PGmm (\PGg)$.
Additional selection cuts based on the photon reconstruction cannot improve the sensitivity, as the spectra of radiated photons is expected to be the same for the considered signal and the dominant background. 

\begin{figure}[htbp]
    \centering
\begin{subfigure}{0.45\textwidth}
    \includegraphics[width=\linewidth]{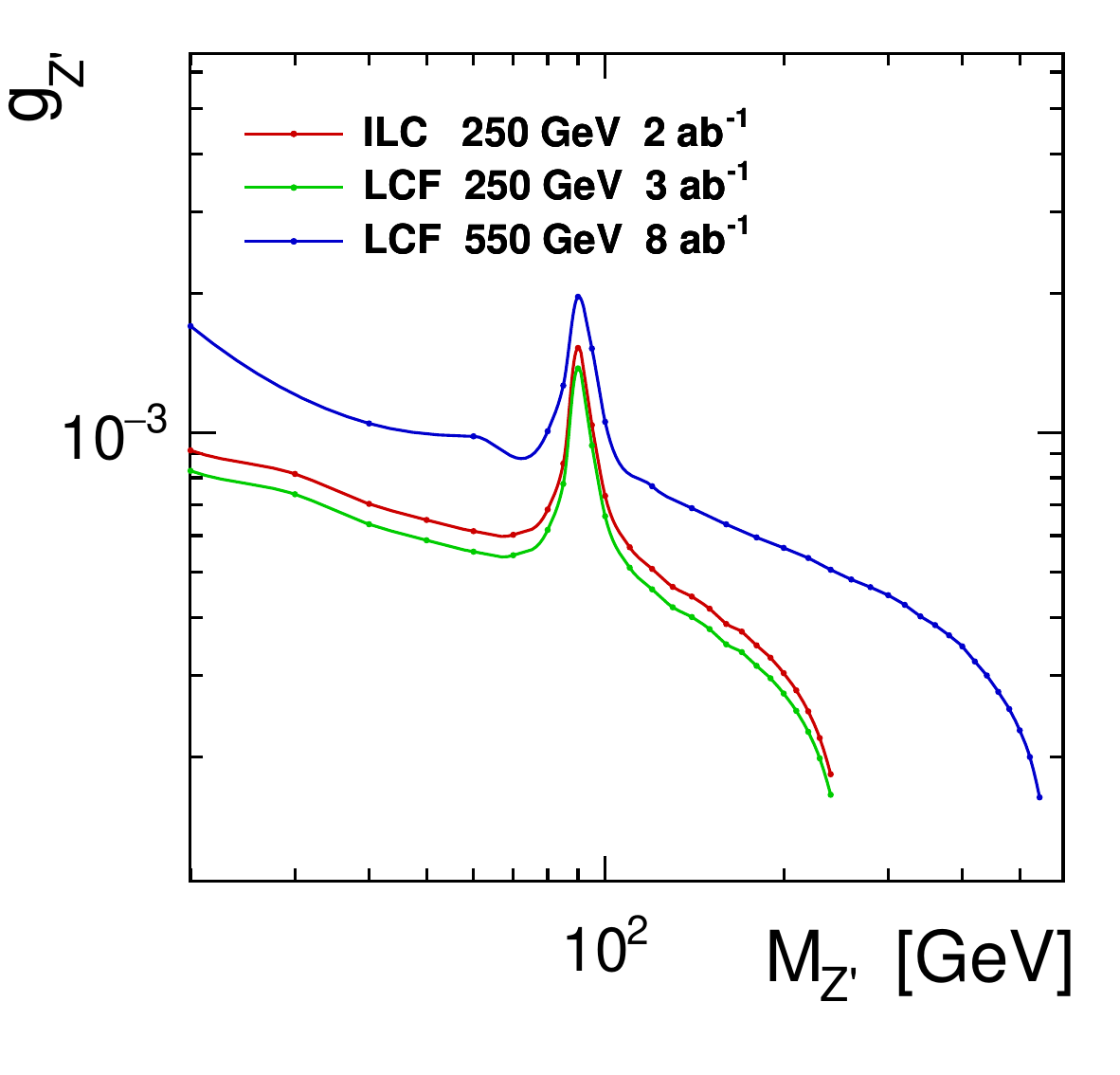}
\caption{}
\label{fig:zprime:a}
\end{subfigure}
\hspace{0.0001cm}
\begin{subfigure}{0.45\textwidth}
    \includegraphics[width=\linewidth,trim=0 -0.45cm 0 0.45cm]{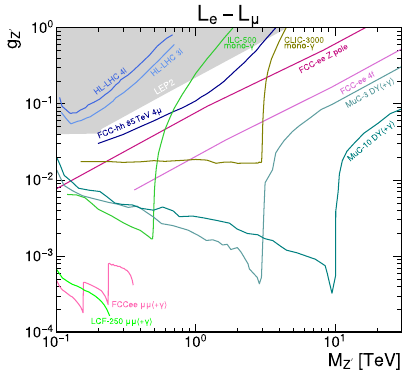}
\caption{}
\label{fig:zprime:b}
\end{subfigure}

    \caption{Expected \SI{95}{\%} CL exclusion limits on the leptophilic
      \PZpr boson coupling. (a) Limits expected from the di-muon
      invariant mass distribution for $\ee \to \PGmp \PGmm (\PGg)$
      events,  based on fast detector simulation for for the
      ILC and LCF running at \SI{250}{GeV}, and LCF running at \SI{550}{GeV}. (b) Comparison of LCF limits at \SI{250}{GeV} with existing
      limits from LEP2 and projected exclusion curves for HL-LHC and
      other future facilities (taken from~\cite{deBlas:2025gyz}).} 
    \label{fig:zprime}
\end{figure}

In the coupling value range considered in this study, the intrinsic width of the \PZpr is expected to be in ${\cal O}$(\SI{10}{keV}) range, negligible compared to the experimental resolution.
For LCF running at \SI{250}{GeV}, the invariant mass of the di-muon from the \PZpr decay can be reconstructed with a resolution of \SI{0.7}{GeV} in the best cases, which worsens to about \SI{2}{GeV} for LCF running at \SI{550}{GeV} due to the higher momenta of the produced muons.
The signal of \PZpr production should be clearly visible as a narrow resonant peak in the di-muon invariant mass distribution on top of the smooth, structure-less (except for the \PZ peak) SM background.
Expected limits on the coupling of the \PZpr boson to SM leptons, assuming no deviations from the SM predictions are observed, are presented in Fig.~\ref{fig:zprime:a} for ILC and LCF running at \SI{250}{GeV}, with a total integrated luminosity of \SI{2}{\abinv} and \SI{3}{\abinv}, respectively, and LCF at \SI{550}{GeV} (\SI{8}{\abinv}).

Also included in  Fig.~\ref{fig:zprime:b} is the comparison of the expected limits from LCF at \SI{250}{GeV} with current LEP2 constraints, HL-LHC projections and prospects for other future facilities~\cite{deBlas:2025gyz}.
One should note that, despite the lower integrated luminosity, the limits for LCF at \SI{250}{GeV} are better than those for FCC-ee at \SI{240}{GeV}.
Multiple factors contribute to this effects. Some are not intrinsic to the collider properties, but to the analysis, i.e.\ the benefit from a higher signal cross-section estimate with multi-photon emission included and the improved selection strategy not requiring a reconstructed photon in the final state. 
Others, however, are intrinsic to the choice of collider, in particular the beam polarisation, helping with signal-to-background ratio, as well as the better muon momentum resolution resulting from higher magnetic field of detectors at linear colliders, which is not possible at circular colliders as it would have too destructive effects on the circulating beam.


\subsubsection{BSM Higgs bosons and triple Higgs couplings at linear colliders}
\label{sec:phys:bsm:BSMHiggs}

As discussed already in Sec.~\ref{sec:gravwaves}, one of the most intriguing open questions
is the origin of the matter-antimatter asymmetry of the universe, which cannot be explained within the SM~\cite{Kajantie:1996mn}.
A broad class of models~\cite{Alves:2018jsw,Li:2019tfd} imply that the required first-order electroweak phase transition (FOEWPT) can only be realized if the mass of an additional Higgs boson, $\PSh$, is not too large, $m_{\PSh} \le $\SI{900}{-}\SI{1000}{GeV}. This makes the boson $\PSh$ a prime target for the HL-LHC, but also for future high-energy \ee\ colliders.
The related physics questions are two-fold. 
Firstly, can such a new, heavier Higgs boson, as favoured by a FOEWPT, be detected at (the HL-LHC and/or) a linear collider? 
Secondly, can we gain access to BSM triple Higgs couplings (THCs) through the processes $\Pep\Pem \to \PZ\PH\PH/\,\PGnl\PAGnl\PH\PH$?

\begin{figure}[!htbp]
\centering

\begin{subfigure}{0.49\textwidth}
\includegraphics[width=\linewidth]{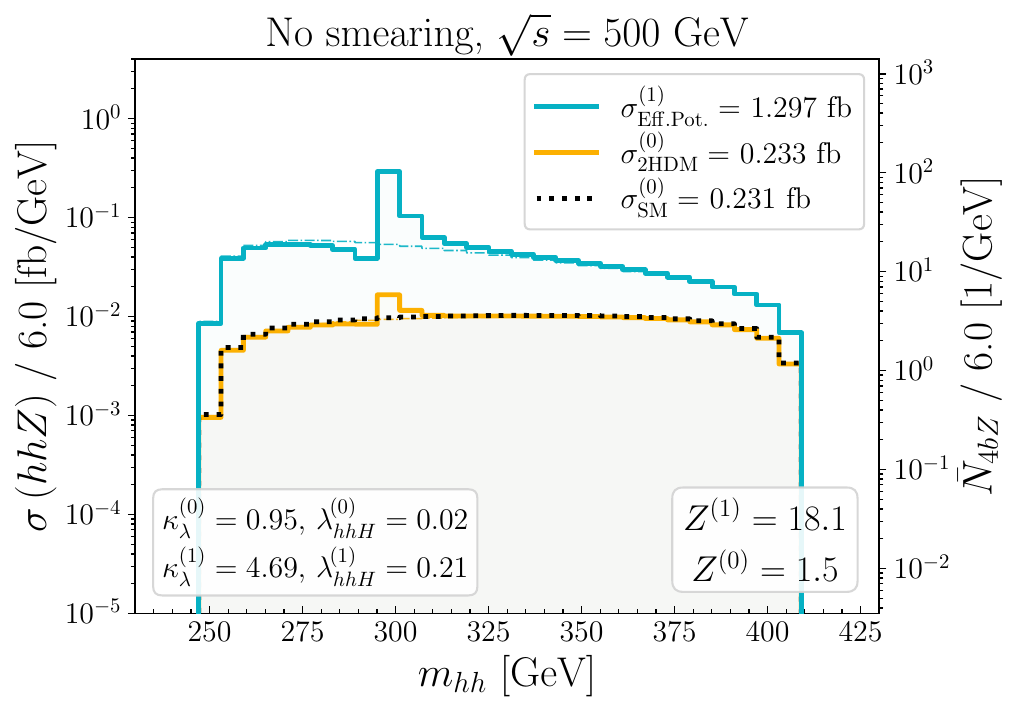}
\caption{}
\label{fig:mHH-2HDM:a}
\end{subfigure}
\begin{subfigure}{0.49\textwidth}
\includegraphics[width=\linewidth]{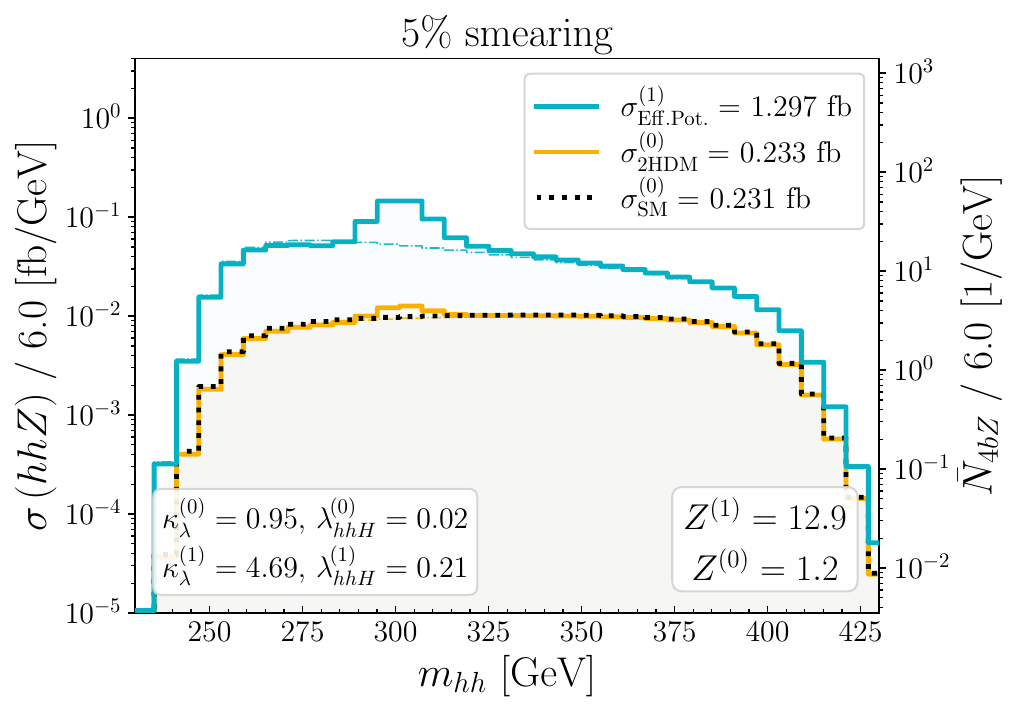}
\caption{}
\label{fig:mHH-2HDM:b}
\end{subfigure}

\caption{Differential cross section distribution w.r.t.\ $m_{\PH\PH}$ in the 2HDM type~I (for the parameters see text) at \SI{500}{500} for $P_{\Pem} = $\SI{-80}{\%} and $P_{\Pep} = $\SI{+30}{\%}. 
(a) without smearing and (b) with \SI{5}{\%} smearing. The colour coding is described in the text.
Note that in the figure $\PSh$ denotes the detected 125-GeV Higgs boson. Adapted from~\cite{Arco:2025nnn}. 
}
\label{fig:mHH-2HDM}

\end{figure}

A case study carried out in the frame of the 2HDM is presented in~\cite{Arco:2025nnn}.\footnote{For a recent analysis in the Higgs-singlet extension of the SM, see~\cite{Arco:2025nii}.}
In this work several benchmark points are evaluated, and the access to the THC $\lambda_{\PH\PH\PSh}$ (i.e.\ involving two CP-even Higgs bosons, assumed to correspond to the Higgs bosons discovered at the LHC, and one additional CP-even Higgs boson) is analysed. 
In Fig.~\ref{fig:mHH-2HDM}~\cite{Arco:2025nnn} results for one benchmark point in the 2HDM type~I are given.
The parameters are chosen as: $m_{\PH} = $\SI{125}{GeV}, $m_{\PSh} = $3\SI{300}{GeV}, $m_{\PSA} = m_{\PSHpm} = $\SI{650}{GeV},
$\tan\beta = 12$, $\cos(\beta-\alpha) = 0.12$ and $m_{12}^2/(\sin\beta\,\cos\beta) = $\SI{400}{GeV}.
This yields $\kappa_\lambda^{(0)} = 0.95$, $\kappa_\lambda^{(1)} = 4.69$, $\lambda_{\PH\PH\PSh}^{(0)} = 0.02$ and 
$\lambda_{\PH\PH\PSh}^{(1)} = 0.21$, where $(i)$ denotes the loop order.
Shown are the differential cross sections of the process $\Pep\Pem \to \PZ\PH\PH \to \PZ\PQb\PAQb\PQb\PAQb$ w.r.t.\ $m_{\PH\PH}$, the invariant mass of the di-Higgs system. The cross section is evaluated at $\sqrt{s} =$ \SI{500}{GeV} for $P_{\Pem} = $\SI{-80}{\%} and $P_{\Pep} =$\SI{+30}{\%}. 
The blue lines include the one-loop values of $\kappa_\lambda^{(1)}$ and $\lambda_{\PH\PH\PSh}$ (evaluated via the effective potential approach), 
the yellow lines show the 2HDM tree-level prediction and the dotted black lines show the SM tree-level prediction.
The number of events is calculated assuming an integrated luminosity of ${\cal L}_{\mathrm int} = $\SI{1600}{\fbinv} for each choice of polarisation ($P_{\Pem} =$ \SI{-80}{\%}, $P_{\Pep} =$\SI{+30}{\%} and $P_{\Pem} =$\SI{+80}{\%}, $P_{\Pep} =$\SI{-30}{\%} (not shown in the plots)).
The $m_{\PH\PH}$ distribution is binned with a bin size of \SI{6}{GeV}, ensuring a minimum of at least 2~events per bin for each polarisation in the case of no smearing (indicated by the vertical lines), as shown in Fig.~\ref{fig:mHH-2HDM:a}. Figure~\ref{fig:mHH-2HDM:b} shows the same curves, but with a smearing of \SI{5}{\%}, which could be seen as a realistic value for the ILC detectors~\cite{Durig:2016jrs}. 
To estimate the significance of the peak-dip structure caused by the resonant $\PSh$-exchange (and thus the presence of a BSM THC, $\lambda_{\PH\PH\PSh}$) cuts on the $\PQb$-jet energy and momentum are introduced, together with a $\PQb$-jet acceptance (see~\cite{Arco:2025nnn} for details). 
The significance, $Z$, is then calculated from the number of events expected with or without the peak-dip structure (and quadratically summed over the two polarisations). The results are given for tree-level THCs ($Z^{(0)}$) and one-loop THCs ($Z^{(1)}$ in the plots.
One can observe that the significance in this benchmark point becomes very large only because of the large one-loop correction to $\lambda_{\PH\PH\PSh}$, and degrades somewhat by the inclusion of a ``realistic smearing'' of \SI{5}{\%}. However, even after including the smearing the significance is found at $Z^{(1)} \sim 13$, i.e.\ at a very high level.
Clearly, a full experimental analysis is needed to realistically determine the precision with which $\lambda_{\PH\PH\PSh}$ can be measured at the ILC500. However, this example clearly indicates that BSM THC's can be accessible, if the BSM Higgs-boson mass is within the kinematical reach. 

\subsubsection{Linear collider sensitivity to unexpected new physics}
\label{sec:phys:bsm:novelBSM} 

While the above examples provide a panorama of well-motivated scenarios that address open questions of the SM, they remain speculative: we do not know what form BSM physics will take or at what scale it will manifest. Indeed, despite our theoretically well-founded ideas about BSM, it is possible that new physics shows up in an unexpected way. It is therefore important to search for new phenomena with an open mind, and with maximum experimental flexibility. 

Not all is completely open, however. As stated in the introduction, the priority for the next collider is to gain understanding of the Higgs sector and EWSB. And, dynamical explanations of the puzzles of the Higgs sector must come from new interactions and particles that couple to the Higgs. 
For new particles with electroweak interactions, the sensitivity of a linear collider essentially goes up to the kinematic limit. 
The extensibility in energy of a linear machine is thus a huge advantage, as it allows to adapt future upgrades to the findings at the HL-LHC and/or early stages of linear collider operation. Moreover, one can exploit beam polarisation to maximize sensitivity to whatever new physics may show up. 
Indeed, as already advocated in section~\ref{sec:phys:bsm:light_susy}, the clean experimental environment with well defined initial state, precisely known collision energy, electron and positron polarisation, hermetic detectors and trigger-less operation make an $\ee$ linear collider an ideal machine for looking
for unexpected and possibly very subtle new phenomena. Runs at different centre-of-mass energies (see section~\ref{sec:operating}) and the
$\PGg\PGg$ and $\Pe\PGg$ options (see section~\ref{sec:phys:altmodes}) further increase this potential.

It is also important to stress here that indirect measurements, e.g.\ in the context of EFT fits discussed in the next section, may provide evidence of new physics, but will not be able to pinpoint its exact nature. For this, direct observation and precise measurements of on-shell processes are vital,  which makes the possibility of energy upgrades and beam polarisation even more important.     


\subsection{Global interpretations}
\label{sec:glob}

Without any particular guidance in terms of what type of new physics signal we are looking for, and given the vast number of BSM deformations that could appear in physical observables with the expected precision at future $\ee$ colliders, global explorations covering as many different types observables as possible in new physics searches become a necessity.
Indeed, while a significant deviation from the SM prediction in a single observable would be sufficient as evidence of new physics, characterizing its origin requires us to study the pattern of deviations that could appear in correlation with such a signal across different observables.      

From the point of view of indirect searches, and assuming that new physics is somewhat heavier than the electroweak scale, as seem to be suggested by LHC direct searches,\footnote{However, many scenarios with light new physics remain uncovered by LHC and can be probed at \ee\ linear colliders; see Secs.~\ref{sec:directBSM} and~\ref{sec:phys:PBC:beamdumps}.}
the low-energy effects of the new physics in the precision measurements that will be possible at future electroweak, Higgs and Top factories can be adequately described by an {\em Effective Field Theory} (EFT). In this section, we adopt the framework of the so-called {\em SM Effective Field Theory} (SMEFT), which describes new physics in a generic way {\em under the assumptions} that at low energies the fields and symmetries are those of the SM, and the effects of new states decouple as these become heavy.  
While some of these assumptions could be questioned, and one could use other EFTs,\footnote{For instance, including extra light degrees of freedom or, staying with the SM spectrum at low energies, assuming that the electroweak symmetry is non-linearly realized, leading to what is known as the HEFT, where the Higgs described by a singlet field.} this is a minimal set that is consistent with current observations, it is well-motivated phenomenologically and has an ample coverage of BSM scenarios. 
Because the SMEFT Lagrangian describes all processes at all energies where the EFT is valid, the same parameters are constrained by measurements of a variety of observables. This  allows us to explore the complementary of the measurements performed at \ee\ colliders, and to quantitatively assess the importance of beam polarisation and running at different energies.  Finally, compared to other alternative EFTs, the SMEFT is a significantly more mature framework in terms of formal and practical developments.

SMEFT studies at future colliders prepared for the 2020 European Strategy Update~\cite{deBlas:2019rxi,EuropeanStrategyforParticlePhysicsPreparatoryGroup:2019qin} and the 2021 Snowmass process~\cite{deBlas:2022ofj} were based on global fits to projections of measurements of EW and Higgs observables, while the top sector was either minimally described~\cite{deBlas:2019rxi} or treated separately~\cite{deBlas:2022ofj}. 
Very recently, in the context of the preparations for the upcoming 2026 European Strategy Update, a global SMEFT fit to observables in the EW/Higgs/top sectors including the available results in the literature for next-to-leading order effects has been developed~\cite{deBlas:2025gyz}.

All the above-mentioned studies were performed truncating the SMEFT Lagrangian to dimension six:
\begin{equation}
{\cal L}_{\mathrm{SMEFT}}^{(6)}={\cal L}_{\mathrm{SM}} + \frac{1}{\Lambda^2}\sum_{i} C_i {\cal O}_i,
\end{equation}
where $\Lambda$ is the cut-off of the EFT, ${\cal O}_i$ are Lorentz and SM gauge-invariant operators of canonical mass dimension six, with $C_i$ the corresponding Wilson coefficients, and we have implicitly assumed lepton (and baryon) number conservation, thus removing the contribution from the dimension-five Weinberg operator. All the new physics effect are encoded in the $C_i$, which can be obtained matching the SMEFT to a given particular scenario. The different dimension-six operators will
induce observable effects suppressed by $q^2/\Lambda^2\ll 1$, with $q=v$ or the energy scale $E$ of the process of interest. Effects from higher-dimensional operators will carry extra suppressions
in $q/\Lambda$ and are therefore expected to be sub-leading, unless the new states are relatively light.

In this section, we present studies of the sensitivity to dimension-six SMEFT effects, focusing our attention on the LCF scenario described in this document. In Sec.~\ref{sec:glob:LO} we discuss the results in terms of the SMEFT setup used in previous ILC reports (e.g.~\cite{ILCInternationalDevelopmentTeam:2022izu}), focused on the characterization of the Higgs boson properties at leading order. 
These results are extended in Sec.~\ref{sec:glob:NLO}, combining the information of measurements of the Electroweak, Higgs and Top sectors, and going beyond leading order, including some of the most relevant NLO effects. The study in that section follows the setup used in the Electroweak chapter in \cite{deBlas:2025gyz}, with some differences that we briefly discuss before presenting the results. 

\subsubsection{SMEFT studies of the Higgs sector at leading order}
\label{sec:glob:LO}
In this section we supply results from a global SMEFT fit carried out with the assumptions
of \cite{Barklow:2017suo} used in previous ILC  reports~\cite{Bambade:2019fyw,LCCPhysicsWorkingGroup:2019fvj,ILCInternationalDevelopmentTeam:2022izu}.  This includes the expected uncertainties from  precision electroweak measurements in the programme discussed in Sec.~\ref{sec:Zpole} and from measurements of $\PW$ pair production discussed in Sec.~\ref{sec:ewhigh}, as well as the Higgs measurements from the various stages of a linear collider programme described in Sec.~\ref{sec:singleHiggs}.  The fit assumes lepton flavour universality and ignores CP-violating dimension-6 operators. It allows for exotic Higgs decays, parametrising these by the branching fraction for Higgs decay to invisible final states and, separately, the branching fraction for Higgs decay to unclassified exotic modes.  For a more detailed explanation of the assumptions of this fit, see \cite{Barklow:2017suo}.

This fit includes only tree-level effects.    The triple-Higgs coupling affects single-Higgs production only through loop effects, which should be considered together with SM 1-loop diagrams, including effects from CP-violating operators and operators involving the top quark.   The full 1-loop SMEFT corrections to $\ee\to \PZ\PH$ have now been computed~\cite{Asteriadis:2024xuk,Asteriadis:2024xts} and are included in the analysis presented in the next section.

Within this context, we present the Higgs coupling uncertainties predicted by this fit for the proposed LCF run plans.
In principle, we could quote the results of the fit in terms of bounds on SMEFT coefficients.  However, it is more illustrative to quote them in terms of fractional uncertainties in the Higgs couplings to various final states. We present these in Table~\ref{tab:fullprogram}. For the purpose of that table,
we define the relative error in the coupling for $\PH\to \PXXA \bar{\PXXA}$ as
\beq
\delta g(\PH \PXXA\PXXA) = {1\over 2} {\Delta \Gamma(\PH\to \PXXA\bar{\PXXA})\over \Gamma(\PH\to \PXXA\bar{\PXXA})}\ .
\eeq{Higgssigmadef}
Note that, while each Higgs decay depends on several separate dimension-6 SMEFT operators, the uncertainties in these operator coefficients are conflated in this metric.  This is rather different from the commonly used but oversimplified  $\kappa$ fit to  Higgs couplings, where the separate effects of different SMEFT operators are ignored and also constraints from processes other than Higgs production and decay are not included.

In the LCF run plan, the initial stage will take \SI{3}{\abinv} of luminosity at \SI{250}{GeV} in the centre of mass and carry out a separate run at the $\PZ$ resonance, as described above. 
Longitudinally polarised beams are used, assuming \num{80} electron polarisation and \SI{30}{\%} positron polarisation. 
The total luminosity is assumed to be divided among the four possible polarisation sign configurations as (\SI{10}{\%}, \SI{40}{\%}, \SI{40}{\%}, \SI{10}{\%}), c.f.\ Sec.~\ref{sec:RunScenarios}.  
The like-sign modes contribute little to the rate of Higgs boson production but are important to measure 2-photon and  beam-related backgrounds.  
All of the processes included in the fit have been studied in full simulation using the ILC beam conditions and model detectors, as described above.  
We use these results to obtain the measurement uncertainties associate with this data sample.  
Estimated systematic errors are included.  The results for $\delta g(\PH \PXXA\PXXA)$ as defined in \leqn{Higgssigmadef} are given in Table~\ref{tab:fullprogram}.

\begin{table}[htb]
    \centering
    \begin{tabular}{lcccc}
  \qquad\qquad  
  \qquad\qquad\qquad\qquad\qquad  LCF,    & 250\,GeV      & 250 + 550\,GeV & 250 + 550\,GeV    & 250 +550 +1000\,GeV     \\ \hline
   \qquad\qquad  Higgs couplings\qquad   L$[\mathrm{ab}^{-1}]$& 3, pol.      & 3+4, pol. & 3+8, pol.    & 3+4+8, pol.     \\ \hline   
  $g(\PH \PQb\PAQb)$                          & 0.72     &    0.46         &     0.38  &     0.36    \\
   $g(\PH \PQc\PAQc)$                         & 1.45      &    1.04         &    0.87  &     0.72     \\
  $g(\PH \Pg\Pg)$                               & 1.31           & 0.86    &      0.70  &     0.60    \\
  $g(\PH \PGt\PGt)$                         & 0.83      &  0.61      &    0.53  &     0.52    \\
  $g(\PH \PW\PW)$                           & 0.34        & 0.24      &     0.22 &     0.22     \\
  $g(\PH ZZ)$                               & 0.34         &  0.24      &    0.22  &     0.22     \\
  $g(\PH \PGg\PGg)$                          & 1.02        &  0.98       &    0.94   &     0.90 \\
  $g(\PH \PGg \PZ) $                        & 7.51      &  5.75    &    5.41  &    5.64   \\
      $g(\PH \PGm\PGm)$                     & 3.87       &  3.68     &    3.53 &   3.36    \\ 
   \qquad\qquad \qquad  Higgs widths \\  \hline 
  total width                            & 1.39          &  0.97     &  0.85 &    0.83  \\
 width to invisible  (\SI{95}{\%} CL)           & 0.21       & 0.20     &   0.20 &   0.20   \\
      width  to unclassified    (\SI{95}{\%} CL)    &  1.53  &  1.23 &  1.11 &  1.19   \\
      \end{tabular}
    \caption{Expected measurement precision of Higgs boson coupling and width determinations, in percent, according to the SMEFT fit  described in the text. Higgs coupling uncertainties are defined by \leqn{Higgssigmadef}.
We show the results for the initial-stage programme of the LCF at \SI{250}{GeV} (first column of results), and 
for the full program, considering different possible upgrades to higher energies: an extension to \SI{550}{GeV} collecting a total of \num{4} or \SI{8}{\abinv} (columns 2 and 3, respectively); an extension to \SI{550}{GeV} collecting \SI{4}{\abinv}, followed by a \SI{1}{TeV} run with a total of \SI{8}{\abinv} (fourth column).
}
    \label{tab:fullprogram}
  \end{table}

\begin{figure}[htb]
\begin{center}
    \includegraphics[width=0.90\hsize]{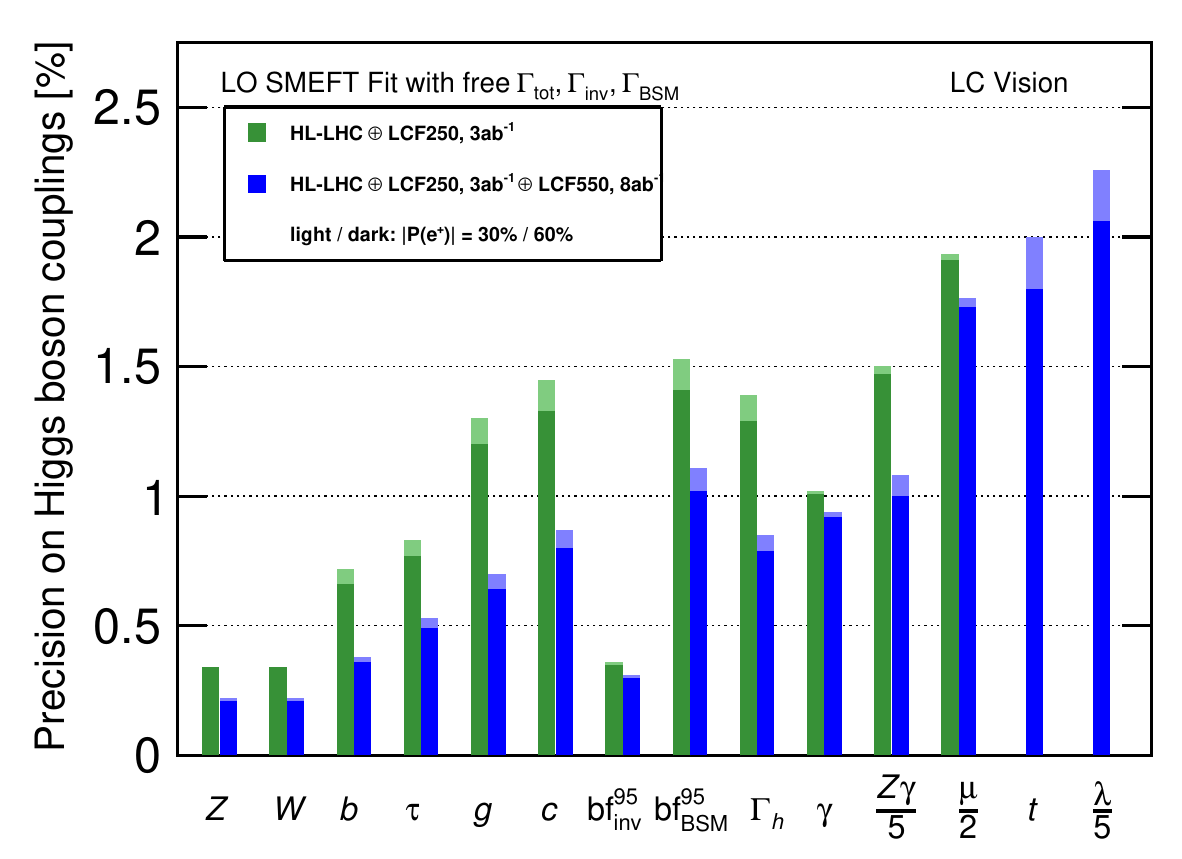}
\end{center}    
\caption{\label{fig:smeft:3ab8ab} Projected uncertainties on Higgs boson couplings for \SI{3}{\abinv} at \SI{250}{GeV} and \SI{8}{\abinv} at \SI{550}{GeV}; the bars for the $\PH\PZ\PGg$ coupling, the muon Yukawa and the trilinear Higgs self-coupling $\lambda$ have been scaled by $1/5$, $1/2$ and $1/5$, respectively. On the invisible and unclassified partial widths the expected \SI{95}{\%} CL limit is shown.}
\end{figure}

Figure~\ref{fig:smeft:3ab8ab} illustrates the projected uncertainties for \SI{3}{\abinv} at \SI{250}{GeV} and \SI{8}{\abinv} at \SI{550}{GeV}, i.e.\ the 3rd column of Table~\ref{tab:fullprogram}. The projected precisions on the top Yukawa and the trilinear Higgs self-coupling are from $\PQt\PAQt\PH$ and di-Higgs production, as discussed in Sec.~\ref{sec:Higgstop} and Sec.~\ref{sec:phys:selfcoup}, respectively.

  In Table~\ref{tab:fullprogram}, we also give a similar set of LCF results
  for the full programme of Higgs studies.  For LCF, the full programme would consist of \SI{3}{\abinv} at \SI{250}{GeV} plus either \SI{8}{\abinv} at \SI{550}{GeV} or \SI{4}{\abinv} at \SI{550}{GeV} and \SI{8}{\abinv} at \SI{1}{TeV}, plus additional short runs at the $\PZ$ resonance and the top-quark threshold.
  We also include the results for a programme with only \SI{4}{\abinv} at \SI{550}{GeV} for comparison.  

With either extension at high energies, the Higgs programme at the LCF would be comparable to that of any of the other Higgs factories that have been proposed, e.g.\ at circular colliders, bringing permil-level sensitivity to potential new physics effects in Higgs couplings. 
It should be noted that the LCF program, more balanced between Higgs production via $\PZ\PH$ and $\PW\PW$ fusion, also offers non-trivial checks of anomalies found in these two separate reactions.  
Of course, as we have discussed in the previous sections, the higher-energy programme of the LCF also includes many  reactions not accessible to circular colliders, including the measurements of the $\PQt\PAQt\PH$ and $\PH\PH$ production processes, where one could test at tree-level the top-quark Yukawa coupling and the Higgs self-coupling, respectively.

\subsubsection{Global SMEFT interpretation of the electroweak, Higgs and top-quark sectors}
\label{sec:glob:NLO}

The baseline of the SMEFT fit that will be used for comparison in this section is similar to that in~\cite{deBlas:2025gyz}, including current Electroweak Precision Observables from LEP/SLD, as well as the HL-LHC projections for Higgs and top-quark measurements.
This baseline will be combined with the information obtained from the observables  that would be measured at the different centre-of-mass energies that could be  accessible at a future $\ee$ linear collider. 
In this regard, we consider three operating scenarios, following mostly the stages presented in Table~\ref{tab:LCF-runplan}:
\begin{itemize}
    {\item Scenario {LCF}$_{\PZ/250}$ includes only the $\PZ$-pole run and the first stage as a Higgs factory at \SI{250}{GeV}. The polarisation and luminosities assumed are exactly the ones reported in Table~\ref{tab:LCF-runplan}.}
    {\item Scenario {LCF}$_{\PZ/250/550}$ consists of an extension of the previous scenario up to \SI{550}{GeV}, including also the small run at \SI{350}{GeV}. For the \SI{550}{GeV} run, we assume the same \SI{8}{\abinv} reported in Table~\ref{tab:LCF-runplan}, but all the projections used for this stage correspond to a polarisation setup were the positron polarisation is only \SI{30}{\%}.}
    {\item Finally, scenario {LCF}$_{\PZ/250/550/1000}$ considers a further extension to high-energies up to \SI{1}{TeV}, collecting a total of \SI{8}{\abinv} with the same distribution among polarisations described in Table~\ref{tab:LCF-runplan}.}
\end{itemize}

In terms of the LCF observables included in the fit, we include the following list:

\begin{itemize}
    {\item The full set of measurements for single-Higgs processes in Table~\ref{tab:HiggsBRPrec} in Sec.~\ref{sec:singleHiggs}.  
    We use the same projections in \cite{deBlas:2025gyz}, which were obtained from a common set of inputs for all existing projects for future \ee\ colliders, extrapolated to the luminosities and polarisations indicated above.
    }
    {\item The Higgs mass determination at \SI{250}{GeV} with a precision of \SI{12}{MeV}, as explained in Sec.~\ref{sec:Higgs_mass_xs}.}
    {\item We use the same projections for $\ee\to \PWp\PWm$ processes as in~\cite{deBlas:2022ofj}, but again updating the luminosity to the scenarios discussed above.}
    {\item The projections for electroweak precision measurements at linear $\ee$ colliders. These include the measurements taken during the \PZ-pole run and the properties of the $\PW$ boson, following again the input set used in the studies in \cite{deBlas:2025gyz}. These also include the electroweak precision observables that could be measured at \SI{250}{GeV}, e.g.\ the $\PZ$-pole observables via radiative return, taken from~\cite{deBlas:2022ofj} and scaling the uncertainties with the luminosity change from \num{2} to \SI{3}{\abinv}.
    }
    {\item On the electroweak side, we extend the set of observables with the projected measurements of $2\to 2$ fermion processes above the $\PZ$ pole, which were treated separately in~\cite{deBlas:2022ofj}. For the leptonic channels we use the same information in~\cite{deBlas:2022ofj}, scaled to the appropriate luminosities, whereas for the hadronic final states we use the results in~\cite{Irles:2024ipg,ECFA-HF-Report}.
    }
    {\item The top-quark mass measured in the $\PQt\PAQt$ threshold scan, see Sec.~\ref{sec:Top_mass}. This is of particular relevance for the fit to electroweak precision observables. For the {LCF}$_{\PZ/250}$ scenario, we assume a $m_{\PQt}$ precision from HL-LHC measurements of \SI{200}{MeV}~\cite{ATLAS:2025eii}.
    }
    {\item Similarly, we include the top sector directly in the global combination, using the results presented in Sec.~\ref{sec:Top_physics}. (Note however that the luminosities used in that section are different than the ones considered here.) The LCF inputs are combined with the HL-LHC top-quark projections used in~\cite{deBlas:2025gyz}.
    }
\end{itemize}

The projected precision for some of the above-mentioned inputs, in particular for the electroweak precision observables, depend strongly on the precision of theory QED, EW and QCD calculations. Here, for the theory systematics of the different experimental measurements, as well as for the uncertainties of the SM predictions of the different observables, we follow the assumptions of the ``aggressive'' scenario for theory uncertainties used as the baseline in the studies of~\cite{deBlas:2025gyz}. Finally, for some of the SM input parameters, namely $\alpha(m_{\PZ}^2)$ and $\alpha_s(m_{\PZ}^2)$, we assume their experimental precision will be improved according to future lattice QCD projections~\cite{deBlas:2025gyz}.

On the technical side of the EFT interpretation, we use the so-called {\em Warsaw basis} for the dimension-six SMEFT Lagrangian~\cite{Grzadkowski:2010es},  we restrict the analysis to CP-even operators and, to be consistent with flavour constraints, we assume a $U(2)^5$ flavour symmetry (see~\cite{deBlas:2025xhe} for the notation used here).
In the quark sector, this flavour symmetry is defined choosing the third family along the direction of the up-quark mass eigenstates. 
We consider all the dimension-six SMEFT operators that contribute to the LCF precision observables at leading order or via NLO effects proportionally to $g_3$ or $y_{\PQt}$. 
The full NLO effects for the resulting set of 84 operators is included for the electroweak precision observables and the Higgs\-strahlung cross section, whereas for other observables, the logarithmic one-loop corrections are included via the RG-evolution of the coefficients contributing at LO. 
Compared to~\cite{deBlas:2025gyz}, where the previous assumptions are defined at a scale of \SI{10}{TeV}, taken as the cut-off scale of the EFT, here a value of $\Lambda=$ \SI{1}{TeV} is used instead for such cut-off. 
The results for the different operators discussed in this section are therefore presented in terms of quantities evaluated at a renormalization scale equal to \SI{1}{TeV}. 
Note that, unlike in the previous section, here we work exclusively within the SMEFT framework, and therefore do not include the possibility of decays of any SM particle into BSM states, which are assumed to be heavier than the cut-off scale of the EFT. 
Finally, the contributions from the different operators are kept at order $1/\Lambda^2$, given by the interference between the new physics amplitudes and the SM ones. 
All the fits presented in what follows were obtained using the \texttt{HEPfit} code~\cite{DeBlas:2019ehy,deBlas:2025xhe}.

\begin{figure}[t!]
\includegraphics[width=\textwidth]{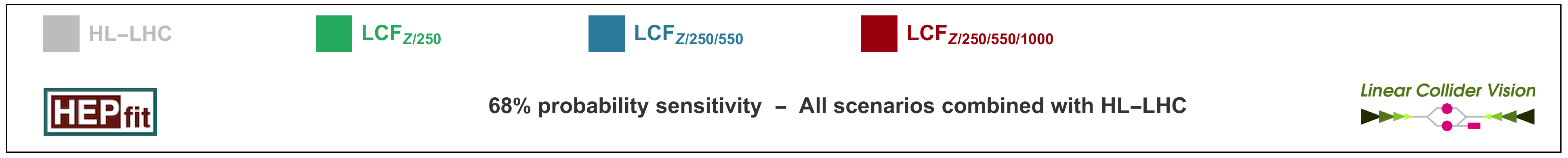}\\
\includegraphics[width=\textwidth]{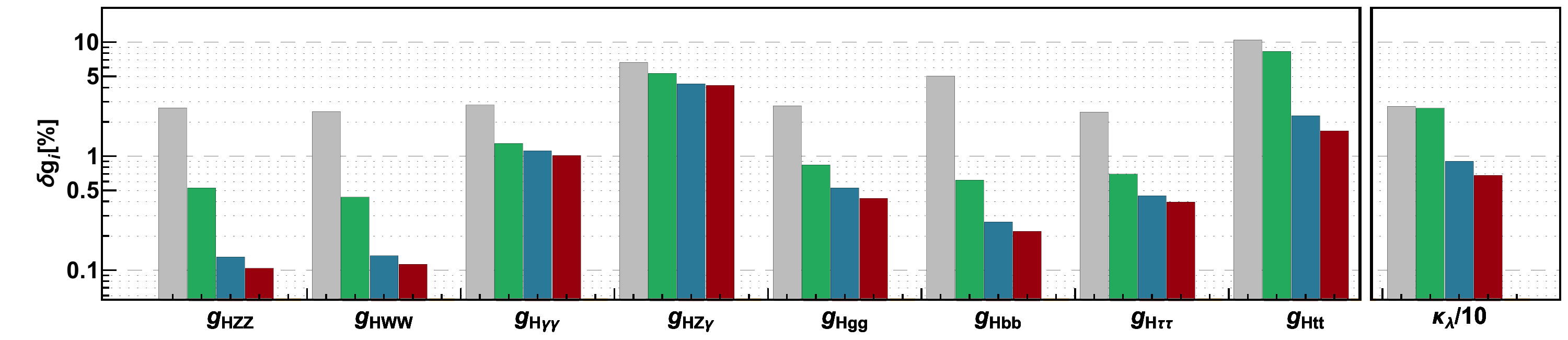}\\
\caption{\SI{68}{\%} probability sensitivity to modifications of the effective single Higgs couplings in the dimension-six $U(2)^5$-symmetric SMEFT framework. The last set of bounds indicate the expected sensitivity via modifications of the triple Higgs coupling, $\kappa_\lambda = \lambda_3/\lambda_3^{\mathrm{SM}}$. 
All numbers are derived assuming the experimental observations are centred around the SM prediction.
}
\label{fig:eft:gheff}
\end{figure}

Similar to the results presented in the previous section, in Figure~\ref{fig:eft:gheff} we show the precision on the effective single Higgs couplings, defined similarly to Eq.~(\ref{Higgssigmadef}). The numerical values for the precision of the different couplings are given in Table~\ref{tab:eft:Heff}. 
Note that, compared to the fit in the previous section, the $U(2)^5$ symmetry imposed here only allows for independent new physics contributions to the Higgs couplings to the fermions of the third generation. 
While a direct comparison of the results of the two fits is not straightforward, due to the differences in terms of theory assumptions, inclusion of NLO effects, and the differences in the set of inputs used, the overall conclusions of both studies in terms of Higgs precision are similar, with the first stage at \SI{250}{GeV} bringing a substantial improvement in most couplings with respect to the HL-LHC, in several cases reaching sub-percent precision, and the combination with high-energy runs further pushing this precision near the permil-level, in particular for the Higgs couplings to $\PZ$ and $\PW$ bosons.  
For the effective $\PH\PQt\PAQt$ coupling, which cannot be defined in an on-shell manner as the ones for other SM particles via Eq.~(\ref{Higgssigmadef}), we report in Fig.~\ref{fig:eft:gheff} the uncertainty obtained from the operator $({\cal O}_{u\phi})_{33}$, evaluated at the EW scale, via $\delta g_{\PH\PQt\PQt} = -\frac{v^2}{\Lambda^2} (C_{u\phi})_{33}$. The results for $g_{\PH\PQt\PQt}$ are consistent with the findings in sections~\ref{sec:Higgstop} and the LO fit discussed above for the top-quark Yukawa coupling, reaching precisions at the level of \SI{1}{-}\SI{2}{\%} when combining the information of the high-energy runs. 
Finally, the Higgs self-coupling, $\lambda_3$, is also shown in Figure~\ref{fig:eft:gheff}. The sensitivity to new physics modifying this interaction is derived using the information both from Higgs pair production at \SI{550}{GeV} and \SI{1}{TeV}, and from single-Higgs measurements, where $\lambda_3$ enters at the 1-loop level. While the precision of $\lambda_3$ is controlled by the Higgs pair production determination (taken here to be $\sim$ \SI{11}{\%} and \SI{7}{\%} at $\sqrt{s}=$ \SI{550}{GeV} and \SI{1}{TeV}, respectively), one still observes a slight improvement in the determination from the combination with single-Higgs measurements, in particular in the \SI{550}{GeV} run. 

\begin{figure}[t!]
\includegraphics[width=\textwidth]{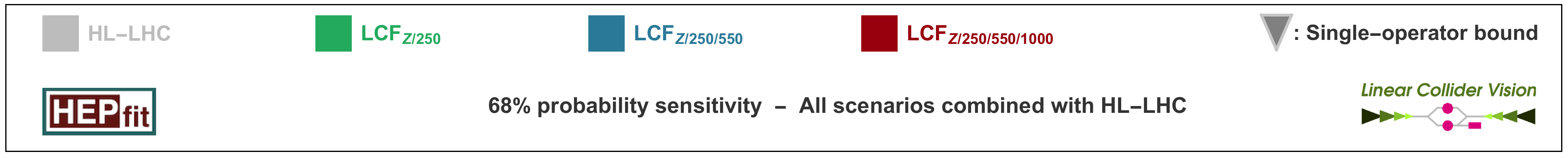}\\
\includegraphics[width=\textwidth]{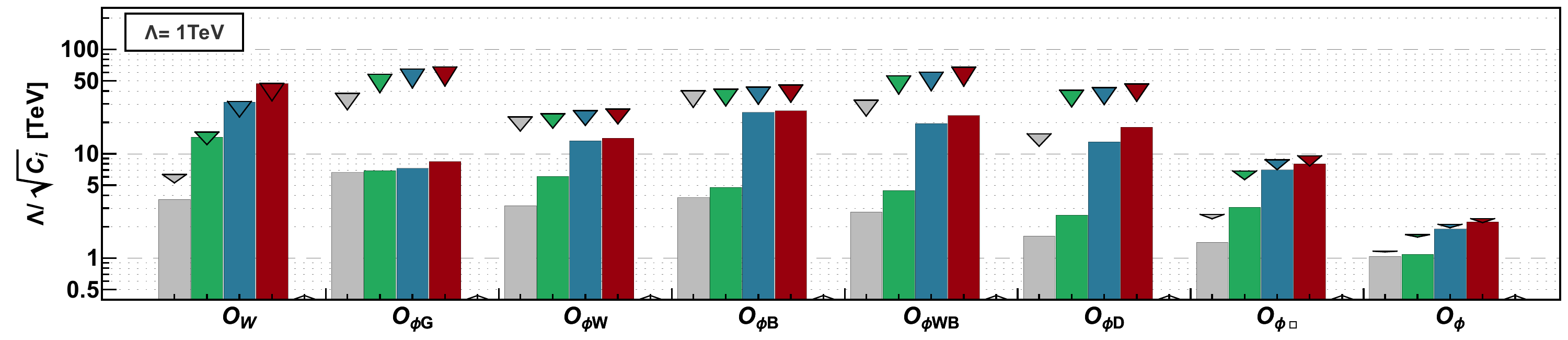}\\
\includegraphics[width=\textwidth]{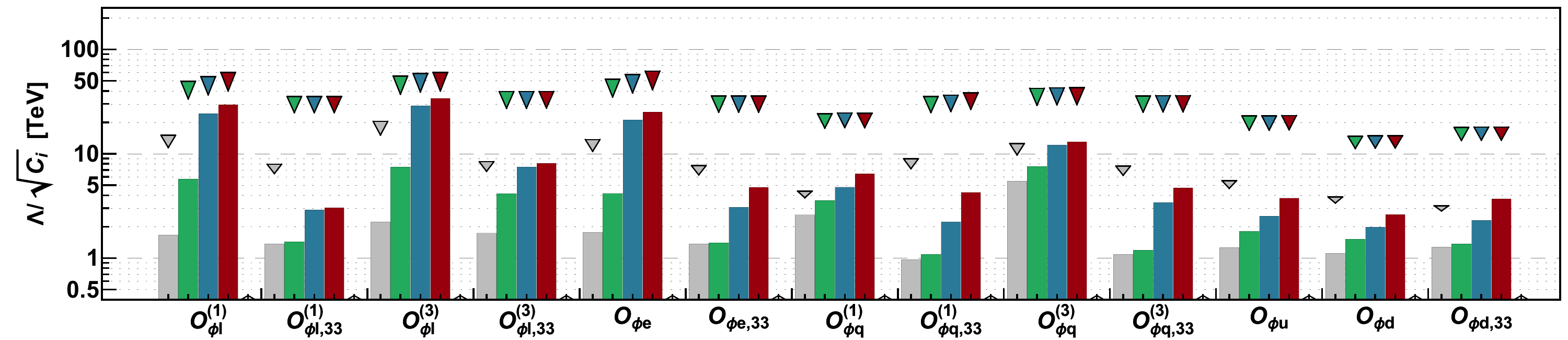}\\
\includegraphics[width=\textwidth]{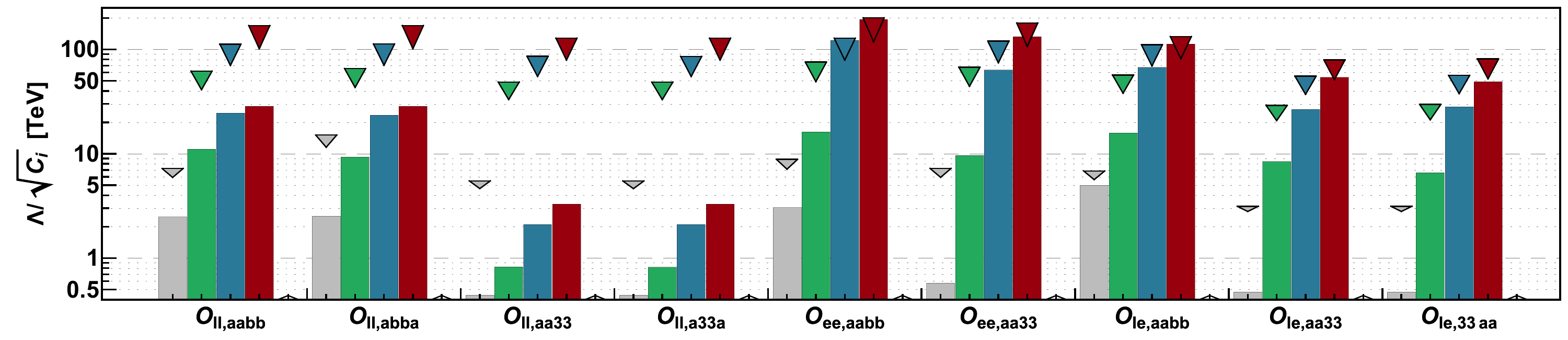}\\
\includegraphics[width=\textwidth]{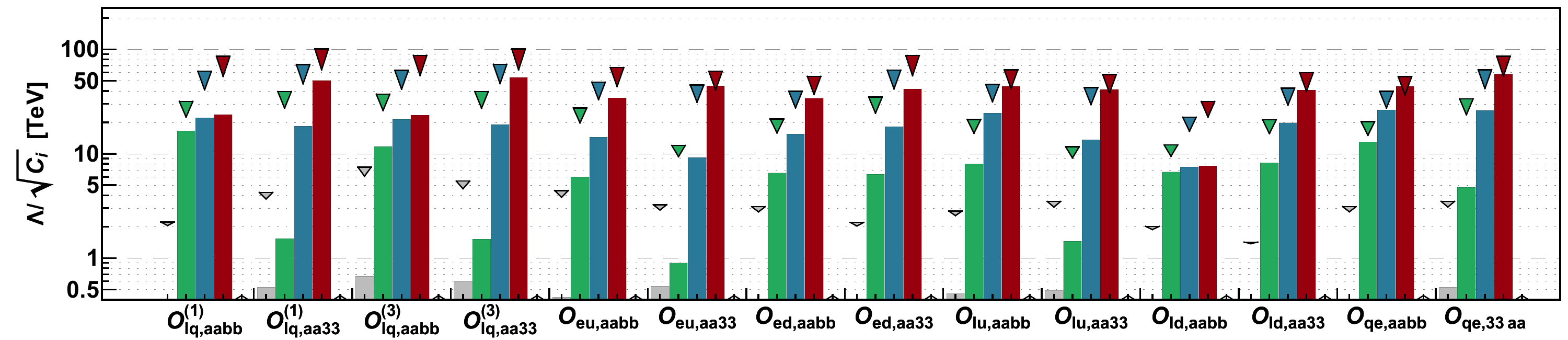}
\caption{\SI{68}{\%} probability limits on a representative sample of dimension-six bosonic operators (first panel), operators involving two-fermion fields contributing to EWPO (second panel), and four fermion operators (third panel for four-lepton operators and fourth panel for lepton-quark operators). Latin indices aa,\,bb denote the fermion indices of the first two generations, which are related by the $U(2)^5$ flavour symmetry. 
The inverted triangles indicate the single operator limits.
See text for details.
}
\label{fig:eft:operators}
\end{figure}

\begin{table}[htb]
    \centering
    \begin{tabular}{lccc}
    \hline
  \qquad\qquad  
  \qquad\qquad\qquad  HL-LHC +     & LCF$_{\PZ/250}$     & LCF$_{\PZ/250/550}$ & LCF$_{\PZ/250/550/1000}$     \\ \hline
   \qquad SMEFT effective Higgs couplings  \qquad       &                   &                   &                            \\ \hline   
  $g_{\PH \PZ\PZ}$                                    & 0.53            & 0.13    &  0.11       \\
  $g_{\PH \PW\PW}$                                & 0.44            & 0.13    &  0.11      \\
  $g_{\PH \PGg\PGg}$                              & 1.3             & 1.1     &  1.0    \\
  $g_{\PH \PZ \PGg }$                             & 5.4             & 4.3     &  4.2    \\
  $g_{\PH \Pg\Pg}$                                & 0.84            & 0.53    &  0.43    \\
  $g_{\PH \PQb\PAQb}$                             & 0.62            & 0.26    &  0.22      \\
  $g_{\PH \PGt\PGt}$                              & 0.70            & 0.45    &  0.40     \\
  \qquad SMEFT Higgs self-coupling                &                 &         &        \\ \hline   
  $\kappa_\lambda = \lambda_3/\lambda_3^{\mathrm{SM}}$ & 27 (HL-LHC)     &  9.1    &  6.8    \\
  \hline
      \end{tabular}
    \caption{\SI{68}{\%} probability sensitivity to modifications of the effective single Higgs couplings in the dimension-six $U(2)^5$-symmetric SMEFT framework, see Figure~\ref{fig:eft:gheff}.}
    \label{tab:eft:Heff}
  \end{table}


Figure~\ref{fig:eft:operators} focuses on the reach in terms of the interaction scale, $\Lambda/\sqrt{C_i}$, for subset of the operators entering in the fit. The results of the global fit are presented by the solid bands, while the bounds obtained assuming the new physics generates only one operator at a time are given by the inverted triangles. The difference between the two results can be used a sign of how strong the correlations between the different operators are in the global fit. 
The first panel focuses on bosonic interactions, which are tested mainly by Higgs and electroweak processes. 
The latter are also affected by the operators involving fermionic currents shown in the second panel of the figure, and that induce fermion-specific modifications of the EW neutral and charged currents. 
The interactions in the last two panels correspond to four-fermion operators. The best bounds in this case come from two-to-two fermion processes, and they exploit two of the main advantages at linear colliders: the access to high energies and the use of polarised beams. Indeed, the interference of a given four-fermion interaction, e.g.\ $(\bar{\Pe}\gamma_\mu \Pe) (\bar{\mathrm{f}}\gamma^\mu \mathrm{f})$, with the SM amplitudes of the process $\ee\to \mathrm{f}\bar{\mathrm{f}}$, generates a relative contribution proportional to $E^2/\Lambda^2$. Therefore, going to higher energies brings a quadratic increase in the sensitivity, which can overcome the one from more precise experimental measurements at low energies. This effect is observed for almost all operators in the lower two panels, with the LCF bounds becoming stronger as the centre-of-mass energy grows, and eventually reaching scales of the order of \SI{100}{TeV}.~\footnote{A notable exception is $({\cal O}_{ld})_{aabb}$, due mainly to the absence of projections for $\ee\to \PQs\PAQs$ at energies above $\sqrt{s}=$ \SI{250}{GeV}.} On the other hand, the use of polarisation helps to separate contributions from operators with the same fields but different chiralities for the \ee\ initial states.
The relevance of the high energy reach of linear colliders is also observed in the leptonic operators ${\cal O}_{\phi l}^{(1),(3)}$ and ${\cal O}_{\phi \Pe}$ in the second panel, which induce similar growing-with-energy effect in di-boson processes.

\begin{figure}[ht]
\includegraphics[width=\textwidth]{figures/global/LegGeneral_1op.pdf}\\
\includegraphics[width=\textwidth]{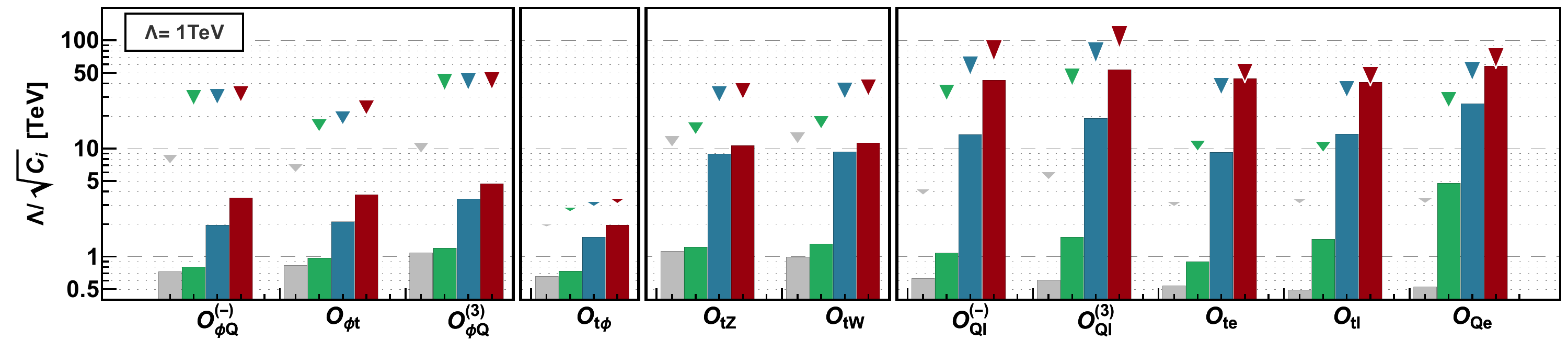}
\caption{\SI{68}{\%} probability limits on dimension-six operators entering in top-quark processes at \ee\ colliders. The different sub-panels separate operators modifying the top-quark EW couplings, Yukawa interactions, dipolar couplings and four-fermion $\ee \PQt \PAQt$ operators.
The inverted triangles indicate the single operator limits.
}
\label{fig:eft:top-eft}
\end{figure}

The bounds from top-quark observables are shown separately in Figure~\ref{fig:eft:top-eft}.
Following the conventions of the LHC Top WG~\cite{Aguilar-Saavedra:2018ksv}, also used in Figure~\ref{fig:smeft_top_sector} in Section~\ref{sec:Top_physics}, we present in this figure the results for the combinations of operators involving the quark top, e.g.\ $C_{\phi Q}^{-}\equiv (C_{\phi q}^{(1)})_{33} - (C_{\phi q}^{(3)})_{33}$. 
The qualitative aspects of the global fit results are similar to the ones presented in Sec.~\ref{sec:Top_physics}. 
Quantitative differences can be observed in some cases, but these are expected due to different setup in terms of energy/luminosity between the two fits, the different set of top-quark physics inputs from the HL-LHC, plus the fact that the analysis in this section is performed including NLO effects of the different top operators in other observables, and the results are presented for the operators evaluated at the renormalization scale of \SI{1}{TeV}. 
The impact of the NLO effects is more visible in the single-operator bounds, which in some cases are significantly improved by the precision measurements taken at the \SI{250}{GeV} stage (including the \PZ-pole run). 
Similarly to what was discussed for the effect of four-fermion operators in $\ee \to \mathrm{f}\bar{\mathrm{f}}$ above,  the access to high energies is of particular relevance to test new physics mediating interactions between electron and top-quark currents.

\begin{figure}[htbp]
\centering
\begin{subfigure}{0.45\textwidth}
\includegraphics[width=\linewidth]{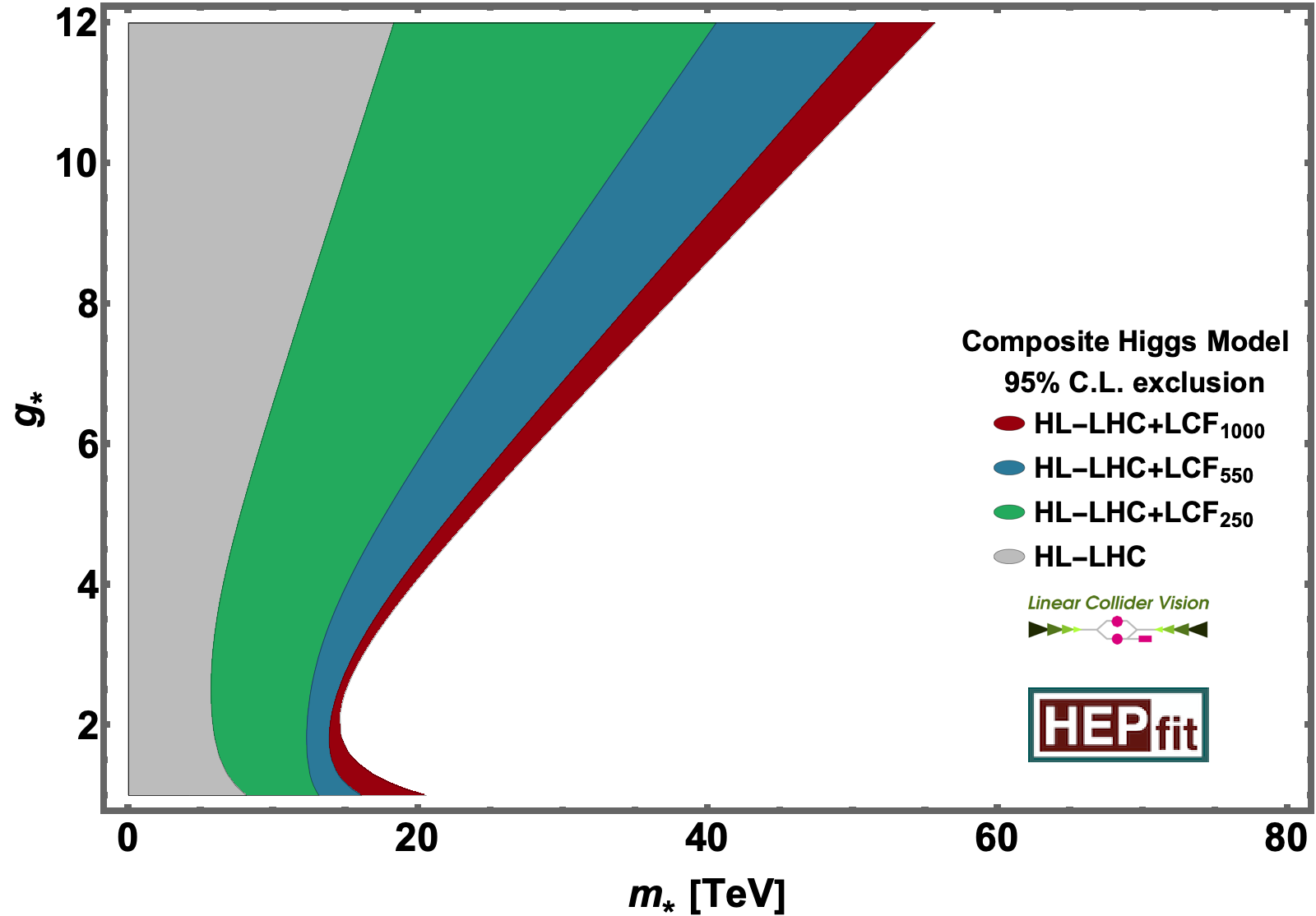}
\caption{}
\label{fig:eft:CompHiggsYp:a}
\end{subfigure}
\begin{subfigure}{0.45\textwidth}
\includegraphics[width=\linewidth]{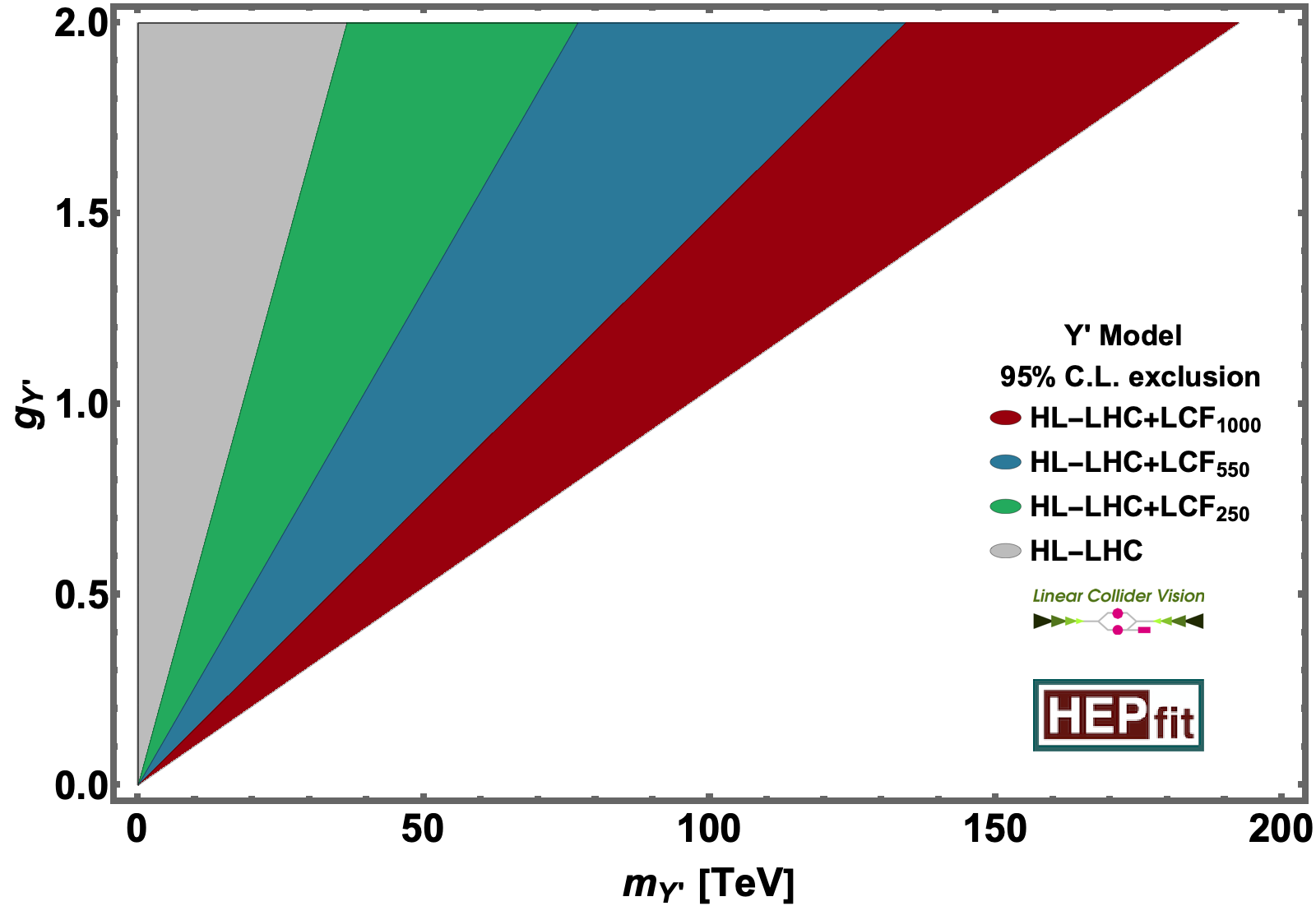}
\caption{}
\label{fig:eft:CompHiggsYp:b}
\end{subfigure}

\caption{(a) \SI{95}{\%} CL exclusion region in the $(g_\star,m_\star)$ plane for the class of composite Higgs scenarios described in~\cite{Giudice:2007fh}. (b) \SI{95}{\%} CL exclusion region in the $(g_{Y^\prime},m_{Y^\prime})$ plane for an SM extension with a new vector boson, $Y^\prime$, coupling proportionally to hypercharge current. The bounds displayed here are based on the new global SMEFT fit described in this section. (See appendix A in \cite{deBlas:2025xhe} for details on the implementation of both scenarios.)}
\label{fig:eft:CompHiggsYp}
\end{figure}

The implications of the top-quark constraints on composite models were discussed briefly in Section~\ref{sec:Top_physics}, see e.g.\ Fig.~\ref{fig:LCbounds}, highlighting the effects on the parameter space of different operators. To conclude this section, we show in Fig.~\ref{fig:eft:CompHiggsYp} the global bounds combining all the recent information for the case of composite Higgs models and new vector resonances. In Fig.~\ref{fig:eft:CompHiggsYp:a} we consider the general scenario in~\cite{Giudice:2007fh}, where the mass and interactions of the resonances of the strongly-interacting sector are parametrized by a single mass scale and coupling, $m_\star$ and $g_\star$, respectively. 
These models predict sizeable deviations on the Higgs couplings, proportional to $g_\star^2/m_\star^2$, which highlight the importance of a programme for precision Higgs physics. The LCF measurements would improve the capabilities of testing this type of scenarios by up to a factor of $\sim 3$, compared to the HL-LHC. On the other hand, in Fig.~\ref{fig:eft:CompHiggsYp:b} we choose a model that illustrates the benefits of high-energy measurements to set constraints on new physics. This is an scenario with an additional vector boson, $Y^\prime$, that couples to the SM particles proportionally to the hypercharge current, with coupling $g_{Y^\prime}$. While this simple scenario also generates corrections to electroweak precision observables, the strongest bounds in the LCF case come from the $Y^\prime$ contributions to four-fermion operators. As explained above, the indirect sensitivity to these effects via $\ee\to \mathrm{f}\bar{\mathrm{f}}$ grows quadratically with the energy, highlighting the benefits of the high energy runs at the LCF. These are just two examples of the physics opportunities to test new physics indirectly via a combination of measurements of high precision and at high energies. 

\subsection{Photon-photon, electron-photon and electron-electron collisions}
\label{sec:phys:altmodes}

High-energy $\PGg\PGg$, $\Pe\PGg$  and $\Pem\Pem$ collisions offer a rich physics programme, complementary to  
$\ee$ collisions both in kinematic as well as physics aspects~\cite{ECFADESYPhotonColliderWorkingGroup:2001ikq,Yokoya:2000cg,Boos:2000ki,Watanabe:1998mn,Ginzburg:1981vm}. 
In the following subsections, we will first introduce key properties like the luminosity spectra of $\PGg\PGg$ colliders (c.f.\ Sec.~\ref{sec:phys:altmodes:gg_intro}), before discussing selected physics opportunities, including property determinations of the 125-GeV Higgs boson and possible additional ones via s-channel production (c.f.\ Sec.~\ref{sec:phys:altmodes:gg_H}), a novel study of di-Higgs production indicating intriguing potential to measure the Higgs self-coupling (c.f.\ Sec.~\ref{sec:phys:altmodes:gg_HH}) as well as further examples (c.f.\ Sec.~\ref{sec:phys:altmodes:gg_other}). Opportunities in $\Pem\PGg$ and $\Pem\Pem$ collisions will be briefly summarised in Secs.~\ref{sec:phys:altmodes:eg} and~\ref{sec:phys:altmodes:ee}, respectively. 

\subsubsection{Photon-photon colliders}
\label{sec:phys:altmodes:gg_intro}

\begin{figure}[htbp]
    \centering
    \includegraphics[width=0.6\linewidth, trim=0 1cm 0 3cm, clip]{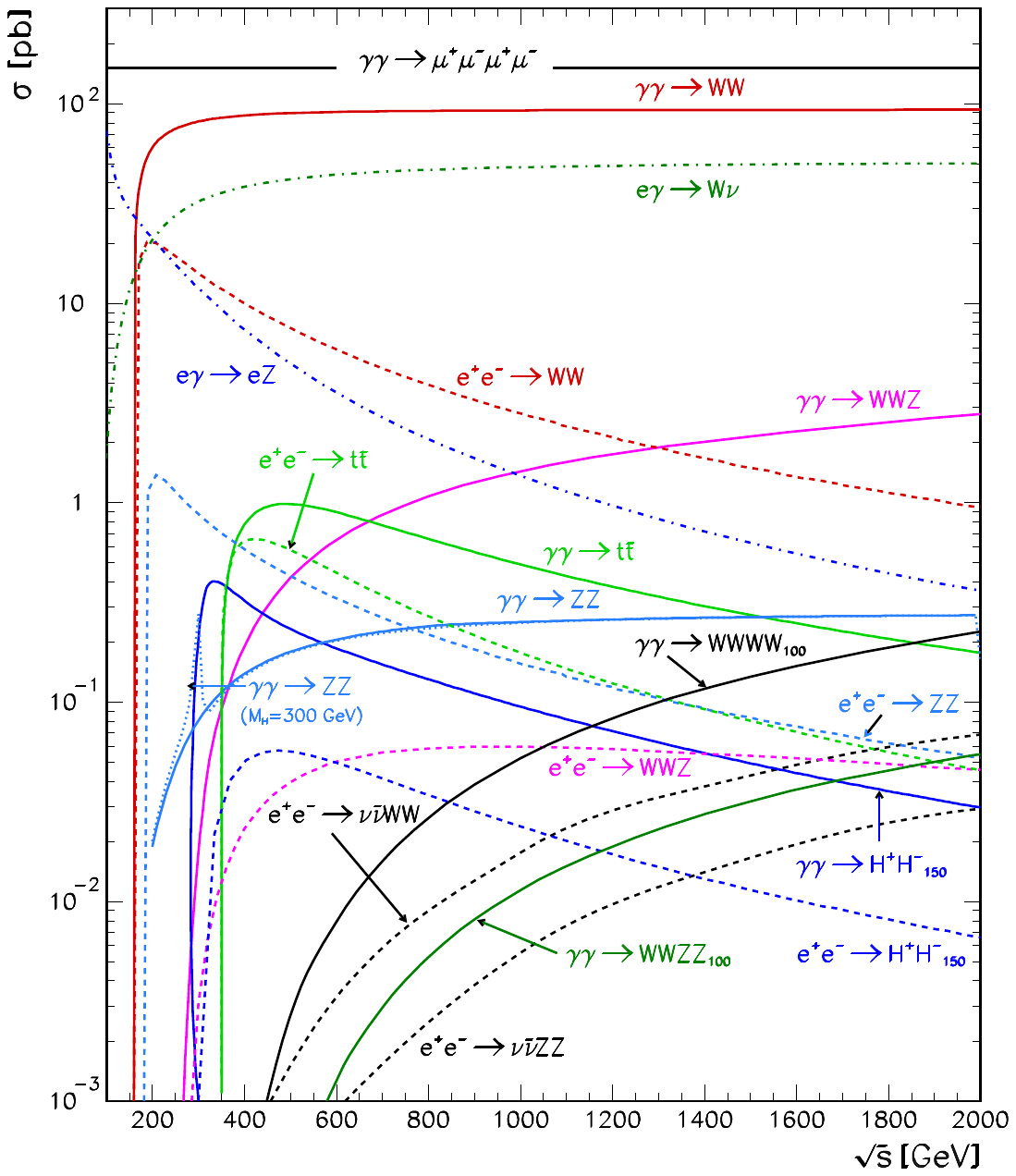}
    \caption{Fermion, gauge boson and Higgs production in $\ee$ and $\PGg\PGg$ and as a function of $\sqrt{s}$, including production of a hypothetical charged Higgs boson $\PH^{\pm}$ as well as some $\PGg\Pem$ processes. Modified from~\cite{Boos:2000ki}. 
    \label{fig-gg-overview}}
\end{figure}

An overview about the most important SM processes in $\PGg\PGg$ collisions compared to their counterparts in \ee\ collisions is shown in Fig.~\ref{fig-gg-overview}. It shows that typical pair production processes in $\PGg\PGg$ collisions have significantly larger cross sections than their \ee\ counterparts. In the particular case of the Higgs boson, $\PGg\PGg$ colliders do not have to rely on associated production with a $\PZ$ boson, but can directly produce Higgs bosons in the s-channel via $\PW$-boson and top-quark loops.

Implementation scenarios for $\PGg\PGg$ colliders will be discussed more detail in Sec.~\ref{sec:acc:altmodes}. 
However it is important to note already here that $\PGg\PGg$ collisions can be created via Compton backscattering of either optical laser or X-ray  free-electron lasers (XFELs) off $\Pem\Pem$  or off $\epem$ beams. These choices, as well as the polarisation configurations, result in different luminosity spectra.

\begin{figure}[htb]
 \centering
 \begin{subfigure}{0.44\textwidth}
     \includegraphics[width=\linewidth]{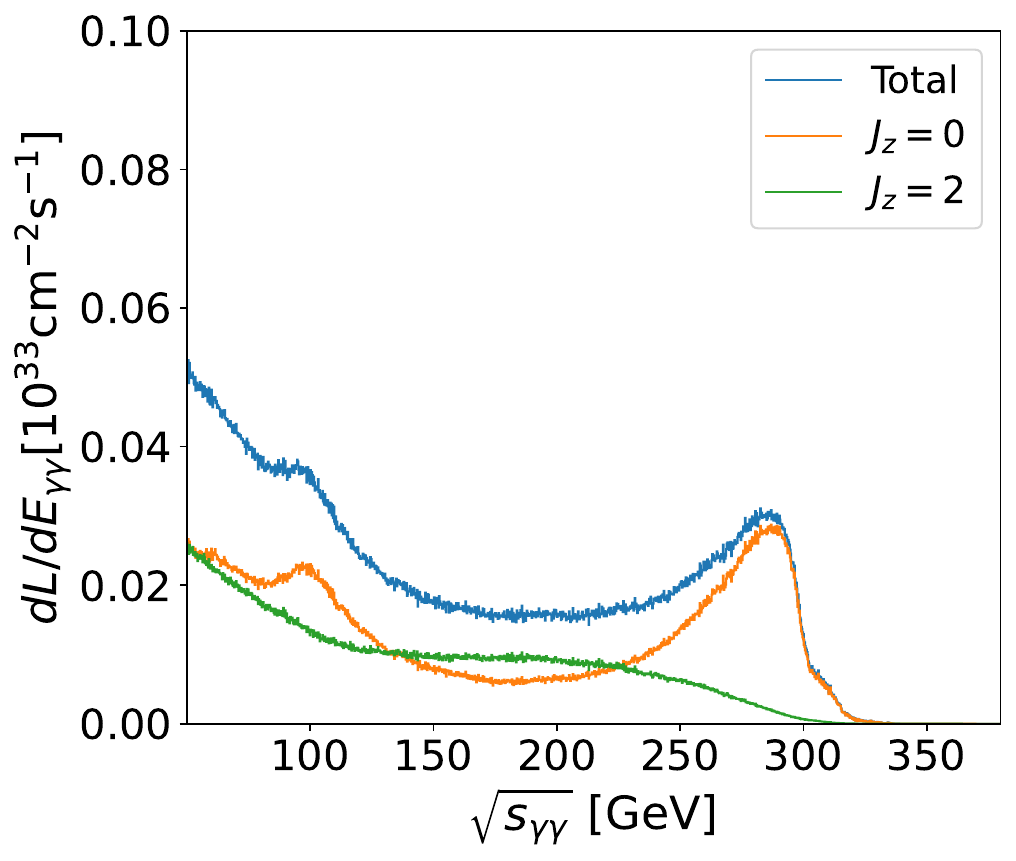}
\caption{}
\label{fig:phys:aalumi:optical}
\end{subfigure}
\begin{subfigure}{0.45\textwidth}
 \includegraphics[width=\linewidth]{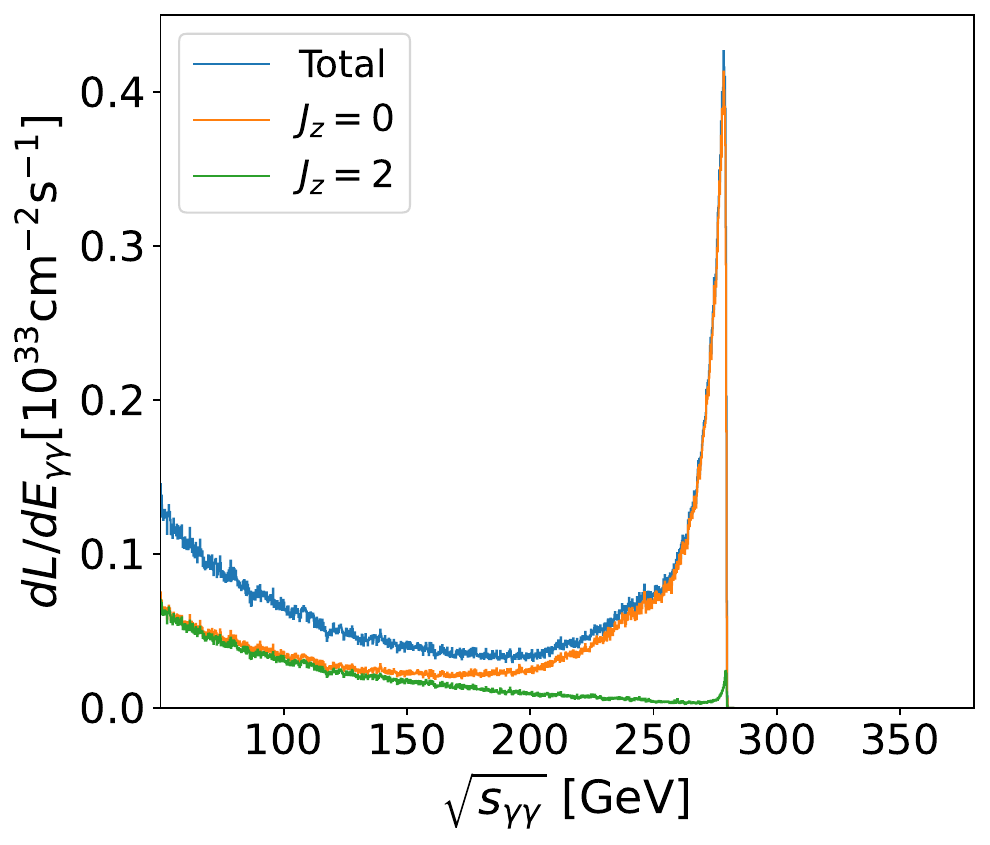}  
\caption{}
\label{fig:phys:aalumi:xcc}
\end{subfigure}
 \caption{$\PGg\PGg$ luminosity spectra in the  $J_z=0$ and $J_z=2$ configurations of a $\PGg\PGg$ collider for backscattering of (a) an optical laser ($\Pem\Pem$ centre-of-mass energy \SI{380}{GeV}) and (b) via an XFEL ($\Pem\Pem$ centre-of-mass energy \SI{280}{GeV}). From~\cite{Berger:2025ijd}.
 \label{fig:phys:aalumi}}
 \end{figure}

As example, Fig.~\ref{fig:phys:aalumi} compares the $\PGg\PGg$ luminosity spectra for employing optical lasers and XFELs. In the optical case, an $\Pem\Pem$ centre-of-mass energy of \SI{380}{GeV} is required to reach a peak $\PGg\PGg$ energy of \SI{280}{GeV}, while in case of an XFEL, this can already be achieved for an $\Pem\Pem$ centre-of-mass energy of \SI{280}{GeV}. The $J_z=0$ and $J_z=2$ cases are shown separately.  

Given the broad luminosity spectra, the integrated luminosity is typically defined in regard to the of the upper \SI{30}{\%} of the spectrum, while physics projections take into account the full spectrum to correctly account for backgrounds from lower energy processes. Early studies typically assumed a few hundred \fbinv, while modern implementations (c.f.\ Sec.~\ref{sec:acc:altmodes}) justify the assumption of \SI{4}{\abinv} being collected at a nominal centre-of-mass energy \SI{280}{GeV} in a 10-year operation of e.g.\ one of the interactions regions at LCF in a $\PGg\PGg$ mode. At \SI{380}{GeV} even \SI{5}{\abinv} could be collected a similar running time.

\subsubsection[Single-Higgs production in the s-channel at a $\gamma\gamma$ collider]{Single-Higgs production in the s-channel at a $\PGg\PGg$ collider}
\label{sec:phys:altmodes:gg_H}

One of the most interesting opportunities in $\PGg\PGg$ collisions is to test the properties of the 125-GeV Higgs boson as well as of additional ones~\cite{Boos:2000ki}: Higgs bosons can be singly produced as s-channel resonances through one-loop triangle diagrams and observed in their subsequent decays into $\PQb\PAQb$, $\PGtp\PGtm$, $\PW\PW^*$, $\PZ\PZ$, etc.
Choosing the initial photon polarisations to produce spin-zero resonant states simultaneously reduces the cross section for important background processes and is thus ideal for precision measurements of scalars.

Table~\ref{table:ggHparameters} compares key parameters and the produced number of Higgs bosons for $\PGg\PGg$ colliders using optical lasers and XFELs. 
The XFEL option produces a luminosity profile that is sharply peaked around the Higgs mass (see Fig.~\ref{fig:phys:aalumi}), leading to a much higher number of signal events and slightly smaller backgrounds compared to the optical variant. A full analysis of single Higgs production comparing both laser options is ongoing.

\begin{table}
    \centering
    \small 
    \begin{tabular}{l c c}
        \toprule
        & \textbf{Optical} & \textbf{XFEL} \\
        \midrule
        Laser $E_{\PGg}$ (eV) & 1.17 & $1.04 \times 10^3$ \\
        $\sqrt{s_{ee}}$ (GeV) & 216.00 & 125.72 \\
        $w_{\max}$ (GeV) & 218.16 & 126.97 \\
        $N_{\text{bunch}}$/train & 2820 & 165 \\
        Rep Rate (Hz) & 5 & 120 \\
        $L_{\PGg\PGg}$ (cm$^{-2}$s$^{-1}$) & $3.618 \times 10^{34}$ & $3.570 \times 10^{34}$ \\
        $N_{\PH}/1.0\times10^7$s & 25,000 & 72,200 \\
        $N_{\PQb\PAQb}/1.0\times10^7$s & 64,390 & 58,220 \\
        $N_{\PQb\PAQb}/N_H$ & 2.58 & 0.81 \\
        $\langle \mu \rangle$ & 0.90 & 0.63 \\
        \bottomrule
    \end{tabular}
    \caption{Comparison of $\PGg\PGg$ collider parameters as well number of produced single-Higgs events and $\PQb\PAQb$ background events with $m_{\PQb\PAQb} \ge$ \SI{100}{GeV} and $\beta_z \le 0.1$ for an optical laser and an XFEL.}
    \label{table:ggHparameters}
\end{table}

\paragraph[Higgs effective coupling to photons]{Higgs effective coupling to photons}
The effective coupling of the Higgs boson to photons, $g_{\PH\PGg\PGg}$, is particularly sensitive to new heavy charged particles in different BSM scenarios and beyond the kinematic range via their loop contributions~\cite{Moortgat-Pick:2015lbx, Godbole:2006eb, Gunion:1996kn, Kramer:1993jn, Grzadkowski:1992sa}. As shown previously in Figs.~\ref{fig:smeft:3ab8ab} and ~\ref{fig:eft:gheff}, the \ee\ mode of a linear collider can improve the constraints on the effective $g_{\PH\PGg\PGg}$ coupling by about a factor 2-3 w.r.t.\ HL-LHC, to a precision of about \SI{1}{\%}. At a $\PGg\PGg$ collider, this coupling determines the Higgs boson production rate.  

By combining the measurement of the $\PGg\PGg \to \PH \to \PQb\PAQb$ rate with $BR(\PH \to\PQb\PAQb)$ as measured in the \ee\ mode, the partial width $\Gamma_{\PGg\PGg}$ can be determined. 
For the measurement of $\sigma(\PGg\PGg \to \PH) \times BR(\PH \to\PQb\PAQb)$, earlier studies found a precision of \SI{2}{\%} for on basis of ${\cal L}_{\PGg\PGg} =$ \SI{410}{\fbinv}~\cite{Moortgat-Pick:2015lbx,Abe:2001rdr}, which scales to a precision of \SI{0.7}{\%} for \SI{4}{\abinv}. As can be seen in Fig.~\ref{fig:eft:gheff}, $g_{\PH\PQb\PQb}$ can be determined to about \SI{0.25}{\%} from the \SI{550}{GeV} \ee\ run of LCF, leading to a combined uncertainty of about \SI{0.9}{\%} on the total cross section $\sigma(\PGg\PGg\to\PH)$, corresponding to \SI{0.45}{\%} on $g_{\PH\PGg\PGg}$. This not only a significant quantitative improvement w.r.t.\  the \ee\ collider measurements, by more than a factor of two, but also offers an important independent cross check with very different systematic uncertainties. 

\paragraph[CP quantum numbers]{CP quantum numbers}
The possibility to control of the polarisations of backscattered photons provides a powerful tool for exploring the CP properties of any single neutral Higgs boson that can be produced with reasonable rate at the photon collider:  CP-even Higgs bosons couple with maximal strength to linearly-polarised photons with parallel polarisation vectors whereas CP-odd Higgs bosons prefer linearly-polarised photons with perpendicular polarisation vectors, or in other words, parallel (perpendicular) polarised $\PGg \PGg$ beams generate Higgs particles with quantum numbers $J^{PC}=0^{++}$ $(0^{-+})$, respectively~\cite{Beloborodov:2022byx,Telnov:2020gwc,Telnov:2016lzw,Ginzburg:2013hoa,Ginzburg:2000rm,Kramer:1993jn}. 
This means that not only the SM-like Higgs boson, but also scalar and pseudo-scalar Higgs bosons of extended Higgs sectors could be produced in the s-channel, and its properties measured with significant precisions.

\paragraph{BSM Higgs production}
In \ee\ collisions, Higgs bosons (or other new scalar or pseudo-scalar particles) either need to be produced together with a $\PZ$ boson, or in association with another Higgs boson, e.g.\ $\ee\to\PH\PSA$. This limits the kinematic reach, in particular if an additional CP-even $\PH$ and an CP-odd $\PSA$ are both heavy. 
In $\PGg\PGg$ collisions, scalar and pseudo-scalar bosons can be produced in the s-channel, significantly increasing the kinematic reach by up to a factor of two compared to \ee\ collisions of the same centre-of-mass energy. 
This is particularly interesting for practically all BSM models with an extended Higgs sector, since they usually offer both CP-even as well CP-odd heavy Higgs bosons that become accessible
in the $\PGg\PGg$ option, even allowing a precise determination of  the two-photon width of those particles, respectively~\cite{Boos:2000ki}.

\subsubsection[Di-Higgs production and measurement of trilinear couplings at a $\gamma\gamma$ collider]{Di-Higgs production and measurement of trilinear couplings at a $\PGg\PGg$ collider}
\label{sec:phys:altmodes:gg_HH}

\begin{figure}[htb]
 \centering
 \begin{subfigure}{0.50\textwidth}
    \includegraphics[width=\textwidth, trim=0 0.1cm 0 0,clip]{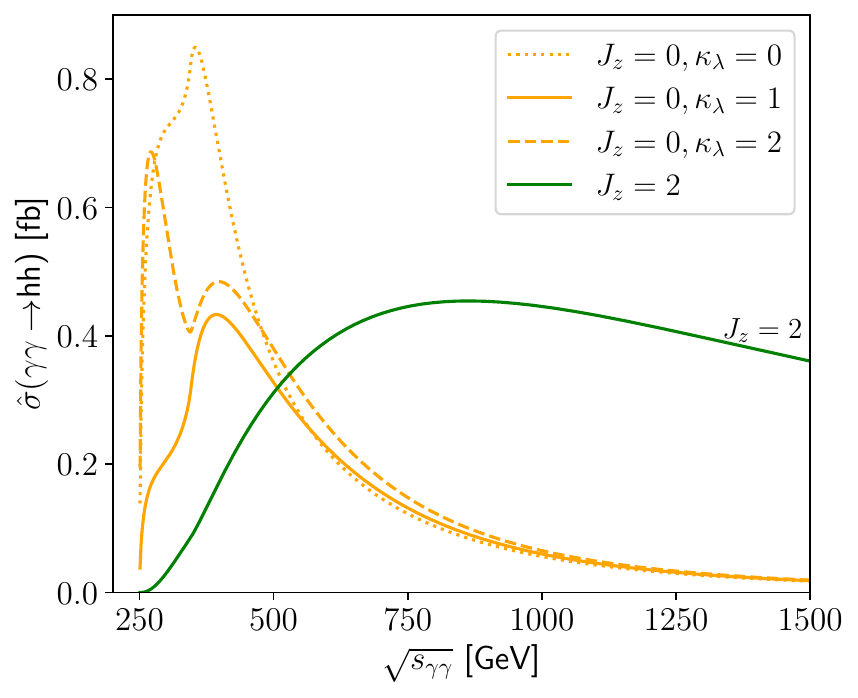}
    \caption{}
    \label{fig:phys:aa2HH:xsecECM}
 \end{subfigure}
 \begin{subfigure}{0.48\textwidth}
   \includegraphics[width=\textwidth]{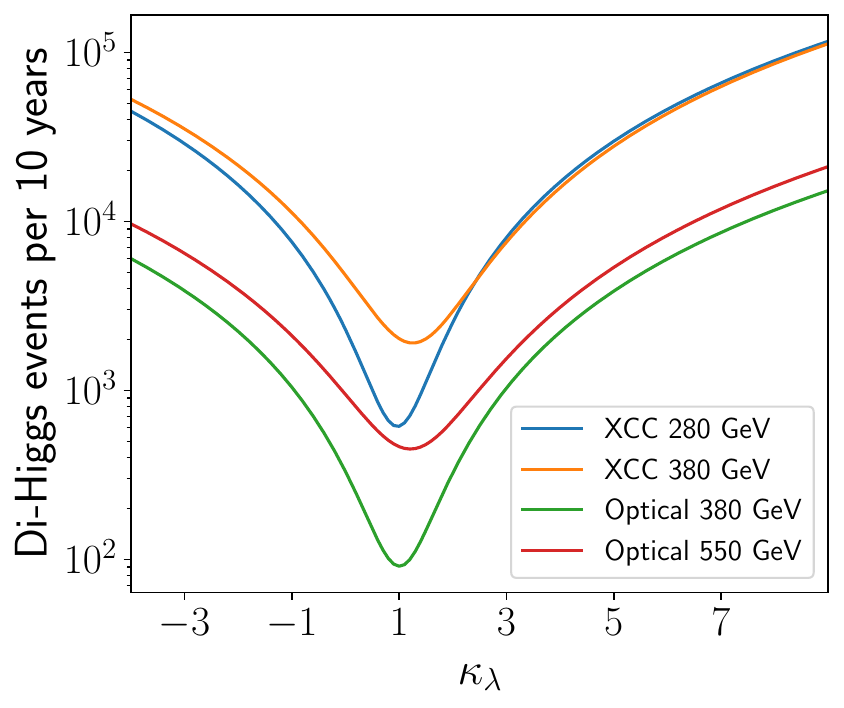}
    \caption{}
    \label{fig:phys:aa2HH:xseckala}
 \end{subfigure}
 \caption{Di-Higgs production cross section (a) as a function of $\sqrt{s_{\PGg\PGg}}$ for different $\kappa_{\lambda}$ in the $J_z=0$ configuration, as well as for $\kappa_{\lambda}=1$ in the $J_z=2$ configuration, and (b) as a function of $\kappa_{\lambda}$ for different $\PGg\PGg$ collider configurations. Both from~\cite{Berger:2025ijd}.
 \label{fig:phys:aa2HHxsec}}
\end{figure}

The case for di-Higgs production in $\PGg\PGg$ collisions, $\PGg\PGg \to \PH\PH$ is even more intriguing~\cite{Berger:2025ijd,Bharucha:2020bhy,Takahashi:2009sf,Belusevic:2004pz,Jikia:1992mt}. 
In the $J_z =0$ configuration, this process opens at a centre-of-mass energy of about \SI{280}{GeV}, which is much less than required for $\ee \to \PZ\PH\PH$, c.f.\ Sec.~\ref{sec:phys:highEHiggs}. This can  clearly be seen in Fig.~\ref{fig:phys:aa2HH:xsecECM}, which shows the di-Higgs production cross section as a function of $\sqrt{s_{\PGg\PGg}}$ for different $\kappa_{\lambda}$ in the $J_z=0$ configuration, as well as for $\kappa_{\lambda}=1$ in the $J_z=2$ configuration.

The di-Higgs production cross section depends strongly on the value of the trilinear Higgs self-coupling, with a minimum of the cross section very close to the SM case. This is illustrated in Fig.~\ref{fig:phys:aa2HH:xseckala} for different $\PGg\PGg$ collider configurations, also indicating the number of di-Higgs events produced in a 10-year run, based on the implementations discussed in Sec.~\ref{sec:acc:altmodes}. The dependence on the value of the Higgs self-coupling differs from the situation at the LHC as well as from the two di-Higgs production modes in \ee\ collisions (c.f.\ Sec.~\ref{sec:phys:highEHiggs}).

In order to estimate the achievable precision, the process $\PGg\PGg \to \PH\PH \to \PQb\PAQb \PQb\PAQb$ has recently been analysed~\cite{Barklow:2025} based on fast simulation of the SiD detector concept, assuming the  $J_z =0$ of an XFEL-based 
$\PGg\PGg$ collider with an $\Pem\Pem$ centre-of-mass energy of \SI{380}{GeV}, including all sources of physics backgrounds from $\PGg\PGg$, $\Pem\PGg$, and $\ee$, properly accounting for the luminosity profile predicted by CAIN~\cite{Chen:1994jt}.  The $\ee$ background, not present in previous optical photon collider concepts, arises from Compton photons interacting with laser photons producing $\ee$ pairs. The process $\ee \rightarrow \PZ\PH \rightarrow\PQb\PAQb\PQb\PAQb$ is in fact the dominant background to $\PGg\PGg \rightarrow \PH\PH \rightarrow\PQb\PAQb\PQb\PAQb$ after final selection. Pileup backgrounds were not considered but preliminary studies show that the $\PGg\PGg \rightarrow$hadrons pileup background is predominantly forward and can be suppressed at the analysis level.  

\begin{figure} 
    \centering
    \fbox{ 
        \begin{minipage}{0.8\linewidth}
            \centering
            \includegraphics[width=\linewidth]{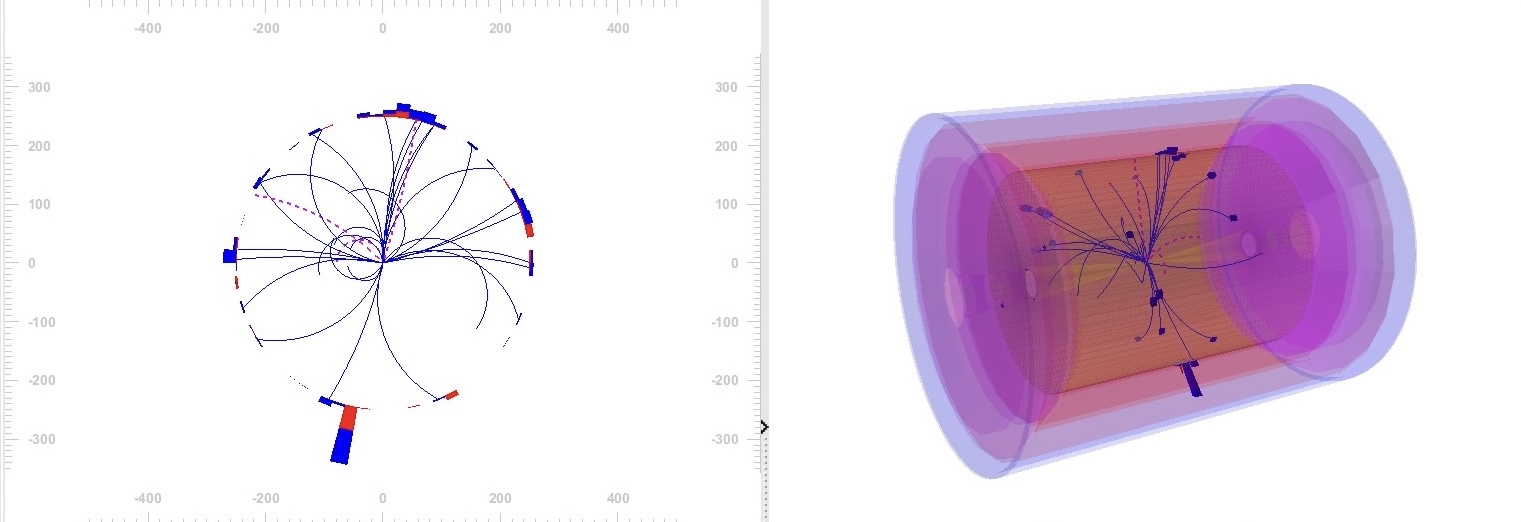}
            \vspace{5pt} 
            \includegraphics[width=\linewidth]{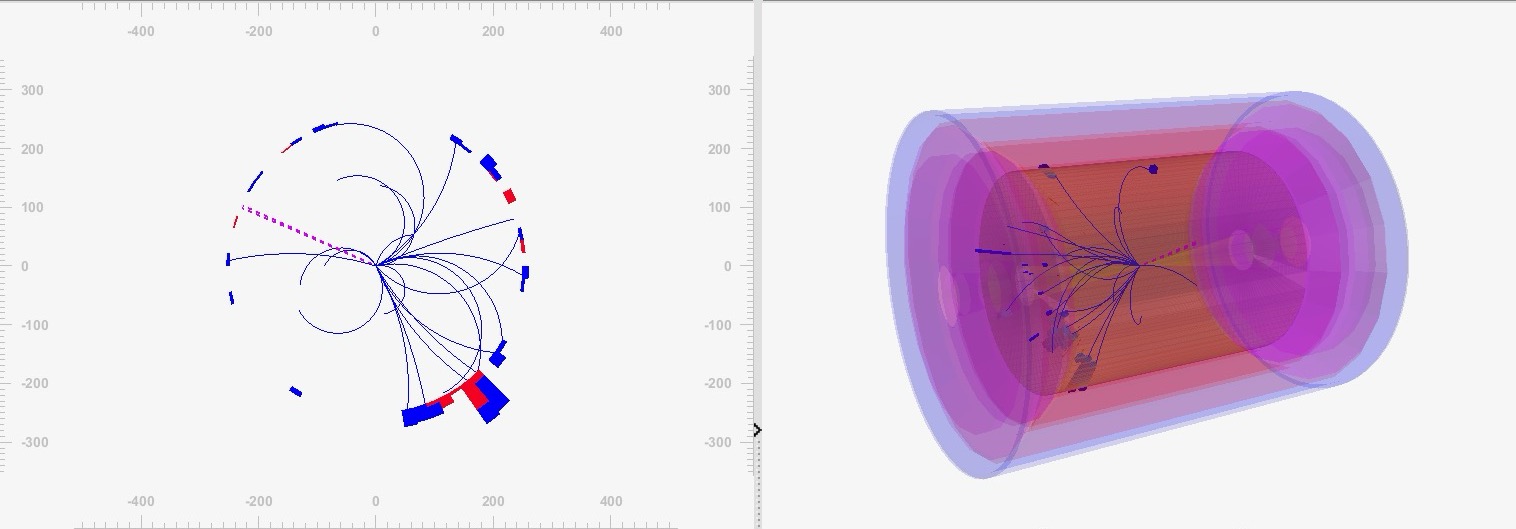} 
        \end{minipage}
    }
    \caption{$\PGg\PGg \rightarrow \PH\PH \rightarrow\PQb\PAQb\PQb\PAQb$ and $\Pem\PGg \rightarrow \PQq\PAQq\PH +$X at $\sqrt{s}$ = \SI{380}{GeV} event displays. The $\Pem\PGg$ background events are highly boosted and distinct from the di-Higgs signal.}
    \label{fig:HH_eGamma_event_display}
\end{figure}

Figure~\ref{fig:HH_eGamma_event_display} shows an event display of $\PGg\PGg \rightarrow \PH\PH \rightarrow \PQb\PAQb\PQb\PAQb $ at $\sqrt{s}$=\SI{380}{GeV}, highlighting the central production of the four $\PQb$ jets and the less busy final state compared to $\ee \to \PZ\PH\PH$, where the decay products of the $\PZ$ boson would be present in addition to the four $\PQb$ jets. The display of an $\Pem\PGg \rightarrow qqH$ event with $\PQb$-jets illustrates that although  $\Pem\PGg$ backgrounds have the largest cross sections, their final states are highly boosted and easily suppressed.

Several Boosted Decision Trees (BDT) were trained to separate signal from the dominant backgrounds. A genetic algorithm was then used to pick the output cuts of each BDT that maximize sensitivity. A likelihood fit analysis yields a significance of \SI{8.5}{\sigma} (statistical only) and a statistical uncertainty of \SI{12}{\%} on the di-Higgs cross section in the $\PGg\PGg \rightarrow\PQb\PAQb\PQb\PAQb$ channel for an integrated $\PGg\PGg$ luminosity of \SI{5}{\abinv} for \SI{260}{GeV} $< \sqrt{s_{\PGg\PGg}} <$ \SI{380}{GeV}  i.e., the top \SI{30}{\%} of the luminosity spectrum. When  eventually all $\PH\PH$ decay modes would be analysed, the statistical uncertainty on the cross section is expected to reduce to \SI{7}{\%}.

\begin{figure}[htbp]
    \centering
    \includegraphics[width=0.65\linewidth]{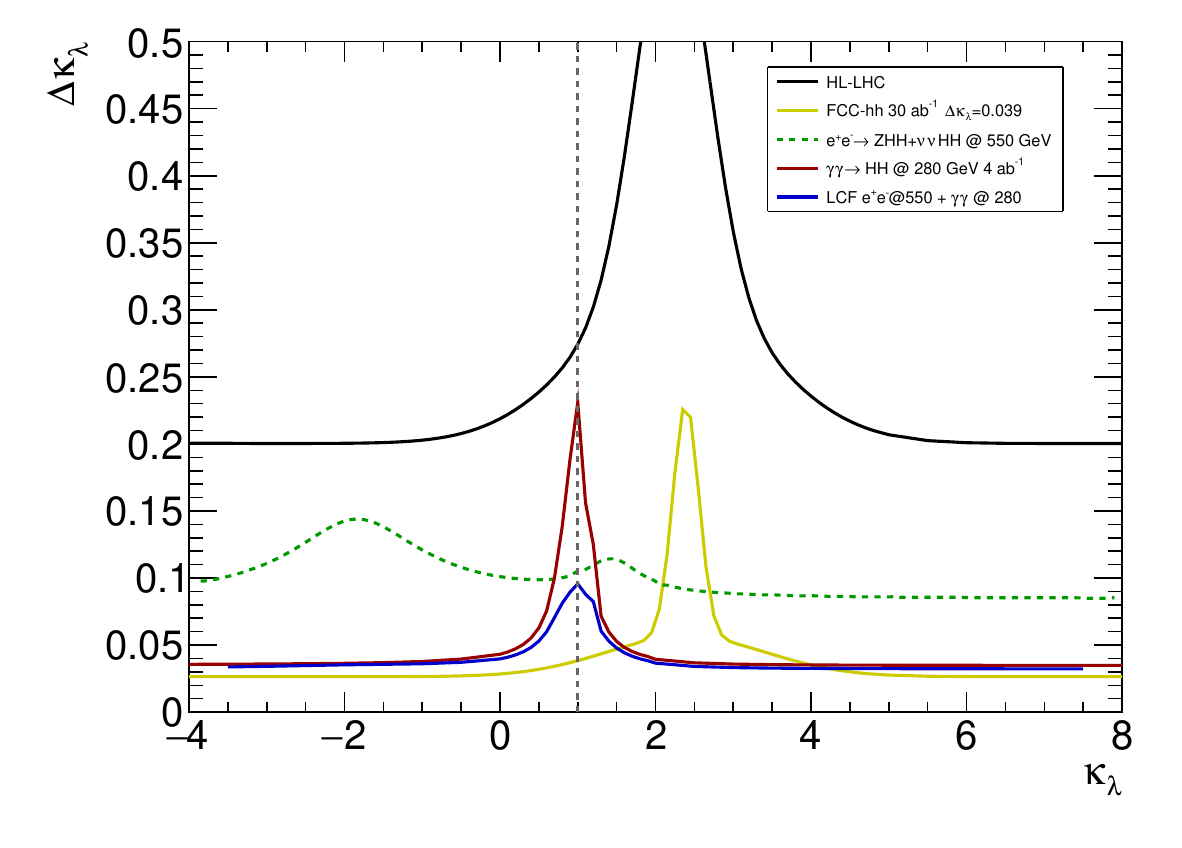} 
    \caption{$\Delta\kappa_\lambda$ for various colliders including HL-LHC, FCC-hh, LCF at $\sqrt{s}$ =\SI{550}{GeV} and the XFEL $\PGg\PGg$ collider at $\sqrt{s}$ = \SI{280}{GeV}. $\Delta\kappa_\lambda$ for the XFEL $\PGg\PGg$ collider was calculated assuming \SI{12.4}{\%} cross section error at \SI{280}{GeV}.  This \SI{12.4}{\%} error is an extrapolation of a \SI{7}{\%} error at \SI{380}{GeV} which in turn is an extrapolation of an \SI{12}{\%} error for $\PGg\PGg\rightarrow \PH\PH\rightarrow \PQb\PAQb\PQb\PAQb$ to an analysis of all $\PH\PH$ decay topologies.}
    \label{fig:delklamcomparison}
\end{figure}

The error on the Higgs self-coupling parameter $\kappa_\lambda$ is proportional to both the di-Higgs cross-section measurement error and the inverse of the absolute value of the slope of the di-Higgs cross section versus $\kappa_\lambda$ curve. 
As shown in Fig.~\ref{fig:phys:aalumi:xcc}, the slope of this curve is much larger at \SI{280}{GeV} than at \SI{380}{GeV}. 
Future double Higgs studies at $\PGg\PGg$ colliders will therefore concentrate on $\sqrt{s}$ = \SI{280}{GeV}. 
The double Higgs cross section error for $\PH\PH\rightarrow \PQb\PAQb\PQb\PAQb$ at \SI{380}{GeV} can be extrapolated to \SI{280}{GeV} assuming  an integrated $\PGg\PGg$ luminosity of \SI{4.0}{\abinv} for \SI{250}{GeV} $< \sqrt{s_{\PGg\PGg}} <$ \SI{280}{GeV} (10 years running time with the XFEL collider described in Sec.\ref{sec:acc:altmodes}), yielding an expected precision of \SI{12.4}{\%} on the total $\PH\PH$ cross section. From this number, ultimate error on $\kappa_\lambda$ can be estimated assuming all $\PH\PH$ event topologies have been analysed, as shown in Fig.~\ref{fig:delklamcomparison}.

\subsubsection[Opportunities beyond Higgs physics  at a $\gamma\gamma$ collider]{Opportunities beyond Higgs physics  at a $\PGg\PGg$ collider}
\label{sec:phys:altmodes:gg_other}

\paragraph{Multi-quark resonances}
In general, the $\PGg\PGg$ option allows a precise spectroscopy of C-even resonances in multi-quark states or  glueballs~\cite{Beloborodov:2022byx}.
These information might lead to new insights into the strong sector. 

\paragraph[Photon structure functions]{Photon structure functions}
The $\PGg\PGg$ as well as the $\Pe\PGg$ options allow to measure the hadronic and electromagnetic structure of photons~\cite{Aurenche:1996mz,Vogt:1999mu}. These measurements can be performed with different polarisation settings.

\paragraph[Di-photon resonances]{Di-photon resonances}
Motivated by the former excess at \SI{750}{GeV} in the 2015 LHC data, the implications for future colliders of such excesses
that show up in the invariant mass distribution of a pair of high-energy photons were discussed in~\cite{LCCPhysicsWorkingGroup:2016szh}.
Any resonance decaying into two photons can be directly produced at a $\PGg\PGg$ collider. Via the polarisation choice, the spin of the resonance can be directly tested, and likewise the CP quantum numbers using transversely polarised photons.

\paragraph[ALPS and dark matter]{ALPS and dark matter}
The $\PGg\PGg$ option also allows explores the dark sector, axion-like particles and non-linear QED. For instance, the
direct access to light-by-light scattering opens a new plethora of ``old'' and ``new'' physics as Born-Infeld, constraints for non-linear extensions of QED, ALPS and dark matter candidates, complementary to the $\Pep\Pem$ mode~\cite{Ellis:2022uxv,Ginzburg:1982bs}.

\subsubsection[Physics opportunities at an $\mathrm{e}\gamma$ collider]{Physics opportunities at an $\Pe\PGg$ collider}
\label{sec:phys:altmodes:eg}

The $\Pem\PGg$ collider option is ideal for single production of  heavy charged particles like heavy Higgs bosons or supersymmetric fermions, beyond the kinematical limit in direct production at $\ee$. Similar as for the $\PGg\PGg$ case, both beams $\Pem\PGg$ are available highly polarised.

\paragraph[Higgs and electroweak physics]{Higgs and electroweak physics}
Anomalous top couplings are well accessible via single-top production in $\Pem\PGg$ collisions. The rate of single-top production is directly proportional to the $\PW\PQt\PQb$ coupling. Its anomalous part contains terms $f_{L,R}\sim 1/\Lambda$, where $\Lambda$ is the scale of new physics. At $\roots=$ \SI{500}{GeV}, rates for single-top production of $\ge$ \SI{100}{fb} are predicted, leading to 
increased sensitivity in $f_{L,R}$ compared to the LHC~\cite{Boos:2000ki}.

\paragraph[Supersymmetry]{Supersymmetry}
The $\Pem\PGg$ mode extends and complements the kinematic range of $\ee$ colliders in particular for heavy sfermions~\cite{Kon:1993aw,Kon:2001hm} 
which might be absolutely substantial for the direct discovery of Supersymmetry!

\subsubsection[Physics opportunities at an $\mathrm{e}^-\mathrm{e}^-$ collider]{Physics opportunities at an $\Pem\Pem$ collider}
\label{sec:phys:altmodes:ee}

Electron-electron collisions extend the physics potential of $\epem$ colliders. They offer an extreme cleanliness since background processes are usually strongly suppressed. For instance, the annihilation processes from $\ee$ collisions are absent. Since in case of $\Pem$ beams, both beams can be highly-polarised ($|P(\Pem)|$>0.85), the weak quantum numbers of the initial state spin $S_z$, weak isospin $I^3_w$ and hypercharge $Y$ can be nearly completely specified. The beam-beam interaction is much lower due to the absence of the pinch effect, and implications for the luminosity tuning have been studied~\cite{AlabauPons:2006xm}.
Background processes involving $\PW$ bosons can be substantially suppressed using right-handed  electrons. 

\paragraph[SM Higgs physics]{SM Higgs physics}
In $\Pem\Pem$ collisions, single Higgs production dominantly proceeds via $\PZ\PZ$-fusion, a process very  sub-dominant in $\ee$ collisions.  A cross section of about \SI{10}{fb} is expected at $\sqrt{s}=$ \SI{500}{GeV}. Combination with either branching fraction  measurements in \ee\ -- or from considering the decay via the same coupling, i.e.\ $\PH\to\PZ\PZ^*$ leads to an independent measurement of the $\PH\PZ\PZ$ coupling~\cite{Barger:1994wa,Gunion:1998jc,Minkowski:1997cv,Abe:2001rdr}, with  systematic uncertainties complementary to the recoil technique in \ee. 

\paragraph[Doubly-charged exotic particles]{Doubly-charged exotic particles}
The initial state with lepton number $L=2$,  electron number $L_e=2$ and electric charge $Q=-2$ opens a window to new classes of BSM models, for instance to bi-lepton models or models with doubly-charged Higgs bosons as in Georgi-Machacek models~\cite{Cao:2018xwy, Yu:2016skv}.

\paragraph[Majorana neutrinos]{Majorana neutrinos}
The process $\Pem\Pem \to \PWm\PWm$, which violates lepton number conservation, is the inverse of neutrino-less double beta decay. It has  clear kinematical signatures, e.g.\ $\PWm\PWm \to$ 4 jets or the semi-leptonic channel. With the detectors and reconstruction algorithms under development for future colliders, reliable charge information will not only be available for charged leptons, but also for jets, c.f.\ Sec.~\ref{sec:hlreco}. The observation of this process would indicate a model-independent evidence for the Majorana nature of the neutrino, and the measurement of the cross section could serve as input to distinguish between theoretical models~\cite{Asaka:2015oia, Banerjee:2015gca, Greub:1996ct}.


\subsection{Beyond-collider physics}
\label{sec:phys:PBC}

The beam dumps of a linear collider absorb a huge number of electrons or positrons.
Also, small fractions of the beams can be extracted before the main interaction point, providing high-energy and high-quality particle bunches with well-known properties for experiments or R\&D facilities.
The pre-collision beams and the particles produced in the beam dumps can be utilized for exploring BSM, QED in extreme conditions, material science, or future accelerator studies. This section summarizes the physics opportunities 
Considerations for the implementation as well as R\&D and irradiation facilities will be discussed in Sec.~\ref{sec:beyond}.


\subsubsection{Physics at the main beam dumps}
\label{sec:phys:PBC:beamdumps} 

Unlike at circular colliders, experiments using the main beam dumps can be conducted entirely in parallel with collider experiments. The number of particles impinging on each dump for various collider configurations are shown in Table~\ref{tab:ILCbaseline}. For a beam energy of \SI{125}{GeV} and a power of \SI{2.6}{MW}, corresponding to the ILC250 option, the main beam dumps see $N_{\mathrm{EOT}}= \num{4e21}$ electrons on target in about \SI{2.5}{years}.  The ILC design adopted cylindrical, water-based beam dumps with a diameter of \SI{1.8}{m} and a length of \SI{11}{m}, which can handle up to \SI{17}{MW} beam power~\cite{Satyamurthy:2012zz}. While alternative designs are possible, we assume ILC-like water dumps for quantitative results in this section. Space requirements for shielding and background reduction to be considered for the implementation of such experiments will be discussed in Sec.~\ref{sec:beyond:maindump}. 

\begin{figure}[htbp]
    \centering

\begin{subfigure}{0.43\textwidth}
        \includegraphics[width=\linewidth]{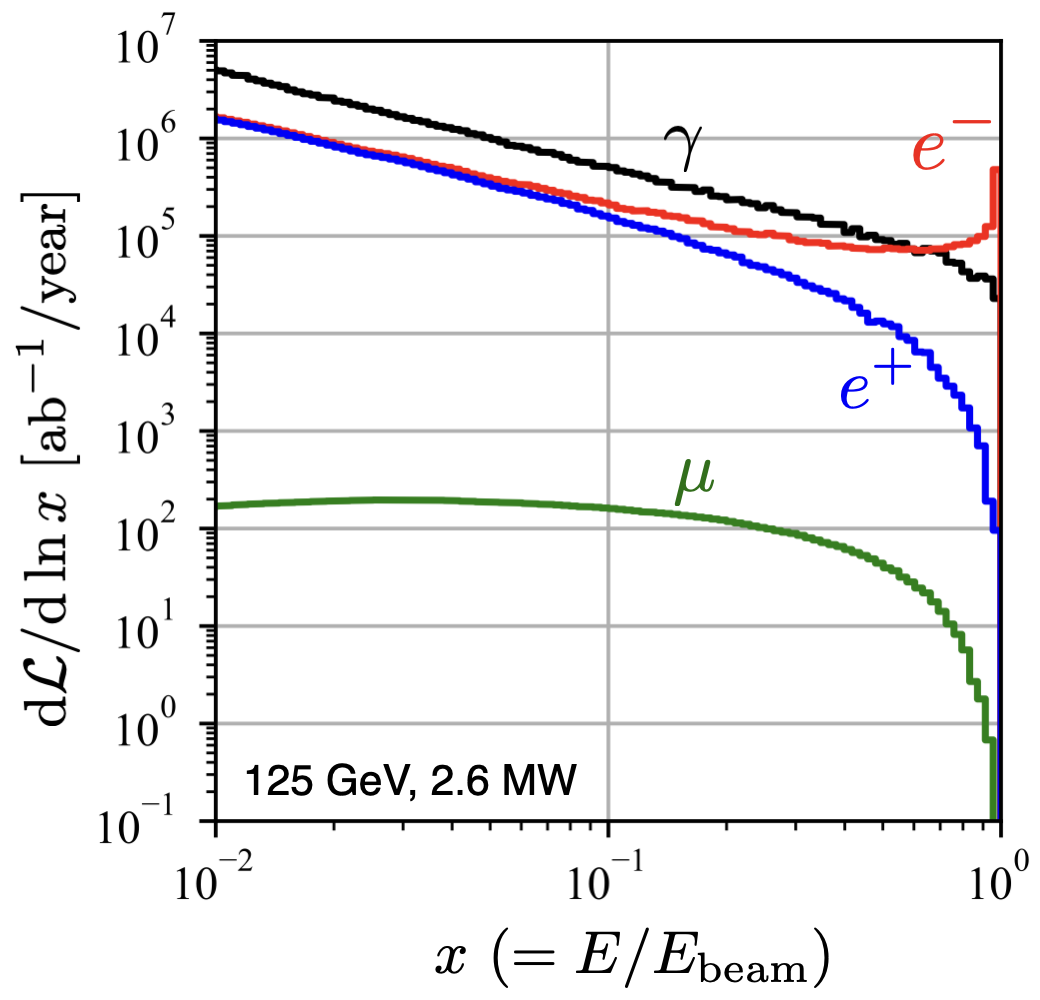}
\caption{}
\label{fig:lumi}
\end{subfigure}
\hspace{0.0001cm}
\begin{subfigure}{0.46\textwidth}
        \includegraphics[width=\linewidth]{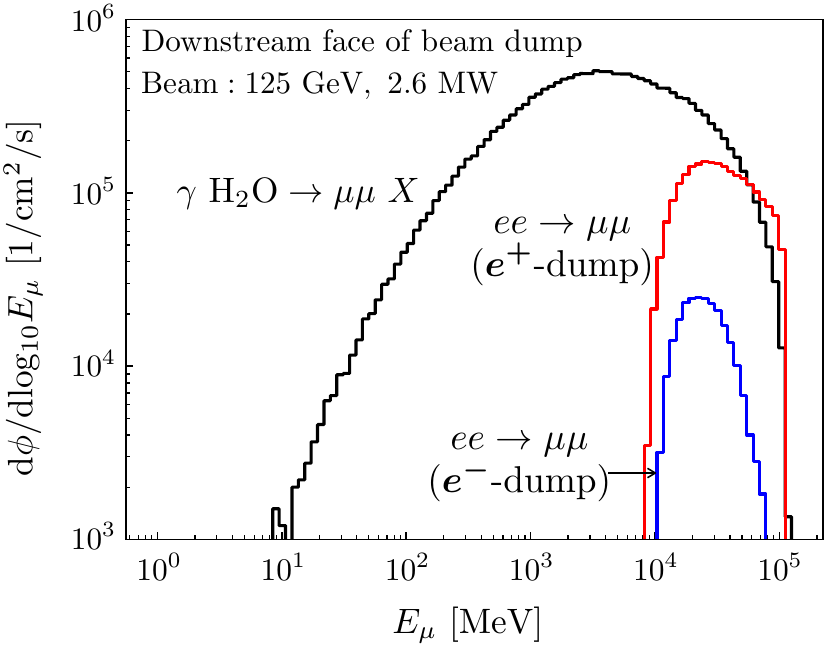}
\caption{}
\label{fig:2process_E}
\end{subfigure}
    \caption{(a) The luminosity of  $\Pepm$, $\PGg$, $\PGmpm$ to nucleons on water target inside the electron beam dump as the function of $\ln x$, where $x=E/E_{\mathrm{beam}}$. (b) The energy distribution of muons downstream of the beam dump. The contribution of muons from the annihilation process in the positron beam dump (red curve) is higher than the one in the electron beam dump (blue curve)  and is dominant in the high energy region~\cite{Sakaki:2022udd}.}
    \label{fig:overall}
\end{figure}

\paragraph[Rates and spectra of produced particles]{Rates and spectra of produced particles} In the beam dump, $\Pepm$, $\PGg$, and $\PGmpm$, are produced through electromagnetic showers, and they interact further with nucleons. 
Staying with the ILC250 parameters as example, high integrated luminosities of $10^5$ to $10^6$\,\abinv\ per year for $\Pepm$ and $\PGg$, and approximately \SI{100}{\abinv} per year for muons can be achieved. This result considers only interactions with water inside the beam dump. If shielding is placed downstream of the beam dump, interactions with these materials can further increase the luminosity of muons. New particles that couple feebly to SM particles may also be created from electromagnetic shower particles. Figure~\ref{fig:lumi} shows the effective luminosity of $\Pepm$, $\PGg$, $\PGmpm$ on nucleons of water target inside the electron beam dump as a function of $\ln x$, where $x=E/E_\mathrm{beam}$. 

Positrons and electrons behave similarly in electromagnetic showers. Important differences arise from the annihilation of the positrons with the target electrons. For example, pairs of muons are produced more efficiently in the positron beam dump through the annihilation process between the $\Pep$ beam and atomic electrons ($\ee \to \PGmp\PGmm$). Figure~\ref{fig:2process_E} shows the energy distribution of muons downstream of the beam dump. The contribution of muons from the annihilation process in the positron beam dump (red curve) is higher than the one in the electron beam dump (blue curve) and is dominant in the high-energy region. 

\begin{figure}
    \centering
    \begin{subfigure}{0.45\textwidth}
        \includegraphics[width=\linewidth]{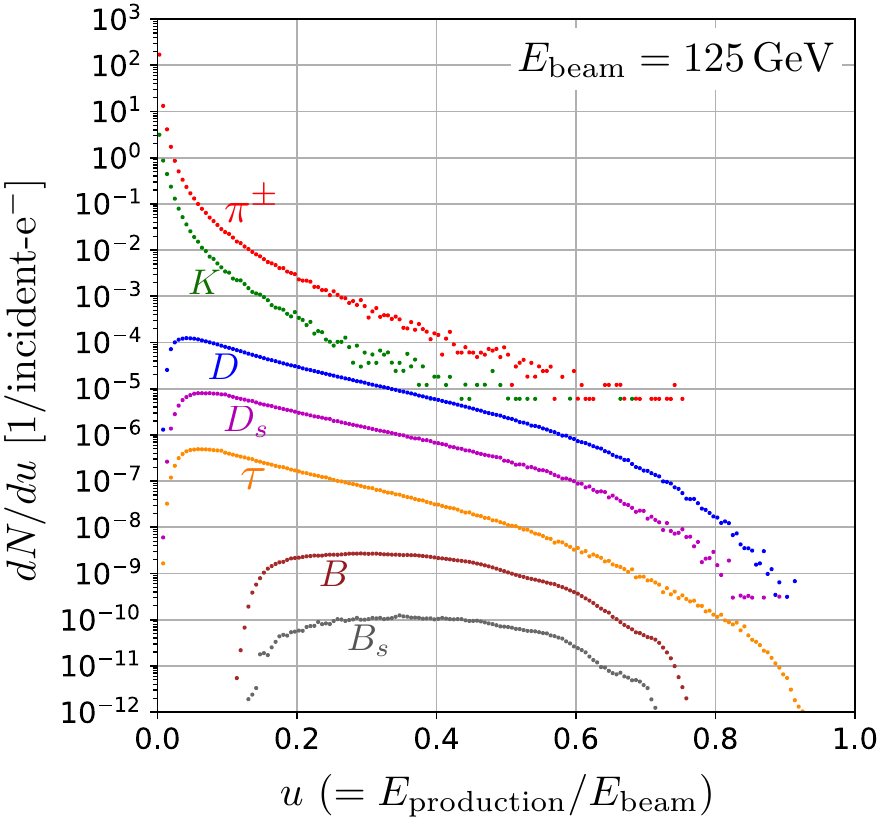}
\caption{}
\label{fig:production:a}
\end{subfigure}
\hspace{0.0001cm}
\begin{subfigure}{0.45\textwidth}
    \includegraphics[width=\linewidth]{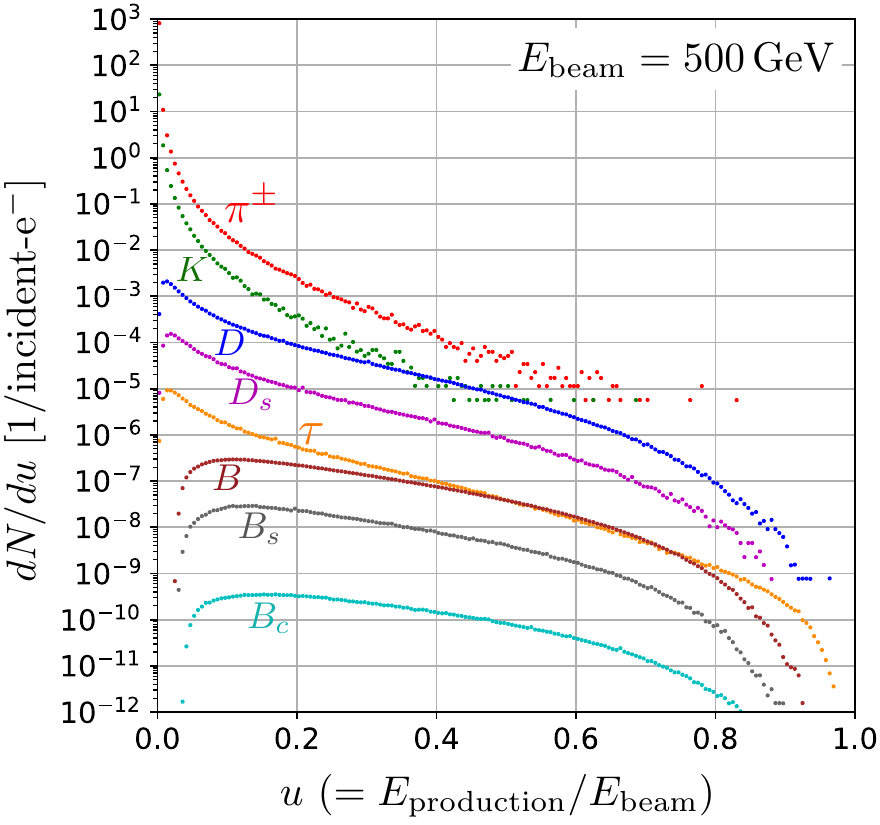}
\caption{}
\label{fig:production:b}
\end{subfigure}
    \caption{The energy distribution of mesons and $\PGt$ leptons at  production inside the linear collider beam dump for beam energies of (a) $E_\mathrm{beam}=$ \SI{125}{GeV} and (b) $E_\mathrm{beam}$ = \SI{500}{GeV}~\cite{Nojiri:2022xqn}.}
    \label{fig:production}
\end{figure}

\paragraph[Heavy meson and $\tau$ lepton production]{Heavy meson and $\PGt$ lepton production}
High-energy electron or positron beam dumps produce not only electromagnetic shower particles but also a large number of hadrons and $\PGt$ leptons. Figure~\ref{fig:production:a} shows the energy distribution of mesons and tau leptons at their production inside the linear collider beam dump for $E_\mathrm{beam}=$\SI{125}{GeV}~\cite{Nojiri:2022xqn}. For this evaluation, meson photoproduction has been considered for hadron production because photons are dominant secondary particles in electromagnetic showers. In contrast, $\PGt$-lepton production cross section includes both 
the contributions from \PDs decay $(\PDs \to \PGt \nu_{\PGt} )$ and pair production from real photons $(\PGg \PN \to \PGtp\PGtm X)$. In the high-energy region where $u>0.65$, the pair production process becomes dominant, a characteristic feature of electron beam dumps. In Fig.~\ref{fig:production:b}, we show the same distribution of  $E_\mathrm{beam}=$\SI{500}{GeV}. At the higher energy, the contributions of  \PD and \PB are more prominent.  The total number of \PB mesons, \PD mesons and $\PGt$ leptons  per $N_{\mathrm{EOT}}=4 \times 10^{21}$ are 
$4\times 10^{12}$, $7\times 10^{16}$ and $3\times 10^{14}$, respectively, for $E_{\mathrm{beam}}=$\SI{125}{GeV}, and \num{3e14}, \num{5e17} and \num{2e15}, respectively,
for $E_{\mathrm{beam}}=$\SI{500}{GeV}. 

\paragraph[Dark sector particle searches]{Dark sector particle searches}

Many extensions of the SM contain particles that do not carry SM gauge charges. They are called ``dark sector'' particles and often arise in models that address the hierarchy problem (e.g., relaxion models), the strong CP problem (models with axions or axion-like-particles (ALPs)), and very importantly Dark Matter (DM) models.

In the past decade, new ideas to search for dark sector particles at accelerator-based experiments (colliders, meson factories, fixed target experiments, ...) have been flourishing. For a recent review, see \cite{Antel:2023hkf}. Dark sector particles typically have a long lifetime, making them a target for beam dump experiments.

\paragraph[Axion-like particles (ALPs)]{Axion-like particles (ALPs)}
ALPs, \PXXA, 
are pseudoscalar particles whose Lagrangian has an approximate shift symmetry. ALPs can couple to the SM gauge bosons, $\PV$, with interactions of the type 
\begin{equation}
    {\cal L}_{VV}=g_{\PXXA VV} ~\PXXA V_{\mu\nu}\tilde V^{\mu\nu} /4 \ ,
\end{equation}
where $V_{\mu\nu}$ is the field strength tensor and $\tilde V_{\mu\nu}=\epsilon_{\mu\nu\alpha\beta}V^{\alpha\beta}/2$. 
ALPs can also couple to SM fermions with derivative couplings $\propto \partial_\mu \PXXA (\bar f\gamma^\mu \gamma_5 f)$. 

These ALPs could be produced in the electron beam dump through the bremsstrahlung process initiated by the secondary photon beam. The ALPs could decay to two photons after the dump. Their lifetime can, in fact, be macroscopic and is given by 
\begin{equation}
    \tau_{\PXXA} = 4\pi/(c_{\PXXA\PGg\PGg}^2 m_{\PXXA}^3)\simeq 4\pi\left(\frac{{\mathrm{GeV}}}{m_{\PXXA}}\right)^3\left(\frac{1}{g_{\PXXA\PGg\PGg}{\mathrm{GeV}}}\right)^2\frac{\mathrm{m}}{5\times 10^{15}} \ .
\end{equation}

\begin{figure}
\centering
   \begin{subfigure}{0.45\textwidth}
    \includegraphics[width=\linewidth]{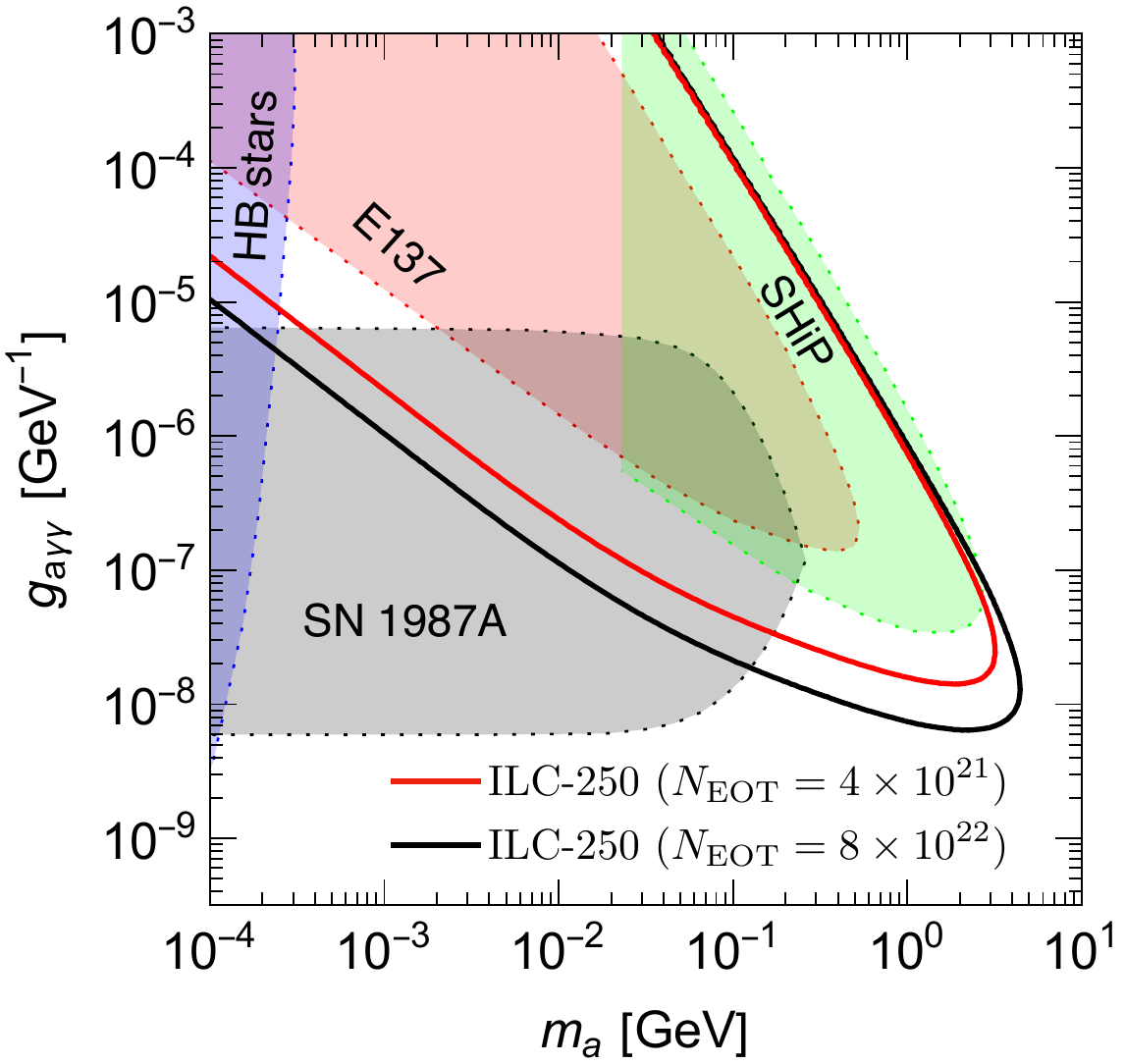}
    \caption{}
    \label{Fig:ILCALPs}
    \end{subfigure}
\hspace{0.0001cm}
    \begin{subfigure}{0.45\textwidth}
    \includegraphics[width=\linewidth]{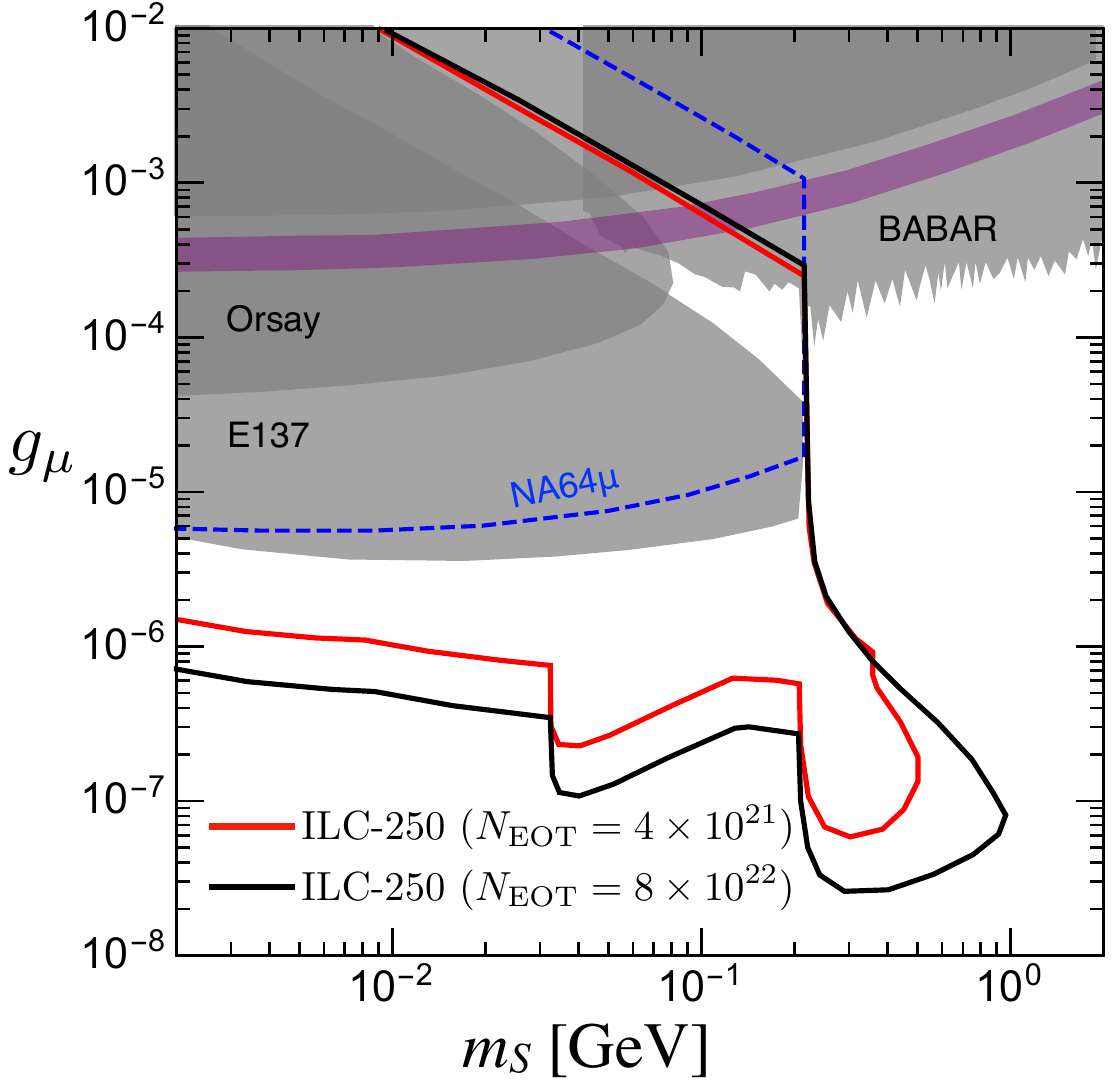}
    \caption{}
    \label{Fig:lepscalar}
    \end{subfigure}

    \caption{(a) Sensitivity reach of the beam dump experiment to ALPs coupled to photons \cite {Sakaki:2020mqb}. The SHiP limit is taken from~\cite{Lantwin:2024cfl}.  (b) Phase-space coverage of leptophilic scalars coupling to muons. Figures modified from~\cite{Asai:2021ehn}.}
    \label{}
\end{figure}

Due to the setup of the ILC-like beam dumps, this experiment can be sensitive to GeV-scale ALPs with coupling to photons of the order of $\mathcal O(10^{-8}$\,GeV$^{-1})$. This is demonstrated in Fig.~\ref{Fig:ILCALPs} where we show the sensitivity for the $E_\mathrm{beam}=\SI{125}{GeV}$ option at \SI{95}{\%} CL with \num{4e21} and \num{8e22} electrons on target. Notably, already \SI{2.5}{years} at a dump of such a facility could extend the SHiP reach by roughly an order of magnitude to smaller couplings.

Apart from dumping charged particles on a fixed target to search for ALPs, high-energy photons can also be used. Such photons can be produced via Compton backscattering when colliding high-energy electrons with a high-intensity laser pulse, such as in setups of strong-field quantum electrodynamics experiments described in Sec.~\ref{sec:phys:PBC:sfqed}. Fixed-target searches using photons have been shown to have significantly lower background rates in certain cases~\cite{Bai:2021gbm}. Simulations showed that a linear collider facility can probe new phase space of the ALP-photon coupling compared to existing constraints and projections of current experiments in such a configuration~\cite{Schulthess:2025tct}. Additionally, the approach remains highly complementary since this method leverages photon interactions via the Primakoff production mechanism directly.

\paragraph[Leptophilic scalars (LS)]{Leptophilic scalars} 
The beam dump of a linear collider is a good place to search for light scalar particles coupled to electrons and muons, so-called leptophilic scalars. The Lagrangian of a simplified model can be expressed as
\begin{equation}
    {\cal L}=\frac{1}{2}(\partial_{\mu} S)^2 
    -\frac{1}{2}m^2_S S^2
    - \sum_l g_l S \bar{l}{l}
    -\frac{1}{4} S F_{\mu\nu} F^{\mu\nu} \ .\footnote{The mixing of $S$ to the SM Higgs bosons also has to be considered.} 
\end{equation}

Such leptophilic scalars could be produced from the secondary muons in the dump. They can be detected if they decay back to a pair of charged leptons before the detector. Figure~\ref{Fig:lepscalar} shows the expected reach on the leptophilic scalar coupling to muons.  The peculiar shape of the excluded region originates from competing production processes for such scalars (bremsstrahlung, Primakoff and pair annihilation).\footnote{This case is not explored in the SHiP report. } 

\paragraph[Heavy neutral leptons (HNL)]{Heavy neutral leptons (HNL)} 
The neutrino sector can be extended to include the right-handed neutrino. In that case, the HLN is predicted, and the constraint of the HLN is weak enough to allow this sector to play a pivotal role in the leptogenesis by introducing the CP violation to the sector. 
The charged current interaction between a charged lepton and a HNL is expressed as 
\begin{equation}
    {\cal L}= -\lambda_{lI}(\bar{L}_l \tilde{H} ) N_I  
    -\frac{1}{2}M_I\bar{N}^c_I N_I + h.c. \ ,
\end{equation}
where $N_I$ express the extra neutral lepton that mixes with the SM neutrinos. 
When the coupling $\lambda_{lI}$ is small, 
the mixing angle is approximated as 
\begin{equation}
    U_{lI}=\frac{v\vert\lambda_{lI}\vert}{M_I}. 
\end{equation}
The interaction of a charged lepton is already strongly constrained as low as $10^{-10}<U^2<10^{-4}$ depending on the mass $m_{HNL}$ for mixing with the electron. 

\begin{figure}
\begin{center}
\includegraphics[width=0.5\textwidth]{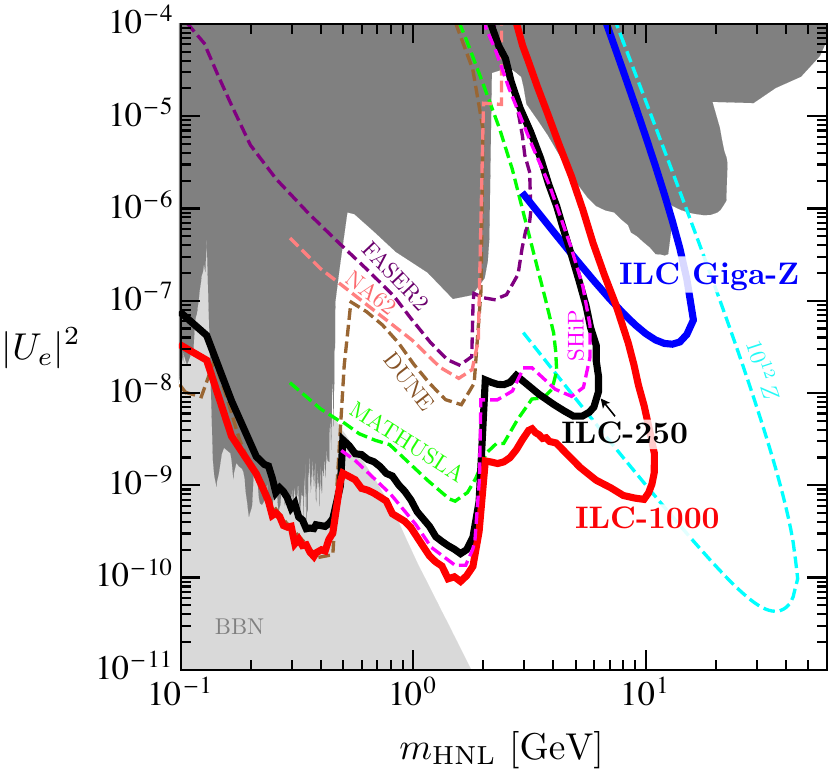}
\end{center}
\caption{Sensitivity reach of the beam dump experiment to HNLs mixing with the electron neutrino in the mass and mixing plane, assuming \num{4e22} electrons on target with a centre-of-mass energy of \SI{250}{GeV} (black solid) and \SI{1000}{GeV} (red solid). Figure taken from \cite{Nojiri:2022xqn}. }
\label{fig:Ue}
\end{figure}

The phenomenology of HNLs at a beam dump experiment is complex because various production processes contribute. The charged current interaction of electron or muon produces the HNL directly and dominant production cross section of the high mass region. The production cross section is proportional to the square of the mixing angle of $U^{2}$, while the probability of decaying inside the decay volume reduces if the mixing angle is too large or too small. Figure~\ref{fig:Ue} compares the various limits of HNLs mixing with $\nu_e$.   
The region accessible via direct production in the dump and decay in the decay volume corresponds to the nose-like region above a few GeV in the black and the red lines for ILC-250 and ILC-1000, respectively. 

Heavy neutral leptons can also be produced by heavy-meson decay. Any weak decay process of mesons contributes to HNL production through neutrino mixing. As already discussed, at the beam dumps of a linear collider, heavy mesons are copiously produced by photon-nucleon interaction, and the leptonic decay of the heavy mesons is an additional source of HNLs. The effect can be seen in Fig.~\ref{fig:Ue} in the edge-like structure of the excluded region at \SI{500}{MeV}, \SI{2}{GeV} and around \SI{3}{GeV}, hidden behind the direct production limit. 

Note that the expected reach of ILC250 in Fig.~\ref{fig:Ue} closely follows the SHiP projection. 
This is because the total flux of neutrinos at the beam dump is roughly equal to the expected number of neutrinos at SHiP. As we discuss in the Sec.~\ref{sec:beyond:maindump:exp}, the SHiP limit assumes an active muon veto system. 
In contrast, the ILC limit assumes a rather simple detector setup and kinematical reduction of the background, and is thus expected to improve substantially if an active veto is considered. 

While the sensitivity of the beam dump experiment to HNLs has been estimated, the effect of the beam polarisation has not been carefully studied yet. The electron beam is highly polarised at the injection of beam dumps so that the highest charged current interactions depend on the beam polarisation, while the effect is diminished for the electron and positron at the end of the EM showers. This should allow us to study the interaction of HNLs or other BSM particles that are produced at the beam dump.

The beam-dump search for HNLs is highly complementary to the direct search described in Sec.~\ref{sec:phys:bsm:hnl} as it covers a lower mass range. 

\begin{figure}
    \centering
    \begin{subfigure}{0.49\textwidth}
        \includegraphics[width=\linewidth]{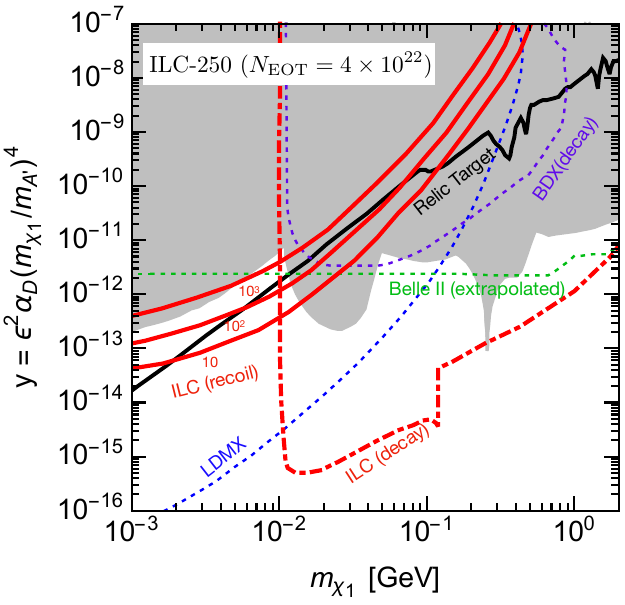}
    \caption{}
    \label{Fig:DM:a}
    \end{subfigure}
\hspace{0.0001cm}
    \begin{subfigure}{0.49\textwidth}
        \includegraphics[width=\linewidth]{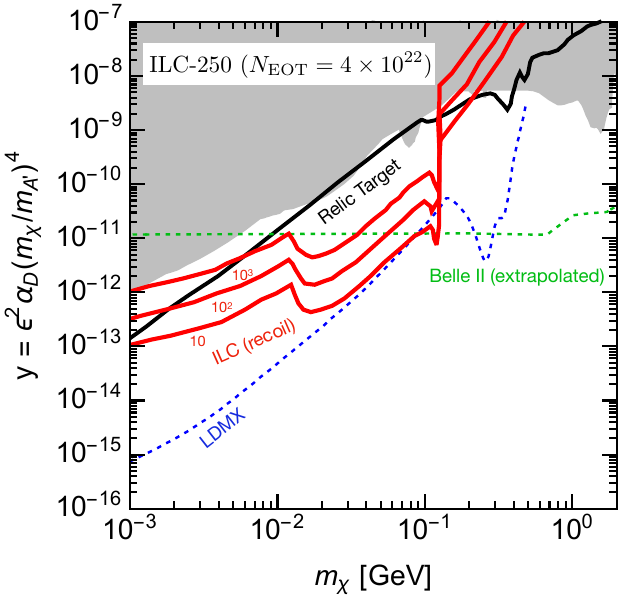}
    \caption{}
    \label{Fig:DM:b}
    \end{subfigure}

    \caption{Projected sensitivity reach of ILC250 with \num{4e22} positrons on target (a) in the pseudo-Dirac DM model  and (b) in the scalar inelastic DM model. See~\cite{Asai:2023dzs} for the details on the specific benchmark chosen for the two models. The solid lines correspond to $10,~10^2,~10^3$ signal events in the case of DM scattering. The dotted-dashed line in the left panel shows the reach of the decay signal search. In both panels, we show the reach at the positron beam dump.}
    \label{Fig:DM}
\end{figure}

\paragraph*{Dark matter} 
Light dark matter particles can be produced at the beam dumps via different processes: Depending on the specific model, DM particles can be produced via bremsstrahlung at both electron and positron beam dumps --- or via pair-annihilation of a positron on an atomic electron at the positron beam dump. The resulting DM particles can scatter off electrons in a detector placed after the muon shield and yield observable electron-recoil events. Several models predict the existence of heavier dark particles in addition to DM and the decay of heavy particles to dark matter can produce a signal at beam dump experiments. This is the case of e.g.\ inelastic DM models or models with strongly interacting massive particles.
In Figs.~\ref{Fig:DM:a} and~\ref{Fig:DM:b}, respectively, we show the reach in the case of pseudo-dirac inelastic dark matter, which is detected via decay, and for scalar inelastic dark matter detected by scattering. In both cases, the ILC-like setup will be able to extend the probed regions of parameter space if compared to past and current experiments (in gray in the figure). Particularly, it will have access to the ``thermal lines'' where DM acquires the measured relic abundance (shown in black in the figure). A linear collider beam dump reach will be complementary to the reach at the proposed LDMX experiments (blue lines in the figure). In the case of scalar inelastic DM, the ILC250 bound is expected to be weaker than the LDMX one. However, the beam dump experiment can potentially give additional information, since it can characterise the DM interaction in detail by identifying the recoil objects. The reach in the figure was computed in the case of production at the positron beam dump. The ILC250 reach varies if the search is done at the electron beam dump or at the positron beam dump, with generally a broader reach in the case of positron beam dump, due to the increase in production coming from the pair-annihilation of positron on an atomic electron.

\subsubsection{Strong-field QED}
\label{sec:phys:PBC:sfqed}

Strong-field quantum electrodynamics describes the behaviour of charged particles and photons in strong electromagnetic fields, where non-linear effects become significant. These effects emerge when field strengths approach the QED field strength scale, which is sometimes also called the critical field or the Schwinger limit $E_\mathrm{qed} = \frac{m_{\Pe}^2 c^3}{e \hbar} \approx 1.32 \times 10^{18}~\mathrm{V/m}$. Processes like non-linear Compton scattering and non-linear Breit-Wheeler pair production, as well as effects of vacuum birefringence, can be observed. With advancements in high-intensity laser technology, strong-field QED is now experimentally accessible. A future linear collider facility with beam energy above \SI{100}{GeV} will provide a unique opportunity to probe this largely unexplored regime of QED. 

The strong-field QED phase space can be defined via two parameters. One is the classical non-linearity parameter $\xi = e E \lambda_c / \hbar \omega$ also called the intensity parameter, although proportional to the square root of the laser intensity. $\lambda_c = \hbar / (m_e c)$ is the Compton wavelength and $\omega$ the angular frequency of the laser photons. The other one is the quantum non-linearity parameter which in plane-wave approximation is $\chi = \gamma \, E / E_\mathrm{qed}$, where $\gamma$ is the Lorentz boost of the electron and $E$ the electric field strength of the laser pulse. In the context of electron-laser collisions it is convenient to also define the linear quantum parameter $\eta = \chi / \xi$ which is proportional to the momentum of the electron.

\paragraph*{Electron-laser interaction}

So far, tests of strong-field QED by colliding electrons with a high-intensity laser have only been conducted with the E-144 experiment at SLAC. They collided electrons with an energy of \SI{46.6}{GeV} with a \SI{1}{TW} laser, resulting in a value of $\chi=0.3$. They observed a total of $106 \pm 14$ positrons from the non-linear Breit-Wheeler process~\cite{Burke:1997ew}. Currently, the E-320 experiment at SLAC is testing strong-field QED with the FACET-II electron beam with an energy of \SI{10}{GeV} and \SI{10}{TW} available laser power~\cite{Reis:2024ep}. At DESY, the LUXE experiment is planned to be measuring in the near future at the \SI{16.5}{GeV} electron beam of the European XFEL and also \SI{10}{TW} available laser power~\cite{Abramowicz:2021zja, LUXE:2023crk}. Both experiments are considering laser upgrades in a second phase. 

At the point when a linear collider facility becomes operational, advances in high-power laser systems will likely result in the availability of laser systems capable of delivering \SI{100}{PW} or more~\cite{Danson:2019pec}. Electron-laser collisions could be performed at a dedicated facility in the tune-up dump area, described in Sec.~\ref{sec:beyond:tunedump}. 

Various measurements are of particular interest. With a slow-extraction beamline, the elementary electron-photon interaction can be investigated by colliding single electrons with a laser pulse at normal incidence~\cite{Blackburn:2018tsn}. Changing the configuration to head-on collisions would allow to measure the coherent electron-positron plasma creation~\cite{Qu:2020gpy} and with an electron bunch instead of single electrons the incoherent plasma creation can be tested, which is of astrophysical interest as such strong fields can appear in magnetars, neutron stars, and black holes~\cite{Lai:2014nma, Kim:2019joy}. These cases have been discussed in the ILC report to Snowmass 2021~\cite{ILCInternationalDevelopmentTeam:2022izu}. 

\begin{figure}
   \centering  
\includegraphics[width=0.8\linewidth]{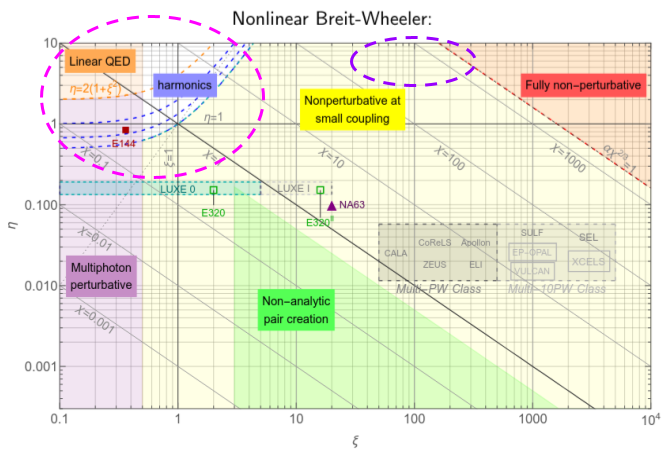}
    \caption{Phase space of the non-linear Breit-Wheeler effect, adapted from~\cite{Fedotov:2022ely}, defined by the classical non-linearity parameter $\xi$ and the linear quantum parameter $\eta$. The different regimes are indicated by the coloured areas. Two particularly interesting regimes that can be tested at a linear collider facility are the harmonic structure of the Breit-Wheeler process and the transition into the fully non-perturbative regime with no reliable QED theory, indicated by the pink and purple dashed lines, respectively. }
    \label{fig:strongFieldQEDLandscape}
\end{figure}

A linear collider facility has two peculiarities that makes it particularly suited to test strong-field QED. The high degree of electron polarisation of the electron beam of more than \SI{80}{\%} allows for the measurements of the spin-dependent rates of the strong-field QED processes~\cite{Seipt:2020diz}. Furthermore, future energy upgrades can enable a beam energy of $\mathcal{O}$(\SI{1}{TeV}). This will allow to thoroughly test the Breit-Wheeler harmonic structure~\cite{Blackburn:2021cuq, Seipt:2024tim} at low laser intensity and delve into the fully non-perturbative regime where there is no reliable QED theory at all. Both cases are indicated as dashed ellipses in Fig.~\ref{fig:strongFieldQEDLandscape} which shows the different regimes of the strong-field QED landscape of the Breit-Wheeler process. 

\paragraph*{Beam-beam interaction}

Beamstrahlung is a relevant process in future electron-positron colliders that is usually tried to be minimised. The particles in each beam radiate photons due to interaction with the electromagnetic fields generated by the opposite beam~\cite{Yokoya:1991qz, Yokoya:2000bv}. In future Higgs factories the effects are in the linear regime, and the radiation is mitigated by having very flat beams. At higher beam energy, such as in a future energy upgrade of a linear collider facility, the beam-beam effects will become more significant because of the non-linear effects. This is especially the case at energy of \SI{1}{TeV} and above, where more than \SI{10}{\%} of the energy is lost~\cite{Adolphsen:2013kya}. This was studied for CLIC at CERN for energies up to \SI{5}{TeV}~\cite{Schulte:1999xb} and for future ultra-high energy linear colliders for energies up to \SI{15}{TeV}~\cite{Barklow:2023iav}. However, the standard tools of the community to model beam-beam interaction, i.e.\ GUINEA-PIG~\cite{Schulte:1997nga} and CAIN~\cite{Chen:1994jt}, are not believed to accurately model all physics processes. Therefore, they should be tested and validated at low energy where the tools are reliable and possibly extended to properly model the processes at higher energy. 

The beamstrahlung is characterised by the beamstrahlung parameter
\begin{equation}
    \Upsilon_\mathrm{avg} = \frac{5 r_{\Pe}^2 \gamma N_{\Pe}}{6 \alpha \sigma_z (\sigma_x + \sigma_y)} \ ,
\end{equation}
where $r_{\Pe}$ is the classical electron radius, $N_{\Pe}$ the number of electrons, $\alpha$ the fine structure constant, and $\sigma$ is the Gaussian sigma in the corresponding direction. From a strong-field QED perspective, this is the average quantum non-linearity parameter that the particles experience in the collision $\Upsilon_\mathrm{avg} \equiv \chi$. For a systematic investigation, various parameters can be tuned, such as the energy, the shape of the beam, and also the misalignment/offset between the two beam axes~\cite{Filipovic:2021xge}. Such tests could be done at the second interaction point of the linear collider facility. The main diagnostics required is a detection system in the forward direction for the produced photons and, at higher energy, the coherent electron-positron pairs. Detailed studies are required. 

The fully non-perturbative regime of strong-field QED, indicated in Fig.~\ref{fig:strongFieldQEDLandscape}, can also be tested in the beam-beam interaction with highly focused beams of energy above \SI{100}{GeV}. This approach may even be more easily accessible since it does not require a high-power laser facility. Simulations showed that in particular with slightly misaligned beams, very high values of $\chi \gtrsim 1000$ can be probed~\cite{Yakimenko:2018kih, Filipovic:2021xge}.

\section{A Linear Collider Facility -- baseline and roadmap}
\label{sec:acc}
This section takes a fresh look at various accelerator concepts and technologies proposed for linear colliders, aiming to develop a coherent picture of providing $\ee$ collisions at the earliest possible date, while remaining open to the many exciting R\&D ideas which could gain maturity on short-, medium- and long-term timescales.

The timing here is important. 
The next collider should plan for data-taking on a schedule that will motivate young scientists involved in the LHC programme to participate in the design and construction of Higgs factory detectors in parallel with late-stage HL-LHC data analysis. 
Delay and uncertainty will discourage young people and sap the vitality of collider physics. On the other hand, our field must be able to react in a flexible way to new scientific developments, be it from HL-LHC, Belle-II, other ongoing experiments ---  or the first stage of a linear collider itself.

Over the last year, the LCVision team has developed a compelling scenario to address this apparent contradiction: 
A Linear Collider Facility (LCF), with a first stage that is affordable and based on the technology which promises earliest realisation, implemented within an underground structure compatible with upgrading the facility with other, more advanced technologies in future. 
Following this guiding paradigm, this section is structured as follows: 
We will first discuss the two most mature technologies, which can be considered for a first stage, namely ``ILC-like'' superconducting radio frequency (SCRF) (Sec~\ref{sec:acc:SCRFbase}) and ``CLIC-like'' warm copper (Sec~\ref{sec:acc:CLICbase}) accelerating structures. 
These sections heavily rely on the mature designs of the ILC~\cite{ILCInternationalDevelopmentTeam:2022izu, Evans:2017rvt, Adolphsen:2013kya} and CLIC~\cite{Brunner:2022usy, CLIC:2016zwp} projects which also have been submitted to the ESPPU~\cite{ILC-EPPSU:2025, CLIC-EPPSU:2025, LCF:EPPSU}. 
Currently the civil engineering studies for a linear collider at CERN are compatible with both starting options described in this document, as they use the same footprint. 
Linear collider proposals have often been criticised for offering only one interaction region, usually a choice made for cost reasons. 
We will address this issue by discussing key aspects of the beam delivery system (BDS) in Sec.~\ref{sec:acc:BDS}, in particular covering the existing designs with a double BDS catering two interaction regions and how they would need to be optimised for a linear collider facility. 
Section~\ref{sec:acc:upgrade}  discusses opportunities to upgrade an initial linear collider facility to higher energies by exploiting technologies under development today, including advanced super-conducting RF, cool and warm copper RF as well as plasma wake-field acceleration. 
Upgrade opportunities leading to higher luminosity will be discussed in Sec.~\ref{sec:acc:lumiup} ---  spanning from straightforward upgrades to the minimal ILC scenario to concepts for energy and particle recovery.  
Section~\ref{sec:acc:scenarios} introduces some examples for possible operation and upgrade timelines for Linear Colliders, followed by a discussion of  alternative collider modes like in particular $\PGg\PGg$ collisions in Sec.~\ref{sec:acc:altmodes}. 

Opportunities for beyond-collider extensions of the facility will be discussed in Sec.~\ref{sec:beyond}.


\subsection{First stage based on superconducting radio-frequency (SCRF) cavities}
\label{sec:acc:SCRFbase}

A linear collider based on SCRF has been studied in great detail in the context of the International Linear Collider (ILC), as documented in a Technical Design Report~\cite{Behnke:2013xla, ILC:2013jhg, Adolphsen:2013jya, Adolphsen:2013kya, Behnke:2013lya} and several updates since~\cite{ILCInternationalDevelopmentTeam:2022izu, Evans:2017rvt}. 
The ILC TDR considered three different sample sites for construction, at Fermilab, at CERN and in Japan. Following the discovery of the Higgs boson at the LHC~\cite{ATLAS:2012yve, Bhat:2012zj} in 2012, the Japanese HEP community expressed interest to host the ILC in Japan, and siting, civil construction and costing studies focused on a minimal version of the ILC in Japan~\cite{Evans:2017rvt}. 
In the last years, the ILC design has been guided by the International Development Team (IDT~\cite{IDT_mandate}), and important R\&D aspects have been advanced by the International Technology Network (ITN~\cite{ITN}). 
In particular for the ESPPU, the cost estimate for the ILC in Japan has been updated~\cite{Dugan:2025utp}. 
For about 2/3 of the costs, the update is based on new quotes from industry, whereas the remaining items have been escalated to 2024 costs.
The cost update as well as the work of the ITN are summarised in the IDT's submission to the ESPPU~\cite{ILC-EPPSU:2025}.

Here, we discuss a SCRF-based linear collider in a more general context, for a project anywhere in the world, including opportunities to modernise the ILC design. 
We will refer to such a generalised and modernised SCRF-based linear collider as Linear Collider Facility (LCF). 
The underlying TESLA SCRF technology is by now well-established with several free electron laser facilities in operation (FLASH and European XFEL in Hamburg, Germany and LCLS-II at SLAC, USA) or under construction (energy upgrade of LCLS-II at SLAC and SHINE in Shanghai, China). 
We summarise the experience with cavity production for the European XFEL as part of Sec.~\ref{sec:acc:SCRFbase:specs}, and discuss the experience from its operation in Sec.~\ref{sec:acc:SCRFbase:XFEL}, both underpinning the maturity and reliability of the technology. 

For an SCRF linear collider, the two main baseline options under consideration are 1) a cost-minimising option and 2) a day-1 operation at \SI{550}{GeV}. 
The first option has a centre-of-mass energy reach of \SI{250}{GeV} within \SI{20.5}{km}, following~\cite{ILCInternationalDevelopmentTeam:2022izu}. 
The second option, retained in~\cite{LCF:EPPSU} is similar to a collider presented in the ILC TDR~\cite{Behnke:2013xla, ILC:2013jhg, Adolphsen:2013jya, Adolphsen:2013kya, Behnke:2013lya} but with a slightly higher centre-of-mass energy of \SI{550}{GeV} fitting into a \SI{33.5}{km} long site. 
Both options can accommodate two interaction points as will be discussed in section~\ref{sec:2IPs}. 
An option of a \SI{27}{km} collider with a centre-of-mass energy between \SI{350}{} and \SI{380}{GeV} could be considered, however is not favoured from the point of view of an optimal science programme (apart from the threshold mass determination, most top quark measurements achieve much better results at higher centre-of-mass energies of at least \SI{550}{GeV}, which also enable the study of $\PQt\PAQt\PH$ and $\PZ\PH\PH$ production, c.f.\ Sections~\ref{sec:Higgstop} and~\ref{sec:phys:highEHiggs}). 
If desired, a path towards the \SI{550}{GeV} collider could include construction of a full-length tunnel and other civil infrastructure but with an initial installation of linacs sufficient to run at a centre-of-mass energy of \SI{250}{GeV} only for the first physics runs. 
Subsequent installation of additional SCRF cryomodules and other accelerator hardware would increase the machine energy to the full \SI{550}{GeV} centre-of-mass energy, or beyond.

Constructing the \SI{33.5}{km} full length tunnel required for the desired final energy (of \SI{550}{GeV}) provides attractive features at reasonable additional cost: 
It avoids the complications and overheads created by excavating more tunnel at a later stage.
In a geology that lends itself to the use of a tunnel boring machine (e.g., in the Geneva area), drilling a full length tunnel in a single go is considerably more economic than inserting and extracting a tunnel boring machine again at a later time. 
Cryomodules can be installed in several shorter shutdowns as they are being produced, avoiding very long interruptions of the physics programme. 
In addition, the relocation of the turn-around arcs, the spin rotators and the bunch compressors, with a total length of about \SI{1.2}{km} on each side, after a tunnel extension  can be avoided.

Even the \SI{20.5}{km} long tunnel includes some reserve length, which is partially necessitated by a timing constraint between damping ring circumference and main linac length. 
Such a reserve is an important risk mitigation, as it would permit, if necessary, to install more cryomodules and thus reduce the gradient requirements, should producing cavities at the design gradient turn out to be economically less advantageous than producing and installing more cavities at slightly lower gradient.

\begin{table}[!htb]
\centering

\begin{tabular}{l l c c c c}
\toprule
Parameter & Unit &   ILC250 & ILC500 & LCF250 & LCF550\\
\midrule
Centre-of-mass energy & GeV & 250 & 500  & 250 & 550\\
Luminosity & $10^{34}$cm$^{-2}$s$^{-1}$ & 1.35/2.7/5.4 & 1.8/3.6 & 2.7/5.4  & 3.9/7.7\\
Polarisation, $P(e^-) \big/ P(e^+)$ &  \% & 80 / 30 & 80 / 30 & 80 / 30  & 80 / 60 \\ \hline 
Number of interaction points &  & 1 & 1 & 2 & 2 \\
Repetition frequency & Hz & 5/5/10 & 5 & 10 & 10 \\
Number of bunches per train & & 1312/2625/2625 & 1312/2625  & 1312/2625 & 1312/2625\\
Bunch spacing & ns & 554/366/366 & 554/366  & 554/366 & 554/366\\
Bunch train duration & $\mu$s & 727/961/961 & 727/961 & 727/961 & 727/961 \\
Cavity quality factor & $10^{10}$ & 1 & 1  & 2 & 2 \\
Klystron efficiency & \% & 65 & 65 & 80 & 80 \\
\midrule
Bunch population & $10^{10}$ & 2 & 2 & 2 & 2 \\
Number of particles per dump & $10^{21}$yr$^{-1}$ & 1.57/3.15/6.3 & 1.57/3.15  & 1.57/3.15 & 1.57/3.15\\
Accelerating gradient & MV/m & 31.5 & 31.5  & 31.5 & 31.5\\
\midrule
Length of 2 SCRF linacs & km & 10 & 22.3  & 10 & 24.1\\
Total facility length & km & 20.5 & 33.5 & 33.5 & 33.5 \\
Site power consumption & MW & 111/138/198 & 173/215 &  143/182 & 250/322\\ 
\bottomrule
\end{tabular}
\caption{Key parameters for the updated superconducting Linear Collider Facility (LCF) compared to the ILC baseline options. Values for ILC250 and ILC500 are taken from Table~4.1 in~\cite{ILCInternationalDevelopmentTeam:2022izu}.
\label{tab:ILCbaseline}
}
\end{table}

Previously reported ILC baseline options assumed that SCRF linacs operate at an accelerating gradient of \SI{31.5}{MV/m}, see~\cite{Behnke:2013xla, ILC:2013jhg, Adolphsen:2013jya, Adolphsen:2013kya, Behnke:2013lya}. 
The worldwide R\&D efforts continue to improve performance of SCRF cavities and efficiency of klystrons. 
Some of the recent advances can benefit the baseline of a SCRF linear collider already and are discussed in section~\ref{sec:acc:SCRFbase:specs}. 
Taking into account the effect of an improved SCRF cavity performance with a quality factor of \num{2E10} at \SI{31.5}{MV/m} and a klystron efficiency of \SI{80}{\%}, Table~\ref{tab:ILCbaseline} shows the main parameters for an updated ILC-like Linear Collider Facility in comparison to the ILC itself. 
Notable differences comprise:
\begin{description}
    \item[ Centre-of-mass energy:] the LCF targets operation at \SI{550}{GeV} instead of \SI{500}{GeV} for a much better precision on the top quark Yukawa coupling. 
    Note that this does not increase the overall facility length as already the ILC main linac tunnel provides a sufficient length.
    \item[ Repetition rate and instantaneous luminosity:] the LCF proposes to operate at a repetition rate of \SI{10}{Hz}, i.e.\ twice the repetition rate, at full gradient. 
    For ILC, \SI{10}{Hz} operation has only been considered when operating the physical \SI{500}{GeV} machine at \SI{250}{GeV}, i.e.\ when operating the main linacs at half their maximal gradient. 
    Due to the improved $Q_0$ and the higher klystron efficiency of a modernised design, the impact of the \SI{10}{Hz} operation at full gradient on the total site power is visible, but not outrageous, and significantly improves the luminosity-to-power ratio. 
    For instance at \SI{550}{GeV}, the power increase by about \SI{50}{\%} more than doubles the luminosity. For more information on the luminosity upgrade options we refer to Sec.~\ref{sec:acc:lumiup:benno}.
    \item[ Beam polarisation:] for the \SI{500}{GeV} ILC, an upgrade of the positron polarisation from \SI{30}{\%} to \SI{60}{\%} was foreseen in the TDR~\cite{Adolphsen:2013kya} by installing a longer helical undulator and a collimation system. 
    Due to more challenging positron production with a lower electron drive-beam energy, this was never considered for the \SI{250}{GeV} ILC. 
    Since the ILC TDR, there has been significant progress both in terms of the design of the positron source (c.f.~\cite{Moortgat-Pick:2024fcy} for a recent summary) as well as in terms of operation experience with long undulator systems, e.g.\ at the European XFEL. 
    Thus, after several years of experience with operating the positron source in the more challenging \SI{250}{GeV} configuration, the positron polarisation can be increased to $|P(e^+)|=60\%$ for the \SI{550}{GeV} stage.
    \item[Number of interaction points:] the LCF foresees two interaction points, while the original second interaction point was removed from the ILC design during the TDR phase in order to reduce the construction cost.
    \item[Number of particles per dump:] one of the advantages of a linear collider facility is the continuous dumping of the beam, which can be used in a fixed-target experiment to search for new physics. We assumed an uptime of the facility of $\SI{1.2E7}{s/yr} \approx 38\%$. This topic is described in detail in Secs.~\ref{sec:phys:PBC:beamdumps} and~\ref{sec:beyond:maindump}.
    \item[Total facility length:] for the LCF, a \SI{33.5}{km} tunnel is foreseen from the beginning. 
    For an initial \SI{250}{GeV} stage, the turn-arounds between the ring to main linac system (RTML) and the main linac (ML) will be built at the ends of the facility, and the outer half of the ML will be equipped with cryomodules, followed by a transport line (e.g.\ with a simple FODO lattice) to the beginning of the BDS. 
    While this adds about \SI{0.65}{BCHF} to the initial construction cost, such a scenario enables a seamless upgrade to \SI{550}{GeV}. 
    Given sufficient resources, cryomodule production could continue after the installation of the \SI{250}{GeV} machine, and cryomodules could be added in the annual or bi-annual shutdowns as they become available, or stored and installed in one longer shutdown. 
    Alternatively, one can consider a scenario in which a host country or region can decide to move ahead with the construction of the \SI{250}{GeV} machine (in a \SI{33.5}{km} tunnel) independently of ---  or only with a small number of ---  other contributors. 
    Should, after a positive decision, other parties decide to join the project, providing additional cryomodules produced in  local industry as in-kind contribution offers a potentially attractive route, directly augmenting the energy reach of the facility. 
    In any case the fact that no additional civil construction will be needed for the upgrade to \SI{550}{GeV} is considered a decisive advantage. 
    \item[More efficient components:] given the necessarily longer preparation time for a new site in the Geneva area, more ambitious R\&D goals for klystron efficiency (\SI{80}{\%} instead of \SI{65}{\%}) and cavity quality factor ($Q_0=\num{2E10}$ instead of \num{1E10}) have been formulated, based on recent technological advances.
\end{description}

While Table~\ref{tab:ILCbaseline} focuses on \num{250} and \num{500}/\SI{550}{GeV}, a SCRF linear collider can always be operated at lower energies, for instance at the $\PZ$ pole or at the $\PW\PW$ and $\PQt\PAQt$ production thresholds as physics demands. 
The overall accelerator parameters for these lower energies have not been optimised in as great detail as for the main energies yet. 
Some recent, preliminary ideas on how to improve the luminosity at the $\PZ$ pole and $\PW\PW$ threshold quite significantly will be discussed in Sec.~\ref{sec:acc:lumiup:zpole}.

\subsubsection{SCRF specifications for the baseline}
\label{sec:acc:SCRFbase:specs}

The ILC TDR~\cite{Behnke:2013xla, ILC:2013jhg, Adolphsen:2013jya, Adolphsen:2013kya, Behnke:2013lya} specifies an accelerating gradient of \SI{31.5}{MV/m} and an intrinsic quality factor $Q_0$ of \num{1E10}. 
This level of performance is by now firmly established, with recent R\&D results pointing towards the possibility of increased performance, potentially combined with lower production costs, through novel surface treatments.

A statistical model, based on the European XFEL data, which accurately includes the gradient increase due to re-treatment of low performing cavities, shows that the ILC specification of an yield of \SI{94}{\%} at \SI{35}{MV/m} for the maximum gradient can be achieved after re-treatment of underperforming cavities.

Taking into account other limitations (field emission and low quality factor), the usable gradient is lower, \SI{33.4}{MV/m} with a \num{82}(\num{91})\SI{}{\%} yield after one (two) re-treatments. 
The re-treatment would mostly be limited to a simple high-pressure rinse (HPR) rather than a more expensive electropolishing step. 
Thus, with a reasonable extrapolation, the European XFEL cavity production data demonstrate that it is possible to mass-produce cavities in industry meeting the ILC TDR specifications with the required yield~\cite{Walker:LCWS2017}.

Further, the ILC-like performance with beam was demonstrated in two laboratories. 
A cryomodule at Fermilab has reached total accelerating voltage of \SI{267.9}{MV}, equivalent to an average gradient of \SI{32.2}{MV/m} for eight cavities, thus exceeding the ILC goal. 
A beam with \SI{3.2}{nC} bunch charge has obtained an energy gain of $>$\,\SI{255}{MeV} after passing through the cryomodule~\cite{FermilabCM}. 
At KEK, nine (including one nitrogen-infused cavity) out of twelve cavities in a demonstration cryomodule at the STF test facility exceeded the ILC specification, achieving an average accelerating gradient of \SI{33}{MV/m} during beam operation~\cite{KEK_STF_CM}. 
Three of these cavities exceeded \SI{36}{MV/m}.

The results discussed above are obtained with the SCRF cavities subjected to the ILC-TDR cavity processing recipe, which include electrochemical polishing (EP), outgassing at $800^\circ$C in a furnace under high vacuum, HPR, and vacuum baking at $120^\circ$C. 
The cost estimate for ILC, for the TDR and the recent update for the ESPPU~\cite{ILC-EPPSU:2025}, is based on quotes from companies around the world which have demonstrated that they can produce cavities following the ILC recipe.

Recent advancements in surface tailoring techniques have led to groundbreaking results, as laboratories have successfully introduced nitrogen, oxygen, and other impurities into the niobium lattice. 
These innovative procedures have yielded increasing efficiency of cavities by a factor of 2 to 3 at medium accelerating fields, \num{20} to \SI{30}{MV/m}. 
The best experiments employing medium temperature, or mid-T, vacuum bake produced \SI{1.3}{GHz} SCRF cavities with a quality factor of \num{5E10} at \SI{30}{MV/m} and \SI{2}{K}~\cite{Posen_mid-T_baking,Pan:PRAB2024}. 
The world's first SCRF cryomodule with nine \SI{1.3}{GHz} mid-T baked cavities was developed and successfully tested at IHEP, China~\cite{Pan:PRAB2024}.

\begin{figure}[htb]
    \centering
    \includegraphics[width=0.75\hsize]{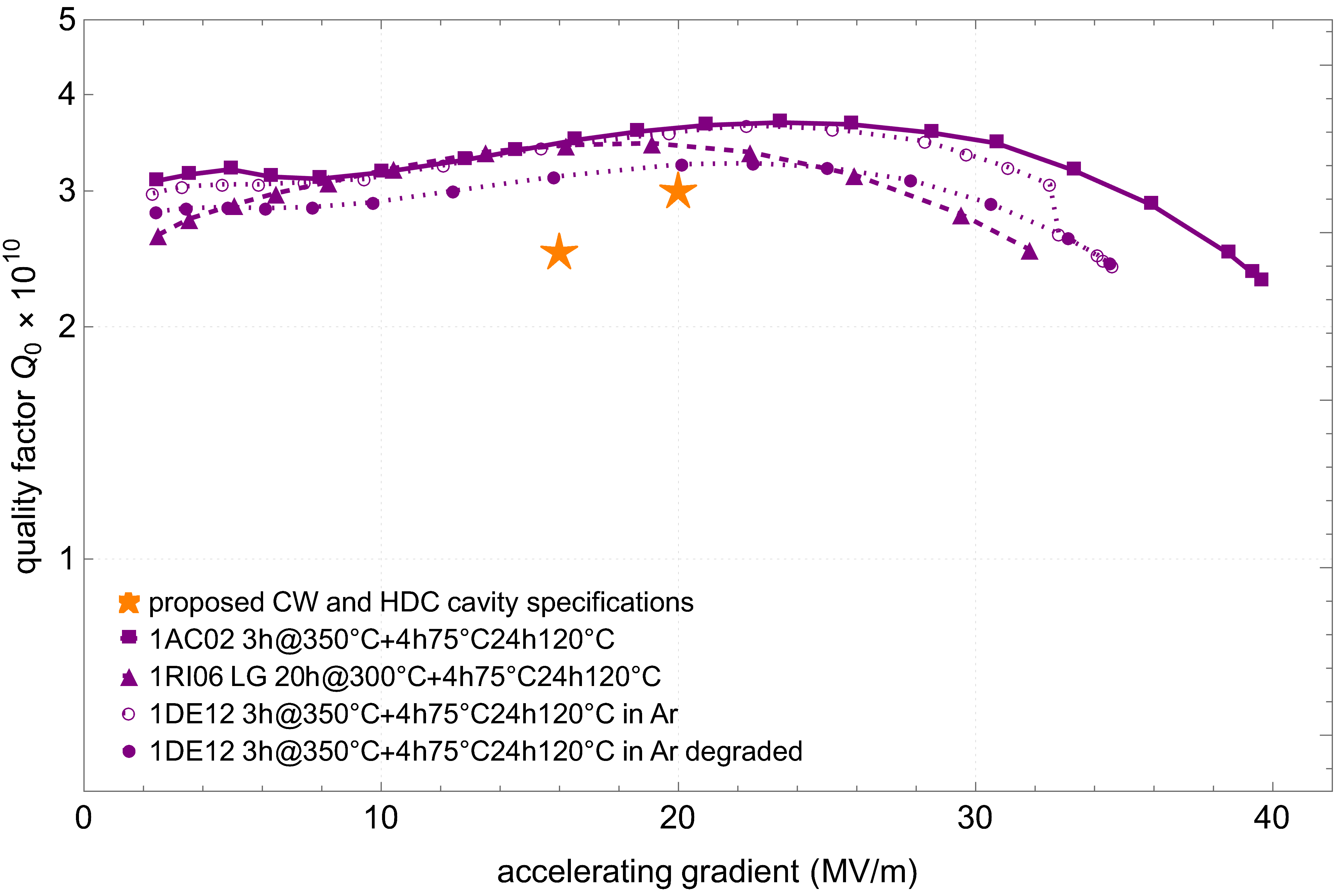}
    \caption{$Q_0 (E_{acc})$ curves for three single-cell cavities at 2~K after the mid-T--low-T heat treatment combination. The stars indicate the two suggested specification points for EuXFEL upgrade cavities~\cite{Steder}. \label{fig:QvsE}    
}
\end{figure}

On the high-gradient side, new advances pushed accelerating gradients towards \SI{50}{MV/m} in vertical tests of single-cell TESLA cavities, and demonstrated that with some additional R\&D it should be practically achievable to operate TESLA cavities in cryomodules at $>$\,\SI{35}{MV/m} while maintaining $Q_0>\num{2E10}$. 
Examples of the new recipes are 1) a combination of cold electropolishing and two-step baking (75/120$^\circ$C)~\cite{2-step_baking} and 2) medium temperature (mid-T) baking followed by low-T baking~\cite{Steder}. 
Figure~\ref{fig:QvsE} shows curves of an intrinsic quality factor vs.\ accelerating gradient for three \SI{1.3}{GHz} cavities after a combination of mid-T and low-T treatment. 
The performances of all three cavities are outstanding and very promising, indicating that large accelerating gradients accompanied by high quality factors are in reach. Following a short-term R\&D to refine the cavity processing procedure, a well-funded industrialisation programme ($\sim$\,\SI{5}{years}) should be launched to demonstrate the required cavity production yield. 
While global availability of high-purity niobium remains a concern, there is progress with increasing SCRF cavity production capacity worldwide. Europe has traditionally been a cavity production hub with two strong companies qualified to make cavities per ILC specifications. 
With several SCRF-based accelerator projects under way, China has developed industrial capabilities to produce \SI{1.3}{GHz} cavities. 
More recently, South Korea's industry has become interested in building up such a capability as well.

In parallel to industrialising the cavity processing advances, better strategies for mitigating the effect of field emissions should be deployed in order to minimise the difference between maximal and usable gradient and maximise the yield of high gradient cavities. Examples comprise a robot-assisted automation of the cryomodule assembly and plasma processing after installation~\cite{Snowmass2021:Martinello}.

Significant technical progress has been made in improving the efficiency of high-power klystrons~\cite{Syratchev:2022a}. 
This will make L-band klystrons with $>$\,\SI{80}{\%} efficiency a reality in the near future if current R\&D efforts will continue with sufficient funding. 
Then the total RF source efficiency would be $\sim$\,\SI{70}{\%}, taking into account losses in a high-voltage modulator and waveguide distribution system. 
Another area where short-term R\&D would greatly benefit the ILC SCRF performance is the waveguide distribution system. 
To avoid an issue when one low-performing cavity in an RF unit can limit the performance of the whole unit, a flexible, remotely-controlled, low-loss design must be developed.

None of these recent advances have been incorporated in the ILC baseline, which ---  in terms of SCRF specifications ---  remained effectively frozen since the TDR. 
Today, about 12 years later, it is time to modernise the design. A most conservative update maintains the ILC average gradient of \SI{31.5}{MV/m}, but assumes a higher quality factor $Q_0$ of \num{2E10}, to be achieved e.g.\ via the mid-T--low-T baking recipe discussed above. 
This recipe so far exhibits an excellent reproducibility, and is actually a slight simplification compared the standard ILC recipe. 
Therefore, after a short optimisation period and industrial test production, it is not expected to invalidate the careful cost evaluation for the ILC~\cite{ILC-EPPSU:2025}.

\subsubsection{Experience with European XFEL}
\label{sec:acc:SCRFbase:XFEL}

The \SI{1.4}{km} \SI{1.3}{GHz} superconducting linac of the European XFEL (Eu-XFEL) is the largest deployment of TESLA technology. 
With \num{100} cryomodules (\num{800} cavities) driven by a total of \num{25} klystrons, the Eu-XFEL can be considered a $\sim\SI{10}{\%}$ prototype of an ILC linac. 
The numbers in this section are based to a large extent on the online monitoring of Eu-XFEL operation; for conference summaries see e.g.~\cite{Branlard:2022gys, Schmidt:2023cqp,Branlard:2024ukm}.  
In operation since 2018, the linac was turned on at \SI{14}{GeV} beam energy, which was then ramped up to the Eu-XFEL maximum design energy of \SI{17.5}{GeV} within the first 18 months of operation. 
The maximum average cavity gradient is \SI{23.4}{MV/m}, with several cavities operating close to \SI{30}{MV/m}. 
The accelerator now routinely operates for photon users with very high availability, accelerating up to 2700 bunches per \SI{10}{Hz} pulse, using an RF pulse structure very similar to ILC. 
Typical  pulse beam currents are \SI{0.24}{-}\SI{1.0}{mA}, with a demonstration of \SI{4.5}{mA}  planned in the near future. 
The \SI{10}{MW} multi-beam klystrons (originally developed for linear collider application) show excellent and robust behaviour (albeit running at typical peak powers of \SI{4}{-}\SI{7}{MW}), as do the solid-state pulse modulators. 
Operationally the linac systems have accumulated close to 40\,000 hours, which is the specified lifetime of the klystrons. 
To date only four klystrons have failed and needed to be exchanged, and only one of these could be considered an end-of-life event. 
The state-of-the-art low-level RF (LLRF) and related controls systems facilitate a high degree of automation and almost turn-key operation. 
Specifically, the LLRF feedback stabilises the electron beam energy to the level of \num{E-4} or better. 
The Eu-XFEL has also demonstrated that cavities can be routinely and robustly operated at gradients within \SI{0.5}{MV/m} of their quench limits over very long periods of time, with quench-related RF trips being quite rare. 
This together with the high beam energy stability is aided by the sophisticated piezo control systems which actively keep the cavities tuned on resonance to within \SI{15}{Hz}.

The exceptionally high availability is in part achieved by sophisticated RF trip recovery which recovers over \SI{50}{\%} of the RF trips with \num{100} seconds. 
The mean time between trips is typically $\sim$12 hours (averaged over a half-year user run), although entire weeks of operation have been observed trip free (\SI{100}{\%} availability).  
The \SI{2}{K} liquid-helium cryoplant in general runs routinely and provides significantly better than specified liquid helium pressure stability at the cavities. 
However, cryoplant failures are a significant source of downtime.
Although trips are rare (few per year), recovery of the liquid helium stability generally takes between \num{8} and \num{48} hours, depending on the nature of the fault. 
Fortunately due to the development of new technology (magnetic bearings in the cold compressors), the longer downtimes (resulting from the need to warm up the the cold compressors) are becoming less frequent, and there is good indication that this particular issue has been resolved. 
Even including these long downtimes, the linac availability over an entire six-month user run routinely achieves \SI{98}{\%}. 


\subsection{First stage based on drive-beam technology}
\label{sec:acc:CLICbase}

CLIC is proposed as a linear collider option at CERN~\cite{Adli:ESU25RDR} and it can also be a starting point for an upgradable linear collider facility as discussed in this document. 
CLIC as proposed would start at \SI{380}{GeV}  and can be upgraded to \SI{1.5}{TeV} in 29 km.
In a 33 km facility one could reach very close to \SI{2}{TeV}. 
In all cases one drive beam is sufficient to provide the RF power needed.
In the following section the main features and parameters of a CLIC technology based accelerator complex are summarised.


\subsubsection{Design overview}

\begin{center}
\begin{figure}[!th]
\includegraphics[width=\textwidth]{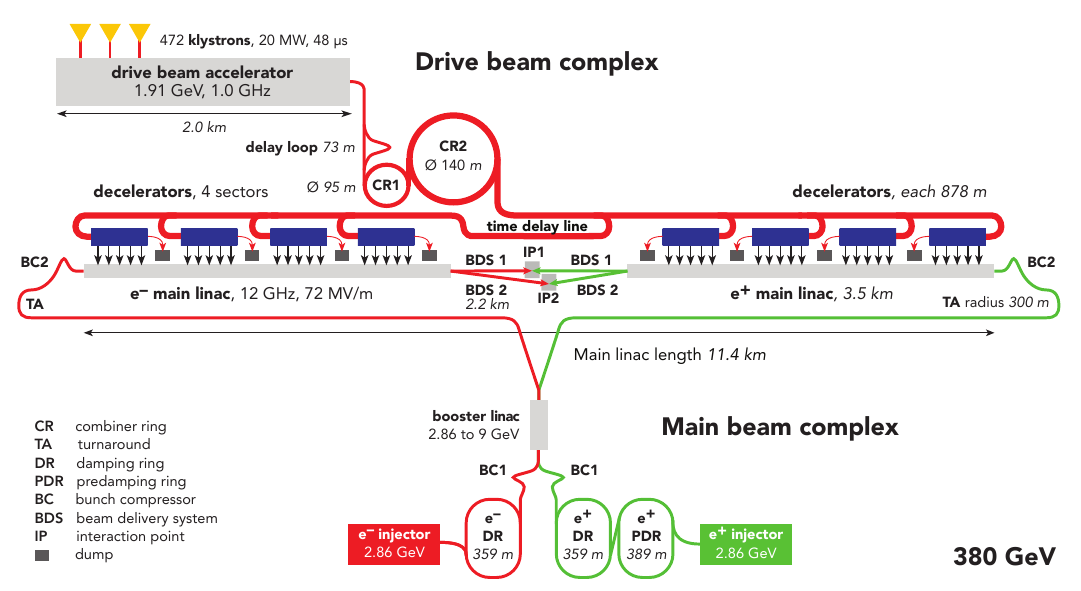}
\caption{Schematic layout of the CLIC complex at \SI{380}{GeV}, with a double beam-delivery system and two detectors~\cite{Adli:ESU25RDR} (image credit: CLIC).}
\label{scd:clic_layout}
\end{figure}
\end{center}

The schematic layout of the baseline CLIC complex for \SI{380}{GeV} operation is shown in Fig.~\ref{scd:clic_layout} and the key parameters are listed in Table~\ref{tab:CLIC380}. 
The schematic layout of the baseline CLIC complex for \SI{380}{GeV} operation is shown in Fig.~\ref{scd:clic_layout} and the key parameters are listed in Table~\ref{tab:CLIC380}. 

\begin{table}[!ht]
\centering
\begin{tabular}{l l l l l}
\toprule
Parameter                  &   Symbol         &   Unit &    \SI{380}{GeV}, \SI{100}{Hz}   & \SI{380}{GeV}, \SI{50}{Hz}      \\
\midrule
Centre-of-mass energy               & $\sqrt{s}$              &GeV                                      & 380      & 380     \\
Number of interaction points        & $N_{\text{IP}}$                &                                         & 2      &  1      \\
Repetition frequency                & $f_{\text{rep}}$        &Hz                                       & 100      &  50      \\
Number of bunches per train         & $n_{b}$                 &                                         & 352      & 352     \\
Bunch separation                    & $\Delta\,t$             &ns                                       & 0.5      & 0.5     \\
Pulse length                        & $\tau_{\text{RF}}$      &ns                                       & 244      & 244     \\
\midrule
Accelerating gradient               & $G$                     & MV/m                                     & 72       & 72     \\
\midrule
Total luminosity                    & $\mathcal{L}$           & \SI{e34}{cm^{-2}s^{-1}}              & 4.5      & 2.3     \\
Luminosity above 99\% of $\sqrt{s}$ & $\mathcal{L}_{0.01}$    &\SI{e34}{cm^{-2}s^{-1}}              & 2.7      & 1.3     \\
Beamstrahlung photons per particle  & $n_{\gamma}$            &                                      & 1.5      & 1.5     \\ 
Total integrated luminosity per year& $\mathcal{L}_{\mathrm int}$ &fb$^{-1}$                                & 540      & 270     \\ 
Power consumption           & $P$                     &MW                                      & 166      & 105     \\ 
\midrule
Main linac tunnel length                  &                   &km                                       & 11.4     & 11.4     \\
Number of particles per bunch       & $N$                     &\num{e9}                                 & 5.2      & 5.2     \\
Bunch length                        & $\sigma_z$              &\SI{}{\mu m}                                   & 70       & 70     \\
IP beam size                        & $\sigma_x/\sigma_y$     &nm                                       & 149/2.0  & 149/2.0     \\
Final RMS energy spread             &                         & \SI{}{\%}                                       & 0.35      & 0.35     \\
\bottomrule
\end{tabular}
\caption{ Key parameters of the baseline machine, a drive-beam based CLIC at $\sqrt{s} =$\,\SI{380}{GeV}, running at \SI{100}{Hz}, operating with two detectors.  
Parameters for an alternative machine, using ca.\ \SI{40}{\%} lower power consumption,  running at \SI{50}{Hz} with a single beam delivery system and one detector, are also given.
The main linac crossing angle is \SI{20}{mrad}. 
The crossing angle at the IP depends on the BDS design, and the number of IPs. 
\label{tab:CLIC380}
}
\end{table}

The main electron beam is produced in a conventional radio-frequency (RF) source and accelerated to \SI{2.86}{GeV}. 
The beam emittance is then reduced in a damping ring.
To produce the positron beam, an electron beam is accelerated to \SI{5}{GeV} and sent into a crystal to produce energetic photons, which hit a second target and produce electron--positron pairs.
The positrons are captured and accelerated to \SI{2.86}{GeV}. Their beam emittance is reduced first in a pre-damping ring and then in a damping ring. 
The RTML accelerates the beams to \SI{9}{GeV} and compresses their bunch length. The main linacs accelerate the beams to the beam energy at collision of \SI{190}{GeV}.
The beam delivery system removes transverse tails and off-energy particles with collimators and compresses the beam to the small sizes required at the collision point.
After the collision the beams are transported by the post collision lines to the respective beam dumps.

The RF power for each main linac is provided by a high current, low-energy drive beam that runs parallel to the colliding beam through a sequence of power extraction and transfer structures
(PETS). The drive beam generates RF power in the PETS that is then transferred to the accelerating structures by waveguides.

The drive beam is generated in a central complex with a fundamental frequency of \SI{1}{GHz}. 
A \SI{48}{\mu s} long beam pulse is produced in the injector and fills every other bucket, i.e.\ with a bunch spacing of \SI{0.6}{m}. 
Every \SI{244}{ns}, the injector switches from filling even buckets to filling odd buckets and vice versa, creating \SI{244}{ns} long sub-pulses.
The beam is accelerated in the drive-beam linac to \SI{1.91}{GeV}. 
A \SI{0.5}{GHz} resonant RF deflector sends half of the sub-pulses through a delay loop such that its bunches can be interleaved with those of the following sub-pulse that is not delayed. This generates a sequence of \SI{244}{ns} trains in which every bucket is filled, followed by gaps of the same \SI{244}{ns} length.
In a similar fashion three of the new sub-pulses are merged in the first combiner ring. Groups of four of the new sub-pulses, now with \SI{0.1}{m} bunch distance, are then merged in the second combiner
ring. 
The final pulses are thus \SI{244}{ns} long and have a bunch spacing of \SI{2.5}{cm}, i.e.\ providing \num{24} times the initial beam current. 
The distance between the pulses has increased to $24 \times \SI{244}{ns}$, which corresponds to twice the length of a \SI{878}{m} decelerator. 
The first four sub-pulses are transported through a delay line before they are used to power one of the linacs while the next four sub-pulses are used to power the other linac directly.
The first sub-pulse feeds the first drive-beam decelerator, which runs in parallel to the colliding beam. When the sub-pulse reaches the decelerator end, the second sub-pulse has reached the beginning of the second drive-beam decelerator and will feed it, while the colliding beam has meanwhile reached the same location along the linac.

The CLIC baseline has $\pm$\SI{80}{\%} longitudinal polarisation for the electron beam, and no positron polarisation, although a polarised positron source could be considered~\cite{Liu:2010bb}. For optimal physics reach an equal sharing of the running time between the two polarisation states is proposed for the initial energy stage; and a sharing in the ratio 80:20 for $-\SI{80}{\%}$ and $+\SI{80}{\%}$ electron-beam polarisation, for centre-of-mass energies above \SI{500}{GeV} ~\cite{robson2018updatedclicluminositystaging}.

\subsubsection{Extension to higher-energy stages}
\label{sect:HE_Intro}

The CLIC \SI{380}{GeV} energy stage can be efficiently upgraded to higher energies.  This flexibility has been an integral part of the design choices for the first energy stage. With a single drive beam and running at \SI{100}{Hz}, the machine can be upgraded to \SI{550}{GeV}. 
With a single drive beam and running at \SI{50}{Hz}, the machine can be upgraded to \SI{1.5}{TeV}, and possibly to \SI{2}{TeV}. 
Parameters for a \SI{1.5}{TeV} stage a given in Table~\ref{tab:CLIC_stages:1}. 
The table also contains numbers for a \SI{250}{GeV}  and \SI{550}{GeV}  CLIC.  The power and luminosity at these energies are scalings, based on the \SI{380}{GeV} and \SI{1500}{GeV} designs.
The key parameters for the different energy stages of CLIC are given in Table~\ref{tab:CLIC_stages:1}.

\begin{table}[!htb]
\caption{Key parameters for 380\,GeV and 1.5\,TeV stages of CLIC.  Parameters for energy options at 250\,GeV and 550\,GeV are also given; for these options the power and luminosity are scalings, based on the 380\,GeV and 1.5\,TeV designs.
\newline
$^*$The luminosity for the 1.5 TeV machine has not been updated to reflect recent alignment studies~\cite{Gohil:2020tzn}. If the same method is applied, the luminosity at 1.5\,TeV is expected to reach $\SI{5.6e34}{\per\centi\meter\squared\per\second}$.
}
\label{tab:CLIC_stages:1}
\centering
\begin{tabular}{llllll}
\toprule 
Parameter  & Unit  & \textbf{380\,GeV}  &\textbf{1.5\,TeV}  & 250\,GeV  & 550\,GeV \tabularnewline
\midrule 
Centre-of-mass energy  & \si{\GeV}  & 380  & 1500  & 250  & 550 \tabularnewline
Repetition frequency  & \si{\Hz}  & 100  & 50 & 100  & 100 \tabularnewline
Nb. of bunches per train  &  & 352  & 312 & 352  & 352 \tabularnewline
Bunch separation  & \si{\ns}  & 0.5  & 0.5 & 0.5  & 0.5 \tabularnewline
Pulse length  & \si{\ns}  & 244  & 244 & 244  & 244 \tabularnewline
\midrule 
Accelerating gradient  & \si{\mega\volt/\meter}  & 72  & 72/100 & 72  & 72 \tabularnewline
\midrule 
Total luminosity  & \SI{e34}{\per\centi\meter\squared\per\second}  & 4.5  & 3.7$^{*}$  & $\sim$3.0  & $\sim$6.5 \tabularnewline
Lum. above \SI{99}{\percent} of $\sqrt{s}$  & \SI{e34}{\per\centi\meter\squared\per\second}  & 2.7  & 1.4  &  $\sim$2.1  &  $\sim$3.2 \tabularnewline
Total int. lum. per year  & fb$^{-1}$  & 540  & 444  & $\sim$350  & $\sim$780 \tabularnewline
Power consumption  & MW  & 166  & 287  & $\sim$130  & $\sim$210 \tabularnewline
\midrule 
Main linac tunnel length  & \si{\km}  & 11.4  & 29.0 & 11.4  & $\sim$15 \tabularnewline
Nb. of particles per bunch  & \num{e9}  & 5.2  & 3.7 & 5.2  & 5.2 \tabularnewline
Bunch length  & \si{\um}  & 70  & 44 & 70  & 70 \tabularnewline
IP beam size  & \si{\nm}  & 149/2.0  & 60/1.5 & $\sim$184/2.5  & $\sim$124/1.7 \tabularnewline
\bottomrule
\end{tabular}
\end{table}

In the \SI{380}{GeV} stage, the linac consists of modules that contain accelerating structures that are optimised for this energy.
At higher energies these modules are reused and new modules are added to the linac.
First, the linac tunnel is extended and a new main-beam turn-around is constructed at its new end. The technical installations in the old turn-around and the subsequent bunch compressor are then moved to this new location.
Similarly, the existing main linac installation is moved to the beginning of the new tunnel.
Finally, the new modules that are optimised for the new energy are added to the main linac. Their accelerating structures have smaller apertures
and can reach a higher gradient of \SI{100}{\mega\volt/\meter}; the increased wakefield effect is mitigated by the reduced bunch charge and length.
The beam delivery system has to be modified by installing magnets that are suited for the higher energy and it will be extended in length. The beam extraction line also has to be modified to accept the larger beam energy but the dump remains untouched.
Alternative scenarios exist. In particular one could replace the existing modules with new, higher-gradient ones; however, this would increase the cost of the upgrade. In the following only the baseline is being discussed.

The design of the first stage considers the baseline upgrade scenario from the beginning. For the luminosity target
at \SI{380}{GeV}, the resulting cost increase of the first stage is \SI{50}{MCHF} compared to the fully optimised first energy stage (without the constraints imposed by a future energy upgrade beyond \SI{380}{GeV}).
To minimise the integrated cost of all stages, the upgrades reuse the main-beam injectors and the drive-beam complex with limited modifications and reuse all main linac modules.

In order to minimise modifications to the drive-beam complex, the drive-beam current is the same at all energy stages. The existing drive-beam RF units
can therefore continue to be used without modification. In addition, the RF pulse length of the first stage is chosen to be the same as
in the subsequent energy stages. This is important since the lengths of the delay loop and the combiner rings, as well
as the spacings of the turn-around loops in the main linac, are directly proportional to the RF pulse length.
Hence, the constant RF pulse length allows the reuse of the whole drive-beam combination complex.
For the upgrade from \SI{380}{GeV} to 1.5\,TeV, only minor modifications are required for the drive-beam production complex.
The drive-beam accelerator pulse length is increased in order to feed all of the new decelerators, and also its beam energy is increased by \SI{20}{\percent}. The energy increase is achieved by adding more drive-beam modules.
The pulse length increase is achieved by increasing the stored energy in the modulators to produce longer pulses.
The klystron parameters in the first energy stage have been chosen to be compatible with the operation using longer
pulses and higher average power. The remainder of the drive-beam complex remains unchanged, except that all magnets after the drive-beam linac need to operate at a \SI{20}{\percent} larger field, which is also foreseen in the magnet design. 

The impact of the upgrades on the main-beam complex has also been minimised by design.
The bunches of the main-beam pulses have the same spacing at all energy stages, while at higher energies the number of bunches per train and
their charge is smaller. Therefore the main linac modules of the first stage can accelerate the trains of the second and third stage without modification. 
Since the drive-beam current does not change, also the powering of the modules is the same at all energies.
The upgrade to \SI{1.5}{TeV} requires an additional nine decelerator stages per side and the \SI{3}{TeV} needs another \num{12}.  An upgrade from \SI{380}{GeV} to \SI{550}{GeV} requires an additional two decelerator stages per side.

Still some modifications are required in the main-beam complex. The injectors need to produce fewer bunches with a smaller charge
than before, but a smaller horizontal emittance and bunch length
is required at the start of the main linac. The smaller beam current requires less RF, so the klystrons can be operated at lower power and the emittance growth due to collective effects
will be reduced. The smaller horizontal emittance is mainly achieved by some adjustment of the damping rings.
The reduction of the collective effects that result from the lower bunch charge will allow to reach the new value with the same risk as in the first energy stage.

The preservation of the beam quality in the main linac is slightly more challenging at the higher energies. 
However, the specifications for the performance of alignment and stabilisation systems for the \SI{380}{GeV} stage are based on the requirements for the \SI{3}{TeV} stage. 
They are therefore sufficient for the high energy stages and no upgrades of these systems are required.

The collimation system is longer at \SI{1.5}{TeV} and \SI{3}{TeV} to ensure the collimator survival at the higher beam energies. Similarly the final focus system is slightly longer to limit the
amount of synchrotron radiation and emittance degradation in the indispensable bending of the beams. The systems have to be re-built using higher field magnets. However, the integration into the
existing tunnel is possible by design. The extraction line that guides the beams from the detector to the beam dump will also need to be equipped with new magnets.

\subsection{Beam delivery system considerations}
\label{sec:acc:BDS}

The beam delivery system (BDS) is the final part of a linear collider which transports the high-energy beam from the linac to the interaction point (IP), and from there to the dump. The primary tasks of the BDS are the focusing of the high-energy particle beams to nanometer sizes at the IP for collision followed by their disposal in the beam dumps. Key challenges include the management of the beam halo for background control (collimation), the mitigation of synchrotron radiation (SR) and beamstrahlung effects, and the provision of sophisticated beam instrumentation for diagnostics and tuning to achieve high luminosity.
From the technical point of view, the beam delivery system is among the most complex systems of a linear collider. 

In the following, we will discuss some of the key aspects to be considered for the BDS of a linear collider facility: the compatibility with energy upgrades, the choice of crossing angle(s), as well as the challenges of double-BDS configurations. To illustrate these aspects, the choices made for the design of the ILC and CLIC beam delivery systems provide important examples.

\subsubsection{Compatibility with energy upgrades: the ILC as example}
\begin{figure}[htb]
    \centering
    \begin{subfigure}{0.49\textwidth}
        \includegraphics[width=\textwidth]{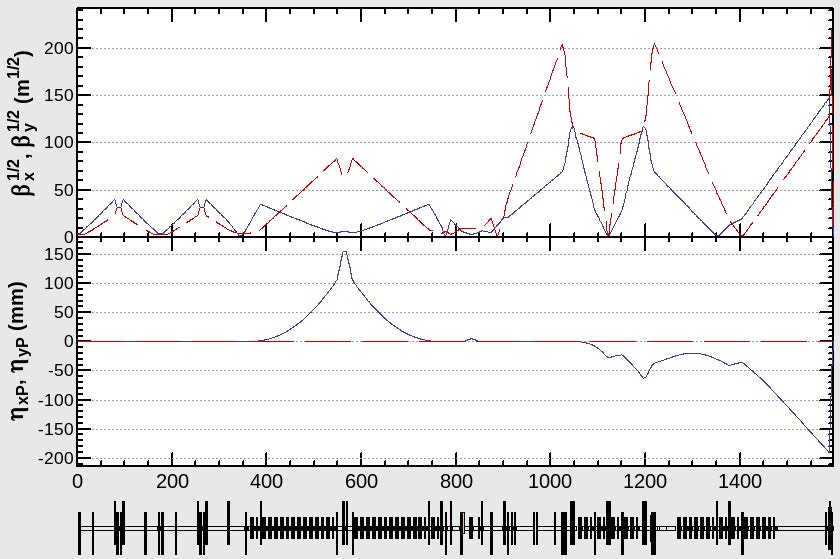}
        \caption{}
        \label{fig:ILC-BDS:250}    
    \end{subfigure}
    \begin{subfigure}{0.49\textwidth}
        \includegraphics[width=\textwidth]{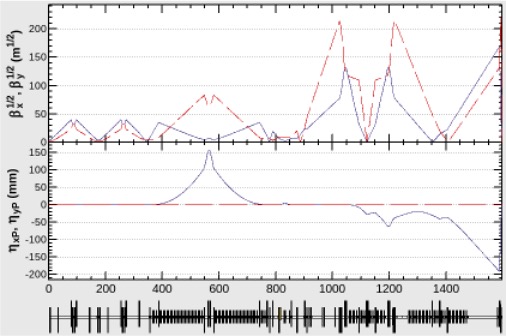}
        \caption{}
        \label{fig:ILC-BDS:500}    
    \end{subfigure}
\caption{ILC BDS optics design~\cite{Okugi:LCWS2024}. (a) BDS for \SI{250}{GeV}  with half-length final-doublet ($\beta_{x}^{*}=\SI{13}{mm}$, $\beta_{y}^{*}=\SI{0.41}{mm}$) (b) BDS for \SI{500}{GeV} ($\beta_{x}^{*}=\SI{11}{mm}$, $\beta_{y}^{*}=\SI{0.48}{mm}$). In both, the blue (red) lines are for the horizontal (vertical) plane.
\label{ILC-BDS}}
\end{figure}

The current baseline ILC BDS at \SI{250}{GeV} (BDS-250) has been designed to be scaled up to \SI{1}{TeV} (BDS-1000) centre-of-mass energy (corresponding to \SI{500}{GeV} beam energy), keeping the same geometry. 
The total length is around \SI{2400}{m} with a crossing angle of \SI{14}{mrad}. 
The optics has been designed taking as starting point the BDS-250 magnet configuration. 
The BDS-500 optics has been matched with the same magnet configuration (location) as the BDS-250 one, but with different quadrupole strengths. 
The field strength of the final-doublet quadrupoles (QD0, QF1) for the BDS-250 is half of the BDS-500. 
The BDS-1000 needs some additional magnets, in dedicated locations that are free magnet areas for the lower energy operation modes, to reduce the horizontal emittance growth from synchrotron radiation, which increases with the fifth power of beam energy for constant bending radius~\cite{Raubenheimer:1998bj,Blanco:2015}. 
The ILC-BDS optics for \num{250} and \SI{500}{GeV} are shown in Fig.~\ref{ILC-BDS}.
Beam dynamics studies taking into account the multi-poles, but not the SR, have been performed for  the full range of beam energies. 
The optimisation of performance and tolerances, including different bending magnets strengths compatible for all the energies, collimation depths and different possible beamline geometries between others, are ongoing. 
More details can be found in~\cite{Okugi:LCWS2024}.

\subsubsection{The crossing angle choice}
The long ILC inter-bunch spacing of \SI{554}{ns} (\SI{366}{ns} for upgrade luminosity), gives the possibility of having a large range of crossing angle configurations from large (\SI{20}{mrad}) to quasi-head-on (\SI{2}{mrad}). With its much shorter bunch distance and higher maximum energies, CLIC prefers larger crossing angles in the range of \SI{16}{mrad} to \SI{25}{mrad}.

Larger crossing angle solutions separate the spent beam early from the incoming beam, with no shared final focus magnets. 
This separation facilitates the beam optics for the spent beam extraction line at the expense of some detector hermeticity in the forward region and the more complex tracking calibration procedures. 
In addition, larger crossing angles are more reliant on a Crab Cavity (CC) system, which rotates the bunches, in order to maximise their overlap and hence the luminosity. 

Small crossing angles are generally favoured from the point of view of detector hermeticity and are less reliant on the CC system. 
However, the fact that the incoming and outgoing beams must share the innermost magnets reduces significantly the design flexibility, e.g.\ to deal with the reduction of the backgrounds and the minimisation of energy losses in the critical elements. 
A detailed study for the \SI{2}{mrad} scheme including the design of the magnets involved in the extraction line has been carried out~\cite{Appleby:pac2007, Appleby:epac2008}.
Detailed studies and discussions about the pros and cons of each crossing scheme can be found in~\cite{ILC-CDR:2006,Blair:epac2006,ILC-RDR:2007}.

The present ILC baseline configuration has a \SI{14}{mrad} crossing angle at the IP. 
The \SI{14}{mrad} option is the minimum angle allowing on one hand independent magnetic channels for ingoing and outgoing beams by means of compact superconducting magnets and on the other hand makes easier the straightforward spent beam extraction while facilitating post-IP diagnostics to serve precision physics studies, such as polarimetry and spectrometry complementing similar devices planned upstream the BDS. 

\subsubsection{The two IP BDS configuration}
\label{sec:2IPs}

\begin{figure}[htb]
    \centering
    \begin{subfigure}{0.54\textwidth}
        \includegraphics[width=\textwidth]{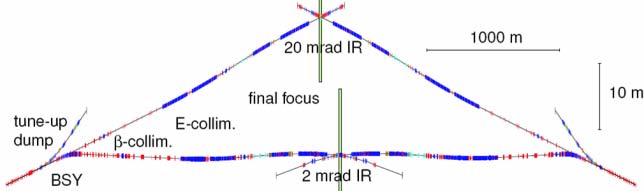}
        \caption{}
        \label{fig:ILC-BDS-2IPs:1}    
    \end{subfigure}
    \vspace{0.001cm}
    \begin{subfigure}{0.41\textwidth}
        \includegraphics[width=\textwidth]{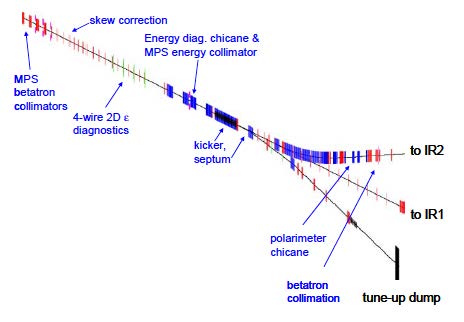}
        \caption{}
        \label{fig:ILC-BDS-2IPs:2}    
    \end{subfigure}
\caption{ILC BDS optics design with two IPs~\cite{Blair:2006ik}. (a) Layout of the two IPs with \num{20} and \SI{2}{mrad} crossing angle; (b) Layout of the BDS section after the Main Linac and the Final Focus section.
\label{ILC-BDS-2IPs}}
\end{figure}

The full realisation of the scientific potential of the ILC argues for the construction and operation of two complementary detectors. The value of having two detectors with different designs, technologies, collaborations and emphasis has proven to be a very effective way to exploit the science. 
In its Baseline Conceptual Design~\cite{ILC-CDR:2006}, the ILC was designed with two BDS sections and two independent IPs with \num{20} and \SI{2}{mrad} crossing angles respectively as shown schematically in Fig.~\ref{ILC-BDS-2IPs} and is described in some detail~\cite{Blair:epac2006}. 
At that time other schemes were under study, including one of \SI{14}{mrad}~\cite{Nosochkov:slac2007} and a head-on scheme~\cite{Napoly:epac2006}. 
Since 2007, the two IPs evolved to only one IP with \SI{14}{mrad} crossing angle and two detectors in push-pull configuration sharing the beam time. 
This configuration is the baseline for the ILC located in Japan~\cite{Behnke:2013xla, ILC:2013jhg, Adolphsen:2013jya, Adolphsen:2013kya, Behnke:2013lya}~\cite{Bambade:2019fyw}.

A detailed study of a dual BDS serving two IPs has been realised in the framework of the CLIC project, this work could be taken as reference for future studies~\cite{Cilento:PRAB2021, Cilento:2021trt}. 
The main objective of this work was the design of a dual BDS system optimised to minimise the luminosity loss due to emittance increase and related widening of the beam sizes caused by the emission of SR as generated in the bending magnet section in the first part of the BDS, added to separate the two IPs. 
This effect is much more significant for CLIC at high energies such as the \SI{3}{TeV} case, since the contribution to the IP beam size scales with the fifth power of energy~\cite{Blanco:2015}.
To mitigate luminosity loss from this contribution, the bending angle needs to be as large as possible, which determines the length of the BDS. 
Another important design parameter is the crossing angle at the IP. 
In fact, the SR in the final quadrupole and the solenoid increases the vertical spot size at the IP. 
Simulations with a \SI{4}{T} solenoid field at CLIC \SI{3}{TeV} give an acceptable growth if the crossing angle is around \SI{20}{mrad}, leading to the current baseline value of \SI{20}{mrad}. 
An interesting consequence of adding a new IP with larger crossing angle is its compatibility with $\PGg\PGg$ collisions~\cite{Telnov:2007}. 
A crossing angle of about \SI{25}{mrad} is considered optimal for this type of collisions using optical lasers~\cite{Telnov:2018}.

\begin{figure}[htb]
\begin{center}
    \begin{subfigure}{0.75\textwidth}
        \includegraphics[width=\textwidth]{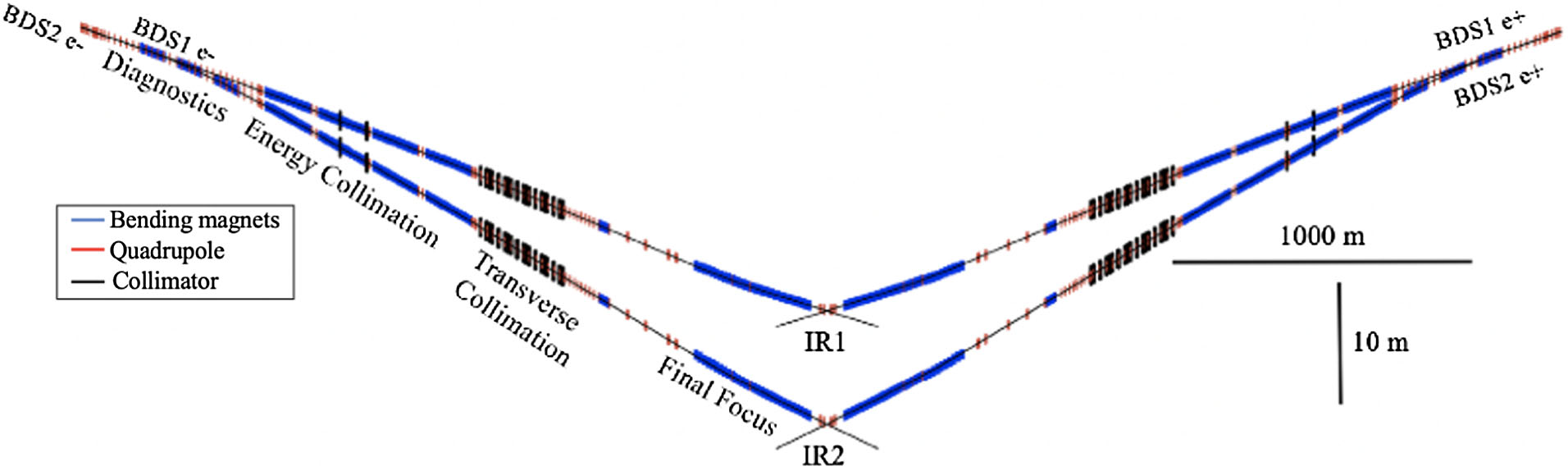}
        \caption{}
        \label{fig:CLIC-BDS-2IPs:1}    
    \end{subfigure}
    \begin{subfigure}{0.75\textwidth}
        \includegraphics[width=\textwidth]{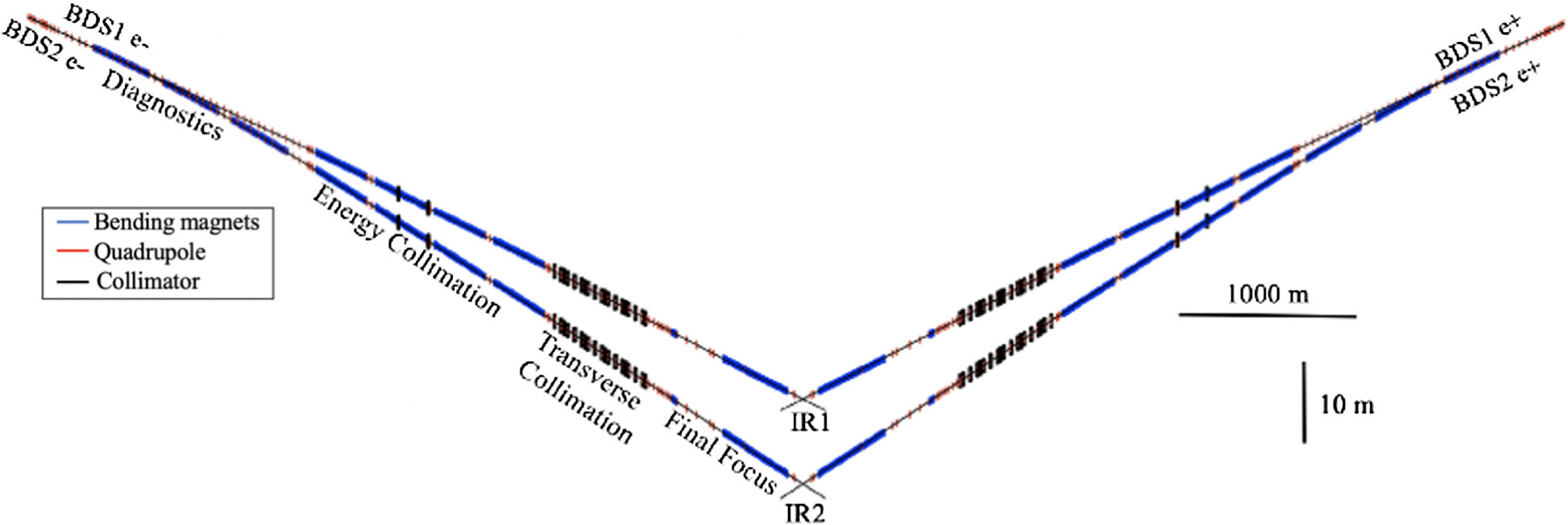}
        \caption{}
        \label{fig:CLIC-BDS-2IPs:2}    
    \end{subfigure}
\end{center}
\caption{CLIC BDS optics design with two IPs~\cite{Cilento:2021trt}. (a) Layout of the two IPs at \SI{380}{GeV} with \num{16.5} and \SI{20}{mrad} crossing angle; (b) Layout of the two IPs at \SI{3}{TeV} with 20 and \SI{25.5}{mrad} crossing angle.
\label{CLIC-BDS-2IPs}}
\end{figure}

Figure~\ref{CLIC-BDS-2IPs} shows the layout of the CLIC dual BDS at \SI{380}{GeV} with \num{16.5}/\SI{20}{mrad} crossing angles and \SI{3}{TeV} with \num{20}/\SI{25.5}{mrad}  crossing angles. 
The BDS layout is constructed by adding FODO cells in the existing Dispersion Suppressor (DS), with an additional total length of about \SI{300}{m} for \SI{380}{GeV} and \SI{1}{km} for the \SI{3}{TeV}.
The four BDS systems at either side of the two IPs need to have different DS lengths to provide the desired longitudinal and transverse separations at the IP.
Further improvements can still be performed for the dual BDS layout in order to recover part of the luminosity performance, especially for the BDS2 of the CLIC \SI{3}{TeV} case. 
Currently all DS bending is placed in BDS2 however this could be distributed between BDS1 and BDS2 (with opposite angle). 
This would reduce luminosity loss in IP2 and increase it in IP1. Another option could be to do a longer BDS to reduce the impact of the SR in the BDS2, followed by optics improvements of the dual BDS.

For the LCF at CERN, the idea of having two IPs is being revisited as baseline. The angle choice has not been finally determined, but the two angles should be different to have physics complementarity between the two detectors, and to optimise the IPs for different upgrade options, e.g.\ photon-photon collisions (c.f.\ Sec.~\ref{sec:acc:altmodes}) or high-energy or high-luminosity upgrades (c.f.\ Secs.~\ref{sec:acc:upgrade} and~\ref{sec:acc:lumiup}).
Preliminary design studies and CFS cost with \SI{20}{mrad} between the Main Linacs (MLs), and  with two IPs  with \num{14} and \SI{26}{mrad} crossing angle with the two detectors in the same longitudinal position, are being performed but no detailed design of the optics is currently being developed. 
Discussions are ongoing to find an optimal solution; some issues have been identified and could be the subject of focused R\&D over the next five years. 
These topics are:

\begin{itemize}
    \item The ``quasi-head on'' (\SI{2}{mrad}) crossing angle configuration will be challenging to implement from the outgoing beams and dumps point of view.
    \item In the same longitudinal position for the two IPs, an optimisation of the two IPs cavern has to be performed to provide a minimum transverse distance. Currently, \SI{30}{m} between two detectors is generally considered to be sufficient. 
    \item The beam dumps have to be reconfigured and special attention has to be paid to the photon dump location. 
    In particular, the photon dump location at the centre (between two tilted beams) becomes a critical issue in the case of the undulator positron source.
    \item Options for different longitudinal positions of the two IPs have the advantage of allowing less separation between the two detectors, but will have timing issues due to the different lengths for the two BDSs. 
    The implementation of these type of options will be complex, and will suppose a global optimisation design for all the accelerator systems as the Damping Rings (DR) (for instance, the length of the BDSs has to be half-integer of the DR length) or the injection system. 
    In this scenario the implementation of two different inter-bunch spacing value, as expected for the the baseline ILC operation with \SI{554}{ns} and with \SI{366}{ns} for high-luminosity, will be very difficult. 
    More in detail, if the path length condition is satisfied by one of the IPs, the other IP can see the bunch collisions only when the distance of the two IPs is a multiple of $\SI{554}{ns}/2$, i.e.\ \SI{83}{m}, (though 1 to 2 collisions among \num{1312} are missed). 
    But this cannot be a multiple of $\SI{366}{ns}/2$ equivalent to \SI{55}{m}.
    \item The angle between the MLs has to be optimised to minimise the SR generated by the bending magnets of the BDS separation region, taking into account the two IPs crossing angle choice.
    \item The idea of having parallel MLs is also being considered and this option could be better for the future implementation of an upgrade by means of ERLs, nevertheless the impact of the increase of total length  and the SR losses in the splitting section should be considered. 
    \item The split location of the BDSs has to studied in detail, as well as the split magnet technology for the high-energy upgrades.
    \item The switching between the two IPs should be at least on a train-by-train basis.  Bunch-by-bunch separation would be an interesting option in case one IP should collect luminosity while the other is used temporarily for R\&D or testing / commissioning, e.g.\ paving the path towards an upgrade. 
    In particular train-by-train switching by pulsed magnets is used already in the ILC design for the \PZ-pole operation and also the polarisation switching of positron beam.
   \item Compatibility with a possible $\PGg\PGg$ collision mode have to be evaluated, with both optical lasers or FEL options. 
   Options with Eu-XFEL lasers are optimal when the crossing angle is "quasi-head on" contrary to the optical ones.
   \item Likewise, the compatibility with all other upgrade options (e.g.\ based on energy- and particle recovery schemes or  higher acceleration gradients) and beyond-collider facilities have to be evaluated.
\end{itemize}

\subsection{Running and upgrade scenarios for a Linear Collider}
\label{sec:acc:scenarios}
The technologies discussed in Sections~\ref{sec:acc:SCRFbase} and~\ref{sec:acc:CLICbase} enable us to pursue a comprehensive programme to study the Higgs boson and its closest relatives in the SM today. At the same time, the energy and luminosity upgrade options discussed in the following sections provide attractive long-term pathways and opportunity for a linear collider facility. 

\begin{figure}[htbp]
    \centering
    \begin{subfigure}{.5\textwidth}
    \centering
        \includegraphics[width=0.95\textwidth]{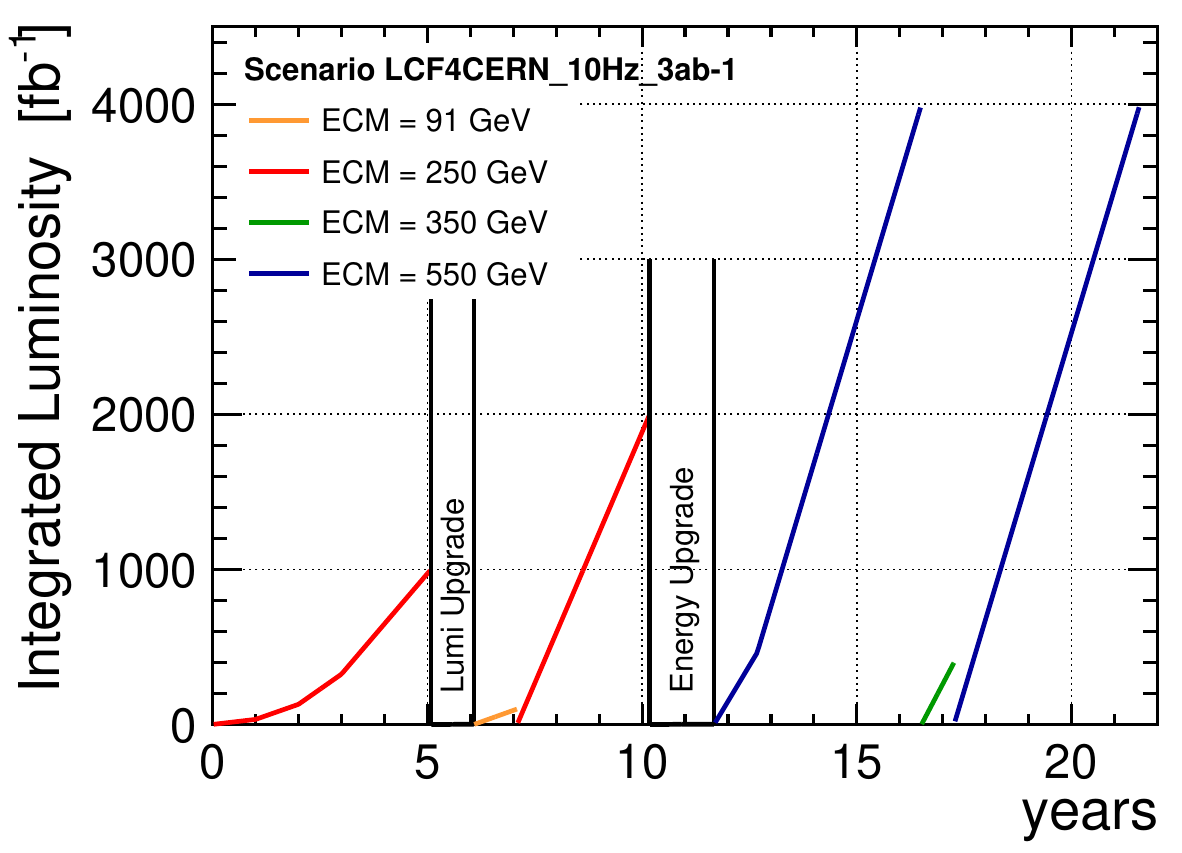}
        \caption{}
        \label{fig:runscen:LCF}    
    \end{subfigure}\hfill%
    \begin{subfigure}{.5\textwidth}
        \centering
        \includegraphics[width=0.95\textwidth]{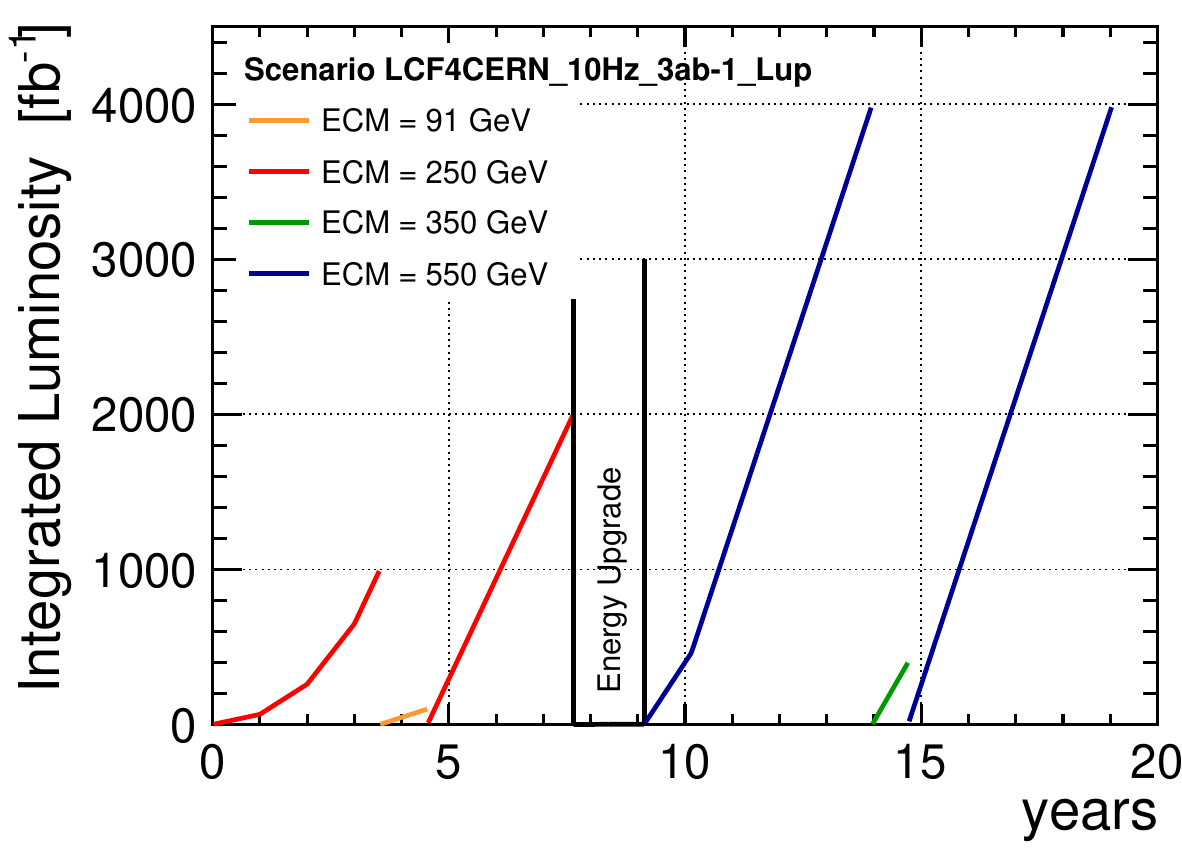}
        \caption{}
        \label{fig:runscen:LCF-LUp}    
    \end{subfigure}%
    \vspace{0.1cm}
    \begin{subfigure}{.5\textwidth}
    \centering
        \includegraphics[width=0.95\textwidth]{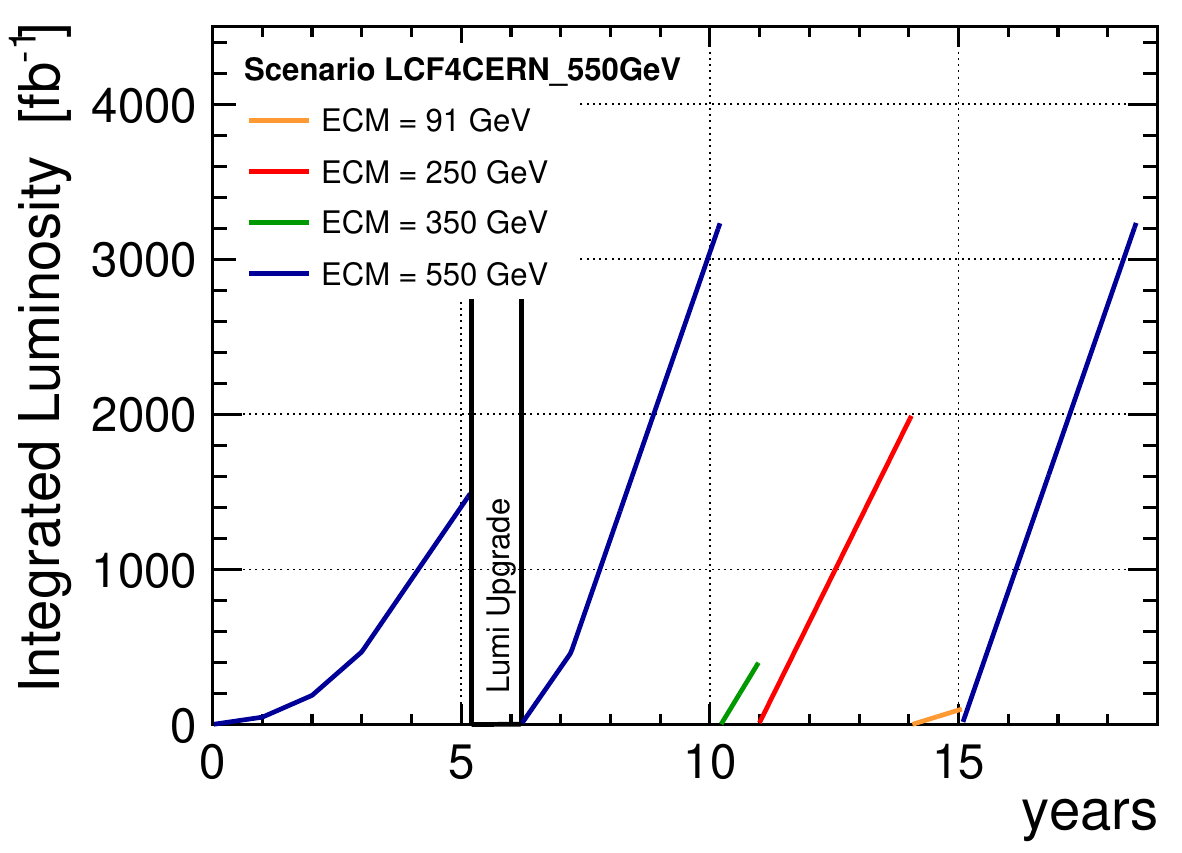}
        \caption{}
        \label{fig:runscen:550GeV}    
    \end{subfigure}\hfill%
    \begin{subfigure}{.5\textwidth}
        \centering
        \includegraphics[width=0.95\textwidth]{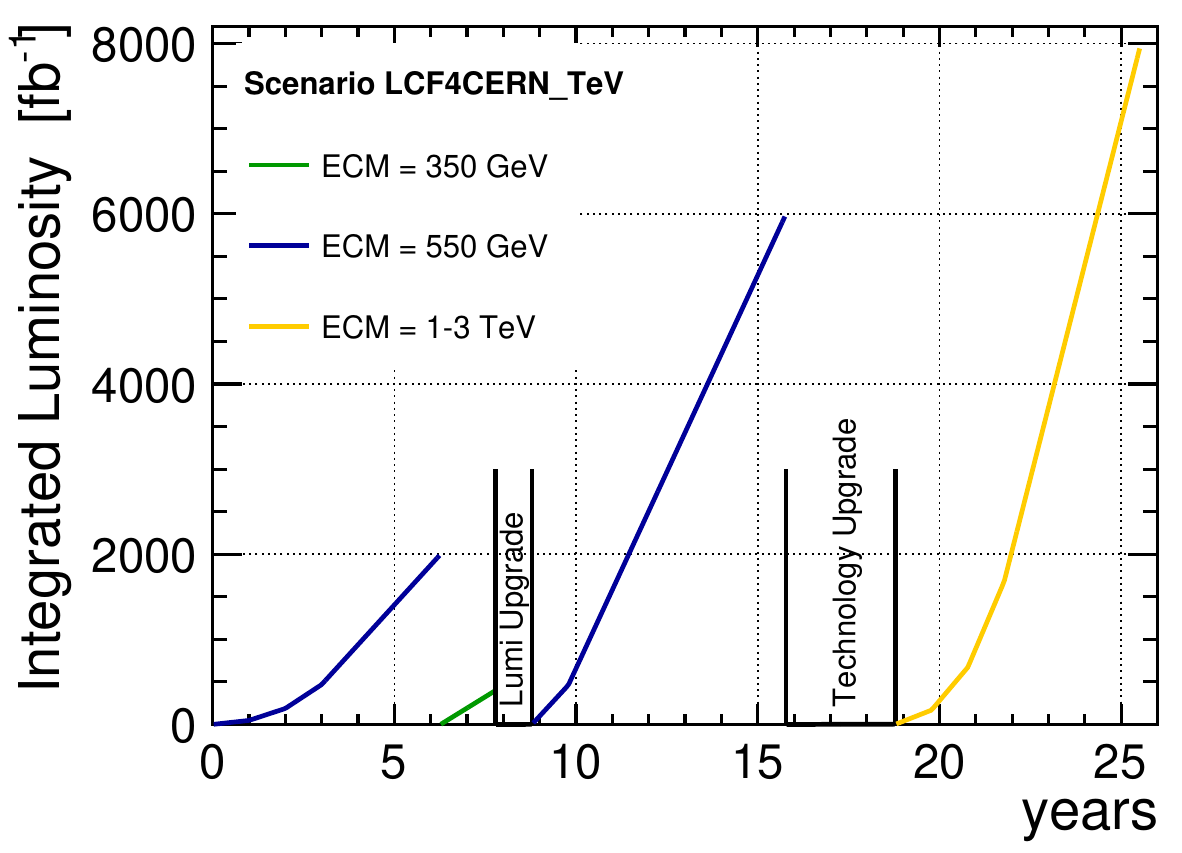}
        \caption{}
        \label{fig:runscen:TeV}    
    \end{subfigure}%
    \vspace{0.1cm}
    \begin{subfigure}{.5\textwidth}
    \centering
        \includegraphics[width=0.95\textwidth]{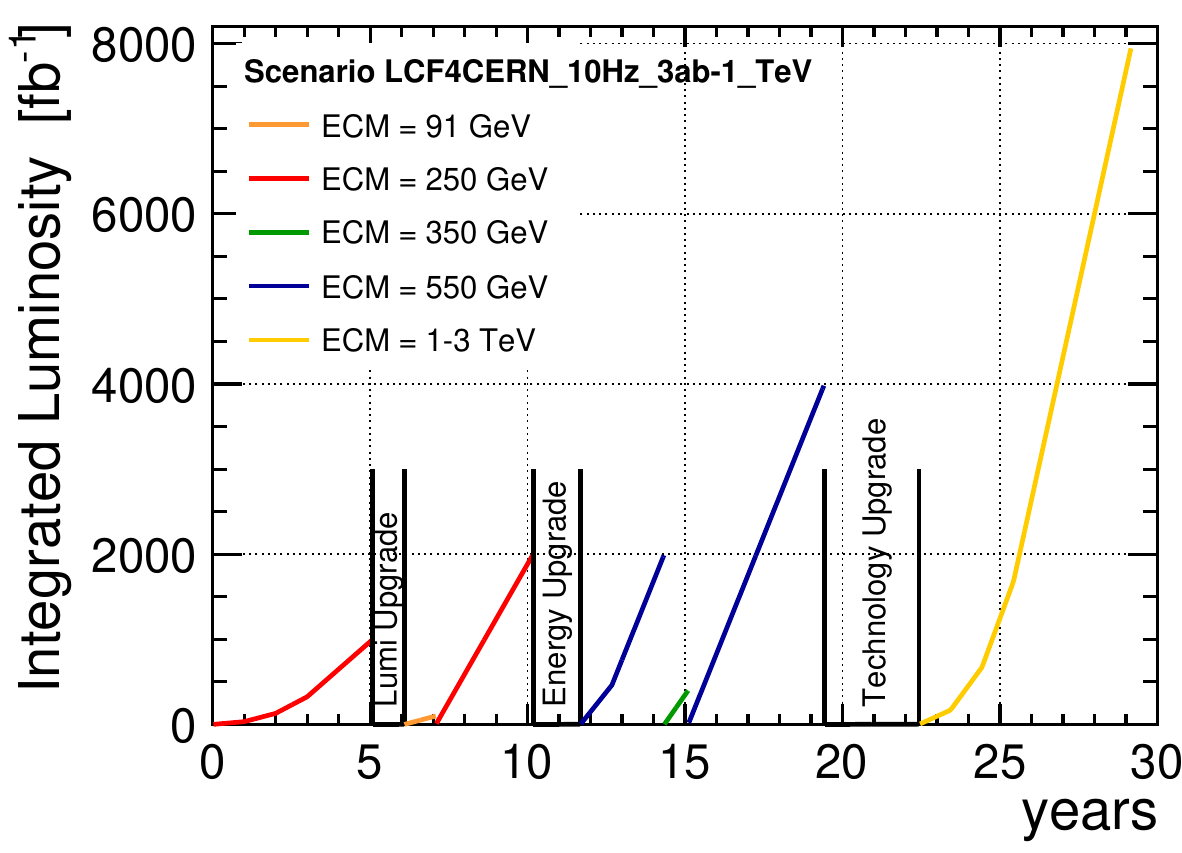}
        \caption{}
        \label{fig:runscen:250-TeV}    
    \end{subfigure}\hfill%
    \begin{subfigure}{.5\textwidth}
        \centering
        \includegraphics[width=0.95\textwidth]{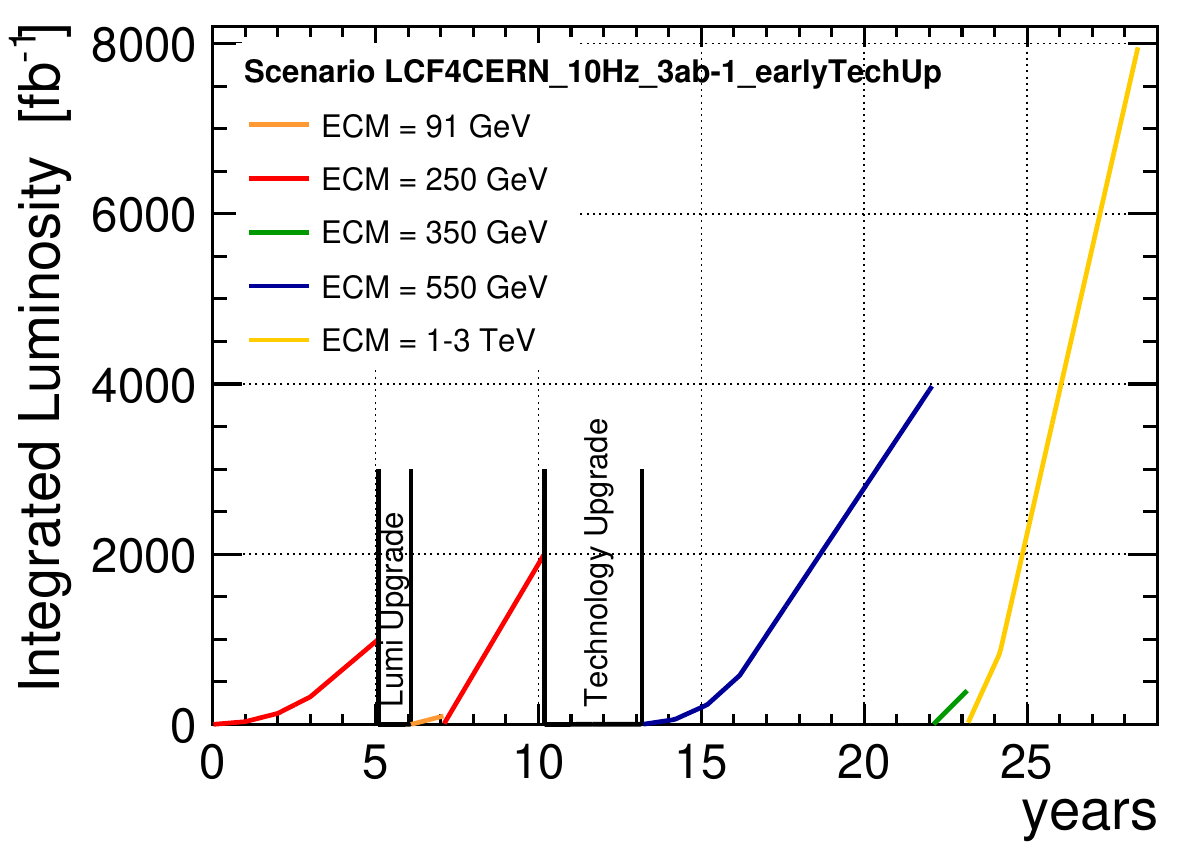}
        \caption{}
        \label{fig:runscen:earlyTechUp}    
    \end{subfigure}%
    \caption{Various examples of operation and upgrade scenarios for a Linear Collider Facility, all starting with a SCRF-based machine, labelled LCF in Fig.~\ref{fig:lep:lumi}. (a) starting at \SI{250}{GeV}, low power and upgrade to full power and then to \SI{550}{GeV}; (b) starting directly with full power at \SI{250}{GeV} and then upgrading to \SI{550}{GeV}; (c) start at \SI{550}{GeV}, taking smaller sets of polarised data at lower energies later; (d) starting at \SI{550}{GeV} and upgrading to \SI{1}{TeV} or beyond; (e) like (a), but with a shortened run at \SI{550}{GeV} in favour of an earlier upgrade to \SI{1}{TeV} or beyond; (f) starting at \SI{250}{GeV}, but then changing technology already for \SI{550}{GeV} (here assuming \CCC\ parameters).}
    \label{fig:runscen}
\end{figure}

Figure~\ref{fig:runscen} illustrates some of the possible choices  optimising the physics programme with several energy stages below \SI{1}{TeV}.  
The instantaneous luminosities assumed follow those for the SCRF-based version of LCF in Fig.~\ref{fig:lep:lumi} up to \SI{550}{GeV}, and for \CCC\ above that energy. 
After each major change to the accelerator, a ramp-up of efficiency is assumed following the prescription developed in~\cite{Barklow:2015tja}. 
Figures~\ref{fig:runscen:LCF} and~\ref{fig:runscen:LCF-LUp} cover the integrated luminosities given in the first four columns of Table~\ref{tab:LCF-runplan}, corresponding to the Higgs coupling results in the third column of Table~\ref{tab:fullprogram}, differing by the initial luminosity (at low power (\num{1312} bunches per train) or directly at full power (\num{2625} bunches per train)). 
Figures~\ref{fig:runscen:250-TeV} and~\ref{fig:runscen:earlyTechUp} cover all stages given in~\ref{tab:LCF-runplan}, corresponding to the fourth column of Table~\ref{tab:fullprogram}, differing in when a new technology would be decision- and production-ready. 
In the case that scientific competition makes a start at \SI{550}{GeV} desirable, it would still be possible to collect (polarised) data at lower energies later in the programme if science requires (Fig.~\ref{fig:runscen:550GeV}), or to directly upgrade the facility to explore the \SI{1}{-}\SI{3}{TeV} regime.

Many other choices are possible. We'd like to re-emphasise that one of the central advantages of a linear collider facility is that the upgrade path remains flexible, and should be chosen according to scientific and technological advances or even revolutions. In the next sections, we discuss several possible avenues towards high-energy and high-luminosity upgrades as well as those enabling $\PGg\PGg$, $\Pem\PGg$ and $\Pem\Pem$ collisions, which all requires more R\&D than the baseline options, but could well become viable on the timescale of an upgrade of an initial facility.


\subsection{Energy upgrades}
\label{sec:acc:upgrade}

The construction of a linear collider facility provides us with the opportunity to consider various pathways for future energy upgrades. 
These upgrades may allow for more advanced technologies presently in development to replace the original technological choices if they provide a significant advantage. 
Ideally, significant portions of the accelerator complex can be re-utilised to leverage the existing common hardware of a linear collider facility and the research and development investment that went into those sub-systems (e.g.\ injectors, positron source, damping ring, beam delivery). 
In the following we will discuss upgrades based on SCRF with higher gradient cavities, an upgrade with warm copper (CLIC-like) cavities, an upgrade based on cool copper (\CCC ) technology as well as upgrades with wakefield acceleration, comprising structures as well as laser- or beam-driven plasma. While upgrades with RF cavities mostly apply if an LCF starts out as a superconducting machine, wakefield acceleration provides intriguing upgrade options also for the CLIC-like baseline.

\subsubsection{Upgrade  of an SCRF-based machine with higher gradient cavities}

The European Strategy for Particle Physics - Accelerator R\&D Roadmap~\cite{ESPP-Accelerator} and Snowmass'2021~\cite{Snowmass2021:SRF-key-directions,Snowmass2021:Nb3Sn,Snowmass:SRF-thin-films} emphasised strategic trends in SCRF development. In particular, the former stated: 
``In SCRF the development is focused in two areas, bulk niobium and thin-film (including high-$T_c$) superconductors. 
In bulk SCRF new treatments are allowing niobium cavities to exceed previous record $Q$ factors and avoiding degradation with increasing gradients. 
This includes nitrogen infusion and doping, and two-step baking processes. 
There is also an emphasis on limiting field emission. For thin-films the community is investigating creating coated cavities that perform as good as or better than bulk niobium (but with reduced cost and better thermal stability), as well as developing cavities coated with materials that can operate at higher temperatures or sustain higher fields. 
One method of achieving this is to use multi-layer coatings. 
Innovative cooling schemes for coated cavities are also being developed. Coupled to the cavity development is improvement in the cost and complexity of power couplers for SCRF cavities.''

The performance limit of SCRF cavities made of pure niobium is determined by the two key parameters: the RF critical, or superheating, magnetic field $B_{sh}$ and the surface resistance $R_s$. 
The RF critical magnetic field limits the maximal achievable accelerating gradient $E_{acc}=B_{sh}/k_m$, where $k_m$ is a coefficient specific to the cavity shape. 
For example, $k_m=\SI{4.26}{mT/(MV/m)}$ for \SI{1.3}{GHz} TESLA-shape cavities. 
The surface resistance determines efficiency of the SCRF cavity: the cavity intrinsic quality factor $Q_0 = G/R_s$ is inversely proportional to the surface resistance. 
The cavity geometry factor $G$ is approximately \num{270}\,$\Omega$. 
Recent advancements in surface tailoring techniques have led to groundbreaking results, as laboratories have successfully introduced nitrogen, oxygen, and other impurities into the niobium lattice in a controlled fashion. 

The superheating field of approximately\footnote{Typically, it is assumed that the RF critical magnetic field is equal to the DC superheating field~\cite{Padamsee_Book_1}, although no definitive theory yet exists that can calculate the RF critical filed.} \SI{210}{mT} at \SI{2}{K} was reached experimentally on single cell cavities, resulting in an accelerating field of about \SI{50}{MV/m} for TESLA-shape cavities~\cite{2-step_baking,Bafia_SRF2019}. 
At $\sim$\SI{50}{MV/m}, the quality factor is still limited to $<\num{2E10}$ at \SI{2}{K}\footnote{Note that a \SI{24}{K} operation of niobium cavities at this frequency is not practical, as the $Q_0$ becomes low, $\sim\num{5E8}$ resulting in an excessive heat dissipation.}.
To date, the experiments show that only low-temperature surface treatments are beneficial for increasing of RF critical field. 
Mid-T bake and nitrogen doping tend to lower the achievable quench field of SCRF cavities.

With additional investments, further developments in the bulk niobium SCRF technology could be impactful for the ILC upgrades on a 5-year time scale. On a longer time scale, about 10 years, travelling wave SCRF structures could be developed to bring the gradient to $>\SI{60}{MV/m}$~\cite{JINST:HELEN}. 
Finally, if there is a breakthrough, thin films and superconductors beyond niobium could potentially bring higher accelerating gradients, higher quality factors at \SI{4}{K}, or both. 
This would open a horizon to more efficient (e.g., ERL-based) and higher energy linear collider options.

\paragraph*{Bulk-niobium and travelling wave SCRF}

The most promising surface treatment of SCRF bulk niobium cavities for high gradient applications is a combination of cold electropolishing and two-step (\num{75}/\SI{120}{^\circ C}) baking. 
It has paved the way for gradients near \SI{50}{MV/m} in single-cell TESLA-shape cavities. 
Applying this advanced treatment to improved shape standing-wave SCRF structures ---  such as Ichiro, Low-Loss (LL), re-entrant, and Low Surface field (LSF) cavities ---  holds the prospect of gradients close to \SI{60}{MV/m} in single-cell cavities and \SI{50}{MV/m} in 9-cell cavities in a cryomodule.

Further improvements in accelerating gradients of bulk Nb structures are possible but require developing an alternative approach. 
Utilising travelling wave (TW) in a resonant ring configuration offers such an approach~\cite{Neal}. 
The SCRF TW structure has several advantages compared to the standing wave structures, see e.g.~\cite{Yakovlev:PAC07_TW,JINST:HELEN,Belomestnykh:IPAC2023_HELEN}. 
The most salient advantages are as follows: A substantially lower shape parameter $k_m$ promises achieving higher $E_{acc}$ at the same magnetic quench limit.
Approximately a factor of 2 higher $R/Q$ (even at somewhat lower $G$) would ensure lower cryogenic losses, see~\cite{Shemelin:2021ahc} for more details on the structure optimisation. 
In addition, the TW structure provides high stability of the field distribution along the structure with respect to geometrical perturbations. 
This allows for much longer accelerating structures than TESLA cavities, limited by the manufacturing technology only. 

Assuming a realistically achievable accelerating gradient of \SI{60}{MV/m} (about 20\% reduction from the ultimate gradient of \SI{73.7}{MV/m}) and the quality factor of \num{1.4E10} (for the surface resistance corresponding to $Q_0 = \num{2E10}$ of TESLA cavity), a superconducting linear collider built with TW cavities could reach \SI{500}{\GeV} centre-of-mass collision energy within the \SI{20.5}{km} footprint of ILC in Japan and \SI{1}{\TeV} within the \SI{33.5}{km} footprint proposed for LCF. 
Some parameters of SCRF travelling-wave based linear collider (TWLC) are presented in Table~\ref{tab:TWLCparameters}.

While there is progress towards proof-of-principle experimental demonstration on a 3-cell structure and developing a longer structure~\cite{Furuta:LINAC2024}, it is slow due to intermittent funding. 
Proper investment in this technology could bring it to ``prime time'' readiness ---  demonstrated performance in a full-scale cryomodule ---  on a 10-year time scale. After that TW-based linear collider should be feasible.\par

\begin{table}[!htb]
\caption{Key parameters for a linear collider based on travelling-wave SCRF.}
\label{tab:TWLCparameters}
\centering

\begin{tabular}{l l l l}
\toprule
Parameter & Unit &   TWLC550 & TWLC1000 \\
\midrule
Centre-of-mass energy & GeV & 550 & 1000 \\
Luminosity & $10^{34}$cm$^{-2}$s$^{-1}$ & 3.9/7.7 & 5.1 \\
Number of interaction points &  & 2 & 2 \\
Repetition frequency & Hz & 10 & 4 \\
Bunch population & $10^{10}$ & 2 & 1.74 \\
Number of bunches per train & & 1312/2625 & 2450 \\
Bunch spacing & ns & 554/366 & 366 \\
Bunch train duration & $\mu$s & 727/961 & 897 \\
\midrule
Accelerating gradient & MV/m & 60 & 60 \\
Cavity quality factor at 60~MV/m & $10^{10}$ & 1.4 & 1.4 \\
Klystron efficiency & \% & 80 & 80 \\
\midrule
Length of 2 SCRF linacs & km & 11.9 & 21.8 \\
Total machine length & km & 20.5 & 33.5 \\
Site power consumption & MW & 225/299 & 223 \\ 
\bottomrule
\end{tabular}
\end{table}

\paragraph*{Advanced SCRF superconductors (\ce{Nb3Sn}, thin films, multilayers)}

Over the years, many high-temperature superconductors have been investigated. 
Some of them (cuprates, bisthmuthates, etc.) have not been found to be useful for high-field accelerating cavities~\cite{Padamsee_Book_1}. 
Others, such as A15 compounds, \ce{MgB2} are still under investigation. 
The most progress to date in advanced SCRF superconductors has been made with \ce{Nb3Sn} using the vapour diffusion process~\cite{Posen_2021}. 
Superconducting accelerating cavities using \ce{Nb3Sn} are expected to be a game changer, bringing higher gradients with more efficient cooling, because \ce{Nb3Sn} has a higher transition temperature ($T_c$) of \SI{18.3}{K}, compared to the \SI{9.2}{K} transition temperature of Nb. 
Development of \SI{1.3}{GHz} accelerating cavities has been progressing since the mid-2010s at Cornell University, Fermilab, and Jefferson Laboratory in the United States~\cite{Posen_PoP-demonstration, Posen_2017, Posen_2021}, and now is being actively pursued worldwide~\cite{PosenSnowmass2021}. 

Normally, \SI{1.3}{GHz} Nb cavities operate at \SI{2}{K} to reduce the surface resistance and achieve high intrinsic quality factors $Q_0$ of greater than \num{1E10}. 
Large cryogenic systems, like the one required for ILC, are very inefficient, with about \SI{800}{W} of wall-plug power needed for every watt dissipated at \SI{2}{K}. 
On the other hand, \ce{Nb3Sn} can achieve the same surface resistance as Nb, but at at \SI{4.2}{K}, which increases the cooling efficiency and substantially simplifies the cryogenic system by removing the need for cold compressors. 
For example, a single-cell cavity having surface resistance of \num{20}\,n$\Omega$ and operating at an accelerating gradient of \SI{10}{MV/m} would dissipate approximately \SI{1}{W} at \SI{4.2}{K}, where the efficiency of cryogenic systems is almost a factor of 3 better. 
Such cavities would not only benefit future colliders but are also suitable for compact conduction-cooled accelerators, as commercial cryocoolers have capacity of \num{1} -- \SI{2}{W} at \SI{4.2}{K}. 

For achieving higher gradients, \ce{Nb3Sn} has a potential advantage due to its high DC superheating field ($B_{sh}$), approximately twice that of Nb~\cite{Padamsee_Book_1}. 
This means that it could be expected that \ce{Nb3Sn} SCRF cavities might reach accelerating gradients as high as \num{90} -- \SI{100}{MV/m} providing good system efficiency at the same time. However, despite continuous efforts to improve the gradient performance of \ce{Nb3Sn} SCRF cavities, the maximum accelerating gradients achieved to date remain quite modest, $\leq\SI{24}{MV/m}$~\cite{Posen_2021}. 
Processes other than vapor diffusion are being investigated at different institutions: electroplating, CVD deposition, magnetron sputtering, just to mention a few. 
As these techniques mature, we should expect further progress to be made. 
If there is a breakthrough, one could potentially apply \ce{Nb3Sn} to travelling wave structures, making even higher accelerating gradients reachable.

Another approach was proposed by Gurevich and further developed by Kubo~\cite{Gurevich_2006,Gurevich_2015,Kubo_2017}, taking advantage of superconductors with higher $T_c$ than niobium without being penalised by their lower $B_{sh}$. 
The idea is to coat SCRF cavities with alternating superconducting and insulating layers (SIS structures) with a thickness smaller than the London penetration depth $\lambda_L$. 
The surface resistance will be strongly reduced because those superconducting materials  (\ce{Nb3Sn}, \ce{NbTiN}, \ce{NbN}, ...) have a higher energy gap $\Delta$ than \ce{Nb} and the surface resistance depends inverse exponentially on $\Delta$.
Theory suggests that by employing such structures, the performance of superconducting cavities can be significantly increased by achieving $Q$-values of a factor of \SI{2}{-}\num{5} higher than values typically obtained for pure niobium. 
Since induced currents and fields in the bulk material are reduced as well by additional shielding provided by the interfaces between the layers, an application of SIS multilayers would in addition enable a significant increase of the maximum achievable accelerating fields above \SI{70}{MV/m}. 
This would allow an operation of cavities above 50~MV/m at 4\,K with a $Q_0$ of greater than \num{1E10}. 
The benefit would then be, that a \SI{4}{K} operation would allow an overall higher cooling efficiency and reduced operational cost of an accelerator the same way as for \ce{Nb3Sn} and the higher gradient would increase the energy reach of a machine.

The SIS multilayer research is in its initial stage, at the technological readiness level 2. 
Teams in America, Asia, and Europe are developing coating processes, investigating the parameter phase space, trying various materials, etc. 
A sustained investment into the basic R\&D studies, infrastructure, analytical instruments is needed before we can expect first practical multilayer SCRF cavities.
Yet, sample studies have shown already, that a field enhancement and a higher external magnetic field can be sustained compared to bulk Nb~\cite{Ito2019}, proving the underlying principle~\cite{Gonzalez2023, Asaduzzaman2023, Antoine2019, Antoine2010}. 
Very first RF tests on prototypes showed weak to no improvement or even deterioration of the surface resistance compared to baseline measurements~\cite{Tikhonov2019, Keckert2019, Kalboussi2023}, which can be explained by magnetic flux pinning during cooldown or a non-optimal, faulty post-processing of the films after deposition. 
Hence, the major focus of the ongoing R\&D is to transfer the promising sample results to cavities, and optimise the post-processing strategy which heavily relies on the deposition technique itself. 
First decisive results can be expected within a 5- to 10-year time frame.   

\paragraph*{Upgrading the initial facility} 
In case of a replacement of the original main linac with better SCRF technology, most of the other infrastructure could be retained, including the cryogenic system. R\&D results expected in the coming ten years will form an important basis for a more detailed design study for such an upgrade.
~\par

\subsubsection{Upgrade using CLIC technology}

In this section we discuss the possible upgrade of the SCRF-based baseline option described in Sec.~\ref{sec:acc:SCRFbase} by replacing the SCRF linac with the CLIC high-gradient RF cavities. In the \SI{33.5}{km} footprint, replacing the \SI{22}{km} of main linac would result in an energy reach up to \SI{2}{TeV}.

The current performances of CLIC are  described in detail in~\cite{Adli:ESU25RDR} and is summarised in section~\ref{sec:acc:CLICbase} above. 
The key parameters of the baseline machine, a drive-beam based CLIC at $\sqrt{s} = \SI{380}{GeV}$, running at \SI{100}{Hz}, operating with two detectors, are listed in Table~\ref{tab:CLIC380}. 
Accelerating gradients of up to \SI{145}{MV/m} were reached with the two-beam concept at CTF3, and breakdown rates of the accelerating structures well below the limit of \SI{3E7}{m^{-1}} are stably achieved at X-band test platforms. 

For the first energy stage, an alternative scenario, with X-band klystrons powering the main-beam accelerating structures has also been studied~\cite{Aicheler:2018arh}. 
When considering using CLIC X-band technology as an upgrade, the klystron-powered solution is the easiest, as it minimises changes to the infrastructure and in particular does not require construction and integration of a drivebeam complex. 
For the higher energies, a drive-beam option may be cost-optimal. In this case, the L-band klystron and modulators from the SCRF-linac may be reused for the CLIC drive beam linac.

As shown in Table~\ref{tab:CLIC_upgrade}), a \SI{550}{GeV} CLIC would fit inside a \SI{15}{km} tunnel a 1500  CLIC would fit inside a \SI{29}{km} tunnel, for both upgrade options. For a \SI{33.5}{km} footprint, even \SI{2}{TeV} could be reached, however a detailed design for this energy needs to be worked out. 

\begin{table}[!htb]
\caption{Main parameters for a CLIC based upgrades showing the different possible energy versions.}
\label{tab:CLIC_upgrade}
\centering
\begin{tabular}{l l l l}
\toprule
Parameter                  &  Unit     & \SI{550}{GeV}   &   \SI{1500}{GeV}  \\
\midrule
Centre-of-mass energy  & \si{\GeV}                                           & 550      & 1500 \\
Repetition frequency   & \si{\Hz}                                            & 100         & 50\\
Nb. of bunches per train  &        & 352      & 312\\
Bunch separation        & \si{\ns}            & 0.5       & 0.5\\
Pulse length  & \si{\ns}                                            & 244          & 244\\
\midrule
Accelerating gradient  & \si{\mega\volt/\meter}          & 72     & 100 \\ 
\midrule
Total luminosity & \SI{e34}{\per\centi\meter\squared\per\second}     &  $\sim$6.5         & 3.7 \\
Lum. above \SI{99}{\percent} of $\sqrt{s}$ & \SI{e34}{\per\centi\meter\squared\per\second}         & (n/c)      & 1.4 \\
Total int. lum. per year     & fb$^{-1}$      &  $\sim$782      & 444 \\ 
Average power consumption  & MW       & $\sim$209 & 287  \\ 
\midrule
Main linac tunnel length   &  \si{\km}   & $\sim$15          & 29.0\\
Nb. of particles per bunch       & \num{e9}       & 5.2       & 3.7\\
Bunch length   & \si{\um}   & 70      & 44\\
Final RMS energy spread & \si{\percent}  & 0.35 & 0.35  \\
\bottomrule
\end{tabular}
\end{table}

\paragraph*{Timeline and R\&D needed}

Based on the desired energy reach for the upgrade, it should be studied whether a klystron-based upgrade  option or a drive-beam based upgrade is optimal.   Furthermore, how to maximise the reuse of  each subsystem of the SCRF-machine (Injectors, BDS etc) needs to be studied.

The key technologies needed for a klystron-based option are related to the powering and in particular 
high-efficiency X-band klystrons: Efficiencies between \SI{80}{-}\SI{90}{\%}  and more than \SI{70}{\%} are being achieved currently in L-band and S-band klystrons respectively~\cite{Syratchev:2022a}. 
For X-band achieving such high efficiency is very difficult, due to the small distances involved. 
A focused R\&D effort is needed to overcome the issues and get higher efficiencies to be able the reach of high-energies a reality. 
These klystrons and associated modulators are also cost-drivers and require reliabilities beyond what is needed for the drivebeam-based CLIC.

\subsubsection{Upgrade using \CCC\  technology} 

\label{sec:acc:upgrade:c3}

\CCC\ is a concept for a linear collider that aims to achieve a compact, high-gradient facility with a normal-conducting distributed-coupling accelerator that is operated at cryogenic temperatures~\cite{vernieri2023cool}.
\CCC\ is designed  
with a wholistic approach that aims to optimise the main linac, collider subsystems, and beam dynamics to deliver the required luminosity at the overall lowest cost. 
\CCC\ was originally optimised for physics programme at centre-of- mass energies of \num{250} and \SI{550}{GeV} to allow for a Higgs physics programme that could also perform measurements of the Higgs self-coupling. 
A total facility length of \SI{8}{km} is sufficient for operation at \num{250}  and \SI{550}{GeV}. 
The initial operation at \SI{250}{GeV} is upgraded to \SI{550}{GeV} simply with the addition of RF power sources to the main linac. 
This is possible because the increase in gradient and resulting increased power requirement is balanced with an adjustment to the beam format to maintain a constant beam-loading, or fraction of the RF power delivered to the beam, which preserves the RF efficiency of the system. 
By optimizing the cavity geometry, RF distribution, and operating at liquid nitrogen temperatures where the conductivity of copper is increased significantly, the peak-power requirement from the high-power RF sources are significantly reduced. 
This allows for a high beam loading, nearly \SI{50}{\%}, making for a compact and efficient machine. 
It is important to emphasise that for a normal conducting machine the cost driver for the main linac is the RF, and through the adoption of distributed-coupling and cold-copper technology peak RF power requirements are reduced by a factor of six.

The present optimised parameters for \CCC\ are listed in Table~\ref{tab:C3baseline}. 
The original and sustainability update parameters (s.u.) are shown. 
The s.u.\ parameters have now been adopted as the baseline after careful analysis~\cite{breidenbach2023sustainability}.

\begin{table}[h!]
\begin{center}
\begin{tabular}{|c | c | c | c | c |} 
 \hline
  Scenario &  \CCC-250 & \CCC-550  & \CCC-250 s.u. & \CCC-550 s.u. \\
   \hline\hline
     Luminosity [x10$^{34}$] & 1.3 & 2.4 & 1.3 & 2.4  \\
Gradient [MeV/m] & 70 & 120  & 70 & 120 \\
Effective Gradient [MeV/m] & 63 & 108 & 63 & 108 \\   %

Length [km] & 8 & 8 & 8 & 8 \\
  Num. Bunches per Train  & 133 & 75 & {266} & {150} \\
  Train Rep. Rate [Hz] & 120 & 120 & {60} & {60} \\
  Bunch Spacing [ns] & 5.26 &  3.5 & {2.65} &  {1.65} \\
  Bunch Charge [nC]  & 1 & 1  & 1 & 1 \\
    Crossing Angle [rad] & 0.014 &  0.014 & 0.014 &  0.014\\
Single Beam Power [MW]  & 2 & 2.45 & 2 & 2.45\\
  Site Power [MW] & $\sim$150 & $\sim$175 & {$\sim$110} & {$\sim$125}  \\ 
\hline
 \end{tabular}
\end{center}
\caption{Baseline \CCC\ parameters.}
  \label{tab:C3baseline}
\end{table}

\CCC\ was optimised based on the incredible advances that were made as part of the ILC and CLIC programmes in the development of electron and positron sources, injector linacs, damping rings, bunch compressors, beam transport, beam dynamics, beam delivery and detectors. 
These highly complex and optimised systems serve as the baseline for what is technologically achievable in the linear collider complex and allow for the optimisation of the main linac using the \CCC\ approach with minor modifications to the relevant sub-systems. 
Indeed, a critical point is that \CCC\ is compatible with the injector complex and beam delivery system that will be installed in the first phase of a linear collider facility with an SCRF technology that reaches a centre-of-mass energy of \SI{250}{GeV} in \SI{20.5}{km}. 
The present design of the damping ring and bunch compressors for \CCC\ will be presented below. 
As is shown there, the size and beam format for the ILC damping ring is compatible with the size of the damping ring of \CCC. 
The extraction kicker for the damping ring would need to be replaced from a fast to a slow kicker. 
The bunch spacing at injection into the damping ring, which is set by the fast kicker into the ring, would need to be optimised to match the bunch format required for the \CCC\ upgrade, in the range of \SI{1}{-}\SI{5}{ns}. 

Of a \SI{20.5}{km} Linear Collider Facility, \SI{15}{km} consist of the main linacs for the collider. 
After completing the initial run of a Linear Collider Facility, we envision the possibility of replacing these \SI{15}{km} with a cold-copper distributed-coupling linac with an energy reach of up to \SI{2}{TeV} in the same footprint. 
Analogously, a centre-of-mass energy of up to \SI{3}{TeV} is possible in a \SI{33.5}{km} facility.  
No additional civil construction (i.e.\ additional tunnel length) would be required. 

The linacs could operate at any energy below the maximum energy (\SI{2}{TeV} or \SI{3}{TeV} depending on the facility length) by a combination of bypass lines or by adjusting the gradient and beam format to maintain a constant beam loading and providing a linear scaling in luminosity with energy and beam power.
The parameters for 1 and \SI{2}{TeV} operation are shown in Table~\ref{tab:TeVmainlinacparam}, and would be accomplished only with the addition of RF sources, upgrades to the beam delivery system and modifications to the bunch format. 
More details on upgrading an SCRF-based LCF with \CCC-technology can be found in~\cite{CCC:ESPPU}.

\begin{table}
\begin{center}
\begin{tabular}{|c | c | c | c |  c |} 
 \hline
  Parameter  & Unit & Value  & Value & Value \\
 \hline\hline
Centre-of-Mass Energy   & GeV & 1000 & 2000 & 3000 \\
 \hline\hline
Site Length & km & 20 & 20 & 33 \\
Main Linac Length (per side) & km & 7.5 & 7 & 10.5\\
Accel. Grad. & MeV/m & {75} & {155} & {155}\\
Flat-Top Pulse Length  & ns & {500} &  {195}  &  {195} \\
Cryogenic Load at 77 K & MW & 14 &  20 &  30 \\
Est. AC Power for RF Sources  & MW  & 68 &  65 &  100 \\
Est. Electrical Power for Cryogenic Cooling  & MW  & 81 & 116 & 175  \\
Est. Site Power & MW  & 200 & 230  & 320 \\
RF Pulse Compression & & {N/A} &  {3X} &  {3X} \\
RF Source efficiency (AC line to linac) & $\%$  & {50} & {80} & {80}\\
Luminosity  & x$10^{34}$ cm$^{-2}$s$^{-1}$  & $\sim$4.5 & $\sim$9 & $\sim$14 \\ 
Single Beam Power  & MW & 5 & 9 & 14 \\ 
Injection Energy Main Linac & GeV  & 10 & 10& 10    \\
Train Rep. Rate & Hz & 60 & 60 & 60  \\
Bunch Charge & nC & 1 & 1 & 1 \\
Bunch Spacing & ns  & 3 &  1.2  &  1.2 \\
\hline
 \end{tabular}
\end{center}
\caption{Main Linac parameters for \CCC\ at \SI{1}{TeV} to \SI{3}{TeV} centre-of-mass energy.}
 \label{tab:TeVmainlinacparam}
\end{table}

Significant technical progress has been made recently in the development of \CCC\ technology.
In particular, the accelerating structures that can achieve this gradient have been built and demonstrated. 
Further research and development is needed for the integration of higher order mode damping, cryomodule integration and demonstration of alignment and vibration tolerances. 
A \CCC\ demonstrator plan was developed to identify the required metrics and experiments that would be needed to advance the technical maturity of the main linac to a level required for a collider design~\cite{nanni2023status}. 
Since this demonstrator plan was published, significant progress has been made on structure design~\cite{schneider2024high}, high gradient testing~\cite{liu2025high}, structure vibration measurements~\cite{dhar2024vibration}, damping materials~\cite{xudesign}, alignment system~\cite{van2024alignment}, low-level RF~\cite{Liu:2025cku} and cryomodule design. 
Ongoing efforts are focused on integration of higher order mode damping, cryomodule design and integration, alignment and vibration measurements, and ultimately the construction and testing of a ``quarter cryomodule (QCM)'' with RF power, beam and full instrumentation for alignment and vibrations. 
The QCM represents an essential step in the process as it contains all of the repeating element that make up the full cryomodules in the main linac. 
On the intermediate timescale a demonstration is needed on the few GeV scale to provide technical maturity for the full scale cryomodules (9~m) that would be used \CCC. 
Such an R\&D programme as proposed at Snowmass is technically achievable before 2030 allowing for \CCC\ technology to be integrated into plans for an upgrade to the Linear Collider Facility. 

The injector complex and the damping rings represent a major capital investment for the LCF. 
Fortunately, with \CCC\ we can reutilise this infrastructure for an energy upgrade of the LCF. 
This is because the damping ring offers an opportunity to adjust the bunch train format from the electron and positron source linacs to match the bunch structure needed for \CCC.

In the original configuration for a SCRF-based LCF the damping ring operates with fast kickers for both the injection and extraction of the electron and positron bunches. 
This is due to the fact that the bunch spacing (hundreds of nanoseconds) is too large to maintain in the damping ring without and excessively large circumference. 
The fast kickers place the bunches with a spacing on the order of five nanoseconds into the damping ring. 
This spacing is well suited for the use of a normal conducting technology such as \CCC. 
Small adjustments to the timing of the injection system and RF system of the damping ring would allow the bunch spacing in the damping ring to match the desired spacing of \CCC. 
The bunch spacing for \CCC\ is nominally in the \num{3} -- \SI{5}{ns} range and possibly as small as \SI{1.5}{ns}. 
The fast kicker in the extraction system would be replaced with a slow kicker to extract the full bunch train which is accelerated in the main linac.

When considering this scheme, it is essential that the circumference of the damping ring be sufficient to provide enough space for the \CCC\ bunch train and enough time for the beam to be damped. 
This also requires that the emittance at the exit of the damping ring be low enough and preserved through the bunch compressors prior to acceleration in the main linac. 
The footprint of the bunch compressor could also limit the energy reach of the collider and should be accounted for. 
In the following sections we show that the size needed for the damping ring is smaller than the SCRF baseline for the LCF. 
A larger damping ring would be lower risk and is therefore acceptable. 
We also show that the emittance is low enough through bunch compression, and the length of the bunch compressors and associated RF is small and on the order of \SI{100}{m}. 
With this preliminary study we can conclude that there is a significant amount of infrastructure that can be reutilised in the injector complex and damping rings of the LCF for a \CCC\ upgrade. 
Future studies should investigate a configuration that matches the circumference of the LCF damping ring. 
Furthermore, R\&D is needed in fast kickers to get down to the single nanosecond timescale for the most tightly spaced bunch formats being considered.

\paragraph*{Damping ring}
\label{sec:DR}

The purpose of the damping ring is to produce a flat beam with as small an emittance as possible subject to the constraint of having a sufficiently fast damping time to match the collider repetition rate. 
Generally, multiple damping cycles are required to allow the injected beam to reach its equilibrium emittance. 
This requirement along with the collider bunch structure which informs the choice of ring circumference, sets roughly the beam energy:
\begin{equation}
\tau_y=\frac{3C}{r_ec\gamma^3I_2}
\end{equation}
where $C$ is the ring circumference, $\gamma$ is the relativistic factor, $r_e$ is the classical electron radius and $I_2$ is the second radiation integral.
\begin{figure}[h!]
   \centering
   \includegraphics*[width=0.5\columnwidth]{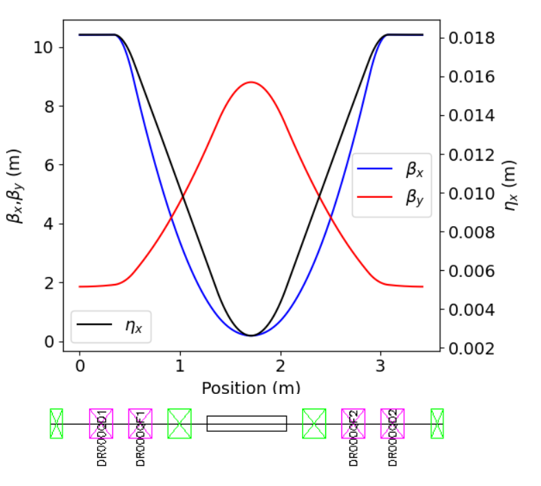}
   \caption{Beta functions and dispersion for a TME cell~\cite{CCC:ESPPU}.}
   \label{TME}
\end{figure}

For the \CCC\ collider the main electron and positron damping rings will have a racetrack layout. 
The arcs are comprised of Theoretical Minimum Emittance (TME) cells while the straight sections are used for injection and extraction and to provide space for damping wigglers. 
An analytic study varying the major parameters (beam energy, number of stored bunches, wiggler strengths and lengths) was performed and a choice of storing two bunch trains was made as a trade-off between size of the machine and achievable emittance resulting in an energy of \SI{4.6}{GeV}. 
Each bunch train is stored for approximately \num{7.9} damping times. 
While this number is relatively large, lowering it by reducing the beam energy results in a larger final emittance when IBS is accounted.

The zero-charge horizontal emittance is largely determined by the number of TME cells in the ring:
\begin{equation}
\gamma\epsilon_x=\frac{C_q\gamma^3\theta^3}{12\sqrt{15}}
\end{equation}
where $\gamma\epsilon_x$ is the normalised horizontal emittance, $C_q$ is a numerical constant and $\theta$ is the bend angle per TME cell. 
The above implies \num{248} cells are required for a baseline zero-charge normalised emittance of \SI{65}{nm}. 
However, at optimal emittance TME cells require tight focusing which introduces a large chromaticity in both planes. 
We therefore moved off of the optimum settings until the normalised chromaticity in each plane was $<\num{3}$ resulting in a zero-charge emittance of \SI{170}{nm}. 
For a cell length of \SI{3.4}{m} the total arc length is \SI{843}{m}. 
Each straight section is \SI{172}{m} long with the majority of that length used for wiggler magnets, but additional space is also included for beam extraction/injection, RF cavities to replenish energy loss from synchrotron radiation and optics for matching the straight sections and arcs and zeroing dispersion. 

In the current design the ring circumference is \SI{1190}{m} which fits \num{2.8} bunch trains. 
It may be possible to reduce the ring circumference by making the TME cells more compact but for an initial study a conservative estimate of \SI{3.4}{m} was used. 
Note that reducing the number of cells would increase the emittance which is not preferable. 
Another possibility may be to increase the circumference to fit 3 bunch trains.

\begin{figure}[h!]
   \centering
   \includegraphics*[width=0.5\columnwidth]{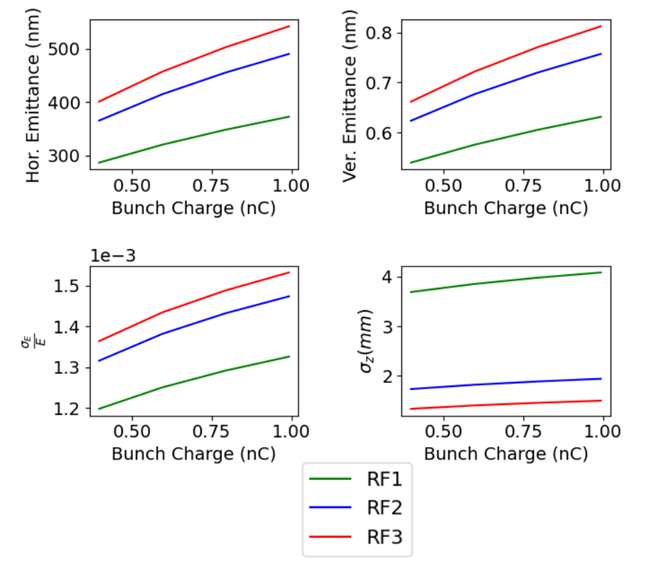}
   \caption{Horizontal, vertical normalised emittances, RMS relative energy spread and bunch length as a function of charge due to IBS. 
   Each color refers to a different RF voltage in the ring~\cite{CCC:ESPPU}.}
   \label{IBS}
\end{figure}

With the major design parameters selected a model of the ring was produced in BMAD. 
A code available in the BMAD ecosystem to model IBS based on the Completely Integrated Modified Piwinski (CIMP) formalism was used (see Fig.~\ref{IBS}).
Final normalised emittances range from \SI{300}{-}\SI{500}{nm} in the horizontal plane depending on RF voltage and bunch charge, while the vertical emittance remains below \SI{1}{nm} for all RF settings and bunch charges considered. 
It is important to note that the vertical emittance value will likely be dominated by magnet tolerances and not IBS, which is an ongoing study.

\paragraph*{Bunch compressors}
\label{sec:BC}

Particle bunches leaving the damping ring (DR) have been damped strongly in the transverse plane but are too long to be accepted into the main linac where they will be boosted to collision energy. 
Therefore, a bunch compression scheme between the damping ring and main linac is required. Because the beam energy leaving the DR is \SI{4.6}{GeV}, ballistic compression is not an option. 
Instead, the compressor consist of two parts: an RF linac which imparts an energy-position correlation (energy chirp) on the bunch, and a magnetic chicane which has a path length that depends strongly on a particle's energy. 

Leaving the damping ring, the particle bunch is uncorrelated in position with a bunch length of the order of a few mm and a relative energy spread of approximately \SI{0.1}{\%} (depending on bunch charge and RF setting in the DR). 
The compression goal is \SI{100}{\mu m}. 
In the absence beam intensity effects (in particular Coherent Synchrotron Radiation (CSR)) the longitudinal emittance is conserved which implies an energy spread on the order of \SI{1}{\%} is expected after compression.

Analytic formulae derived from linear transfer matrices of an RF cavity, bend magnet and drift were used to determine the required $M_{65}$ of the RF system and $M_{56}$ of the chicane for the case of operating at the RF zero-crossing. 
To compensate for a non-negligible  $T_{566}$ from the chicane the beam is moved slightly off the zero-crossing to take advantage of the curvature present in the RF field. 
This results in a slight deceleration of the beam from \SI{4.6}{GeV} to about \SI{3.9}{GeV}. 

An RF frequency of \SI{2.85}{GHz} was chosen since it is a sub-harmonic of the main linac. 
As a reasonable starting point a 15-cell cavity with gradient of approximately \SI{9}{MV/m} was assumed. 
To achieve the required energy chirp for a \SI{100}{\mu m} final bunch length, \num{80} total RF cavities are needed resulting in an approximately \SI{130}{m} long RF section. 
Dipole magnets where chosen to be \SI{2}{m} in length separated by \SI{4}{}m drifts. 
Typical strengths are  \SI{0.5}{T} resulting in a bending angle of \SI{4.5}{^\circ} per dipole. 
For focusing, a FODO cell comprising the length of 8 RF cavities is used. 
The chicane does not contain any quadrupoles. 
Four quadrupoles on each side of the chicane were tuned to keep the beta functions under \SI{30}{m} with relatively weak divergence through it.
\begin{figure}[h!]
   \centering
   \includegraphics*[width=0.8\columnwidth]{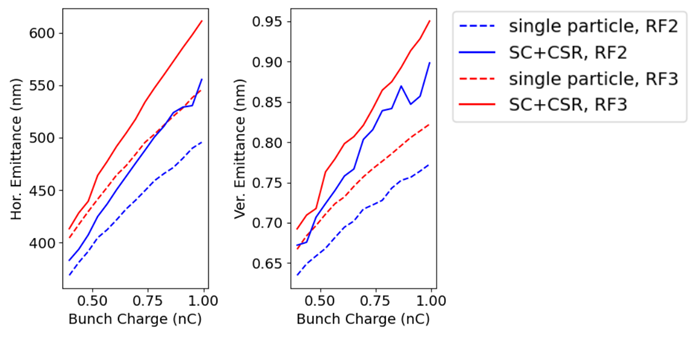}
   \caption{The horizontal and vertical normalised emittances at the end of the bunch compressor~\cite{CCC:ESPPU}. 
   Red and Blue are for two different RF settings in the DR while solid lines include a 1D-CSR calculation while dashed lines do not.}
   \label{CSR}
\end{figure}
The effect of CSR was modeled using BMAD's built-in 1D CSR calculation. 
Tracking was done with \SI{10}{k} particles and \num{100} longitudinal bins (see Fig.~\ref{CSR}). 
We are currently looking into 3D CSR effects. 
However, we note that due to the small transverse emittance and well controlled beta functions in the chicane, the common rule-of-thumb for when 3D effects become non-negligible $\sigma_{x,y} < \rho^{1/3}\sigma_{Z}^{2/3}$ suggest that the 1D model should be accurate.

\paragraph*{Environmental considerations}

Due the the presence of liquid nitrogen in the tunnel, the presence of humans during the cold state of the machine will require proper safety measures. 
These include adequate ventilation, oxygen deficiency monitors, etc. 
Requirements will need to be considered concerning relevant regulations.


\subsubsection{Upgrade using plasma wakefield accelerator technology }

A particular strength of a linear collider at the electroweak scale is that it can be a first step towards a discovery machine that can reach \SI{10}{TeV} parton-centre-of-mass (pCM) collisions. 
Such a machine must necessarily be based on novel accelerator technology with GeV-per-meter accelerating gradients in order to fit within a reasonable facility footprint~\cite{Lindstrom:2025yng}. 
The 2023 P5 Report~\cite{P5:2023} calls for ``vigorous R\&D toward a cost-effective \SI{10}{TeV} pCM collider based on ... wakefield technologies'' and goes on to state ``Wakefield concepts for a collider are in the early stage of development. 
A critical next step is the delivery of an end-to-end design concept, including cost scales, with self-consistent parameters throughout.'' 
The \SI{10}{TeV} Wakefield Collider Design Study has been formed in response to the P5 Report~\cite{10TeV_AAC}.

There are several design studies on future wakefield accelerator (WFA) collider concepts currently underway. 
The HALHF concept~\cite{Foster:HALHF2023} proposes an asymmetric collider where high-energy, plasma-accelerated electrons collide with low-energy positrons accelerated in conventional RF linacs, thus avoiding the challenge of positron acceleration in plasma~\cite{Cao2024:Plasma}. 
The HALHF design addresses previous critiques of WFA colliders~\cite{Adli:2025qzu}.  
ALIVE~\cite{Farmer:ALIVE2024} considers the proton beam-driven plasma wakefield acceleration where the enormous energy stored in the LHC proton beam is harnessed to accelerate electrons to the TeV scale in a single plasma stage. 
The \SI{10}{TeV} pCM Wakefield Collider Design Study aims to develop a self-consistent design of an energy-frontier linear collider based on beam-driven plasma, laser-driven, or structure-based wakefield acclerators~\cite{10TeV_AAC}.

\paragraph*{Plasma booster}
The idea of upgrading a linear collider using newer technology with a higher accelerating gradient has long been discussed~\cite{Raubenheimer:2004} and is a major advantage of a linear compared to a circular collider. 
The difficulty of such an upgrade depends on the similarity of the technologies involved. 
Beam-driven plasma-wakefield acceleration (PWFA), the only WFA technology currently capable of achieving the efficiencies required for collider applications, is rather dissimilar to that of the RF cavities of the baseline facility. 
Implementing it with minimal changes to a running Higgs factory is complex but seems at least in principle feasible. 

Because no efficient and tested scheme to accelerate positrons in a plasma exists, the booster is only used in the electron arm, which leads to a boost to the centre-of-mass energy of the electron-positron system as in the HALHF concept. 
However, the boost required to upgrade for instance a symmetric \SI{250}{GeV} collider to an asymmetric \SI{550}{GeV} one is much smaller than in HALHF, since \SI{550}{GeV} is achieved with a positron energy of \SI{137.5}{GeV}  and an electron energy of \SI{550}{GeV}, which corresponds to a boost of $\approx 1.25$. 
This is sufficiently small to not require significant changes to the detector, except for the luminosity monitoring. 
Should efficient positron acceleration be developed in future, then boosters could be used in both arms, restoring symmetric collisions.  

For this exercise, it is assumed that the Higgs factory to be boosted is based on the ILC TDR design at the baseline  luminosity. 
The centre-of-mass energy is increased from  \SI{250}{GeV} to \SI{550}{GeV}, appropriate for top-pair and Higgs-pair production. 
This requires a slight increase in the positron energy to \SI{137.5}{GeV}, achieved by assuming that the cavities can be run at a gradient of \SI{35}{MV/m},  and an electron energy of \SI{550}{GeV}, produced by the PWFA booster. 
The major complication in using PWFA is the necessity to produce the drive beams required to ``blow-out'' the electrons to form the accelerating cavity in the plasma. 
The number of drive beams required depends on the number of ``stages'', or plasma cells, in the design. 
This depends on a detailed optimisation of costs, operational feasibility, cooling etc.\ for the required application~\cite{Foster:HALHF2025} that is beyond the scope of this exercise. 
To be concrete, the number of stages used in the original HALHF design, \num{16}, is used. 
Thus, \num{16} drive beams must be manufactured, fed to the PWFA cells and synchronised so as to accelerate the electron bunch coherently. 
The most cost-efficient way to do this is to repurpose the original cavity-based electron arm to produce the required number of low-energy drive bunches  in addition to the high-quality bunch destined for collision at the IP. 
Overall, the cavities are run at lower gradient but accelerate higher currents; an additional factor of \num{2.7} in the RF power in the electron arm is required because of the addition power required for the drive beams. 
This is somewhat higher than the doubling that was forseen for the ILC luminosity upgrade in the TDR. 
It can be achieved by providing six klystrons per nine cryomodules rather than the baseline of two; sufficient space is available, and the segmentation of the power distribution system makes this modification easy. 
The number of colliding bunches per bunch train is halved in order to keep the power/colliding bunch approximately constant; the bunch charge for the positrons can then be doubled for the same RF power.
If \SI{5}{Hz}  running is assumed, then the produced luminosity would be half that assumed for a \SI{500}{GeV} standard ILC upgrade. Running at a \SI{10}{Hz} repetition rate, which may be possible, would restore the luminosity. 

The area in which the booster is situated is assumed to be that containing the undulator that produces polarised positrons. 
Since additional space perpendicular to the beam line is required for the drive-beam distribution, this must either have been planned during the original construction or the cavern must be extended somewhat, since polarised positrons will still be required in the upgrade. 

Complications arise from the bunch-compressor design for the ILC, which also accelerates from 5 to \SI{15}{GeV}. 
This is unable to cope with the necessary drive-beam current. 
It is therefore necessary to construct a separate drive-beam injector. 
However, injecting at \SI{15}{GeV} would be too expensive. 
A possible solution is to reduce the injection energy for the drive beams to \SI{2}{GeV} and to construct a bypass for the \SI{15}{GeV} colliding beam from the bunch compressor, returning it to the main beam axis just before the PWFA cells. 
Such a bypass requiring only a simple FODO lattice would be inexpensive and would fit inside the main tunnel. 

Other complications include the bunch length of the colliding bunch. 
This must be sufficiently short to fit within the plasma blow-out bubble, whose length is approximately the plasma wavelength, which is inversely proportional to the plasma density. 
The relatively low plasma densities used in HALHF, which facilitate the operation of the PWFA arm, around \SI{6E14}{cm^{-3}}, still require much shorter bunch lengths than typical for ILC, of around \SI{40}{\mu m} rather than \SI{300}{\mu m}. 
The ILC TDR bunch length is mostly determined by the electron-damping-ring characteristics. 
The HALHF baseline assumption is that polarised sources of  characteristics such as to obviate the necessity for a damping ring will be available before construction. 
This would make the production of short pulses much easier. 
Additionally, the shape of the drive bunches, which should provide a transformer ratio of around 2, are asymmetrical and therefore lead to an increase in high-order mode losses (HOMs) which must be damped, leading to a concomitant increase in the cryoload, which is power hungry. 

In summary, there are a number of challenges that need to be addressed if a detailed design of a PWFA-based energy booster were to be pursued. It should be remarked that CLIC technology is much more compatible with a PWFA booster, in terms of bunch patterns, bunch lengths etc., than is ILC. However, it seems possible that either technology can be made compatible with a PFWA booster. 
There do not appear at this stage to be show stoppers; such a booster could provide a cost-effective way to increase the energy of an initial linear-collider Higgs factory.
It also would provide an important proof of principle for a much higher energy linear collider, as discussed in the next section.

\paragraph[Fully wakefield-based accelerator]{Fully wakefield-based accelerator}
The alternative to a booster-upgrade is a multi-TeV upgrade (up to \SI{10}{TeV}) of a linear collider facility based on wakefield accelerators (either beam-driven plasma, laser-driven plasma, or structure-based wakefield accelerators, depending on technological developments). 
The accelerating gradient in the plasma or structure may well-exceed \SI{10}{GV/m}. 
One of the primary challenges the \SI{10}{TeV} Wakefield Collider faces is the operation of multiple wakefield stages in succession~\cite{Lindstrom:2020pzp}. 
In the case of beam-driven plasma acceleration, the wakefield stages are separated by magnetic chicanes and focusing lattices, which may be longer than the accelerating plasma in each stage. 
This results in a reduction of the \emph{geometric gradient}. 
If the LCF \SI{550}{GeV} SCRF option is built, each linac will be \SI{11}{km} in length. 
Therefore, the minimum geometric gradient to achieve \SI{5}{TeV} beam energy in this \SI{11}{km} tunnel is \SI{450}{MV/m}. 
It is important to examine  trade-offs between the local maximum gradient in the plasma, the length of the interstage optics, and the number of plasma stages in order to achieve this geometric gradient. 
Such problems would be solved if the ALIVE concept~\cite{Farmer:ALIVE2024, ALIVE:EPPSU} of proton-driven PWFA can be successfully developed, since this is envisaged to achieve high energies in a single stage. 
It does however also require a high-energy proton accelerator on site, which would be available if the facility were built at CERN or Fermilab. 
In addition to the staging challenge, the \SI{10}{TeV}  Wakefield Collider must either solve the problems of accelerating positrons in plasma efficiently~\cite{Cao2024:Plasma}, or further develop the $\gamma\gamma$ collider concept for ultra-high energy collisions.

The \SI{10}{TeV}  Wakefield Collider Design Study is currently being formed to address these challenges~\cite{10TeV_AAC}. 
The design study has been initiated by the 2023 P5 Report~\cite{P5:2023}, which calls for ``vigorous R\&D toward a cost-effective \SI{10}{TeV}  pCM collider based on proton, muon, or possible wakefield technologies.'' 
Specifically, the P5 Report requests ''the delivery of an end-to-end design concept, including cost scales, with self-consistent parameters throughout.'' 
The design study is working in close collaboration with the ongoing HALHF~\cite{Foster:HALHF2025}, ALIVE~\cite{Farmer:ALIVE2024}, and ALEGRO~\cite{Cros:2019tns, Proceedings:2024ncv} efforts to design colliders based on wakefield technologies. 

The \SI{10}{TeV}  Wakefield Collider Design Study must deliver solutions to the following challenges in order to achieve high-energy, high-luminosity collisions.
\begin{itemize}
    \item Full acceleration of beams with charge $> \SI{100}{pC}$ and emittance $< \SI{100}{nm}$.
    \item Emittance preservation and stability of staged plasma acceleration.
    \item Beam Delivery Systems for \SI{5}{TeV}  beams.
    \item Mitigation of beam-beam effects at \SI{10}{TeV}  pCM.
\end{itemize}
In addition, if an electron-positron collider is required, the efficient acceleration of positrons in plasmas must be solved.

Assuming that these challenges can be addressed both in theory and in proof-of-principle demonstrations, the next step is to understand their implementation in the context of an existing linear collider facility. 
For example, beam-driven plasma staging requires multiple parallel beamlines to deliver drive and main beams to each plasma stage. 
Are the LCF tunnels wide enough to accommodate the necessary separation between the beamlines? 
In the case of a collider based on laser-driven accelerators, large laser rooms must be built near the beamline. 
Both of these design choices impact the civil engineering of a linear collider facility, but current wakefield collider designs are not mature enough to make specific recommendations for facility enhancements that might accommodate a \SI{10}{TeV}  upgrade. 
The \SI{10}{TeV}  Wakefield Collider Design Study aims to address questions related to civil engineering and deliver its report by 2028, well in-time to provide input to the construction process.

\paragraph*{Timeline and R\&D needed}
\begin{itemize}
   \item \textbf{Electron beam-driven plasma}
    Table~\ref{tab:PWFAreadiness} indicates the timescale necessary for the required development of PWFA technology to produce a Technical Design Report for the HALHF project and its current technological readiness. 
    The technology and design thus developed would have far-reaching implications and be applicable to all plasma-wakefield accelerator facilities.
\item \textbf{Proton beam-driven plasma}
The ideas for proton-driven acceleration for a collider are being explored in the ALIVE project. 
The major gain of such a facility would be the possibility of accelerating to TeV energies in a single stage, leading in principle to significantly higher geometric gradients because interstage optics inserts which grow in length with energy is avoided. 
However, significant R\&D is required, in particular in areas such as rapid-cycling proton accelerators, if the luminosity of such machines is to be large enough to be interesting. 
    \item \textbf{Laser-driven plasma}
    There is very substantial activity in laser-driven plasma acceleration, which, not needing powerful accelerators, is more widely spread worldwide than is the case for PWFA. 
    Investigations with LWFA can illuminate many of the areas that require development in plasma-acceleration facilities, particularly single-stage properties, emittance preservation, etc. 
    The implementation of a laser-driven collider must however await an orders-of-magnitude improvement in the efficiency and repetition rate of the high-power lasers required. 
    The good news is that such developments are required in a much broader scientific field than particle physics and are driven by major industrial actors. 
    There is therefore hope that by the time that a plasma-wakefield accelerator could be seriously contemplated, the required advances will have been made. Scientists inside the ALEGRO framework~\cite{Cros:2019tns, Proceedings:2024ncv} are pursuing research in this area.
    \item \textbf{Structure wakefield acceleration}
    The idea of using dielectric structures to maintain high accelerating fields has been developed in recent decades. 
    Very interesting results have been obtained for a variety of configurations~\cite{Structure_review}. 
    Two of these have been investigated in the context of upgrades to HALHF~\cite{Foster:HALHF2025}. 
\end{itemize}

\begin{table}
\footnotesize
    \begin{tabular}{>{\raggedright}p{0.11\linewidth}>
    {\raggedright}p{0.03\linewidth}>
    {\raggedright}p{0.09\linewidth}>
    {\raggedright}p{0.08\linewidth}>
    {\centering}p{0.13\linewidth}>
    {\centering}p{0.08\linewidth}>
    {\centering}p{0.09\linewidth}>
    {\raggedright}p{0.2\linewidth}}%
    \hline
Critical parameters                                               & TRL    & R\&D time (y) (Design/Total)                                                      & R\&D current (MSF) & R\&D needed (M\euro) (Design/Total)& FTE current & FTE needed & Comments                                 \cr
    \hline
Electron beams \textgreater{}~\SI{100}{GeV} & 1   & 7-8/11-13 &                                                                       & 7/100                                                & 1  & 40  & No PWFA-test facilities have produced \textgreater \SI{100}{GeV} beams                                                      \cr
    \hline
Acceleration in one stage ($\sim$ \SI{10}{GeV})   & 5    & 5/9-10 &                                                                       & 10/100                                              & 3                                & 50                                                                   & AWAKE demonstration but  technology may not be suitable                                             \cr
    \hline
Plasma uniformity (long \& trans.)                                & 4      & 5/9-10    &                                                                       & 2/100                                                & 2                                                                       & 15                                                                   & AWAKE demonstration but  technology may not be suitable                                                              \cr
    \hline
Preserving beam quality/emittance                      & 3.5      & 7-8/11-13 & 0.5 (ERC + Oslo national )                                & 3/100                                              & 5                      & 25                                                                   & Normalised emittance preserved at \textless 3 um levels with small currents\cr
    \hline
Spin, polarisation                                                & 2                                                                       & 5/9-10    &      0.1 (DESY)                                                                 & 3/100                                                & 1                                                 & 16                                                                   & Technology concept formulated                  \cr
    \hline
Stabilisation (active and passive)                                & 3                                                                        & 7-8/11-13 &                                                                       & 1/100                                               & 1                                                                       & 10                                                                   & Studies at AWAKE and LWFA, but not at HALHF requirements                                                          \cr
    \hline
Ultra-low-emittance beams                                         & 2                                                                           & 7-8Y/11-13&                                                                       & 3/100                                                & 0                                                                       & 20                                                                   & Not yet at collider emittance. Better test facilities required.                                                                                                                   \cr
    \hline
External injection and timing                                     & 4                                                                       & 7-8/11-13 &                                                                       & 1/100                                               & 0                                                                       & 10                                                                   & Precise timing for external injection demonstrated at AWAKE                                                                         \cr
    \hline
High rep-rate targetry with heat management                       & 2                                                                      & 5/9-10   &                                                                       & 7/100                                                & 3                                               & 40                                                                   & Heat modification of plasma properties/profile and  target cooling requires inew concepts                 \cr
    \hline
Temporal plasma uniformity/stability                           & 4                                                                    & 5/9-10    &                                                                       & 3/100                                               & 0                                                                       & 10                                                                   & AWAKE demonstration but technology may not be suitable                        \cr
    \hline
Driver removal                                          & 2                                                                         & 7-8/11-13 &                                                                       & 2/100                                               & 0.5                                                                     & 10                                                                   & HALHF concept exists                                                     \cr
    \hline
Drivers @ high rep.\ rate \& wall-plug eff.                 & 5                                                                      & 5/9-10   &                                                                       & 5/100                                                & 0.5                                                                     & 10                                                                   & Similar to CLIC driver, demonstrated in CTF3         \cr
    \hline
Interstage coupling                              & 2                                                                & 7-8/11-13 & 1 (ERC)                                                        & 3/100                                                & 3                                                                       & 10                                                                   & HALHF concept exists              \cr
    \hline
Total system design with end-to-end                               & 3                                                                      & 3-4                                                       & ~0.5 ( Oslo, Oxford)                                           & 3                                                     & 2                                                                 & 20                                                                   & Not yet at pre-CDR level. Aim for pre-CDR document early in 2026.    \cr
    \hline
Simulations                                                       & 5                                                                   & part of above                                                   & 0.5 (ERC + Oslo national)                                & part of above                                                                 & 4                                                 & 5                                                                    & Single-stage simulation (HIPACE++)  well developed - dedicated framework (ABEL) for start-to-end                \cr
    \hline
Self-consistent design                                            & 4                                                                    & part of above                                                   & part of above                                                         & part of above                                                                 & in previous 2 rows                                            & 5                                                                    & Plasma linac start-to-end simulations performed using HIPACE++/ABEL \cr          
     \hline
\end{tabular}
\caption{HALHF Plasma Arm R\&D: Technology Readiness Levels (TRL), required resources and timescales to produce TDR. ``FTE current'' means currently in place; ``needed'' is integrated total requirement. 
``R\&D current'' is only non-zero if dedicated to HALHF. 
(Adapted from LDG Accelerator Roadmap review submission, Feb. 2025).}
      \label{tab:PWFAreadiness}
\end{table}

\paragraph*{Facilities and timelines}
There are several facilities world-wide are carrying out work relevant for wakefield-based colliders. 
A summary of research goals and capabilities for Beam Test Facilities in the US was created for the recent Snowmass process~\cite{Power:2022}, and a summary of research goals and capabilities for Beam Test Facilities in Europe was created for the LDG process~\cite{Muggli2022}.
\begin{itemize}
    \item \textbf{AWAKE} ~\cite{Gschwendtner2022} is engaged in a long-term development programme to explore the parameters of proton-driven wakefield acceleration. 
    It has a variety of ``spin-off'' applications for physics, in particular for high-field QED and fixed-target experiments. Developments that could lead to a collider based on these techniques are concentrated in the ALIVE project.
    \item The \textbf{FACET-II} programme at SLAC focuses on high-efficiency, high-quality two-bunch PWFA with parameters relevant for future colliders~\cite{joshi2018plasma}. 
    FACET-II provides \SI{10}{GeV} electron beams at extremely high peak currents~\cite{Emma2025}. Its advantages include the ability to carefully control the drive-witness separation and bunch profiles. 
    FACET-II has some infrastructure creating positron beams and accelerating them in plasma, but further investment is required to enable this capability.
    \item The \textbf{FLASHForward} facility at DESY has been running for several years and produced results, particularly in emittance preservation and plasma recovery time, of direct relevance to collider applications~\cite{DArcy2019}. 
    It is particularly suited to the study of high-power effects as it can run with multiple bunches closely separated to study large instantaneous power deposition in the plasma cell.
    \item The \textbf{Argonne Wakefield Accelerator (AWA)} facility specialises in beam-driven wakefield acceleration, with a focus on the structure wakefield accelerator technology~\cite{Power:2022}. 
    It has pioneered advanced phase space manipulation techniques for temporal bunch shaping, and is capable of producing electron-bunch trains having charges up to $\simeq \SI{400}{nC}$. 
    Operating at beam energies of approximately \SI{70}{MeV}, the facility contributes significantly to developing next-generation compact accelerators through collaborations with institutions worldwide.
    \item The \textbf{BELLA} centre at LBNL has been performing research on laser-plasma accelerators (LPAs) for over two decades. 
    The main research objectives are the development of LPA modules at the \SI{10}{GeV} level~\cite{Picksley2024} and the staging (coupling) of LPA modules, which are two essential R\&D components for a future plasma-based linear collider. 
    BELLA is pursuing research into coherent combination of fiber lasers for high repetition rate LWFA.
    \item \textbf{SPARC-LAB} at INFN Frascati develops plasma source technologies and has demonstrated precision applications of beam-driven plasma accelerators, such as FELs~\cite{Pompili2022}. 
    The R\&D at SPARC-LAB lays the groundwork for EuPRAXIA, the world's first plasma accelerator-based User Facility~\cite{Assmann2020}.
    \item \textbf{High-Intensity Laser Facilities}:
    A number of laser facilities, including ZEUS, ELI, APOLLON, GEMINI and others pursue LWFA research with very high intensity lasers. 
    Goals for these facilities include single-stage very high energy acceleration.
\end{itemize}
If a PWFA facility such as HALHF is to make progress, it will be necessary to build a substantial test facility to test the various parts of the system if a real accelerator. 
Such a test facility, as proposed for example by the SPARTA project~\cite{Lindstrom:SPARTA}, would produce beams of around \SI{100}{GeV}, which can additionally be very useful for e.g.\ strong-field QED studies or beam-dump experiments. 
The time-scale for such a facility, assuming sufficient resources are obtained for the necessary R\&D, could be around a decade. 
In fact, as discussed in Sec.~\ref{sec:beyond:tunedump:PWA}, a linear collider facility in an initial, SCRF-driven configuration, could itself provide opportunities for an R\&D facility with the necessary beams and experimental environment to conduct and finalise this research.

\paragraph*{Necessary civil construction and environmental impact}
The strength of PWFA colliders is precisely in the area of civil construction and environmental impact. 
Since the effective accelerating gradient is at least an order of magnitude greater than for metallic cavities, the length for a particular application is greatly reduced, and thereby the construction cost and the environmental impact. 
The latter is usually dominated by the carbon footprint of the concrete required for the tunnel. 
For the case of the HALHF Higgs factory, for example, the length of the \SI{250}{GeV} centre-of-mass version is around \SI{5}{km}, much shorter than ILC. 

For many of the PWFA upgrades to an existing ILC discussed in this section, the additional construction and environmental impact is negligible. 
For example, for the PWFA booster discussed above, the only major impact would be the requirement for a larger alcove/cavern at the site of the undulator in order to make space for the drive-beam distribution system. 
This would preferably be part of the initial construction, or a subsequent minor excavation.


\subsubsection{Upgrade using structure wakefield accelerator technology} 

As mentioned in the previous section, structure wakefield acceleration (SWFA), the acceleration of beams in fixed dielectric or metal structures at very high frequency, has also been explored as an approach to high gradient electron acceleration~\cite{Structure_review}. 
Though this cannot reach such high gradients are are possible in Plasma Wakefield Acceleration, SWFA is $\Pem/\Pep$-symmetric and makes the problem of staging somewhat simpler.  

In this section, we explore two applications of SWFA to a future $\sqrt{s} = \SI{10}{TeV}$ pCM linear collider housed in the LCF tunnel. 
The first application is (i) a fully SWFA-based linear collider, while the second option is (ii) a hybrid-WFA that uses SWFA for the positron ($\Pep$) linac and plasma-WFA for the electron ($\Pem$) linac. 
By assessing the feasibility of these options, we explore how SWFA can offer a viable quasi-conventional high-gradient solution alongside PWFA and LWFA. 
This assessment draws on collaborative efforts from the newly formed \SI{10}{TeV} pCM Wakefield Collider Design Study~\cite{10TeV_AAC} and updated beam-delivery system (BDS) scaling from the CLIC \SI{7}{TeV} design report at CERN while leveraging recent SWFA progress at Argonne National Laboratory (ANL).

SWFA includes two configurations: collinear wakefield accelerator (CWA), where the drive and main beams share the same path (similar to PWFA), and two-beam acceleration (TBA), pioneered at CERN's CLIC test facility, where the two beams follow parallel paths. 
For the two linear collider applications considered here, we focus on TBA-SWFA, given its maturity compared to CWA-SWFA. Specifically, we consider the short-pulse TBA-SWFA scheme under development at ANL that employs structures driven by short RF pulses to achieve high gradients.

The short-pulse TBA-SWFA approach is motivated by the CLIC programme's empirical scaling law,
\beq
    \text{BDR} \propto E^{30} \times \tau^5 \ ,
\eeqn
where $E$ is the electric-field gradient [MV/m] and $\tau$ is the RF pulse duration~\cite{Grudiev2009} and $E \propto G$, where $G$ is the average-energy gradient [MeV/m]. 
This equation indicates that shorter pulses enable higher gradients. 
While the CLIC \SI{3}{TeV} collider~\cite{Aicheler:2012bya} is based on $\tau = \SI{240}{ns}$ pulses to achieve $G = \SI{100}{MV/m}$, the Argonne Flexible Linear Collider (AFLC) uses $\tau = \SI{22}{ns}$ pulses to operate at $G = \SI{267}{MV/m}$~\cite{Jing2013}. 
This trend of shorter pulses enabling higher gradients continues with recent short-pulse TBA research at ANL that has demonstrated $G \simeq $\SI{300}{MV/m} in a 3-cell travelling-wave structure and $E \simeq $\SI{400}{MV/m} (corresponding to $G \simeq $\SI{200}{MV/m}) in a standing-wave RF gun, both driven by 9 nsec X-band (\SI{11.7}{GHz}) RF pulses ~\cite{Tan2022}. 
The ongoing research at ANL is targeting $E \simeq \SI{1}{GV/m}$.

\begin{enumerate}
\item \textbf{Fully SWFA-based linear collider.} This application proposes a \SI{10}{TeV}  linear collider employing short-pulse TBA-SWFA for both the electron ($\Pem$) and the positron ($\Pep$) linac, each operating with an average energy gradient of $G \simeq \SI{500}{MeV/m}$ over \SI{10}{km}. 
This configuration yields \SI{5}{TeV}  per linac with a total linac length of \SI{20}{km}. 
The CERN BDS analysis for a 7-TeV collider, which requires \SI{7}{km} per linac, scales to approximately \SI{8}{km} for \SI{10}{TeV}, leaving \SI{5}{km} for the injector complex and IP. This layout fits within the 33.5-km ILC tunnel. Note, the proposed short pulse TBA-SWFA \SI{10}{TeV}  design is an extension of the previous short pulse TBA-SWFA \SI{3}{TeV} , AFLC strawman by ANL~\cite{Jing2013}.

\item \textbf{Hybrid-WFA.} This application proposes a hybrid \SI{10}{TeV}  linear collider that uses short-pulse TBA-SWFA for the $\Pep$ linac and plasma-WFA for the $\Pem$ linac. The advantage of TBA-SWFA is that it is charge-independent and can easily serve as the $\Pep$ linac, which remains a challenge for plasma-WFA. This application can be considered for both LCF baseline options based on tunnel lengths of \SI{20.5}{km} and \SI{33.5}{km}. While the \SI{33.5}{km} option was already outlined above, the case for a \SI{20}{km} tunnel is similar to the asymmetric Higgs Factory, HALHF~\cite{Foster:HALHF2023}. As a motivating example, one could use a \SI{5}{km} e+ TBA-SWFA linac (\SI{2.5}{TeV}) and a \SI{5}{km} e- plasma-WFA linac (\SI{10}{TeV} ) to give $\sqrt{s} = \SI{10}{TeV}$. Assuming \SI{8}{km} for BDS this leaves \SI{2.5}{km} for the injector and IP; further optimisation is needed.

\end{enumerate}

These  two applications toward a \SI{10}{TeV}  pCM collider~\cite{10TeV_AAC} based on short-pulse TBA-SWFA with average-energy gradient of 500~MeV/m appear feasible within both the \SI{33.5}{km} and \SI{20.5}{km} LCF tunnels. 
They are encouraged by current designs and ongoing experimental progress in TBA-SWFA. 
This approach enables effective reuse of the ILC tunnel once its Higgs-factory programme has concluded, thus offering an opportunity to extend the programme to the energy frontier. 
Continued validation of the gradient will strengthen the Linear Collider Facility's contribution to CERN, advancing the energy-frontier programme through collaborative wakefield research.

\subsection{Luminosity upgrades}
\label{sec:acc:lumiup}

Linear Colliders can also be upgraded to higher ---  or even much higher ---  luminosities. In this section, we will first show paths towards higher luminosity specifically at the \PZ pole and the $\PW\PW$ threshold, before discussing the general option to increase the collision rate for all energies.
Finally, we give two examples of how the concepts of energy and particle recovery could boost the luminosity of a Linear Collider by up to two orders of magnitude. 

\subsubsection{Higher luminosity at lower energies: Z pole and WW threshold}
\label{sec:acc:lumiup:zpole}

   The initial accelerator configuration of a Linear Collider can 
also run at the \PZ pole. For the ILC, a conservative estimate yields  
an instantaneous luminosity of \SI{2.05E33}{cm^{-2}s^{-1}} with polarised electron and positron beams as described in~\cite{ILCInternationalDevelopmentTeam:2022izu} and in detail in~\cite{Yokoya:2019rhx}. 
It can also run around the $\PW\PW$ threshold if desired; 
interpolating between the \PZ pole and \SI{250}{GeV}  
to running at $\sqrt{s}=$\SI{161}{GeV} leads to 
an instantaneous luminosity\footnote{This assumes a repetition 
rate of \SI{3.0}{Hz} specifically for $\PW\PW$ threshold operation with the initial ILC-250 configuration.} of \SI{3.65E33}{cm^{-2}s^{-1}}.
The luminosity at the \PZ pole exceeds that of LEP by two orders of magnitude and together with the polarisation is sufficient to produce five billion Z's and accumulate \SI{100}{fb^{-1}} in a couple of years. 
Given the likely investment in a facility that can reach to much higher energy including to \SI{550}{GeV} and above, it makes sense to consider how to efficiently address the physics goals at lower energy as part of the integrated physics programme of a TeV-scale linear collider.
Two of the primary physics drivers discussed in Sec.~\ref{sec:Zpole} that showcase unique linear collider capabilities are the measurement of $A_{\mathrm{LR}}$ at the \PZ pole~\cite{Hawkings:1999ac} and the option for a dedicated $\mw$ measurement from a polarised threshold scan~\cite{Wilson:2016hne}.
The two relevant aspects are the potential for higher luminosity at lower centre-of-mass energies, and the provision of the highest possible polarisation of both beams, thus enabling target physics precisions to be reached in modest running times. 

At lower energies, especially at the \PZ pole, beamstrahlung is naturally suppressed. 
The characteristic centre-of-mass energy spread at the \PZ pole, including the effects of beamstrahlung, $\sigma_{\mathrm{eff}}$\footnote{Defined as half of the range covered by the central 68\% of the centre-of-mass energy luminosity distribution normalised to the nominal centre-of-mass energy.}, is \SI{0.25}{\%}; this is much less than the related fractional half-width of the \PZ (\SI{1.4}{\%}). 
Therefore, the physics constraints leave considerable scope for adopting collision conditions that lead to higher luminosities per bunch crossing, while modestly broadening the luminosity spectrum.
A number of potential improvements have been investigated that may lead to combined luminosity increase factors in the range of \SI{2}{-}\num{10}. 

These include both, straightforward increases of repetition rate, $f_{\mathrm{rep}}$, and/or bunch number, $n_{\mathrm{b}}$, and changes that increase the luminosity per bunch crossing, $\mathcal{L}_{\mathrm{BX}}$, by accepting increased beamstrahlung leading to higher luminosities. 
The latter involves both increases to the geometric luminosity, $\mathcal{L}_{\mathrm{geo}}$, that is a function of the number of electrons/positrons per bunch, $N$, and the beam sizes at the interaction point ($\sigma_{x,y}^{*}$), and  additional increases from the luminosity enhancement factor associated with the pinch effect, $H_{D}$.
The instantaneous luminosity is given by 
\[ \mathcal{L} = f_{\mathrm{rep}} \,  n_{\mathrm{b}} \, \left(\frac{N^{2}}{4 \pi \sigma_{x}^{*} \sigma_{y}^{*}}\right) H_{D} =  f_{\mathrm{rep}} \,  n_{\mathrm{b}} \,  \mathcal{L}_{\mathrm{BX}}  =  f_{\mathrm{rep}} \,  n_{\mathrm{b}} \,  \mathcal{L}_{\mathrm{geo}} \, H_{D}   \: \: ,\]
where $\mathcal{L}_{\mathrm{BX}}$ and consequently $H_{D}$ are evaluated 
with beam-beam simulations. 
The vertical disruption parameter, $D_{y}$, which is the ratio between the bunch length, $\sigma_{z}$, and the focussing length for the vertical focussing from the electromagnetic forces acting like a lens upon the colliding bunch, and is given by  
\[   D_{y} = \frac{2\, r_{\mathrm{e}} \, N \, \sigma_{z} } {\gamma \, \sigma_{y}^{*} \, ( \sigma_{x}^{*} + \sigma_{y}^{*} ) } \: \: ,\] 
plays a key role in determining $H_{D}$~\cite{Yokoya:1991qz}.

The straightforward changes are:
\begin{itemize}
\item Doubling the bunch number to 2625 as already assumed feasible in~\cite{Yokoya:2019rhx}. 
\item If sufficient cryogenic cooling capacity is available, the repetition rate can be increased to permit \num{5}+\SI{5}{Hz} 
operation\footnote{Where the electron linac uses one cycle for undulator-based positron production at \SI{125}{GeV} or above, and one cycle for \SI{45.6}{GeV} or \SI{80.5}{GeV} beams for physics collisions.}. 
Consequently the collision repetition rate can increase to \SI{5.0}{Hz} for both \PZ and $\PW\PW$ threshold operation (from \num{3.7}/\SI{3.0}{Hz} for \PZ pole/$\PW\PW$ threshold). 
\end{itemize}
Combined, these changes lead to luminosities of \num{5.54e33}\,$\mathrm{cm}^{-2} \mathrm{s}^{-1}$ 
and \num{12.2e33}\,$\mathrm{cm}^{-2} \mathrm{s}^{-1}$ at the \PZ pole and $\PW\PW$ threshold respectively.
The LCF proposal includes these changes.

Individual changes were investigated affecting the luminosity per bunch crossing. The studies were done for a centre-of-mass energy of \SI{91.2}{GeV} 
using Guinea-PIG++~\cite{Rimbault:2007wfy} 
starting from the ILC-250 baseline \PZ-pole operation beam parameters with $D_{y}=31.4$.
\begin{itemize}
\item Increasing the bunch length ($\sigma_z$): Increasing  $\sigma_{z}$ from 410\,$\mu$m to e.g.\ 615\,$\mu$m by reducing the compression in the bunch compressor also reduces the linac beam energy spread from \SI{0.30}{\%} to \SI{0.20}{\%}.
The \SI{50}{\%} increase in $D_{y}$ leads 
to a modelled \SI{4.6}{\%} increase in luminosity and a \textit{decrease} in $\sigma_{\mathrm{eff}}$ by a factor of \num{1.34}.
\item Decreased horizontal and vertical normalised emittances: 
For instance, if these could be decreased from \SI{6.2}{\mu m}/\SI{48.5}{nm} to \SI{5.6}{\mu m}/\SI{42}{nm}, the modelled $\mathcal{L}_{\mathrm{BX}}$ increases by \SI{20}{\%} with a \SI{2}{\%} increase in $\sigma_{\mathrm{eff}}$.
\item Decreased horizontal beta function: 
Decreasing $\beta_x^{*}$ from \SI{18}{mm} to e.g.\ \SI{12}{mm} leads to a modelled overall increase of \SI{53}{\%} in $\mathcal{L}_{\mathrm{BX}}$ with a factor of \num{1.26} increase in $\sigma_{\mathrm{eff}}$. 
\item Increased bunch population ($N$) at fixed beam power:
Increasing $N$ from \num{2.0e10} to e.g.\ \num{2.5e10} while decreasing $n_{\mathrm{b}}$ by a factor of \num{1.25} leads to a \SI{56}{\%} increase in $\mathcal{L}_{\mathrm{geo}}$, a modelled overall increase of \SI{79}{\%} in $\mathcal{L}_{\mathrm{BX}}$, and \SI{43}{\%} in instantaneous luminosity, with a factor of \num{1.16} increase in $\sigma_{\mathrm{eff}}$. 
\end{itemize}

Overall, if all four types of change could be implemented, with the discussed magnitudes, the 
net effect is simulated to be a factor of \num{3.1} increase in facility luminosity with 
a tolerable increase of $\sigma_{\mathrm{eff}}$ to \SI{0.62}{\%}. 
Together with the ``straightforward'' luminosity upgrades, there is thus potential for luminosities 
of \SI{1.7E34}{cm^{-2}s^{-1}} at 
the \PZ pole and an estimated \SI{3.8E34}{cm^{-2}s^{-1}} near $\PW\PW$ threshold. 
These correspond to factors of \num{8.5} and \num{10} beyond 
the ILC-250 baseline, and would open up the potential for integrated luminosities of \SI{800}{\fbinv} in \num{4} years at the \PZ pole 
and \SI{1.4}{\abinv} in \num{3} years near the $\PW\PW$ threshold. 
Further study is needed to assess the feasibility of such changes, to verify the fidelity of the simulation estimates in the presence of machine imperfections and offsets in the high disruption regime, and to characterise fully the effects on the physics observables.
There are a number of inter-related issues for the overall accelerator/detectors design:
\begin{enumerate}
\item Beam dynamics (i.e.\ less emittance dilution) is facilitated if beams are accelerated at 
full gradient. This would need a bypass line (and corresponding tunnel area) 
and may be essential for operating well an energy-upgraded machine at the \PZ pole.
\item Study of the beam control at the IP for the high disruption case including the feedback system.
\item Beam backgrounds in the detectors will increase, especially 
for smaller $\beta_x^{*}$, and need to be assessed. 
\item Increasing the bunch populations has wide-ranging challenges across 
the accelerator complex and may not be practical. 
The design bunch charge at the sources is \num{3.0E10}$e$. This includes a 
\SI{50}{\%} margin over the nominal \num{2.0E10}$e$ in collision.
\end{enumerate}

The baseline scenario for ILC is \SI{80}{\%} electron polarisation and \SI{30}{\%} positron polarisation 
leading to an effective polarisation (see Sec.~\ref{sec:polarisation}) of \SI{88.7}{\%}. 
Higher polarisation values of up to \SI{90}{\%} for electrons  
and \SI{60}{\%} for positrons are conceivable, leading to an effective polarisation of up to \SI{97.4}{\%}. 
A \SI{40}{-}\SI{45}{\%} positron polarisation is already believed to be feasible with the current ILC-250 baseline~\cite{Moortgat-Pick:2024fcy}. 

\subsubsection{Higher luminosity at all energies by increasing the collision rate}
\label{sec:acc:lumiup:benno}

There are two ways to increase the collision rate of a Linear Collider: the number of bunches per pulse can be increased, and/or the number of pulses per second, i.e.\ the repetition rate.

\paragraph[Doubling the number of bunches per pulse]{Doubling the number of bunches per pulse}

The doubling of the number of bunches per pulse is the classic mean of choice for the ILC luminosity upgrade.
L-band SCRF cavities as employed at the ILC store a considerable amount of energy in the oscillating accelerating field, which is lost at the end of each pulse and must be replenished before the next beam pulse; in the ILC baseline design, as discussed in Sect.~\ref{sec:acc:SCRFbase:specs}, approximately one third of the total RF pulse of 1.65 ms duration is required for the filling of the cavity.
Therefore, accelerating more bunches, or overall charge, per pulse improves the ratio of RF power transferred to the beam over the RF power lost in the filling, and thus improves the overall efficiency. 
This is also reflected in the fact that increasing the number of bunches from \num{1312} to \num{2625}  bunches per pulse requires only \SI{50}{\%} more klystrons in the Main Linac, not twice as many.

Increasing the number of bunches per pulse (or the bunch charge) further would require overcoming several limitations:
\begin{itemize}
    \item The beam current in the damping rings, which is given by the total accelerated charge per pulse times the DR revolution frequency, is limited to about \num{400} (\num{800}) \SI{}{mA} for positrons (electrons) by electron and ion cloud instabilities. 
    More accelerated charge per pulse would thus require a larger damping ring circumference or stacking several damping rings in the same tunnel (which is already required for positrons when operating at 2625 bunches).
    \item The pulse length of existing klystron designs is limited to about \SI{1.65}{ms}. 
    Operation at significantly longer pulses would require changes to the klystron design. 
    Alternatively, the bunch separation could be reduced to accelerate more bunches within the \SI{1.65}{ms} pulse duration, requiring more klystrons to provide the necessary power, as done for 2625 bunches operation.
    \item Increasing the overall amount of accelerated positrons per second raises the requirements on the positron source's production rate.  
\end{itemize}

\paragraph[10\,Hz operation at full gradient]{10\,Hz operation at full gradient}

The ILC baseline design calls for a repetition rate of \SI{5}{Hz}, while the Eu.XFEL operates at \SI{10}{Hz}. 
To permit the \num{5}+\SI{5}{Hz} operating mode that is necessary for operation below the minimum energy of about \SI{120}{GeV} needed by the undulator positron source to work effectively, a number of key systems have been specified for \SI{10}{Hz} operation, notably: The electron source, the damping rings, the bunch compressors, the HLRF system (modulators, klystrons, and couplers) of the main linac, and the BDS. 
\SI{10}{Hz} operation for physics operation at reduced gradient has also been studied, which is challenging for the undulator positron source, but deemed feasible.
Operating the damping rings at \SI{10}{Hz} repetition rate requires more damping wigglers in order to reduce the damping time, and more RF power to replenish the increased synchrotron radiation power; this is already foreseen in the damping ring design as an option. 
The increased RF power requirements of the sources, bunch compressors and main linacs would lead to a doubling of the electric power demand of the modulators, which needs to be taken into account in the dimensioning of the modulator electric supply, but is not a problem.

The limiting factor for \SI{10}{Hz} operation at full gradient in the Main Linac is the increased cryogenic load from the dynamic heating of the cavities, which dominates the cryogenic load and doubles at a doubled repetition rate. Dynamic heating is dominated by the heat losses in the cavity walls, which are proportional to $1/Q_0$. 
An doubling of $Q_0$ (from the ILC design value of \num{1E10} to \num{2E10}, which appears feasible given recent advances in cavity performance, would make it easier to provide the necessary cooling capacity to go to \SI{10}{Hz}.
However, as not all dynamic cryogenic losses scale with the cavity wall losses, doubling the repetition rate still requires an increase in cryogenic capacity.

This increase from \SI{5}{Hz} to \SI{10}{Hz} operation is the baseline chosen for the LCF proposal for CERN, leading to an initial luminosity twice as high as for the ILC baseline. The increase of bunches per pulse is retained as additional luminosity upgrade, allowing another doubling of the luminosity.

Designing the Main Linac for operation at full gradient with a repetition rate is possible.
Increasing the cryogenic cooling capacity involves an increased throughput of helium through the helium pipes of the cryomodules, increased capacity of the cold compressor system serving one supply area (\num{2} to \SI{2.5}{km} of cryo-strings), and an increased helium plant capacity. 
Helium plants are commercially available up to approximately \SI{20}{kW} equivalent cooling power at \SI{4.5}{K}, and the ILC design is optimised to fully utilise such a plant. 
Therefore, more helium plants would be required to supply a given supply area.
It requires a reoptimisation of the overall cryogenic layout concerning the number of cryomodules in a cryo-string (connected to a single feed cap) and the number of strings supplied by one helium plant, and possibly to increase the helium gas pipe in the cryomodule for increased helium flow, as has been successfully done for the LCLS-II cryomodule.
Provided that the overall distance between access shafts with helium plants stays the same, the cost impact of such a design change is dominated by the increased number of cryo-plants.

This requirement is to be taken into account during the dimensioning of the relevant components and systems, considering the assumed cavity operating gradient and quality factor. 
Revisiting these design assumptions would be part of the value engineering part of a design study.

\subsubsection{Energy recovery technologies}
\label{sec:acc:lumiup:erl}

Increasing luminosity by orders of magnitude from $\mathcal{O}$(\SI{E34}{cm^{-2}s^{-1}}) requires significant modifications in the overall approach to the beam parameters and to some of the system choices for the linear collider. 
A simple-minded increase of the collision frequency would require increased AC power and particle production. 
Hence, an efficient recycling of both the beam energy and of the particles in the bunches is required for any significant luminosity boost. 
All these requirements naturally lead to CW operation of Energy Recovery Linacs. 
After an overview on the status of ERL technology and R\&D facilities, we present two possible upgrade of the proposed linear collider to an average luminosity far beyond  \SI{E34}{cm^{-2}s^{-1}}. 

\paragraph[Status of energy recovery technology]{Status of energy recovery technology}
Energy-recovery linacs (ERLs) are a highly efficient technique for accelerating high-average-current electron beams. 
In an ERL, an electron beam is accelerated to relativistic energies in (typically) a superconducting RF continuous-wave (CW) linear accelerator. 
The beam is then used for its intended application, such as serving as a gain medium for a free-electron laser, producing synchrotron light, acting as a cooling source for ion beams, or generating a beam to collide with ions. 
These applications typically lead to a significant increase in energy spread or emittance of the electron beam, while the majority of the beam power is retained. 
To recover this power, the beam is sent back through the accelerator, but this time it is approximately $180^\circ$ out of phase with the accelerating RF field. 
As a result, the beam is decelerated, transferring its energy back into the RF fields, allowing a new beam acceleration. 
Eventually, the beam's energy drops to a point where further transport becomes impractical, and the beam is dumped with a small residual energy. 
In the case of \epem\ ERL-based colliders, all the beams are recycled and not dumped.

In high-energy physics, ERLs could be used to produce an intense, low-emittance electron beam for interactions with hadrons (\Pe-hadron), positrons (\epem), or photons (\Pe\PGg). For \epem colliders, both beams could be accelerated and decelerated in ERLs as it will be described in the following proposals. 
These large-scale facilities aim to operate at power regime of gigawatts to tens of gigawatts, surpassing current state-of-the-art systems (the \SI{1}{MW} achieved at JLab’s FEL: \SI{200}{MeV}, \SI{8}{mA}) by 3 to 4 orders of magnitude. 
The primary technical challenges in reaching these power levels include ensuring SRF cavities that can sustain high-current beams (tens to hundreds of mA) with manageable cryogenic loads, exceptional beam stability, and efficient energy recovery. 
Current R\&D efforts to address these challenges are being pursued through compact, high-performance ERL demonstrators.

The 2020 update to the European Strategy for Particle Physics (ESPP) emphasised that ERLs represent one of the most innovative technologies for future high-energy physics accelerator infrastructures. 
It also highlighted that the R\&D roadmap for critical technologies needed for future colliders should include research on high-intensity, multi-turn ERL machines. 
Thus, two projects were identified as essential pillars for future ERL development:
\begin{itemize}
    \item The \textbf{bERLinPro} project in Helmholtz-Zentrum Berlin (HZB)- Germany: A single turn ERL (\SI{50}{MeV}) to demonstrate high current (up to \SI{100}{mA}), low-emittance continuous wave (CW) electron beams using SCRF technology. 
    \item The \textbf{PERLE} project at Ir\`ene Joliot-Curie Laboratory (IJCLab) at Orsay-France: A three-pass ERL facility designed to operate ultimately at \SI{10}{MW} beam power ($\SI{20}{mA} \times \SI{500}{MV}$), aiming at testing the technologies increasing the efficiency of the accelerators and necessary for future high-energy physics applications, including SCRF development.
\end{itemize}
The two projects have the ambitions to explore a new power range (5 to \SI{10}{MW}) never reached by ERLs, paving the way to a new generation of powerful machines and appending new milestones toward future large-scale, ERL-based colliders. 
Several challenges have to be addressed along the way: 
\begin{description}
    \item[Development of high-Q$_{\mathrm{0}}$ SCRF systems for high beam currents:] 
    The development of high-$\mathrm{Q_0}$ SCRF systems capable of handling high beam currents (typically \SI{100}{mA} or higher) and continuous wave (CW) mode operation while mitigating higher-order mode (HOM) excitations is a key focus. 
    In addition to the widely-used Tesla cavities (\SI{1.3}{GHz}) adapted for CW operation, which have been implemented in projects such as bERLinPro, CBETA, and cERL, the PERLE project has led to the design and optimisation of a new \SI{802}{MHz}, 5-cell Nb cavity. 
    This cavity has been optimised to facilitate the efficient extraction of parasitic HOMs through features like fewer cells, larger iris apertures, and an optimised end-group, all while enabling the handling of multi-bunch operations.
    \item[Innovations in HOM couplers and beam line absorbers (BLA):]
    Recent advancements include the development of new HOM damping waveguides, which were implemented in the bERLinPro project to address the challenges of high-current operations. 
    Additionally, for the PERLE project, three coaxial HOM coupler designs (Hook, Probe, and DQW) were studied and optimised. 
    An optimised damping scheme was selected for the end-group to efficiently extract harmful modes. For the next generation of high-current ERLs, HOM couplers alone are often insufficient to eliminate all critical modes. 
    To address this, Beam Line Absorbers (BLAs) are employed between cavities to capture and dissipate the modes that propagate along the beamline.
    \item[Enhancing ERL efficiency:] 
    Research is focused on advancing the performance of Energy Recovery Linacs (ERLs) by developing \ce{Nb3Sn} coatings. 
    These coatings have shown higher superconducting transition temperatures and potential for reduced surface resistance compared to traditional niobium cavities. 
    Additionally, efforts are being made to develop multilayered Superconductor-Insulator-Super\-conductor (SIS) structures, utilizing materials like NbN and MgB$_2$. 
    These innovations aim to overcome the performance limitations of bulk niobium, offering the possibility of more efficient and higher-performing SCRF cavities. 
    ERLs using these advancements could operate at higher cryogenic temperatures (\SI{4.2}{K} vs.\ \SI{2}{K}), reducing the need for costly cryogenic cooling systems. 
    Another promising technology under development is the Ferroelectric Fast Reactive Tuners (FE-FRTs). 
    These innovative devices are designed to rapidly adjust the resonant frequency of SCRF cavities, compensating for frequency shifts caused by microphonics. 
    Their integration could significantly improve the performance and efficiency of ERLs and other accelerator systems.
\end{description}

Many of the developments mentioned above are part of the European programme ``Innovate for Sustainable Accelerating Systems'' (iSAS)~\cite{isas}.
This programme is dedicated to enhancing accelerator efficiency, with a particular emphasis on Energy Recovery Linacs (ERLs). 
It focuses on optimising energy consumption and improving the performance of various systems, including cryogenics, RF power, and beam energy recovery.

\paragraph[ReLiC]{ReLiC}

\begin{figure}[htb]
\includegraphics[width=\textwidth]{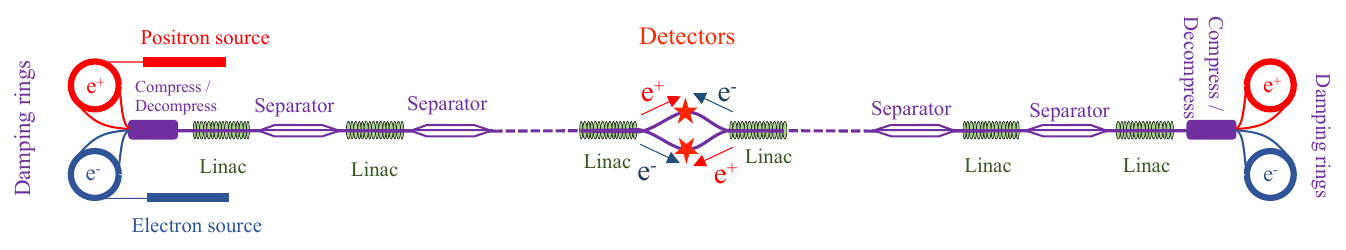}
\caption{Schematic of ReLiC with two detectors~\cite{Litvinenko:2022qbd}. 
One the most important features of the ReLiC design is its capability to collide highly polarised electron and positions beams. 
Beam polarisation is maintained by the Sokolov-Ternov effect in the damping rings.}
\label{fig:relic:footprint}
\end{figure}

This option for a high-luminosity upgrade of a linear collider facility explores and extends ideas of recycling energy and colliding particles that were developed in Refs.~\cite{Litvinenko:2022qbd, litvinenko:lcws2024, litvinenko:erl2024} for a concept called ReLiC, illustrated in Fig.~\ref{fig:relic:footprint}. 

\begin{figure}[htb]
\includegraphics[width=\textwidth]{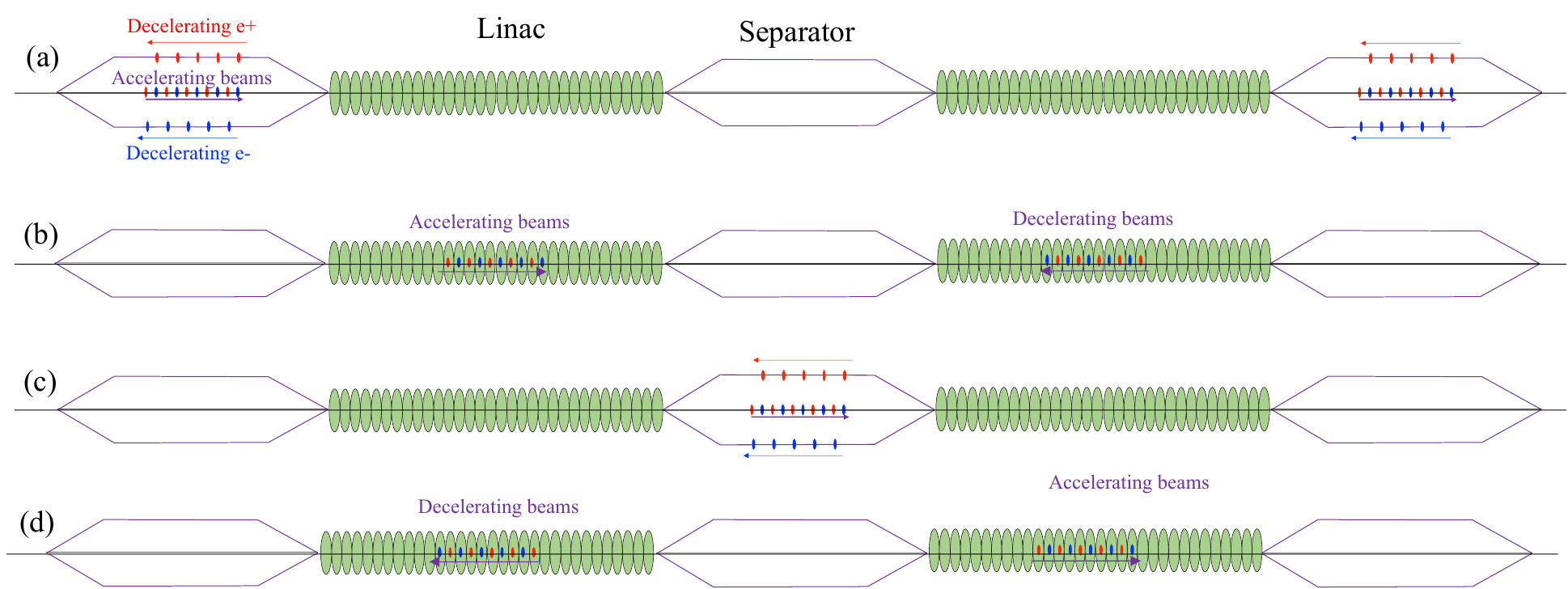}
\caption{Details of separating accelerating and decelerating electron and positron bunches which repeats with the bunch train frequency $f_{\mathrm{train}} =1/T_{\mathrm{train}}$: 
(a) at $t=0$ trains of  $N_{\mathrm{bunch}}$ accelerating and decelerating bunches are separated horizontally in two separators; 
(b) at $t= T_{\mathrm{train}}/4$ beams propagating towards (coming from) the IP accelerate (decelerate) on-axis in corresponding linacs; 
(c) at $t=T_{\mathrm{train}}/2$, the counter-propagating beams are separated again; (d) at $t=3T_{\mathrm{train}}/4$ beams accelerate and decelerate on-axis in opposite linacs. 
Then the process repeats itself. 
Note that scales are greatly distorted for visibility.}
\label{fig:relic:separation}
\end{figure}

ReLiC is based on conventional standing wave (SW) single-axis SCRF linacs which are used both for accelerating and decelerating electron and positron beams. 
The linacs are split into sections where separators are used to avoid parasitic collisions of the accelerating and decelerating beams. 
It is worth noting that a similar concept, but without separators, was discussed by Gerke and Steffen in 1979~\cite{Gerke:1979zs}, however, the separators have been added to achieve high luminosity. 
Using a combination of electric and magnetic fields in the separators provides for on-axis propagation of the accelerating beams, preserving their emittance, while  only the decelerating beams are being deflected~\cite{Litvinenko:2022qbd}, as shown in Fig.~\ref{fig:relic:separation}.

The high luminosity is attained by a relatively high collision frequency $f_c = N_{\mathrm{bunches}} \times f_{\mathrm{train}}$ of flat beams with large vertical disruption parameter, which are typical for linear colliders. For ReLiC, the beams are assumed to be even much flatter than for ILC or LCF, with a horizontal beam size about one order of magnitude larger and a vertical beam size about one order of magnitude smaller.
As shown in~\cite{Litvinenko:2019txu}, such violent collisions can result in a significant ($\sim 2$ fold) increase in vertical emittance, which requires cooling of the decelerated beam in a damping ring for one to two damping times.  Nevertheless, the maximal beamstrahlung parameter at ReLiC would be more than one order of magnitude smaller than at conventional linear colliders, c.f.\ Table~\ref{tab:relic}.

ReLiC operation provides for recycling not only of the beam energy but also of the  particle bunches. 
Efficient particle recovery requires the damping ring to accept particles with deviating energy caused by synchrotron radiation in the beam separators and IP optics and beamstrahlung during beam collision. 
An analysis shows that for high-energy colliders at the centre-of-mass energy of interest beamstrahlung plays the dominant role and requires increasing both horizontal beam size (and horizontal $\beta^*$) and bunch lengths when compared with the LCF baseline parameters (see Table~\ref{tab:relic}). 
While this results in a reduction of luminosity, it also significantly reduces the energy spread in colliding beams and should provide for improved energy resolution. 

The energy efficiency of such a collider depends on the energy of the damping ring: as mentioned above, we expect that each particle spends two damping times in the damping rings and thus loses about twice the damping ring beam energy per round-trip through the whole machine. 
\footnote{After one damping time for the emittance, the total energy loss through synchrotron radiation equals the beam energy in the damping ring~\cite{Wolski:2023pci}.}.
For a damping beam energy of \SI{2}{GeV} and \SI{4}{GeV} for centre-of-mass energies of \SI{250}{GeV} and \SI{500}{GeV}, respectively, we would need a \num{20}-fold decompression of the beams and \SI{10}{\%} energy acceptance of the damping rings to provide lossless operation of the collider. 
In this case only particles lost (``burned'') at the IP and in collisions with residual gas would be replaced by top-off injection from electron and position injectors. 

Current technology provides us with limited choices for \SI{1.3}{GHz} SCRF linacs: the quality parameter is limited to $Q_0$ = \num{4E10} at \SI{2}{K} temperatures~\cite{Grassellino:2013nza} where the coefficient of performance (COP) of liquid Helium refrigerators is $\sim 6$-fold below the theoretical Carnot limit. 
While we list in Table~\ref{tab:relic} possible ReLiC parameters with existing technology, we also explore options of engineering development of improved of \SI{2}{K} liquid Helium refrigerator COP or a breakthrough with \ce{Nb3Sn} SCRF cavities operating at \SI{4.5}{K} temperature with high $Q_0$.

\begin{figure}[htb]
\begin{center}
\includegraphics[width=0.6\textwidth]{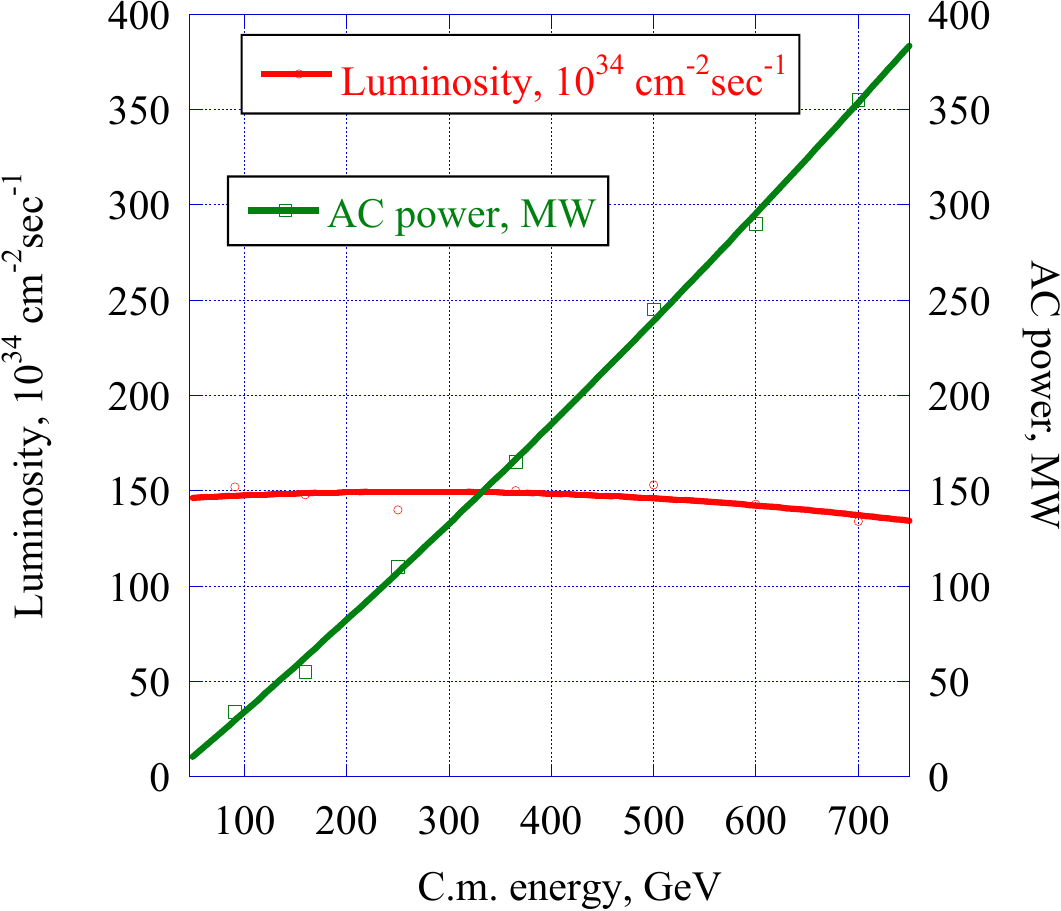}
\end{center}
\caption{Estimated luminosity and AC power consumption for ReLiC operating with futuristic \ce{Nb3Sn} SCRF linac with fixed linac length. 
At high centre-of-mass energies above \SI{500}{GeV}, the AC power consumption is nearly equally split between that used by the cryogenic system and by the damping rings.}
\label{fig:relic:power}
\end{figure}

\begin{table}
\begin{threeparttable}[htb]
\begin{tabular}{|l |c | c | c | c | c |} 
\hline
    & LCF250FP & ReLiC\tnote{1} & ReLiC\tnote{2} & ReLiC\tnote{3} & ReLiC\tnote{3} \\ \hline
Centre-of-mass energy [ \SI{}{GeV} ]  & 250 & 250 & 250 & 250 & 550 \\ \hline
Accel. Grad. [ \SI{}{MeV/m} ] & 31.5 & 12.55 & 12.55 & 12.55 & 27.6 \\  \hline
Cavity $Q_0$ [ $10^{10}$ ] & 2 & 4 & 4 & 4 & 4 \\  \hline
Liquid He temperature [ \SI{}{K} ] & 2 & 2 & 2 & 4.5 & 4.5 \\  \hline
Bunch population [ $10^{10}$ ] & 2 & 2.5 & 2.5 & 2.5 & 2.5 \\  \hline
Collision frequency [ \SI{}{MHz} ] & 2.73 & 1.5 & 1.5 & 1.5 & 1.5 \\
Duty cycle  & 0.0096 & cw & cw & cw & cw \\ \hline
Beam current, all beams [ \SI{}{mA} ] & 0.084 & 12 & 12 & 12 & 12 \\
 \\  \hline
Normalised emittance hor.[ $\mu$m ] / vert.[nm]  & 5/35 & 4/1 & 4/1 & 4/1 & 4/1 \\  \hline
$\beta_x$ / $\beta_y$ [ \SI{}{m}] / [\SI{}{mm}]  & 0.013/0.41 & 2.2/0.19 & 2.2/0.19 & 2.2/0.19 & 4/0.36 \\  \hline
$\sigma_x$ / $\sigma_y$ at IP [$\mu$m] / [\SI{}{nm}]  & 0.52/7.7 & 6/0.9 & 6/0.9 & 6/0.9 & 6/0.9 \\  \hline
$D_x$ / $D_y$ & 0.5/34.5 & 0.01/87 & 0.01/87 & 0.01/87 & 0.01/88 \\  \hline
$ \Upsilon_{\mathrm{ max}}$ & 0.068 & 0.0028 & 0.0028 & 0.0028 & 0.0031 \\ \hline
Luminosity  [$10^{34}$ cm$^{-2}$s$^{-1}$] & 5.4 & 140 & 140 & 140 & 153  \\ \hline
AC Site Power [ \SI{}{MW} ]  & 182 & $\sim$135 & $\sim$105  & $\sim$95 & $\sim$ 250 \\ \hline
\end{tabular}
\caption{\label{tab:relic}
Parameters of ReLiC options with two 14.25 km linacs}
\begin{tablenotes}
      \item [1] ReLiC based on current ``conventional" Nb SW SCRF technology ~\cite{Grassellino:2013nza}. Cryogenics system: \SI{80}{MW}, damping rings: \SI{70}{MW} + others 
       \item [2] ReLiC based on Nb SW SCRF technology, but COP of \SI{2}{K} LiHe refrigerators improved 2-fold. Cryogenics system: \SI{45}{MW}, damping rings: \SI{70}{MW} + others 
       \item [3] ReLiC based on futuristic SCRF technology for \ce{Nb3Sn} SW cavity with $Q_0 =4E10$. 
       \\{250}{GeV}: Cryogenics system: \SI{35}{MW}, damping rings: \SI{70}{MW}; 
       \\{500}{GeV}: Cryogenics system: \SI{95}{MW}, damping rings: \SI{140}{MW}
\end{tablenotes}
\end{threeparttable}
\end{table}
Table~\ref{tab:relic} shows total the luminosity of the collider, which is split 50/50 between two detectors. 
The collision time structure in each detector is determined by that of the bunch trains. Using five electron and five position bunches per train requires the linacs to be split in one kilometre sections. 
In addition, we assume that electron and positron bunches are separated by ten \SI{1.3}{GHz} RF cycles. 
In this case, each detector will see five collisions, separated by \SI{7.7}{ns}, repeated with frequency of \SI{150}{kHz}. 
Separating counter-propagating bunch trains by \SI{1}{cm} would require \num{62}-meter-long separators, which reduces the real-estate gradient of the linacs by  $\sim$\SI{6}{\%}. 
While Table~\ref{tab:relic} lists example parameters for centre-of-mass energies of \SI{250}{GeV}  and \SI{500}{GeV}, ReLiC would provide luminosities above \SI{e36}{cm^{-2}s^{-1}} in polarised beam collision for a large range of energies from \SI{45}{GeV} to \SI{700}{GeV}. 
Fig.~\ref{fig:relic:power} shows the estimations for luminosity and AC power consumption.

The fact that ReLiC can operate with a single cavity is an important advantage of this upgrade option making it cost effective. Doubling the number of linac cavities, in the manner similar to ERLC, would allow to eliminate separators~\cite{litvinenko:erl2024}, reduce collider length by $\sim$\SI{6}{\%} while doubling the cryo-plant power and cost of the linac.

For the power estimation we assume that electron and positron beams circulate in the damping rings for two emittance damping times (i.e.\ will lose energy equal to two times the damping ring beam energy). 
Such operation will provide for \SI{7.4}-fold reduction of transverse emittances and \num{54.6}-fold reduction of the longitudinal emittance growths accumulated during each collision pass. 
Our simulations~\cite{Litvinenko:2019txu} indicated that head-on collisions with vertical disruption parameters below \num{150} results in less than factor two of vertical emittance growth, which could require only a single damping time if the orbit feedback system will maintain nearly head-on collisions.

\paragraph*{Challenges for a ReLiC-like upgrade}  Challenges for upgrading an ILC-like linear collider with a ReLiC-like scheme comprise:
\begin{enumerate}
 \item the ReLiC design does not require a large crossing angle as in the ILC design, which assumes an angle of \SI{14}{mrad} between the main linacs. Since energy recovery requires returning beams into the opposite linac for deceleration, a non-zero crossing angle between the linac arms necessitates a bending of the beams.
 Such bending and aligning of the beam's trajectory would result in synchrotron radiation and growth of energy spread and horizontal emittances. 
 In the case of ReLiC operating at \SI{500}{GeV} centre-of-mass energy an ILC-like main linac angle will result in \SI{15.4}{MeV} loss of energy with a critical photon energy of \SI{0.7}{MeV}, and \SI{3.3}{MeV} growth of the RMS energy spread per pass. 
 This level of energy loss and increase in energy spread is fully acceptable for ReLiC.
 \item The strong two-fold damping in the ReLiC system is sufficient to counteract additional contributions of the energy spread and emittance growth resulting from synchrotron radiation and intra-beam scattering. 
 Still, self-consistent simulations are required to establish the level of losses caused by beamstrahlung, Touschek scattering, scattering on residual gas and particle burn-off in collisions.
\end{enumerate}

\paragraph*{Time and R\&D needed} The following R\&D priorities have been identified:
\begin{enumerate}
\item 	The development of efficient SCRF and cryogenic technology (efficient LiHe refrigerators, development of high $Q_0$ SCRF cavities, including but not limited to \ce{Nb3Sn}) is critically important for reducing the energy consumption. While ERLC and ReLiC can be build using existing SCRF and cryogenic technology, the power consumption would be close to \SI{500}{MW} for \SI{500}{GeV} collider.
\item The demonstration of flat beams with \num{2000} to \num{4000} emittance ratios. 
\item 	The development of high rep-rate and accurate kickers
\item 	Start-to-end simulations of the beam and spin dynamics need to be performed.
\end{enumerate}

In summary, ReLiC offers one of the potential paths towards increasing the luminosity of a linear collider by up to two orders of magnitudes. 

\begin{figure}[htp]
\includegraphics[width=\textwidth]{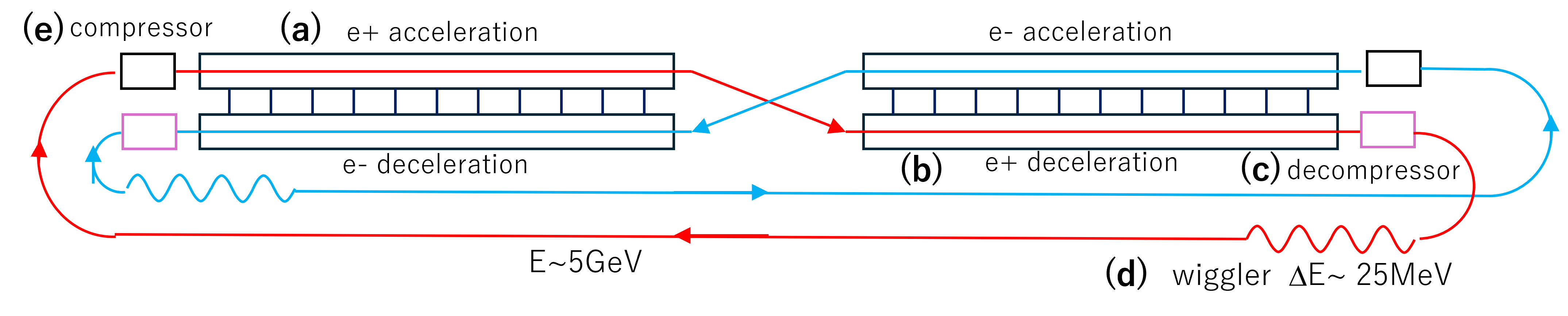}
\caption{Schematic plot showing the concept of ERLC. 
Some systems such as injectors, beam dumps, etc. are omitted for simplicity.
}
\label{fig:ERLCconcept}
\end{figure}

\paragraph[ERLC]{ERLC}
Another idea to profit from energy and particle recovery is the ERLC (Energy Recovery Linear Collider) concept~\cite{Telnov_2021}. 
Figure~\ref{fig:ERLCconcept} illustrates the main idea of ERLC. 
In its steady-state, e.g.\ the positron beam propagates as follows:
It is accelerated in the linac (a) and, after collision with the electron beam, it is 
decelerated in the linac (b) down to about \SI{5}{GeV}. After the bunch length is decompressed in (c), the beam loses about \SI{25}{MeV} of energy in the wiggler section (d). 
Then the beam is compressed in (e) and again accelerated in (a). 

During deceleration, the electromagnetic field generated by the positron beam is transferred to the electron linac running in parallel and accelerates the electron beam. 
Thus, the two linacs are connected electromagnetically by twin-axis cavities.

The beam-beam collision at the IP is not as strong as in linear colliders like the ILC, but is quite moderate so that the beam quality is restored by the weak synchrotron radiation in the wiggler section. 
The whole system works like  a circular rather than a linear collider.

\paragraph*{Optimisation} From the beam dynamics point of view the following three issues must be considered:
\begin{enumerate}
\item  The beam-beam tune-shift limit must be obeyed as in circular colliders.
\item  The average energy spread coming from the turn-by-turn effects of the beamstrahlung must be taken into account.
\item  The low-energy tail of the beam due to the high-energy tail of the single beamstrahlung demands large energy acceptance of the beam line. 
The beam life time is limited by this effect.
The bunch compressors and decompressors are required to relax this third point.
\end{enumerate}

The AC power consumption is dominated by two effects: by the RF losses in the cavities and by HOM (Higher Order Modes). 
The RF loss is proportional to $G\cdot E/((r_{\mathrm{s}}/Q) \cdot Q_0)$ where $E$ is the collision energy, $G$ the accelerating gradient, $(r_{\mathrm{s}}/Q)$ is the $R_{\mathrm{s}}/Q$ per unit length with the shunt impedance $R_{\mathrm{s}}$, and $Q_0$ is the cavity quality parameter. 
This is independent of the beam current.
To compensate for this heating the efficiency of the cryogenics system (COP) is important.

The beam energy loss by HOM is proportional to $N\langle I \rangle(E/G)/a^2$, where $N$ is the number 
of particles per bunch, $\langle I \rangle$ is the average beam current and $a$ the aperture radius of the cavity.
The HOM loss itself must be compensated by the RF system. 
However, a larger contribution to the power consumption comes from the heat load, in particular the HOM component lost in the low (helium) temperature region. 
The design of the cavity and cavity module is important.

Thus, the total power consumption from these two loss mechanisms must be optimised. 

\paragraph*{Twin-axis cavity} In this ERLC scheme, twin-axis cavities are indispensable, since the beam to be accelerated and the beam to be decelerated are running along opposite directions. 
Such cavities have already been proposed and some experiments are going on. 

On the other hand, as an improvement of ILC accelerating cavity, a new idea called HELEN~\cite{Belomestnykh:2023uon, Belomestnykh:2023naf} has been proposed. 
A standing wave (SW), like the RF wave in ILC, consists of two travelling waves (TW) propagating in opposite directions. 
Only the TW travelling along the direction of the beam contributes to the acceleration. In the HELEN cavity, a wave guide is attached such that only the TW to the `right' direction goes through the waveguide and returns to the input port.
In this way, the maximum surface field and the RF heat loss are reduced at any given accelerating gradient.  
In principle the accelerating gradient can thus be increased if it is limited by the maximum field. 

\begin{figure}[htp]
\centering \includegraphics[width=0.67\textwidth]{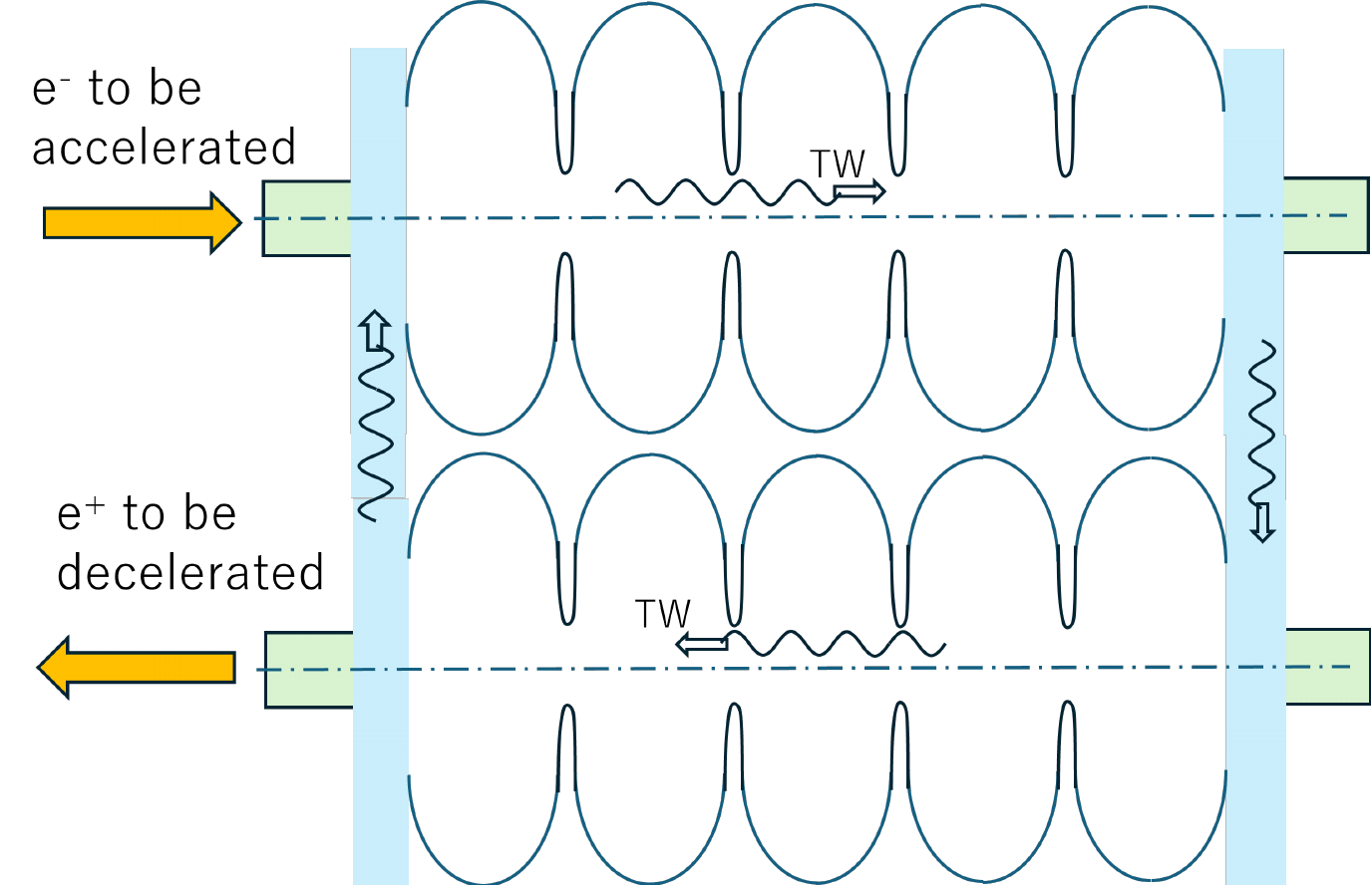}
\caption{Concept of a twin-axis cavity of TW type. The TW propagates only in the clockwise direction in this figure. 
The upper cavity accelerates the electron beam going to the right and the lower cavity decelerates the 
positron beam going to the left.}
\label{fig:ERLCTW-TwinAxisCavity}
\end{figure}

Starting from this idea, a new type of twin-axis cavity has been proposed~\cite{Yokoya:2024vfr}, where the waveguide in the HELEN scheme is replaced with another cavity which decelerates the outgoing beam (see Fig.~\ref{fig:ERLCTW-TwinAxisCavity}). 

An optimisation of the HELEN cavity is described in~\cite{Shemelin:2021ahc}. 
It aims at an accelerating gradient as high as possible and therefore adopts an iris diameter of \SI{50}{mm}, narrower than the \SI{70}{mm} of ILC-type cavities. 
For our purpose, however, an iris diameter \SI{70}{mm} is appropriate for lower HOM generation and presumably easier absorption. 
The shunt impedance in Table~\ref{tab:ERLCparam} is taken from~\cite{Shemelin:2021ahc}.

\paragraph*{Challenges as an ILC upgrade} 
The original idea of ERLC was proposed independently from conventional linear collider plans such as ILC. 
Once it is considered as an upgrade of an ILC-like linear collider, several practical problems must be taken into account. 
An obvious problem is whether the twin-axis cavities can be accommodated in the tunnel.
Here, we have to assume yes, while the actually required tunnel diameter should be worked out in the future. 
This, however, requires a detailed design of  cavity and cryomodule. 

Another problem is the emittance preservation. 
As is mentioned in Fig.~\ref{fig:ERLCconcept}, the \SI{5}{GeV} beam loses \SI{25}{MeV} in the wigglers. 
This means the transverse damping time is $\sim$\num{400} turns. 
ILC assumes a vertical emittance increase of O(\SI{10}{nm}) in a single pass from the damping ring to the IP. 
The ERLC system would not work if this increase were to be accumulated over \num{400} turns, but the major part of this increase will be coherent turn by turn. 
On the other hand, the effects of the synchrotron radiation on the horizontal emittance will certainly be accumulated. 
There are several places in the ILC such as the doglegs in the bunch compressor and the final focus system with significant bending and synchrotron radiation emission leading to an increase of horizontal emittance. 
To upgrade the ILC to ERLC, the beam after the IP  must be bent back into the other arm of the main linac tunnel according to the \SI{14}{mrad} crossing angle. 
A rough evaluation shows the resulting emittance increase is marginal for $E_{CM}=$\SI{500}{GeV} 
(The energy loss must also be taken into account as a source of total power consumption.)  
If this is severe, it is possible to shorten the damping time to $\sim 200$ turns and/or increase the horizontal design emittance by factor of \num{2}, for example.

\paragraph*{Possible parameters} The first issue for the choice of the parameters is the choice of the available technology: \ce{Nb} or \ce{Nb3Sn}, $Q$, cavity type (SW or TW), etc. Here, we choose the most demanding technology, TW twin-axis, \ce{Nb3Sn} cavity. 
Some tentative parameter sets for ERLC are shown in Table~\ref{tab:ERLCparam}, which shows the parameter sets for $E_{CM}=$\num{250} and \SI{500}{GeV} along with ILC-250 for comparison.
Reference~\cite{Telnov_2021} suggests an advantage of a frequency of \SI{650}{MHz} over \SI{1.3}{GHz}, but here we adopt \SI{1.3}{GHz} because twin-axis cavities for \SI{650}{MHz} may not be accommodated in the ILC tunnel.

The accelerating gradient is chosen here so as to accommodate the linacs in the ILC tunnel of the same energy, taking into account the possible space for HOM absorption. 

In this table CW operation is assumed. 
Reference~\cite{Telnov_2021} discusses the possibility of pulsed operation (duty cycle operation) which allows more a flexible choice of parameters. 
For the case $Q_0=$\num{1E10}, the luminosity in the pulsed mode would be some \SI{60}{\%} of the CW luminosity with the same AC power.

\begin{table}[htp]
\centering
\begin{tabular}{|l|r|r|r|l|}
\hline
     & ILC & ERLC & ERLC &    \\
\hline
centre-of-mass energy   &  250  &  250  &  500  &  GeV  \\
Accelerating gradient     &  31.5  &   40   &  40    &  MV/m \\
Cavity $Q_0$               &  1       &   3     &   3     & $\times 10^{10}$ \\
Aperture radius             &  35     &   35    &   35    &  mm \\
Shunt impedance per unit length    &   996   &   1690   &    1690   &   Ohm/m   \\
Operating temperature   &   2       &   4.5   &  4.5    &  K   \\
Bunch population          &     2      &   0.075    &   0.081    &        $\times 10^{10}$ \\
Bunch distance              &    166   &  0.23   &    0.23   &  m  \\
Average beam current     &  0.021   &    157    &   169  &   mA   \\
Beam energy in the return line  &      &    5      &     5     &  GeV \\
Total HOM power                   &   0.014    &    2.9       &   5.85    &   MW   \\
Energy acceptance of the return line   &    &    3    &   3   &  \% \\
Radiation loss in the wiggler      &      &     25      &     25    & MeV \\
Bunch length in main linac  and IP        &   0.3     &   0.31      &   0.89      &   mm  \\
Normalised emittance at IP (x/y)    &    5/35      &    10/35   &  10/35  &      $\mu$m / nm \\
Beta function at IP(x/y)       &     13/0.41       &  12/0.31   &   40/0.89    &   mm  \\
Beam size at IP(x/y)           &    515/7.66        &   700/6.2    &  900/7.4       &   nm  \\
Disruption parameter (x/y)   &  0.5/34.5     &    0.011/1.14    &  0.010/1.14      &           \\
Beam-beam tune shift (x/y)    &          &    0.033/0.097         &    0.036/0.098        &            \\
Upsilon (max)                &    0.068     &  0.00182    &   0.00106    &     \\  
Luminosity                     &   1.35      &    135     &   102    &   $10^{34}/\mbox{cm}^2/\mbox{s}$ \\
AC power for RF heat cooling   &    5     &     91     &    181      &    MW  \\
AC power for HOM cooling    &    1    &     35     &     71     &     MW  \\
Total site power             &    111       &    170    &    320    &   MW  \\
\hline
\end{tabular}
\caption{Tentative Parameters of ERLC for $E_{CM}=$\num{250} and \SI{500}{GeV}.
\label{tab:ERLCparam}
}
\end{table} 

\paragraph*{Timeline and R\&D needed} 
The following R\&D priorities have been identified:
\begin{enumerate}
\item The development of twin-axis cavities of SW or TW type as the key technology for ERLC. TW-type cavities are desired for their higher shunt 
impedance and lower surface field. 
The first step is the HELEN cavity, which may need more than \SI{5}{-}\num{10} years. 
Obviously, the TW twin-axis cavity requires even more R\&D. 
It is difficult to estimate the  required R\&D time, however, we mention here some of the known challenges: 
The structure fabrication is more complex than for single-axis cavities and the corresponding surface polishing method must be developed. 
The tuning of the two cavities can be a serious problem. 
Possible trapped modes must be examined. 
Transverse deflection of the beam must also be investigated since the cavity is quite asymmetric and the beam current is very high.
\item  Development of the cavity material, in particular \ce{Nb3Sn} which can be operated at \SI{4.5}{K}. 
A high Q$_0$ value $\gsim$\num{3E10} is desired.
\item  Development of appropriate HOM absorbers/dampers. 
As shown in Table~\ref{tab:ERLCparam}, the HOM loss per unit length is a somewhat higher than in the latest designs of ERL for light sources (the bunch charge and the average current). 
Table~\ref{tab:ERLCparam} assumes \SI{1}{\%} of the HOM power is deposited in the helium temperature region. 
\item  High efficiency cryogenics system. 
Table~\ref{tab:ERLCparam} assumes a COP of 230 at \SI{4.5}{K}.
\item  Beam dynamics issues to be investigated comprise:
\begin{itemize}
  \item Vertical emittance increase in the main linac
  \item  Horizontal emittance increase due to synchrotron radiation in various parts, in particular the bending section after IP.
  \item  Background in the BDS section (the average beam current is four orders of magnitude higher than in ILC.)
\end{itemize}
\end{enumerate}
The items (2) to (4) are more or less common to all energy recovery colliders.

\paragraph*{Acknowledgement} This section on ERLC has been written after discussions with V.~Telnov.


\subsection{Upgrades to alternative collider modes}
\label{sec:acc:altmodes}

A collider based on the Compton backscattering of laser light off the  high-energy
electron beams provides $\PGg\PGg$, $\Pem\PGg$, and $\Pem\Pem$ collisions with centre-of-mass energies and luminosities comparable to those of the $\ee$ mode. 
As discussed in Sec.~\ref{sec:phys:altmodes}, these collider modes nicely complement the $\ee$ physics programme and are a natural fit for the second IP of a linear collider facility.

While of course a $\PGg\PGg$ collider can be created from one electron and one positron beam, we assume in the following that two electron beams will be employed. 
This has the advantage that neither the intensity nor the degree of polarisation are limited by the capabilities of the positron source. 
In combination with high-energy upgrades based on plasma wakefield acceleration the advantage of using electron beams on both sides is even more obvious. 
However, this requires that positron arm of the initial $\ee$ can be operated with electrons, including a second electron gun and the ability to reverse the polarity of all bending magnets, in particular those in the ``positron'' damping ring.

Three $\PGg\PGg$ collider designs could be considered:
\begin{enumerate}
    \item SCRF accelerator + \SI{1.2}{eV} optical laser
    \item SCRF accelerator + \SI{1.0}{keV} X-ray free electron laser (XFEL)
    \item \CCC\ accelerator + \SI{1.0}{keV} XFEL
\end{enumerate}

Design (1) is well developed and was seriously considered for the second interaction region of the ILC before the second interaction region was dropped in order to reduce the costs of the machine.
The XFEL laser in designs (2) and (3) leads to a greater physics reach, but significant R\&D on the production and focusing of $\sim$~Joule per pulse XFEL light is required, among other issues.  
While design (2) is very recent, design (3) has been assumed in most of the literature on XFEL Compton colliders~\cite{Barklow_2023, Barklow:2023ess, Barklow:2022vkl}. 
Thus we will present example parameters for designs (1) and (3) in the following, while emphasising that design (2) is very worthwhile to be substantiated in the future.

\begin{figure}[htbp]
    \centering
    \begin{subfigure}{.5\textwidth}
    \centering
        \includegraphics[width=0.95\textwidth]{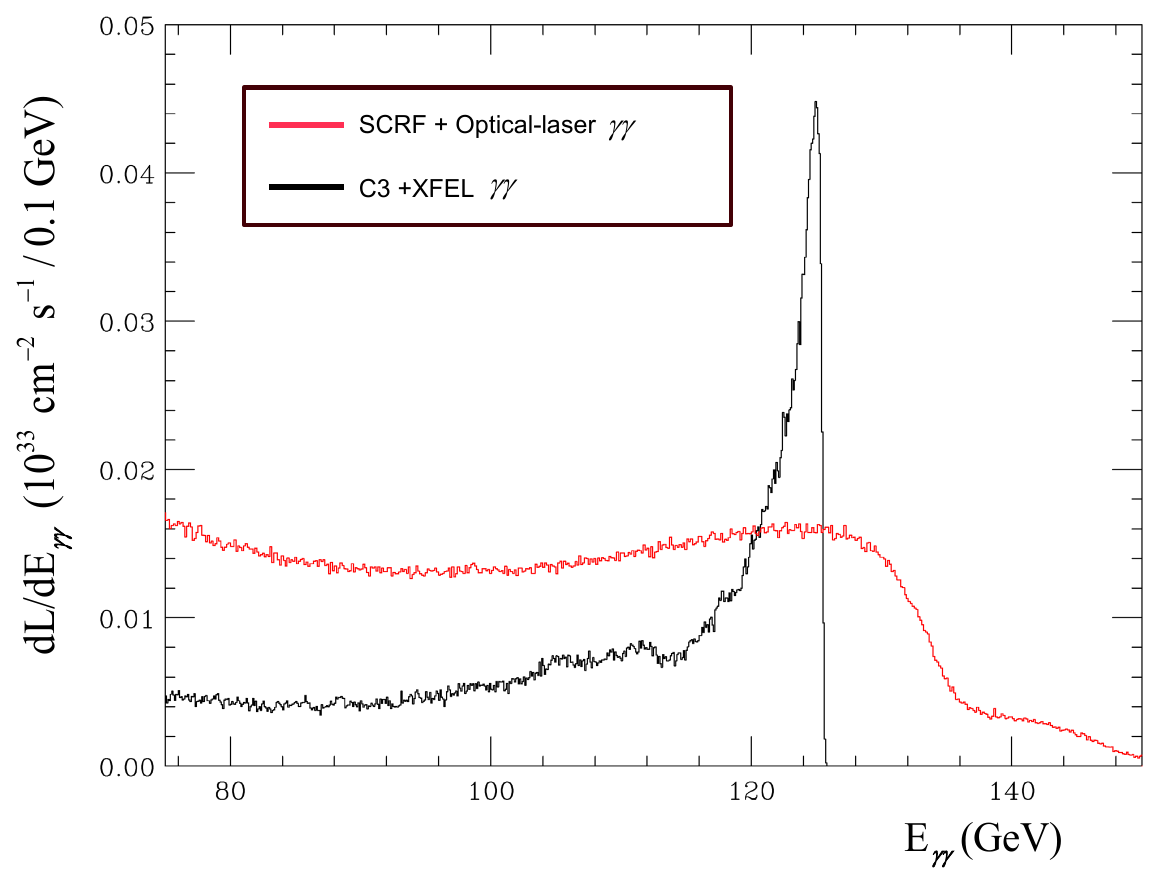}
        \caption{}
        \label{fig:ggLumi125:gg}    
    \end{subfigure}\hfill%
    \begin{subfigure}{.5\textwidth}
        \centering
        \includegraphics[width=0.95\textwidth]{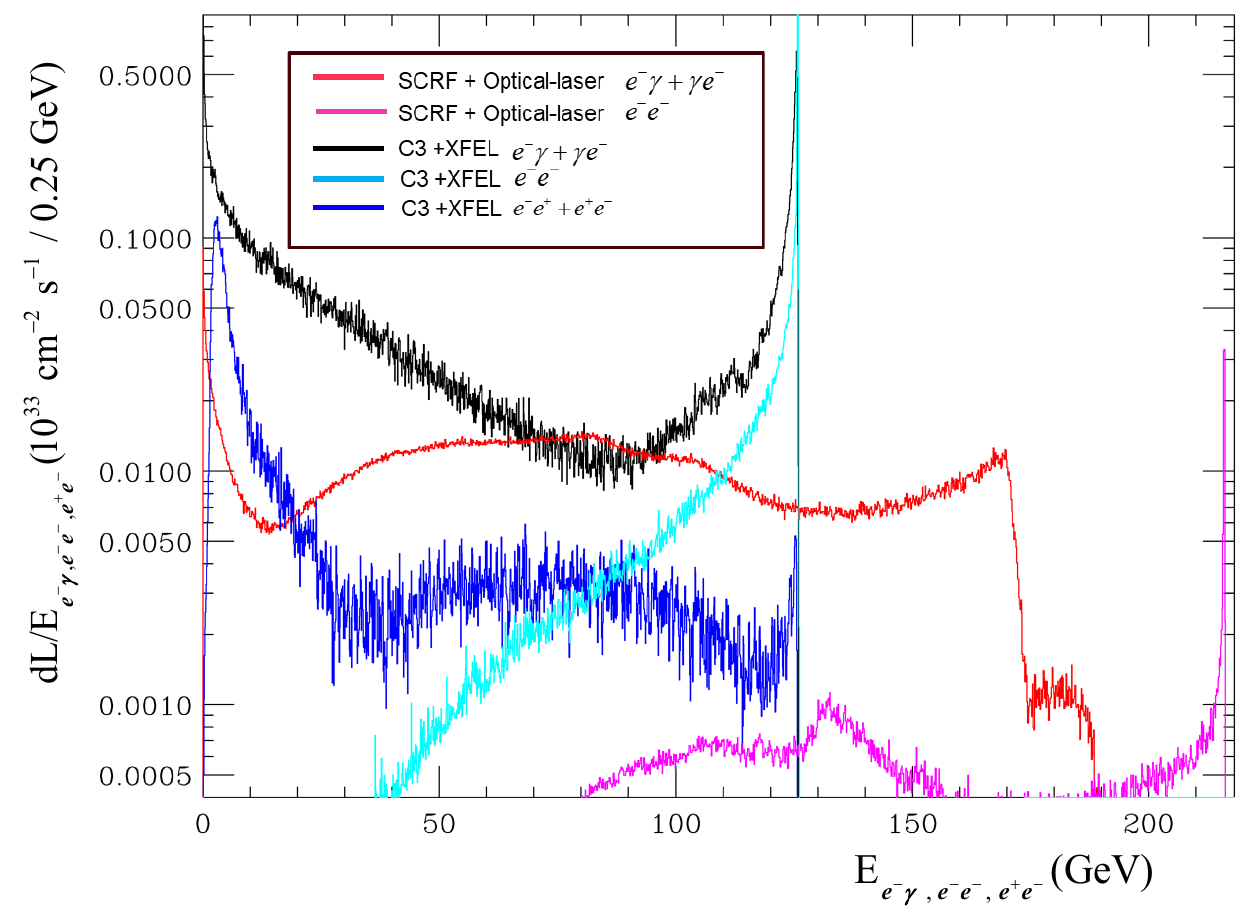}
        \caption{}
        \label{fig:ggLumi125:eg}    
    \end{subfigure}%
    \caption{Luminosity spectra for (a) $\PGg\PGg$ collisions and (b) $\Pem\PGg$, $\Pem\Pem$, and $\ee$ collisions at a $\PGg\PGg$ collider with a peak of the luminosity spectrum near \SI{125}{GeV}. 
    In the XFEL-based design, the peaks of the $\PGg\PGg$ and $\Pem\PGg$ are much sharper, and coincide with the peak in the $\Pem\Pem$ spectrum, while in the case of an optical laser a much higher $\Pem\Pem$ centre-of-mass energy is required.}
    \label{fig:ggLumi125}
\end{figure}

\begin{figure}[htbp]
    \centering
    \begin{subfigure}{.5\textwidth}
    \centering
        \includegraphics[width=0.95\textwidth]{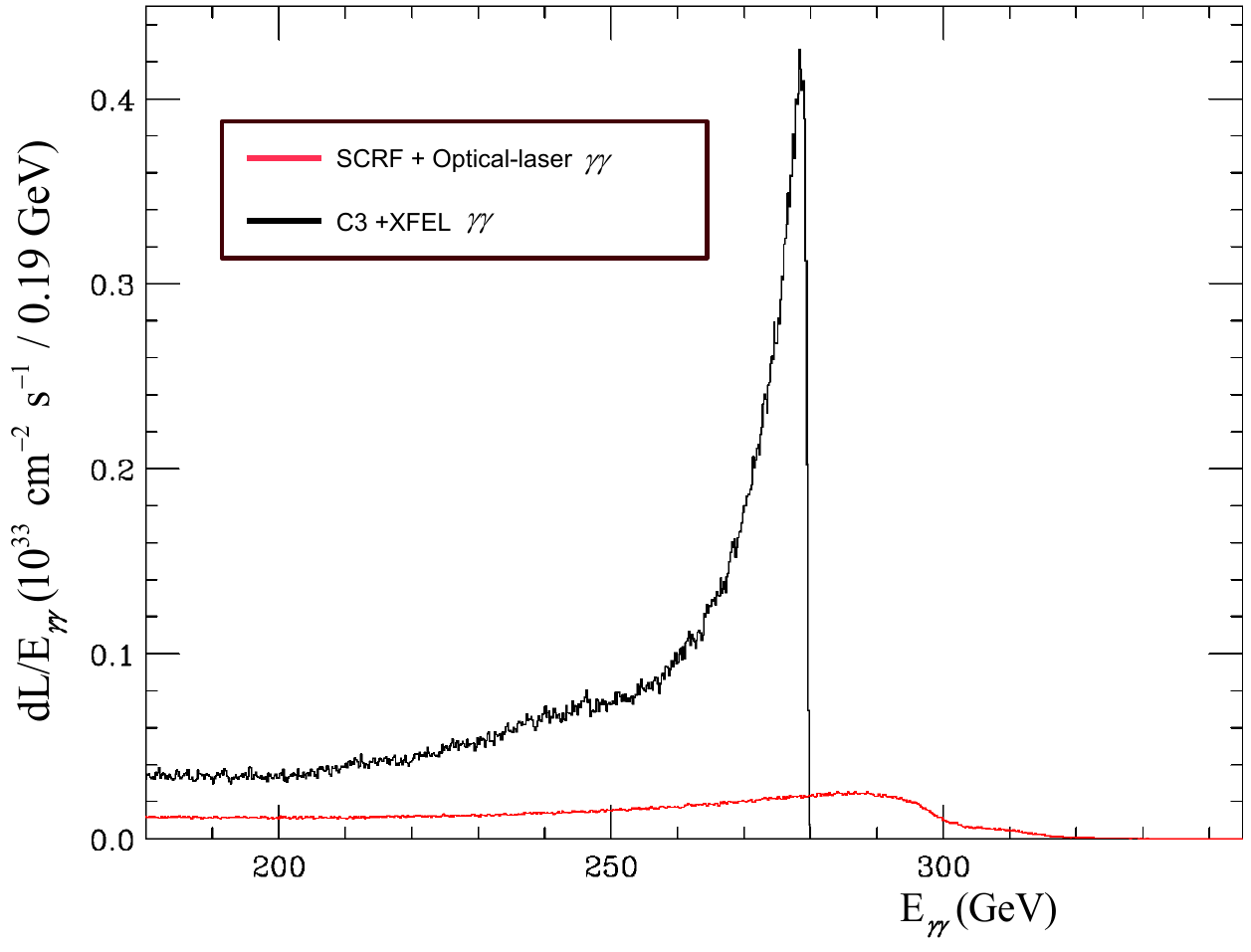}
        \caption{}
        \label{fig:ggLumi280:gg}    
    \end{subfigure}\hfill%
    \begin{subfigure}{.5\textwidth}
        \centering
        \includegraphics[width=0.95\textwidth]{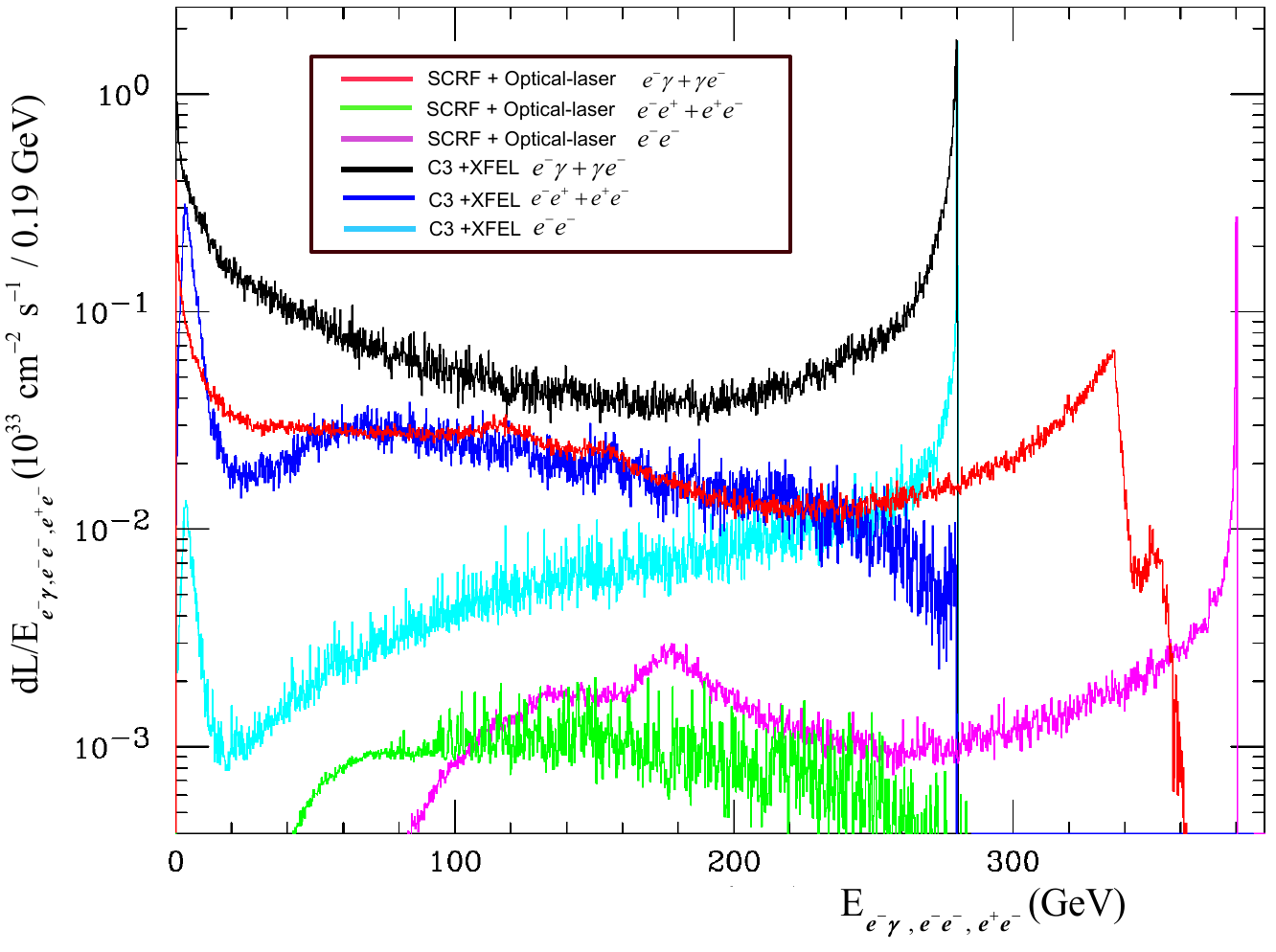}
        \caption{}
        \label{fig:ggLumi280:eg}    
    \end{subfigure}%
    \caption{Luminosity spectra for (a) $\PGg\PGg$ collisions and (b) $\Pem\PGg$, $\Pem\Pem$, and $\ee$ collisions at a $\PGg\PGg$ collider with a peak of the luminosity spectrum near \SI{280}{GeV}. In the XFEL-based design, the peaks of the $\PGg\PGg$ and $\Pem\PGg$ are much sharper, and coincide with the peak in the $\Pem\Pem$ spectrum, while in the case of an optical laser a much higher $\Pem\Pem$ centre-of-mass energy is required. }
    \label{fig:ggLumi280}
\end{figure}

The electron beam parameters for a $\PGg\PGg$ collider with a luminosity spectrum peaking at $\sqrt{s}\approx$ \SI{125}{GeV} are given in Table~\ref{tab:ggeminusbeam}, and the corresponding laser parameters are shown in Table~\ref{tab:gglaser}. 
The corresponding luminosity spectra of the different colliding particle species are shown in Fig.~\ref{fig:ggLumi125}.
The corresponding electron beam and laser parameters as well as the resulting luminosity spectra for a $\PGg\PGg$ collider with a luminosity spectrum peaking at $\sqrt{s}\approx$ \SI{280}{GeV} are given in Tables~\ref{tab:ggeminusbeam280} and~\ref{tab:gglaser280} as well as in Fig.~\ref{fig:ggLumi280}, respectively.

\begin{table}
\begin{center}
\begin{tabular}{ |l|l|l| }
\hline
Final Focus parameters  &  SCRF + Optical Laser  &  \CCC\ + XFEL  \\
\hline
\hline
Electron energy [GeV] & 108   & 62.8  \\ 
Electron beam power [MW] & 4.5   & 1.2  \\
$\beta_x/\beta_y$ [mm] & 1.5/0.3   & 0.03/0.03  \\
$\gamma\epsilon_x/\gamma\epsilon_y$ [nm] & 2500/30   & 120/120  \\
$\sigma_x/\sigma_y$ at $e^-e^-$ IP [nm] & 133/6.5   & 5.4/5.4  \\
$\sigma_z$ [$\mu\mathrm{m}$] & 300  & 20  \\
Bunch charge [$10^{10} e^-$ ] & 2   & 0.62  \\
Bunches/train at IP & 2625  & 165 \\
Train Rep. Rate at IP [Hz] & 5  & 120  \\
Bunch spacing at IP [ns] & 366  & 4.2  \\
$\sigma_x/\sigma_y$ at IPC [nm] & 266/56.7   & 12.1/12.1  \\
$\mathcal{L}_\textrm{geometric}$ [$10^{34}\textrm{cm}^2\ \textrm{s}^{-1}$] &  4.8  & 21 \\
$\delta_E/E$ [\%] &    & 0.1 \\
$L^*$ (QD0 exit to $e^-e^-$ IP) [m] & 3.8  &  1.5 or 3.0 \\
$d_{cp}$ (IPC to IP) [$\mu$m ] & 2600  & 60 \\
crossing angle [mrad] & 20  & 2 or 20  \\
\hline
\end{tabular}
\caption{\label{tab:ggeminusbeam} $e^-$ beam parameters for $\PGg\PGg$ collider with a luminosity spectrum peaking at $\sqrt{s}\approx$ \SI{125}{GeV}.}
\end{center}
\end{table}

\begin{table}
\begin{center}
\begin{tabular}{ |l|l|l|}
\hline
Laser parameters  &  SCRF + Optical Laser  &  \CCC\ + XFEL  \\
\hline
\hline
 Electron energy [GeV] & n.a.  & 31  \\ 
 Normalised emittance [nm] & n.a.  & 120  \\ 
 RMS energy spread $\langle\Delta\gamma/\gamma\rangle$ [\%] & n.a.   &  0.05\\ 
 Bunch charge [nC] & n.a.   & 1  \\ 
  Undulator B field [T] & n.a.  & $\gtrsim$ 1 \\ 
 Undulator period $\lambda_u$ [cm] & n.a.   &  9  \\ 
 Average $\beta$ function [m] & n.a.   &  12 \\
 \hline
$\gamma\gamma$ collider $x$ & 1.9   &  1000  \\
  photon $\lambda$ (energy) [ nm (keV)] & 1056 (0.0012)   & 1.2 (1.0) \\
 photon pulse energy [J] & 2.4   &  0.7  \\
 RMS pulse length [$\mu\mathrm{m}$] & 450    &  20   \\
 $a_{\gamma x}$/$a_{\gamma y}$ (x/y waist) [nm] & 5000/5000   & 21/21\\
 non-linear QED $\xi^2$  & 0.21  & 0.10  \\ 
\hline
\end{tabular}
\caption{\label{tab:gglaser} Laser parameters for $\PGg\PGg$ collider with a luminosity spectrum peaking at $\sqrt{s}\approx$ \SI{125}{GeV}.}
\end{center}
\end{table}

\begin{table}
\begin{center}
\begin{tabular}{ |l|l|l| }
\hline
Final Focus parameters  &  SCRF + Optical Laser  &  \CCC\ + XFEL  \\
\hline
\hline
Electron energy [GeV] & 190   & 140  \\ 
Electron beam power [MW] & 10.5   & 2.1  \\
$\beta_x/\beta_y$ [mm] & 1.5/0.3     & 0.01/0.01  \\
$\gamma\epsilon_x/\gamma\epsilon_y$ [nm] & 2500/30    & 60/60  \\
$\sigma_x/\sigma_y$ at $e^-e^-$ IP [nm] & 88/4.3   & 1.5/1.5  \\
$\sigma_z$ [$\mu\mathrm{m}$] & 300   & 10  \\
Bunch charge [$10^{10} e^-$ ] & 2   & 0.62  \\
Bunches/train at IP & 2625    & 121 \\
Train Rep. Rate at IP [Hz] & 5    & 120  \\
Bunch spacing at IP [ns] & 366  & 5.2  \\
$\sigma_x/\sigma_y$ at IPC [nm] & 176/37.5   & 6.1/6.1  \\
$\mathcal{L}_\textrm{geometric}$ [$10^{34}\textrm{cm}^2\ \textrm{s}^{-1}$] &  12   & 209 \\
$\mathcal{L}_{\gamma\gamma\ \textrm{ for }\sqrt{s}\ >\ 250\ \textrm{GeV}}$ [$10^{34}\textrm{cm}^2\ \textrm{s}^{-1}$] &  0.56   & 2.6 \\
$\delta_E/E$ [\%] &    & 0.1 \\
$L^*$ (QD0 exit to $e^-e^-$ IP) [m] & 3.8   &  1.5 or 3.0 \\
$d_{cp}$ (IPC to IP) [$\mu$m ] & 2600   & 40 \\
crossing angle [mrad] & 20  & 2 or 20  \\
\hline
\end{tabular}
\caption{\label{tab:ggeminusbeam280} $e^-$ beam parameters for $\gamma\gamma$ collider with peak luminosity at $\sqrt{s}$\SI{280}{GeV}}
\end{center}
\end{table}

\begin{table}
\begin{center}
\begin{tabular}{ |l|l|l| }
\hline
Laser parameters  &  SCRF + Optical Laser   & \CCC\ + XFEL  \\
\hline
\hline
 Electron energy [GeV] & n.a. & 31  \\ 
 Normalised emittance [nm] & n.a.  & 60  \\ 
 RMS energy spread $\langle\Delta\gamma/\gamma\rangle$ [\%] & n.a. &  0.05\\ 
 Bunch charge [nC] & n.a. & 1  \\ 
  Undulator B field [T] & n.a. & - \\ 
 Undulator period $\lambda_u$ [cm] & n.a.  &  -  \\ 
 Average $\beta$ function [m] & n.a.  &  - \\
 \hline
  $\gamma\gamma$ collider $x$ & 4.5   &  1300  \\
  photon $\lambda$ (energy) [ nm (keV)] & 809 (0.0015)   & 2.0 (0.61) \\
 photon pulse energy [J] & 9.0   &  1.0  \\
 RMS pulse length [$\mu\mathrm{m}$] & 450    &  20   \\
 $a_{\gamma x}$/$a_{\gamma y}$ (x/y waist) [nm] & 7150/7150   & 21/21\\
 non-linear QED $\xi^2$  & 0.21 & 0.82  \\ 
\hline
\end{tabular}
\caption{\label{tab:gglaser280} Laser parameters for $\PGg\PGg$ collider with peak luminosity at $\sqrt{s}\approx $\SI{280}{GeV}.}
\end{center}
\end{table}

\paragraph*{Timeline and R\&D needed}
While the physics potential of a $\PGg\PGg$ collider is intriguing (c.f.\ Sec.~\ref{sec:phys:altmodes}), the accelerator effort is so far limited to conceptual studies. 
However, first important experimental insights could be gained from the next generation experiments aiming to probe strong-field QED by colliding high-intensity lasers with electron beams. 
For instance at the LUXE experiment~\cite{LUXE:2023crk} planned at the ELBEX facility at the Eu.XFEL in the early 2030s, the creation of one backscattered photon beam could be exercised, tuned and characterised and thus could serve to validate the simulations. 
The actual collision of two photon beams created from laser back-scattering, however, would likely need to wait until a linear collider facility is operational. 
With a flexible sharing of the luminosity between the two IPs, one IP could routinely collect $\ee$ data, while the other could be used as a test facility served by a permil-level fraction of the bunches.

\clearpage

\section{Beyond-collider facilities}
\label{sec:beyond}

\begin{figure}[htb]
    \centering
    \includegraphics[width=1.0\linewidth]{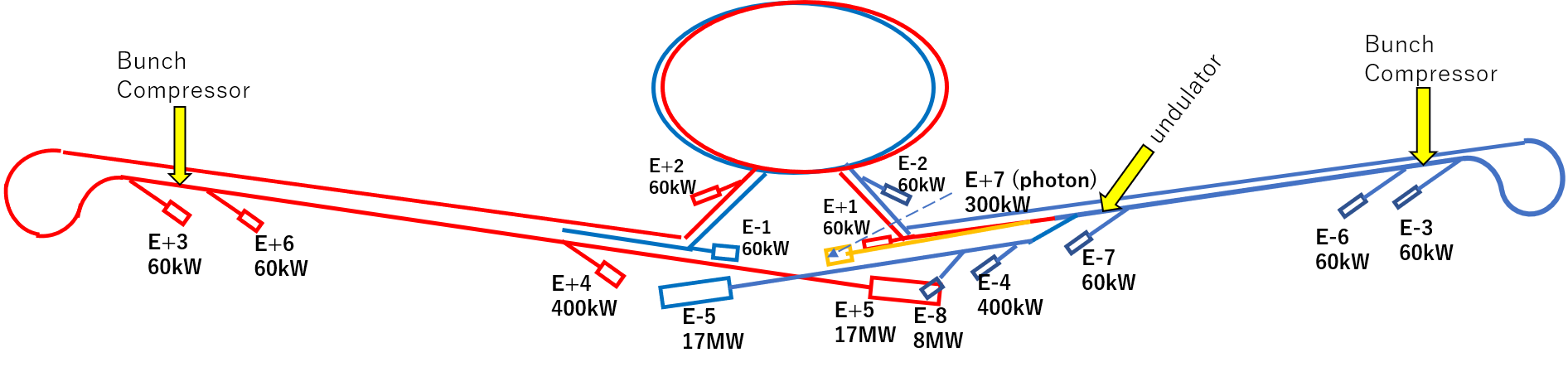}
    \caption{The layout of the ILC and its beam dumps~\cite{ILCInternationalDevelopmentTeam:2022izu}. The main beam dumps (E$-$5 for electrons and E$+$5 for positrons) are designed to absorb \SI{17}{MW} each, the tune-up dumps (E$-$4 and E$+$4) for \SI{0.4}{MW} each. For a Linear Collider Facility with two interaction points, c.f.\ Sec.~\ref{sec:acc:BDS}, the arrangement of beam dump facilities still needs to be optimised.}
    \label{fig:ILCbeamDumps}
\end{figure}

The Linear Collider Vision team aims to plan for a facility that exploits all of its opportunities as well as possible, and fully integrates experiments that are not directly related to the collider physics programme.
A key difference of a linear collider compared to a circular collider is the continuous dumping of the electron and positron beams in the main beam dumps. 
In addition, other auxiliary dumps, such as the tune-up dumps, are foreseen in the collider complex for setup and commissioning, and at these locations beam could be extracted for experiments and R\&D facilities. 
The locations of the various beam dumps and the power they are designed to absorb are indicated in Fig.~\ref{fig:ILCbeamDumps} for the example of the ILC~\cite{ILCInternationalDevelopmentTeam:2022izu}.

The physics opportunities with dumped or extracted beams have been discussed in Sec.~\ref{sec:phys:PBC}.
In this section, we elaborate on the corresponding facilities and provide some preliminary assessments of the requirements. 
We first describe in Sec.~\ref{sec:beyond:maindump} the infrastructure required at the main beam dumps to investigate beyond-collider physics as well as to create neutron and muon facilities for irradiation and nuclide production. 
In Sec.~\ref{sec:beyond:tunedump}, we discuss facilities for experiments and R\&D with beams extracted before collisions, including a possible plasma-wakefield accelerator R\&D facility.

\subsection{Main beam dump facilities}
\label{sec:beyond:maindump}

The main beam dumps of a linear collider absorb a few $10^{21}$ high-energetic electrons or positrons per year, c.f.\ Table~\ref{tab:ILCbaseline}, corresponding to $\mathcal{O}$(10)\,MW. The ILC main beam dumps, labelled E-5 and E+5 in Fig.~\ref{fig:ILCbeamDumps}, are planned as water dumps designed for a maximum power of \SI{17}{MW}~\cite{Satyamurthy:2012zz}. 

As discussed in Sec.~\ref{sec:phys:PBC:beamdumps}, the dumped beams will create a plethora of secondary particles, which can be used for particle physics experiments as well as for irradiation facilities. In the following sections, we discuss some considerations for the implementation of such facilities in Secs.~\ref{sec:beyond:maindump:exp} and~\ref{sec:beyond:maindump:irrad}, respectively.

\subsubsection{Beam dump experiments}
\label{sec:beyond:maindump:exp}

\begin{figure}[htbp]
   \centering
   \begin{subfigure}{0.95\textwidth}
      \includegraphics[width=\linewidth]{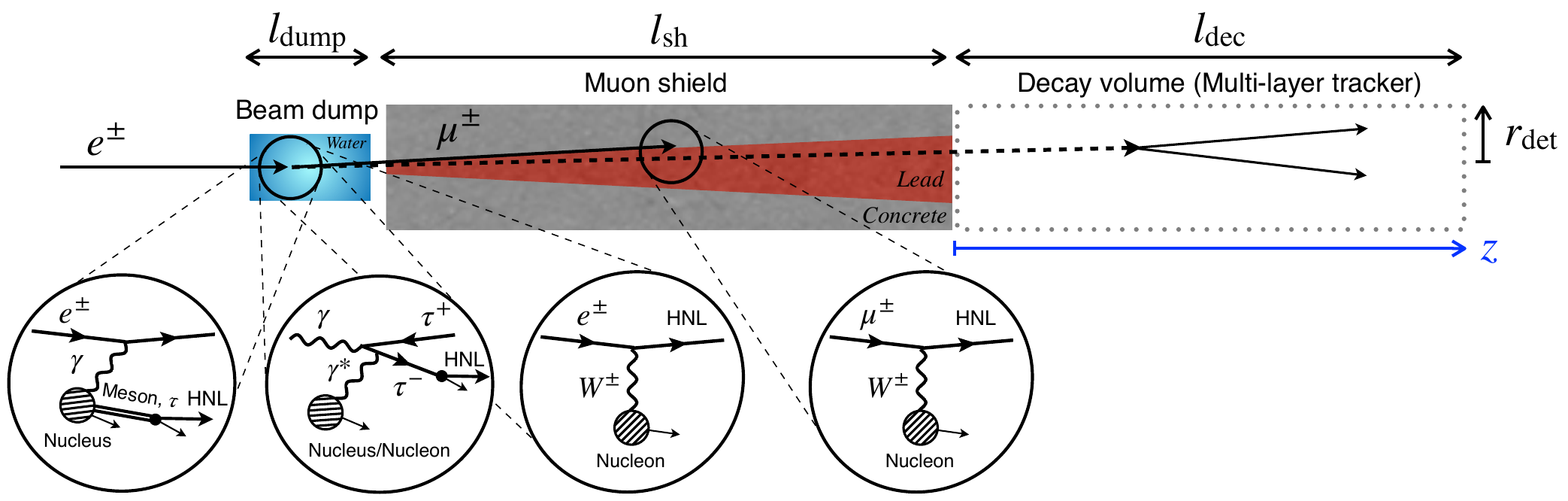}
      \caption{}
      \label{fig:dump:exp:lowE}
   \end{subfigure}
   \begin{subfigure}{0.95\textwidth}
   \vspace*{0.5cm}
      \includegraphics[width=\linewidth]{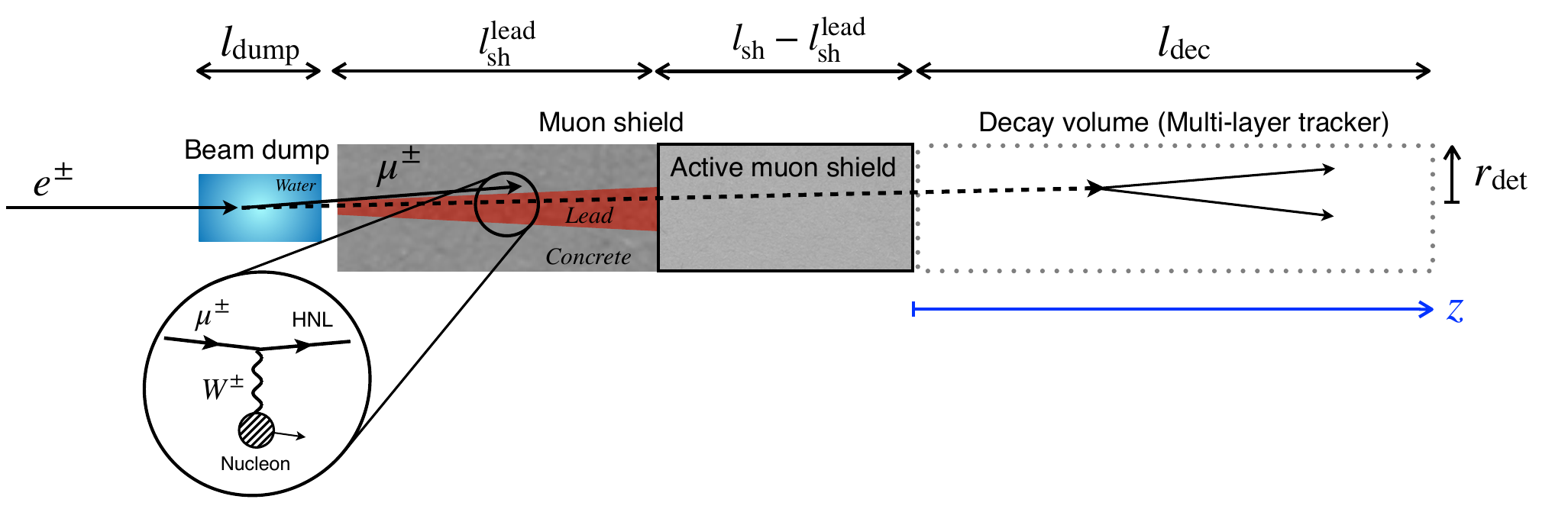}
      \vspace*{-0.5cm}
      \caption{}
      \label{fig:dump:exp:highE}
   \end{subfigure}
   \caption{Schematics of an experiment searching for exotic particles (here with heavy neutral leptons, HNL, as example) produced in the beam dumps of a linear collider. (a) At beam energies of \SI{125}{GeV}, a passive muon shield is sufficient. (b) At higher beam energies, e.g.\ \SI{500}{GeV}, active shielding is required. Both figures from~\cite{Nojiri:2022xqn}.}
   \label{fig:dump:exp}
\end{figure}

The generic setup of an experiment searching for the production of exotic particles in the beam dump is shown in Fig.~\ref{fig:dump:exp}. 
A long muon shield aims to reduce the background from secondary muons, followed by a decay volume instrumented with e.g.\ tracking layers to detect charged decay products of the ``invisible'' exotic particles and/or identify an veto background leaking from the shield, and further downstream detectors. 
In previous studies, a muon shield of length \SI{70}{\meter} was assumed~\cite{Sakaki:2020mqb,Asai:2021ehn}, as shown in  Fig.~\ref{fig:dump:exp:lowE}. An active veto, illustrated in Fig.~\ref{fig:dump:exp:highE}, could reduce the background levels further, as we will discuss below.

Neutrino interactions near the end of the muon shield induce additional background: 
High-energy neutrinos can produce neutral hadrons, such as \PKzS or \PKzL that decay to charged pions. 
The tracking system in the decay volume is essential to identify and veto them. 
This type of background has been estimated with \textsc{Geant4} for an early setup of the SHiP experiment~\cite{Bonivento:2013jag,SHiP:2015vad}. 
The relevant hadrons were found to always be produced at the edge of the muon shield and can be significantly reduced by requiring charged tracks consistent with the kinematics of e.g.\ heavy neutral leptons (c.f.\ Sec.~\ref{sec:phys:PBC:beamdumps}). 

The background in the beam dump experiments at a linear collider can be inferred by comparing the number of produced neutrinos with SHiP. The SHiP experiment expects $7\times 10^{17}$ neutrinos for $2\times 10^{20}$ protons on target. Many of the background events with charged tracks can be reduced by simple topological cuts. Similar reduction factors using the topology cuts are assumed in~\cite{Nojiri:2022xqn}. At a beam energy of \SI{125}{GeV} (\SI{500}{GeV}), we estimate \num{8e15} (\num{3e16}) neutrinos for \num{4e21} electrons on target. 
The active veto system of SHiP reduces the muons pointing to the decay volume by a magnet, thereby reducing the neutrino background further by a factor of $\mathcal{O}(10^{-4})$. This technique, sketched in Fig.~\ref{fig:dump:exp:highE}, should achieve an analogous reduction of backgrounds at a linear collider beam dump, and the physics projections in Sec.~\ref{sec:phys:PBC:beamdumps} should be updated accordingly in the future.

Another important consideration is the required lateral space. 
The location of the beam dump in the ILC design, with its \SI{14}{mrad} crossing angle, is about \SI{300}{m} away from the IP, where the transverse separation between the incoming and outgoing beams is roughly \SI{4}{m}. 
If one of the LCF IPs has a small crossing angle of e.g.\ \SI{2}{mrad}, a location \SI{2}{km} from the IP would be needed to provide the same separation, which we consider close to the minimum required to install an experiment.
Integrating a beam dump detector into a linear collider facility would require careful consideration by accelerator and experimental physicists.
If the second IP has a larger crossing angle, e.g.\ \SI{20}{mrad}, a reasonable transverse separation can be achieved in a much shorter distance. 
Then, the constraints are much less severe, and the experimental site can be planned independently of the main accelerator system. 
Since it is planned to share the luminosity equally between the two IPs, it is advantageous to use the beam dump of the second IP for the beam dump facilities.

\subsubsection{Neutron and muon facilities}
\label{sec:beyond:maindump:irrad}

At the main dump of the linear collider facility, a high number of high-energy neutrons and muons are produced, see Sec.~\ref{sec:phys:PBC:beamdumps}. 
They can be used for auxiliary purposes, such as irradiation tests of integrated circuits or to produce radionuclides, as described in the following. 
In both cases, remote access and changes of the devices under test in the radiation cavern must be possible during beam operation. 
Furthermore, the facility must be large enough to accommodate entire devices, which will require a metre-sized cavern. 

\paragraph*{Irradiation facility}

Our daily lives increasingly depend on larger integrated circuits with ever-smaller transistor structures. 
The demand for verifying their reliability grows. 
One possible source of malfunctions are soft errors induced by cosmic neutrons and muons~\cite{Iwashita:2020er}.
They can induce a change of state (a bit flip) that can be reset by a power cycle (a cold reboot). 
In a soft error, the device itself is not damaged, but only the data on it. 
Assessing the soft error rate of an entire circuit requires a high-intensity radiation field with an atmospheric-like energy spectrum over a large area to irradiate a system as a whole. 
The vicinity of a linear collider beam dump meets all these requirements~\cite{Sakaki:2022udd}. 
Other irradiation facilities exist, e.g.\ at RCNP in Osaka (Japan) or at TRIUMF in Vancouver (Canada). 
However, they can only provide neutrons with energy up to a few hundred MeV in small irradiation areas. 

\begin{figure}
    \centering
    \includegraphics[width=1.0\linewidth]{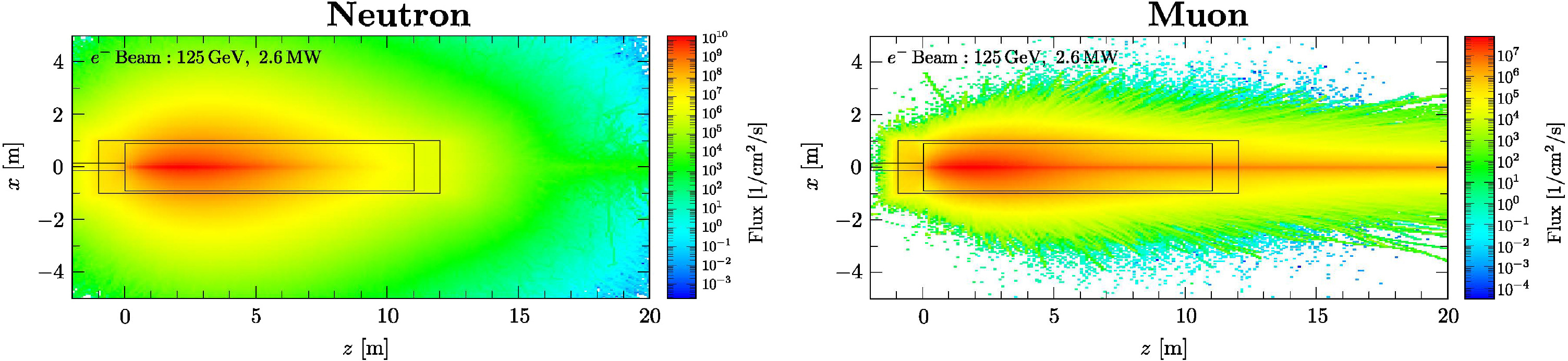}
    \caption{Neutron and muon fluxes around the electron beam dump of an LCF. The water beam dump (inner rectangle) is in a cavern (outer rectangle) which is surrounded by a uniform concrete shield. Figure from~\cite{Sakaki:2022udd}. }
    \label{fig:beamDumpNeutronsMuons}
\end{figure}

Neutrons and muons have different production characteristics in a beam dump as shown in Fig.~\ref{fig:beamDumpNeutronsMuons}. 
The neutrons are mostly produced through photo-nuclear reactions of the photons from the electromagnetic shower.
They are emitted almost isotropically from the dump, with the core of the production being \num{2}-\SI{4}{m} in the beam direction from the beam injection point. 
Therefore, high-intensity neutrons can be obtained close to the neutron emission region on the side of the beam dump. 
The muons are mainly produced by pair production from bremsstrahlung and tend to be scattered in the beam direction. 

\begin{figure}
  \centering
  \begin{subfigure}[b]{0.49\textwidth}
    \includegraphics[width=\textwidth]{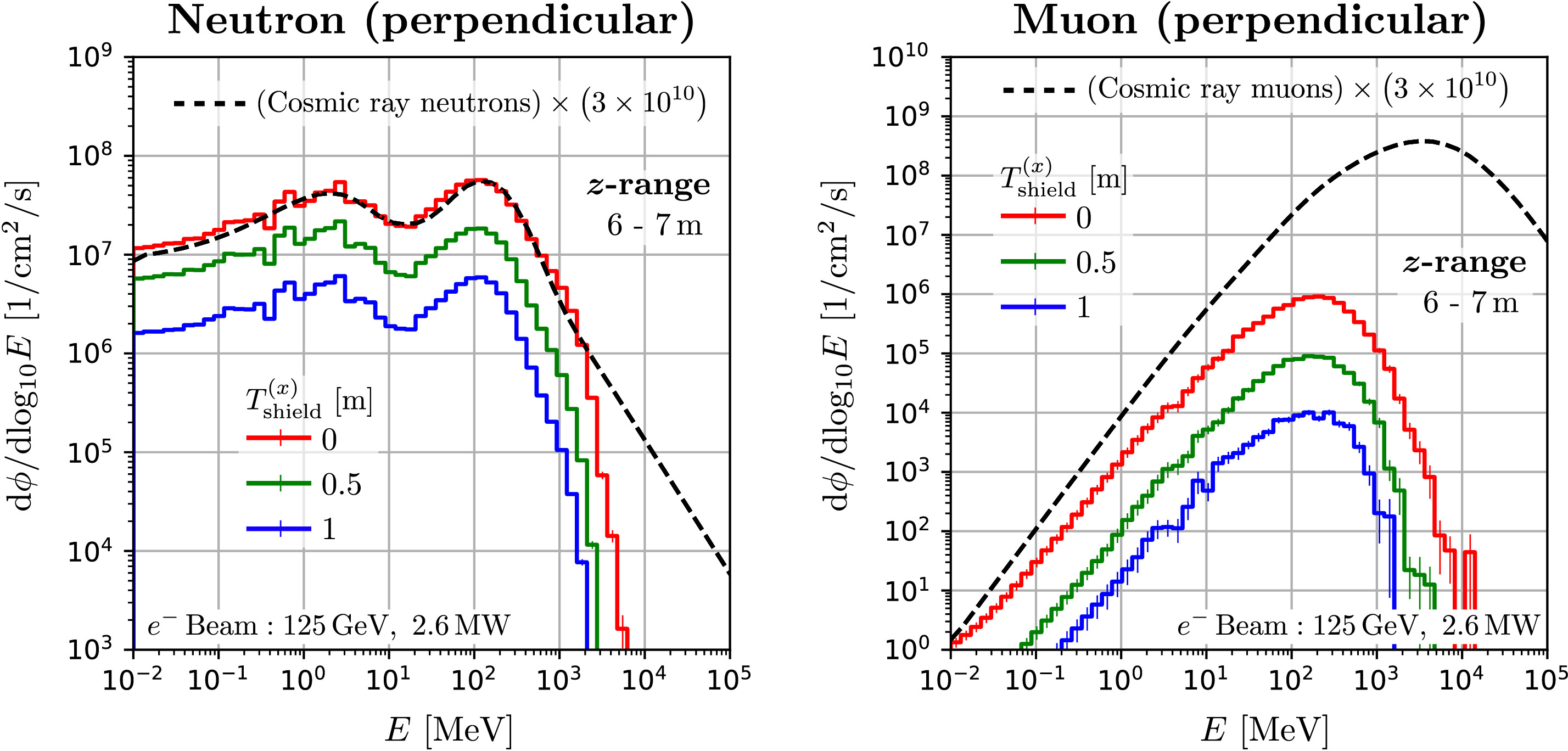}
    \caption{~}
    \label{fig:beamDumpPerpendicularSpectra}
  \end{subfigure}
  \hfill
  \begin{subfigure}[b]{0.49\textwidth}
    \includegraphics[width=\textwidth]{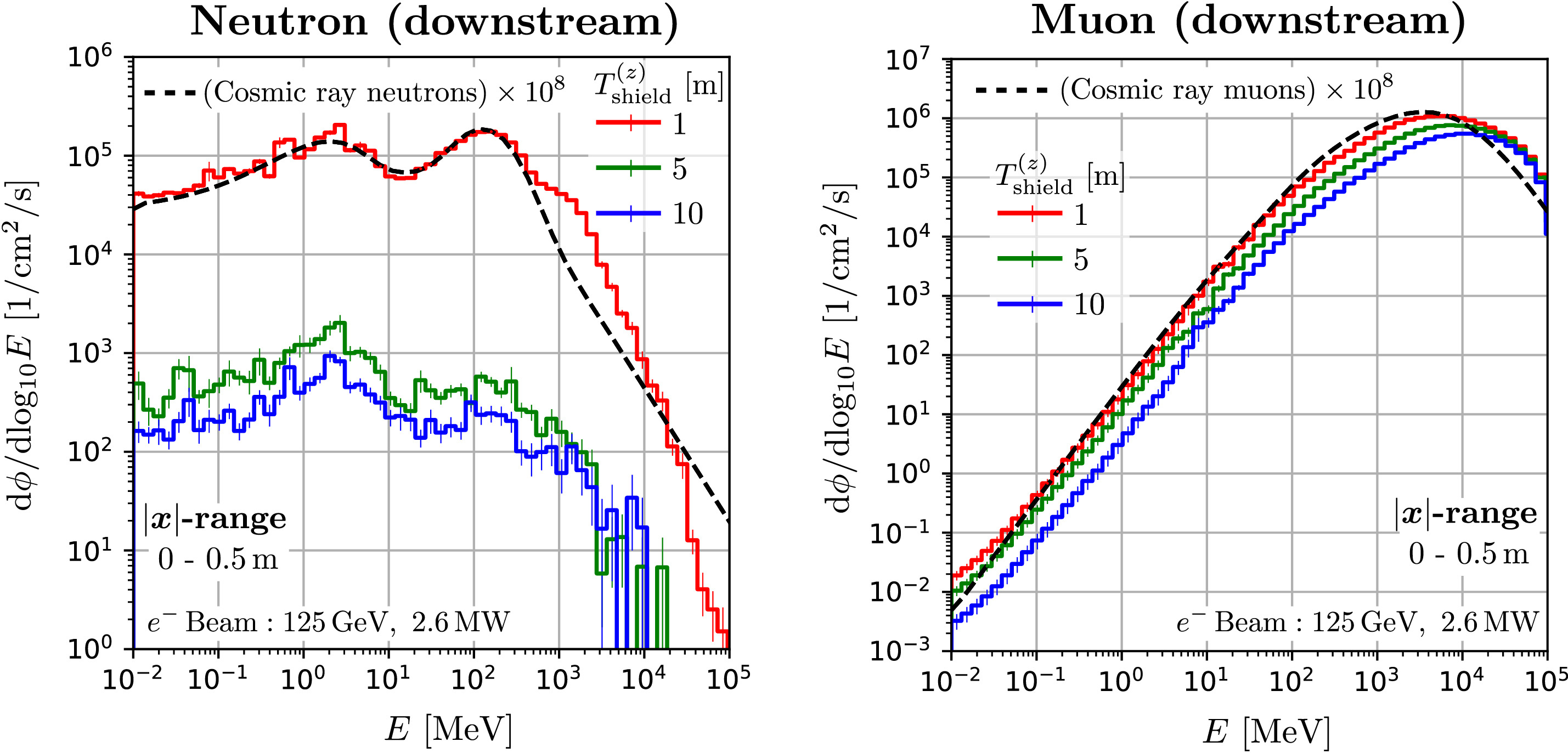}
    \caption{~}
    \label{fig:beamDumpDownstreamSpectra}
  \end{subfigure}
  \caption{Energy distribution of neutrons and muons for various concrete shield thicknesses (colors) and in comparison to cosmic rays (dashed line). (a) Spectra on the side of the dump at $z =$ \num{6}-\SI{7}{m}. (b) Spectra downstream of the beam dump at $|x| \leq$ \SI{0.5}{m}. Figure from~\cite{Sakaki:2022udd}. }
  \label{fig:beamDumpSpectra}
\end{figure}

As described in~\cite{Sakaki:2022udd}, there are two possible locations for the irradiation facilities at the beam dump of a linear collider. 
A cavern on the side of the dump at $z =$ \num{6}--\SI{7}{m} is mainly suitable for neutron irradiation. 
The neutron spectrum is atmospheric-like up to an energy of about \SI{2}{GeV} with a flux more than a factor $10^{10}$ higher than cosmic rays at sea level. 
The muon flux in this area is much lower and can be further reduced by increasing the thickness of the concrete shield between the dump and the irradiation cavern. 
A cavern downstream of the dump on-axis is well suited for a combined neutron and muon irradiation facility. 
With a concrete shielding of the order of \SI{1}{m}, both spectra are atmospheric-like but with a flux higher by a factor of $10^8$. 
Since the shield thickness has almost no effect on the muons, but reduces the neutron flux significantly, a thicker concrete shield can be used if only muons are desired for the irradiation facility. 
The neutron and muon energy spectra for both locations are shown in Fig.~\ref{fig:beamDumpSpectra}. 

\paragraph*{Facility for radionuclide production}

Radionuclides or radioactive isotopes find applications in many fields such as nuclear physics, but most notably in nuclear medicine for diagnosis, treatment, and research. 
These are primarily produced through the exposure of stable targets to neutrons or charged particles. 
The result of such irradiation is an activated target material that has to undergo further processing to be usable. 
As part of the PRISMAP project network, European facilities provide a wide variety of radionuclides for medical research~\cite{Prismap:2021mr}. 

As described above, the beam dump of a linear collider facility can provide a high flux of high-energy neutrons that can possibly also be used for radionuclide production. 
Furthermore, a study for the ILC showed that the photon dump after the undulator for the generation of positron (E+7) could also be used, especially for $^{99}$Mo production~\cite{Morikawa:2018ilc}. 
The detailed potential and competitiveness must be studied with a suitable tool such as ActiWiz~\cite{Vincke:2018mpj}.

\subsection{Facilities at tune-up and auxiliary dumps}
\label{sec:beyond:tunedump}

Complementary to the extreme intensity but broad energy spectrum of the disrupted main beam after collisions, beam extraction at selected points before collision provides opportunities for experiments and R\&D with highest-quality bunches. As shown in Fig.~\ref{fig:ILCbeamDumps}, several extractions and dumps are foreseen at the ILC. In particular the location of the so-called tune-up dumps E$+$4 and E$-$4, located after the main linac and thus providing beams at full energy, as well as the extractions before and after the bunch compressors (E$\pm$3 and E$\pm$6, respectively) providing \SI{5}{GeV} are of interest. 

While the main physics programme is running, single or few bunches could be extracted at these locations without noticeably harming the collider luminosity. In particular during commissioning of the accelerator, the tune-up dumps are foreseen to be used continuously and thus could serve high-rate experiments during these times. Due to the full beam energy being available, the tune-up dump is particularly interesting for thin-target beam dump searches as described in the following, but also for tests of strong-field QED (Sec.~\ref{sec:phys:PBC:sfqed}). For R\&D facilities, e.g.\ for plasma-wakefield acceleration, the pre-linac extractions with \SI{5}{GeV} could be more suitable as will be discussed Sec.~\ref{sec:beyond:tunedump:PWA}.  The size of the facilities may vary depending on the usage, and the details have to be worked out. For strong-field QED experiments, an energy spectrometer is needed which leads to a length requirement of about \SI{100}{\meter} for the maximum energy of \SI{1}{TeV} after the upgrade. For plasma-wakefield acceleration R\&D at \SI{100}{GeV}, this requirement may increase by a factor of ten.

\subsubsection{Slow extraction}
\label{sec:beyond:tunedump:slow}

For many applications, having a low bunch charge even down to quasi-single electrons or positrons has advantages. Two examples are fixed-target experiments with a thin target and strong-field QED experiments with electron-laser collisions. The former allows access to bigger couplings in new-physics searches since these particles do not decay within the target and can therefore be detected, due to the short target length. This option was studied for the linear accelerator facility at KEK~\cite{Ishikawa:2021qna}. The latter allows for studying the elementary electron-photon interaction in a strong field and coherent electron-positron plasma creation, as described in more detail in Sec.~\ref{sec:phys:PBC:sfqed}. 

Slow extraction usually refers to the gradual extraction of particles in a circulating beam. At a linear collider facility, the problem is more delicate due to the different bunch structure. However, similar approaches can be applied. One of those is the extraction of particles from the beam halo with the channelling effect in a bent crystal, as successfully demonstrated at CERN's Large Hadron Collider LHC with \SI{6.5}{TeV} protons~\cite{Scandale:2016krl}. This could be done at the location of the tune-up dump, where a single bunch can first be extracted from each bunch train with the full energy. Afterwards, the crystal can be brought to close proximity of the beam such that the desired number of particles can be extracted. The rate can be tuned by the distance from the crystal to the beam. Depending on the charge of the particle, different orientations of the crystal planes are advantageous. For energy of \SI{120}{GeV}, deflections of \num{100}--$\SI{150}{\mu rad}$ are possible for electrons and positrons with axial channelling and even more in the case of planar channelling of positrons~\cite{Bagli:2016mbf, Bandiera:2020rwx}. 

\subsubsection{Laser facility}
\label{sec:beyond:tunedump:laser}

To test the fundamental physics of strong-field QED described in Sec.~\ref{sec:phys:PBC:sfqed}, a high-power laser is required. 
Usually, laser facilities are built above ground, and the laser pulse is guided through a laser beam pipe to the electron-laser interaction point. 

Nowadays, systems with \SI{200}{TW} peak power can be bought off the shelf and have a small, container-sized footprint of approximately ${4 \times 4~\mathrm{m^2}}$~\cite{Thales:2020qut}. 
They can deliver a pulse energy of about \SI{5}{J} which, depending on the waist at the beam focus, leads to an intensity parameter of $\xi \approx 20$. 
The laser systems with the highest intensities currently available have a peak power of \SI{10}{PW} and can reach intensity parameters of $\xi \gtrsim 200$~\cite{Papadopoulos:2016tjf, Radier:2022ppf}.
However, they require hall-sized clean rooms of roughly ${20 \times 50~\mathrm{m^2}}$.

\subsubsection{Plasma-wakefield accelerator R\&D facility}
\label{sec:beyond:tunedump:PWA}

Plasma-Wakefield Acceleration (PWA) shows considerable promise for the future of high-energy physics because it can provide exceptionally high acceleration gradients and focussing capabilities~\cite{Joshi:2003fj}. Also within the framework of the Linear Collider Vision, it is considered one of the main upgrade options for the future energy upgrade of the collider as described in Sec.~\ref{sec:acc:upgrade}. 
The higher acceleration gradient compared to classic acceleration with superconducting cavities leads to a reduction in accelerator length.

In a PWA, a driver is shot into a plasma, creating a wakefield that can be used to transfer the energy of the driver into a witness bunch of charged particles. There are three options for the driver, namely a laser, a proton, or an electron pulse. For the applicability of this new technology to a high-energy electron-positron collider, only electrons are currently suitable as a driver. So far, lasers are not efficient enough and protons cannot be provided at high enough rates to be feasible.  

For electron-driven PWA, one of the main problems today is staging~\cite{Lindstrom:2020pzp}. Since a single cell offers only limited capability of transferring energy into the witness bunch, many shorter acceleration cells have to be added/staged together. This issue is being addressed and will probably be solved by the time a linear collider facility is built. However, other challenges may persist and need to be overcome before a collider can be upgraded to PWA technology. They may include the acceleration of positrons~\cite{Cao:2023ksg} and the achievement of the required high beam currents to reach the luminosity goals~\cite{Litos:2014yqa}. 

A linear collider facility offers various options for the R\&D of electron-driven PWA. For many applications, a low-energy beam is sufficient and easier to handle. Therefore, the beam before the bunch compressor with an energy of about \SI{5}{GeV} could be used. In the layout of the linear collider facility, an extraction beamline with a low-power beam dump is planned (called E-3 in the ILC design). This cavern could be enlarged to host a facility for the PWA R\&D. A tunnel of a few hundred metres is required. For applications requiring high energy, the beam after the entire accelerator is needed. A possible location is the area of the tune-up dump (E-4). Since the length of a PWA facility scales roughly with $\sqrt{E}$, a km-long complex would be necessary. Such a beam could also be used for high-field QED experimentation, c.f.\ Sec.~\ref{sec:phys:PBC:sfqed}.

\section{Detectors -- opportunities for the community}
\label{sec:det}

Detectors at a collider are an integral part of the facility and need to be part of the planning from the beginning. The detectors should be able to extract science as efficiently and as easily as possible from the enormous amounts of data produced by the collider facilities. They rely on technologies that in many cases go well beyond the currently established state-of-the-art, and they utilise advanced concepts and methodologies to handle and curate (and eventually extract scientific results from) the data.  Detectors need to be tightly integrated into the design of the collider, in particular taking into account the characteristics of the interaction region.

Detectors also need to be optimised towards the particular environment of the collider at hand. Linear colliders operate in a pulsed mode, at very high energies, but in a rather low-rate environment. The intrinsically clean events, i.e.\ the absence of very large hadronic backgrounds, offer a great opportunity to the community to concentrate on the development of technologies that stress precision and on the extraction of the complete collision information. The detectors are, therefore, not only very powerful tools for the science at a linear collider facility, but also platforms for the development and demonstration of cutting-edge technologies. 

The main requirements for linear collider detectors have been introduced in Sec.~\ref{sec:expenv}; they are recalled here for convenience: 
\begin{description}
    \item[Impact parameter resolution:] An impact parameter resolution of 
   $ \SI{5}{\mu m} \oplus \SI{10}{\mu m} / [ p~[{\mathrm{GeV}/c}]\sin^{3/2}\theta]$ has been defined as a goal, where $\theta$ is the angle between the particle and the beamline. 
    \item[Momentum resolution:] An inverse momentum resolution of $\Delta (1 / p) = 2 \times 10^{-5}~\mathrm{[(GeV/c)}^{-1}]$ asymptotically at high momenta should be reached. Maintaining excellent tracking efficiency and very good momentum resolution at lower momenta will be achieved by an aggressive design to minimise the detector's material budget.
    \item[Jet energy resolution:] Using the paradigm of particle flow, a jet energy resolution $\Delta E/ E = 3-4\%$ or better for light flavour jets should be reached. The resolution is defined in reference to light-quark jets, as the RMS of the inner \SI{90}{\%} of the energy distribution. 
    \item[Readout:] The detector readout will not use a hardware trigger, ensuring full efficiency for all possible event topologies.
    \item[Powering:] The power of major systems will be cycled between bunch trains. This will reduce the power consumption of the detectors and minimise the amount of material in the detector. The corresponding smaller need of services will also be beneficial for the acceptance and hermeticity of the detectors.
\end{description}

More recently, the ability to identify different kinds of charged hadrons, like pions, kaons and protons has been recognised as a further asset~\cite{Radkhorrami:2024bbf, Irles:2024ipg, Okugawa:2024dks}. Another interesting development not yet fully considered in the typical linear collider detector concepts are fast timing detectors, which offer a wide range of potential improvements from time-of-flight particle identification to 5D-calorimetry~\cite{Dudar:2024quz}.

The high granularity of detectors optimised for particle-flow reconstruction cannot be fully exploited with advanced reconstruction algorithms. In fact, the optimisation of the detectors is intimately intertwined with the sophistication of the reconstruction techniques. These currently undergo a change of paradigm towards machine learning, which has only started to reveal its impact on the detector design and performance, c.f.\ Sec.~\ref{sec:RunScenarios}.

\subsection{Concepts}

The linear collider community has developed several integrated detector concepts, adapted to the different linear accelerator options under discussion. ILD, SiD and CLICdet~\cite{ILDConceptGroup:2020sfq, Breidenbach:2021sdo, CLICdp:2018cto} have all undergone intense optimisation efforts and have been subjected to a series of international reviews. Nevertheless, they all evolve continuously, and remain eager to integrate new ideas and technologies. 

These detectors are all conceived as multi-purpose devices, optimised for precision measurements in the area of Higgs and electroweak physics. They should be able to cover a collision energy range from \SI{90}{GeV} up to several \SI{}{TeV}, in the case of CLICdet, or \SI{1}{TeV} for a detector at the ILC. Over the course of the development of these systems, the concept of particle flow as a method to optimally reconstruct the event properties has been developed and has been integrated into the detector design for all concepts. Particle flow has profound consequences for the design of the detector, as it stresses the need for (i) high granularity throughout the detector (in particular in the calorimeter system) and (ii) very high reconstruction efficiency in the tracker, combined with an excellent capability to link objects in the tracker with objects in the calorimeter.

Regardless of the concrete linear collider option, all linear accelerator technologies together define a set of boundary conditions that have a significant impact on the way these detectors are designed, in addition to the requirements coming from the science programme at these facilities. 

Most prominently, linear colliders are pulsed machines. Collisions will happen during bunch trains, followed by relatively long periods without collisions. The detailed timing structure, the length of the bunch trains and, most importantly, the time differences between pulses within a train vary a lot, leading to  somewhat differing constraints on the related detailed detector designs. 

Due to the pulsed operation, the front-end electronics of the detectors stays mostly in idle mode at shut-down bias current levels of the power consuming parts like the pre-amplifiers in the calorimeter. This significantly reduces the average power consumption of the detectors and, with proper design, reduces very much the need for active cooling of detector components. This has very advantageous consequences on the overall material budget and reduces, particularly in the inner part of the detector, the amount of inactive material needed to thermally control the detectors. 

Furthermore, the pulsed mode of operation makes it much easier to locally collect data during the collision part of the train and then ship the data out during the inter-train times. This makes a trigger-less operation of these detectors much easier to achieve. 

\subsubsection{Vertexing system}
The system closest to the interaction region is a pixel detector designed to reconstruct decay vertices of short-lived particles with great precision, as well as low-momentum tracks barely reaching the main tracker. Over the past few years, the monolithic active pixel sensor (MAPS) CMOS technology has been adapted as baseline by all three concepts. It has matured to a point where all the requirements (material budget, readout speed, granularity) that a linear collider detector imposes can be met. The technology has experienced a first large-scale use in the STAR vertex detector~\cite{CONTIN20167}; more recently it has been deployed in the upgrade of the ALICE Inner Tracking System (ITS-2)~\cite{ALICE:2013nwm}. 

The detailed system design is the subject of active R\&D. Typically several layers of high-precision pixel detectors form a barrel that is complemented by endcap disks. In the CLICdet design, the endcap disks are arranged in a screw-like pattern, which allows efficient air-cooling of the detectors. In ILD sensors are arranged in pairs, which allows the combination of sensors with high spatial resolution with sensors that are optimised for timing resolution, in one super-layer. A range of different technologies for the sensors and the readout are being explored, in close cooperation with the DRD collaborations recently initiated at CERN. 

With current technology, a spatial resolution around $\SI{3}{\mu m}$ at a pitch of about $\SI{17}{\mu m}$ is in reach, and a timing resolution of around \num{2}-$\SI{4}{\mu s}$, possibly lower, per (super-)layer can be obtained. R\&D is directed towards improving this even further, to a point which would allow hits from individual bunch crossings to be resolved. 

A key challenge for an integrated detector design is the development of an extremely light-weight support structure, which brings the goal of \SI{0.15}{\%} of a radiation length per layer within reach.  Structures that reach \SI{0.21}{\%} X$_0$ in most of the fiducial volume are now used in the Belle~II vertex detector~\cite{PLUME:2011rwc}. At a linear collider, the intrinsically pulsed operation of the system, which opens the possibility to have no liquid cooling at all, is a significant boost for the efforts towards achieving passive-material budgets of around \SI{0.1}{\%} of a radiation length. Particularly promising is the prospect of combining thin MAPS sensors that can be bent into a cylindrical geometry with ultra-thin support structures without the need for active cooling. 

\begin{figure}
    \centering
    \begin{subfigure}{.33\textwidth}
       \includegraphics[width=\textwidth]{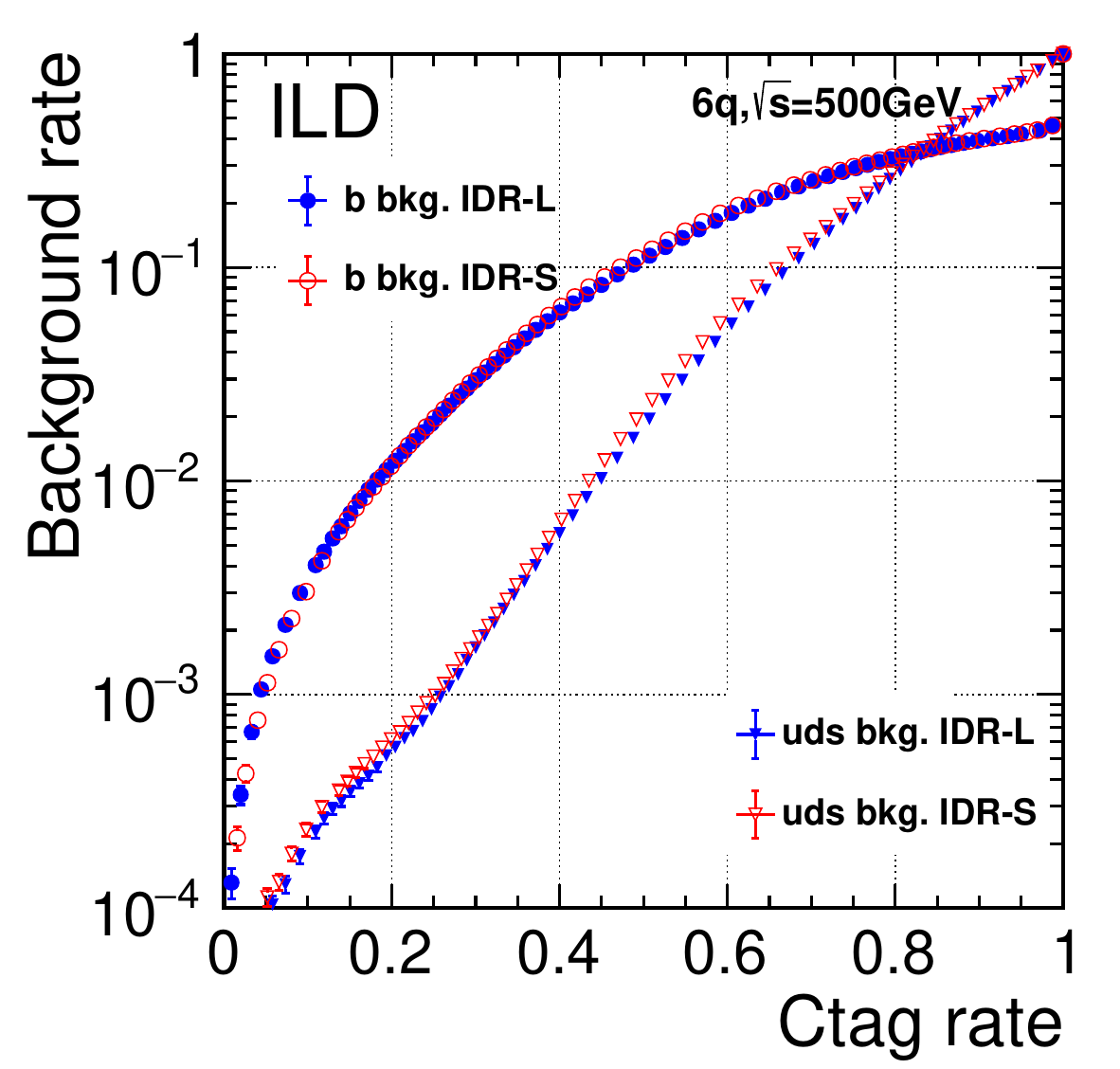}      
        \caption{}
        \label{fig:flavtag:lcfiplus_c}    
    \end{subfigure}%
    \begin{subfigure}{.33\textwidth}
        \includegraphics[width=\textwidth]{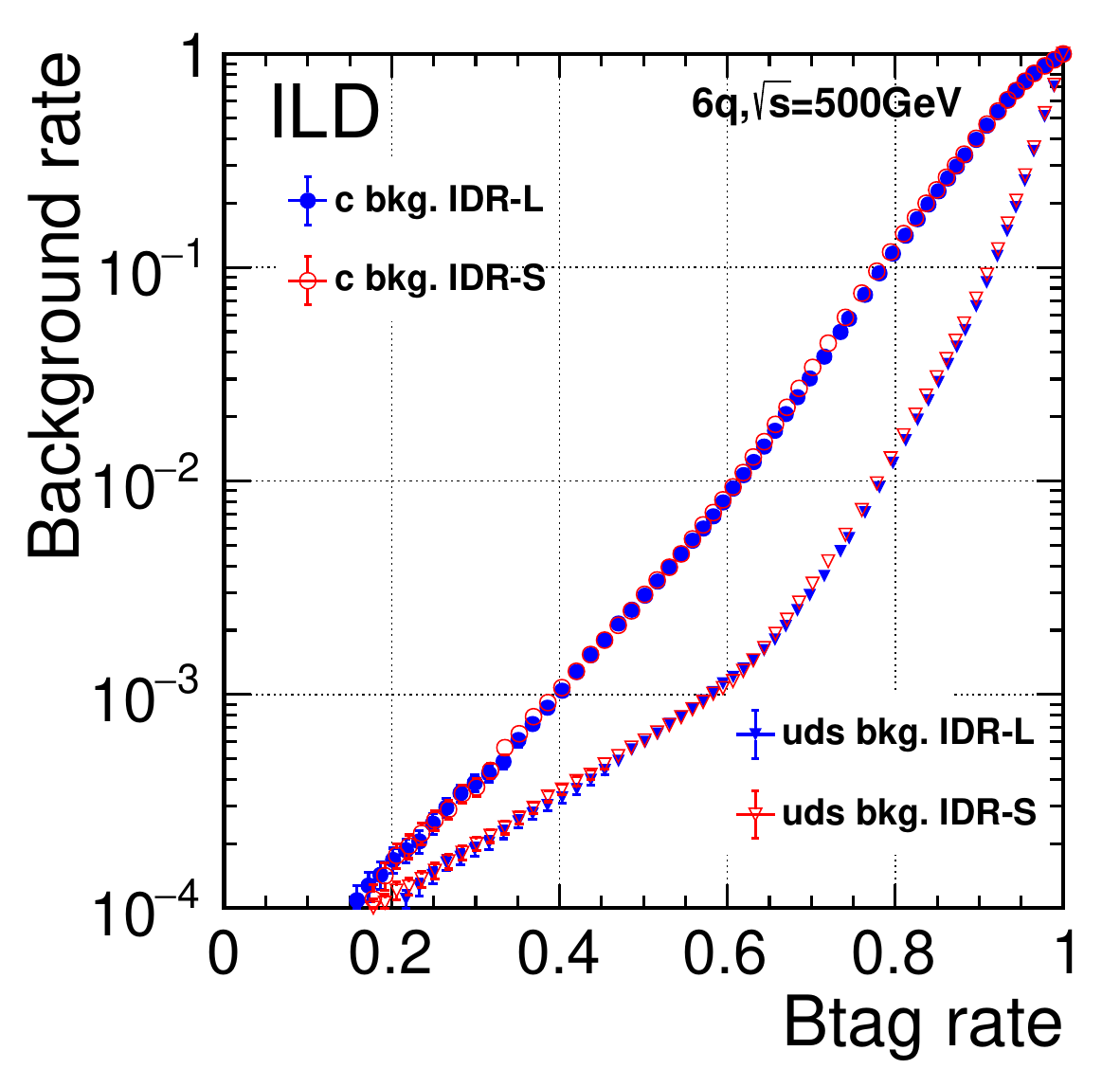}
        \caption{}
        \label{fig:flavtag:lcfiplus_b}    
    \end{subfigure}%
    \begin{subfigure}{.33\textwidth}
        \includegraphics[width=\textwidth,trim=15pt 10pt 25pt 60pt, clip]{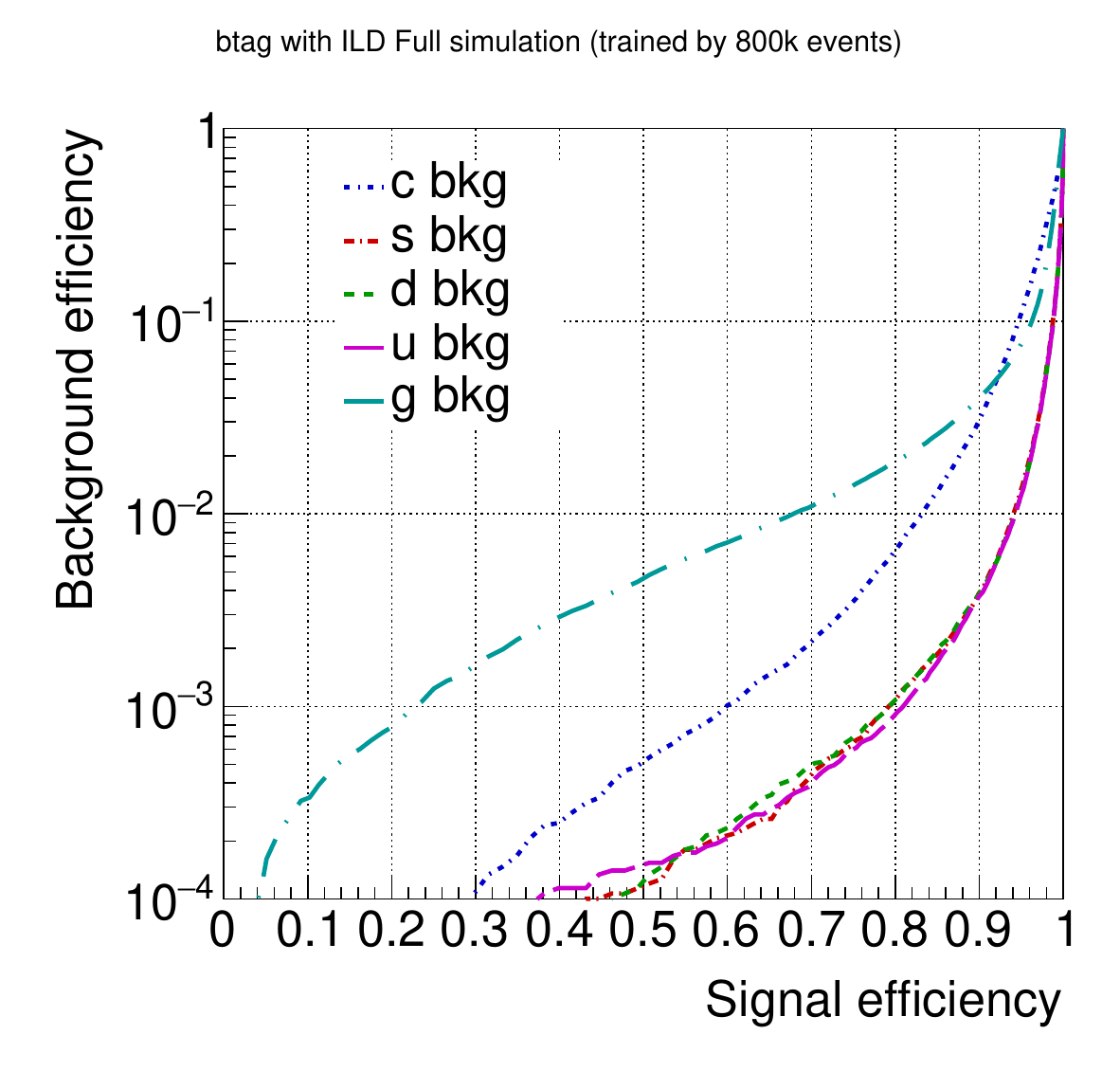}
        \caption{}
        \label{fig:flavtag:parT_11cat}    
    \end{subfigure}%
    \caption{Flavour-tagging performance in ILD in terms of mis-tag rate vs signal efficiency for (a) \PQc-jet  signal against $\PQb$- and light-jet background with the classic LCFIPlus flavour tag~\cite{Suehara:2015ura}, (b)  \PQb-jet signal against $\PQc$- and light-jet background with LCFIPlus and (c) \PQb-jet signal against $\PQc$-, $\PQs$, $\PQd$-, $\PQu$- and gluon-jet background of the classic LCFIPlus  as a function of the efficiency with the new, 11-class particle-transformer-based  ParT algorithm~\cite{Tagami:2024gtc}. All based on full simualtion of the ILD detector concept, (a) and (b) from~\cite{ILDConceptGroup:2020sfq}, (c) from~\cite{List:2024ukv}.}
    \label{fig:flavtag}
\end{figure}

The excellent vertexing capabilities and low material budget of linear collider detectors are key to the identification of jets originating from \PQb and \PQc quarks. The system also allows the accurate determination of the charge of displaced vertices, and it contributes strongly to the low-momentum tracking capabilities of the overall system, down to a few 10's of \SI{}{MeV}. Figure~\ref{fig:flavtag:lcfiplus_c} and~\ref{fig:flavtag:lcfiplus_b} show the performance of the jet flavour identification in ILD based on the  LCFIPlus algorithm~\cite{Suehara:2015ura}, which exploits classic boosted decision trees and is the basis of nearly all physics projections for linear colliders so far. The already excellent performance of LCFIPlus can be improved even further using deep-learning architectures. Corresponding tools are still under development and only start to be used in physics projections. Figure~\ref{fig:flavtag:parT_11cat} shows one of the first examples among the ongoing developments~\cite{List:2024ukv, Tagami:2024gtc}, which can be directly compared to Fig.~\ref{fig:flavtag:lcfiplus_b}, showing $\PQc$- and light-quark miss-tagging rates lower by an order of magnitude from algorithmic improvements only, based on the same assumptions on hardware performance and detector configuration.
 
\subsubsection{Tracking system}

The central tracking system in a linear collider detector is a key ingredient in the quest to optimise the detector towards particle flow. The system needs to deliver excellent momentum resolution, combined with outstanding efficiency, and overall low mass. Two different design routes are being followed: 

ILD has decided to approach the problem of charged particle tracking with a hybrid solution, combining a high resolution time-projection chamber (TPC) with a few layers of strategically placed strip or pixel detectors before and after the TPC. The inner layers could use CMOS detectors; the extended part is still open, including the possibility of using the first ECAL layer instead of extra detectors.

SiD and CLICdet feature all-silicon tracking systems. The baseline designs call for strip sensors, arranged in layers surrounding the vertex detectors, and a system of endcap disks to maximise the solid angle coverage. As for the vertex detector, a solution requiring no or only minimal liquid cooling is anticipated. 

The SiD concept has developed an integrated solution, where the readout ASIC is made part of the silicon strip detector, simplifying the design and minimizing the overall material budget. With the rapid technological developments in the area of silicon sensors and readout architectures, the option of a fully pixel-based central tracker is under investigation. This requires detailed studies of the various trade-offs such as hit resolution, power consumption, data volume and physics requirements. 
 
The time-projection chamber of ILD will fill a large volume about \SI{4.6}{m} in length, spanning radii from \SI{33}{} to \SI{180}{cm}. In this volume, the TPC provides up to \num{220} three-dimensional points for continuous tracking with a single-hit resolution of better than $\SI{100}{\mu m}$ in $r \phi$ and about \SI{1}{mm} in $z$. For momenta above \SI{100}{MeV}, and within the acceptance of the TPC, a tracking efficiency larger than \SI{99.9}{\%} has been found in events simulated realistically with full backgrounds. The complete TPC system will introduce about \SI{10}{\%} of a radiation length into the detector~\cite{Diener:2012mc}. 

Inside and outside of the TPC volume, a few layers of silicon detectors provide high-resolution points, at a point resolution of $\SI{10}{\mu m}$. Combined with the TPC track, this will result in an asymptotic momentum resolution of $\delta p_t / p_t^2 = 2 \times 10^{-5}$ ((GeV/c)$^{-1}$) for the complete system. 

\begin{figure}
    \centering
    \begin{subfigure}{.525\textwidth}
    \centering
        \includegraphics[width=0.95\textwidth]{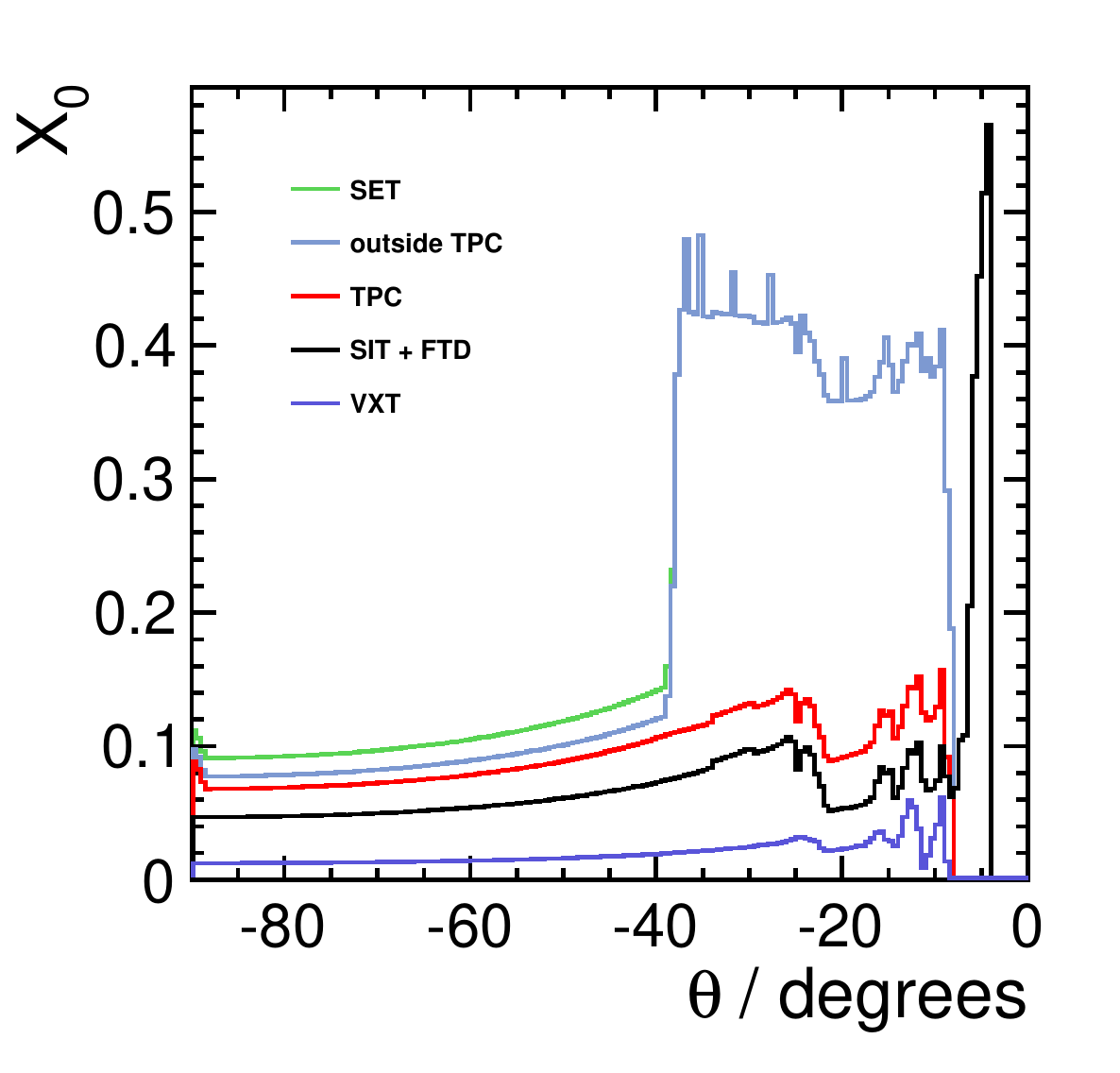}
        \caption{}
        \label{fig:tracking:ptres}    
    \end{subfigure}\hfill%
    \begin{subfigure}{.475\textwidth}
    \centering
        \includegraphics[width=0.95\textwidth]{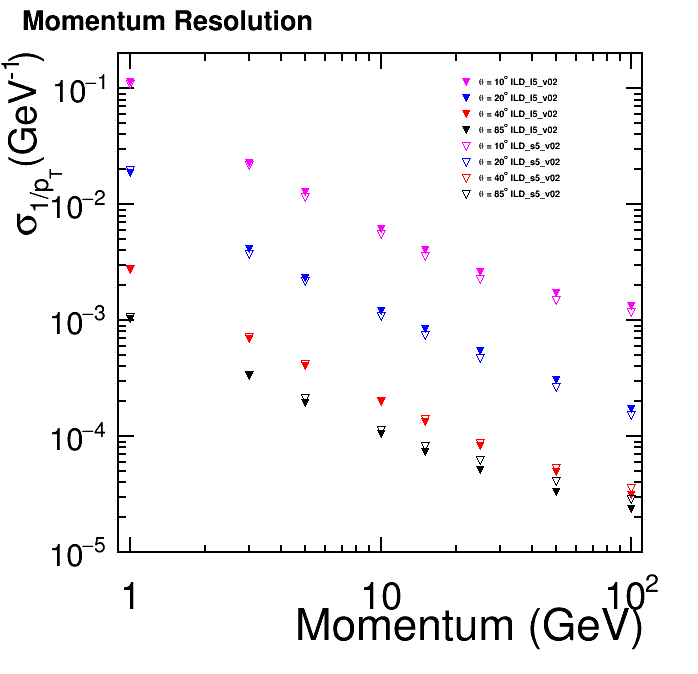}
        \caption{}
        \label{fig:tracking:material}    
    \end{subfigure}\hfill%
    \caption{(a) Cumulative material budget in ILD up to the calorimeter, in fraction of a radiation length (b) Simulated resolution in $1/p_t$ as a function of the momentum for single muons. The different curves correspond to different polar angles (both from \cite{ILDConceptGroup:2020sfq}).}
    \label{fig:tracking}
\end{figure}

Both options have demonstrated excellent momentum resolution over a large range of momenta. Differences between the options are mostly visible at lower momenta, where the lower material of the TPC tracker offers some advantages. The achievable resolution is illustrated in Fig.~\ref{fig:tracking:ptres}, where the $1/p_t$ resolution is shown as a function of the momentum of the charged particle. 

\begin{figure}
    \centering
    \includegraphics[width=.99\hsize]{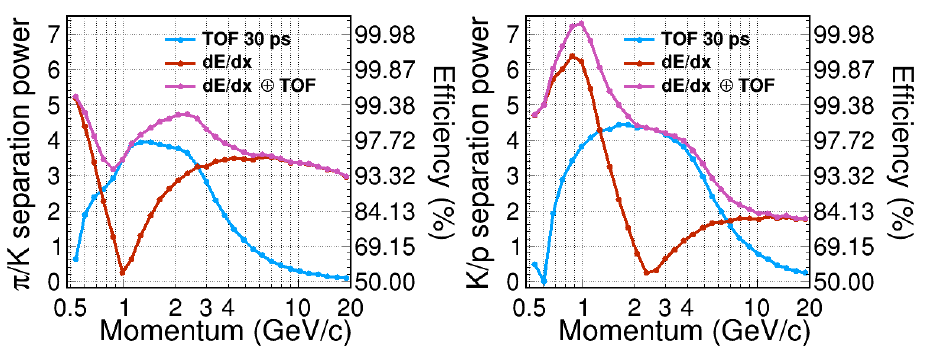}
    \caption{Simulated separation power between pions and kaons (left) and between kaons and protons (right), from dE/dx in the ILD TPC and from time-of-flight, assuming a \SI{30}{ps} timing resolution per particle, which can be achieved e.g.\ via \SI{100}{ps} hit-time resolution in the first 10 ECAL layers, or  via \SI{50}{ps} hit-time resolution in the two SET layers (from~\cite{Dudar:2024quz}).}
    \label{fig:PID}
\end{figure}

Particle ID is an important additional role the tracking system should be able to deliver for dedicated studies within the overall physics programme. The performance of time-of-flight- and $dE/dx$-based approaches are shown in Fig.~\ref{fig:PID}. Time-of-flight measurements alone (blue) provide kaon-pion separation at more than $\SI{3}{\sigma}$ for momenta between \SI{0.9}{GeV} and \SI{2.8}{GeV}, assuming a timing precision per particle at the outer radius of the tracker of \SI{30}{ps}, which could be extended to momenta between \SI{0.9}{GeV} and \SI{4.7}{GeV} with \SI{10}{ps} per particle. The TPC-based option can extend the $\SI{3}{\sigma}$ range for a kaon-pion separation down to lowest reconstructable momenta and up to \SI{18}{GeV}, based on the specific energy loss dE/dx~\cite{Dudar:2024quz}.

In order to compensate for the absence of dE/dx-based PID in an all-silicon tracking system, SiD is investigating the option of inserting a gaseous RICH detector for PID in front of the ECAL~\cite{Basso:2023zuq}. The full impact on the particle flow performance still needs to be understood.

\subsubsection{Calorimeter and muon system}
A very powerful calorimeter system is essential to the performance of a detector designed for particle flow reconstruction. Particle flow stresses the ability to separate the individual particles --- both charged and neutral --- in a jet. This puts the imaging capabilities of the system at a premium, and pushes the calorimeter development in the direction of systems with very high granularity in all parts, both transverse to and along the shower development direction. A highly granular sampling calorimeter is the chosen answer to this challenge~\cite{Sefkow:2015hna}. The detectors comprise around \SI{2500}{m^2} of active volume, subdivided into up to 100~million readout cells. 

Baselines for the electromagnetic calorimeter of all three linear collider detector concepts are silicon-tungsten (SiW) sampling calorimeters. ILD and CLICdet consider CALICE designs based on diodes with pads of about  \num{5}\,$\times$\,\SI{5}{mm^2} to sample a shower up to \num{40} times in the electromagnetic section. 

\begin{figure}[htbp]
  \centering
  \begin{subfigure}{.525\textwidth}
  \centering
      \includegraphics[width=0.95\textwidth]{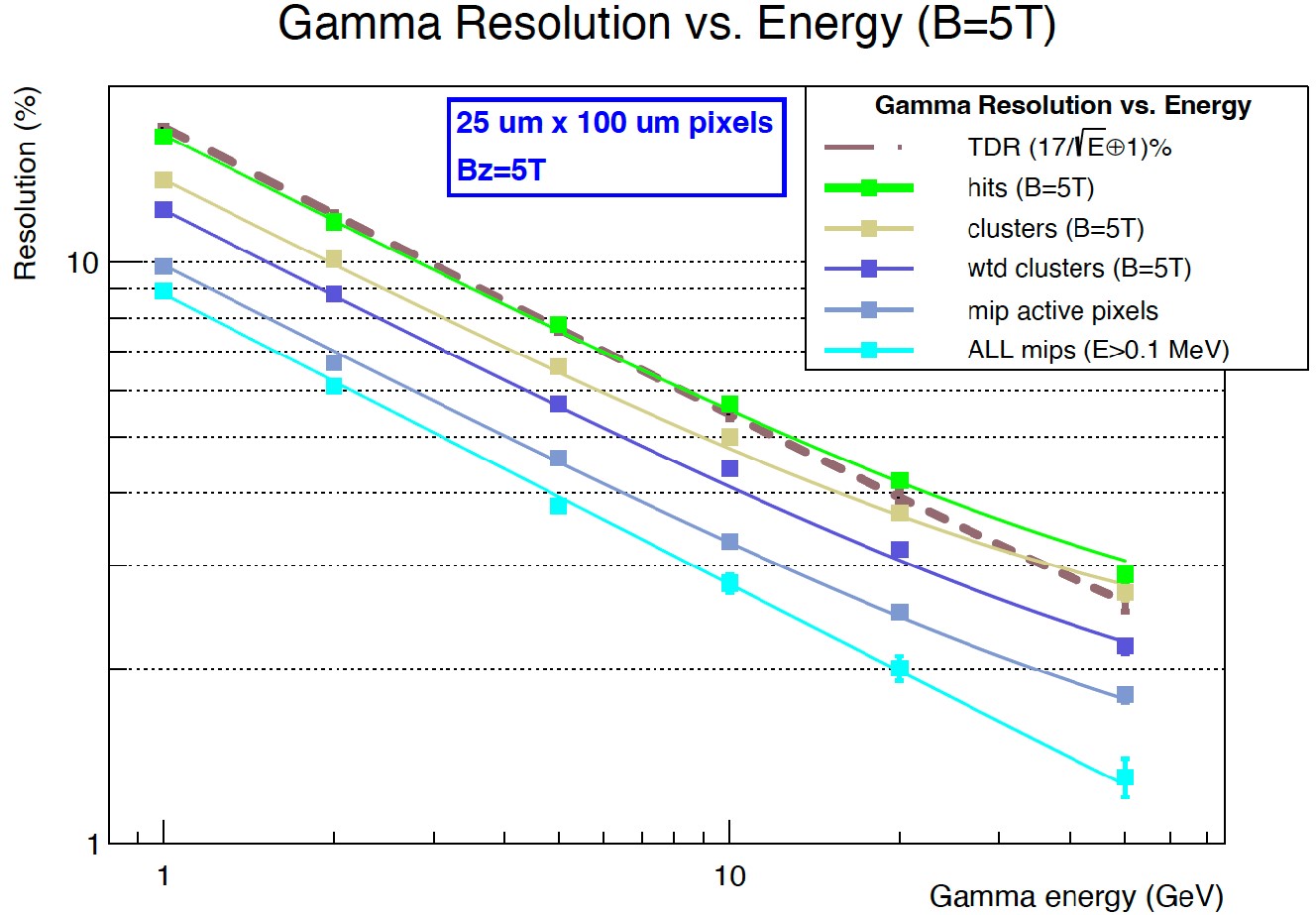}
	  \caption{}
      \label{fig:MAPS-ECal}
  \end{subfigure}\hfill%
  \begin{subfigure}{.475\textwidth}
  \centering
       \includegraphics[width=0.95\textwidth]{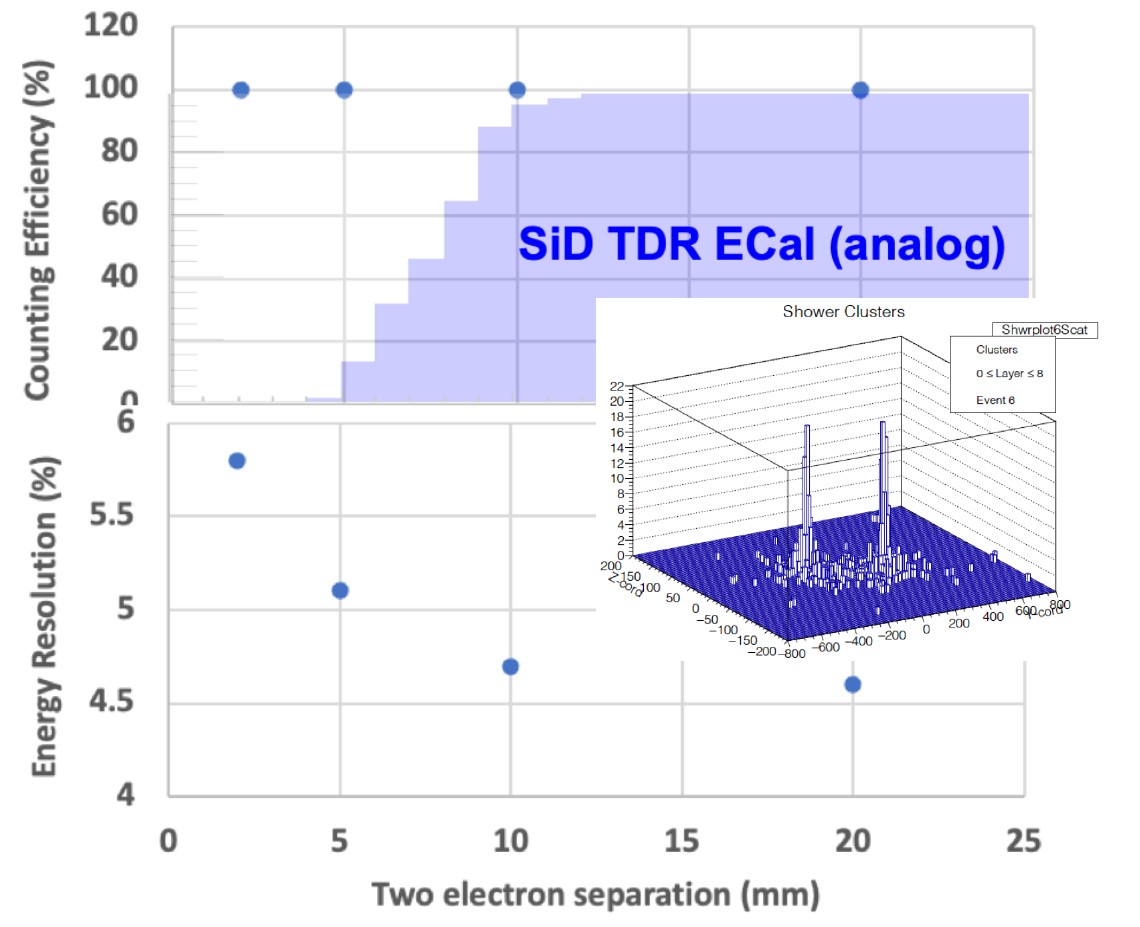}
	   \caption{} 
       \label{fig:Two_electrons}
  \end{subfigure}\hfill%
  \caption{(a) Photon energy resolution versus energy~\cite{Brau:2024jrq}, and (b) efficiency for distinguishing two nearby \SI{10}{GeV} electron showers and the degradation of energy resolution versus separation~\cite{Brau:ILCX2021}.}
\end{figure}

The SiD ECAL design has been upgraded to a digital SiW sampling calorimeter employing MAPS sensors. Detailed studies of a design with a pixel area of $\SI{2500}{\mu m^2}$ 
($\SI{50}{\mu m} \times \SI{50}{\mu m}$ or $\SI{25}{\mu m} \times \SI{100}{\mu m}$) have demonstrated excellent performance. Results of the energy resolution from these studies~\cite{Brau:2024jrq}, based on this fine, digital configuration,
are shown in Figure~\ref{fig:MAPS-ECal}.  

The performance for two showers versus their separation is summarised in Figure~\ref{fig:Two_electrons}. With the fine granularity, showers can be identified at the millimeter scale of separation, and the energy resolution of each of the showers does not degrading much at this level of shower separation. 

As cost effective alternative, ILD is considering scintillator strips as active material. Extensive prototyping work and test beam experiments have demonstrated the feasibility of all these approaches.

A self-sustaining structure holding tungsten plates in a carbon fiber reinforced polymer (CFRP) supports the detector elements while minimizing non-instrumented spaces. A very similar system has been adopted by the CMS experiment for the upgrade of the endcap calorimeter. This system will deliver invaluable information on the scalability and engineering details of the technology.

In recent years, the capabilities of the system to also provide precise timing information on the shower has been studied. The most obvious application of this is the instrumentation of the first layers of the ECAL with very good timing capability, to provide time-of-flight measurements on charged particles, hitting the face of the calorimeter.

As an alternative to the silicon-based system, sensitive layers made from thin scintillator strips are also investigated. Orienting the strips perpendicular to each other has the potential to realise an effective cell size of \num{5}\,$\times$\,\SI{5}{mm^2}, with the number of read-out channels reduced by an order of magnitude compared to the all silicon case. Also here, extensive prototyping work has been performed, though not as complete as for the silicon ECAL option. 

For the hadronic part of the calorimeter, several technologies are studied, based on either silicon photo diode (SiPM) on scintillator tile technology~\cite{CALICE:2022uwn} or resistive plate chambers~\cite{Baulieu:2015pfa}. The SiPM-on-tile option has a  moderate granularity, with \num{3}\,$\times$\,\SI{3}{cm^2} tiles, and provides an analogue readout of the signal in each tile (AHCAL). The RPC technology has a better granularity, of \num{1}\,$\times$\,\SI{1}{cm^2}, but provides only 2-bit amplitude information (SDHCAL). For both technologies, significant prototypes have been built and operated. As for the ECAL, the SiPM-on-tile technology has been selected as baseline for part of the upgrade of the CMS hadronic endcap calorimeter; it will thus see a major application in the near future.

SiD is considering extensions and optimisations of the hadron calorimeter
design to improve performance. Ideas include the inclusion of timing layers to assist the particle flow algorithm in separating the delayed shower components from slow neutrons from the prompt components, inclusion of precision tracking layers to assist the PFA. and exploration of the benefits of using on-board intelligence. This last item might include features ranging from simple zero suppression to interlayer
communication to assist tracking through the HCAL and or PFA jet reconstruction.

The calorimeters comprise around \SI{2500}{m^2} of active volume, subdivided into about 100\,million readout cells (or even three orders of magnitude more in case of a MAPS design). Therefore, a key challenge for any of these technologies is the demonstration that the chosen technology scales to the large areas and large channel counts needed for a detector at a linear collider facility. Here great progress has been made through coordinated test beam campaigns and engineering efforts, to develop and demonstrate cost scaling to large systems. 

For the SiW ECAL of ILD, an engineering study has been carried out on the power consumption under pulsed operation of the accelerator~\cite{rpoeschl:2019}. In idle time, power needed by the front-end electronics is charged by small, $\mathcal{O}$(\num{100}-\SI{200}{mA}) in-rush currents. The power is stored locally in capacitors and released during the bunch trains. This allows avoiding large peak currents during the actual data taking that would otherwise be as large as \num{10}-\SI{15}{A} for a complete SiW ECAL layer with consequences for the mechanical stability when operated in a magnetic field of several Tesla. The chosen concept for the operation of the front-end electronics allows for limiting the power consumption of the entire SiW ECAL of ILD to around \SI{20}{kW} for an assumed duty cycle of \SI{1}{\%}. This number may be compared to the power consumption of the CMS HGCAL  which is around \SI{110}{kW} per endcap. The SiW ECAL of ILD features around 80\,million cells while the CMS HGCAL features around 6\,million cells per endcap\footnote{We are aware that this comparison is rather simplistic, still it gives an impression about the savings.}. 
The low power consumption favours a hermetic design of the calorimeter system of ILD. The gap between the ECAL and HCAL barrel and the calorimeter endcaps can be as small as \SI{62.5}{mm}~\cite{hvideau:2019}.  

\begin{figure}[th]
    \centering

    \begin{subfigure}{0.45\textwidth}
        \includegraphics[width=\linewidth]{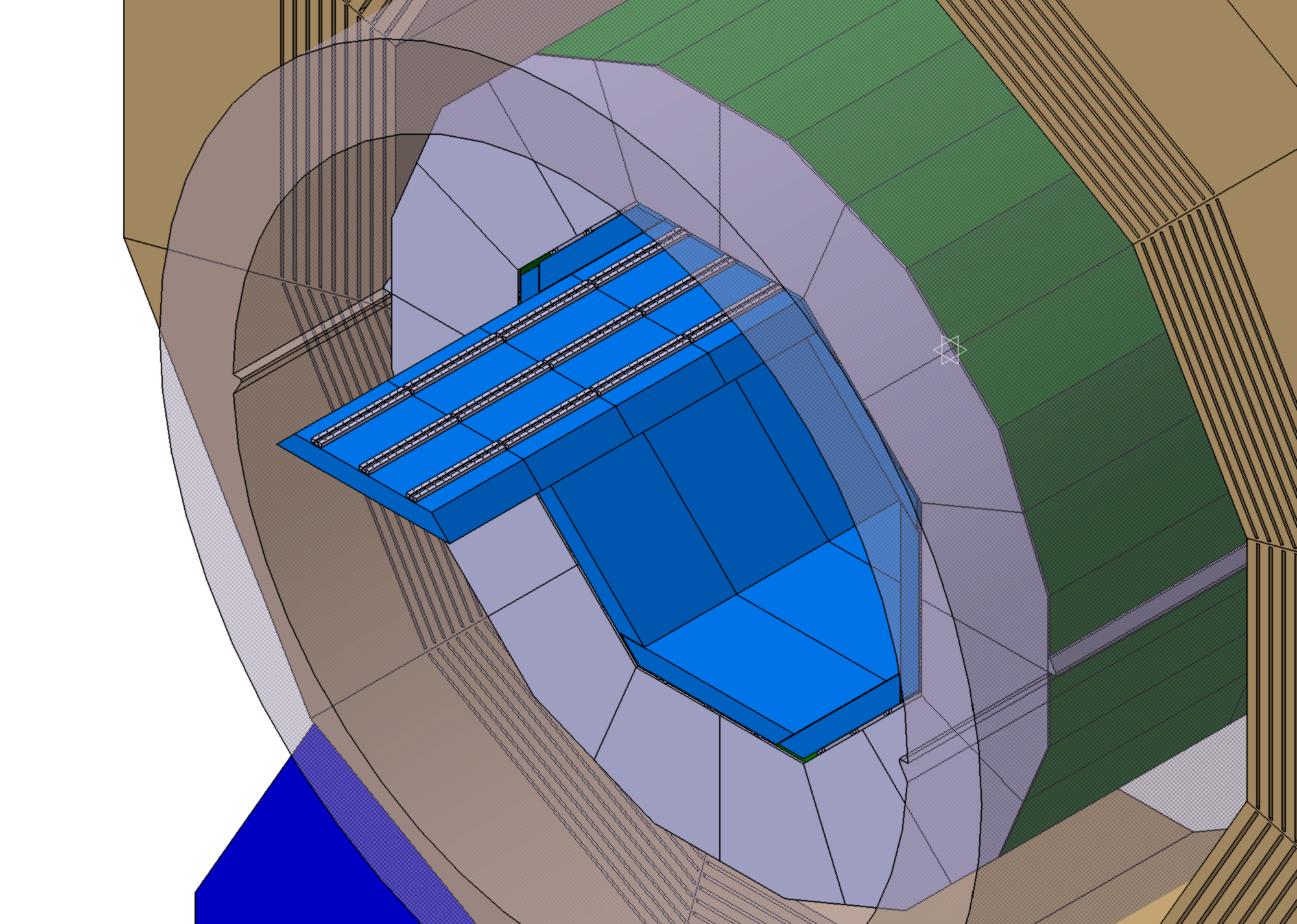} 
    \caption{}
    \label{ild-fig-CALO:a}
    \end{subfigure}
\hspace{1.cm}
    \begin{subfigure}{0.24\textwidth}    
    \includegraphics[width=\linewidth]{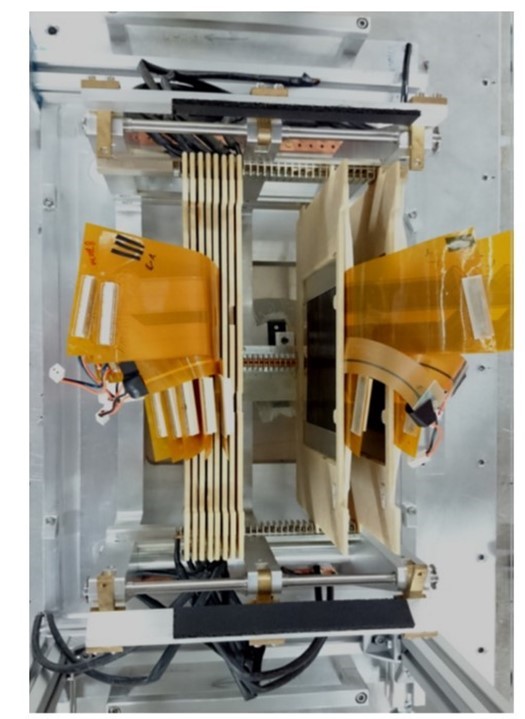}
    \caption{}
    \label{ild-fig-CALO:b}
    \end{subfigure}
    
    \caption{(a) Three-dimensional rendering of ILD's barrel calorimeter, solenoid and yoke, with one ECAL module partially extracted  (from~\cite{Behnke:2013lya}). (b) Prototype module of the LumiCal calorimeter (from~\cite{ILDConceptGroup:2020sfq}).}
    \label{ild-fig-CALO}
\end{figure}

A rendering of ILD's barrel calorimeter, solenoid and yoke is shown in Fig.~\ref{ild-fig-CALO:a}. 

The iron return yoke of the detector, located outside of the coil, is instrumented to act as a tail catcher and as a muon identification system. The size and thickness of the iron return yoke scales with the ultimate energy for which the detector is designed. Several technologies are under study for the instrumented layers. The main challenge here is, however, primarily one of finding a cost efficient, robust technology, and not so much one of pushing technological limits. 

The current ILD and SiD designs in addition were optimised for the so-called push-pull scheme, in which both detectors are housed in one common experimental hall, but share one interaction point. Servicing one detector while the other takes data requires additional shielding of the detector, most efficiently realised through the combination of a thick yoke and additional inactive shielding walls. In case, as anticipated for the LCF, two separate interaction regions are available, this additional complexity of the detector is not needed. 

\subsubsection{The interaction region}
In a linear collider, highly collimated electron and positron beams are brought into collision. This creates a volume of extremely high energy density, which is the source of a very intense background radiation, the so-called beamstrahlung. This effect is more pronounced at the currently considered linear collider options than it is at an FCC-ee, since the focus of the linear colliders is more intense. 
The interaction region thus needs to be optimised to be able to extract as much as possible the background from the detector, without interacting with active or passive detector elements. 

Compared to circular colliders, the final focus length L$^*$ can be larger at linear colliders (i.e. ${\mathrm L^*}\approx\SI{4.1}{m}$ in case of ILC compared with ${\mathrm L^*}\approx\SI{2}{m}$ in case of FCC-ee). This allows for more space for instrumentation and background mitigation. In particular, no need exists in the linear collider case to protect the inner region with a massive mask reaching inside the detector acceptance. 

The region around the beam-pipe is instrumented with a series of smaller calorimeters, optimised to measure the luminosity, and to close the solid angle coverage of the electromagnetic calorimeter. Compared to the main calorimetric detectors, technologies need to be employed in this region which are significantly more radiation hard, to survive the intense electromagnetic background from the beam-beam interaction. These areas thus require rather special instrumentation.

The area surrounding the beam pipe is equipped with a set of compact calorimeters, including LumiCal and BeamCal, which are based on sampling calorimeter technologies. These detectors are designed to measure luminosity and extend the solid-angle coverage of the electromagnetic calorimeter. While both calorimeters utilise tungsten planes as absorbers, LumiCal employs silicon sensors as the active material, whereas BeamCal uses gallium arsenide sensors to withstand the intense electromagnetic background generated by beam-beam interactions. Figure~\ref{ild-fig-CALO:b} shows a 10-layer LumiCal prototype used during a test beam campaign.

\subsubsection{Experimental magnet}
The three linear collider detector concepts all utilise a solenoid to provide strong magnetic fields between \num{3.5} and \SI{5}{T} in the central region. This field is needed to measure charged tracks with high precision, and to focus and then remove charged background created in the beam beam interaction from the detector. 

Technologically these magnets are based on the CMS experiment at CERN. A special challenge will be to build these magnets as thin as possible, to introduce minimal inactive material into the detector system. Cable-in-Conduit Conductors (CICC) is a specialised technique that could be well-suited for the construction of these superconducting magnets. In this approach, a large number of superconducting wires are bundled together into a cable and enclosed within a metallic conduit or jacket. The conduit serves multiple purposes, such as structural support, thermal stabilisation, and protection. Cooling channels are present within the conduit to allow cryogenic fluids to flow directly through the cable bundle to maintain low temperatures. The CICC designs allow the conductor to carry extremely high currents, and the cable bundle can accommodate many superconducting strands, providing a large cross-sectional area for current flow keeping the magnet thin. Cooling is very efficient, with liquid helium flowing directly through the strands. The design also can offer better quench protection through the inclusion of copper strands alongside the superconducting wires (see~\cite{cicc}). 

For all three concepts the current baseline calls for these magnets to be positioned outside of the hadronic calorimeter, to minimise the interference with the calorimeter measurement. Recent technological developments promise a significant reduction in thickness of the coils. Whether or not this then would allow to move the coil to smaller radii, at significant savings, is an area of active investigation. 

\subsection{Achievements}

Ideas for the construction of a linear collider and its associated detectors have been around for a very long time. The first workshop for a Japan Linear Collider (JLC) took place in 1989 and a zeroth-order design report for a next linear collider (NLC) at SLAC was published in 1996. 
A conceptual design report of a superconducting linear accelerator (TESLA) was published in 1997~\cite{TESLA-CDR}. 
Concurrent with the development of the accelerator design, new detector technologies were proposed for the experiments at these new facilities. 
In November 2003 an International Technology Recommendation Panel was formed to decide on the accelerator technology for a future linear collider. 
This was followed in 2005 by the formation of the Global Design Effort by an international team of physicists to develop the Technical Design for a linear collider based on SCRF with its associated detectors. 
A global effort emerged for the development of new technologies for a linear collider detector, building on the strong foundation laid by the NLC, the JLC and the TESLA efforts. 

In 2003 and following the proposal of the TESLA accelerator a number of R\&D collaborations formed, with the goal to advance the state of detector technologies to be used in any linear collider. The collaborations were organised under the auspices of ECFA, but were drawing on a fully international membership and participation. 

The CALICE (Calorimeter for Linear Collider Experiment) collaboration has pursued a broad spectrum of novel calorimeter technologies to address the critical issue of jet energy resolution to discriminate between $\PW$ and $\PZ$ bosons in the jet final state at the ILC. 
A focus of the collaboration has been the study of particle flow, where the combined tracker and calorimeter is used to identify charged and electromagnetic particles in a hadronic shower and the calorimeter is used to match energy deposits with each track and the remaining energy is associated with neutrals. To achieve the best energy resolution, a very fine grained readout was required. 
The collaboration has pioneered the silicon-tungsten and SiPM-on-tile calorimeter and demonstrated that is was a viable technology. The R\&D programme comprised the test of power pulsing~\cite{Poschl:2015yqj} and included hardware that could, in terms of performance but also due to its compact design, more or less readily be used for the ILD detector~\cite{Poschl:2022lra}. Detector units that implement the powering scheme of the SiW ECAL described above are being examined in beam test campaigns in 2025 and 2026. 
The concepts of the silicon-tungsten and SiPM-on-tile calorimeter have been adopted for the HL-LHC upgrade of the CMS experiment where the technology is being implemented in the High-Grained Endcap Calorimeter. The collaboration between CALICE and CMS has lead to a common testbeam campaign that is described in~\cite{CMS:2022jvd}. 
The technology is currently also being implemented for the forward hadron calorimeter of the ePIC detector for the EIC. The SiW calorimeter design has also been adopted by the ALICE experiment for the forward calorimeter (FoCal), that will enable a new and unique programme at the LHC focused on small-x gluon distributions of hadrons and nuclei. Similar designs are being considered to instrument the forward region for the ePIC detector. 

A demonstration of a full digital hadron calorimeter was also demonstrated by the CALICE collaboration. A \SI{1}{m^3} digital calorimeter was built, with one million readout channels using iron absorber plates interspersed with RPC readout chambers. The technique was been refined by developing the semi-digital hadronic calorimetry, where the deposits from a hadronic shower are recorded with high granularity with only a coarse energy information (2-3 thresholds). Different readout technologies were explored besides RPCs including Micromegas chambers.

The linear collider community has been deeply engaged in the advancement of technologies for tracking detectors with major contributions in three distinct areas. The linear collider TPC (LCTPC) collaboration has explored novel technologies to improve the performance of a TPC. A large prototype TPC was built and tested deploying four different readout technologies. A triple GEM design minimised the inactive area and targeted an improvement in the operational stability of a triple GEM stack. An alternative design was based on a double GEM where the dead area pointing towards the interaction point was minimised. Using thicker GEM stacks that favour higher gas gains, only two GEM chambers were needed. To cover larger areas with fewer and thus larger pads without degrading the performance a new readout concept was developed using a resistive layer on the pads, spreading the narrow charge distribution of a Micromegas gas amplification stage over several pads thus enabling a more precise charge interpolation than otherwise possible. An inverted HV scheme was developed in which the amplification mesh was on ground potential and the resistive layer on a positive high voltage. This configuration showed a reduction of field distortions between the modules by one order of magnitude. A fourth technology studied used the GridPix technology. A GridPix uses a highly pixelised readout ASIC. For the linear collider studies the Timepix ASIC was used. The ASIC was covered with a resistive layer to protect it from discharges and a Micromegas was mounted on top using photolithographic post-processing techniques. It was shown that this concept offers the best possible resolution only limited by the diffusion in the drift region. Many of these studies have informed the construction of new TPCs and the development of MPGDs. In particular it is noted that the design of the new TPC for the ND280 near detector of the T2K experiment was informed by studies within the framework of a linear collider. 

The readout of MPGDs was based on the AFTER and ALTRO ASICs. The Alice TPC Readout (ALTRO) chip was an existing chip that was later adapted to the SAMPA chip to read out the ALICE GEM-based upgraded tracker. The AFTER chip, that was heavily exercised during the linear collider studies is currently used in the readout of the ND280 near detector of the T2K experiment.  

Given the stringent requirements on the material budget the PLUME (Pixelated Ladder with Ultralow Material Embedding) collaboration advanced the development of low-mass support structures for vertex detectors. In parallel, there was a very strong effort on the development of CMOS sensors for vertex detectors that has resulted in opening a new paradigm for vertex detectors with the introduction of monolithic active pixel sensors (MAPS). The Mimosa series of pixel sensors were candidate sensors for a vertex detector. The telescope for the beam test facility at DESY was equipped with Mimosa sensors under the EUDET programme. The linear collider detector R\&D programme provided a fertile ground for the development of this new technology that is now being implemented at scale in the ALICE experiment. The Chronopixel design was developed to provide for single bunch-crossing time stamping in a CMOS process. These efforts targeted pixel sizes of \num{20}\,$\times$\,$\SI{20}{\mu m^2}$ or smaller. 

One of the first silicon-based pixel detectors that reached very low mass budget by thinning the sensor itself was the DEPFET sensor. This technology provides very high signal-to-noise hits while only requiring $\SI{50}{\mu m}$ of silicon per layer in the tracking volume. The device can store signals over the readout cycle. Through very active development as a possible ILC vertex detector, the technology became mature enough to form the basis of the vertex detector for the upgrade of the Belle experiment. 

The community has also been at the forefront in developing alternative bonding technologies. With the advent of the through-silicon-via (TSV) technology an intense effort on developing a direct-bond interface techniques was initiated. With this approach, different functionalities of the readout can be separated in different wafers, that then can be thinned resulting in very thin readout addressing the mass budget of vertex detectors. Initial devices have been produced, but given the cost, the early stages of the development of this technology and the promise of MAPS detectors, it never took hold for large-scale implementation. Capacitive coupled pixel detectors in which a sensor is capacitively coupled to the readout through, for example, a glue layer that eliminated bump bonding is another example of an approach pioneered by the linear collider community. 

The challenges to instrument the forward detector region are significant in an \ee environment. Novel ideas have been proposed within this community to advance forward calorimetry with fast readout in high occupancy and high radiation  environments to measure the luminosity. Three-tiered detector systems have been proposed and tested. A LumiCal, covering the region from about \num{40}-\SI{140}{mrad} from the beam axis, would provide a measurement of the luminosity with a precision of 10$^{-3}$ using Bhabha events, c.f.\ Sec.~\ref{sec:LEPmeas}. A BeamCal (\num{5}-\SI{40}{mrad}) would be used for beam diagnostics and a GamCal even closer to the beam axis would be used for beam diagnostics using beamstrahlung photons. Sensors studied are PCVD diamond, GaAs and SiC
sensors. 

The concept of particle flow calorimetry has been a central notion for linear collider detectors. This type of calorimetry fully integrates tracking with calorimetry and requires very high granularity in the calorimeter. If tracking provides the best possible energy or momentum resolution, that measurement is taken and the hits associated with the track in the calorimeter are removed. This has led to designing detectors not as a sum of individual subsystems, but the detector is viewed as a well choreographed integrated system where there is a mutual dependence of the overall performance of the integrated performance of each sub-detector. Optimisation of the overall geometry and technologies was emphasised from the inception of a detector concept and is an idea that is now being adopted, where applicable, by other new detector designs. 

\subsection{New ideas} 
Particle physics detectors are the eyes of the scientists to reveal the underlying laws of nature. The precision with which processes can be measured is limited by the precision with which the measurements can be performed. The experimentalist always needs to be prepared for the unexpected. History has many examples of detector systems discovering new phenomena for which the experiment was not originally designed. Key to these breakthroughs is building detectors that provide measurements with the highest precision. The pace of technology development is astounding and holds great promise for the next generation of detectors. The timescale from invention to application is sometimes remarkably short. A striking example is the origination of the time projection chamber (TPC). The idea arose in 1974 with a very small prototype~\cite{NYGREN201822} and it took a mere two years before it was adopted as the tracking technology of choice for the PEP-4 experiment~\cite{Marx:1976js}. In 1983 the ALEPH experiment at LEP chose the TPC as its main tracking detector. 

One of the workhorses of current detectors is the silicon photomultiplier (SiPM), sometimes also called the multi-pixel photon counter (MPPC). A precursor to the current SiPM was the visible light photon counter (VLPC) developed by the company Rockwell. The devices had to be operated at cryogenic temperatures and their implementation in detectors was far from trivial. The development was significantly advanced in Russia in the late nineties and early twenties with first excellent devices produced by the Moscow Engineering and Physics Institute~\cite{Dolgoshein}. 

Given the challenges of the experimental environment of the LHC detectors, where events have many interactions per bunch crossing, the impact of high timing resolution was quickly realised and addressed with the development of the low-gain avalanche diode (LGAD) technology. The idea for the implementation of a low-gain amplification region in a silicon detector was first proposed in 2012 and the technology has been advanced to the level where it can be implemented at scale at an impressive pace~\cite{Sadrozinski:2014dkk}. Both the ATLAS and CMS experiments now have large-area timing layers as integral part of their upgrades to help distinguish the multiple interactions in an event. Within a timespan of seven years or so this technology has gone from a mere idea to an essential element of new detectors. 

Another astounding technological revolution has been the development of the monolithic active pixel sensor (MAPS) CMOS technology. Although the hybrid pixel technology was recognised as a most powerful technology for vertex detectors, where the separation of readout and sensor allowed the devices to be made radiation hard for the very harsh environments at the LHC, integration of the two was considered the next logical step with potential access to commercial foundries for the production of the sensors. The MIMOSA (Minimum Ionizing MOS Active pixel sensor) series of sensors was the first venture into this area of sensor development that started around 1999. Using ever more advanced technologies, moving from AMS $\SI{0.6}{\mu m}$ to TSMC \SI{180}{nm}, the technology was matured with the first deployment of a MAPS vertex detector for the STAR experiment at RHIC in 2014~\cite{CONTIN20167}. The technology has advanced unabatedly and a major milestone was reached with the construction of the large MAPS-based vertex detector for the ALICE experiment, ITS2, that has 12.5\,billion pixels of \num{27}\,$\times$\,$\SI{29}{\mu m}$, a spatial resolution of $\SI{5}{\mu m}$ and a material budget of \SI{0.36}{\%} X$_0$~\cite{RAVASENGA2024169311}. It has one of the lowest material budgets of any vertex detector and intense efforts aim at reducing that even further. It is the most promising candidate technology for future collider vertex detectors. 

With sensors being thinned to be flexible, embedding sensors in thin flexible circuits will become feasible. This opens up the possibility to study geometries that have been out of reach to date. Current geometries have by necessity, due to the constraints imposed by the sensors, relatively large uninstrumented areas. With flexible sensors, continuous cylindrical or even spherical geometries could be considered, which could allow for significant advantages. An example of a study in this direction is the idea to embed a Timepix4 chip with through-silicon vias laminated onto a polyimide microRWELL flexible PCB, with the connection to the readout system on one side and the amplification stage on the other side, leading to a detector with full active coverage. 

The emergence of quantum sensors could also enable the development of new detectors for particle physics to reach unimaginable precision in the measurement of certain quantities. A long-held goal of the community, for example, has been to construct calorimeters with superb energy resolution for hadronic jets. Quantum dots could be a first step in this direction. One-dimensional materials offer the capability to tune the fluorescence spectrum. Both crystals and scintillators can be doped with the appropriate quantum dots to enhance the properties of the light emission and improve the measurement of the energy resolution. Using different doping schemes for separate longitudinal sections in the calorimeter, chromatic calorimetry can developed and the calorimeter response tuned to different particles. A further example for improving vertex detectors is DoTPiX technology~\cite{Hallais:2023gmq}, which can be described as a technology for vertexing to in a semi-classical way, using quantum wells.

As these few examples show, new technologies can be developed on a relatively short time scale that truly have a transformative impact on the next generation of experiments. The timescale for the development of new techniques is not many decades as the previous examples have shown, but can happen within a decade. The opportunity exists now to introduce a complete new paradigm for the next-generation experiments that will have capabilities that transcend the current state of the art significantly. This is a very exciting prospect. Given that the experiments will need to operate for many decades, at different centre of mass energies where the physics topologies will be different, it provides an opportunity to develop an approach to detector design, potentially using advanced AI and ML techniques, that integrates not only all detector technologies but also the full lifetime of the experiments with its full physics programme.

\section{Summary and conclusion}
\label{sec:conclusions}
In this report, we have surveyed the physics opportunities offered by linear $\ee$ colliders.   The most urgent problem in particle physics is to understand the nature of the Higgs boson, its structure and couplings to quarks, leptons, and gauge bosons, and the physics responsible for their mass generation.  As the energy of an $\ee$ collider is raised from the Higgs boson threshold, a series of different reactions allow us to measure the Higgs boson's properties in different ways.  This is the way to obtain the most complete and most precise data to address the Higgs' mysteries. Precision electroweak measurements, measurements of the weak interaction bosons, and measurements of the top quark contribute to this story.  Only a linear $\ee$ collider capable of observing the full evolution up to an energy of \SI{1}{TeV} can see the full picture.

We have reviewed the technologies available for construction of a linear $\ee$ collider.  For experiments at the peak of the cross section for $\ee\to \PZ\PH$ at \SI{250}{GeV}, there is a well-understood technology based on superconducting RF cavities that can provide a first-stage collider readily and with little risk. This can solve the problem of assuring a continuous programme of  collider experiments beyond the lifetime of the LHC.  But also, new technologies with higher accelerating gradients and luminosities contribute new possibilities for the future of a linear collider program.   These will make it possible to reach higher energies in an economical way.  Ideas such as the $\PGg\PGg$ collider will add new observables.  A linear collider can also provide a testbed for ideas such as plasma wakefield acceleration that someday will deliver colliders with parton collisions at \SI{10}{TeV} and above.

The low-background environment of an $\ee$ colliders allows the development of new particle detector technologies that could not survive in the intense setting of high-energy proton colliders.   These new technologies are needed to supply measurements of the Higgs boson at the highest possible precision.  Some ideas for highly capable trackers and calorimeters are already included in the current  designs for linear collider detectors.  But we are looking forward to new and powerful ideas to emerge as young scientists educated at the LHC turn their creativity to this new challenge.

In all of these aspects --- compelling physics, novel accelerators, high-precision detectors --- construction of a linear collider facility will lead to a fresh and exciting era of experimental particle physics.

As for applications from our field to other areas of science, the future  is linear.  Spallation neutron sources and free-electron lasers are based on  technologies invented to build linear colliders, and now these have become the workhorses of materials science, structural biology, and ultrafast chemistry. Our highly capable detectors and our experience with massive data sets are already needed, and are already being applied,  to make sense of the results of new experiments that these facilities host.
As we advance along this line, our colleagues will advance along with us.

Linear $\ee$ colliders promise a bright future for physics, for technology, and for the advance of science along many fronts.  Members of our field, and our colleagues in physical science, need this future as soon as possible. The next step to that era is the construction of a linear collider Higgs factory.

\section*{Acknowledgements}

This work was supported by EAJADE, a Marie Sklodowska-Curie Research and Innovation Staff Exchange (SE) action, funded by the EU under Horizon-Europe Grant agreement ID: 101086276; by AIDAinnova, a project within the European Union's Horizon 2020 Research and Innovation programme under GA no. 101004761; 
by the CNRS/IN2P3, France;
by the Deutsche Forschungsgemeinschaft (DFG, German Research Foundation) under grant 491245950 and under Germany's Excellence Strategy --- EXC 2121 ``Quantum Universe'' --- 390833306, 
and the DFG Emmy Noether Grant no.\ BR 6995/1-1; 
by a Department of Science and Technology and Anusandhan National Research Foundation Government of India Startup Research Grant, grant agreement no. SRG/2022/000363 and Core Research Grant, grant agreement no. CRG/2022/004120; 
by the National Science Centre (Poland) under the OPUS research project no. 2021/43/B/ ST2/01778  
and 
MAESTRO grant no. 2023/50/A/ST2/00224; 
by the Spanish Ministry of Science under grant agreements PID2021-122134NB-C21 and PID2021-122134NB-C22, by the Generalitat Valenciana under CIPROM/2021/073 and ASFAE2022/013 and 015, by the Severo Ochao excellence program,  
and by the Atracci\'on de Talento Grant no. 2022-T1/TIC-24176 of the Comunidad Aut\'onoma de Madrid, Spain; 
by the Swiss National Science Foundation under grant no. 214492; 
by the Science and Technology Facilities Council, United Kingdom; 
by the US Department of Energy under contracts No.~DE–AC02–76SF00515,  No.~89243024CSC000002,  
No.~DE-SC0010107, and
No.~DE-SC0010107,  
by the US National Science Foundation through the award NSF2310030, and by the Los Alamos National Laboratory LDRD programme. 

\section*{Editors}
We thank all colleagues who contributed to the actual editing of this document:
H.~Baer, T.~Barklow, T.~Behnke, S.~Belomestnykh, M.~Berger, J.~de~Blas, J.~Braathen, G.~Durieux, M.~Demarteau, A.~Faus-Golfe, B.~Foster, S.~Gessner, S.~Gori, S.~Heinemeyer, A.~Irles, M.~Ishino, D.~Jeans, W.~Kaabi, S.~Kanemura, W.~Kilian, P.~Koppenburg, S.~Kraml, K.~Kr\"uger, B.~List, J.~List, V.~Litvinenko, K.~Mekala, K.~Mimasu, A.~Miyazaki, G.~Moortgat-Pick, M.M.~M\"uhlleitner, N.~Nagata, E.~Nanni, M.~Nojiri, M.E.~Peskin, J.R.~Reuter, A.~Robson, R.~P\"oschl, H.~Sakai, Y.~Sakaki, T.~Sch\"orner-Sardenius, I.~Schulthess, A.~Schwarzman, S.~Stapnes, J.~Strube, T.~Suehara, G.~Taylor, K.~Tsumura, J.~Tian, C.~Vernieri, M.~Vos, G.~Weiglein, M.~Wenskat, G.W.~Wilson, A.~White, K.~Yagyu, K.~Yokoya, A.F.~\.Zarnecki, D.~Zerwas, J.~Zhang.


\section*{References}
\printbibliography[heading=none]

\end{document}